\documentclass[english,12pt]{article}
\usepackage{graphicx}
\usepackage{amsmath}
\usepackage{amssymb}
\usepackage{bm}
\usepackage{slashed}
\usepackage{dcolumn}
\usepackage{epsfig}
\usepackage{times}
\usepackage{cite}
\usepackage{afterpage}
\usepackage{color}
\usepackage[T1]{fontenc}
\usepackage[latin9]{inputenc}
\usepackage{babel}
\usepackage{lipsum}
\usepackage{ulem} 
\usepackage{url}
\usepackage[breaklinks]{hyperref}
\usepackage{hyperref}
\hypersetup{
    colorlinks,
    citecolor=black,
    filecolor=black,
    linkcolor=black,
    urlcolor=black
}

\allowdisplaybreaks[1]

\newcommand{\ord}[1]{{\cal{O}}\left( #1 \right)}
\newcommand{\Lagr}{\mathcal{L}}
\newcommand{\no}{\nonumber}
\newcommand{\Slash}[1]{\ooalign{\hfil/\hfil\crcr$#1$}}

\def\lambdabar{\lambda\kern-1ex\raise0.65ex\hbox{-}}
\def\bra#1{\left\langle #1\right|}
\def\ket#1{\left| #1\right\rangle}

\usepackage[margin=1in]{geometry}
\begin{document}
\hfill{\bf Date:} {\today}
\begin{center}
{\huge\bf Workshop on Physics with Neutral Kaon Beam at JLab \\
	(KL2016) \\
	Mini-Proceedings} 
\end{center}
\hspace{0.1in}
\begin{center}
{\large 1st - 3rd February, 2016}
{\large Thomas Jefferson National Accelerator Facility, Newport 
	News, VA, U.S.A.}
\end{center}
\begin{center}
M.~Albrow,
M.~Amaryan,
E.~Chudakov,
P.~Degtyarenko,
A.~Feijoo,
C.~Fernandez-Ramirez,
I.P.~Fernando,
A.~Filippi,
J.L.~Goity,
H.~Haberzettl,
B.C.~Jackson,
H.~Kamano,
C.~Keith,
M.~Kohl,
I.~Larin,
Wei-Hong~Liang,
V.K.~Magas,
M.~Mai,
D.M.~Manley,
V.~Mathieu,
F.~Myhrer,
K.~Nakayama,
H.~Noumi,
Y.~Oh,
H.~Ohnishi,
E.~Oset,
M.~Pennington,
A.~Ramos,
D.~Richards,
E.~Santopinto,
R.~Schumacher,
A.~Szczepaniak,
S.~Taylor,
B.~Wojtsekhowski,
Ju-Jun~Xie,
V.~Ziegler, and
B.~Zou
\end{center}
\begin{center}
\textbf{Editors}: M.~Amaryan, E.~Chudakov, C.~Meyer, M.~Pennington,
        J.~Ritman, and I.~Strakovsky
\end{center}
\noindent

\begin{center}{\large\bf Abstract}\end{center}

The KL2016 Workshop is following the Letter of Intent LoI12--15--001
\textit{Physics Opportunities with Secondary $K_L^0$ beam at JLab} 
submitted to PAC43 with the main focus on the physics of excited 
hyperons produced by the Kaon beam on unpolarized and polarized 
targets with GlueX setup in Hall~D. Such studies will broaden a 
physics program of hadron spectroscopy extending it to the strange 
sector. The Workshop was organized to get a feedback from the 
community to strengthen physics motivation of the LoI and prepare 
a full proposal.

Further details about the Workshop can be found on the web page of 
the conference:\\ 
http://www.jlab.org/conferences/kl2016/index.html . \\

We acknowledge the support of The George Washington University, 
U.S.A., Institute for Kernphysik \& J\"ulich Center for Hadron 
Physics, J\"ulich, Germany, Jefferson Science Associates, U.S.A., 
Old Dominion University, U.S.A., Thomas Jefferson National 
Accelerator Facility, U.S.A.


\newpage
\tableofcontents

\newpage
\pagenumbering{arabic}
\setcounter{page}{1}
\section{Preface and Summary}
\addtocontents{toc}{\hspace{2cm}{\sl M.~Pennington}\par}
\setcounter{figure}{0}
\halign{#\hfil&\quad#\hfil\cr
\large{Michael Pennington}\cr
\textit{Thomas Jefferson National Accelerator Facility}\cr
\textit{Newport News, VA 23606, U.S.A.}\cr}

\begin{enumerate}
\item \textbf{Primary Physics with Secondary Beams of $K_L$'s}

Not all of physics can be explored by primary beams of electrons 
and protons, however much these have taught us over the past 60 
years. Electron-positron colliders give access to restricted sets 
of quantum numbers. From proton-proton collisions we have long 
learned about all manner of high energy reactions. However, 
access to excited mesons and baryons has been most universally 
achieved with pion and kaon secondary beams, and latterly photon 
beams both virtual and real. Nowadays the wealth of information 
on polarized photon beams on polarized targets has totally 
revolutionized baryon spectroscopy, especially in the lightest 
flavor sector. The measurement of polarization asymmetries has 
constrained partial wave analyses far beyond anything conceivable 
with pion beams. Nevertheless, these only give access to the 
product of photocouplings of each excited baryon and its coupling 
to the hadron final state, such as $\pi N$, $\pi\pi N$, $\eta N$, 
\textit{etc.}  Contemporaneously, in the meson sector, the COMPASS 
experiment with pion beams on protons at 190~GeV/$c$, together 
with heavy flavor decays in $e^+e^-$ annihilation, have given 
access to multi-meson final states, like 2$\pi$, 3$\pi$, with 
greater precision than ever before. This has given hints and 
suggestions of new resonances, like the $a_1(1420)$.

To understand the constituent structure of hadrons requires 
information on the relationship of each meson and baryon to those 
with different flavors, but the same $J^P$ quantum numbers. At 
its simplest, this is to understand the quark model multiplet 
structure, where this is appropriate.  What is more, so 
ubiquitous are $\pi\pi$, $\pi K$, $K{\overline K}$, $\eta\pi$, 
$\eta K$, ... final states as the decay products of almost every  
hadron, that knowledge of the properties of these is critical to 
every analysis. Unitarity colors and shapes the universality
of such final state interactions in each set of quantum numbers.
This means that however precise our measurements of $\gamma N\to
\pi\eta^{(\prime)} N$, with GlueX for instance, we cannot really 
determine the fine resonant structure of such a process without 
some information on $\pi\eta$ and $\pi \eta^\prime$ scattering 
too. Nor can we determine any flavor partners without 
information on the corresponding $K \eta^{(\prime)}$ channels. 
For that secondary kaon beams are essential. 

While J-PARC has a whole program of charged strange particle and 
hypernuclear reactions, photon beams allow unique access to other 
channels. It was realized long ago that intense photon beams like 
that about to be delivered to Hall~D at JLab, could produce 
secondary beams. The charged particles can readily be bent away, 
leaving a neutral particle beam dominated by $K_L$: long lived 
kaons being produced far more copiously than neutrons above
3--4~GeV momentum. Such a facility provides access to a whole 
range of physics that is the subject of this meeting.

The reactions that can be studied cover the meson spectrum and 
dynamics, the baryon spectrum and dynamics, and final state 
interactions that link mesons and baryons together. This includes 
particularly the channels $K\pi$, $K\eta$, $K\eta^\prime$ in the 
meson sector.  These explore the very limitations of chiral 
dynamics: the strange quark's current mass is 30-50 times that of 
the average $u,d$ mass, so corrections from explicit breaking are 
likely significant. Moreover, are there eight or nine Goldstone 
bosons? In a world of large $N_c$, a nonet of light pseudoscalars
is natural, but is the $N_c=3$ world close to this or not? At the 
same time, information on $K M$ final states, with $M\,=\,\pi,\eta,
\eta^\prime$, is vital in understanding excited meson multiplets. 
This also provides links to $J/\psi\pi\pi$ dynamics and its 
relation to $\phi\pi\pi$ and $K^*{\overline K}\pi$ with hidden 
strangeness.  In the baryon sector, while earlier seemingly 
``missing" states are appearing with $N^\ast$'s and 
$\Delta^\ast$'s, few of the related $\Sigma^\ast$'s and $\Xi^\ast$'s 
are known.

The exact connection of the Constituent Quark Model to QCD is not 
really understood. Nevertheless calculations in Lattice QCD with 
heavier than physical pions give a baryon spectrum much like that 
of the Constituent Quark Model. This is perhaps not surprising 
since with only single hadron operators, the excited states are 
almost stable. Though dynamical quarks allow additional $q\overline 
q$ pairs to be produced, with a heavy pion (and hence heavy $u,d$ 
quarks) the Fock space of excited $N^\ast$'s are dominated by 
$qqq$ configurations. Computations including gluonic operators 
allow hybrid states of $qqqg$ that are a GeV or so heavier than 
$qqq$ configurations with the same $J^P$. The spectrum from 
Lattice QCD can be regarded as  the 21st century version of the 
Constituent Quark Model. However, it is naturally improvable, not 
just because by lowering the pion mass the phase space for decays 
increases. This  is not sufficient to reproduce experiment. More 
efficacious is the inclusion of multiparticle scattering states. 
Hadrons become resonances by coupling to these scattering states. 
For mesons like the $\rho$ and $K^\ast$, rather precise 
calculations have now been made, but for baryons like those in 
Fig.~\ref{fig-1}, which are much more computationally intensive, 
these are still to come. 
\begin{figure}[ht!]
\centering
\includegraphics[width=11.cm]{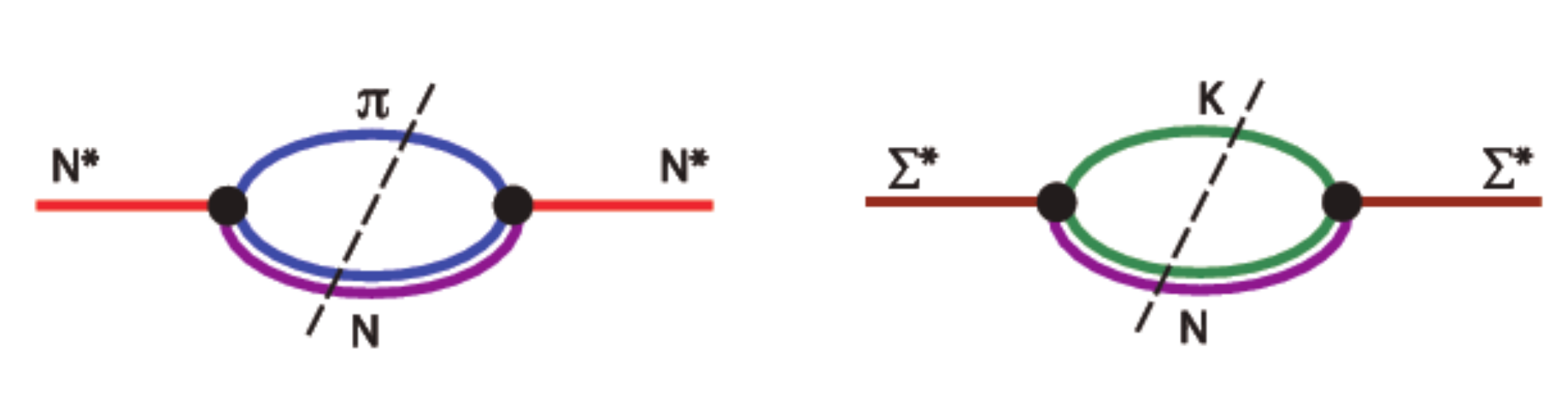}
\centerline{\parbox{0.80\textwidth}{
 \caption{The propagator of an unstable particle is affected 
	by the coupling to hadronic intermediate states. The 
	imaginary part of these contributions (signified by 
	the dashed line that places the intermediate state 
	particles on-shell) gives the resulting resonance a 
	width, as well as changing the real part of its mass 
	function.} \label{fig-1}}}
\end{figure}
\begin{figure}[ht!]
\centering
\includegraphics[width=16.cm]{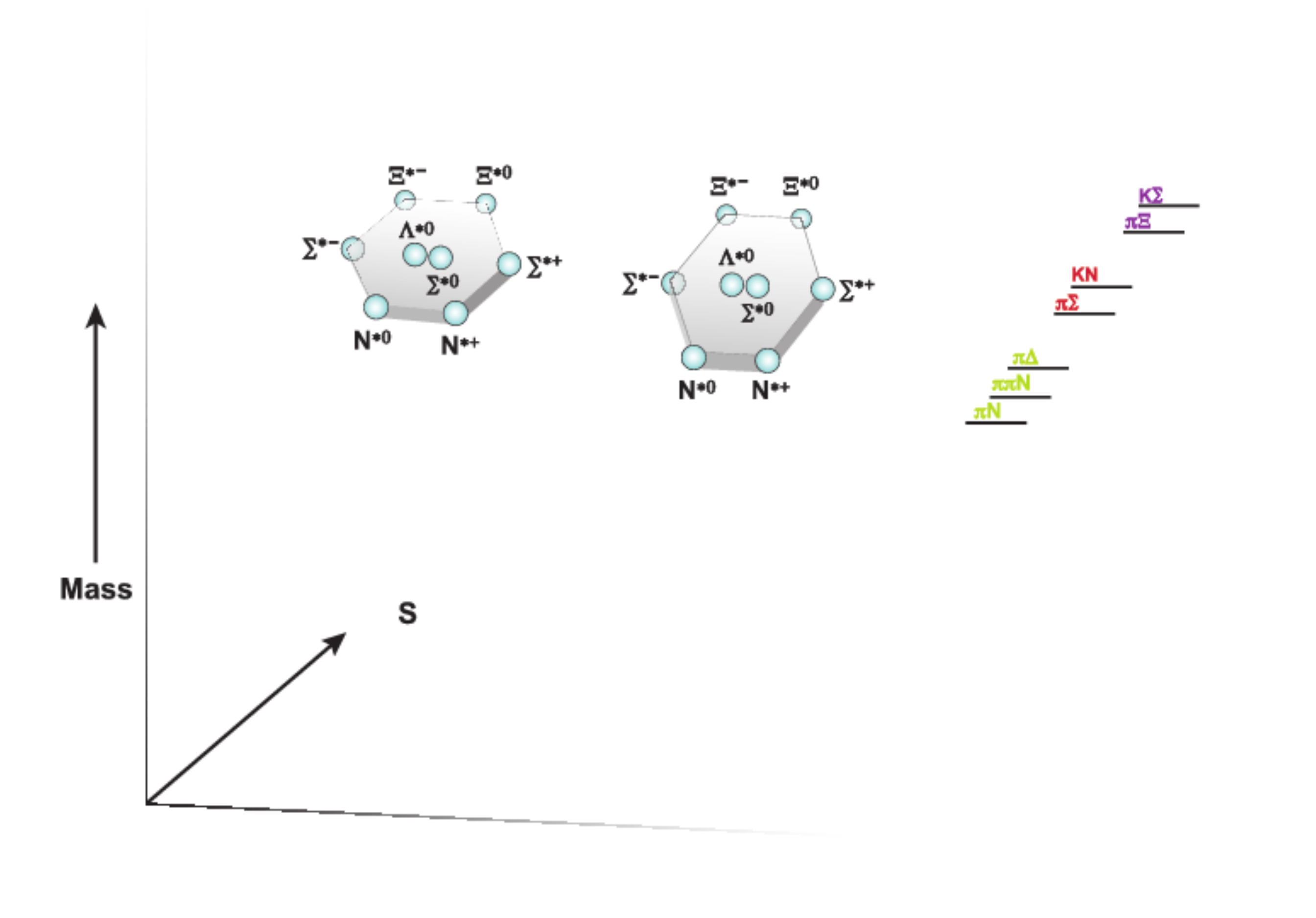}
\centerline{\parbox{0.80\textwidth}{
 \caption{An example of a flavor octet of excited baryons. The 
	multiplet on the left is that expected for stable states 
	in the Constituent Quark Model with the strange quark 
	having a 150~MeV heavier mass than the $u,d$ quarks. The 
	coupling to nearby hadronic channels, indicated on the 
	right, through graphs like that in Fig.~\protect\ref{fig-1}, 
	affects the mass of the unstable states. This might leave 
	the $\Xi^\ast$'s least disturbed, while significantly 
	shifting the masses of the less strange members of the 
	multiplet.} \label{fig-2}}}
\end{figure}

Scattering states do not just allow decays and give the states a 
width, but these in turn shift the masses of the resonances, 
particularly when the decays are $S$-wave. For their strange 
partners, with fewer open channels, the shifts may be less, as 
illustrated in Fig.~\ref{fig-2}. Thus one might expect the 
strange, doubly strange and even more the triply strange baryons 
to be closest to where our new Lattice version of the Constituent 
Quark Model may predict, while the $N^\ast$'s and  $\Delta^\ast$'s 
may be shifted significantly.

Indeed in some cases they may be indistinguishable from the 
continuum, in others coventional and hybrid baryons may have 
distinct decay patterns. So where the $\Sigma^\ast$'s, 
$\Lambda^\ast$'s and $\Xi^\ast$'s are that partner the Roper, 
$N^\ast(1440)$ with $J^P =(1/2)^+$, may teach us about how this 
dynamics works. It is likely much of this same dynamics with its 
interplay of  both colored (quark and gluon) and color singlet 
degrees of freedom underlies the appearance of the $X,Y,Z$, and 
$P_c$ states. Indeed such interplay is integral to generating 
their very existence. Strange baryon studies enabled by $K_L$ 
beams may provide unique insights into this newly exposed world.

Since the primary focus of the GlueX experiment is to study the 
photoproduction of mesons, it is unrealistic to expect its 
replacement by a secondary beam of $K_L$'s before a few years of 
photon data-taking. Indeed, there would be little point until the 
DIRC bars, and even a RICH for kaon identification, are in place. 
Then while the GlueX analysis is concentrating on gluonic 
excitations with millions of events on many channels, from which 
to extract physics, perhaps an opportunity arises to show that a 
secondary $K_L$ beam can provide unique data on the very strange 
baryons. A short program that successfully identifies $\Xi^\ast$'s 
and $\Omega^\ast$'s  not just as bumps in cross-sections, but with 
quantum numbers determined from their decays, may spur the demand 
for longer running and make the $K_L$ beam a facility for hadron 
physics unique in the world.

A long march starts with the rst stride: this is the step taken 
by this workshop. The second step is to plan for the minimal Be 
target, sweeping magnet and pair spectrometer required for a 
feasibility run. As always this requires a lead time of several 
years. Consequently whether to proceed, to even this limited 
stage within the Hall~D schedule, demands a timely decision.

\item \textbf{Acknowledgements} 

This material is based upon work supported by the U.S. 
Department of Energy, Office of Science, Office of Nuclear 
Physics under contract DE--AC05--06OR23177.
\end{enumerate}

\newpage
\section{Summaries of Talks}

\subsection{Photoproduction of $K^0$: Early History}
\addtocontents{toc}{\hspace{2cm}{\sl M.~Albrow}\par}
\setcounter{figure}{0}
\setcounter{footnote}{0}
\halign{#\hfil&\quad#\hfil\cr
\large{Michael Albrow}\cr
\textit{Fermi National Accelerator Laboratory}\cr
\textit{P.O.Box 500, Wilson St.}\cr
\textit{Batavia, IL 60510, U.S.A.}\cr}

\begin{abstract}
I discuss the first measurements of photoproduction of neutral 
Kaons in the 1960's, and early experiments on $K^0$ decays and 
interactions carried out with such beams.
\end{abstract}

\begin{enumerate}
\item \textbf{Motivation for $K^0$ Photoproduction}

In July 1964 Christenson, Cronin, Fitch and Turlay 
announced~\cite{cpviolW} the observation of the long-lived $K^0$ 
decaying to two pions, which was the discovery of CP-violation. 
It was already realised that $K_0 - \bar{K^0}$ physics was 
extremely interesting~\cite{pkkabirW}, with particles and their 
antiparticles being distinct under strong interactions, but able 
to mix through weak interactions to form short-lived and 
long-lived states, now called $K^0_S$ and $K^0_L$. That was the 
year I started my postgraduate studies at Manchester University. 
Prof.~Paul~Murphy led the group from 1965, and the nearby 
Daresbury Nuclear Physics Laboratory (DNPL) was constructing a 
5~GeV electron synchrotron, NINA (Northern Institutes National 
Accelerator). This would be a local laboratory for the northern 
universities to balance the Rutherford Laboratory's NIMROD 7~GeV 
proton accelerator in southern England. The possibility that a 
useful $K^0$ beam could be made at an electron synchrotron by 
photoproduction was being considered, and a 1965 prediction for 
SLAC by Drell and Jacob~\cite{kodrelljacobW} was optimistic. They 
expected the dominant mechanism to be virtual $K^\ast$-exchange 
in the $t$-channel, $\gamma + K^\ast\rightarrow K$, with a cross 
section, peaking at $\theta\sim 2^\circ$, at least 20~$\mu$b/sr, 
for 15~BeV (now GeV !) photons. We now know that $K^\ast$-exchange
is not the dominant process. 

In 1965, the Manchester Group decided to measure $K^0$ 
photoproduction at NINA with a view to using the beam for decay 
measurements.

\item \textbf{First Observations}

While our Manchester experiment was being carried out the first 
``observation" of photoproduced $K^0$ was published~\cite{kocbcW} 
by the Cambridge Bubble Chamber Group; after scanning 865,000 (sic !) 
hydrogen bubble chamber photographs they found about 50 examples of 
$K^0 \rightarrow \pi^+\pi^-$. Of these, 35 were classified as 
$\Lambda^0K^0\pi^+$ events, and a few were possibly photoproduction 
of $\phi\rightarrow K^0_SK^0_L$, but the evidence for that was 
``weak". This was followed by an experiment~\cite{koceaW} in 
1967, using a multi-gap optical spark chamber, still taking 
photographs but now with the ability to trigger on a neutral 
particle entering the fiducial volume (with a veto counter) and a 
pair of charged particles in a scintillation counter hodoscope 
behind. As in our Manchester experiment, a $K^0_S$-regenerator was 
placed in front of the decay volume. There was no magnetic field, 
but coplanar V's were selected, and assuming $K^0 \rightarrow 
\pi^+\pi^-$ the Kaon momentum is known from kinematics. The 
Cambridge Electron Accelerator produced 5.5~GeV electrons, and the 
$K^0$ were produced in Al- and Be-targets at polar angles from 
0$^\circ$ - 10$^\circ$. Since an intermediate photon beam was not 
used, the flux is a combination of electro- and photo-production 
in the targets. With about 400 events, they reported a yield of 
1.3 $\times 10^{-5}~K^0$/electron for $1.0 < p_K < 5.5$~GeV/$c$ 
at 3.5$^\circ$. The spectrum peaks at 2~GeV/$c$, which they 
interpreted as due to $\phi$-photoproduction and decay, 
incorrectly as it turns out. Absence of $K^0$ with momenta closer 
to the beam momentum gave an upper limit on the two-body processes: 
$\gamma + p\rightarrow K^0+ \Sigma^+, \gamma + n\rightarrow K^0 + 
\Sigma^0,$ and $ \gamma + n\rightarrow K^0 +\Lambda^0$. 
Schivell~\textit{et al.} said it was~\cite{koceaW}: \textit{an 
extremely clean beam which appears quite free of neutrons}, 
although they did not elucidate. Of course for most ``decay" 
experiments neutron background in unimportant, but for scattering 
and interaction experiments it can be serious. 

A major hydrogen bubble chamber study~\cite{koabbhhmW} (1.7 million 
pictures !) using a bremsstrahlung beam up to $E_\gamma$ = 5.8~GeV 
at DESY reported a few $\gamma + p\rightarrow K^0 + \Sigma^+$ 
events with $\sigma$ = 0.68$\pm$0.48~$\mu$b, with a considerably 
larger cross section for $\gamma + p \rightarrow \Lambda^0 + K^0 
+ \pi^+ (+\pi^0 ...)$. So $K^0$ photoproduction at these energies 
is usually accompanied by pions.

\item \textbf{The Manchester Experiment at NINA (Daresbury, UK)}

A key innovation of the Manchester experiment was to use 
``automatic" (electronic) spark chambers, to progress beyond 
the prevalent scanning and measuring of bubble chamber or optical 
spark chamber pictures. We built small prototypes of three types: 
(a) sonic, with three microphones around the edge of the spark gap, 
and timing the sound of the sparks, 
(b) magnetostrictive, with one electrode consisting of wires 
crossing a magnetostrictive ribbon, and 
(c) ferrite core memory, in which each wire of an electrode plane 
was threaded through a $\sim$1~mm diameter ferrite core. 
The spark current-pulse flipped the magnetisation of the core, 
inducing a pulse on a ``read" wire connected to the data 
acquisition, before being reset by a pulse on a third wire. (So our 
core memory board had about 1~mm$^3$/bit; compare with memory 
density today, 50 years later !) The ferrite core technique was 
chosen for the experiment, and we made 12 planes, each 70 $\times$ 
121~cm. I believe that was the largest system of electronic spark 
chambers in operation in 1968. 
\footnote{With the lights off, we could see through the wires 
	the pions making tracks of sparks; that was exciting !} 
But that was the very year that Charpak invented the multi-wire 
proportional chamber that superceded spark chambers, with a steady 
high voltage, no sparks, more gentle discharges, and high rate 
capability. 
 
The method of measuring the $K^0$ flux after a 40~m flight path, 
where there are only $K^0_L$ left ($c\tau(K^0_L)$ = 15.34~m), was 
to insert a $K^0_S$ regenerator made of 14~cm of iron. Because $K^0$ 
and $\bar{K^0}$ have different strong interactions and are 
differently attenuated in the iron, a $K^0_S$ component appears, and 
since $c\tau (K^0_S)$ = 2.684~cm they conveniently decay in a short 
fiducial region, 69.2\% of the time to $\pi^+\pi^-$. With two tracks 
to be detected from the $K^0_S$ decays we needed to resolve the 
$x,y$-ambiguity (if only two coordinates had been measured), so the 
chambers immediately following the decay region were in $u,v,x$ 
triplets, with wires at 60$^\circ$. The sides ratio 70/121 = 
tan(30$^\circ$) was conveniently chosen so the inclined $u$- and 
$v$-wire planes were parallel to the diagonals, and the wires 
emerged at the top and bottom with exactly twice the spacing of the 
$x$-wires. The sum of the coordinates $u+v+x$ is a constant for a 
real track, simplifying analysis on our PDP8 computer (programmed 
with punched paper tape !).
 
Tha main improvements over the previous experiments~\cite{kocbcW,koceaW} 
were the first use of ``electronic" wire chambers (no more 
scanning and measuring photographs !), using a bremsstrahlung photon 
beam (from a 0.1 $X_0$ tungsten target), using a 60~cm hydrogen 
target~\cite{koalbrow1W} as well as nuclear targets (Be, Al, 
Cu)~\cite{koalbrow2W}, and using a magnet to measure the momentum of 
at least one pion. The acceptances and efficiences were calculated 
with Monte Carlo simulations. After fitting to $K^0_S \rightarrow 
\pi^+\pi^-$, the reconstructed $K^0_S$ lifetime showed that the 
sample had no background. The yields showed a strong increase with 
$E_\gamma$, see Fig.~\ref{fig:kobeam}(a), rising from threshold to 
$d\sigma/d\Omega = 15\pm3 \: \mu b/sr/e.q.$ 
\footnote{e.q., means ``equivalent quantum", a measure of the photon 
	flux using a quantameter.} 
on hydrogen for $p_K >$ 1.5~GeV/$c$ at $\theta$ = 3$^\circ$. The 
momentum spectra of the $K^0$ peak at low values, $\sim$ 1.5~GeV/$c$; 
two-body reactions (see above) do not dominate. The yield is 
significantly less than that predicted by Drell and Jacob, implying 
a $K^\ast K\gamma$ coupling much smaller than they expected. 
Photoproduced $\phi\rightarrow K\bar{K}$ can only account for a small 
fraction of the data; the total cross section for $\gamma\rightarrow 
\phi$ being small ($\sim 0.4 \: \mu b$).
\begin{figure}[ht!]
\centering
\includegraphics[width=0.8\textwidth]{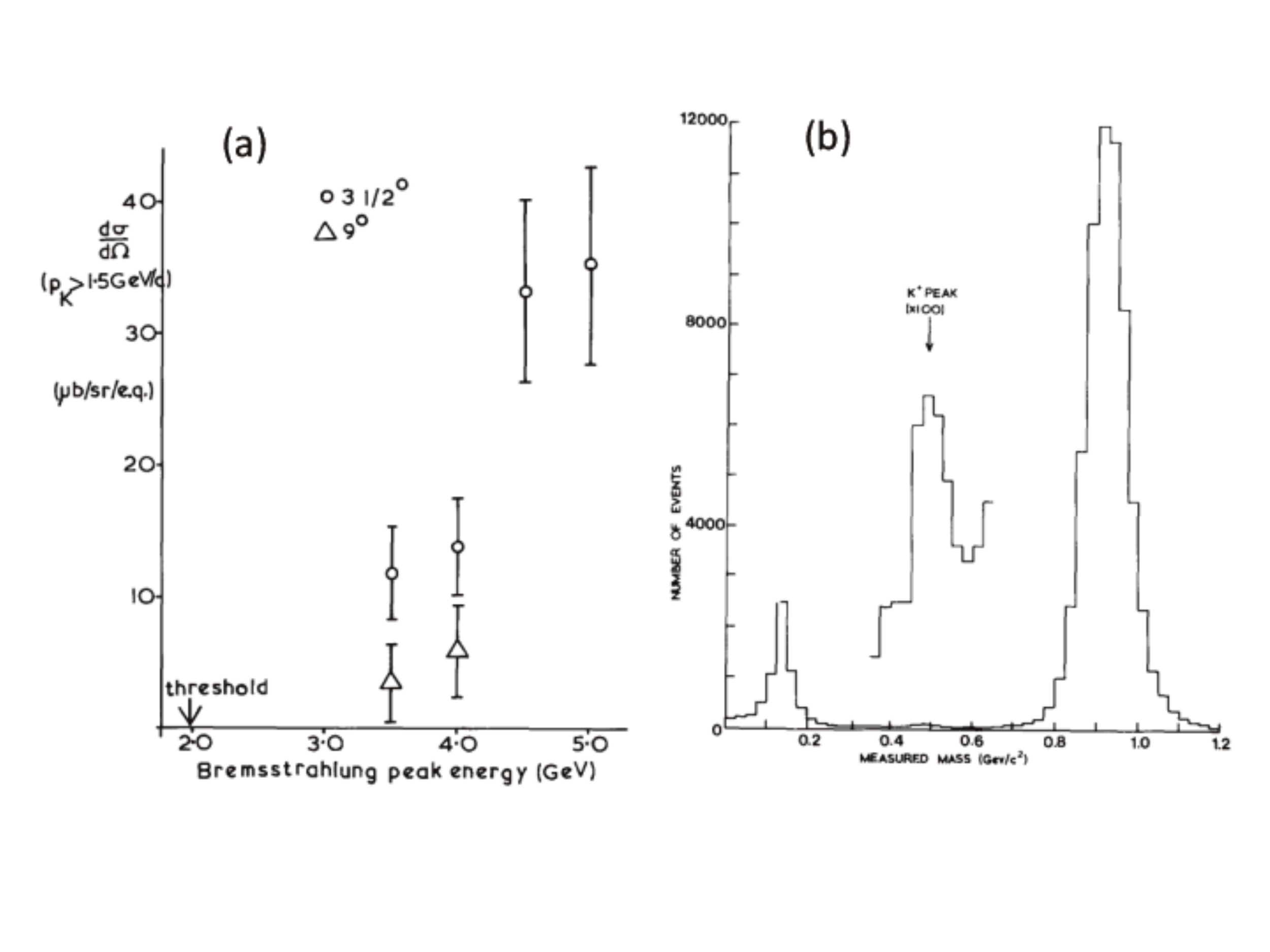}
\centerline{\parbox{0.80\textwidth}{
 \caption{(a) $K^0$ yields~\protect\cite{koalbrow2W} from hydrogen,
        above 1.5~GeV/$c$, as a function of the peak
        bremsstrahlung energy; (b) Mass spectrum of charged
        particles~\protect\cite{kocexW} from time-of-flight 
	of the beam particle ($K^0$ or $n$) plus final 
	charged particle ($K^+$ or $p$), knowing the momentum 
	of the latter. Note the $\pi^+$ background.} 
	\label{fig:kobeam}}}
\end{figure}

The data with nuclear targets (0.45 $X_0$ of Be, Al, and Cu) required 
corrections for the absorption of the photon beam in the target and of 
the $K^0$ leaving it. The photoproduction cross section ``per 
nucleon" shows a very small $A$-dependence : $d\sigma/d\Omega \propto
A^{1.09 \pm 0.03}$ (this is after subtracting the few \% contribution 
from coherent $\phi$-photoproduction). The residual small rise could 
result from another coherent process, or from a difference in the 
production off neutrons and protons. Quoting Ref.~\cite{koalbrow2W} 
\textit{It seems probable that most of the $K^0_L$ yield is produced 
in association with pions in multiparticle production events. This 
would account for the large yield of low momentum $K^0_L$ mesons.} 

\item \textbf{$K_L^0$ Decays}

Having established a photoproduced $K^0$ beam, the $K^0_S$-regenerator 
was removed to measure $K^0_L$ 3-body decays. (The CP-violating $K^0_L 
\rightarrow\pi^+\pi^-$ decay has a probability $\sim 2 \times 10^{-3}$ 
and was not on our menu.) At the time these studies were ``state 
of the art". An excellent review of the contemporary theory is given 
in Ref.~\cite{gaillardW}.

\begin{enumerate}
\item \textbf{$K^0_L \rightarrow \pi^+ \pi^- \pi^0$}

Our first decay measurement~\cite{ko3piW} was the $\pi^+\pi^-\pi^0$ 
mode $K_{3\pi}$, which has a branching fraction $BF$ = 12.5\%~\cite{pdgW}. 
The $\pi^0$ was not detected, and the $K_{3\pi}$ mode was distinguished 
from the $\pi^\pm \mu^\mp \nu$ mode using kinematics. In a particular 
reference frame (the Astier~\cite{astierW} frame) the kinetic energy of 
the $K^0_L$ is positive for the 3$\pi$ mode, but it is mostly negative 
for the semileptonic modes~\cite{ko3piW}. From 660,000 triggers 70,000 
two-track decays in the fiducial region were selected, of which 29,000 
were classified as $K^0_L\rightarrow\pi^+\pi^-\pi^0$, with 17\% 
semileptonic background. The main thrust of the analysis was to study 
the decay matrix element using the distribution of decays over the 
triangular ``Dalitz plot". (The energies of the three pions in 
their c.m. frame have to add up to $M(K)$, and the normals to the sides 
of an equilateral triangle have to add up to a constant. So each event 
can be plotted as a point on 60$^\circ$-triangular graph paper.)

Reconstructing the $K^0_L$ momentum from the two pions involves a 
quadratic ambiguity, which has to be considered in the analysis. The 
results can be presented in the form of the $\pi^0$ kinetic energy, 
$T_0$, spectrum, for which Weinberg had proposed~\cite{weinbergW} a 
general form with linear, quadratic, cubic, \textit{etc.} terms. 
Including a cubic term improved the, basically linear, fit.

\item \textbf{$K^0_L \rightarrow \pi^\pm \ell^\mp \nu$}

Studies were made of both the muon decay~\cite{kopimunuW} $K_{\mu 3}$ 
(BF = 27\%) and the electron decay~\cite{kopienuW} $K_{e3}$ (BF = 
40.6\%). This was a decade before the discovery of the real $W$, but 
the term ``strangeness-changing hadronic vector current"
described in V -- A theory was the current language.  The $K^0$ emits 
a highly off-shell (!) $W$ with 4-momentum-squared $q^2 = (p_K - 
p_\pi)^2$ (4-vectors), that couples to $e\nu$ or $\mu\nu$. The aim is 
to investigate the structure of this weak current; one writes down 
``form factors" $f_+(q^2)$ and $f_-(q^2)$ that can be 
determined by fitting the Dalitz plot of the decays. Scalar and/or 
tensor exchanges could show up in these distributions, so it was a 
test of the V -- A theory.
 
For the $K_{\mu 3}$ measurements an iron absorber and a scintillation 
counter hodoscope were added behind the spark chambers. The 
distribution of 9,066 events over the Dalitz plot was measured, after 
correcting for acceptance and efficiencies using 10$^5$ Monte 
Carlo-generated events. I will not discuss the fits using form factor 
parametrisations, except that the $f_+$ form factor, which should 
depend only on $q^2$, shows a clear linear increase from $q^2/m(\pi)^2$ 
= 1.5 to 5.0. In contrast the $f_-$ form factor is negative and did not 
show significant $q^2$-dependence. Limits were put on scalar and tensor 
couplings.
 
The $K_{e3}$ decays were distinguished by kinematics and the absence 
of a muon penetrating the Pb + Fe wall; this still left 13\% $K_{\mu3}$ 
background (there was no electromagnetic calorimeter). The $f_+(q^2)$ 
distribution is similar to that for $K_{\mu 3}$, but $f_-$ is
suppressed by the kinematic factor $(m_e/m_K)^2$. Again the data did 
not show any evidence for scalar or tensor couplings, but the 
sensitivity was not very high: $f_S < 0.19 \: f_+(0)$ and $f_T < 
1.0 \: f_+(0)$ separately, assuming no destructive inteference between 
them.
\end{enumerate}

\item \textbf{$K^0_L$ Interactions: $K^0 + p\rightarrow K^+ + n$}

Although photoproduced $K^0$ beams are not neutron-free, at least 
in the conditions of our DNPL experiment, the $n:K^0$ ratio is much 
less than in proton-produced beams, and the Manchester-DNPL group 
took advantage of that~\cite{kocexW} to measure $K^0 + p \rightarrow 
K^+ + n$ from 0.6 to 1.5~GeV/$c$. The quark-model was still relatively 
new, and while the known baryons could all be accommodated as 
\{$qqq$\}, a baryon with positive strangeness, then called $Z^\ast$, 
would have to be \{$qqqq\bar{q}$\}, a ``pentaquark", although 
that name came much later ($\sim$1987). A $K^+p$ state or resonance, 
having B = +1 and S = +1 would have to be \{$uuud\bar{s}$\} and a 
$K^+n$ state \{$uudd\bar{s}$\}, these were called ``exotics". 
(I had left the Manchester group by this time, but had looked for a 
$Z^\ast$ in $K^+p$ elastic scattering (pure I = 1) with a polarized 
target~\cite{albrowkpW} at CERN.)  The reactions $K^+ + n\rightarrow 
K^+ + n$ and $K^+ + n\rightarrow K^0 + p$ required deuterium targets;  
they have isospin amplitudes $\frac{1}{2}(f_1 + f_0)$ and $\frac{1}{2} 
(f_1 - f_0)$, respectively. The inverse reaction $K^0 + p \rightarrow 
K^+ + n$ avoids the neutron target complication, but has a neutron in 
the final state. This was detected in a large scintillator block, but 
there would be a large background from the reaction $n + p \rightarrow 
p + n$. One must distinguish $K^0\rightarrow K^+$ from $n \rightarrow 
p$, knowing the momentum of the outgoing charged track but not its 
identity. The trick used was to measure the time-of-flight over the 
21~m of the beam particle ``plus" the 5m of the charged particle, 
using the RF of the synchrotron (0.5~ns bunch  every 4.908 ns) picked 
up in  a cavity on the circulating beam. See Fig.~\ref{fig:kobeam}(b). 
For our experiment, the bunch spacing in NINA was doubled, since 5~ns 
caused ambiguities. The dominant (by a factor $\sim$200) $n + p 
\rightarrow p + n$ events could be used to calibrate the neutron 
counter timing. The $K^0 + p\rightarrow K^+ + n$ cross sections 
$d\sigma/d\Omega(p, cos \: \theta)$ are presented, fit to Legendre 
polynomials and partial and total cross sections derived. For 
details, see Ref.~\cite{kocexW}, but the last sentence is: 
\textit{The evidence for a $Z^\ast_0$ state must therefore be 
considered slender}. Nevertheless, some 40 years later (now), an 
experiment could certainly be done with much higher statistics and 
resolution and less background. Pentaquarks are now in fashion. An 
exercise for this Workshop.
 
One last remark or suggestion. The $K^0_L$ beam is $(K^0 - 
\bar{K^0})/\sqrt{2}$, while a $K^0_S$ is $(K^0 + \bar{K^0})/\sqrt{2}$. 
Normally $K^0$ strong interactions are studied in a (pure) $K^0_L$ 
beam. If the interaction target can be placed very close to the $K^0$ 
production target, before the $K^0_S$ have decayed, the strong 
interactions will have a different mixture of $K^0$ and $\bar{K^0}$. 
By subtraction one can in principle study the I = 0 and I = 1 
amplitudes separately. Alternatively one can study the interactions 
close behind a regenerator when the beam is a mixture of $K^0_S$ and 
$K^0_L$, and by varying the distance between the regenerator and the 
target one can vary the mix in a known fashion.

\item \textbf{Acknowledgments}

I thank Paul Murphy (my professor 1965-1969) and Fred Loebinger 
for ``the good old days". And I thank the organizers, 
especially Igor Strakovsky, and JLab for the opportunity to 
(reminisce and) participate in this Workshop. This work is supported 
by the US DOE.
\end{enumerate}


\newpage
\subsection{Overview of Hall~D Complex}
\addtocontents{toc}{\hspace{2cm}{\sl E.~Chudakov}\par}
\halign{#\hfil&\quad#\hfil\cr
\large{Eugene Chudakov}\cr
\textit{Thomas Jefferson National Accelerator Facility}\cr
\textit{Newport News, VA 23606, U.S.A.}\cr}

\begin{enumerate}
\item 
Hall~D is a new experimental hall at Jefferson Lab, designed for 
experiments with a photon beam. The primary motivation for Hall~D 
is the GlueX experiment~\cite{a1EE,a2EE}, dedicated to meson
spectroscopy. The Hall~D complex consists of:

\begin{itemize}
\item An electron beam line used to extract the 5.5-pass electrons 
	from the accelerator into the Tagger Hall. The designed beam 
	energy is E$_e$ = 12~GeV.
\item The Tagger Hall, where the electron beam passes through a thin 
	radiator ($\sim$0.01\%~R.L.) and is deflected into the beam 
	dump. The electrons that lost $>$30\% of their energy in the 
	radiator are detected with scintillator hodoscopes providing 
	a $\sim$0.1\% energy resolution for the tagged photons. 
	Aligned diamond radiators allow to produce linearly polarized 
	photons via the Coherent Bremsstrahlung. The beam dump is 
	limited to 60~kW (5~$\mu$A at 12~GeV).
\item The Collimator Cave contains a collimator for the photon beam 
	and dipole magnets downstream in order to remove charged 
	particles. The 3.4~mm diameter collimator, located about 75~m 
	downstream of the radiator, selects the central cone of the 
	photon beam increasing its average linear polarization, up to
	$\sim$40\%in the coherent peak at 9~GeV.
\item Hall~D contains several elements of the photon beam line, and 
	themain spectrometer. A Pair Spectrometer consists of a thin 
	converter, a dipole magnet, and a two-arm detector used to 
	measure the energy spectrum of the photon beam. The main 
	spectrometer is based on a 2-T superconducting solenoid, 4~m 
	long and 1.85~m bore diameter. The liquid hydrogen target is 
	located in the front part the solenoid. The charged tracks 
	are detected with a set of drift chambers; photons are 
	detected with two electromagnetic calorimeters. There are 
	also scintillator hodoscopes for triggering and time-of-flight
	measurements. The spectrometer is nearly hermetic in an 
	angular range of $1^\circ < \theta < 120^\circ$. The momentum 
	resolution is $\sigma_p/p\sim 1 -- 3\%$ depending on the 
	polar angle $\theta$. The energy resolution of the 
	electromagnetic calorimeters is about 7\% at 1~GeV.
\end{itemize}

The main spectrometer is designed for photon beam rates below 100~MHz 
in the coherent peak. Such a rate can be provided by a 2.2~$\mu$A beam 
on a 0.02~mm = 0.0125\%~R.L. diamond crystal. The 1-st stage of GlueX 
is planned to run at a lower rate of 10~MHz.

Hall D and the GlueX experiment had 3 commissioning runs in 2014--2016. 
By April 2016, all the systems have been commissioned at some level and 
most of them have reached the specifications. Preliminary results of 
the 2014--2015 commissioning have been reported~\cite{a3EE}. In addition 
to the GlueX experiment, two other experiments (both using Primakoff-type
reactions) have been approved by the Program Advisory Committee (PAC). 
In total, about 500~days of effective running have been approved by the 
PAC.

\newpage
\item \textbf{Acknowledgments}

This material is based upon work supported by the U.S. Department of
Energy, Office of Science, Office of Nuclear Physics under contract 
DE--AC05--06OR23177.
\end{enumerate}


\newpage
\subsection{The $K_L^0$ Beam Facility at JLab}
\addtocontents{toc}{\hspace{2cm}{\sl M.~Amaryan}\par}
\setcounter{figure}{0}
\setcounter{equation}{0}
\halign{#\hfil&\quad#\hfil\cr
\large{Moskov Amaryan}\cr
\textit{Department of Physics}\cr
\textit{Old Dominion University}\cr
\textit{Norfolk, VA 23529, U.S.A.}\cr}

\begin{abstract}
Following a Letter of Intent submitted to PAC43 at JLab 
in this talk we discuss the possibility to create a 
secondary $K_L^0$ beam in Hall~D to be used with GlueX 
detector for spectroscopy of excited hyperons.
\end{abstract}

\begin{enumerate}
\item \textbf{Introduction}

Our current understanding of  strong interactions is 
embedded in Quantum Chromodynamics (QCD). However, QCD 
being a basic theory, extremely successful in explaining 
the plethora of experimental data in the perturbative 
regime, faces significant challenges to describe the 
properties of hadrons in non-perturbative regime.  
Constituent Quark Model (CQM) is surprisingly successful 
in explaining spectra of  hadrons,  especially in the 
ground state; however, CQM appears to be too naive to 
describe properties of excited states. It is natural 
that excited states are not simply explained with spatial 
excitations of constituent quarks, but it is an effective 
representation revealing complicated interactions of 
quarks and gluons inside. Hadron spectroscopy aims to 
provide a comprehensive description of hadron structure 
based on quark and gluon degrees of freedom. Despite 
many successes in observing hundreds of meson and baryon 
states experimentally we haven't succeeded to either 
observe or rule out existence of glueballs, hybrids and 
multi quark systems; although it is tempting to explain 
recently observed  X, Y, Z~\cite{bib1} states as first 
evidences of tetraquarks as well as recently observed 
heavy baryon states at LHCb~\cite{bib2} as charmed 
pentaquarks. 

An extensive experimental program is developed to search 
for hybrids in the GlueX experiment at JLab. Over the 
last decade, significant progress in our understanding of 
baryons made of light $(u,d)$ quarks have been made in 
CLAS at JLab. However, systematic studies of excited 
hyperons are very much lacking with only decades old very 
scarce data filling the world database in many channels. 
In this experiment we propose to fill this gap and study 
spectra of excited hyperons using the modern CEBAF 
facility with the aim to use proposed secondary $K^0_L$ 
beam with physics target of the GlueX experiment in 
Hall~D. The goal is to study $K_L-p$ and $K_L-d$ 
interactions and do the baryon spectroscopy for the 
strange baryon sector.

Unlike in the cases with pion or photon beams, Kaon 
beams are crucial to provide the data needed to identify 
and characterize the properties of hyperon resonances. 

Our current experimental knowledge of strange resonances 
is far worse than our knowledge of $N$ and $\Delta$ 
resonances; however, within the quark model, they are 
no less fundamental. Clearly there is a need to learn 
about baryon resonances in the ``strange sector" to 
have a complete understanding of three-quark bound states.

The masses and widths of the lowest mass baryons were 
determined with Kaon-beam experiments in the 
1970s~\cite{bib1}. First determination of pole positions, 
for instance for $\Lambda(1520)$, were obtained only 
recently from analysis of Hall~A measurement at 
JLab~\cite{bib3}. An intense Kaon beam would open a 
window of opportunity not only to locate missing 
resonances, but also to establish properties including 
decay channels systematically for higher excited states.

A comprehensive review of physics opportunities with 
meson beams is presented in a recent paper~\cite{bib4}. 
Importance of baryon spectroscopy in strangeness sector 
was discussed in Ref.~\cite{bib5}.

\item \textbf{Reactions that Could be Studied with 
	$K_L^0$ Beam}

More details about this chapter could be found in a talk 
by Mark Manley at this workshop.

\begin{enumerate}
\item \textbf{Elastic and charge-exchange reactions}

\begin{eqnarray}
	K_L^0p\to K_S^0p\\
	K_L^0p\to K^+n
\end{eqnarray}

\item \textbf{Two-body reactions producing $S=-1$ hyperons}

\begin{eqnarray}
	K_L^0p\to \pi^+\Lambda\\
	K_L^0p\to \pi^+\Sigma^0
\end{eqnarray}

\item \textbf{Three-body reactions producing $S=-1$ hyperons}

\begin{eqnarray}
 	K_L^0p\to \pi^+\pi^0\Lambda \\
 	K_L^0p\to \pi^+\pi^0\Sigma^0 \\
 	K_L^0p\to \pi^0\pi^0\Sigma^+  \\
 	K_L^0p\to \pi^+\pi^- \Sigma^+  \\
 	K_L^0p\to \pi^+\pi^- \Sigma^- 
\end{eqnarray}

\item \textbf{Two- and three-body reactions producing $S=-2$ 
	hyperons} 

\begin{eqnarray}
	K_L^0p\to K^+\Xi^0 \\
	K_L^0p\to \pi^+K^+\Xi^-\\
	K_L^0p\to K^+\Xi^{0^\ast} \\
	K_L^0p\to \pi^+K^+\Xi^{-\ast}
\end{eqnarray}

\item \textbf{Three-body reactions producing $S=-3$ hyperons}

\begin{eqnarray}
	K_L^0p\to K^+K^+\Omega^-\\
	K_L^0p\to K^+K^+\Omega^{-\ast}
\end{eqnarray}

Reactions 10--15 will be discussed in more details below.
\end{enumerate}

\item \textbf{The $K_L^0$ Beam in Hall~D}

\begin{figure}[htb!]
\centerline{\includegraphics[width=450pt]{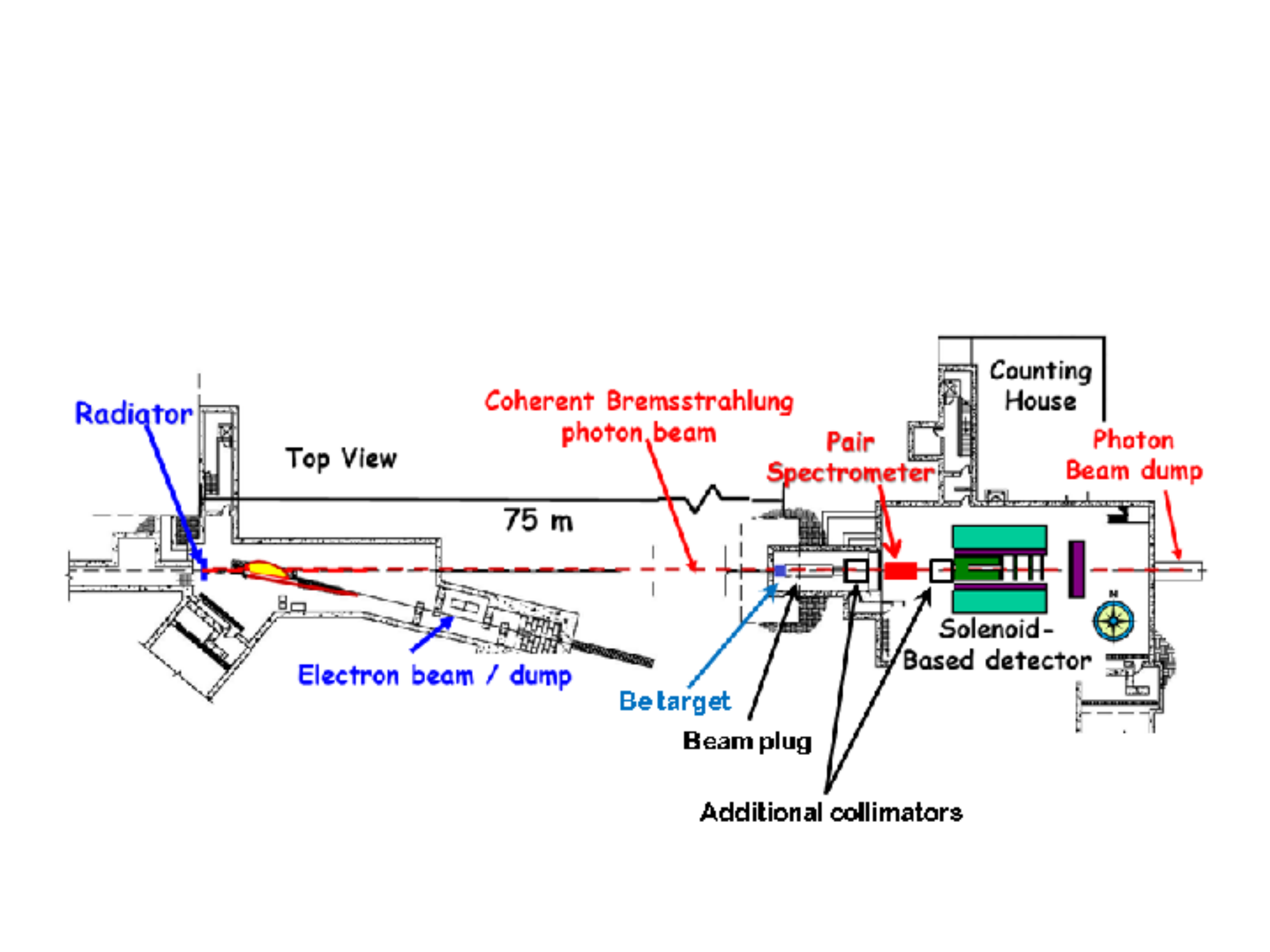}}
\vspace{-1.0cm}
\centerline{\parbox{0.80\textwidth}{
 \caption{Schematic view of Hall~D beamline. See a
        text for explanation. \label{fig:setup} } } }
\end{figure}
In this Section, we describe photo-production of secondary 
$K_L^0$ beam in Hall~D. There are few points that need to 
be decided. To produce intensive photon beam one needs to 
increase radiation length of the radiator up to 10$\%$ 
radiation length.  In a first scenario $E_e=12$~GeV, 
electrons produced at CEBAF will scatter in a radiator in 
the tagger vault, generating intensive beam of 
bremsstrahlung photons. This may will then require removal 
of all tagger counters and electronics and very careful 
design of radiation shielding, which is very hard to 
optimize and design. In a second scenario one may use 
Compact Photon Source design (for more details see a talk 
by Pavel Degtiarenko at this workshop) installed after the 
tagger magnet, which will produce bremsstrahlung photons 
and dump electron beam inside the source shielding the 
radiation inside. At the second stage, bremsstrahlung 
photons interact with Be target placed on a distance 
16~m upstream of liquid hydrogen ($LH_2$) target of GlueX 
experiment in Hall~D producing $K_L^0$ beam. To stop 
photons a 30 radiation length  lead absorber will be 
installed in the beamline followed by a sweeping magnet 
to deflect the flow of charged particles. The flux of 
$K_L$ on LH$_2$ target of GlueX experiment in Hall~D 
will be measured with pair spectrometer upstream the 
target. Details of this part of the beamline (for a 
details see a talk by Ilya Larin at this workshop). 
Momenta of $K_L$ particles will be measured using the 
time-of-flight between RF signal of CEBAF and start 
counters surrounding  $LH_2$ target. Schematic view of 
beamline is presented in Fig.~\ref{fig:setup}. The 
bremsstrahlung photons, created by electrons at a 
distance about 75~m upstream, hit the Be target and 
produce $K_L^0$ mesons along with neutrons and charged 
particles. The lead absorber of $\sim$30 radiation 
length is installed to absorb photons exiting Be target. 
The sweeping magnet deflects any remaining charged 
particles (leptons or hadrons) remaining after the 
absorber. The pair spectrometer will monitor the flux  
of $K_L^0$ through the decay rate of Kaons at given 
distance about 10~m from Be target. The beam flux could 
also be monitored by installing nuclear foil in front 
of pair spectrometer to measure a rate of $K^0_S$ due 
to regeneration process $K_L +p \to K_S +p$ as it was 
done at NINA (for details see a talk by Michael Albrow at 
this workshop).  

Here, we outline experimental conditions and simulated 
flux of $K_L^0$ based on GEANT4 and known cross sections 
of underlying subprocesses~\cite{bib6,bib7,bib8}.
\begin{itemize}
\item An electron beam with energy $E_e$ = 12~GeV and 
	current $I_e = 5~\mu A$ (maximum possible, limited 
	by the Hall~D beam dump).
\item A thickness of radiator 5~$\%$ radiation length.
\item Primary Be target with $R = 4$~cm, $L = 40$~cm.
\item $LH_2$ target with $R = 2$~cm, $L = 30$~cm.
\item Distance between Be and $LH_2$ targets $24$~m.
\end{itemize}

The expected flux of $K_L^0$ mesons integrated in the range 
of momenta P = 0.3 -- 10~GeV/$c$  will be $\approx 2\times 
10^3$~$K_L^0$/sec on the physics target of the GlueX setup.\\

In a more aggressive scenario with 
\begin{itemize}
\item A thickness of radiator 10$\%$.
\item Be target with a length L = 60~cm.
\item $LH_2$ target with R = 3~cm.
\end{itemize}

The expected flux of $K_L^0$ mesons integrated over the same 
momentum range  will increase to $\approx 10^4~K_L^0$/sec.

In addition to these requirements it will require lower 
repetition rate of electron beam with $\sim 40$~ns spacing 
between bunches to have enough time to measure time-of-flight 
of the beam momenta and  to avoid an overlap of events 
produced from alternating pulses. Lower repetition rate was 
already successfully used by G0 experiment in Hall~C at 
JLab~\cite{bib9}.

The radiation length of the radiator needs further studies 
in order to estimate the level of radiation and required 
shielding in the tagger region. During this experiment all 
photon beam tagging detector systems and electronics will 
be removed.

The final flux of $K_L^0$ is presented with 10$\%$ radiator, 
corresponding to maximal rate .

In the production of a beam of neutral Kaons, an important 
factor is the rate of neutrons as background. As it is well 
known, the ratio $R=N_n/N_{K_L^0}$ is on the order $10^3$ 
from primary proton beams~\cite{bib10}, the same ratio with 
primary electromagnetic interactions is much lower. This is 
illustrated in Fig.~\ref{fig:ratio}, which  presents the rate 
of Kaons and neutrons as a function of the momentum, which 
resembles similar behavior as it was measured at 
SLAC~\cite{bib11}.
\begin{figure}[htb!]
\centering{
\includegraphics[width=4.0in]{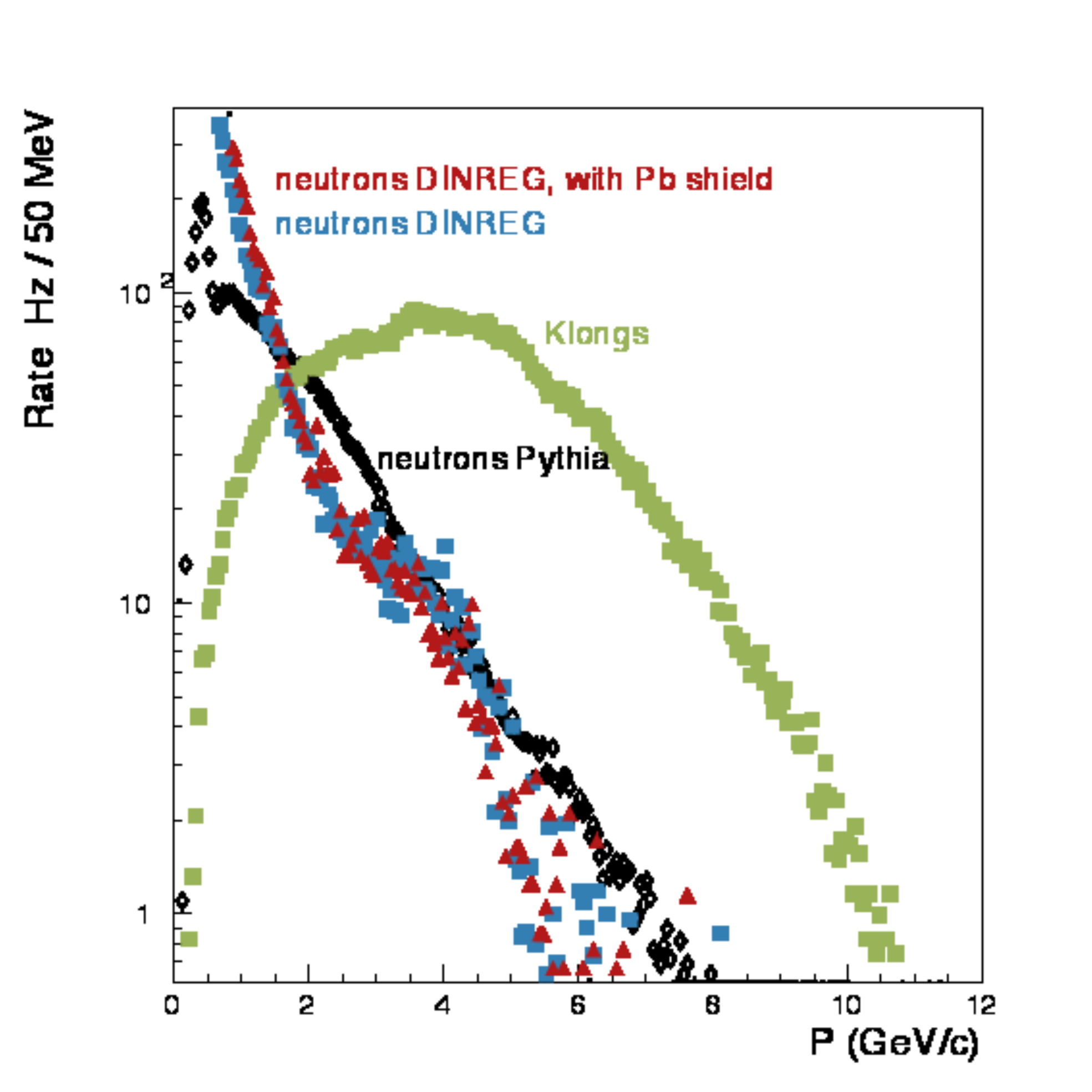}
\centerline{\parbox{0.80\textwidth}{
 \caption{\baselineskip 13pt The rate of neutrons (open 
	symbols) and $K_L^0$ (full squares) on $LH_2$ 
	target of Hall~D as a function of their momenta 
	simulated with different MC generators with 
	$10^4~K^0_L$/sec. \label{fig:ratio} } } } }
\end{figure}

Shielding of the low energy neutrons in the collimator cave 
and flux of neutrons has been estimated to be affordable, 
however detailed simulations are under way to show the level 
of radiation along the beamline. 

Th response of GlueX setup, reconstruction efficiency and 
resolution are presented in a talk by Simon Taylor at this 
workshop.

\item \textbf{Expected Rates}

In this Section, we discuss expected rates of events for 
some selected reactions. The production of $\Xi$ hyperons 
has been measured only with charged Kaons with very low 
statistical precision and never with primary $K_L^0$ beam. 
In Fig.~\ref{fig:xi_prod} panel a) shows existing data for 
the octet ground state $\Xi$'s with theoretical model 
predictions for $W$ (the reaction center of mass energy)  
distribution, panel b) shows the same model 
prediction~\cite{bib12}  presented with expected 
experimental points and statistical error for 10 days of 
running with our proposed setup with a beam intensity 
$2\times 10^3~K_L$/sec using missing mass of $K^+$ in the 
reaction $K_L^0+p\to K^+\Xi^0$ without detection of any of 
decay products of $\Xi^0$ (for more details on this topic 
see a talk by Kanzo Nakayama at this workshop).
\begin{figure}[htb!]
\begin{center}
\includegraphics[angle=0, width=0.70\textwidth ]{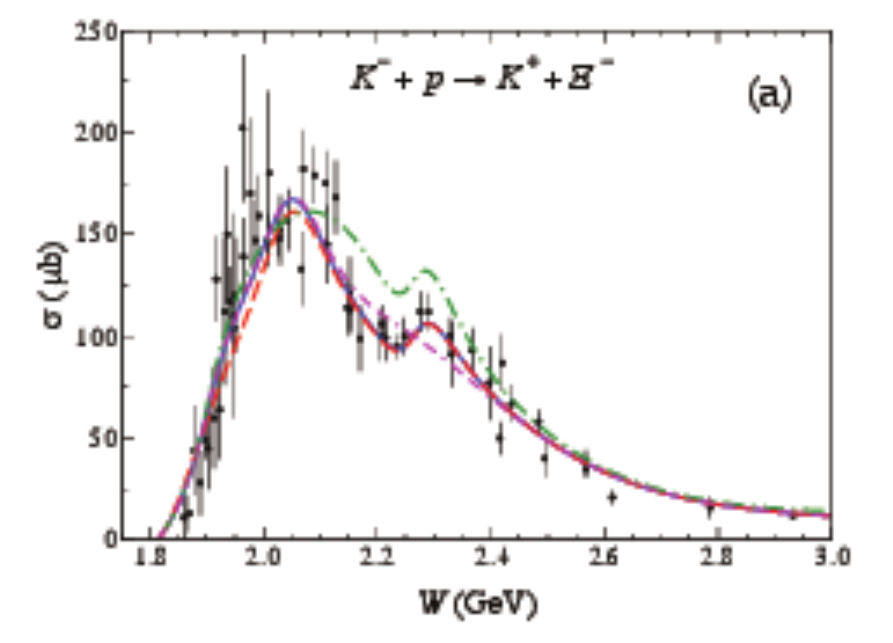}
\includegraphics[angle=0, width=0.55\textwidth ]{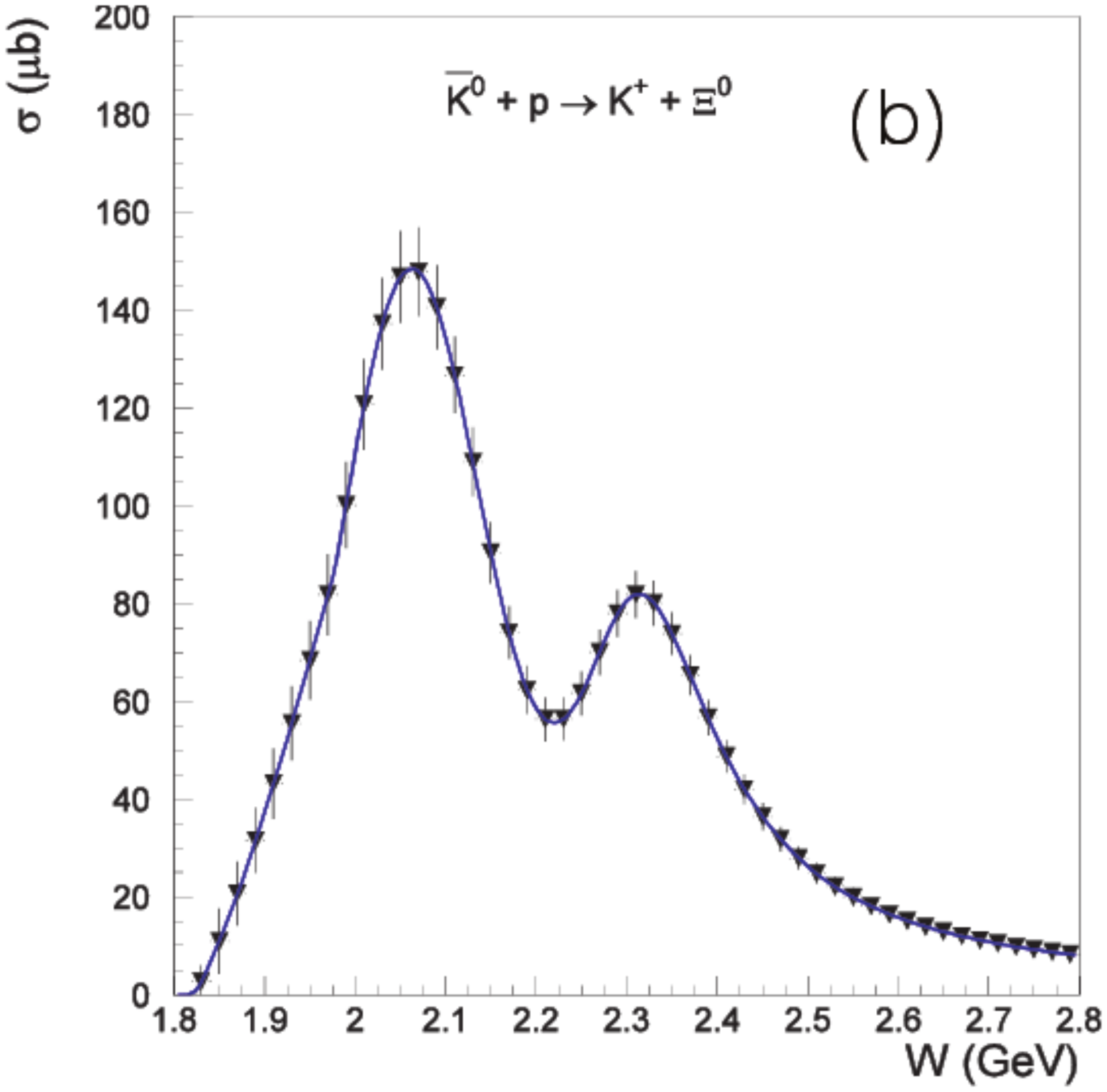}
\end{center}
\vspace{-2.0cm}
\centerline{\parbox{0.80\textwidth}{
\caption{\baselineskip 13pt a) Cross section for existing 
	world data on $K^-+p\to K^+\Xi^-$ reaction with model 
	predictions from~\protect\cite{bib12}; 
	b) expected statistical precision for the reaction 
	$K_L^0+p\to K^+\Xi^0$ in 10 days of running with a 
	beam intensity $2\times 10^3~K_L$/sec overlaid on 
	theoretical prediction~\protect\cite{bib12}. 
	\label{fig:xi_prod} } } }
\end{figure}

The physics of excited hyperons is not well explored, remaining 
essentially at the pioneering stages of '70s-'80s. This is 
especially true for $\Xi^\ast(S=-2)$  and $\Omega^\ast(S=-3)$ 
hyperons. For example, the $SU(3)$ flavor symmetry allows as many 
$S=-2$ baryon resonances, as there are $N$ and $\Delta$ resonances 
combined ($\approx 27$); however, until now only three [ground 
state $\Xi(1382)1/2^+$, $\Xi(1538)3/2^+$, and $\Xi(1820)3/2^-$] 
have their quantum numbers assigned and few more states have been 
observed~\cite{bib1}. The status of $\Xi$ baryons is summarized 
in a table presented in Fig.~\ref{fig:table1} together with quark 
model predicted states~\cite{bib13}.
\begin{figure}[htb!]
\centering{
\includegraphics[width=7.5in]{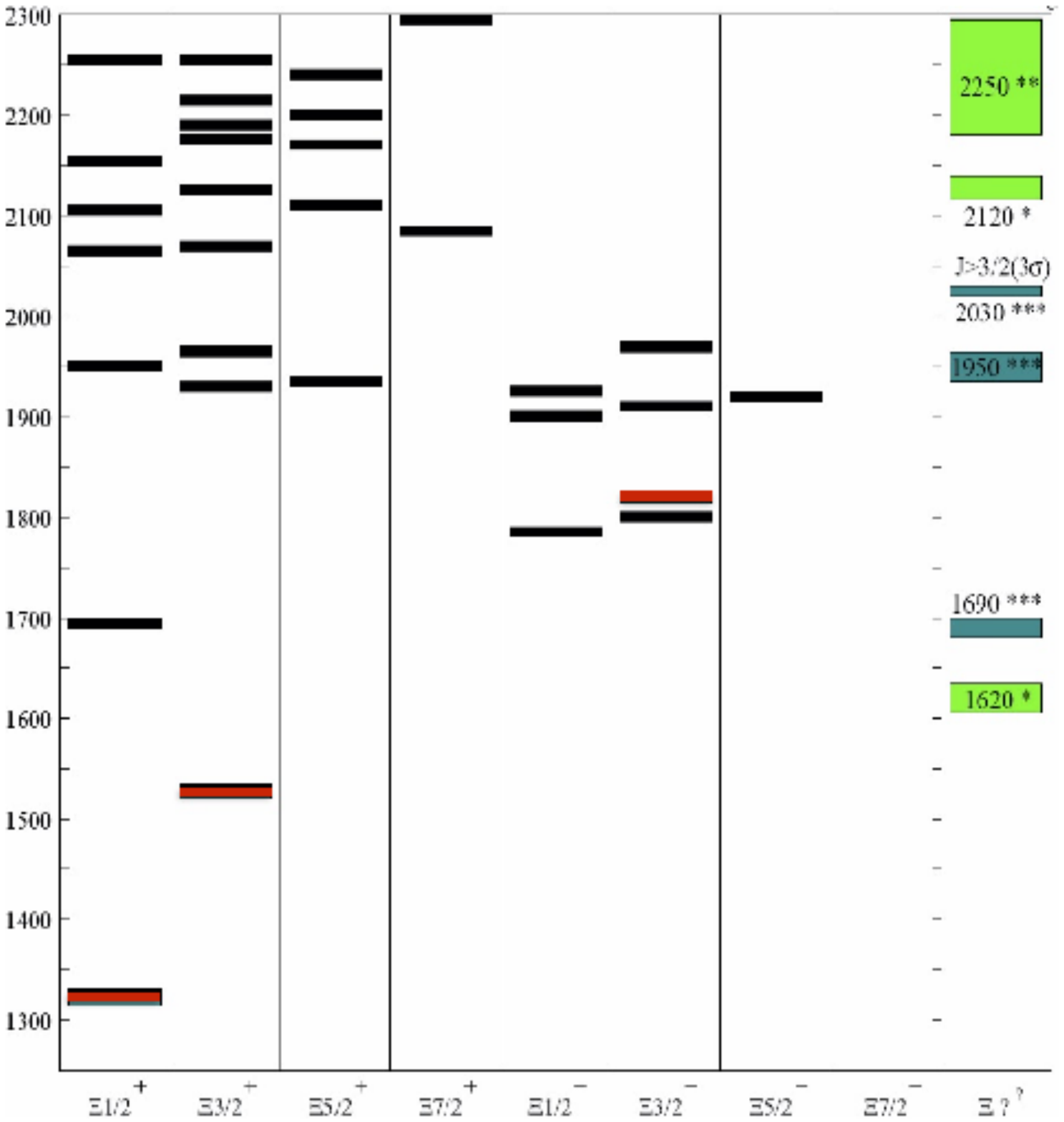}
\centerline{\parbox{0.80\textwidth}{
 \caption{\baselineskip 13pt Black bars: Predicted $\Xi$ 
	spectrum based 	on the quark model 
	calculation~\protect\cite{bib13}. Colored bars: 
	Observed states. The two ground octet and decuplet 
	states together with $\Xi(1820)$ in the column 
	$J^P=3/2^-$ are shown in red color. Other observed 
	states with unidentified spin-parity are plotted in 
	the rightest column. \label{fig:table1} } } } }
\end{figure}

Historically the $\Xi^\ast$ states were intensively searched for 
mainly in bubble chamber experiments using the $K^-p$ reaction in 
'60s--'70s. The cross section was estimated to be on the order of 
1--10~$\mu b$ at the beam momenta up to ~10~GeV/$c$. In '80s--'90s, 
the mass or width of ground and some of excited states were 
measured with a spectrometer in the CERN hyperon beam experiment. 
Few experiments have studied cascade baryons with the missing mass 
technique. In 1983, the production of $\Xi^\ast$ resonances up to 
2.5~GeV were reported from $p(K^-,K^+)$ reaction from the 
measurement of the missing mass of $K^+$~\cite{bib14}. The 
experimental situation with $\Omega^{-\ast}$'s is even worse 
than the $\Xi^\ast$ case, there are very few data for excited 
states. The main reason for such a scarce dataset in multi strange 
hyperon domain is mainly due to very low cross section in indirect 
production with pion or, in particular, - photon beams. Currently 
only ground state $\Omega^-$ quantum numbers are identified. 
Recently significant progress is made in lattice QCD calculations 
of excited baryon states~\cite{bib15,bib16} which poses a 
challenge to experiments to map out all predicted states (for more 
details see a talk by David Richards at this workshop). The advantage 
of baryons containing one or more strange quarks for lattice 
calculations is that then number of open decay channels is in 
general smaller than for baryons comprising only the light $u$ and 
$d$ quarks. Moreover, lattice calculations show that there are 
many states with strong gluonic content in positive parity sector 
for all baryons.  The reason why hybrid baryons have not attracted 
the same attention as hybrid mesons is mainly due to the fact that 
they lack manifest ``exotic" character. Although it is 
difficult to distinguish hybrid baryon states, there is significant 
theoretical insight to be gained from studying spectra of excited 
baryons, particularly in a framework that can simultaneously 
calculate properties of hybrid mesons. Therefore this program will 
be very much complementary to the GlueX physics program of hybrid 
mesons.

The proposed experiment  with a beam intensity $10^4~K_L$/sec will 
result in about $2\times 10^5$~$\Xi^\ast$'s  and $4\times 
10^3$~$\Omega^\ast$'s per month.

A similar program for $KN$ scattering is under development at 
J-PARC with charged Kaon beams~\cite{bib17}. The current maximum 
momentum of secondary beamline of 2~GeV/$c$ is available at the 
K1.8 beamline. The beam momentum of 2~GeV/$c$ corresponds to 
$\sqrt{s}$=2.2~GeV in the $K^-p$ reaction which is not enough to 
generate even the first excited $\Xi^\ast$ state predicted in the 
quark model. However, there are plans to create high energy beamline 
in the momentum range 5 -- 15~GeV/$c$ to be used with the 
spectrometer commonly used with the J-PARC $E50$ experiment which 
will lead to expected yield of $(3-4)\times10^5$~$\Xi^\ast$'s and 
$10^3$~$\Omega^\ast$'s per month. 

Statistical power of proposed experiment with $K_L$ beam at JLab 
will be of the same order  as that in J-PARC with charged Kaon 
beam.

An experimental program with Kaon beams will be much richer and 
allow to perform a complete experiment using polarized target and 
measuring recoil polarization of hyperons. This studies are under 
way to find an optimal solution for GlueX setup.

\item \textbf{Summary}

In summary, we intend to create high intensity $K_L$ beam using 
photoproduction processes from a secondary Be target. A flux as 
high as $10^4~K_L$/sec could be achieved. Momenta of $K_L$ beam 
particles will be measured with time of flight. The flux of Kaon 
beam will be measured through partial detection of $\pi^+\pi^-$ 
decay products from their decay to $\pi^+\pi^-\pi^0$ by exploiting 
similar procedure used by LASS experiment at SLAC~\cite{bib11}.
Besides using unpolarized $LH_2$ target currently installed in 
GlueX experiment additional studies are needed to find the optimal 
choice of polarized targets. This proposal will allow to measure 
$KN$ scattering with different final states including production 
of strange and multi strange baryons with unprecedented statistical 
precision to test QCD in non perturbative domain. It has a potential 
to distinguish between  different quark models and test lattice QCD 
predictions for excited baryon states with strong hybrid content.

\newpage
\item \textbf{Acknowledgments}

This work is supported, in part, by the U.S. Department of Energy, 
Office of Science, Office of Nuclear Physics, under Award Number 
DE--FG02--96ER40960.
\end{enumerate}


\newpage
\subsection{Hadron Physics at J-PARC}
\addtocontents{toc}{\hspace{2cm}{\sl H.~Ohnishi}\par}
\setcounter{figure}{0}
\setcounter{equation}{0}
\halign{#\hfil&\quad#\hfil\cr
\large{Hiroaki Ohnishi}\cr
\textit{RIKEN, Nishina Center}\cr
\textit{2-1 Hirosawa}\cr
\textit{Wako, Saitama 351-0198, Japan \&}\cr
\textit{Research Center for Nuclear Physics (RCNP)}\cr
\textit{Osaka University, Osaka}\cr
\textit{567-0047, Japan}\cr}

\begin{abstract}
One of the main goals for the hadron physics is to understand 
the effective degrees of freedom (EDoF) in hadron and reveal 
their interactions. Spectroscopy of ground and excited states 
of baryons will give us hints to understand the EDoF of hadron. 
In addition, testing properties of known mesons/baryons, such 
as masses and decay widths, inside nuclear matter, will give 
us unique information on the interaction between EDoF and QCD 
vacuum.

In this paper, I will discuss the goal of the hadron physics 
and summarize experimental programs performed and planned at 
J-PARC. Finally, I will briefly discuss a future project at 
J-PARC, which is now under discussion.
\end{abstract}

\begin{enumerate}
\item \textbf{Introduction}

The strong interaction between elementary particles has been 
described very well by the quantum chromo dynamics (QCD). The 
missing element of a standard model of elementary particles, 
{\it i.e.}, the Higgs boson, has been discovered at CERN/LHC 
in 2012. Therefore the theory of known elementary particles, 
including the strong interaction, is now completed. There are 
many varieties of matter created by QCD, such as hadrons, 
nuclei and very high-density nuclear matter, such as neutron 
stars. Those type of matter must be interpreted by the QCD.
However, due to the complexity of the QCD theory, it is very 
difficult to solve all problems and to understand the 
connection between elementary particles like quarks and gluons 
and hadrons or extremely high-density matter. It should be 
noted that not even the first step, how the hadrons and their 
excited states are created, is clearly understood. Therefore, 
not only more experimental efforts to understand hadron 
phenomena, but also strong theoretical supports for the 
hadron/nuclear physics are still mandatory to understand the 
matter created by QCD.  

Some of the goals of hadron physics could be summarized as in 
the following two questions. First, how the hadrons are created 
via QCD? In other words, what are the effective degrees of 
freedom to describe hadron and excited hadrons? Second, hadrons 
are understood as excitations of QCD vacuum. Therefore, a change 
of vacuum condition should affect directly to the properties of 
hadron, such as mass and width. Thus, we need to know how hadron 
properties change when environmental condition changed, {\it 
i.e.}, vacuum inside nuclear matter.

In normal conditions, the world consists only from light quarks, 
{\it i.e.}, $u$ and $d$ quarks. However, inside the compressed 
QCD matter, creation of hadrons with strangeness is expected, 
theoretically. For such condition, hadrons with strangeness 
cannot be ignored to understand high-density matter. For example, 
anti-Kaon in nucleus is a hot subject in the hadron physics, 
which may give us hints toward physics in high-density nuclear 
matter.  One the other hand, baryons with strangeness themselves 
are also a very important subject. According to the quark model, 
color magnetic interaction between constituent quarks can be 
expressed as follows.
\begin{equation}
	V_{CMI}\sim\frac{\alpha_s}{m_i m_j}(\lambda_i\cdot\lambda_j)
	(\vec{\sigma_i}\cdot\vec{\sigma_j}),
\end{equation} 
where, $m$, $\lambda$ and $\vec{\sigma}$ are mass, color and spin 
of constituent quarks, respectively. The equation tells us that 
if we choose heavy quarks as constituents for hadrons, 
color-magnetic interaction between a light quark and heavy quark 
is going to be zero. 

Therefore, the interaction between light quarks will be dominant. 
In the case of baryons, strong correlation between di-quark will 
be realized. Hints for this type of correlation are expected to 
appear in the excited baryon spectra/decay pattern of hadron. 
Since the strange quark mass is heavier that $u$ and $d$ are, we 
may expect signal for such di-quark correlation in the S=-1 
baryon system. In addition, S=-2 baryon can be treated as an 
analogy of baryon with two heavy quarks. It should be noted that 
baryon with two heavy quarks, such as $\Xi_{cc}$ for example, 
have not been observed. Therefore, S=-2 baryon spectroscopy will 
be a unique doorway to understand the structure of baryons with 
heavy quarks, in other words, the investigation will give us an 
insight to the effective degrees of freedom to describe hadrons. 
Therefore, hadron with strangeness, \textit{i.e.}, baryon with 
strangeness ($\Lambda$/$\Xi$/$\Omega$) and/or Kaon, will be a 
key ingredient to understand the questions mentioned above. 

\item \textbf{J-PARC}

The Japan Proton accelerator Research Complex (J-PARC) is one of 
the key machines to perform hadron physics. Proton beam 
accelerated up to 30~GeV by J-PARC Main Ring Synchrotron (MR) is 
delivered to Hadron experimental facility (HD) and shoot onto the 
production target, which is made by gold, to produce secondary 
hadron beams, such as $\pi^{\pm},K^{\pm}$ and $p,\bar{p}$. 
Typical beam intensity of the primary proton beam is 4.8$
\times$10$^{13}$ proton per spill (pps), where the spill length 
is 2~seconds with a 5.52~seconds repetition cycle.  Inside HD, 
four beamlines are designed, two (K1.8, K1.8BR) are in operation 
and two (K1.1 and High-p) are under construction. The typical 
beam intensities for secondary particle for each beam lines are 
summarized in Table~\ref{table1}. As one can see, particle 
separated beams can be available up to 2~GeV/$c$ and unseparated 
beam is available up to 20~GeV/$c$. In addition, a primary proton 
beam is also available for the experiment. A more detailed 
description can be found elsewhere~\cite{JPARCN}.
\begin{table*}
\centerline{\parbox{0.80\textwidth}{
\caption{\label{tab:table1} J-PARC Beam line specifications.} }}
\vspace{0.3cm}
\begin{tabular}{|c|c|c|c|}
\hline \hline
beamline & paricle & momentum range & typical beam intensity \\ 
         &         &                &  (40~kW MR operation)  \\
\hline \hline
K1.8BR   & $\pi^{\pm},K^{\pm}$ and $p,\bar{p}$ (separated)   & $<$ 1.1~GeV/$c$  & 1.5$\times$10$^5$ K$^-$/spill@  1~GeV/$c$ \\
K1.8     & $\pi^{\pm},K^{\pm}$ and $p,\bar{p}$ (separated)   & $<$ 2.0~GeV/$c$  & 5.0$\times$10$^5$ K$^-$/spill@  2~GeV/$c$ \\
K1.1     & $\pi^{\pm},K^{\pm}$ and $p,\bar{p}$ (separated)   & $<$ 1.1~GeV/$c$  & 1.5$\times$10$^5$ K$^-$/spill@  1~GeV/$c$ \\
High-p   & $\pi^{\pm},K^{\pm}$ and $p,\bar{p}$ (unseparated) & up to 20~GeV/$c$ & $ >   \sim$10$^7 \pi^-$/spill@ 20~GeV/$c$ \\
         &                                                   &                  & $ >   \sim$10$^6   K^-$/spill@  7~GeV/$c$ \\
         & primary proton    & 30~GeV                        & $\sim$ 10$^{11}$ proton / spill                          \\
\hline \hline
\end{tabular}\label{table1}
\end{table*}

\item \textbf{Hadron Physics Performed at J-PARC}

\begin{enumerate}
\item \textbf{Search for Penta-Quark Baryon}

Only color singlet state can exist as hadrons. This is a conclusion 
from QCD. Therefore, hadrons which have 5 quarks ($qqq q\bar{q}$) 
are not forbidden by QCD. Thus, many experimental challenges have 
been performed to search for such exotic states. Strong evidence 
have been reported from photo-production experiment~\cite{LEPSoN}, 
however, also many negative results are reported (mainly from 
hadro-production)~\cite{P1N,P2N,P3N,P4N}.  At J-PARC hadro-production 
of penta-quark state using high intensity pion beam is performed, 
by $(\pi,K)$ reaction on hydrogen target\cite{E19-1N,E19-2N}. No 
signal has been observed so far. To date, still the conclusion has 
not been reached concerning whether a penta-quark state exists or 
not.

\item \textbf{Search for Kaonic Nucleus}

Because strong attractive force exists between anti-Kaon and nucleon, 
the existence of the strongly bound Kaonic-nuclear state has been 
discussed for a long time. It is interesting to note that, 
theoretically, the inside of the Kaonic nucleus could turn into high 
density, much higher than normal nuclear matter density. Therefore, 
the study of Kaonic nucleus will give us some insight on QCD at 
high-density matter. There are many experiments to search for such 
exotic state, which have been performed to date, however, still not 
strong conclusion is made. Two new experiments have been performed at 
J-PARC. Both experiments are focussing on the lightest Kaonic nuclear 
cluster, \textit{i.e.}, $K^-pp$ state. One is the E27 experiment, 
which aims to search for $K^-pp$ cluster via $(\pi,K)$ reaction. The 
result shows some indication for the deeply bound $K^-pp$ bound 
state~\cite{E27N}. The other experiment is the E15 experiment, which 
aims to search for the $K^-pp$ bound state via $^3$He($K^-,n$) 
reaction. The first results from the E15 experiment shows~\cite{E15-1N} 
no clear signal found in deeply bound region, but an interesting 
events enhancement has been observed near the $K^-pp$ threshold 
region. Recently, the E15 experiment reported new results on 
exclusive analysis on $^3$He($K^-,\Lambda p$)n reaction~\cite{E15N}. 
The result is shown in Figure~\ref{fig:E15}. Figure~\ref{fig:E15}(a) 
shows a scatter plot for the invariant mass of $\Lambda p$ versus 
neutrons emitted angle in the center of mass frame. 
Figure~\ref{fig:E15}(b) and (c) are the projection of the plot to 
the axes. Expected contributions are also plotted as histogram in 
Figure~\ref{fig:E15}(b). A clear enhancement with respect to the 
expected contributions are seen just bellow the $\bar{K}NN$ 
threshold. It is interesting to note that as shown in 
Figure~\ref{fig:E15}(c), a clear event concentration at 
$\cos(\theta_{CM})\sim 0$, where slowly moving $\bar{K}$ produced is 
seen. This will be a necessary condition to form $\bar{K}NN$ bound 
state. However, due to the small statistics, it is still hard to 
conclude whether $\bar{K}NN$ bound states are produced or not.  

Because both experiments try to produce $K^-pp$ cluster by different 
production mechanisms further detailed studies are still needed to 
conclude whether $K^-pp$ cluster really exists or not, and to know 
its properties.
\begin{figure}[ht!]
\begin{center}
\includegraphics[width=12cm]{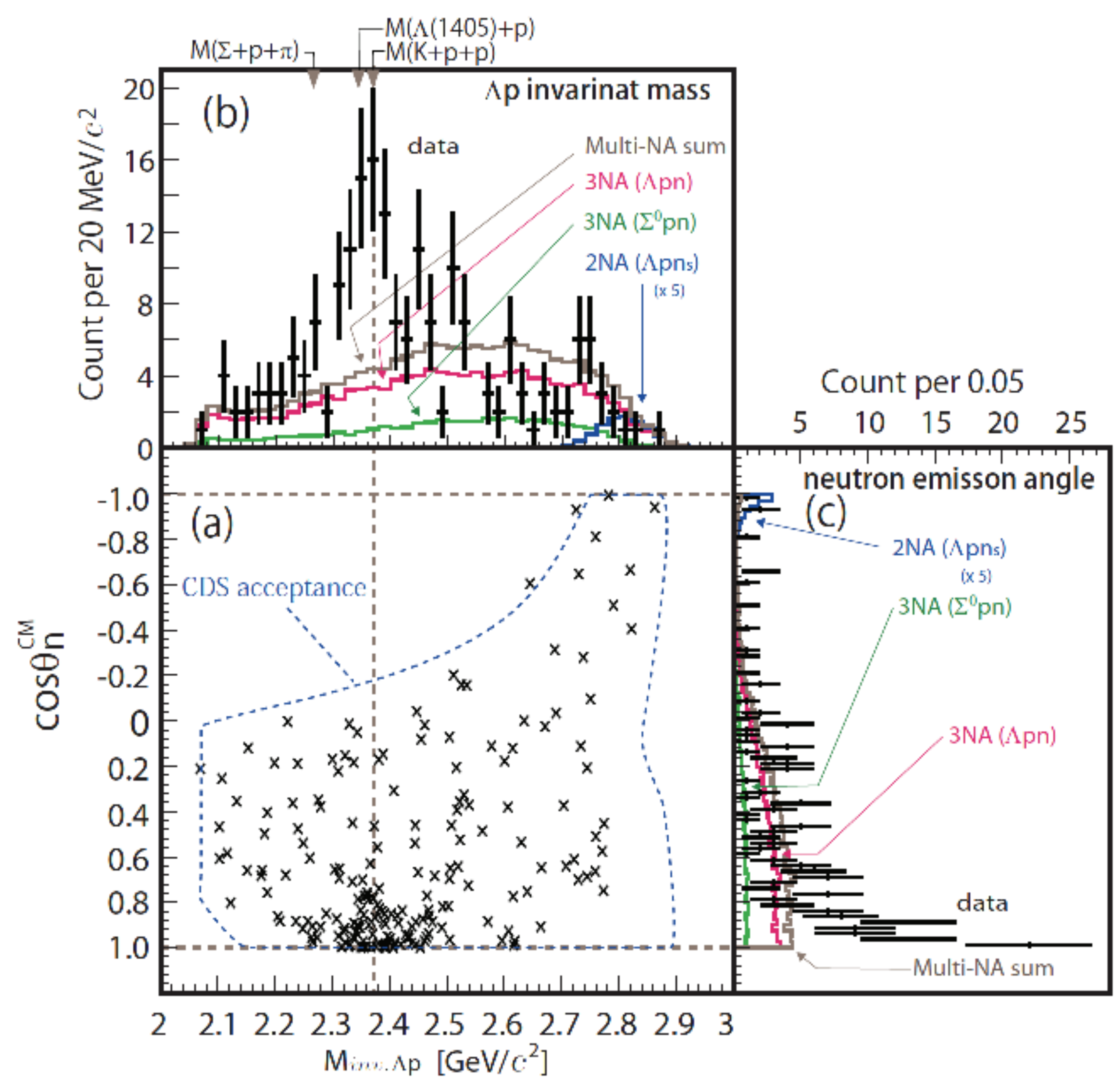}
\end{center}
\centerline{\parbox{0.80\textwidth}{
\caption{\label{fig:E15} Results recently published by the J-PARC 
	E15 Collaboration~\protect\cite{E15N}. } } }
\end{figure}

\item \textbf{Mesons in Nuclei}

Chiral symmetry in QCD vacuum is spontaneously broken. It is now 
understood that mass of the hadrons is generated dynamically by 
the broken symmetry. However, the chiral symmetry will be partially 
restored in high-density matter such as inside nuclei. This 
information can be checked through the measurement of the mass of 
mesons or search for the meson nuclear bound state. Because vector 
mesons have relatively long lifetime, therefore three experiments 
have been proposed to study the vector mesons in nuclear matter: 
E16~\cite{E16N}, E26~\cite{E26N}, and E29~\cite{E29N}. 

The E16 experiment is aiming to measure the line shape of vector 
mesons via measuring $V\rightarrow e^+e^-$ decay inside a nucleus. 
The experiment is planed to be perform using the 30~GeV primary 
proton beam at the high momentum beam line.

The E26 experiment is planed to search for $\omega$ meson nuclear 
bound state. To maximize the formation probability of $\omega$ 
meson nucleus, slowly moving $\omega$ mesons are selectively 
produced via $(\pi,n$) reaction using the 2~GeV/$c$ pion beam at 
K1.8 beamline. Signal of $\omega$ mesic nucleus is identified via 
missing mass spectroscopy of forward going neutron.

The E29 experiment is forccusing on the $\phi$ meson nuclear bound 
state. Very exotic elementary reaction channel, $\bar{p}p\rightarrow
\phi\phi$, has been chosen to produce slowly moving $\phi$ meson. 
The experiment is planning to use the 1.1~GeV/$c$ $\bar{p}$ beam at 
K1.8BR beamline. The signal is identified via missing mass analysis 
using the forward going $\phi$ meson, together with the $K^+$ and 
$\Lambda$ from the target as final state particles, to ensure the 
double strangeness pairs are produced.

\item \textbf{S=-2 and S=-3 Baryon Spectroscopy}

To understand the effective degree of freedom to describe hadrons, 
in other words, what are the DoF to control the excited baryon 
spectra, it is very important to identify the complete spectra of 
S=-2 and/or S=-3 baryons. However, according to the PDG, only a 
small number of S=-2 baryons are established. In case of $\Omega$ 
baryon, only ground state is known. High intensity Kaon beam will 
improve the situation drastically. It should be noted that in case 
of nucleon resonances, the widths are very broad, typically more 
than $\sim$270~MeV, thus it is hard to identify the states easily. 
However, the trend of baryon with strangeness shows widths which 
are much narrower than of nuclear resonances, it is about 
$\sim$40~MeV. Therefore, we have a chance to identify those excited 
multi-strangeness baryon clearly. The experiment to identify $\Xi$ 
baryons are in preparation at High-p beam line where high momentum 
$K^{-}$ beam will be available. The missing mass spectroscopy via 
$(K^{-},K^{+})$ or $(K^{-},K^{+}\pi^{+})$ is planed to establish 
and search for the $\Xi$ baryons~\cite{NarukiLoiN}. The experiment 
is expected to pin down the $\Xi$ baryon spectra up to the baryons 
with mass $\sim$3~GeV/$c^2$. 

\item \textbf{Future projects: Hadron Hall Extension}

To extend the physics cases reachable at J-PARC, an extension of 
the HD facility is under discussion. It is true that high momentum 
K$^-$ is already available at High-p beamline. However, the 
intensity of the beam is rather low, which limits the reach of the 
excited $\Xi$ search. Moreover, excited state for $\Omega$ baryon 
are not possible at High-p beamline, because the expected cross 
section is very small (sub $\mu$b). In addition, at High-p beamline, 
only a cocktail beam of $\pi^-$,K$^-$ and $\bar{p}$ is available, 
but most of it are pions. Therefore, experiments will be facing 
serious problems of pion interactions, which is indeed the main 
background for Kaon interaction studies. Therefore, high intensity 
and particle separated beamline is very important to enhance the 
physics opportunities at J-PARC.
\begin{figure}[ht!]
\begin{center}
\includegraphics[width=14cm]{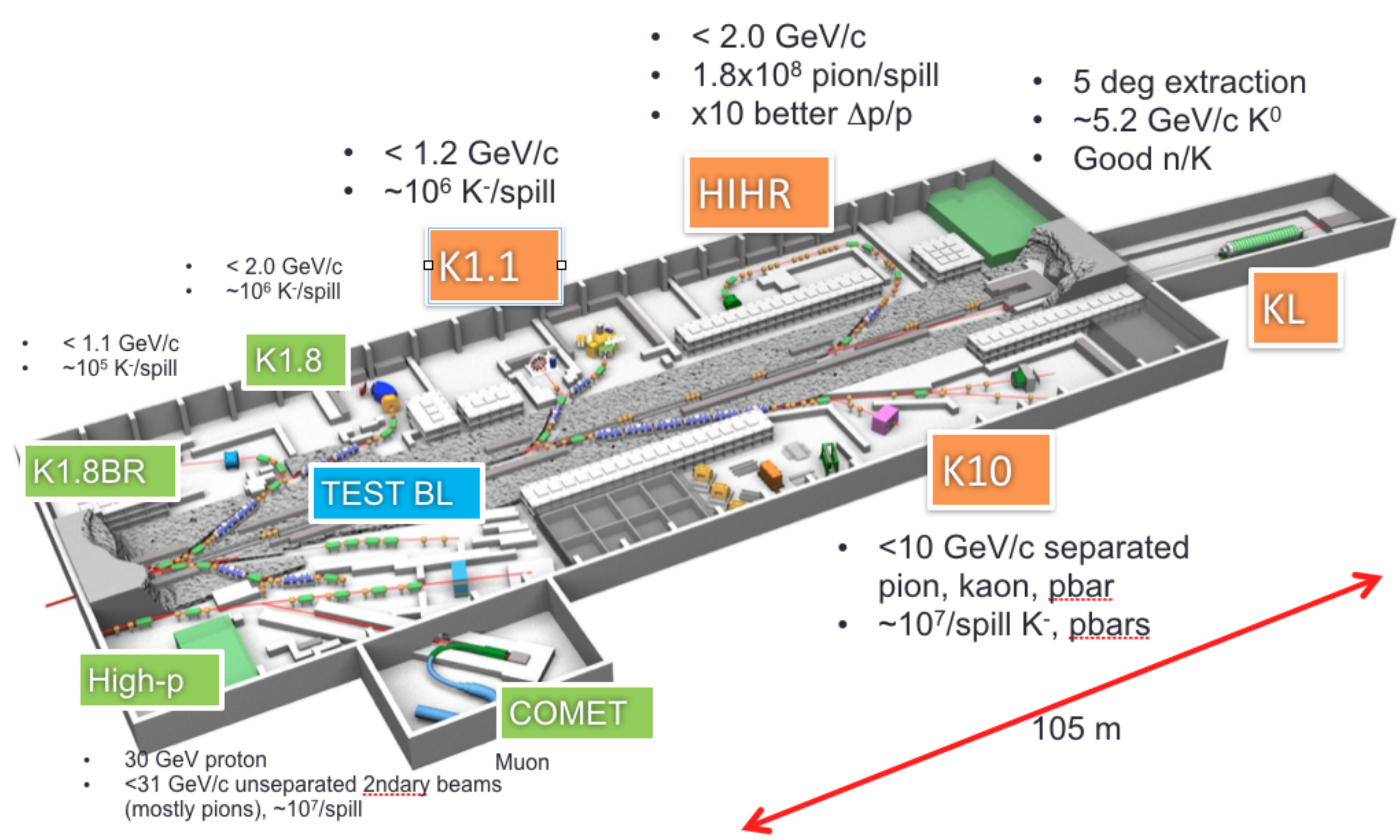}
\end{center}
\centerline{\parbox{0.80\textwidth}{
 \caption{\label{fig:HDext} Conceptual design for extended 
	Hadron Hall.} } }
\end{figure}

Figure~\ref{fig:HDext} shows a conceptual design for the extended 
hadron hall. For this extension, we will construct three new 
charged particle beamlines for hadron/nuclear physics and one new 
$K_0$ beam line to search the $K^0_L\rightarrow\pi^0\nu\bar{\nu}$ 
decay. which has great sensitivity to the beyond the standard model. 
Here, we will concentrate on the K10 beamline, where high intensity 
and high momentum K$^-$ and $\bar{p}$  beam will be available. 
Figure~\ref{fig:K10Int} shows the expected beam intensities at K10 
for K$^-$ and $\bar{p}$. As one can see in Figure~\ref{fig:K10Int}, 
the expected beam intensities will be 10$^7$  per spill at 4 to 
6~GeV/$c$ K$^-$ and 10$^7$ per spill for 10~GeV/$c$ $\bar{p}$.  
Utilizing those beam particles, we are planing to perform $\Omega$ 
baryon spectroscopy which will be possible at J-PARC once the 
hadron hall extension is realized. Moreover, recent lattice QCD 
calculation shows that tha interaction between $\Omega$ baryon and 
nucleon is attractive. If this is true, $\Omega$ baryon and nucleon 
may form $\Omega$-N bound state. Therefore the experiment to search 
for the $\Omega$-N bound state may be very important. 
\begin{figure}[ht!]
\begin{center}
\includegraphics[width=10cm]{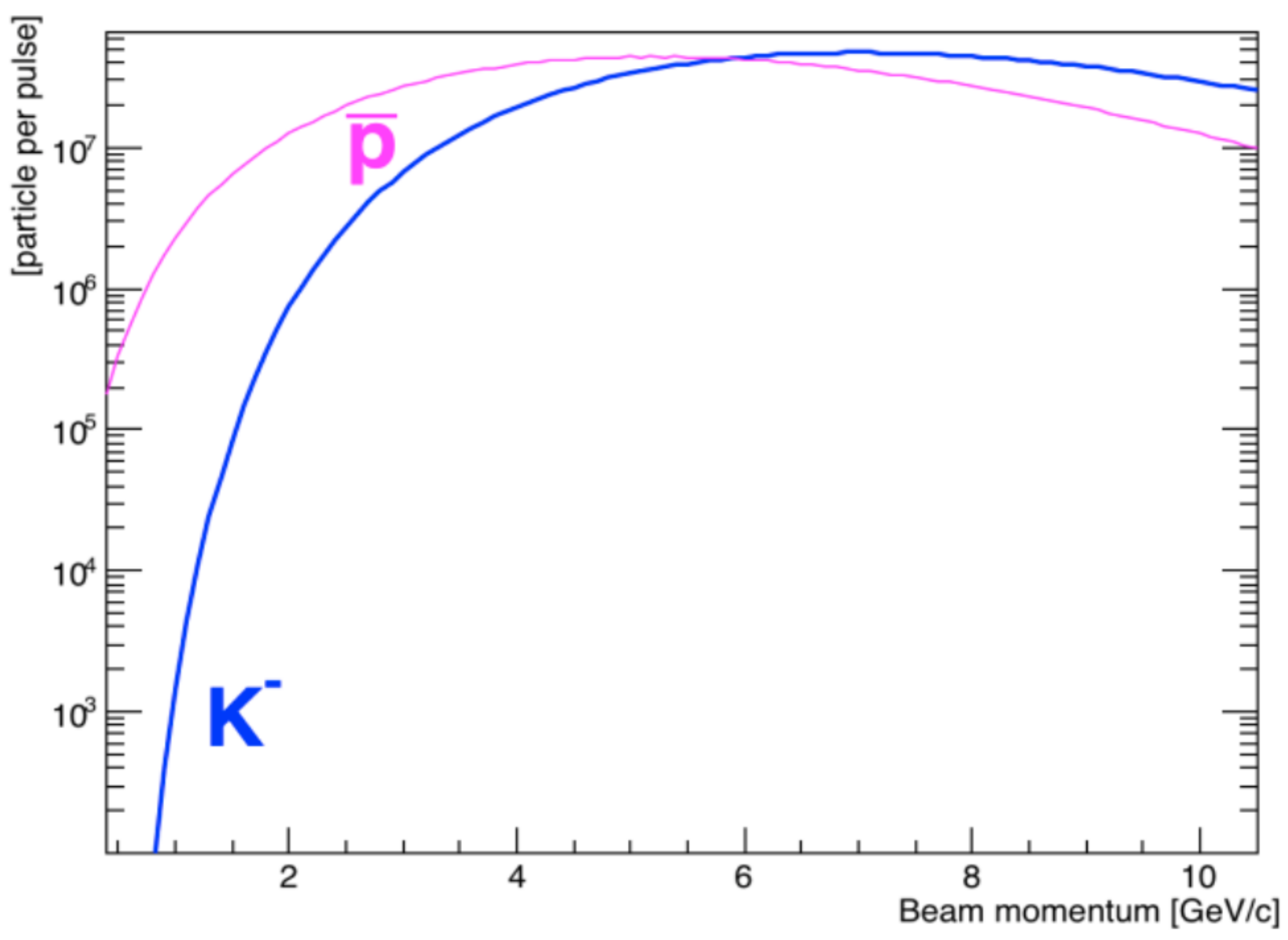}
\end{center}
\centerline{\parbox{0.80\textwidth}{
 \caption{\label{fig:K10Int} Expected beams intensities at K10 
	beamline.} } }
\end{figure}

In addition, high intensity $\bar{p}$ beam will open new opportunity 
to investigate charmed meson properties in nucleus (nuclear matter). 
Since long time, the interaction between D meson and nucleon is 
believed to be attractive, based on the many theoretical 
predictions~\cite{Th1N,Th2N,Th3N}. However, recent QCD sum rule 
calculation shows it is repulsive, in other words, D~meson is getting 
heavier in nuclear matter~\cite{Th4N,Th5N}. Because no experiment was 
performed to investigate the D~meson and nucleon interaction, no 
concrete information is available experimentally. 

Therefore, the D~meson properties in nuclear matter is one of the 
interesting subject to date. At K10, we plan to perform the 
experiment to measure $\bar{D}D$ production with $\bar{p}$ beam on 
proton and on nuclei, which will give us a hints for the $DN$ 
interaction.  It is interesting to note that recently many exotic 
hadrons are reported by collider experiments, such as Belle, BaBar, 
LHCb, BES \textit{etc.} Those measurements provide insight into the 
structure of hadrons, which can be summarize as follows.
\begin{enumerate}
\item Charmonium spectra can be describe very well by the model of 
	constituent quarks acting as effective degree of freedom 
	to describe charmonium, \textit{i.e.}, constituent quark 
	model.
\item Many exotic hadrons are also discovered. It is interesting 
	that those exotic hadrons exist only above the $D\bar{D}$ 
	production threshold. 
\end{enumerate}
Those phenomena indicates that the production cross section of 
$D\bar{D}$ near the production threshold might be sensitive whether 
such exotic states are really produced or not.

Figure~\ref{fig:K10Det} shows conceptual design for the spectrometer 
we are planing to install K10 beamline. The detector consists of 
large volume solenoid detector surrounding the target together with
forward dipole spectrometer. 
\begin{figure}[ht!]
\begin{center}
\includegraphics[width=12cm]{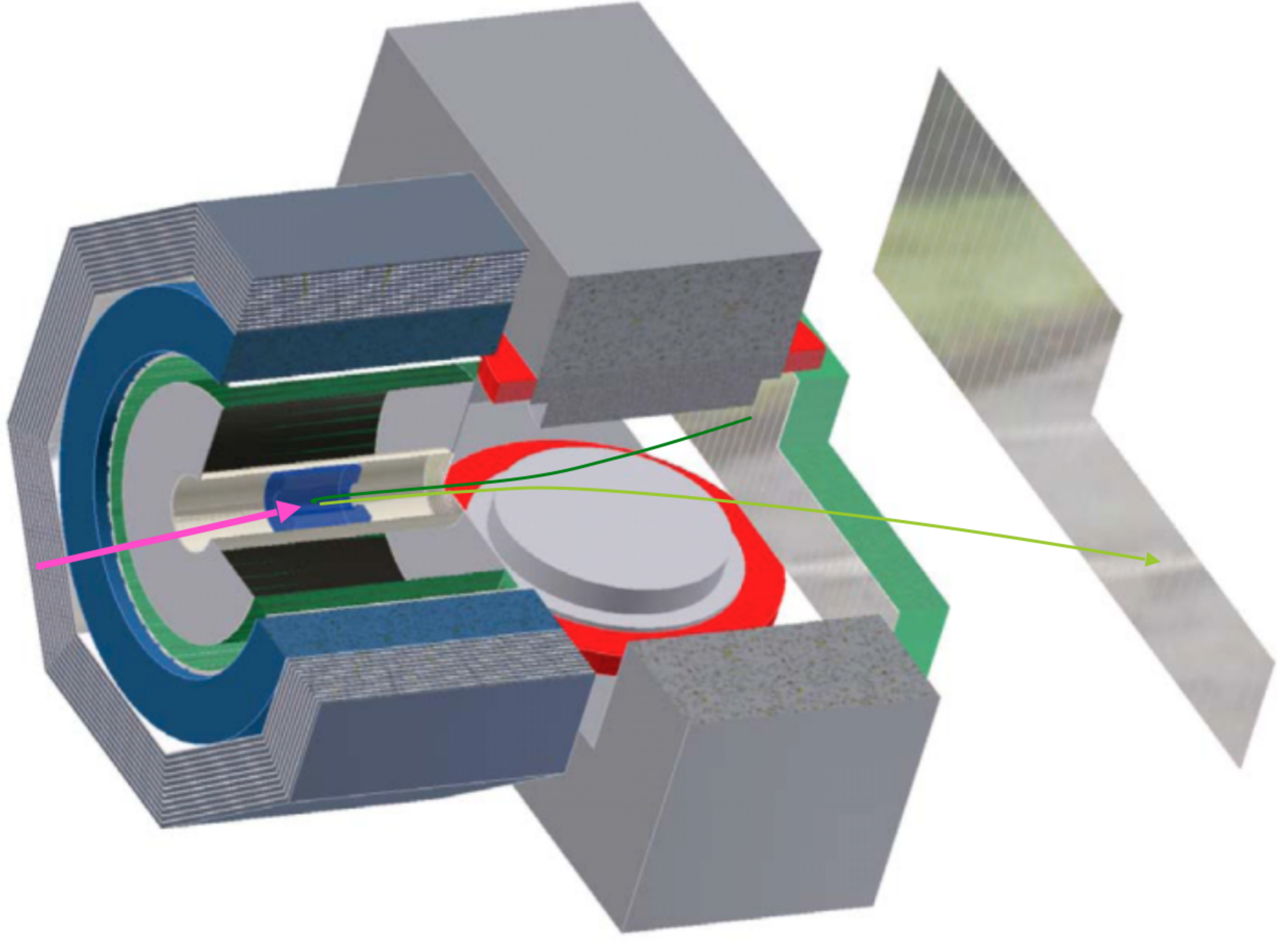}
\end{center}
\centerline{\parbox{0.80\textwidth}{
 \caption{\label{fig:K10Det} Conceptual design for Detector at 
	K10.} } }
\end{figure}
\end{enumerate}

\item \textbf{Summary}

In this paper, physics programs currently performed at J-PARC are 
reviewed. Baryon spectroscopy and mesons in nucleus using high 
intensity pion and Kaons beams are main topics for experimental 
programs at the current J-PARC hadron hall. Many new results are 
coming out. 

At present, investigation for $\bar{K}N$ interaction is performed 
using high intensity low momentum K$^-$. Recently available new 
data from E15 shows strong hint about $\bar{K}NN$ cluster. However, 
to make strong  conclusion, we need to wait the completion of the 
analysis with large data sample. Data have already been taken and 
data analysis is under the way.

A new project at J-PARC, {\it i.e.}, Hadron hall extension, was 
introduced.  Three separated charged secondary beamlines will be 
constructed.  In particular high intensity and high momentum particle 
separated beamline(K10) is very important for the hadron physics.  High 
momentum Kaons beam at K10 will allow to perform multi-strangeness 
baryons spectroscopy. Moreover, the high intensity anti-proton beam will 
open the door to a new physics subject, {\it i.e.}, charmed mesons in 
nuclei. 
\end{enumerate} 


\newpage
\subsection{Low Energy Kaon Scattering: Present Status and Open 
	Possibilities}
\addtocontents{toc}{\hspace{2cm}{\sl A.~Filippi}\par}
\setcounter{figure}{0}
\halign{#\hfil&\quad#\hfil\cr
\large{Alessandra Filippi}\cr
\textit{I.N.F.N. Sezione di Torino}\cr
\textit{Torino 10125, Italy}\cr}

\begin{abstract}
An overview of the experimental results on low energy scattering 
of charged and neutral Kaons is given. Emphasis is put on the
still missing information, which could be essential to provide a 
thorough description of the $\overline KN$ interaction close to 
threshold as well as below it, and that could be achieved by 
exploiting the unique features of a high intensity $K^0_L$ beam.
\end{abstract}

\begin{enumerate}
\item \textbf{Neutral Kaon Scattering: Properties and Cross 
	Sections Measurements at Low Energies}

The low energy cross section data for the interaction of charged 
Kaons with protons or neutrons (in deuterium targets) are rather 
few and imprecise. Below 350~MeV/$c$ incident momentum only old 
measurements exist, which date back to the Eighties and earlier 
years, and were performed in bubble chamber experiments or with 
emulsions~\cite{re:pdgJ}.  This low energy  region could still be 
fruitfully~explored by the DA$\Phi$NE machine in Frascati, and 
proposals were put forward some years ago in this 
respect~\cite{re:ikonJ}.

For neutral Kaons the situation is even worse. Few data exist 
down to 130~MeV/$c$ with a statistical accuracy limited to 10-20\% 
for the $K^0_Lp$ scattering, and a little better for 
$K^0_Ld$~\cite{re:clelandJ}. The trend of low momentum $K^0_Lp$ 
total cross section, from Ref.~\cite{re:clelandJ}, is reported in 
Fig.~\ref{fig:K0lp}: a typical total cross section at low momenta
for $K^0_L$ induced scattering on protons is around 70~mb, and 
twice as large on deuterons.
\begin{figure}[ht!]
\centering
\includegraphics[height=6.5truecm,clip=]{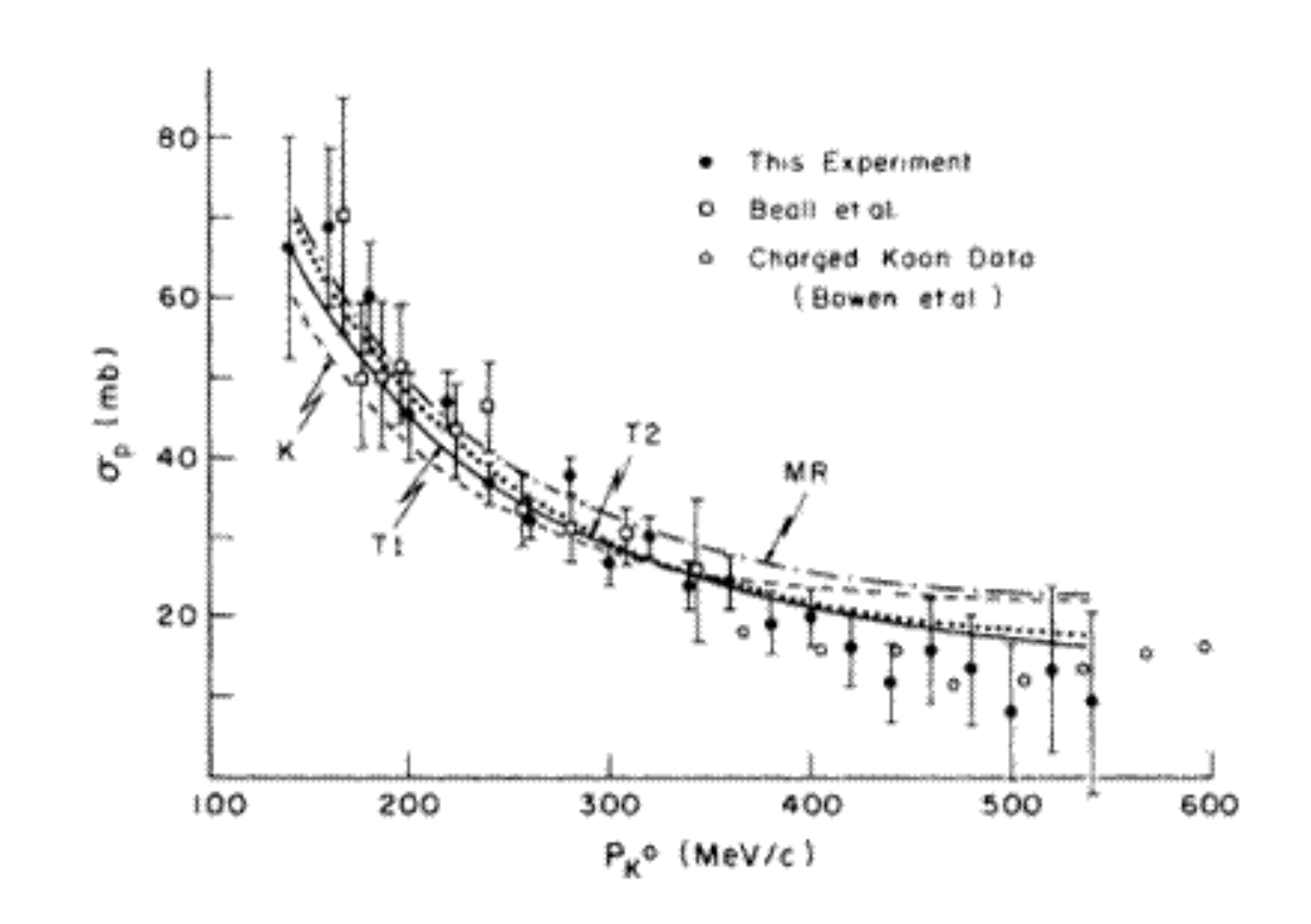}
\centerline{\parbox{0.80\textwidth}{
\caption{Total $K^0_Lp$ cross section. From 
	Ref.~\protect\cite{re:clelandJ}.  The superimposed 
	curves are the trends expected on the basis of 
	different solutions for the $K^-p$ scattering 
	lengths (see text).} \label{fig:K0lp} } }
\end{figure}

It is useful to recall that neutral Kaons behave as two different 
kinds of particles depending on the interaction they are subject 
to. The weakly interating particles, $CP$-eigenstates $K^0_S$ and 
$K^0_L$, are linear combinations (of almost equal strenght, at 
the precision level of scattering measurements) of the strangeness 
eigenstates, $K^0$ and $\overline{K^0}$. These are the relevant 
particles for evaluating the effect of the strong interaction in 
the scattering. According to the strangeness of the meson, 
however, its behavior as hadronic probe is largely different.

The Kaons with strangeness $S=+1$, {\it i.e.}, $K^+$ and $K^0$, 
have just a mild interaction with the medium. The cross sections 
are small, of the order of 10~mb, and are dominated by the 
elastic channel with a small contribution from Charge Exchange. 
The relevance of these scattering processes is mainly related to 
the possible formation of exotic pentaquark systems ($q^4\overline 
q$), that however have never been observed so far (anyway, if 
ever existing, these states are not likely to show up at low 
energies). In the $K^+p$ scattering a sizeable contribution, 
almost as large as the one due to strong interaction, comes from 
the electromagnetic interaction, relevant especially at small 
angles~\cite{re:KplusJ}. The $K^+p$ system is a pure isospin 
$I=1$ state, and its $S$-wave scattering length, which will be 
described in more detail in Sec.~2.5.1.a, has been determined 
with good ($\sim 1$\%) precision.

On the other hand, Kaons with strangeness $S=-1$ are strongly 
absorbed. The interaction cross sections are larger than 50~mb, 
and several baryonic resonances (formerly known as $Y^\ast$), 
both with isospin 0 and 1, may be excited even below threshold. 
The $\overline KN$ system is therefore strongly coupled, via 
these resonances, to several channels, like $\Lambda\pi,\; 
\Sigma\pi,\; Y\eta,\; Y\pi\pi$ \textit{etc}. The different 
behavior of the two $K^0_L$ components implies that the 
interaction of such a beam with dense matter basically kills 
the $\overline{K^0}$ amplitude, which is almost completely 
absorbed. However, if the interaction of $K^0_L$ occurs on 
protons, final states are produced resulting from both $K^0$ 
and $\overline{K^0}$ interactions with different amplitudes: 
from their interference one might extract information on the 
relative sign of the $K^0N$ and $\overline{K^0}N$ potentials. 
While the $K^0p$ system is a mixture of $I=0$ and $I=1$ 
amplitudes, the $\overline{K^0}p$ is in pure $I=1$: the 
information the latter can provide is complementary to what 
can be obtained by the study of $K^+p$, but without any 
Coulomb interaction. Moreover, the final states which can be 
produced in a $\overline{K^0}p$ scattering are the charge 
conjugate of those reachable in a $K^-n$ interaction; 
therefore, they carry the same information but don't require 
the use of  deuterium as a target, which inevitably introduces 
three-body interactions between the target and the projectile 
that need to be properly taken into account.

It is also worthwhile to notice that on the basis of charge 
symmetry one can assume that $\sigma_{tot}(K^0p) = \sigma_{tot}
(K^+n)$ and $\sigma_{tot}(\overline{K^0}p) = \sigma_{tot}
(K^-n)$. These equalities were proved to be valid at least to 
the precision level of old bubble chamber experiments, and were 
often used to indirectly assess unmeasured cross 
sections~\cite{re:chargeSimJ}.

The existence of resonant states prevents the use of 
perturbative theories to describe the $\overline KN$ interaction 
close to threshold. To this purpose, non-perburbative chiral 
based coupled channel approaches are usually applied, adapting 
the models to all the available experimental observations, 
including, besides elastic and inelastic cross sections, also 
measurements of hadronic branching ratios close to threshold, 
resonances lineshapes, and inputs from Kaonic atom levels 
shifts due to strong interaction and their widths. Several 
models have been elaborated in the years to reproduce the 
$K^-N$ experimental data~\cite{re:modelliJ}; more new inputs 
would of course be welcome not only to improve the data 
description, but also to provide a more reliable prediction of 
the below-threshold behavior, that is relevant for the study 
of sub-threshold baryonic resonances and the possible existence 
of multinucleon-antiKaon aggregates, as will be discussed in 
Sec.~2.5.2.a.

\begin{enumerate}
\item \textbf{Low energy scattering parameterizations}

An old fashioned simple but useful way to describe the low 
energy interaction of particles is to parametrize the 
scattering cross sections in terms of $S$-wave {\it 
scattering lengths}~\cite{re:dalitzJ}. Assuming the reaction 
energy to be low enough to allow only the $S$-wave to be 
involved, and the ``zero-effective range" approximation 
to be appliable, the scattering length $A = a+ib$, that in 
general is a complex number, can be used to describe 
univoquely the phase-shift in each channel of given isospin
and strangeness through the relationship  $\cot\delta = 
1/kA$, where $\delta$ is the phase-shift and $k$ the 
projectile wave number. In a definite isospin-strangeness 
channel, the scattering cross section may then be expressed 
by the general formula:
\begin{equation}
	\sigma = \frac{a^2+b^2+b/k}{k^2a^2+(1+kb)^2}.
\end{equation}
The efforts of the first experiments measuring $K^-$ 
scattering was mainly to extract the real and imaginary part 
of the scattering lengths for the two isospin sources from 
the available cross sections~\cite{re:kMinusScatlenJ}. Due 
to the lack of data and the loose constraints provided, 
however, these assessments were far from being precise and 
several equally good solutions were often found, with large 
ambiguities which survived until recently, when precise 
measurements of Kaonic atom levels were performed and could 
be used as precise additional inputs.

The $S$-wave $K^0_Lp$ scattering cross sections may be 
expressed, in zero-range approssimation, through four 
parameters: the isospin I=0 and I=1 real scattering lengths 
$a_0$ and $a_1$ for the $S=+1$ channels, and the complex 
(absorptive) $\overline A = \overline a_1 + \overline b_1$ 
scattering length for the $S=-1$, $I=1$ 
channel~\cite{re:biswasJ}. By means of these parameters the 
low-energy cross sections have the following simple 
expressions:
\begin{description}
\item[total cross section] $$\sigma_{tot} = 2\pi\left[\frac{1}{2}
	\frac{a_0^2}{1+k^2a_0^2}+\frac{1}{2}\frac{a_1^2}{1+k^2a_1^2}
	+\frac{\overline a_1^2+\overline b_1^2
	+ \overline b_1/k}{k^2\overline a_1^2+(1+k\overline b_1)^2}
	\right];$$
\item[elastic cross section] $$\sigma(K^0_Lp\to K^0_Lp) =
	\pi\left|\frac{1}{2}\frac{a_0}{1-ika_0}+\frac{1}{2}\frac{a_1}{1
	-ika_1}+\frac{\overline a_1+i\overline b_1}{k^2\overline a_1^2
	+(1+k\overline b_1)^2}\right|^2;$$
\item[regeneration cross section] $$\sigma(K^0_Lp\to K^0_Sp) =
	\pi\left|\frac{1}{2}\frac{a_0}{1-ika_0}+\frac{1}{2}\frac{a_1}{1
	-ika_1}-\frac{\overline a_1+i\overline b_1}{k^2\overline a_1^2
	+(1+k\overline b_1)^2}\right|^2;$$
\item[one nucleon absorption cross section] $$\sigma(K^0_Lp\to Y\pi) =
	\frac{2\pi}{k}\frac{\overline b_1}{k^2(\overline a_1^2+b_1^2)
	+2k\overline b_1+1}.$$
\end{description}

In the following a short account of the existing measurements of the 
above cross sections at low momenta will be given.

\begin{enumerate}
\item \textbf{$K^0_Lp\to K^0_Sp$ regeneneration cross section}

The main purpose of the first measurements of the regeneration cross 
sections~\cite{re:regeJ} was the investigation of the features of the 
$Y^\ast_1$ resonances (in particular, the $\Sigma(1385)$), and the 
search for the possible existence of exotic $I=0$, $S=+1$ $Z^\ast$ 
states. The amplitude may be written by the sum of the $I=0$ and 
$I=1$ $K^0N$  terms, and the $I=1$ $\overline{K^0}N$ one: $T= 
\frac{1}{4}(Z_0+Z_1)-\frac{1}{2}Y_1$; the resulting cross section 
derives from the interference between the $S=-1$ and $S=1$ 
amplitudes. Regeneration cross sections were measured down to 
300~MeV/$c$~\cite{re:bigiJ}, and at the lowest momenta they amount 
to about 5~mb.  Fig.~\ref{fig:rege} reports the available 
experimental data with, superimposed, a few parameterizations 
deduced from different solutions for the $K^-n$ scattering length 
value (via the application of the charge symmetry assumption).  The 
differential cross sections exhibit moreover a marked backward 
peaked trend as a function of the $K^0_S$ emission angle in the 
reaction center of mass~\cite{re:choJ}.
\begin{figure}[ht!]
\centering
\includegraphics[height=6.5truecm,clip=]{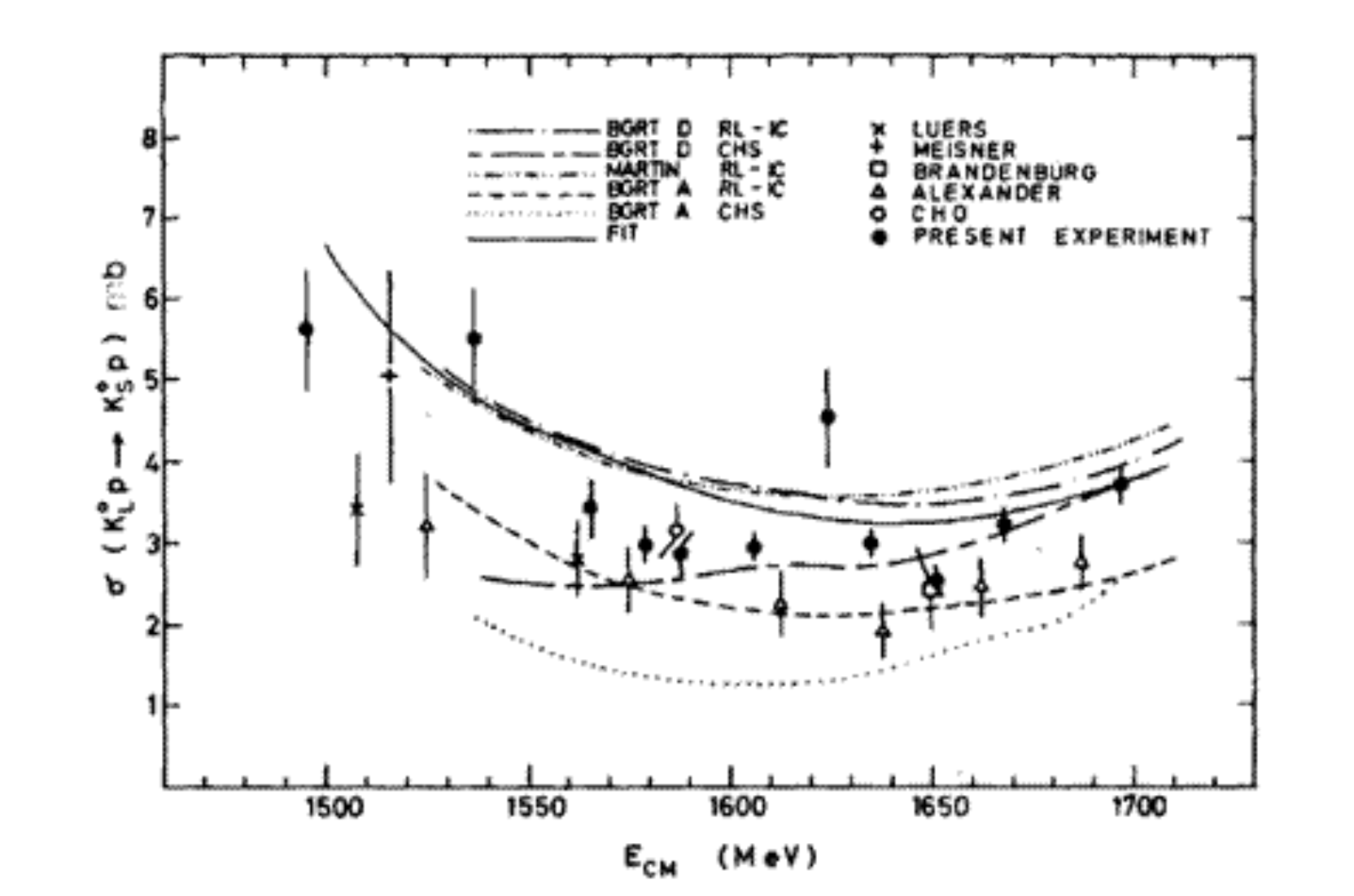}
\centerline{\parbox{0.80\textwidth}{
\caption{Low momentum $K^0_L$ regeneration cross section,
	from Ref.~\protect\cite{re:regeJ}.}
	\label{fig:rege} } }
\end{figure}

\item \textbf{Inelastic Cross Sections and Hyperon Production 
	Yields}

The relevant reactions for the $K^0_L$ induced production of 
baryonic resonances at low energies are $K^0_L p\to \Lambda\pi^+,\; 
\Sigma^0\pi^+$ and $K^0_L p\to \Lambda\pi^+\pi^0$. An assessment of 
the ratio of regeneration to elastic yields, $R = 
\frac{\sigma(K^0_Sp)}{\sigma(\Lambda\pi^+)+2\sigma(\Sigma^0\pi^+)}$, 
was used by early experiments~\cite{re:kadykJ,re:luersJ} to 
discriminate among the expected trends, as a function of $K^0_L$ 
momentum, from different sets of solutions for the $K^-p$ scattering 
length. As shown in Fig.~\ref{fig:Rratio} (left), none of the trends 
expected for $R$ on the basis of different solution sets could 
reproduce in a satisfactory way the observed yields.
\begin{figure*}[ht!]
\centering
\begin{tabular}{ccc}
\includegraphics[height=5truecm,clip=]{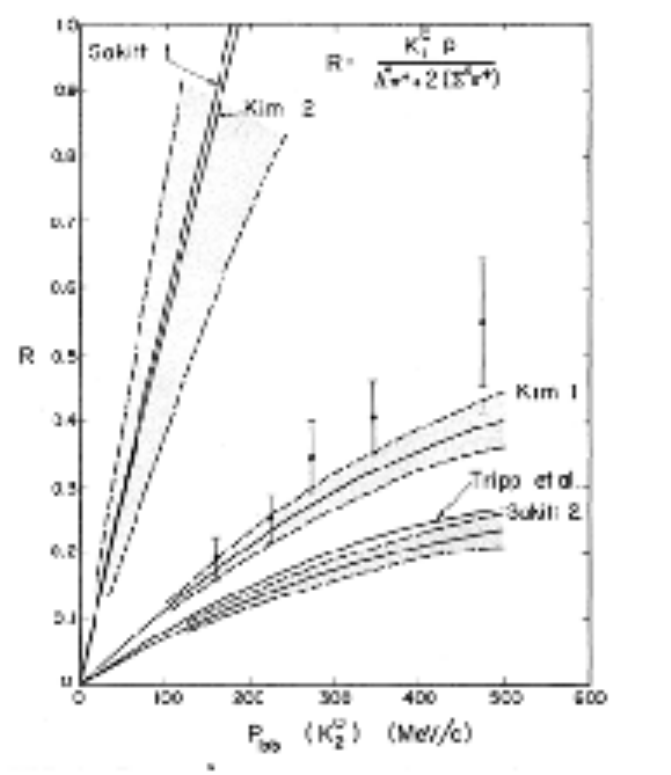} &
\includegraphics[width=6.2truecm, height=5truecm,clip=]{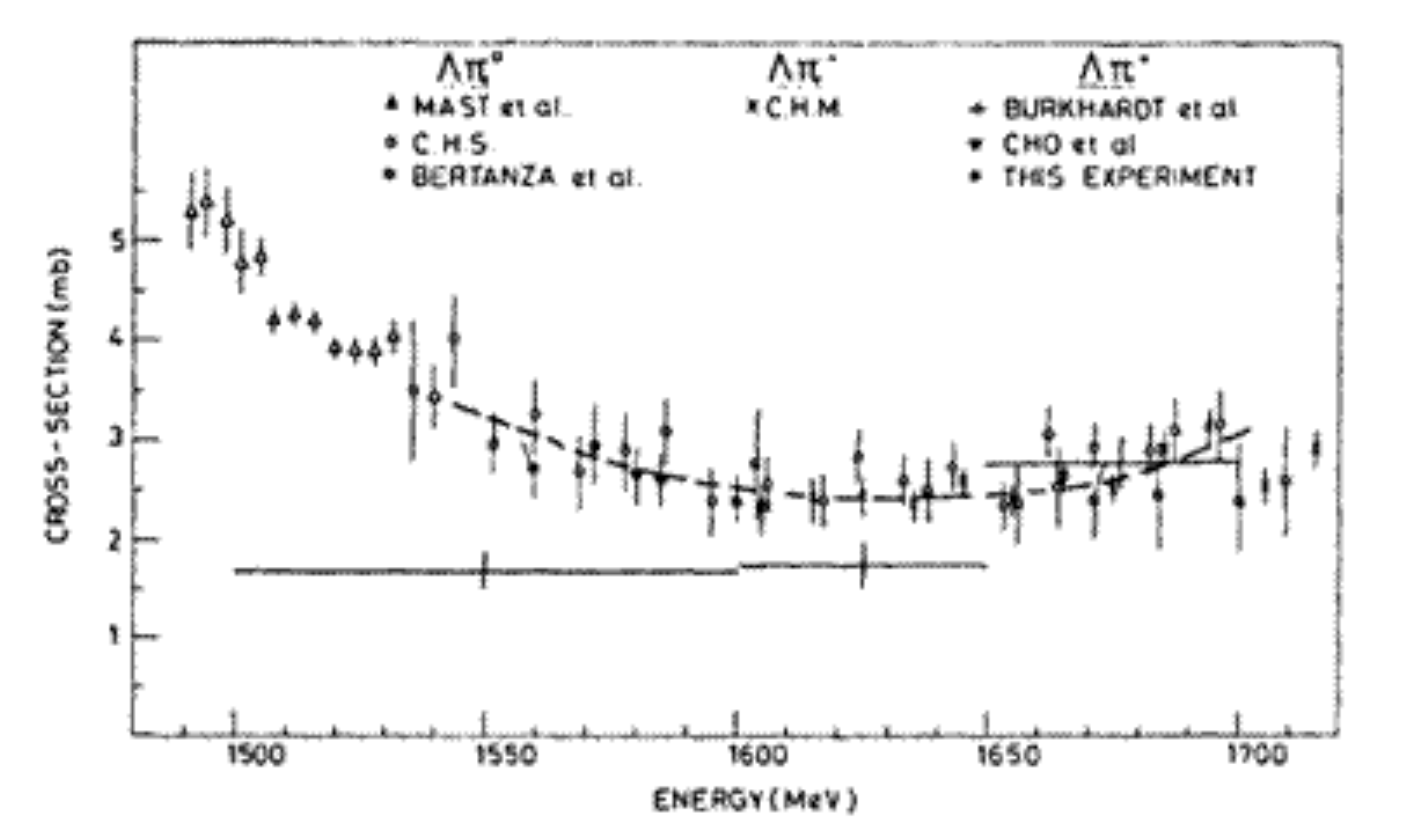} &
\includegraphics[width=6.2truecm, height=5truecm,clip=]{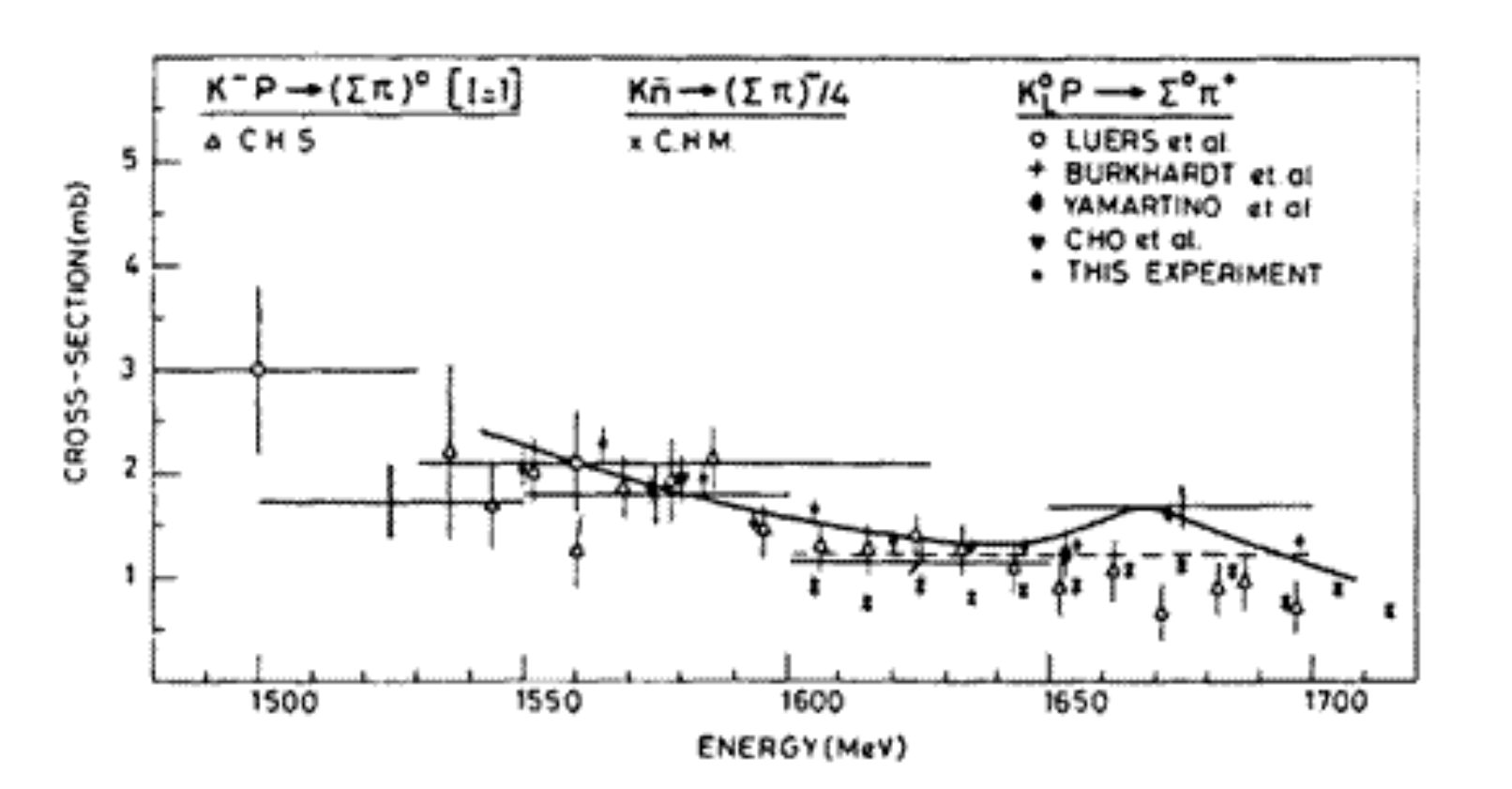} \\
\end{tabular}
\centerline{\parbox{0.80\textwidth}{
\caption{Left: ratio, as a function of $K^0_L$ momentum, of the 
	yields for regeneration to inelastic scattering. The 
	lines represent the expected trends on the basis of 
	different solutions chosen for the $K^-p$ scattering 
	length, from Ref.~\protect\cite{re:kadykJ}.  
	Center: $K^0_L p\to\Lambda\pi^+$ cross section; 
	Right: $K^0_L p\to\Sigma^0\pi^+$ cross section. When 
	existing, the data are compared to cross section 
	measurements in the charge conjugated channels. The 
	last two pictures are from 
	Ref.~\protect\cite{re:cameronJ}.} \label{fig:Rratio} } }
\end{figure*}

Figures~\ref{fig:Rratio} center and right show, respectively, 
the inelastic cross sections for the reactions $K^0_Lp\to\Lambda
\pi^+$ (about 5~mb at 300~MeV/$c$) and  $K^0_Lp\to\Sigma^0\pi^+$ 
($\sim 3$~mb).  The $K^0_Lp\to\Lambda\pi^+\pi^0$ channel is less 
relevant ($< 1$~mb)~\cite{re:cameronJ}, and is mainly dominated by 
the $\Sigma^0(1385)$ production.
\end{enumerate}
\end{enumerate}

\item \textbf{Low Energy $\overline KN$ Dynamics: Open Problems}

The $\overline KN$ interaction still presents some obscure aspects 
which only few more accurate data will be able to shed light on. 
The basic fact is the strong attractiveness of the interaction 
close to threshold and even below it, that manifests itself with 
the existence of a baryonic quasi-bound $\overline KN$ state, the 
$\Lambda(1405)$, embedded in the $\pi\Sigma$ continuum. This means 
that a strong coupled-channel dynamics between $\overline KN$ and 
$\Sigma\pi$ exists; to reproduce this behaviour a below-threshold 
extrapolation of the trend of the $\overline KN$ amplitude based 
on observed data must be exploited. However, the relatively 
scarce precision of the presently available experimental data 
close to threshold has severe drawbacks on the accuracy of the 
sub-threshold extrapolations. For this reason, new experimental 
inputs would certainly be welcome, especially if characterized by 
fixed quantum numbers (like the Coulomb free $I=1$ $K^0_Lp$ 
interaction).

Several chiral inspired coupled-channels models have been 
elaborated over the years~\cite{re:modelliJ}. The most recent 
ones~\cite{re:maximJ} are able to reproduce satisfactorily most 
of the existing data through global fits, especially since when 
the newest measurement of the Kaonic hydrogen $1S$ level 
performed by the SIDDARTHA Collaboration~\cite{re:siddarthaJ} was 
included in the 
data set.  We recall that the energy shift $\Delta E$ and width 
$\Gamma$ of the $1S$ Kaonic hydrogen line are directly related 
to the $a(K^-p)$ scattering length value through the Trueman-Deser 
formula (including second order isospin corrections): $\Delta E 
- i\Gamma/2 = -2\alpha^3\mu_T^2\; a(K^-p)\left[1+2\alpha\mu_T(1
-\log\alpha)a(K^-p)\right]$, where $\alpha$ is the strong coupling 
constant, and $\mu_T$ the reduced mass of the $K^-p$ system. The 
new measurement performed by the SIDDARTHA Collaboration fixes the 
inconsistencies emerging from the previous experiments on Kaonic 
hydrogen, and is 
fully compatible with all the existing scattering data. 
Unfortunately, the experiment was not sensitive enough to perform 
also a measurement of the $1S$ Kaonic deuterium level, that could 
allow the determination of the $K^-n$ scattering length; however, 
an upgrade was proposed to this purpose and is foreseen to run at 
DA$\Phi$NE in the near future.

The Kaonic hydrogen new measurement is very useful to provide much 
more stringent constraints for the determination of the scattering 
lengths in the two different isospin channels~\cite{re:doringJ}. 
Calculations have been performed also to assess the extent of the 
$K^-n$ (fixed $I=1$) scattering length~\cite{re:ikeda2J}, but the 
evaluation is still rather imprecise due to the large uncertainty 
of the experimental inputs (especially of the scattering data in 
the $\Lambda\pi$ channel). As shown in Ref.~\cite{re:alesJ} the 
$I=1$ $\overline KN$ interaction is expected to be weaker as 
compared to the $I=0$ source; therefore, data from Kaonic 
deuterium or from $K^0_Lp$ scattering would be useful in this 
respect.

\begin{enumerate}
\item \textbf{Subthreshold behavior: the $\Lambda(1405)$ case and 
	the case for possible nuclear-Kaonic aggegates} 

The measurement of the Kaonic hydrogen $\Delta E$ and $\Gamma$ 
provides a single experimental point to constrain the behavior 
of the below-threshold real and imaginary part of the $K^-p$ 
elastic scattering amplitude, as shown in 
Fig.~\ref{fig:amplitBelowThreshold} from Ref.~\cite{re:ikedaJ}. 
This result is just one typical snapshot of the outcomes of 
several equivalent high-quality below-threshold extrapolations:
$Re(a(K^-p)) = -0.65\pm 0.10$~fm, and $\Im m(a(K^-p)) = 0.81\pm 
0.15$~fm. In spite of the uncertainty of the prediction, 
represented by the grey band around the best fit result, 
basically all models agree on the existence of the 
$\Lambda(1405)$ resonance, to be interpreted as a $I=0$ 
$\overline KN$ system bound by 27~MeV. This resonance is 
dinamically generated by the interplay of two poles in the 
second Riemann sheet, one at higher mass ($1424-i26$~MeV) 
coupled to the $\overline KN$ channel, and the second at a 
lower mass value ($1381-i81$~MeV) dominated by the $\Sigma\pi$ 
coupling~\cite{re:ikeda2J}.
\begin{figure}[ht!]
\centering
\includegraphics[height=5.5truecm,clip=]{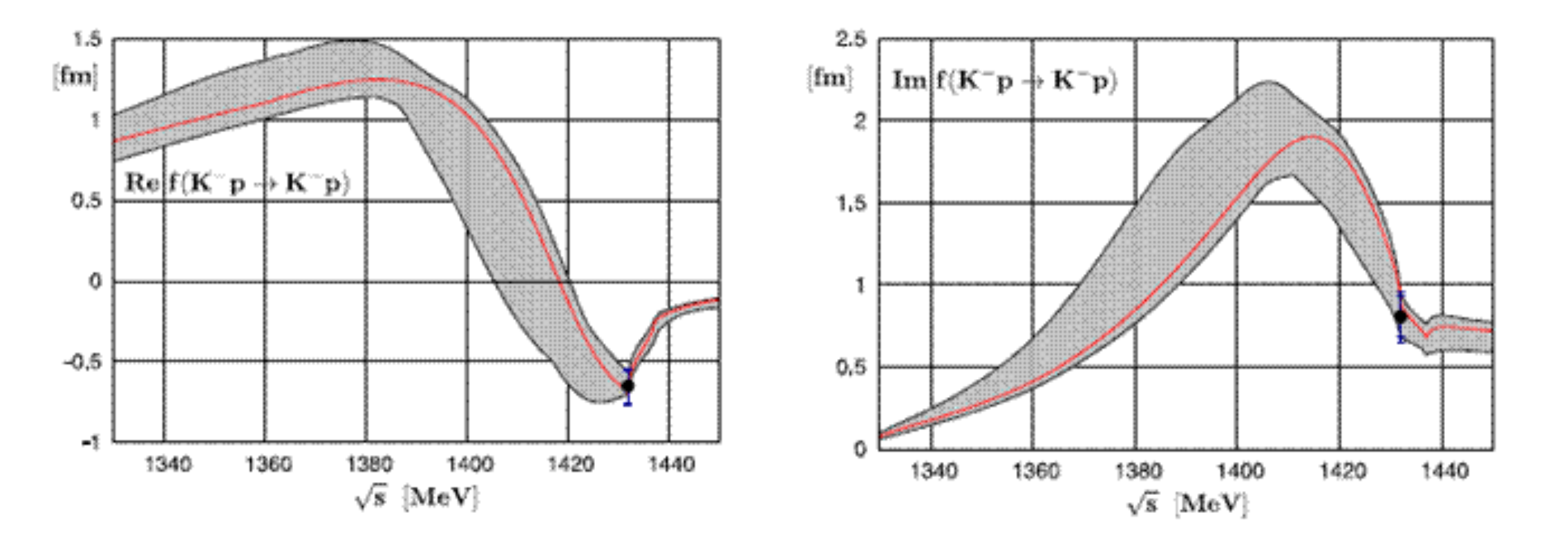}
\centerline{\parbox{0.80\textwidth}{
\caption{Solutions for the real and imaginary part of the $K^-p$ 
	elastic scattering amplitude, based on the chiral inspired 
	model of Ref.~\protect\cite{re:ikeda2J}. The best fit to 
	the experimental data is represented by the continuous 
	line, while the grey area shows the uncertainty of the 
	model, determined by the precision of the available 
	experimental data used for the fit. The two data points 
	correspond to the assessment of the real and imaginary 
	parts of the $K^-p$ scattering length derived from the 
	experimental measurement of the Kaonic hydrogen $1S$ 
	level by the SIDDARTHA Collaboration.} 
	\label{fig:amplitBelowThreshold} } }
\end{figure}

From the experimental point of view the observations of the 
$\Lambda(1405)$ were often hindered by the existence, in the same 
mass region, of the $\Sigma(1385)$ baryon, which shares with the 
$\Lambda(1405)$ the charged $\Sigma\pi$ decay mode. The decays in 
charged $\Sigma\pi$ pairs were studied by early experiments on 
deuterium targets~\cite{re:oldLambda1405J}. The first observations 
were confirmed, with higher statistics, by second generation 
experiments which were also able to measure the $\Sigma^0\pi^0$ 
decay channel, which is mostly important as it is only allowed 
for the decay of the $\Lambda(1405)$, but is prevented to 
$\Sigma(1385)$. The latter, on the other hand, may only decay to 
$\Lambda\pi^0$ (channel excluded for $\Lambda(1405)$). The most 
complete data-set collected so far defining the $\Lambda(1405)$ 
lineshape comes from the CLAS experiment, based on photo- and 
electroproduction of the $\Sigma\pi$ final states~\cite{re:clasJ}. 
The best fit of the data suggests the lineshapes of the three 
charge combinations of the $\Sigma\pi$ invariant mass systems to 
be reproduced by introducing a dominant $I=0$ contribution (at 
$m=1338\pm 10$~MeV/$c^2$, with $\Gamma = 85\pm 10$~MeV), plus two 
$I=1$ amplitudes, one of which is most probably related to the 
$\Sigma(1385)$ broad resonance ($m=1394\pm 40$~MeV/$c^2$, $\Gamma 
= 149\pm 40$~MeV), while the second, narrower and at higher mass 
($m=1412\pm 10$~MeV/$c^2$, $\Gamma = 52\pm 10$~MeV), has a still 
uncertain nature. Its necessity, to provide a good description of 
the data, has been remarked by theoretical models~\cite{re:rocaJ}; 
for its interpretation, the possibility that it might be due to a 
new, exotic pentaquark baryonic state~\cite{re:zouJ} is still open.

Related to the existence of the $\Lambda(1405)$ is the case of the 
so-called (anti)Kaon-nuclear clusters. Following the hypothesis 
suggested by Akaishi and Yamazaki in 2002~\cite{re:ayJ}, the 
$\Lambda(1405)$ could be the founding block based on which more 
complex aggregates, composed by an anti-Kaon deeply bound to two or 
more nucleons, could exist. Even though the existence of such 
states is not ruled out in most of the chiral inspired models 
elaborated so far~\cite{re:modelsDbksJ}, very few of them agree on 
their observability as narrow states mainly decaying via the non 
mesonic channel ($\Sigma\pi$ and $\Lambda\pi$ being prevented by 
their strong binding and by isospin conservation). Most of the 
models, in fact, foresee for the $\overline KN$ potential rather 
shallow wells, so a mild binding. In the initial formulation, on 
the contrary, these states are expected to be narrow, bound by 
more than 100~MeV and forming very compact systems, with a density 
more than three times as large as compared to ordinary nuclear 
matter. The medium in which their formation could more likely 
occur is also a controversial point: while according to the 
starting hypothesis the observation in light targets could be 
easier, other calculations~\cite{re:jiriJ} indicate that heavy 
targets should be preferred. If this were the case, however, 
probably large Final State Interaction effects would spoil 
completely their observability as narrow states.

From the experimental point of view, the situation is still rather 
confused and a few observations claimed so far~\cite{re:dbksEXPJ} 
still need a sound confirmation.  For the latest findings of this 
search using a $^3$He target (E15 experiment running at J-PARC) 
the reader may refer to Ref.~\cite{re:ohnishiJ}.

The search for such states has been performed so far only relative 
to the $K^-NN(N)$ systems; no measurement were ever attempted with 
neutral Kaon beams. Therefore, provided a $^3$He or $^4$He (or even 
heavier) target could be exploited, the search for such states 
could represent a completely new field of investigation to be 
pursued with $K^0_L$ as projectiles (again, free from Coulomb 
interactions and related to the binding properties of $I=1$
$\overline KN$ systems only).
\end{enumerate}

\newpage
\item \textbf{Possible Measurements with $K^0_L$ Beams and 
	Experimental Reach}

With a liquid hydrogen/deuterium target, the following reactions 
could be measured at low momenta, to improve the present knowledge 
on the scattering cross sections:
\begin{description}
\item[elastic scattering] $K^0_L p\rightarrow K^0_L p$, $K^0_Ld
	\rightarrow K^0_L d$ (coherent), $K^0_L d\rightarrow 
	K^0_Lnp$ (quasi-elastic scattering on $n$);
\item[inelastic scattering on protons with $Y$ formation] 
	$K^0_Lp\to\Lambda\pi^+,\; \Sigma^0\pi^+,\; \Sigma^+\pi^0,\;
	\Lambda\pi^+\pi^0$;
\item[inelastic scattering on deuterons for below-threshold $Y$ 
	resonances production] $K^0_Ld\to\Lambda(1405)N$;
\item[charge exchange reactions] $K^0_Lp\to K^+n$;
\item[regeneration reaction] $K^0_Lp\to K^0_Sp$.
\end{description}

One or two measurements of low momentum cross sections below 
350~MeV/$c$ at the 10\% precision level would be highly desireable 
to complement the experimental data set on which close-to-threshold 
$\overline KN$ interaction studies are based. Differential 
information, for instance as a function of the emission angle, 
could be fruitfully explored as well.

A few experimental possibly critical drawbacks have however to be 
taken into account. Among them:
\begin{enumerate}
\item The $K^0_L$ beam intensity at low momentum. As shown by 
	experiments exploiting the $K^0_L$ production by means of 
	photoproduction on a Be target, $K^0_L$'s are produced with 
	a continuum momentum spectrum~\cite{re:KLphotoJ,re:brandeburgJ}. 
	The low momentum portion is roughly some $10^{-3}$ of the 
	total integrated $K^0_L$ momentum spectrum, for a maximum 
	photon energy of around 10~GeV~\cite{re:brandeburgJ}, close 
	to that foreseen for the 12~GeV CEBAF machine. This could 
	still allow to have a fair number of low momentum $K^0_L$ 
	(some Hz), provided they can be effectively discriminated 
	from neutrons even at these low energies;
\item The capability of detecting low momentum particles in the 
	final state. The momentum resolution is not a crucial 
	problem in a few body reaction, but the curling of low 
	momentum particles in a high intensity magnetic field could 
	prevent them from reaching the position sensitive detectors 
	and therefore impair the observation of the mentioned 
	reactions. A careful study on how to increase the apparatus 
	acceptance to low momentum particles would most likely be 
	required in the planning of such measurements.
\end{enumerate}

A tentative yield evaluation, with some optimistic but reasonable 
detection efficiencies (assuming that all the emitted particles 
enter the apparatus acceptance), indicates that for an elastic 
cross section measurement with a precision at the level of 10\% 
some hours of data taking could be enough, while a few days at most 
would be required for the less frequent inelastic channels.

\begin{enumerate}
\item \textbf{Hypernuclei formation studies} 

A completely new research field, that could be explored with a 
$K^0_L$ beam and for which no experimental result exist so far, 
is the production of hypernuclei in $\overline{K^0}$ induced 
reactions. The spectroscopy of the formation pion, in reactions 
on $^A$Z nuclei like $^A\mathrm{Z}(\overline{K^0}, \pi^+)^A_\Lambda
(\mathrm{Z}-1)$ or $^A\mathrm{Z}(\overline{K^0}, \pi^0)^A_\Lambda
\mathrm{Z}$, requires a very high momentum resolution (on the 
order of a few per mil), which, however, is probably out of scope 
for an apparatus conceived for hadron spectroscopy like GlueX. 
This information might be of unprecedented value for the 
investigation of the so-called Charge Symmetry Breaking effect, 
which consists in a sizeable difference between the binding 
energies of the ground states of mirror hypernuclei. So far, 
the effect has been observed in light mirror hypernuclei pairs 
(like $^4_\Lambda$He vs $^3_\Lambda$H), and is supposed to be 
due to a strong $\Lambda\Sigma$ mixing~\cite{re:galJ}. While in 
this case the binding energies differ of about 250~KeV, for 
heavier ($P$-shell) hypernuclei the difference is expected to 
decrease. Studies of mirror light hypernuclei production would 
be welcome to investigate this interesting effect in deeper 
detail.
\end{enumerate}

\item \textbf{Conclusions}

With a beam of low momentum $K^0_L$ new tools to improve the 
knowledge of the $\overline KN$ interaction, never exploited so 
far, could be available. It is important to recall that with a 
$K^0_L$ beam the isospin $I=1$ source of the $\overline KN$ 
amplitude may be selected: its features are largely unknown as, 
with charged Kaons, this information may only be pursued using 
deuterium as a target, which involves a complicated treatment 
due to the inherent few-body interaction. Moreover, the $K^0_Lp$ 
interaction is free from any Coulomb-related effect.

Data on $K^0_L p$ scattering might improve the present knowledge 
of $I=1$ scattering length, providing complementary information 
to the already planned measurements of Kaonic deuterium.  An 
extension of the charged Kaon scattering database to neutral 
Kaon induced reactions would  be important to improve the 
precision of $\overline KN$ models especially regarding their
below-threshold extrapolations, that are crucial to improve the 
understanding of some still critical subjects, like the nature 
of the $\Lambda(1405)$ as a true baryonic resonance.

With targets heavier than deuterium, the study of more complex 
systems like Kaon-nuclear bound states or hypernuclei produced 
in $K^0_L$ induced reactions could potentially be feasible, and 
thorougly yet unexplored research topics could be opened.
\end{enumerate}


\newpage
\subsection{$K^0_Lp$ Scattering to Two-Body Final States}
\addtocontents{toc}{\hspace{2cm}{\sl D.M.~Manley}\par}
\setcounter{figure}{0}
\setcounter{equation}{0}
\halign{#\hfil&\quad#\hfil\cr
\large{D.~Mark Manley}\cr
\textit{Department of Physics}\cr
\textit{Kent State University}\cr
\textit{Kent, OH 44242 U.S.A.}\cr}

\begin{abstract}
Our main interest in creating a high-quality secondary $K^0_L$ beam 
is to investigate hyperon spectroscopy through both formation and 
production processes.  Here we review what can be learned by 
studying hyperon formation processes using $K^0_L p$ scattering 
going to two-body final states.
\end{abstract}

\begin{enumerate}
\item \textbf{Introduction and Formalism}

The mean lifetime of the $K^-$ is 12.38~ns ($c\tau = 3.7$~m) whereas 
the mean lifetime of the $K^0_L$ is 51.16~ns ($c\tau = 
15.3$~m)~\cite{rppW}.  For this reason, it is much easier to perform 
measurements of $K^0_Lp$ scattering at low beam energies compared 
with $K^-p$ scattering. Here, we summarize some of the physics 
issues involved with such processes. The differential cross section 
and polarization for $K^0_Lp$ scattering are given by
\begin{equation}
	\frac{d\sigma}{d\Omega} = \lambdabar^2 (|f|^2 + |g|^2),
\end{equation}
\begin{equation}
	P \frac{d\sigma}{d\Omega} = 2\lambdabar^2 {\rm Im}(fg^\ast),
\end{equation}
where $\lambdabar = \hbar/k$, with $k$ the magnitude of c.m.\ momentum 
for the incoming meson.  Here $f = f(W,\theta)$ and $g = g(W,\theta)$ 
are the usual spin-nonflip and spin-flip amplitudes at c.m.\ energy 
$W$ and meson c.m.\ scattering angle $\theta$.  In terms of partial 
waves, $f$ and $g$ can be expanded as
\begin{equation}
	f(W,\theta) = \sum_{l=0}^\infty [(l+1)T_{l+} 
	+ lT_{l-}]P_l(\cos\theta),
\end{equation}
\begin{equation}
	g(W,\theta) = \sum_{l=1}^\infty [T_{l+} - T_{l-}]P_l^1
	(\cos\theta),
\end{equation}
where $l$ is the initial orbital angular momentum, $P_l(\cos\theta)$ 
is a
Legendre polynomial, and $P_l^1(\cos\theta) = \sin\theta \times
dP_l(\cos\theta)/d(\cos\theta)$ is an associated Legendre function.  
The total angular momentum for the amplitude $T_{l+}$ is $J=l+
\frac{1}{2}$, while that for the amplitude $T_{l-}$ is $J=l-
\frac{1}{2}$.  For hadronic scattering reactions, we may ignore 
small CP-violating terms and write
\begin{equation}
	K_L^0 = \frac{1}{\sqrt{2}} (K^0 - \overline{K^0}),
\end{equation}
\begin{equation}
	K_S^0 = \frac{1}{\sqrt{2}} (K^0 + \overline{K^0}).
\end{equation}

We may generally have both $I=0$ and $I=1$ amplitudes for $KN$ and
$\overline{K}N$ scattering, so that the amplitudes $T_{l\pm}$ can 
be expanded in terms of isospin amplitudes as
\begin{equation}
	T_{l\pm} = C_0 T^0_{l\pm} + C_1 T^1_{l\pm},
\end{equation}
where $T_{l\pm}^I$ are partial-wave amplitudes with isospin $I$ 
and total angular momentum $J = l \pm \frac{1}{2}$, with $C_I$ 
the appropriate isospin Clebsch-Gordon coefficients.

\item \textbf{$KN$ and $\overline{K}N$ Final States}

The amplitudes for reactions leading to $KN$ and $\overline{K}N$ 
final states are
\begin{eqnarray}
	T(K^-p \to K^-p) &=& \frac{1}{2}T^1({\overline K}N \to 
	{\overline K}N) + \frac{1}{2}T^0({\overline K}N \to 
	{\overline K}N), \\
	T(K^-p \to \overline{K^0} n) &=& \frac{1}{2}T^1({\overline 
	K}N\to {\overline K}N) - \frac{1}{2}T^0({\overline K}N \to 
	{\overline K}N), \\
	T(K^+p \to K^+p) &=& T^1(KN \to KN), \\
	T(K^+ n \to K^+ n) &=& \frac{1}{2}T^1(KN \to KN) + 
	\frac{1}{2}T^0(KN \to KN),
\end{eqnarray}
\begin{equation}
	T(K_L^0 p \to K_S^0 p) = \frac{1}{2} \left (\frac{1}{2}T^1
	(KN\to KN) + \frac{1}{2}T^0(KN \to KN) \right ) - \frac{1}{2}
	T^1({\overline K}N \to {\overline K}N),
\end{equation}
\begin{equation}
	T(K_L^0 p \to K_L^0 p) = \frac{1}{2} \left (\frac{1}{2}T^1(KN
	\to KN) + \frac{1}{2}T^0(KN \to KN) \right ) + \frac{1}{2}
	T^1({\overline K}N\to {\overline K}N),
\end{equation}
\begin{equation}
	T(K_L^0 p \to K^+ n) = \frac{1}{\sqrt{2}} \left (\frac{1}{2}
	T^1(KN \to KN) - \frac{1}{2}T^0(KN\to KN) \right ) - 
	\frac{1}{2}T^1({\overline K}N\to {\overline K}N).
\end{equation}
\begin{figure}[ht!]
\begin{center}
\includegraphics[angle=0, width=0.45\textwidth]{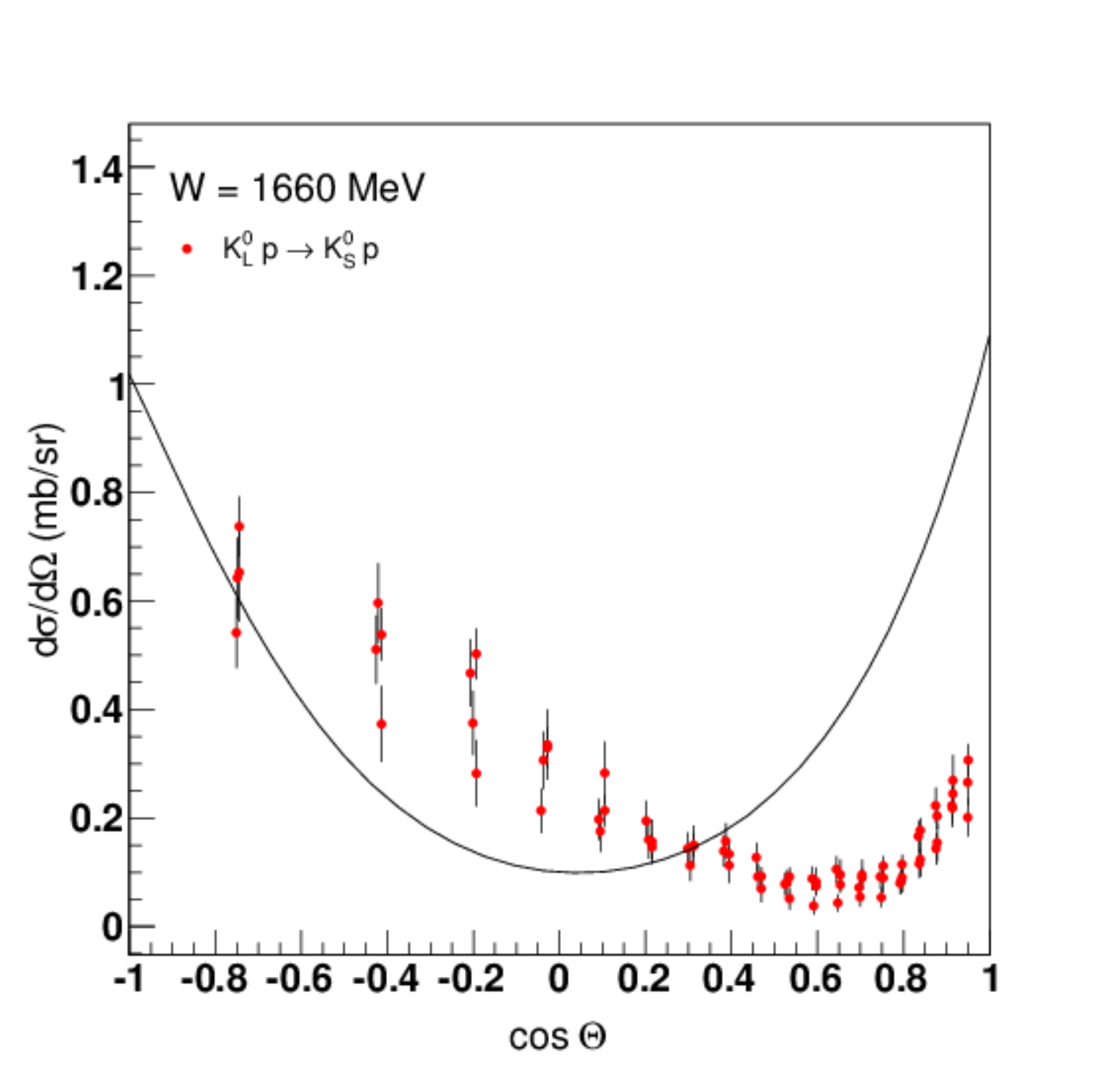}
\includegraphics[angle=0, width=0.45\textwidth]{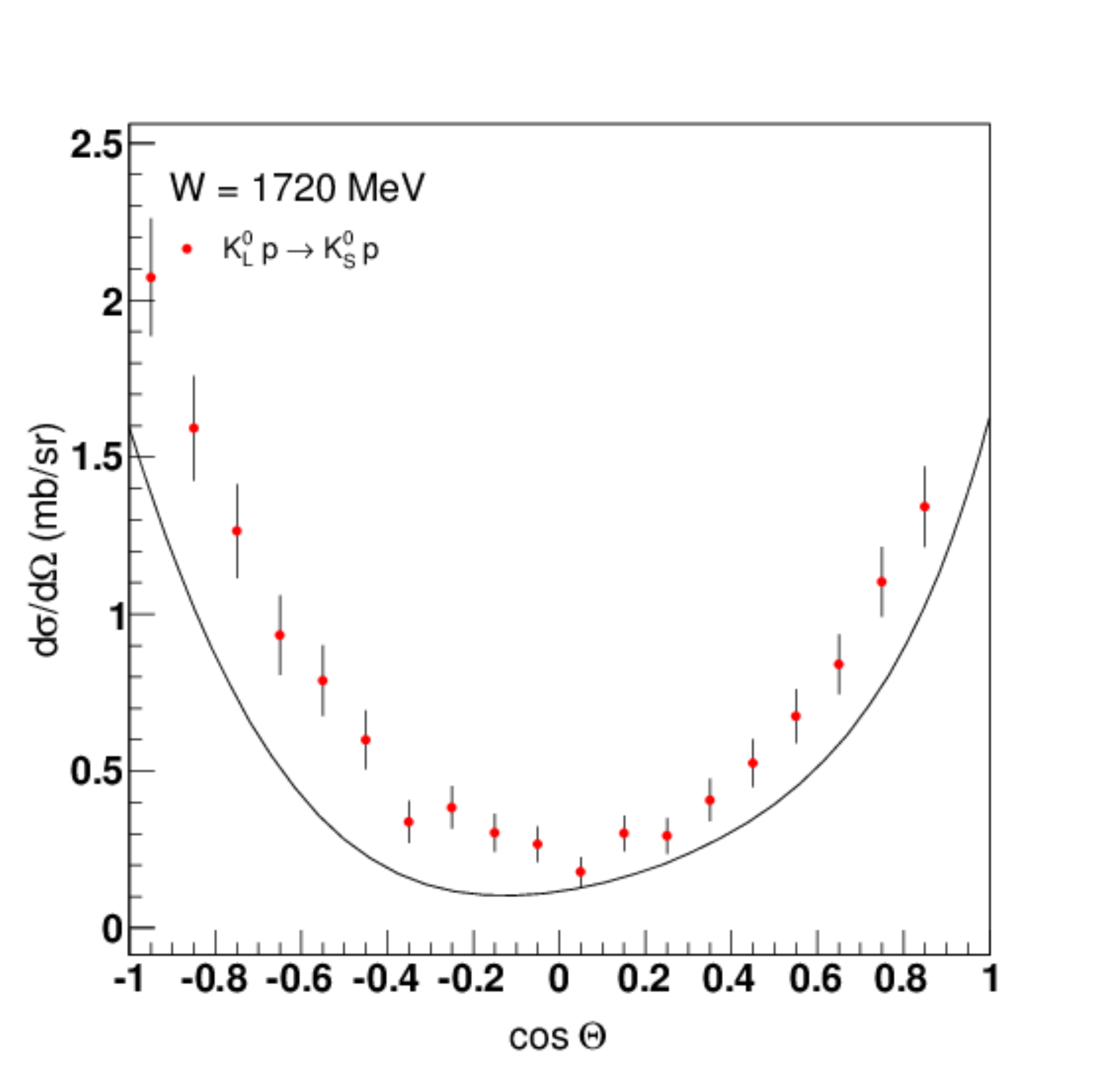}
\includegraphics[angle=0, width=0.45\textwidth]{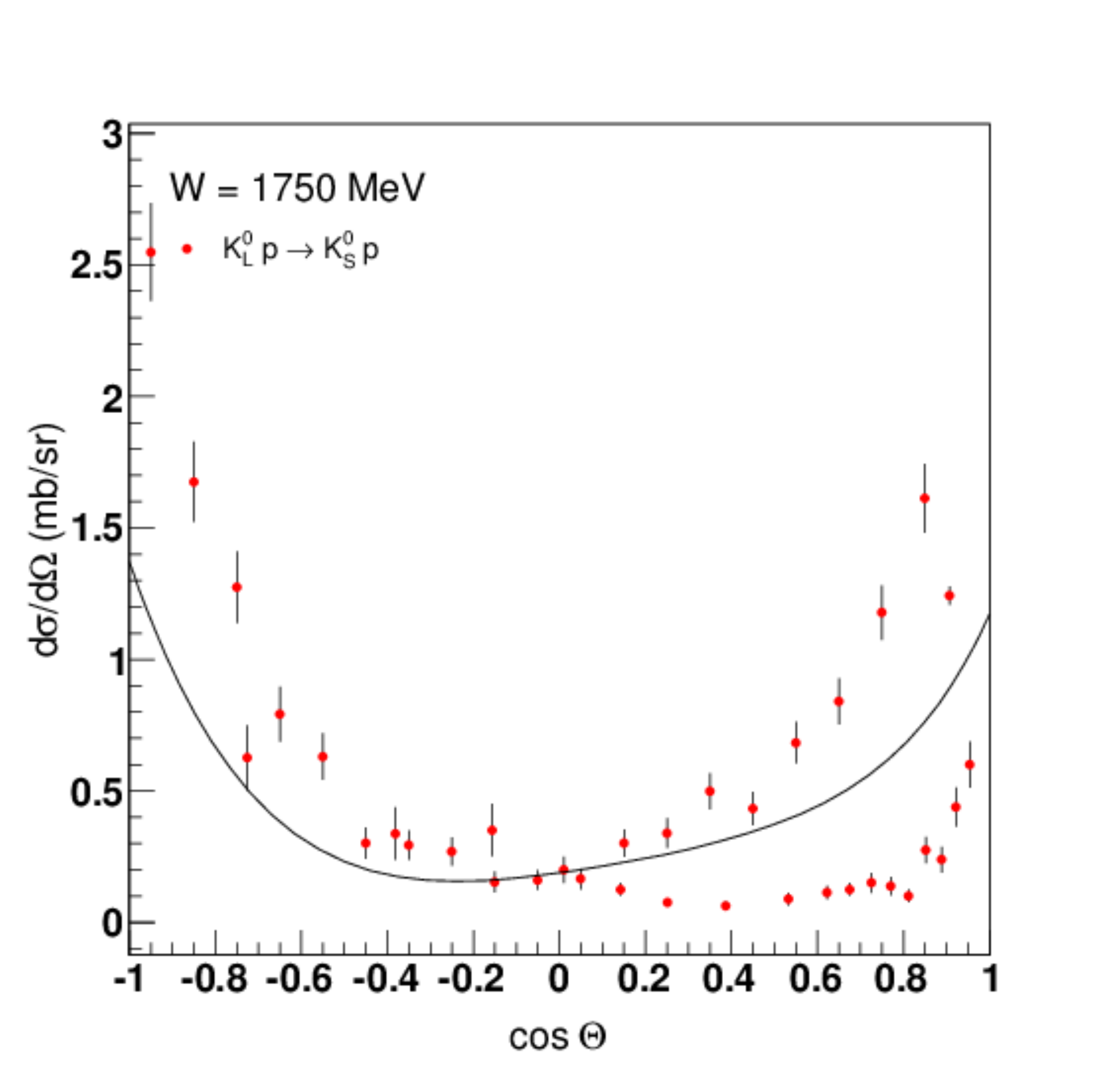}
\includegraphics[angle=0, width=0.45\textwidth]{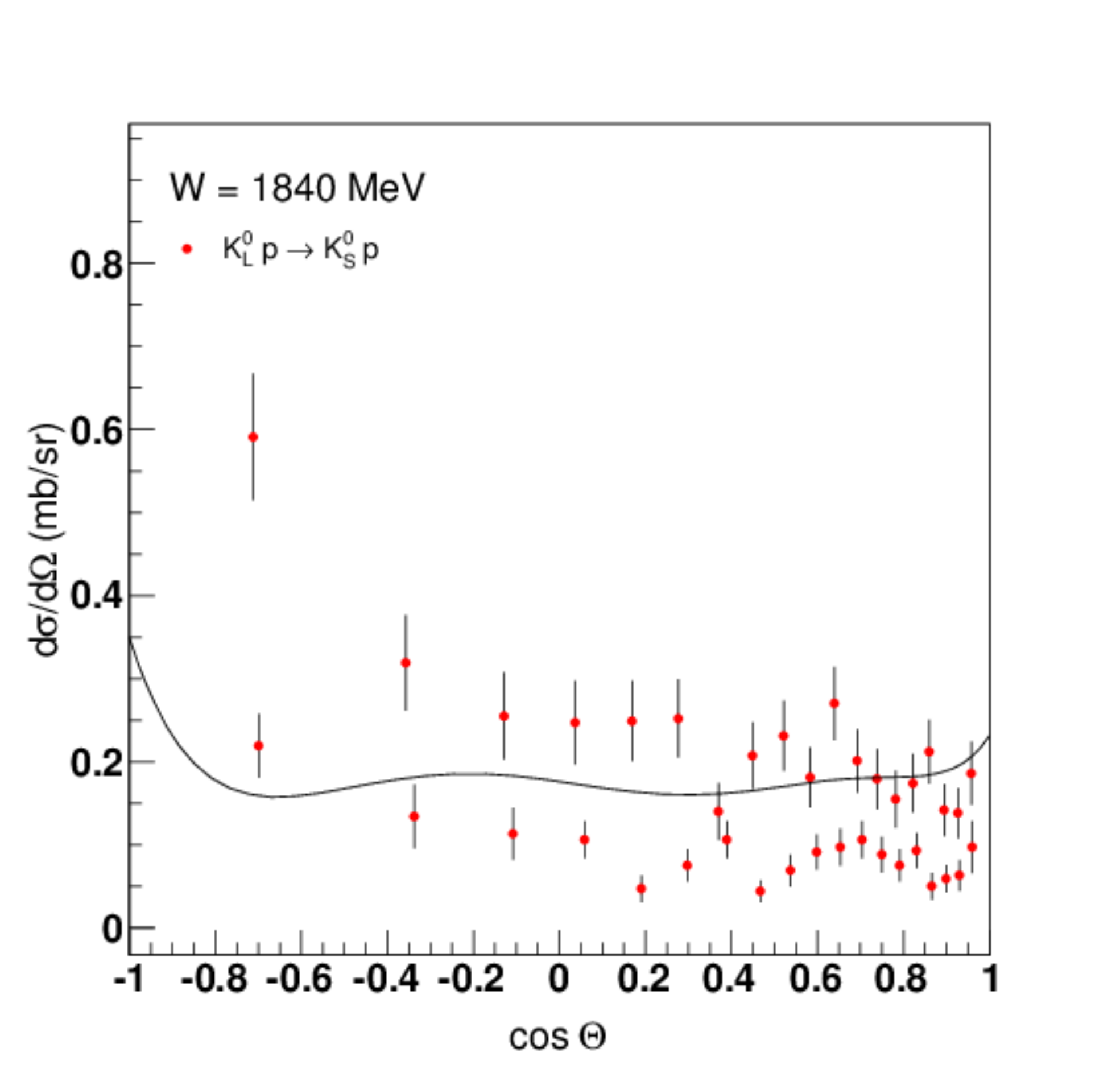}
\end{center}
\centerline{\parbox{0.80\textwidth}{
 \caption[] {\protect\small Selected differential cross section data
        for $K^0_Lp\to K^0_Sp$ at 1660~MeV, 1720~MeV, 1750~MeV, and
        1840~MeV.  The curves are predictions using amplitudes from
        our previous partial-wave analysis of $\overline{K}N\to
        \overline{K}N$ data~\protect\cite{zhang2013aW,zhang2013bW},
        combined with $KN\to KN$ amplitudes from the SAID
        website~\protect\cite{SAID-websiteW}.} \label{fig:KLp_KSp} } }
\end{figure}

No differential cross section data are available for $K_L^0p\to 
K_L^0p$ below $W \sim 2948$~MeV.  A fair amount of data are available 
for the reaction, $K^+n\to K^0p$, measured on a deuterium target.  
Figure~\ref{fig:KLp_KSp} shows a sample of available differential 
cross section data for $K_L^0p\to K_S^0p$ compared with predictions 
determined from our previous partial-wave analysis of $\overline{K}N
\to\overline{K}N$ data~\cite{zhang2013aW,zhang2013bW}, combined with 
$KN\to KN$ amplitudes from the SAID website~\cite{SAID-websiteW}.  
The predictions at lower and higher energies tend to agree less well 
with the data.

\item \textbf{$\pi\Lambda$ Final States}

The amplitudes for reactions leading to $\pi\Lambda$ final states 
are
\begin{eqnarray}
	T(K^-p \to \pi^0 \Lambda) &=& \frac{1}{\sqrt{2}}T^1
	({\overline K}N\to\pi\Lambda), \\
	T(K^0_L p \to \pi^+ \Lambda) &=& -\frac{1}{\sqrt{2}}T^1
	({\overline K}N \to \pi\Lambda).
\end{eqnarray}
The $K^-p\to\pi^0\Lambda$ and $K^0_Lp\to\pi^+\Lambda$ amplitudes 
imply that observables for these reactions measured at the same 
energy should be the same except for small differences due to the 
isospin-violating mass differences in the hadrons. No differential 
cross section data for $K^-p\to\pi^0\Lambda$ are available at c.m.\ 
energies $W < 1540$~MeV, although data for $K^0_Lp\to\pi^+\Lambda$ 
are available at such energies.  At 1540~MeV and higher energies,
differential cross section and polarization data for the two 
reactions are in fair agreement, as shown in 
Figs.~\ref{fig:KLp_piLambda} and \ref{fig:KLp_piLambda_P}.
\begin{figure}
\begin{center}
\includegraphics[angle=0, width=0.45\textwidth]{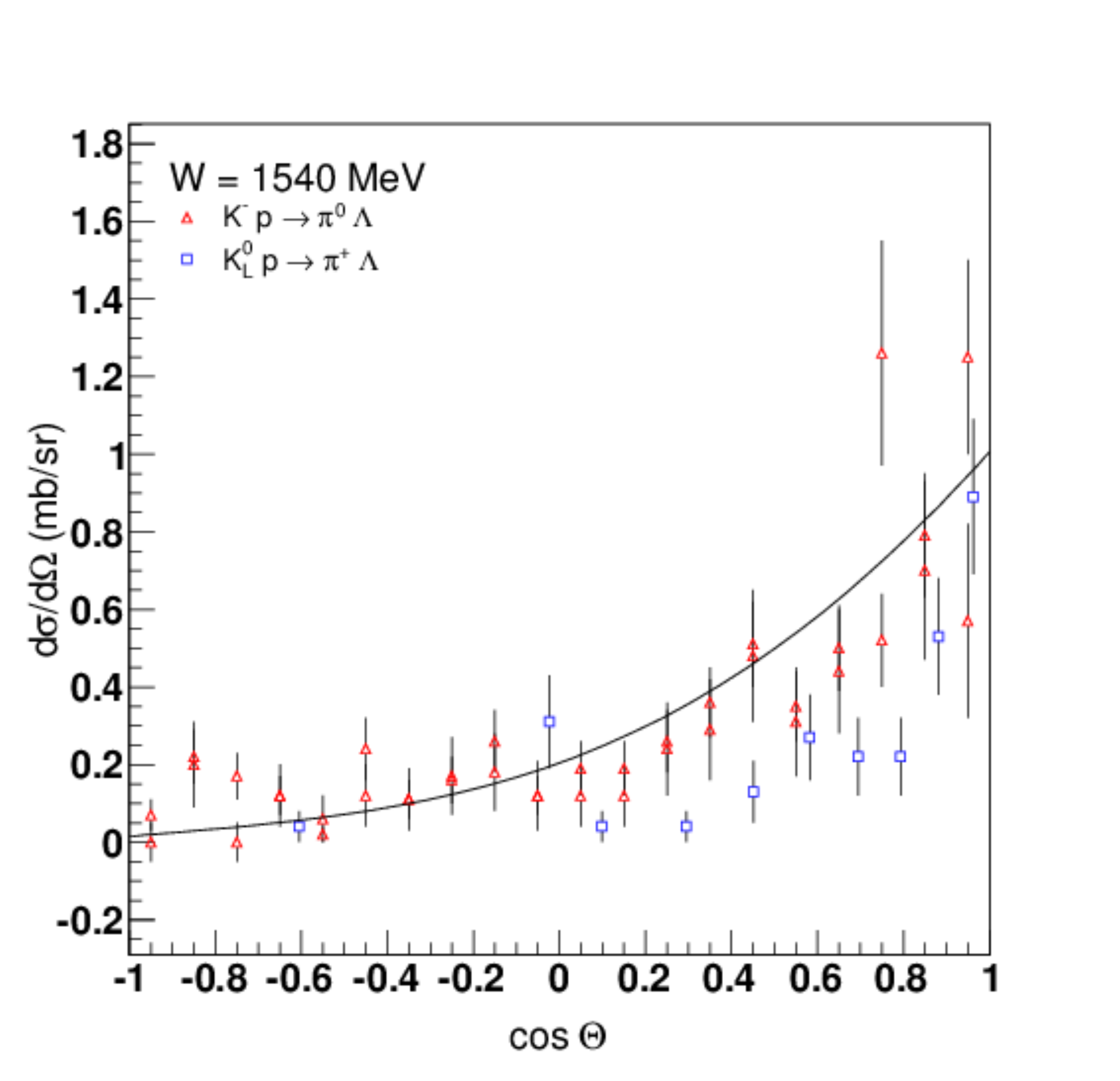}
\includegraphics[angle=0, width=0.45\textwidth]{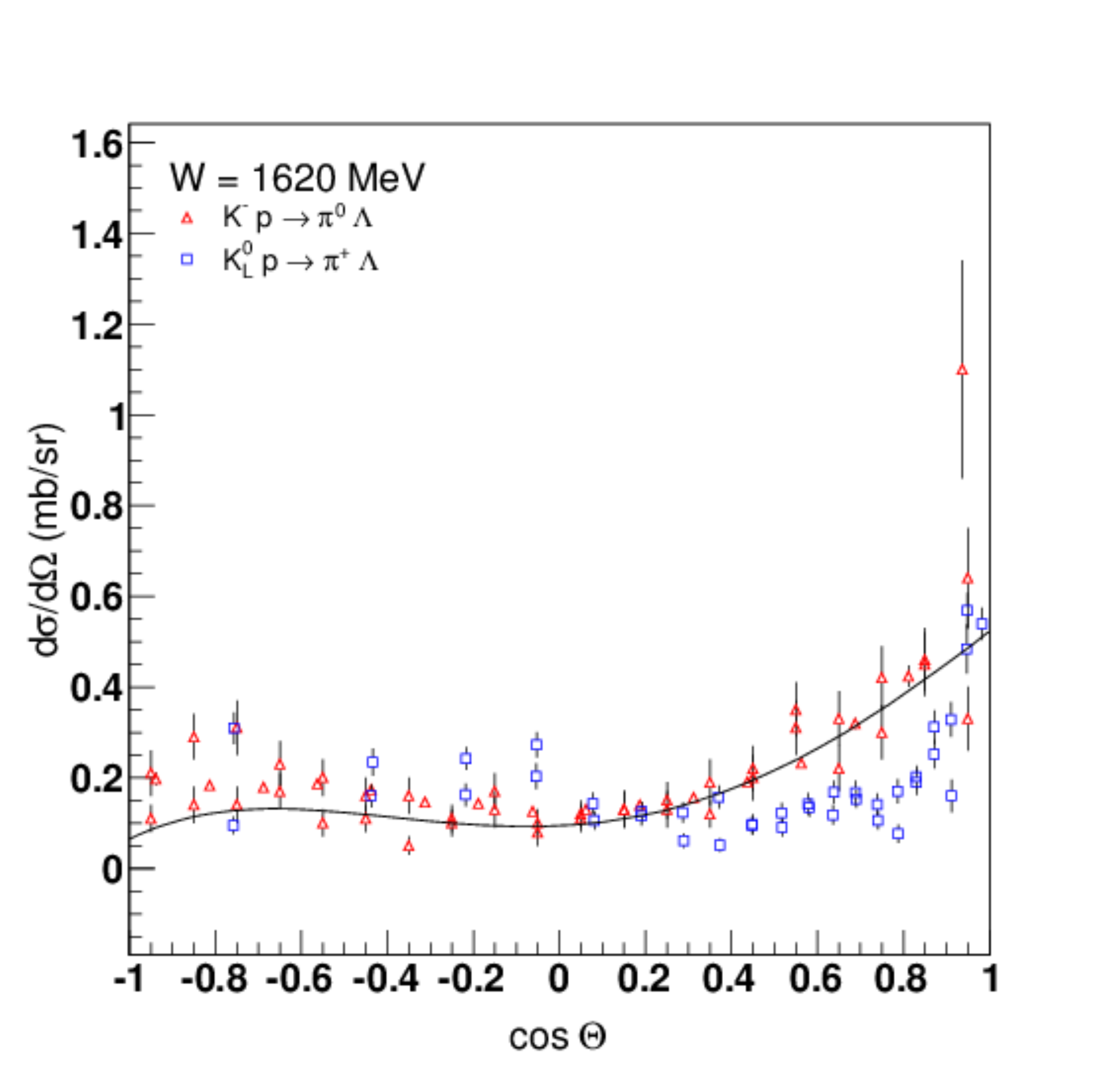} \\
\includegraphics[angle=0, width=0.45\textwidth]{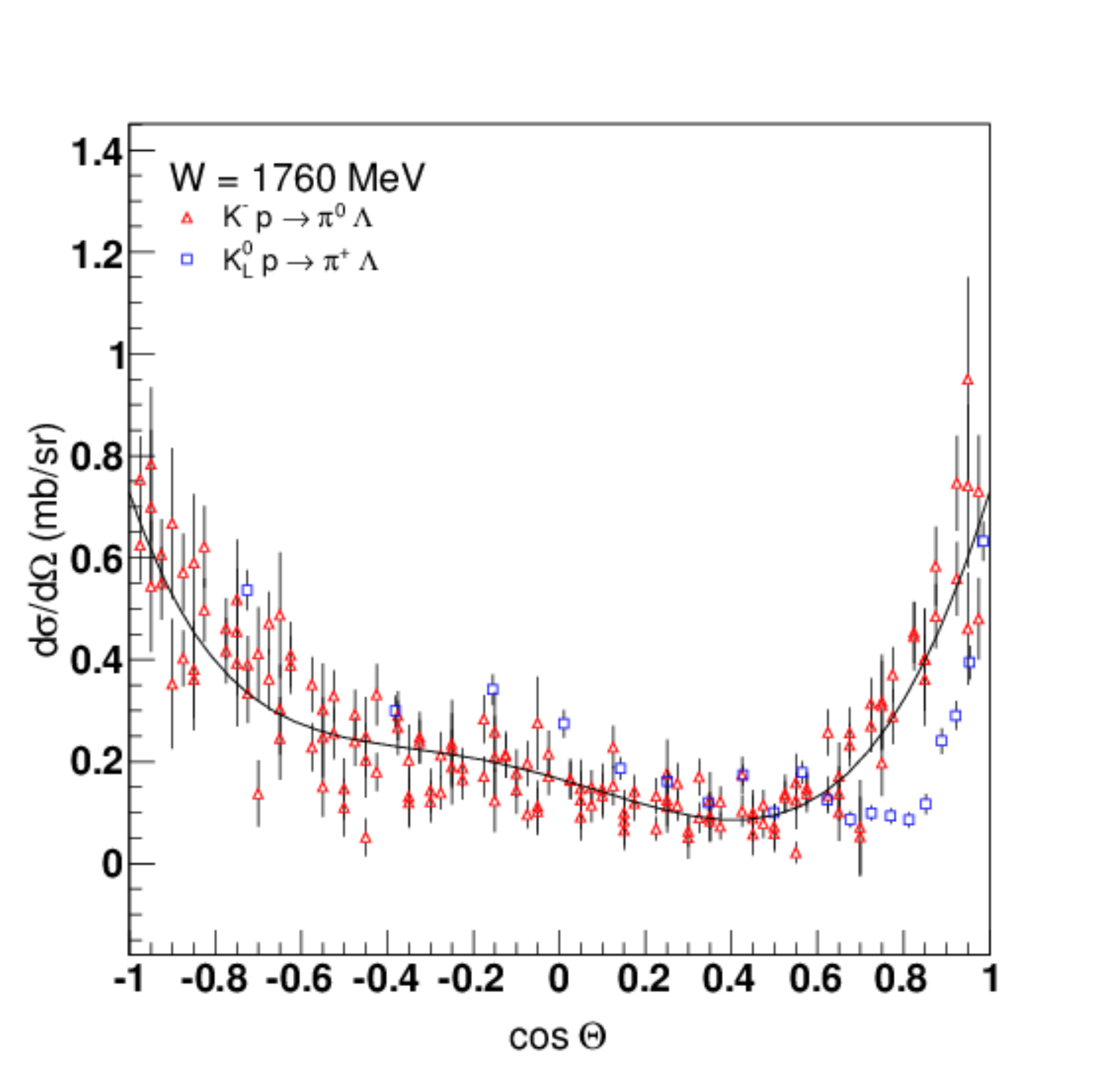}
\includegraphics[angle=0, width=0.45\textwidth]{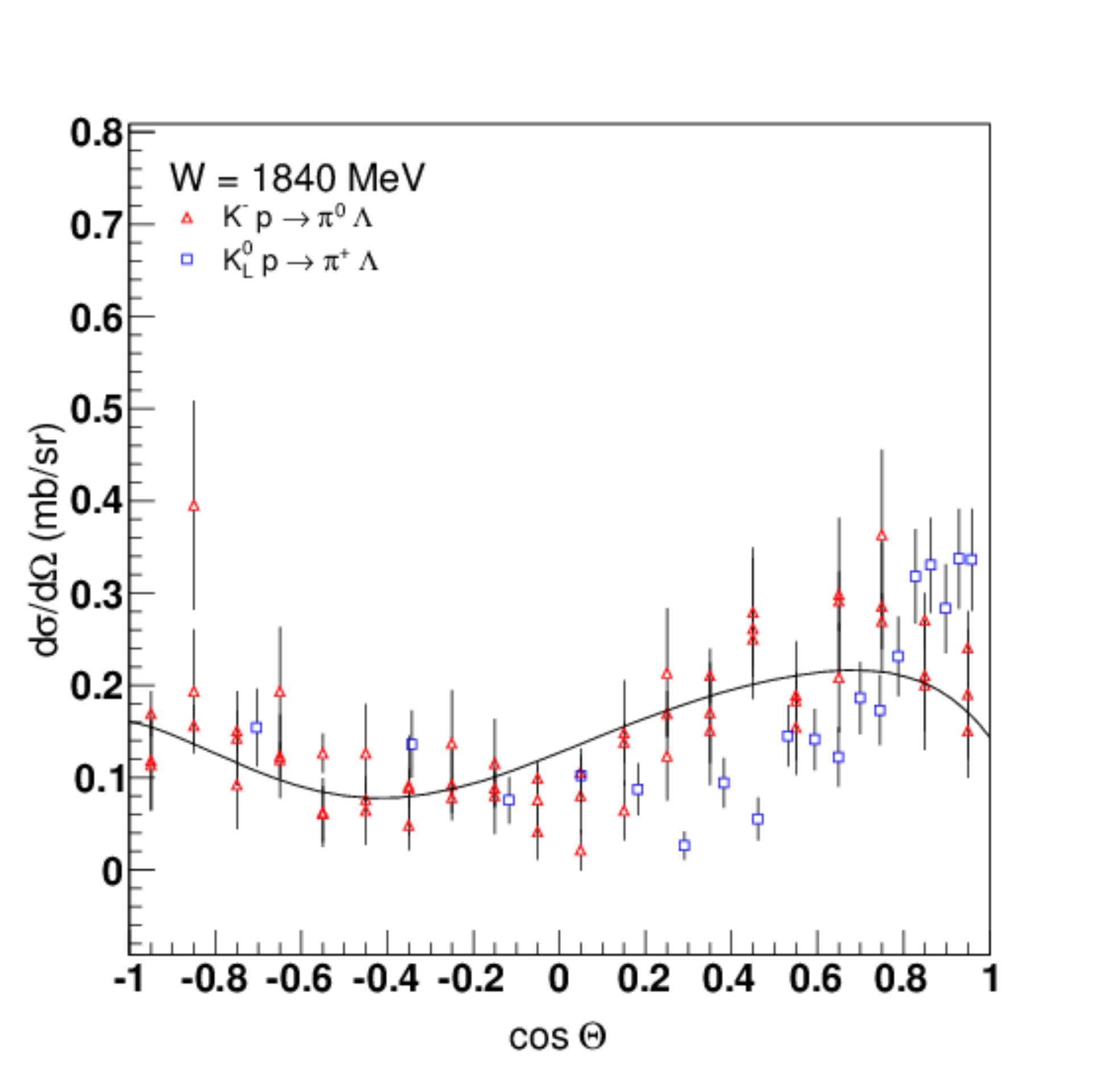} \\
\end{center}
\centerline{\parbox{0.80\textwidth}{
 \caption[] {\protect\small Comparison of selected differential 
	cross section data for $K^-p\to\pi^0\Lambda$ and 
	$K^0_Lp\to\pi^+\Lambda$ at 1540~MeV, 1620~MeV, 1760~MeV, 
	and 1840~MeV.  The curves are from our previous 
	partial-wave analysis of $K^-p\to\pi^0\Lambda$ 
	data~\protect\cite{zhang2013aW,zhang2013bW}.  }
	\label{fig:KLp_piLambda} } }
\end{figure}
\begin{figure}
\begin{center}
\includegraphics[angle=0, width=0.45\textwidth]{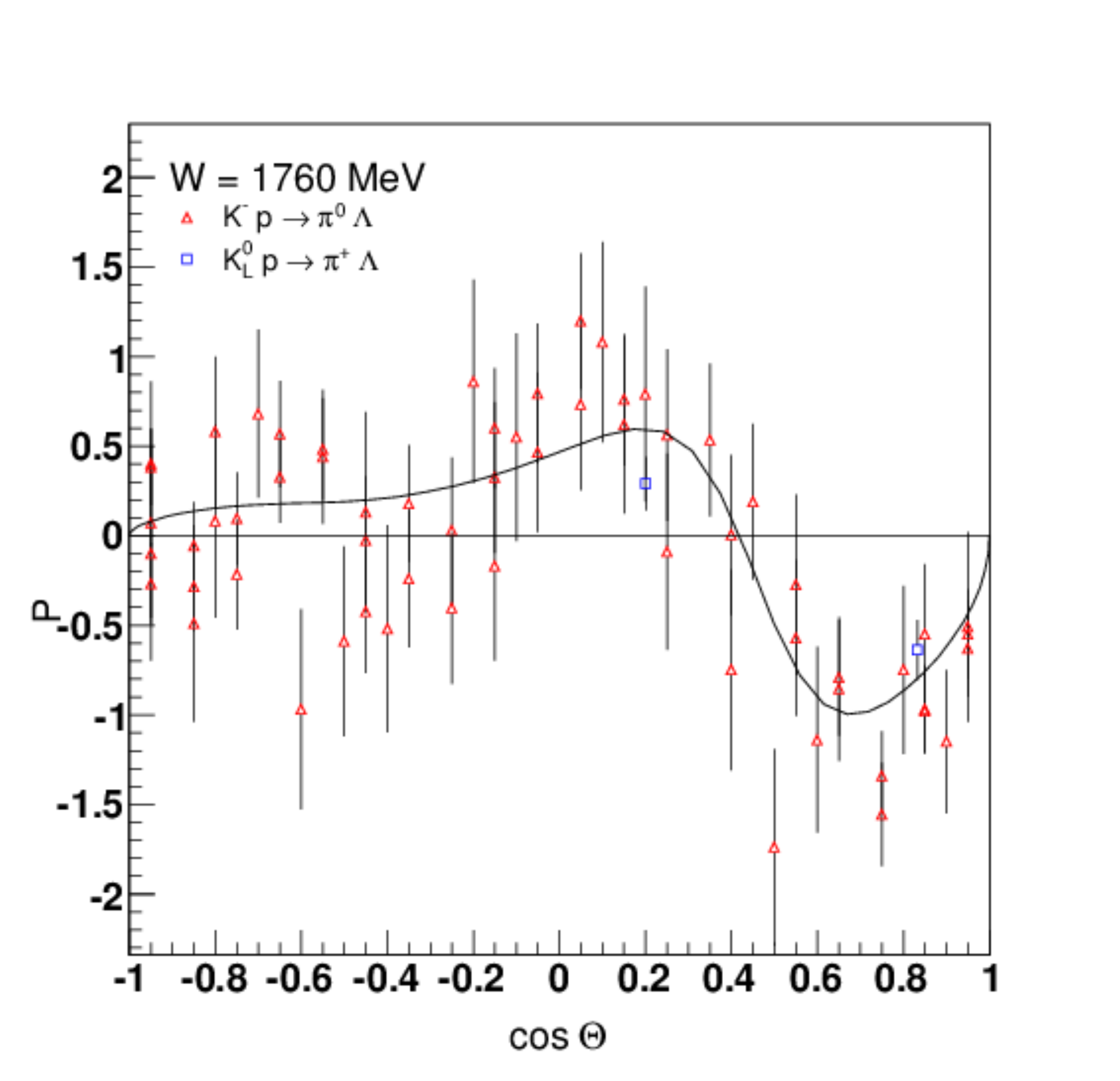}
\includegraphics[angle=0, width=0.45\textwidth]{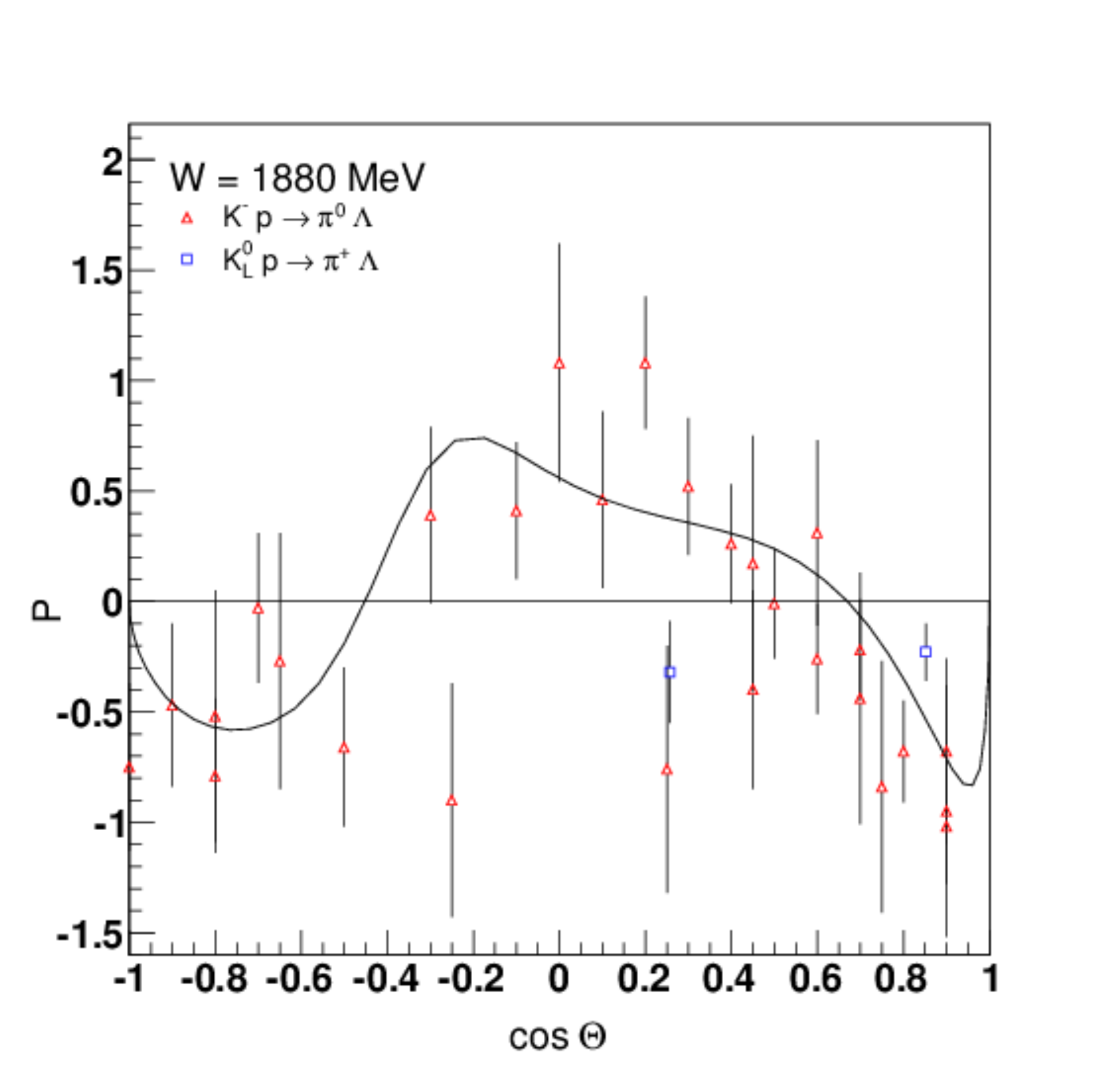} \\
\end{center}
\centerline{\parbox{0.80\textwidth}{
 \caption[] {\protect\small Comparison of selected polarization 
	data for $K^-p\to\pi^0\Lambda$ and $K^0_L p\to\pi^+\Lambda$ 
	at 1760~MeV and 1880~MeV.  The curves are from our previous 
	partial-wave analysis of $K^-p\to\pi^0\Lambda$ 
	data~\protect\cite{zhang2013aW,zhang2013bW}.  }
	\label{fig:KLp_piLambda_P} } }
\end{figure}

\item \textbf{$\pi\Sigma$ Final States}

The amplitudes for reactions leading to $\pi\Sigma$ final states are
\begin{eqnarray}
	T(K^-p\to\pi^-\Sigma^+) &=& -\frac{1}{2}T^1({\overline K}N 
	\to\pi\Sigma)-\frac{1}{\sqrt{6}}T^0({\overline K}N\to\pi
	\Sigma),\\
	T(K^-p\to\pi^+\Sigma^-) &=& \frac{1}{2}T^1({\overline K}N 
	\to\pi\Sigma)-\frac{1}{\sqrt{6}}T^0({\overline K}N\to\pi
	\Sigma),\\
	T(K^-p\to\pi^0\Sigma^0) &=& \frac{1}{\sqrt{6}}T^0({\overline 
	K}N\to\pi\Sigma),\\
	T(K^0_Lp\to\pi^+\Sigma^0) &=& -\frac{1}{2}T^1({\overline K}N 
	\to\pi\Sigma),\\
	T(K^0_Lp\to\pi^0\Sigma^+) &=& \frac{1}{2}T^1({\overline K}N 
	\to\pi\Sigma).
\end{eqnarray}
Figure~\ref{fig:KLp_piSigma} shows a comparison of differential 
cross section data for $K^-p$ and $K^0_Lp$ reactions leading to 
$\pi\Sigma$ final states at $W = 1660$~MeV (or $P_{\rm lab} = 
716$~MeV/$c$).  The curves are based on energy-dependent isospin 
amplitudes from our previous partial-wave analysis~\cite{zhang2013aW,
zhang2013bW}.  No differential cross section data are available for 
$K^0_Lp\to\pi^0\Sigma^+$. As this example shows, the quality of the 
$K_L^0p$ data is comparable to that for the $K^-p$ data.  It would 
therefore be advantageous to combine the $K^0_Lp$ data in a new 
coupled-channel partial-wave analysis with available $K^-p$ data.  
Note that the reactions $K^0_Lp\to\pi^+\Sigma^0$ and $K^0_L p\to
\pi^0\Sigma^+$ are isospin selective (only $I=1$ amplitudes are 
involved) whereas the reactions $K^-p\to\pi^-\Sigma^+$ and $K^-p
\to\pi^+\Sigma^-$ are not.  New measurements with a $K^0_L$ beam 
would amplitudes for $K^-p$ scattering to $\pi\Sigma$ final states.
\begin{figure}
\begin{center}
\includegraphics[angle=0, width=0.45\textwidth]{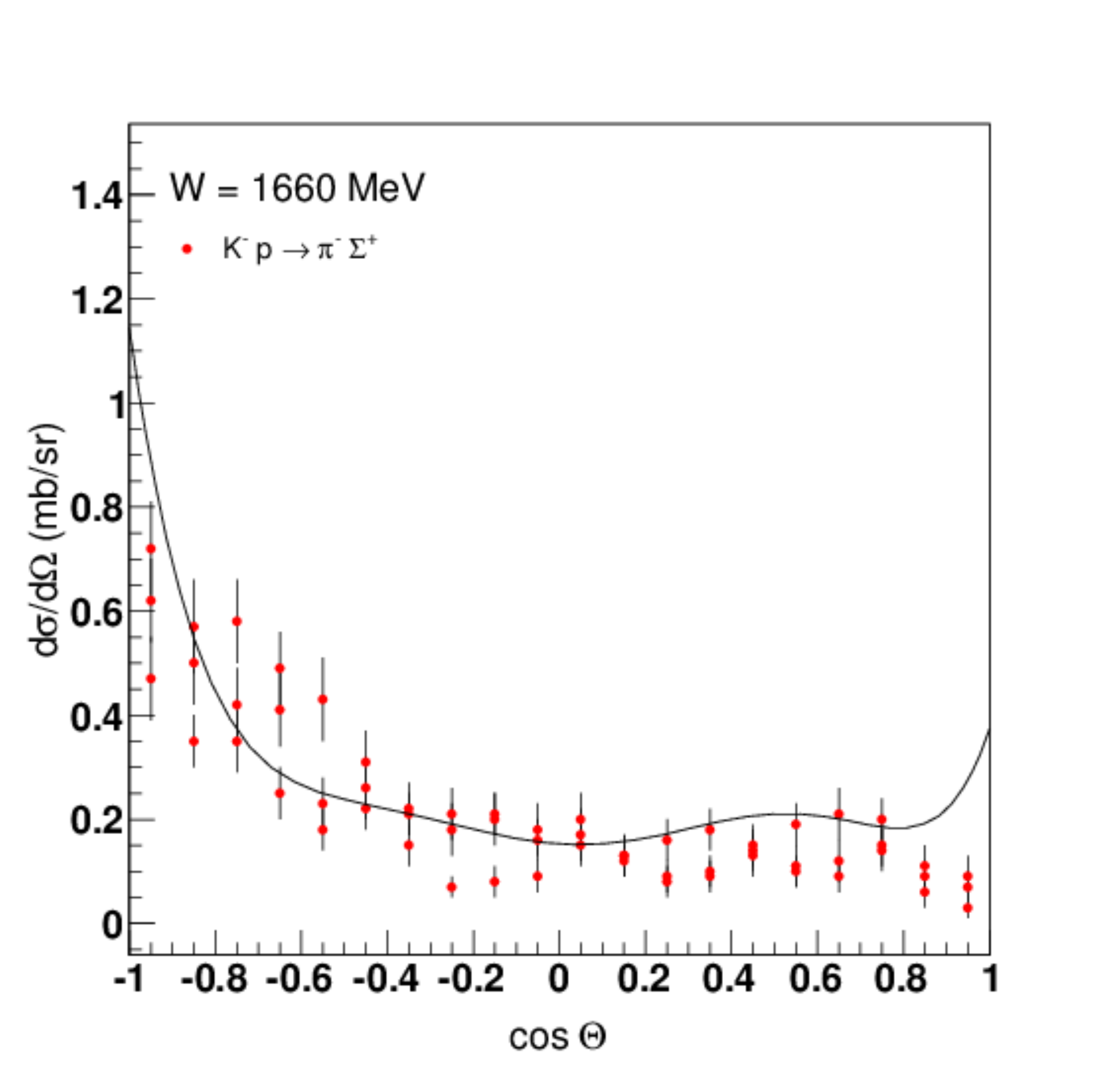}
\includegraphics[angle=0, width=0.45\textwidth]{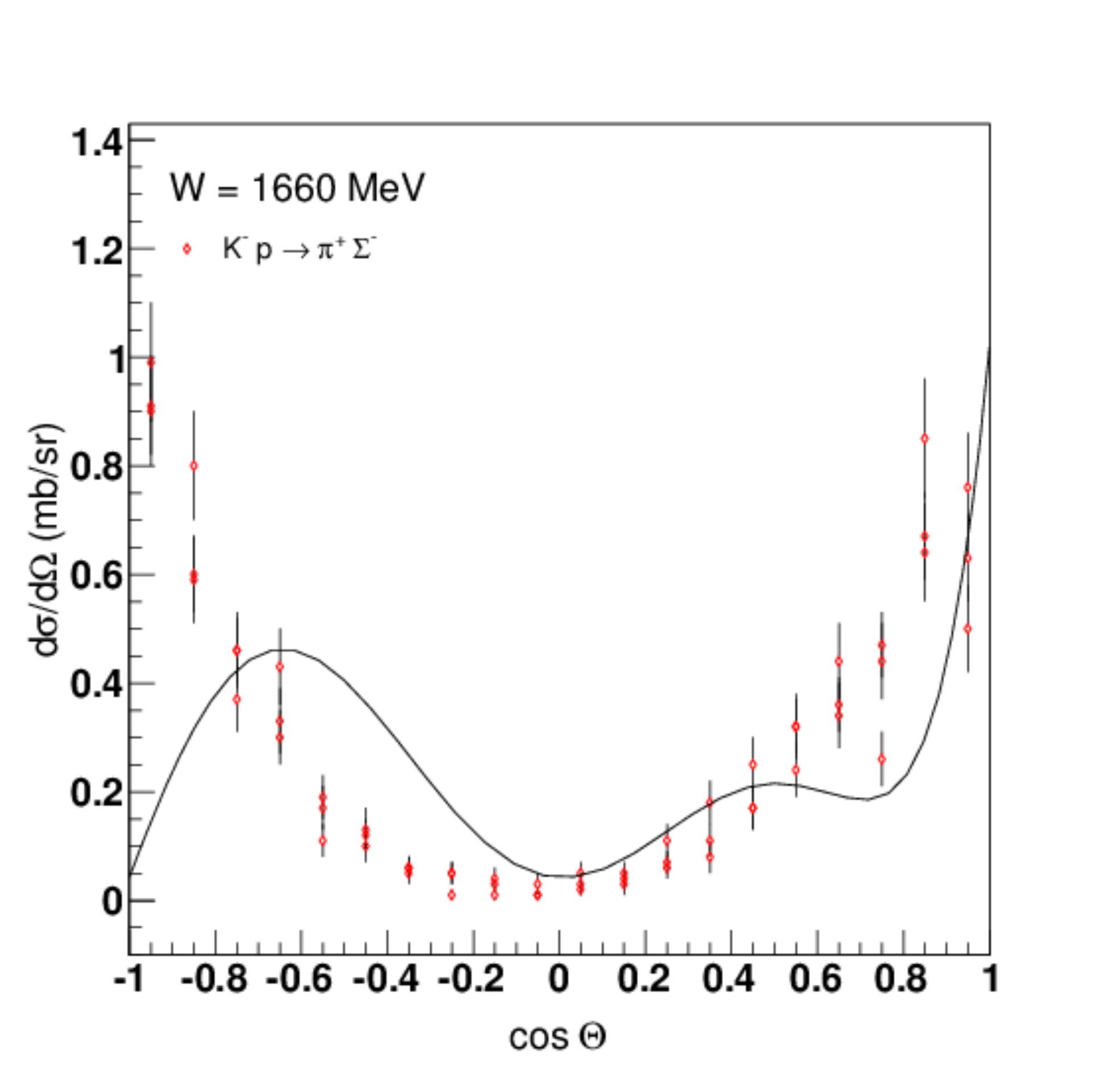} \\
\includegraphics[angle=0, width=0.45\textwidth]{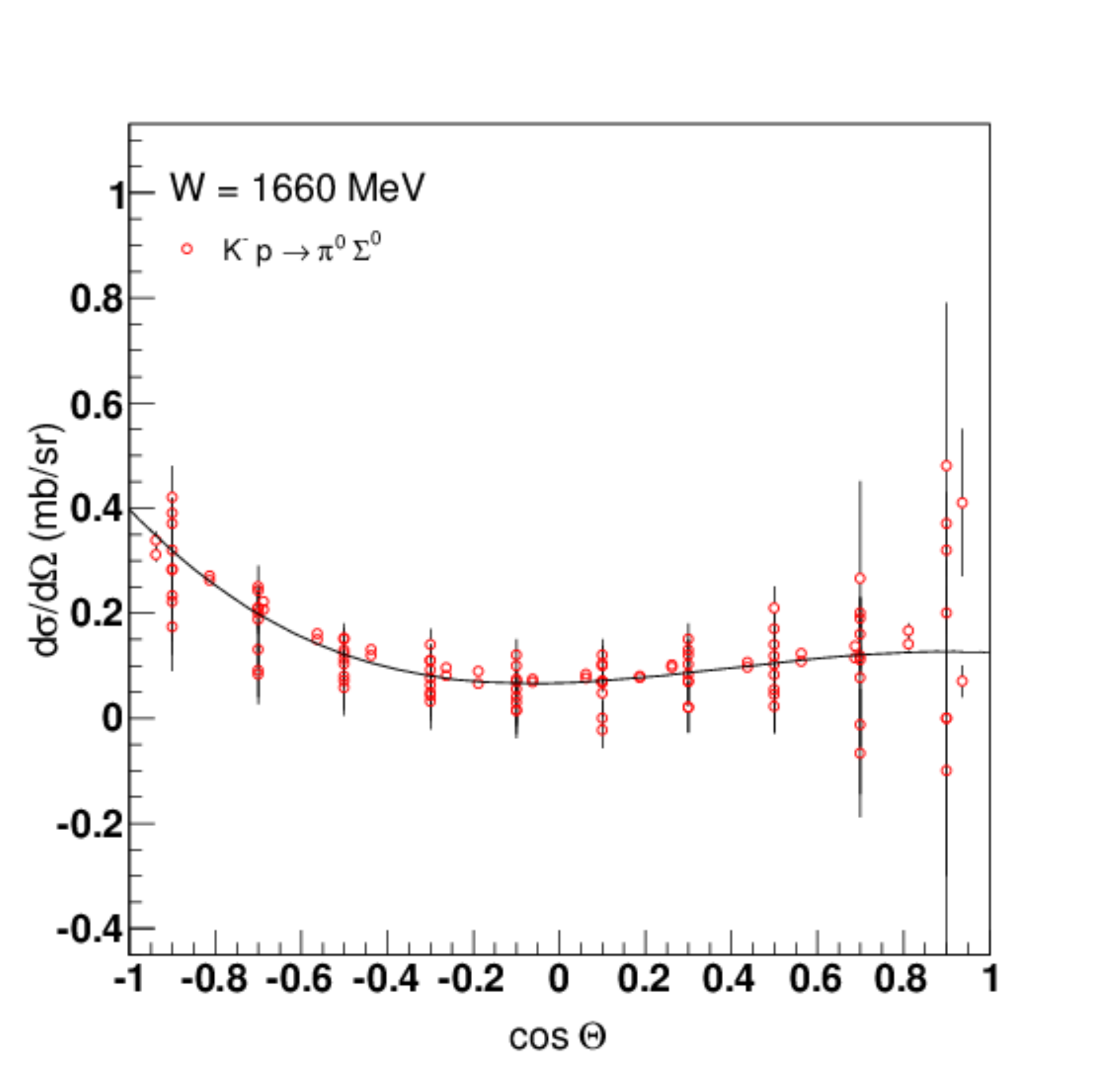}
\includegraphics[angle=0, width=0.45\textwidth]{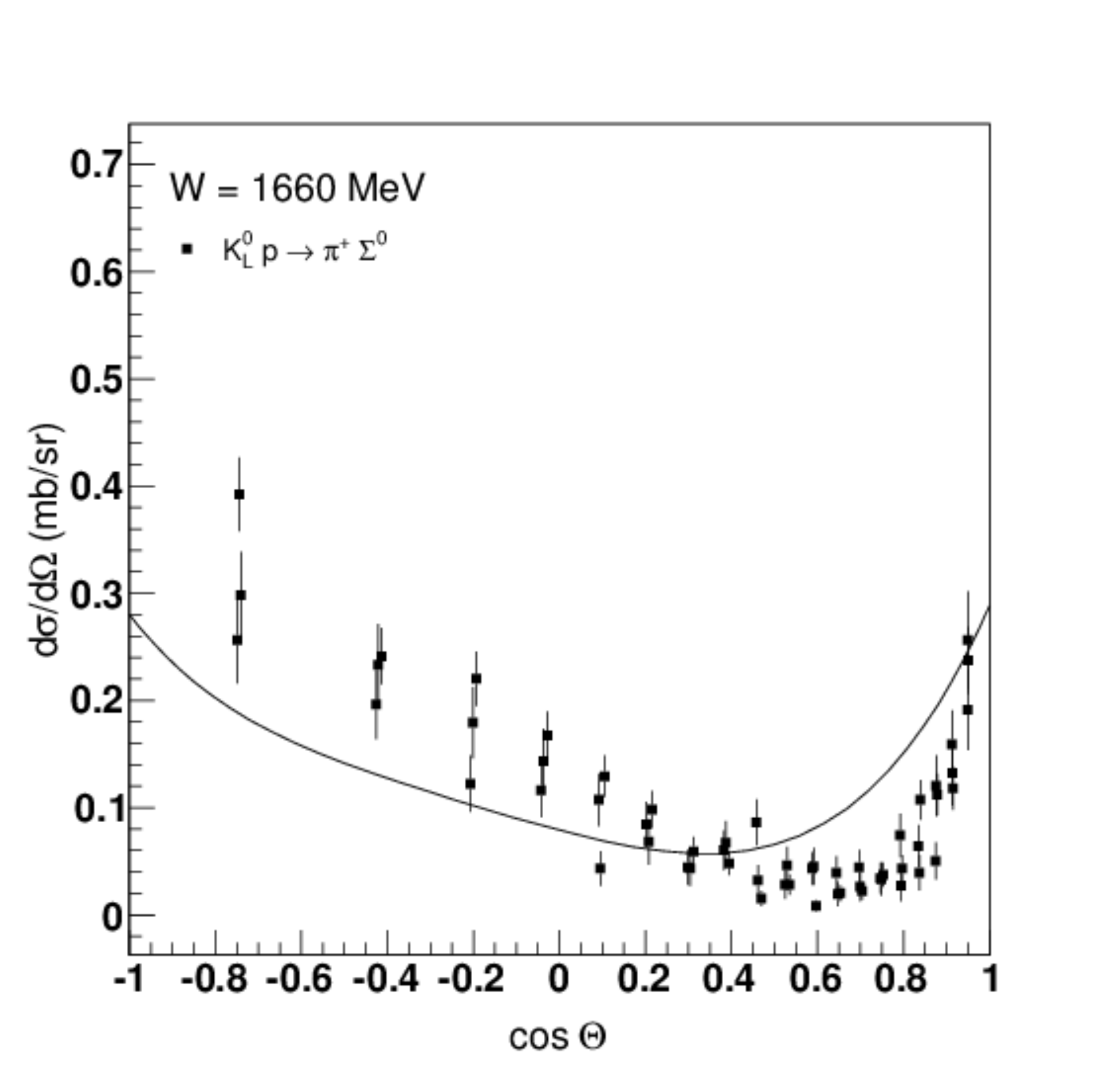} \\
\end{center}
\centerline{\parbox{0.80\textwidth}{
 \caption[] {\protect\small Comparison of selected differential 
	cross section data for $K^-p\to\pi^-\Sigma^+$, $K^-p\to
	\pi^+\Sigma^-$, $K^-p\to\pi^0 \Sigma^0$,  and $K^0_Lp\to
	\pi^0\Sigma^+$ at 1660~MeV.  The curves are from our 
	previous partial-wave analysis of $K^-p\to\pi\Sigma$ 
	data~\protect\cite{zhang2013aW,zhang2013bW}.  }
	\label{fig:KLp_piSigma} } }
\end{figure}

\item \textbf{$K\Xi$ Final States}

The amplitudes for reactions leading to $K\Xi$ final states are
\begin{eqnarray}
	T(K^-p \to K^0 \Xi^0) &=& \frac{1}{2}T^1({\overline K}N 
	\to K\Xi)+\frac{1}{2}T^0({\overline K}N \to K\Xi),\\
	T(K^-p \to K^+ \Xi^-) &=& \frac{1}{2}T^1({\overline K}N 
	\to K\Xi)-\frac{1}{2}T^0({\overline K}N \to K\Xi),\\
	T(K^0_Lp\to K^+\Xi^0) &=& -\frac{1}{\sqrt{2}}T^1
	({\overline K}N\to K\Xi).
\end{eqnarray}
The threshold for $K^-p$ and $K^0_Lp$ reactions leading to 
$K\Xi$ final states is fairly high ($W_{\rm thresh} = 1816$~MeV).  
There are no differential cross section data available for 
$K^0_Lp\to K^+\Xi^0$ and very few (none recent) for $K^-p\to 
K^0\Xi^0$ or $K^-p\to K^+\Xi^-$.  Measurements for these reactions 
would be very helpful, especially for comparing with predictions 
from dynamical coupled-channel (DCC) models.  The {\it Review of 
Particle Physics}~\cite{rppW} lists only two states with branching 
fractions (BF) to $K\Xi$, namely, $\Lambda(2100)\frac{7}{2}^-$ 
(BF $<$ 3\%) and $\Sigma(2030)\frac{7}{2}^+$ (BF $<$ 2\%).

\item \textbf{Summary}

In summary, precise new data for $K_L^0p$ scattering with good 
kinematic coverage could significantly improve our knowledge of 
$\Lambda^\ast$ and $\Sigma^\ast$ resonances.  Although not the 
focus of this talk, a $K_L^0$ beam facility would also be 
advantageous for studying $\Xi^\ast$ and $\Omega^\ast$ states 
via production processes. Polarization data are very important 
to measure in addition to differential cross sections to help 
remove ambiguities in partial-wave analyses.  Unfortunately, the 
current data base for $K_L^0p$ scattering includes very few 
polarization data. As noted here, several $K_L^0p$ reactions are 
isospin-1 selective, which would provide a useful constraint for 
a combined partial-wave analysis of $K_L^0p$ and $K^-p$ 
reactions.  Finally, the long lifetime of the $K_L^0$ compared 
with the $K^-$ would allow a larger beam flux on target, which 
would allow $K_L^0p$ measurements to be made at lower energies 
than easily measurable with $K^-$ beams.

\item \textbf{Acknowledgments}

The author thanks Dr.~Igor~Strakovsky for providing the data 
files used in this work, and Brian~Hunt for providing all the 
figures.  This material is based upon work supported by the U.S. 
Department of Energy, Office of Science, Office of Medium Energy 
Nuclear Physics, under Award No. DE--SC0014323.
\end{enumerate}

\newpage
\subsection{Excited Hyperons and their Decays}
\addtocontents{toc}{\hspace{2cm}{\sl F.~Myhrer}\par}
\setcounter{figure}{0}
\setcounter{table}{0}
\setcounter{equation}{0}
\halign{#\hfil&\quad#\hfil\cr
\large{Fred Myhrer}\cr
\textit{Department of Physics and Astronomy}\cr
\textit{University of South Carolina}\cr
\textit{Columbia, SC 29208, U.S.A.}\cr}

\begin{abstract}
There are several missing states in the mass spectrum of 
the first excited negative parity $\Lambda^\ast$ and 
$\Sigma^\ast$ states. To gain further understanding of 
QCD quark confinement and how QCD should be implemented 
in quark models, it is desirable to establish if these 
missing states exist, and to measure more accurately the 
decay properties of the hyperon states
\end{abstract}

\begin{enumerate}
\item \textbf{Introduction}

The QCD theory describes the forces among the quarks, and 
perturbative QCD has  successfully explained asymptotic 
freedom and  high four-momentum hadronic processes.  At high 
energies QCD predicts the existence of quark-jets and the 
gluon-jets which were observed as predicted~\cite{Hoyer:1979taF,
Ellis1979F}. This means that high energy quarks radiate gluons, 
which  hadronize producing a jet of gluon quantum numbers. 
At low energy the large strong QCD coupling constant requires 
an effective  theory or a model approach to explain the 
structure of baryons. The quark model to organizes the many 
observed colorless meson and baryon states, and the model 
provides  some insights of the structure of these states. The 
quark model was extended to also predict the existence of 
multi-quark (exotic) mesons and baryons including 
{\it glueballs}, {\it e.g.}, QQ$\bar{Q}\bar{Q}$, QQQQ$\bar{Q}$,  
Q$\bar{Q}$glue, where $Q$ \underline{and} $glue$ are both 
treated as building blocks. The question is if such states 
exist. Have we taken the quark and gluon building blocks 
scenario too literary? We need further guidance from what 
QCD would allow. At present the QCD confinement of light 
quarks is not understood. According to the quark model there 
are several states among the first excited hyperon states 
which are missing or are not established. Do QCD require that 
these missing states should exist, or does QCD require further 
model restrictions not implemented in todays quark models?  
Many of the quark model predicted $N^\ast$ and $\Delta^\ast$ 
states, made of the very light $u$ and $d$ quarks, have been 
extensively studied. These states will not be discussed in 
this paper. Instead, I will concentrate only on the 
\underline{first} excited hyperon, negative parity 
$\Lambda^\ast$ and $\Sigma^\ast$ states and make some 
observation on what we could learn about QCD through the use 
of quark model evaluations. In order to answer the questions 
above, it is imperative that we experimentally can establish 
the first excited hyperon mass spectrum of $\Lambda^\ast$ 
and $\Sigma^\ast$. We know that most excited baryon states 
have large decay widths. The decay branching ratios can give 
us further information about the structure of these states.  
In this presentation I will first present a few general quark 
model arguments, which will be used in the discussion to 
follow. 

\item \textbf{The Quark Models}

Due to the very slow experimental progress on strange 
baryon spectroscopy during the last decades, there are 
several of the first negative parity excited hyperon 
states which have not been established. Some of the 
$\Sigma^\ast$ states have at most one star in the 
Particle Data Tables rating~\cite{PDGF}. One possible 
question to ponder is: Does QCD contain dynamical features,  
which are not considered by present day quark models and 
would imply that these states should be absent? The 
$\Lambda^\ast$ and $\Sigma^\ast$ states have one ``heavy" 
$s$-quark and two very light quarks. Could this extra 
feature of having one heavy quark give us some extra 
insight into QCD beyond what $N^\ast$ and $\Delta^\ast$ 
states can provide? These observations and arguments 
would then be used to gain a better understanding of the
$\Xi^\ast$ and $\Omega^\ast$ states. 

The quarks interact via gluon exchanges, which couple the 
quark spins, very similar to what happens to pion- and 
rho-meson exchanges between nucleons in a nucleus. 
Analogous to the three-nucleon states, $^3H$ and $^3He$, 
we should therefore expect the three valence quark baryon 
ground states to have  a three-quark spatial wave function 
which contains a mixture of  S, S$^\prime$ and D quark 
states. The effective pseudo-scalar meson cloud surrounding 
the quark core of the baryons will contribute to this 
spatial mixture of states. As will be presented these 
spatial admixtures affect strongly some excited hyperon 
decays. Much of this talk is based on the extensive work 
of the non-relativistic quark model (NRQM) by Nathan 
Isgur and Gabriel Karl and their coworkers~\cite{IsgurKarlF}. 

\begin{enumerate}
\item \textbf{Quark Model Assumptions}

The generic non-relativistic baryon wave function has the 
following structure     
\begin{eqnarray}
	\Psi =  \Psi_{color}
	\;\Psi_{flavor}\; \Psi_{spin}\;\Psi_{space}\; .
	\label{eq:baryonwf}
\end{eqnarray} 
We assume that isospin is a good symmetry, {\it i.e.},
the masses of the $u$ and $d$ quarks are equal: $m_u$ = $m_d$ 
= $m_q$.  However, the SU$_F$(3) is a broken symmetry since 
the $s$ quark has a mass $m_s > m_q$.  For this reason, we 
will adopt the $uds$ basis when  the baryon wave functions 
are evaluated. [Please note that in the (cloudy or MIT) bag 
model the masses of the $u$ and $d$ quarks are zero.] 

In bag models the effective quark masses are generated by 
the confinement condition, which presumably reflects the 
very soft gluon exchanges between the three quarks. 
The other usual quark model assumptions are the following: 
\begin{itemize}
\item All hadrons are $SU(3)$-color singlets, \textit{i.e.},  
	$\Psi_{color}$ is a totally anti-symmetric wave function 
	under the interchange of any two quarks. 
\item Confinement of quarks is universal and  is the same for 
	all quark flavors. It is presumed to be a Lorentz 
	scalar condition. 
\item The Pauli principle tells us that two identical quarks 
	must have a totally anti-symmetric  wave function. 
	Since $\Psi_{color}$ is anti-symmetric the product of 
	the other components in Eq.(\ref{eq:baryonwf}) must 
	be symmetric under the interchange of any two quarks. 
\item The non-relativistic quarks interact via an effective 
	one-gluon-exchange, a la De Rujula {\it et 
	al.}~\cite{DeRujulaF}. This effective gluon exchange 
	generates a \underline{spin-spin interaction} among 
	the quarks and makes the {\it decuplet} baryons 
	\underline{heavier} than the {\it octet} baryons. 
\end{itemize} 
The non-relativistic effective one-gluon-exchange (OGE) 
between quarks $i$ and $j$ is: 
\begin{eqnarray}
	H_{hyp}^{ij} &=& A_{ij} \left\{ 
	\frac{8\pi}{3} \vec{S}_i\cdot \vec{S}_j \delta^3(\vec{r}_{ij}) 
	+\frac{1}{r_{ij}^3}\left( 
	\frac{ 3 (\vec{S}_i\cdot \vec{r}_{ij})(\vec{S}_j\cdot 
	\vec{r}_{ij}) } { r_{ij}^2 } 
	- \vec{S}_i\cdot \vec{S}_j
	\right) \right\} \, , \nonumber
\end{eqnarray} 
where $A_{ij}$ is a constant which depends on the quark 
masses~\cite{DeRujulaF}. As can be inferred from this 
expression, the spin-spin and the tensor quark-quark 
interactions are closely related, and the tensor component 
will produce a spatial D-state quark wave function. Note 
that  this non-relativistic reduction of the effective OGE 
quark-quark interaction neglects the spin-orbit force. 
Isgur and Karl argue that the spin-orbit force should be 
small. In bag models ($m_q =0$~MeV), the quark P-state with 
j=3/2 has a lower energy than  j=1/2, {\it i.e.}, the bag 
models' Lorentz-scalar confinement condition introduces a 
spin-orbit splitting of the quark states. Fortunately, the 
relativistic OGE introduces an effective spin-orbit force 
of opposite sign.  In cloudy bag model calculations these 
two spin-orbit contributions basically cancel, and what 
remains are the spin-spin and tensor interactions due to 
OGE and  the pseudo-scalar meson (pion, Kaon) cloud 
surrounding the quark core, see for example 
Refs.~\cite{MyhrerWroldsenF,um91F}. The spin-spin and tensor 
interactions strongly affect the decay rates of 
$\Lambda^\ast$ and $\Sigma^\ast$ states to $\bar{K}N$ and 
$\pi\Sigma$. 

\item \textbf{A Decay Rate Observation}

As mentioned we adopt the $uds$ basis and not the SU$_F$(3) 
flavor basis in order to construct the baryon wave functions. 
The spatial wave function of three quarks is given by the two 
relative coordinates between the three quarks (ignoring the 
center of mass motion): 
\begin{eqnarray} 
	\vec{\rho} &=& \left(\vec{r}_1 -\vec{r}_2\right)/\sqrt{2},  
	\label{eq:rho} \\ 
	\vec{\lambda} &=& \left(\vec{r}_1 +\vec{r}_2 -2\vec{r}_3
	\right)/\sqrt{6} .
	\label{eq:lambda}
\end{eqnarray}  
\begin{figure}[ht!]
\begin{center}
\includegraphics[width=6in]{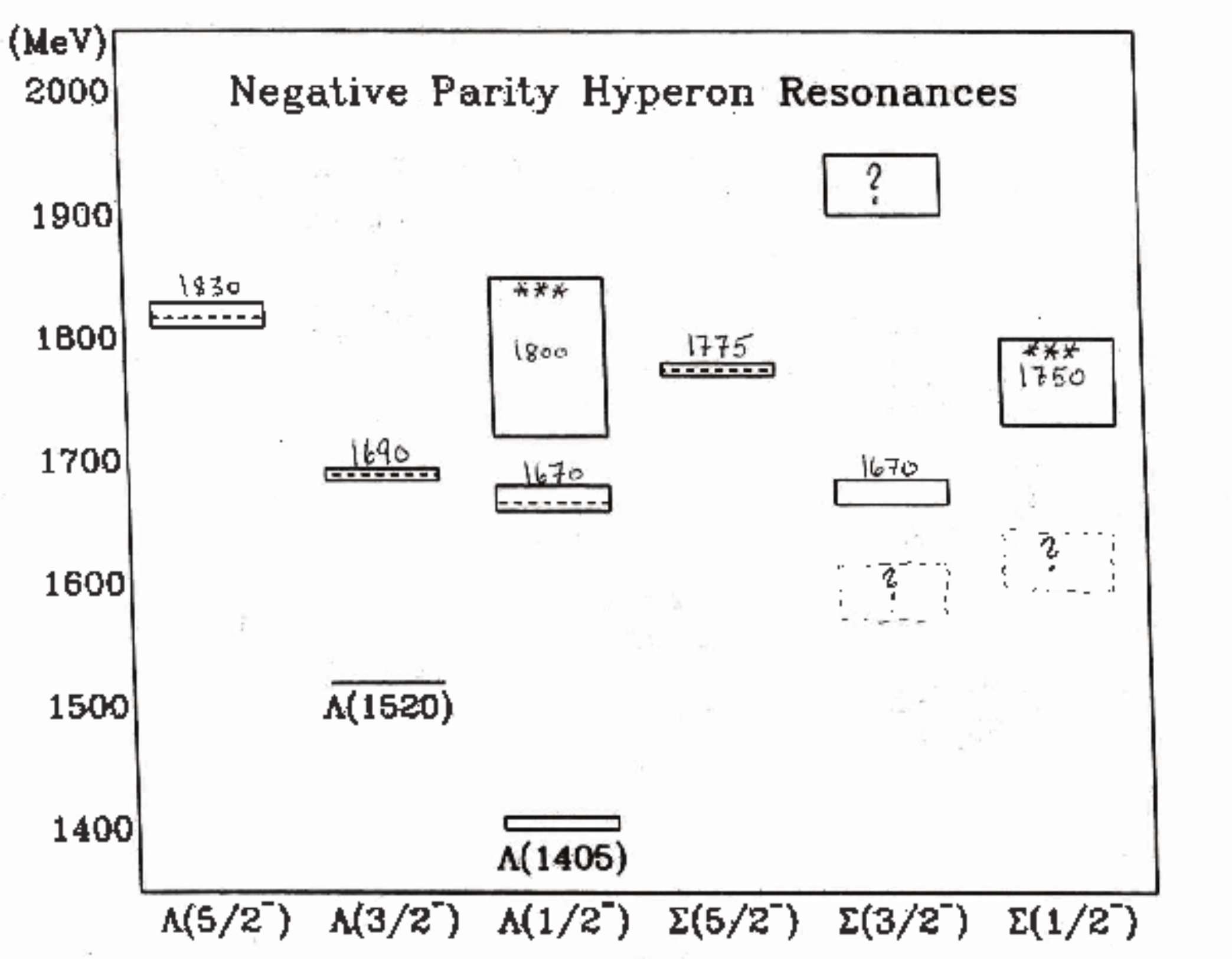}
\end{center}
\centerline{\parbox{0.80\textwidth}{
 \caption{The confirmed mass spectrum of the first excited
        negative parity $\Lambda^\ast$ and $\Sigma^\ast$
        states where we have included the three-star
        states~\protect\cite{PDGF}. The hight of the squares
        illustrate the mass uncertainties of the
        ``established" states. The squares with the
        question marks are states which are controversial.
        According to quark models there are two completely
        missing three-quark states in this figure.}
        \label{fig:masses} } }
\end{figure}

Here quarks 1 and 2 are the $u$ and $d$ quarks, and quark 3 is 
the heavy $s$-quark. The corresponding reduced masses are 
$m_\rho=m_q$ and $m_\lambda= 3m_q m_s/(2m_q+m_s)$. In an NRQM, 
the spatial confinement of the quarks is simulated by an harmonic 
oscillator potential. The harmonic oscillator confinement 
potential with a given flavor independent spring constant gives a 
difference in $\rho$ and $\lambda$ oscillator frequencies, 
$\omega_\rho$ and $\omega_\lambda$ due to the difference in the 
two reduced quark masses~\cite{IsgurKarlF}, 
\begin{eqnarray}
	\omega_\rho-\omega_\lambda = \omega_\rho \left[ 1- 
	\left(\frac{2(m_q/m_s)+1}{3}\right)^{1/2}\right] > 0 \, , 
	\label{eq:frequencies}
\end{eqnarray} 
where the frequency $\omega_\rho$ is the one relevant for 
the nucleon ground state. 

The mass splitting between the $\Lambda^\ast(5/2^-)$ and the 
$\Sigma^\ast(5/2^-)$ states, shown in Fig.~\ref{fig:masses}, 
can easily be understood.  In essence, the masses of these two 
$J^P= 5/2^-$ states differ mainly due to confinement and 
SU$_F$(3) breaking since $m_q <m_s$. The detailed explanation 
goes as follows: The quarks in these excited hyperons have an 
orbital angular momentum $L=1$. Both $J^P= 5/2^-$ states have 
totally symmetric spin wave functions since the total spin of 
the quarks must be $S=3/2$. Furthermore, since 
$\Lambda^\ast(1830)$ is an iso-singlet, due to the Pauli 
principle, it must have a $\vec{\rho}$-dependent spatial wave 
function, which is anti-symmetric  under the interchange of 
quarks 1 and 2,  as seen in Eq.(\ref{eq:rho}). On the other 
hand, $\Sigma^\ast(1775)$, is an iso-triplet, and since 
$\vec{\lambda}$ is symmetric  under the interchange
($1\leftrightarrow 2$), Eq.(\ref{eq:lambda}), 
$\Sigma^\ast(1775)$ must have a $\vec{\lambda}$-dependent 
spatial wave function.  In other words $\Lambda^\ast(1830)$ 
contains the energy $\hbar \omega_\rho$ whereas 
$\Sigma^\ast(1775)$ has $\hbar \omega_\lambda$, and their 
mass difference is $\hbar (\omega_\rho - \omega_\lambda) 
\approx 75$~MeV, which will be  modified by 
$H_{hyp}$~\cite{IsgurKarlF}.  

This difference in the spatial decomposition of the two $J^P= 
5/2^-$ states' wave functions has the following decay 
implications: $\Lambda^\ast(1830)$ couples weakly to $\bar{K} 
N$ since the nucleon spatial wave function is symmetric under 
the interchange $1\leftrightarrow 2$, whereas $\Sigma^\ast(1775)$ 
couples easily to $\bar{K} N$ for the same reason. When  
modifications due to $H_{hyp}$ is included, the difference in 
the  two observed decays widths are easily 
explained~\cite{IsgurKarlF}, as will be detailed in the next 
subsection. This example illustrates the close relations between 
the internal structure of the initial and final baryon and the 
magnitude of the corresponding decay width. With more precise 
experimental data one looks forward to be able to make similar 
arguments for mass differences and decay branching ratios for 
the other excited hyperon states.
 
\newpage
\item \textbf{Spatial Wave Function and the Decay Widths}

As emphasized by Isgur and Karl even the ground state baryons 
have a complicated spatial wave function due to $H_{hyp}$ as 
mentioned earlier.  For example, in their NRQM the nucleon 
state has the following structure: 
\begin{eqnarray}
	| N \rangle &\simeq & 0.90 | ^2S_S\rangle -0.34 | 
	^2S_S^\prime \rangle -0.27 | ^2S_M\rangle -0.06 | 
	^2D_M\rangle \, , 
	\label{eq:N}
\end{eqnarray} 
where $ |S^\prime \rangle $ and $|D\rangle$ are the excited 
S- and D- quark states of the harmonic oscillator. The subscripts 
$S$ and $M$ denote symmetric and mixed symmetry spatial states, 
respectively.  Similarly, Isgur and Karl find the $\Lambda(1116)$ 
state to be: 
\begin{eqnarray}
	| \Lambda \rangle &\simeq & 0.93 | ^2S_S\rangle -0.30 | 
	^2S_S^\prime \rangle -0.20 | ^2S_M\rangle -0.03 | 
	^4D_M\rangle -0.05 |\mathbf{1}, ^2S_M\rangle .
	\label{eq:L}
\end{eqnarray} 
Given present day experimental accuracies of the masses and 
decay branching ratios determinations, Isgur and Karl assume 
that the D-state components of the states have no practical 
consequences and can be  neglected. The expressions in 
Eqs.~(\ref{eq:N}) and (\ref{eq:L}) tell us that the ground 
state baryons are not pure symmetric $|^2S_S\rangle$ states.    
They contain spatially mixed symmetry states, and the $| 
^2S_S^\prime \rangle $ and $| ^2S_M\rangle$ components 
of the ground state baryon octet will modify (sometimes 
strongly) the excited baryon  to ground states decay widths.  
For example, by including the mixed symmetric component 
$|^2S_M\rangle$ of the nucleon state, one finds the ratio of 
decay amplitudes~\cite{IsgurKarlF}:
\begin{eqnarray}
	\frac{{\rm A}(\Lambda^\ast (1830)\to \bar{K}N)}{{\rm A}
	(\Sigma^\ast (1775)\to \bar{K}N)} \simeq -0.28 \, .  
	\label{eq:BR} 
\end{eqnarray}
The excited $\Lambda^\ast$ and $\Sigma^\ast$ states have 
similar mixed spatial states, again due to $H_{hyp}$ and 
also due to the pseudo-scalar meson cloud surrounding the 
quark core in cloudy bag models. These mixed states will 
further affect the relative decay branching ratios of the 
excited hyperon states. 

\item \textbf{Electromagnetic Decays of the $\Lambda^\ast$ 
	and $\Sigma^\ast$ states}

In the first excited states, one quark is in a $P$-state 
relative to the two others which are in a relative $S$-state. 
In an electromagnetic decay to the ground state the $P$-state 
quark couples to the photon (or in strong decays to the 
outgoing meson, {\it e.g.}, $\pi$, $\bar{K}$ or $\eta$). For 
example, the $\Lambda^\ast(1520)$ state is a well established 
state and has the following decomposition in terms of SU$_F$(3) 
multiplets: 
\begin{eqnarray}
	| \Lambda^\ast(1520)\rangle &\simeq & a | ^2\mathbf{1}\rangle 
	+b | ^4\mathbf{8}\rangle + c | ^2\mathbf{8}\rangle .
	\label{eq:Lsymm} 
\end{eqnarray}
Different quark models give different values for the $a$, $b$ 
and $c$ coefficients. Using the harmonic oscillator of the 
NRQM of Isgur and Karl find $a=0.92$, $b= -0.04$, and $c= 
0.39$~\cite{IsgurKarlF}, which results in the following values 
for two electromagnetic widths 
$\Gamma [\Lambda (1520)\to \Lambda \gamma] = 96$~keV and 
$\Gamma [\Lambda (1520)\to \Sigma^0 \gamma] = 74$~keV~\cite{dhk83F}. 
A cloudy bag model calculation produces the values $a= 0.95$, 
$b= -0.09 $ and $c= 0.30 $. The decay widths in the cloudy bag 
model, where the emitted photon also couple to the meson cloud, 
are 
$\Gamma [\Lambda (1520)\to \Lambda \gamma] = 32$~keV,  
$\Gamma [\Lambda (1520)\to \Sigma^0 \gamma] = 49$~keV~\cite{um91F}.  

In Table~\ref{tab:tableF}, we compare several different quark 
model calculations of the decay rate  $\Gamma_\gamma$ for 
$\Lambda(1520)\to\Lambda(1116) + \gamma$. As can be read off 
from the Table~\ref{tab:tableF}, this rate is not only sensitive 
to the coefficients $a$, $b$, and $c$ in Eq.(\ref{eq:Lsymm}), but 
also to the $\Lambda(1116)$ configuration mixing, which may change 
$\Gamma_\gamma$ by 50\% or more. The decay width $\Gamma_\gamma$ 
is very difficult to measure and it is not very well 
determined~\cite{RateF}. The evaluations of  $\Gamma_\gamma$ in 
quark models where SU$_F$(3) is broken are very involved and it 
is desirable to have a more precise experimental determination 
of $\Gamma_\gamma$ before one revisit such a calculation. 
\begin{table*}
\centerline{\parbox{0.80\textwidth}{
\caption{\label{tab:tableF} The evaluation of $\Gamma_\gamma$ 
	by several quark model calculations are given.  The 
	references to the various models can be found 
	in~\protect\cite{Myhrer2006F}. The columns give the 
	coefficients $a$, $b$ and $c$ of the $\Lambda(1520)$ 
	wave functions, Eq.(\ref{eq:Lsymm}), found in the 
	various publications as well as the structure of the 
	ground state wave function $\Lambda(1116)$ used in 
	the different calculations. The ``dash" means that 
	values of the coefficients or the $\Lambda(1116)$ 
	state cannot be ascertained.} } } 
\vspace{0.3cm}
\begin{center}
\begin{tabular}{|c|c||c|c|c|c|}
\hline \hline
Models               &  a   &  b   &  c   &$\Lambda(1116)$& $\Gamma_\gamma$ (keV) \\
\hline
\hline
NRQM                 & 0.91 & 0.01 &  0.40& $|^2S_S>$     &  96 \\
\hline 
NRQM (SU(6)$-$basis) & 0.91 & 0.01 &  0.40& $-$           &  98 \\ 
\hline 
$\chi$QM             & 0.91 & 0.01 & -0.40& $|^2S_S>$     &  85 \\
\hline 
$\chi$QM             & 0.91 & 0.01 & -0.40& mixed         & 134 \\
\hline 
NRQM (uds$-$basis)   & $-$  & $-$  & $-$  & mixed         & 154 \\
\hline 
MIT bag              & 0.86 & 0.34 & -0.37& $-$           &  46 \\
\hline
Cloudy bag           & 0.95 & 0.09 & -0.29& $|^2S_S>$     &  32 \\
\hline 
RCQM                 & 0.91 & 0.01 &  0.40& mixed         & 215 \\
\hline 
Bonn$-$CQM           & $-$  & $-$  &  $-$ & $-$           & 258 \\
\hline 
\end{tabular}
\end{center}
\end{table*}

\item \textbf{A short note on $\Lambda(1405)$}

A question within quark models, which has not been resolved 
satisfactory, is: Why is $\Lambda(1405)$ about 100~MeV below 
$\Lambda(1520)$ in mass?  If we assume that the $\Lambda^\ast$ 
states are mainly three-quark states, quark models have serious 
problems generating this large observed spin-orbit-like mass 
splitting. Could a strong coupling of the lowest three-quark 
state with $J^P = \frac{1}{2}^-$ to the meson-baryon decay 
channels (beyond how this is presently treated in cloudy bag 
models) explain this mass-splitting? Historically, Dalitz and 
Tuan~\cite{DalitzF} proposed that $\Lambda(1405)$ is a $K^-p$ 
bound state, {\it i.e.}, could  $\Lambda(1405)$ be like a 
quark molecule?.  Could it have a large multi-quark 
(pentaquark) state component? A very readable recent paper on 
the arXiv by Molina and D\"{o}ring~\cite{MolinaDoringF} 
discusses the possible pole structure of $\Lambda(1405)$. 
This paper contains an overview of many theoretical 
publications, including lattice evaluations, regarding 
possible structure of $\Lambda(1405)$. Apart from recent 
measurements of the $K^-p$ atom~\cite{KpAtomF}, most data on 
this subject are old. We urgently need better data to settle 
the numerous theoretical discussions regarding $\Lambda(1405)$. 
\end{enumerate}

\item \textbf{Summary and Outlook}

It is imperative that we can experimentally establish the 
mass spectrum of the lowest excited negative parity 
$\Lambda^\ast$ and $\Sigma^\ast$ states in order to enhance 
our understanding of how QCD operates among the three ``light" 
quarks, $u$, $d$ and $s$, and generates the masses and decay 
widths of the  hyperons. The JLab proposed $K_L^0$ beam 
scattering off a hydrogen target can access $\Sigma^\ast$ 
states and could firmly establish some of the missing 
$\Sigma^\ast$ states in Fig.~\ref{fig:masses}.  In order to 
explore the $\Lambda^\ast$ states, the reaction $\gamma +p\to 
K^+\Lambda^\ast$ looks more promising. The $\Sigma^\ast$ 
states decay to $\bar{K} N$ or $\pi\Sigma$ or possibly both, 
as well as $\pi\Lambda$ according to theory estimates. 
Measurements of the branching ratios of these decays will 
further enhance our understanding of these states.  At the 
moment our understanding of light quark confinement in QCD 
is very rudimentary. The suggested measurements would 
contribute to a clarification of this aspect of QCD. 

\item \textbf{Acknowledgments}

This work is supported in part by funds provided by the 
National Science Foundation, Grant No. PHY--1068305.  
\end{enumerate}


\newpage
\subsection{Hadron Physics with High-Momentum Hadron Beams at 
	J-PARC}
\addtocontents{toc}{\hspace{2cm}{\sl H.~Noumi}\par}
\setcounter{figure}{0}
\setcounter{equation}{0}
\halign{#\hfil&\quad#\hfil\cr
\large{Hiroyuki Noumi}\cr
\textit{Research Center for Nuclear Physics (RCNP)}\cr
\textit{Osaka University}\cr
\textit{Osaka, 567-0047, Japan}\cr}

\begin{abstract}
Baryon spectroscopy with heavy flavors provides unique 
opportunities to study internal motions of ingredients 
in baryons, from which we can learn the effective degrees 
of freedom to describe hadrons. We proposed an experiment 
on charmed baryon spectroscopy via the $(\pi^-,D^{\ast -})$ 
reaction at the J-PARC high-momentum beam line. Mass 
spectrum from the ground state to highly-excited states 
of the excitation energy up to more than 1~GeV will be 
observed by means of a missing mass technique. Production 
cross sections and decay branching ratios of these states 
will be measured. It is worthy to compare how nature of 
baryons with different flavors changes. In particular, 
double-strangeness baryons, $\Xi$, are of interest. A 
neutral Kaon beam at JLab is unique to produce $\Xi$ 
baryons. Hadron beams at J-PARC play  complimentary 
roles to the neutral Kaon beam at JLab.
\end{abstract}

\begin{enumerate}
\item \textbf{Baryon Spectroscopy with Heavy Flavors}

``How hadrons are formed from quarks ?" is a fundamental 
question in hadron physics. We know that the quantum 
chromo-dynamics (QCD) is a fundamental theory to describe 
dynamics of quarks and gluons. However, it is still very 
hard to describe hadrons by solving the QCD equation in 
low energy because of its non-perturbative nature of the 
strong interaction, where the coupling constant becomes 
very large when the energy scale is close to the scale 
parameter $\Lambda_{\rm QCD}$. Quarks  drastically change 
their nature below $\Lambda_{\rm QCD}$. Then, constituent 
quarks as effective degree of freedom to describe hadrons 
seem to work rather well. Actually, the constituent quark 
model well describes properties of hadrons in the ground 
state, such as masses, spin-flavor classifications, magnetic 
moment of octet baryons, and so no. However, it sometimes 
fails in excited states. In particular, not only recent 
reports on so-called exotic hadrons,  such as $X$, $Y$, $Z$, 
and pentaquark states in heavy sector but also a 
long-standing puzzle in $\Lambda$(1405) indicate that we 
need a new aspect in describing hadrons. Internal 
correlations among ingredients of hadrons such as diquarks 
and hadron clusters are expected to play an important role. 
Since they are confined in a hadron, hadron spectroscopy to 
look into more details of internal structure or motions of 
the composites in hadrons is necessary.

Since the color magnetic interaction between quarks is 
proportional to the inverse of the quark mass, spin-dependent 
interactions to a heavy quark vanish in the heavy quark mass 
limit. In the heavy quark mass limit, a heavy quark spin and 
the spin of the other system become good quantum numbers. 
This is the so-called heavy quark symmetry of QCD. Let us 
consider a baryon with a heavy quark. A relative motion 
between two light quarks ($\rho$ mode) and a collective 
motion of the light quark pair ($\lambda$ mode) are 
separated in excited states, as illustrated in 
Fig.~\ref{fig:rholam}. This is known as the so-called 
isotope shift. These states are further split due to 
spin-dependent interactions between quarks~\cite{yoshidaU}. 
The spin-correlation between light quarks becomes stronger 
than those to a light quark and a heavy quark due to the 
heavy quark spin symmetry. Internal structure of the baryon 
with a heavy quark is characterized by the two light-quark 
(diquark) correlation. Nature of these baryons is reflected 
in mass, decay width (branching ratio), and production rate.

Therefore, we proposed an experimental study of charmed 
baryons via the ($\pi^-,D^{\ast -}$) reaction on 
hydrogen~\cite{e50U} at the J-PARC high-momentum beam line. 
In the reaction, we reconstruct charmed baryons by means of 
missing mass technique. An excitation spectrum of charmed 
baryon states can be measured independent of their decay 
final states. We could also identify decay modes with 
detecting a decay particle together with scattered $D^{\ast 
-}$ and identifying a daughter particle in a missing mass. 
A branching ratio (partial decay width) of the decay mode 
can be obtained rather easily. Branching ratios provide 
information on diquark motions of excited charmed baryons, 
as described later. This is an advantage of the missing mass 
method.
\begin{figure}[ht!]
\begin{center}
\includegraphics[width=8cm]{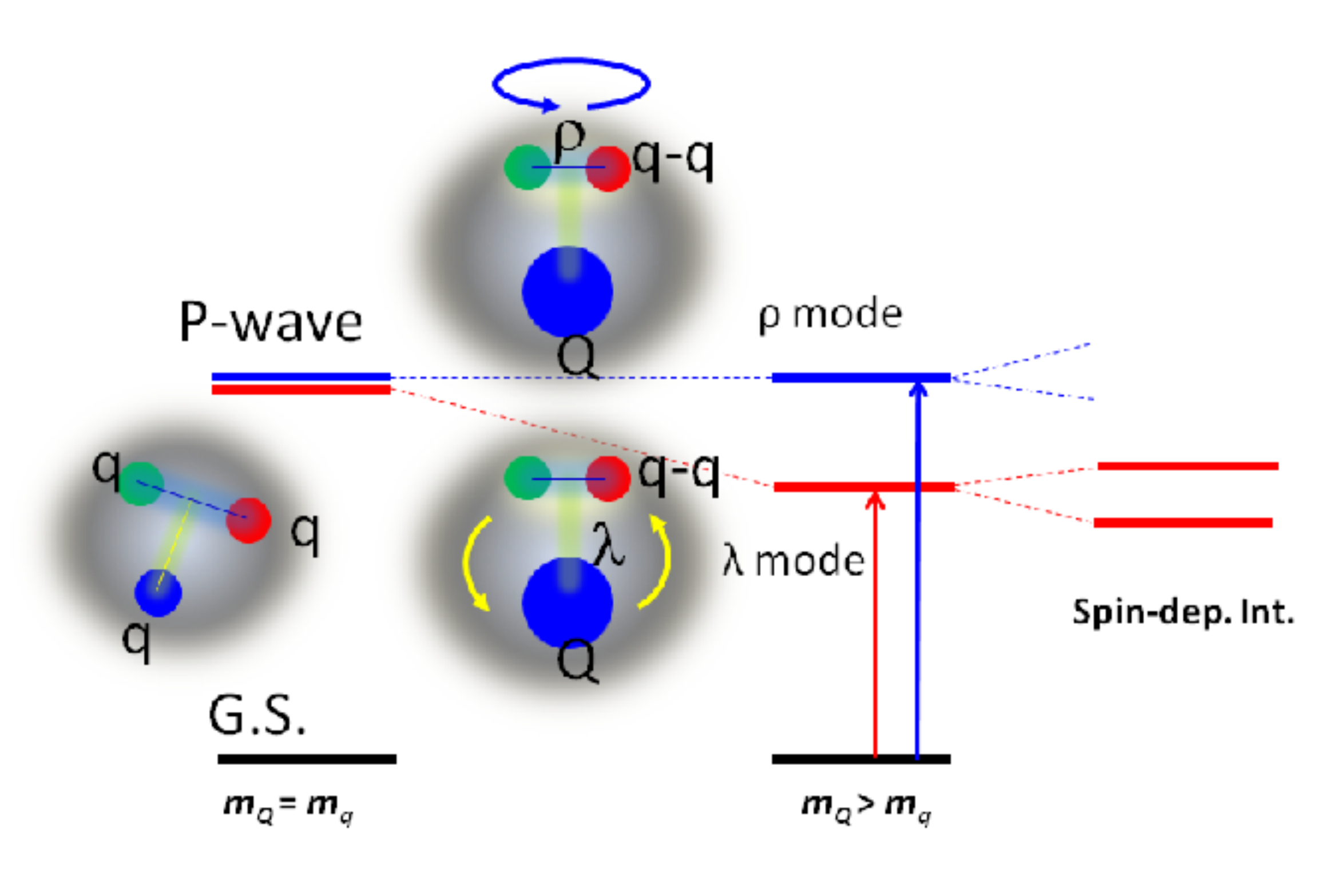}
\end{center}
\centerline{\parbox{0.80\textwidth}{
 \caption{\label{fig:rholam}Schematic level structure of 
	excited baryons with a heavy quark.} } }
\end{figure}
\begin{figure*}[ht!]
\begin{center}
\includegraphics[width=15cm]{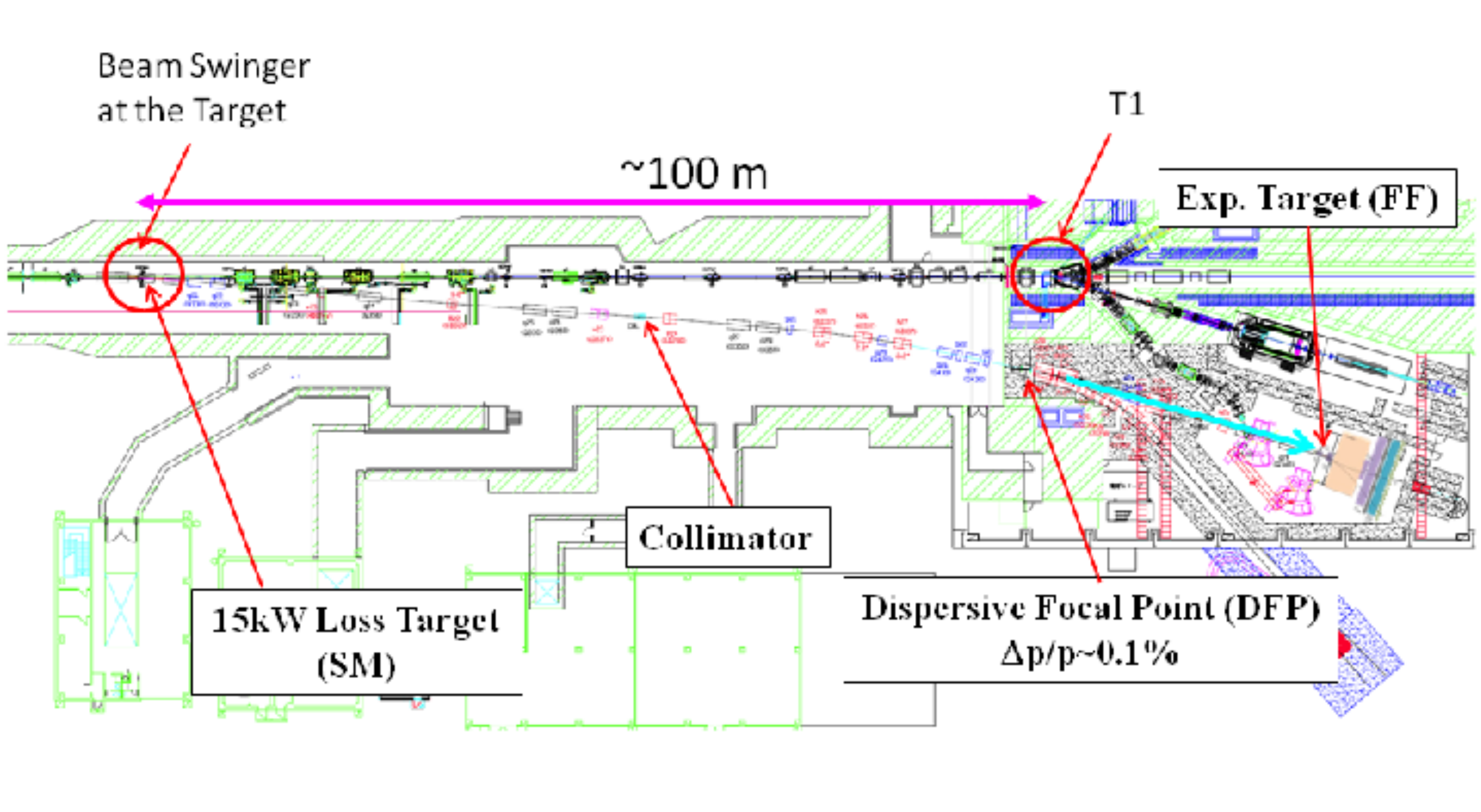}
\end{center}
\centerline{\parbox{0.80\textwidth}{
 \caption{\label{fig:layout}High-momentum beam line at the 
	J-PARC Hadron Experimental Facility.} } }
\end{figure*}

\item \textbf{Beam Line}

A new beam line, called high-momentum beam line, is being 
constructed in the Hadron Experimental Facility of the 
Japan Proton Accelerator Research Complex (J-PARC). It is 
branched from the slow extraction primary beam line in the 
switch yard. A small fraction of the primary beam is 
transported to the experimental area, located about 130~m 
downstream from the branch point, for the E16 experiment 
which aims at measuring spectral changes of vector mesons 
in nuclei~\cite{e16U}. 
 
The high-momentum beam line can deliver secondary beams if 
we install a production target at the branching point. The 
layout of the beam line magnets are arranged so as to 
transport secondary beams up to 20 GeV/$c$, as shown in 
Fig.~\ref{fig:layout}. The beam line is carefully designed 
to realize a dispersive beam at the dispersive focal point, 
where a momentum and a horizontal position of the secondary 
particles are strongly correlated. 
Fig.~\ref{fig:dispersion}-top demonstrates the correlation 
calculated by the DECAY TURTLE~\cite{turtleU}. We place 3 
sextupole magnets to reduce second order aberrations and 
to sharpen the correlation. The calculation tells that a 
momentum resolution of 0.12\% is expected by measuring a 
beam position with a spatial resolution of 1~mm, as 
illustrated in Fig.~\ref{fig:dispersion}-bottom. The 
momentum resolution is mostly determined by a beam size 
at the production target (actually, its image at the 
dispersive focal point). We assume the beam size of 1~mm 
in $\sigma$ in the horizontal direction at the production 
target.
\begin{figure}[ht!]
\begin{center}
\includegraphics[width=8cm]{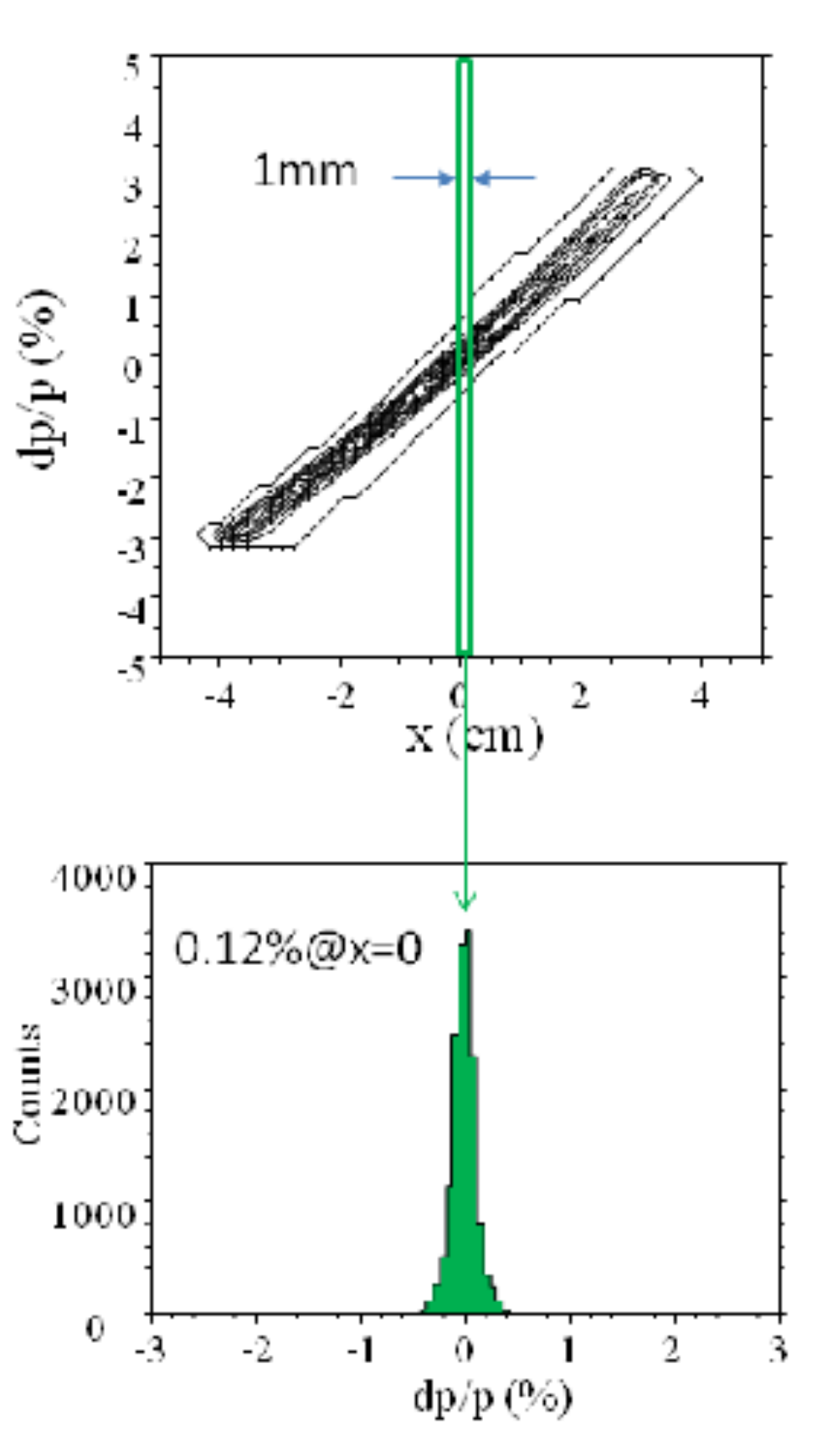}
\end{center}
\centerline{\parbox{0.80\textwidth}{
 \caption{\label{fig:dispersion}Top: correlation between 
	beam momentum and position at the dispersive focal 
	point. 
	Bottom: momentum distribution  within a 1-mm 
	space in horizontal at the beam center.} }}
\end{figure}

The beam line length and acceptance are 133~m and 
1.5~msr$\cdot$\%.  We estimated intensities of secondary 
particles by the so-called Sanford-Wang formula~\cite{swU}, 
assuming that a 6-cm thick platinum target is irradiated 
by a 30-GeV proton beam of 30~kW (15-kW beam loss at the 
target), as shown in Fig.~\ref{fig:intensity}. Here, a 
production angle for negative and positive particles are 
assumed to be 0~degree and 3.9~degrees. We expect that 
the negative pion beam intensity is more than 10$^7$ per 
second at 20~GeV/$c$. 
\begin{figure}[ht!]
\begin{center}
\includegraphics[width=10cm]{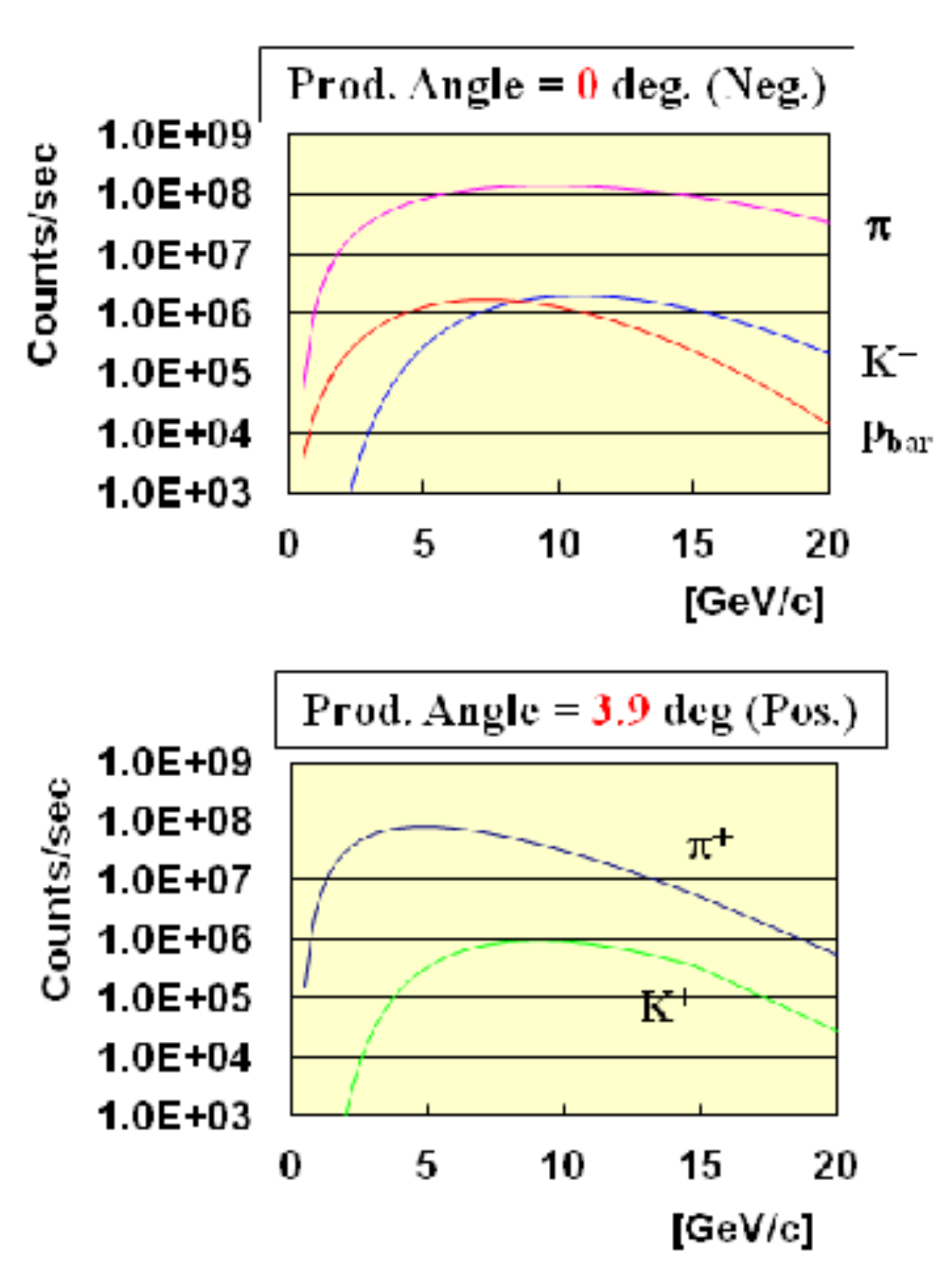}
\end{center}
\centerline{\parbox{0.80\textwidth}{
 \caption{\label{fig:intensity}Intensities of secondary 
	beams calculated by Sanform-Wang's 
	formula~\protect\cite{swU}. See text for assumed 
	conditions.} } }
\end{figure}

\item \textbf{Spectrometer}

An incident pion momentum will be measured at a resolution 
as good as $\sim$0.1\% in the high-momentum beam line. We 
designed a spectrometer system to reconstruct scattered 
$D^{\ast -}$ from its decay chain of $D^{\ast -}\rightarrow 
\bar{D}^0\pi^-$, $\bar{D}^0\rightarrow K^+\pi^-$, as shown 
in Fig.~\ref{fig:spectrometer}. The spectrometer is based 
on a single dipole magnet with a circular pole of 2.1~m in 
diameter and a gap of 1~m. A rigidity of the magnet is 
2.3~Tm. A typical momentum resolution is expected to be 
$\sim$0.5\% at 5~GeV/$c$. A liquid hydrogen target of 57~cm 
in length (4~g/cm$^2$ in thickness) will be placed close to 
the entrance face of the magnet. Fiber trackers with 1~mm 
scintillating fiber will be placed just after the target. A 
set of drift chambers will be placed surrounding the pole 
and after the magnet to detect scattered particles with 
lower and higher momenta, respectively. A ring image 
Cherenkov counter (RICH) with dual radiators of aerogel 
with a reflection index of 1.04 and a C$_4$F$_{10}$ gas 
with an index of 1.00137 will be used for identifying pion, 
Kaon, and proton in a wide momentum range from 2 to 
16~GeV/$c$~\cite{yamagaU}. Time of flight counters will 
be placed to identify scattered particles with lower 
momenta. The spectrometer covers about 60 \% of solid 
angle for scattered $D^{\ast -}$ and about 80 \% for decay 
pions from produced charmed baryons.
\begin{figure}[ht!]
\begin{center}
\includegraphics[width=10cm]{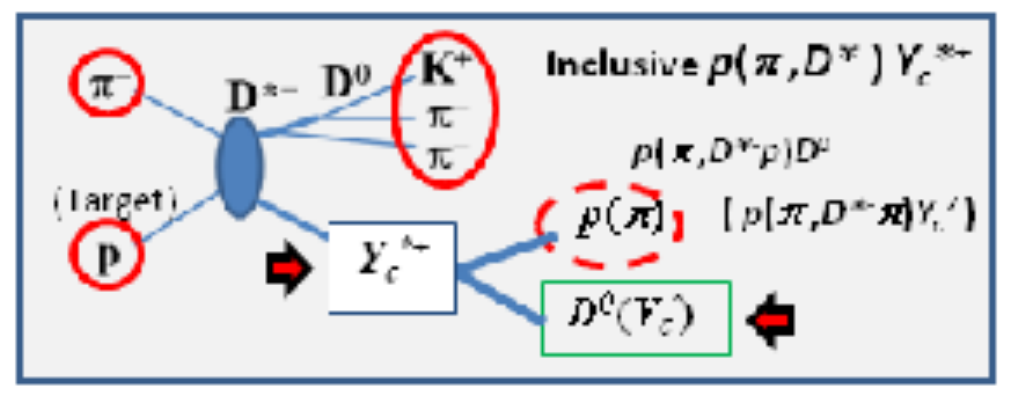}\\
\end{center}
\vspace{0.5cm}
\begin{center}
\includegraphics[width=12cm]{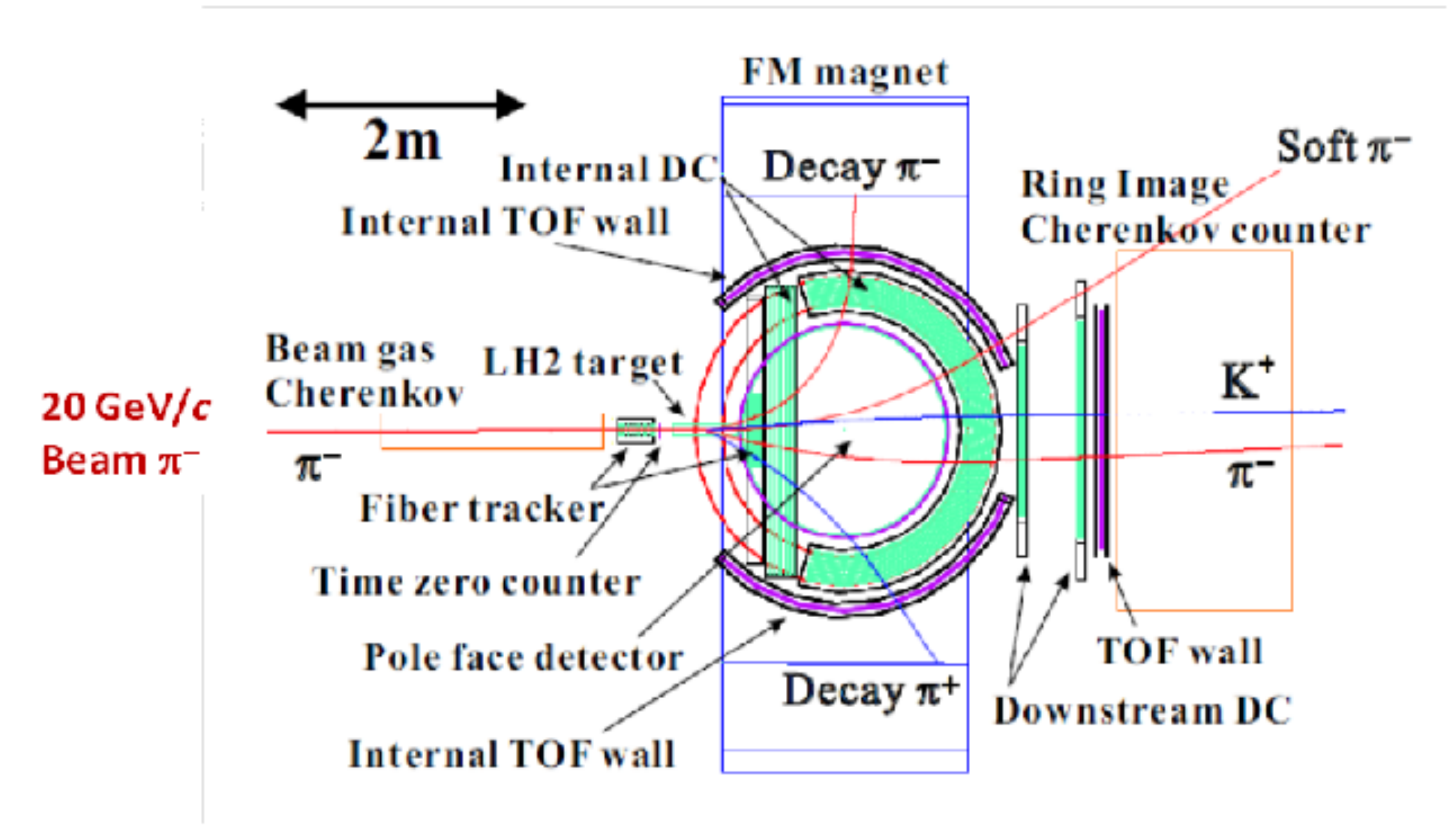}
\end{center}
\centerline{\parbox{0.80\textwidth}{
 \caption{\label{fig:spectrometer}Top: reaction scheme 
	to identify charmed baryons and their decays by 
	means of a missing mass technique. 
	Bottom: designed spectrometer layout.} } }
\end{figure}

Identifying two charmed mesons from the decay final state, 
$K^+\pi^-\pi^-$, we could reduce huge background events of 
$K^+\pi^-\pi^-$ productions by a factor of $\sim$10$^6$. 
Expected charmed baryon spectrum is demonstrated by a Monte 
Carlo simulation in Fig.~\ref{fig:spectrum}. Here, the 
states reported by the Particle Data Group~\cite{pdgU} are 
taken into account. One sees that a series of charmed 
baryons from the ground state to highly excited states 
with higher spins are clearly observed.
\begin{figure}[ht!]
\begin{center}
\includegraphics[width=12cm]{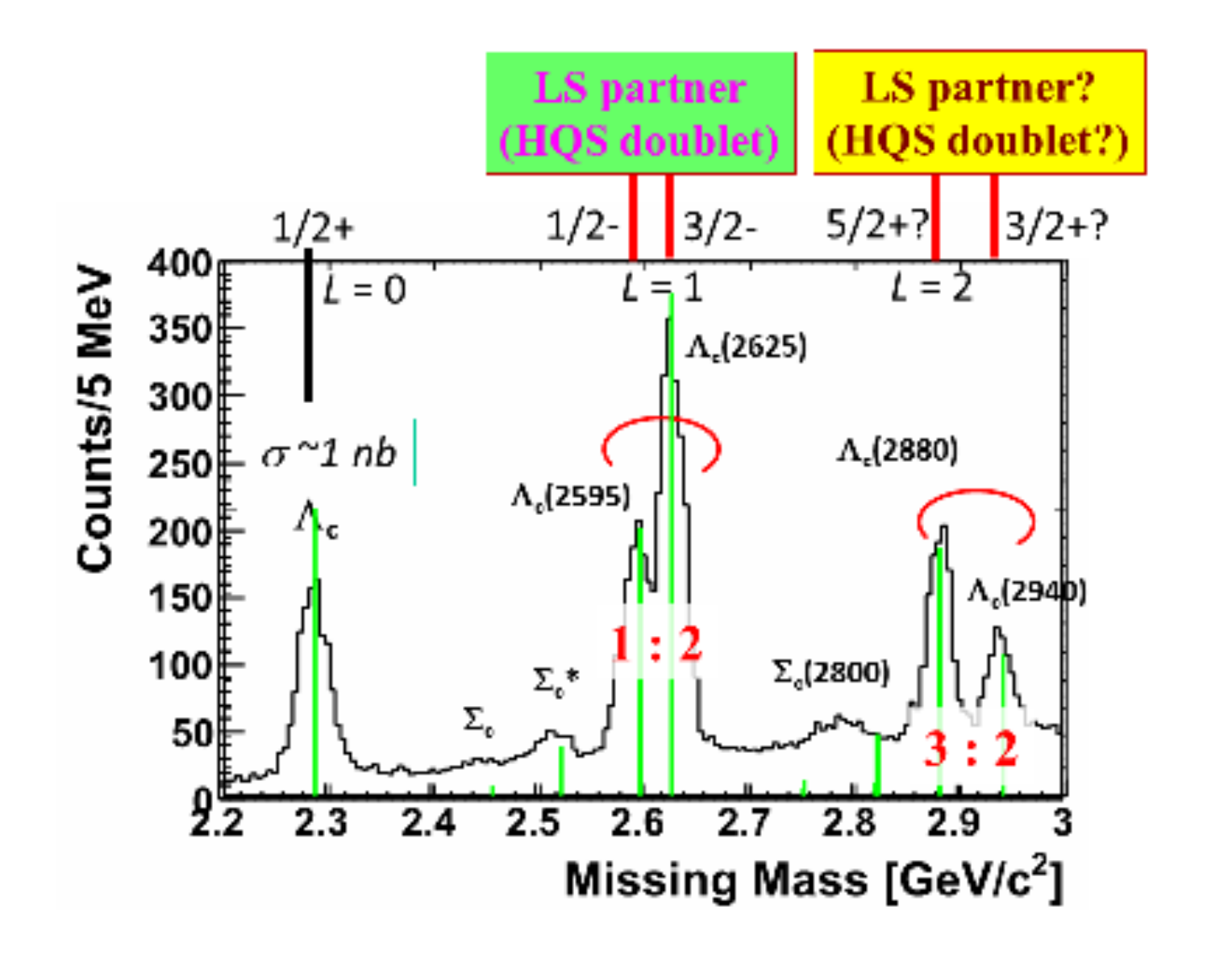}
\end{center}
\centerline{\parbox{0.80\textwidth}{
 \caption{\label{fig:spectrum}Expected missing mass 
	spectrum in the  ($\pi^-,D^{\ast -}$) reaction on 
	hydrogen.} } }
\end{figure}

We found that the production cross sections of the excited 
states relative to that of the ground state do not go down. 
This is an important feature of the $(\pi^-,D^{\ast -}$) 
reaction. We estimated the production rates of the excited 
charmed baryons in the framework of a $D^\ast$ exchange in 
the $t$-channel at a very forward scattering 
angle~\cite{shkimU}. Employing harmonic oscillator wave 
functions of constituent quarks in the initial and final 
baryons, we estimate that the production rate, R, is 
expressed as 
\newpage
\begin{eqnarray}
	R&\sim&<\phi_f|\sqrt{2}\sigma_-\exp(i{\bf q_{eff}r})|\phi_i>,\\
	&\sim&(q_{eff}/A)^L\exp({-q_{eff}^2/2A^2)},
\end{eqnarray}
where $\phi_i$ and $\phi_f$ represent initial and final states 
of baryons. An effective momentum transfer $q_{eff}$, taking a 
recoil effect of the residual $ud$ diquark, is as large as 
1.4~GeV/$c$ in the $p(\pi^-,D^{\ast -})$ reaction at the pion 
beam momentum of 20~GeV/$c$. An oscillator parameter $A$ is 
taken to be $\sim$0.4~GeV, which corresponds to the inverse of 
a typical baryon size. In this reaction, a $u$ quark in a proton 
is converted to a $c$ quark in the final charmed baryon in the 
reaction. The $ud$ diquark behaves as a spectator. Thus, this 
reaction well populates $\lambda$-mode excited states, where an 
angular momentum $L$ is introduced between a $c$ quark and a 
diquark. Due to a large factor of $(q_{eff}/A)$, an absolute 
value of the production cross section is reduced very much. 
On the other hand, the ratio of $R$ for an excited state with 
$L$ to that for the ground state ($L=0$) is $\sim(q_{eff}/A)^L$, 
which does not go down even for $L>0$. The $(\pi^-,D^{\ast -}$) 
reaction is suitable to populate higher spin states.

The $\lambda$-mode $\Lambda_c$ baryons with $L>0$ has two spin 
states coupled to $L\pm1/2$. These states are LS partners. 
Therefore, we find that the production ratio of the two states 
should be $L$:$L$+1. Reversely, we can determine the 
spin-parity of the $\lambda$-mode $\Lambda_c$ baryons by 
measuring the cross sections.
 
So far, a production cross section of $\Lambda_c$ in the 
p($\pi^-,D^{\ast -})$ reaction has not been measured. Only 
upper limit, 7~nb, was reported in 1985~\cite{christensenU}.
We estimate the production cross section to be a few nano 
barn at incident pion momentum of 20~GeV/$c$~\cite{shkim2U} 
by employing a framework of reggeon exchange model, which 
describes binary peripheral reactions well at high energy. 
We expect to observe 1000 events of the ground state 
$\Lambda_c$ production for 100 days.

Decay branching ratios carry information on internal 
structure of a baryon. The ratio of decay into a heavy 
meson and a light baryon to that into a light meson and a 
heavy baryon is of particular interesting. The former is 
expected to be dominant, if it is energetically allowed, 
in $\lambda$-mode excited baryon, which is an orbital 
excitation between a heavy quark and a light diquark.
The situation is to be opposite in $\rho$-mode. One can 
find a suggestion in the case of $\Lambda$(1520), which is 
a P-wave hyperon with spin-parity of 3/2$^-$. In 
$\Lambda$(1520) dominantly decays into a Kaon and a nucleon, 
while a Q-value in the decay is smaller than that in the 
decay into a pion and a $\Sigma$ hyperon. The $\Lambda$(1520) 
hyperon can be classified as a $\lambda$-mode hyperon 
although $\rho/\lambda$ mode classification has yet to be 
established in any baryon excited states. Systematic 
measurements of decay branching ratios for the excited 
charmed baryons are of particular importance.

\item \textbf{Baryon Spectroscopy with Different Flavors}

The above-mentioned discussion on internal structure of 
baryons with a single heavy quark can be extended to 
baryons with double heavy quarks. In the case of double 
heavy-quark baryons, the order of the excitation energy 
for $\lambda$ and $\rho$ modes  interchanges. Here, the 
$\lambda$ mode is a motion of the light quark to the 
heavy-quark pair, and the $\rho$ mode is a relative 
motion between two heavy quarks. One expects that a 
$\lambda$ mode excited state favors a decay into a light 
meson and a double heavy-quark baryon. A $\rho$ mode state 
may dominantly decay into a single-heavy meson and a 
single-heavy baryon.

A several states of cascade hyperons $\Xi$ are 
listed~\cite{pdgU}. Little is known about their spin-parities 
and decay branching ratios. The $(\bar{K},K^+)$ reaction is 
one of promising reactions to produce cascade hyperons.
Since the $(\bar{K},K^+)$ reaction has no single-meson 
exchange process in $t$ channel, $\Xi$ productions at 
backward angles are expected to play a principal role.
$\rho$-mode $\Xi$ hyperons may be populated well through 
$u$-channel process. It is quite worthy to measure 
production rates and decay branching ratios of $\Xi$ 
hyperons.

\item \textbf{Concluding Remark}

\begin{itemize}
\item Masses, decay branching ratios, and production 
	rates of baryons with heavy flavors provide 
	information on internal motions of ingredients, 
	such as diquark correlation. 
\item We proposed an experiment on charmed baryon 
	spectroscopy via the $(\pi^-, D^{\ast -})$ 
	reaction at the J-PARC high-momentum beam line. 
	We will measure a mass spectrum of charmed 
	baryons from the ground state to highly excited 
	states in an excitation energy range of more than 
	1~GeV by means of a missing mass technique. 
	Production cross sections and decay branching 
	ratios of produced charmed baryons will be 
	measured.
\item The present argument on baryon spectroscopy with a 
	charm quark should be extended to those with 
	different flavors. In particular, $\Xi$ baryons 
	are of interest as double-heavy quark system that 
	can be accessible in experiment. Neutral Kaon beam 
	at JLab is unique in hadron spectroscopy and plays 
	a complimentary role to the J-PARC. Constructive 
	collaboration to integrate efforts to realize 
	hadron spectroscopy with hadron beams in JLab and 
	J-PARC is desired.
\end{itemize}
\end{enumerate}


\newpage
\subsection{Cascade Production in $\overline{K}$- and Photon-Induced 
	Reactions}
\addtocontents{toc}{\hspace{2cm}{\sl K.~Nakayama, B.C.~Jackson,
	Y.~Oh, and H.~Haberzettl}\par}
\setcounter{figure}{0}
\setcounter{footnote}{0}
\setcounter{equation}{0}
\halign{#\hfil&\quad#\hfil\cr
\large{Kanzo~Nakayama and B.C.~Jackson}\cr
\textit{Department of Physics and Astronomy}\cr
\textit{University of Georgia}\cr
\textit{Athens, GA 30602, U.S.A.}\cr\cr
\large{Yongseok Oh}\cr
\textit{Department of Physics}\cr
\textit{Kyungpook National University}\cr
\textit{Daegu 702-701, Korea \&}\cr
\textit{Institute for Nuclear Studies and Department of Physics}\cr
\textit{The George Washington University}\cr
\textit{Washington, DC 20052, U.S.A.}\cr\cr
\large{Helmut Haberzettl}\cr
\textit{Institute for Nuclear Studies and Department of Physics}\cr
\textit{The George Washington University}\cr
\textit{Washington, DC 20052, U.S.A.}\cr}

\begin{abstract}
The $\bar{K} + N\to K + \Xi$ and $\gamma + N\to K + K + \Xi$ reactions 
are investigated in a combined analysis within an effective Lagrangian 
approach to learn about the basic features of these reactions. Such a 
study should help construct more complete reaction models within a full 
coupled-channels approach to extract relevant physics information from 
forthcoming experimental data in the multi-strange particle physics 
programs at modern experimental facilities including J-PARC and JLab. 
Among the above-threshold three- and four-star $S=-1$ resonances 
considered in this work, a minimum of three resonances, namely 
$\Sigma(2030)7/2^+$, $\Sigma(2265)5/2^-$, and $\Lambda(1890)3/2^+$, 
are found to be required to reproduce the available data in the 
considered $\bar{K}$- and photon-induced reactions. Among them, the 
$\Sigma(2030)7/2^+$ resonance is shown to play a clear and important 
role in both reactions.
\end{abstract}

\begin{enumerate}
\item \textbf{Introduction}

One of the major interests in baryon spectroscopy in the strangeness 
sector is the possibility to learn about the properties of the so-called
multi-strangeness baryons, \textit{i.e.}, baryons with strangeness 
quantum number $S < -1$. Although the multi-strangeness baryons have 
played an important role in the development of our understanding of 
strong interactions, and thus should be an integral part of any baryon 
spectroscopy program, the current knowledge of these baryons is still 
extremely limited. In fact, the SU(3) flavor symmetry allows as many 
$S=-2$ baryon resonances, called $\Xi$, as there are $N$ and $\Delta$ 
resonances combined ($\sim 27$); however, until now, only eleven $\Xi$ 
baryons have been discovered~\cite{PDG14K}. Among them, only three 
[ground state $\Xi(1318)1/2^+$, $\Xi(1538)3/2^+$, and $\Xi(1820)3/2^-$] 
have their quantum numbers assigned. This situation is mainly due to 
the fact that multi-strangeness particle productions have relatively 
low yields. For example, if there are no strange particles in the 
initial state, $\Xi$ is produced only indirectly and the yield is 
only of the order of nb in the photoproduction reaction~\cite{CLAS07bK}, 
whereas the yield is of the order of $\mu$b~\cite{FMMR83bK} in the 
hadronic $\bar{K}$-induced reaction, where the $\Xi$ is produced 
directly because of the presence of an $S=-1$ $\bar{K}$ meson in the 
initial state. The production rates for $\Omega$ baryons with $S=-3$ 
are much lower~\cite{RHHK13K}. The initiative to having a $K_L$ beam at 
the Thomas Jefferson National Accelerator Facility (JLab) in particular 
to study, among other things, multi-strangeness baryon spectroscopy is, 
therefore, extremely valuable.

The study of multi-strangeness baryons has started to attract renewed
interests recently. Indeed, the CLAS Collaboration at JLab plans to 
initiate a $\Xi$ spectroscopy program through the photoproduction 
reaction using the upgraded 12-GeV machine, and measure exclusive 
$\Omega$ photoproduction for the first time~\cite{VSC12K}. Some data 
for the production of the $\Xi$ ground state, obtained from the 6-GeV 
machine, are already available~\cite{CLAS07bK}.  J-PARC is going to 
study the $\Xi$ baryons via the $\bar{K} + N \to K + \Xi$ process 
(which is the reaction of choice for producing $\Xi$)~\cite{JPARCKbarK,
JPARCbK} in connection to its program proposal for obtaining information 
on $\Xi$ hypernuclei spectroscopy. It also plans to study the $\pi + N 
\to K + K + \Xi$ reaction as well as $\Omega$ production. At the FAIR 
facility of GSI, the reaction $\bar{p} + p \to \bar{\Xi} + \Xi$ will 
be studied by the $\overline{\rm P}$ANDA Collaboration~\cite{PANDA09K}.

In the present work, we concentrate on the production of $S=-2$ $\Xi$, 
in particular, on the production reaction processes of the ground state 
$\Xi$:
\begin{eqnarray}
	\bar{K} + N &\rightarrow K + \Xi ~, \label{reac1}\\
	\gamma +N  &\rightarrow K + K +\Xi ~.
	\label{reac}
\end{eqnarray}

The $\bar{K}$-induced reaction (\ref{reac1}) has been studied 
experimentally mainly throughout the 60's which was followed by 
several measurements made in the 70's and 80's. The existing data 
are rather limited and suffer from large uncertainties. There 
exist only very few early attempts to understand this reaction. 
Recent calculations are reported by Sharov \textit{et 
al.\/}~\cite{SKL11K} and by Shyam \textit{et al.\/}~\cite{SST11K}. 
Although the analyses of both works are based on very similar 
effective Lagrangian approaches, the number of $S=-1$ hyperon 
resonances included as intermediate states are different.  While 
in Ref.~\cite{SKL11K} only the $\Sigma(1385)$ and $\Lambda(1520)$ 
were considered in addition to the above-threshold $\Sigma(2030)$ 
and $\Sigma(2250)$ resonances, in Ref.~\cite{SST11K} eight of the 
3- and 4-star $\Lambda$ and $\Sigma$ resonances with masses up to 
2.0~GeV have been considered. While the authors of 
Ref.~\cite{SKL11K} pointed out the significance of the 
above-threshold resonances, the authors of Ref.~\cite{SST11K} have 
found the dominance of the sub-threshold $\Lambda(1520)$ resonance. 
Reaction~(\ref{reac}) has been also considered by Magas \textit{et 
al.}~\cite{MFR14K} within the coupled-channels Unitarized Chiral 
Perturbation approach when determining the parameters of the 
next-to-leading-order interactions. The authors of Ref.~\cite{MFR14K} 
have added the $\Sigma(2030)$ and $\Sigma(2250)$ resonances into 
their calculation to improve the fit quality to the total cross 
section data. Also the Argonne-Osaka group~\cite{KNLS14K,KNLS15K} 
reported applying their Dynamical Coupled Channels (DCC) approach 
to $\bar{K}$-induced two-body reactions for center-of-momentum 
energies up to $W = 2.1$~GeV. Some of the model-independent 
aspects of the reaction~(1) have been studied recently by the 
present authors~\cite{NOH12K,JOHN14K}.

We note here that the proper identification of resonances and the 
reliable extraction of their parameters require detailed knowledge 
of the analytic structures of the scattering amplitude that, to 
date, can only be obtained through a full coupled-channel treatment, 
such as that of Refs.~\cite{KNLS14K,KNLS15K}.  However, because the 
currently available data in the $K\Xi$ channel are scarce and of 
low quality, they do not provide sufficient constraints for the 
model parameters to permit an in-depth analysis of that channel. In 
this context, we mention that a coupled-channel partial-wave 
analysis of $\bar{K}$-induced reactions up to $W=2.1$~GeV has also 
been performed recently by the Kent State University 
group~\cite{ZTSM13aK,ZTSM13bK} which includes seven reaction channels, 
but not the $K\Xi$ channel.

The available experimental data for the photon-induced reaction (2) 
are also very scarce. In fact, the only data available for this 
reaction in the resonance energy region are those from 
JLab~\cite{CLAS07bK} using the 6-GeV machine. Specifically, the 
total cross sections, both the $K$ and $\Xi$ angular distributions 
and the $KK$ and $K\Xi$ invariant mass distributions are
available. Theoretical studies of this reaction are scarce, too. To 
date, the work of Refs.~\cite{NOH06K,MON11K} is the only one that 
analyzes the JLab data of Ref.~\cite{CLAS07bK}.

One of the purposes of the present work is to search for a clearer
evidence of the $S=-1$ hyperon resonances in reactions (1) and 
(\ref{reac}).  However, we emphasize that our main interest here 
lies not so much in the accurate extraction of $S=-1$ hyperon 
resonance parameters, but in an exploratory study to learn about 
the pertinent reaction mechanisms and, in particular, to identify 
the resonances that come out to be most relevant for the description 
of the existing $\Xi$ production data. In fact, with the exception 
of the $\Sigma(2250)$ resonance, whose mass was adjusted slightly 
to better reproduce the observed bump structure in the total cross 
section in the charged $\Xi$ production, the masses and widths of 
the resonances incorporated here are taken from other sources.
Only the product of the coupling constants and the cutoff parameters 
in the corresponding form factors are adjusted in the present work.

\item \textbf{Formalism}

In the present work, we perform an analysis of the existing data
based on \textcolor{blue}{a} relativistic effective Lagrangian
approach that includes a phenomenological contact amplitude which
accounts for the rescattering contributions and/or unknown
(short-range) dynamics that have not been included explicitly
into the model. For photoproduction, local gauge invariance as
dictated by the appropriate generalized Ward-Takahashi identity is
strictly enforced~\cite{HNK2006K}. Figures~\ref{fig:Kbardiagrams}
and \ref{fig:Photodiagrams} display the Feynman diagrams considered
in the present work for the $\bar{K}$- and photon-induced reactions,
(\ref{reac1}) and (\ref{reac}), respectively. Further details of
the model can be found in Ref.~\cite{JOHN15K} for reaction (1) and,
in Refs.~\cite{NOH06K,MON11K}, for reaction (\ref{reac}). While the
tree-level model used here is not very sophisticated, it captures
the essential aspects of the processes in question.  As such, the
use of a simplified and flexible model is particularly well suited
for a situation, such as for the reactions~(\ref{reac1}) and
(\ref{reac}), where scarce and/or poor data prevent a more detailed
and complete treatment.  The present study is our first step toward
building a more complete reaction model capable of reliably
extracting the properties of hyperons from the forthcoming
experimental data, in addition to providing some guidance for
planning future experiments.
\begin{figure}[ht!]
\begin{center}
\includegraphics[width=0.4\columnwidth,clip=]{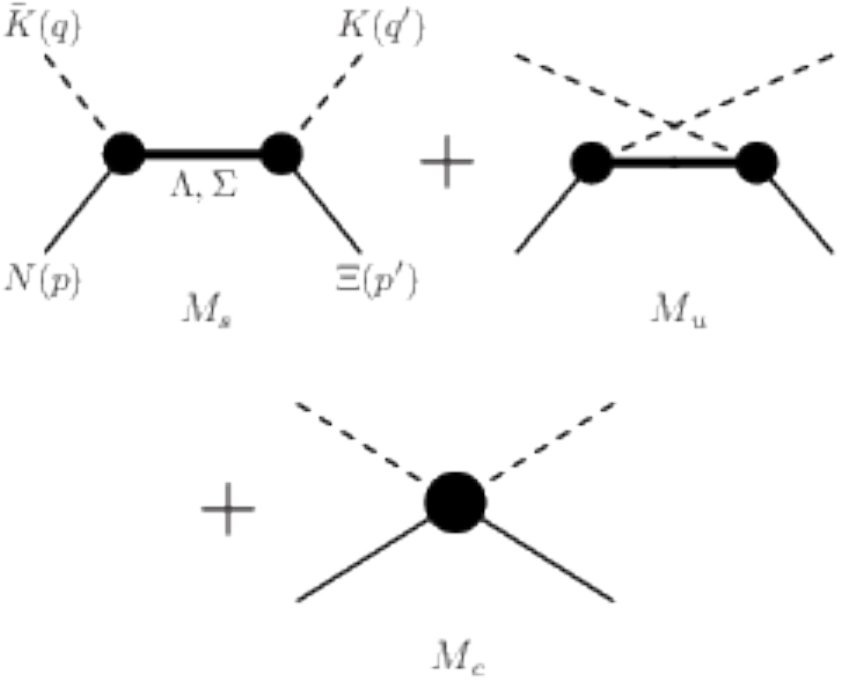}
\end{center}
 \caption{\label{fig:Kbardiagrams}%
	Diagrams describing the amplitude for reaction (1) in 
	the present calculation. The labeling of the external 
	legs of the $s$-channel diagram, $M_s$, follows the 
	reaction equation  (\protect\ref{reac}); the labels 
	apply correspondingly also to the external legs of 
	the $u$-channel diagram, $M_u$, and the contact term 
	$M_c$. The intermediate hyperon exchanges, $\Lambda$ 
	and $\Sigma$, indicated for $M_s$ also appear in $M_u$. 
	The details of the formalism, including the contact 
	amplitude, $M_c$, are given in 
	Ref.~\protect\cite{JOHN15K}.}
\end{figure}
\begin{figure}[ht!]
\centering
\vskip -0.4cm
\includegraphics[width=0.5\textwidth,angle=0,clip]{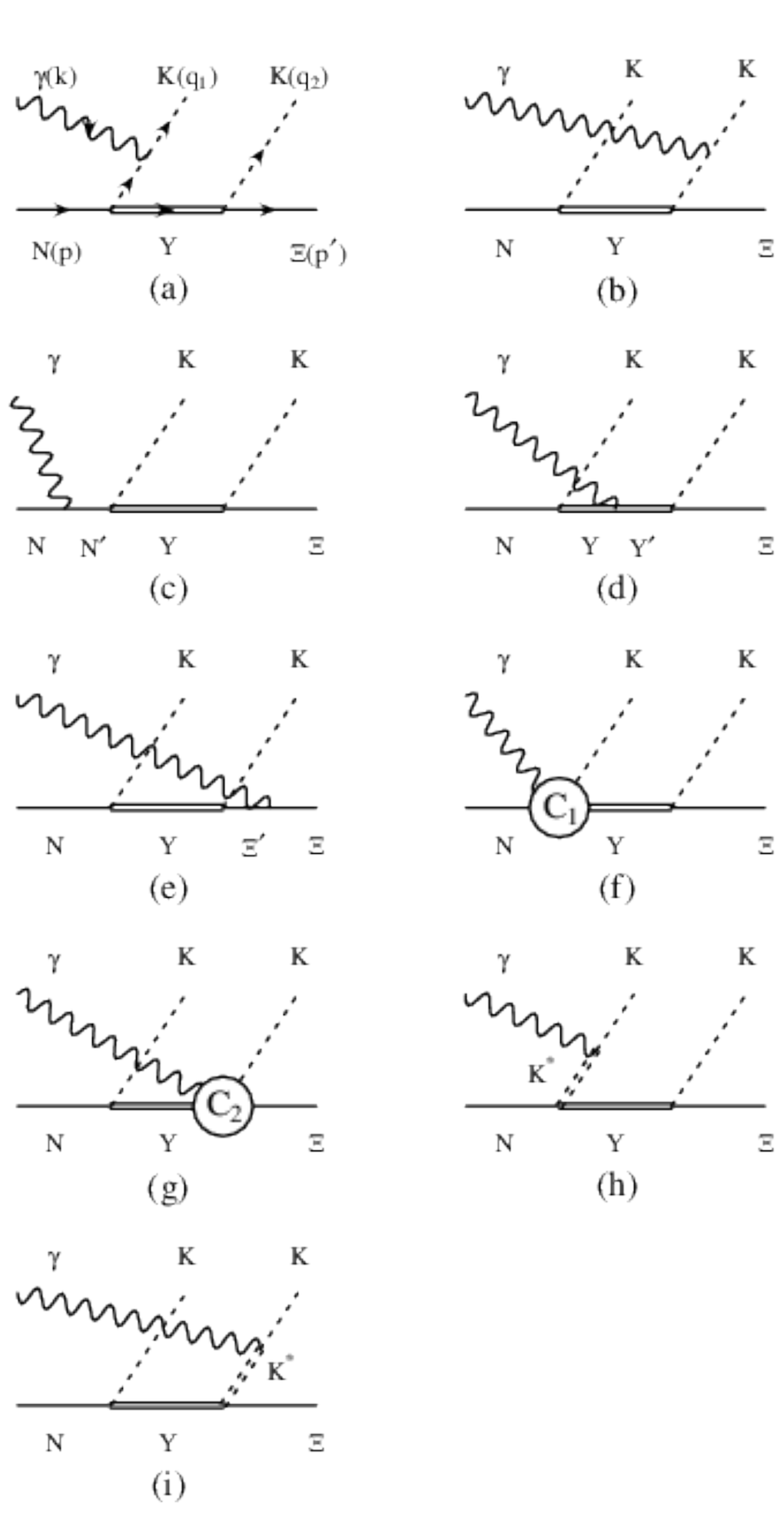}
\centerline{\parbox{0.80\textwidth}{
 \caption{\label{fig:Photodiagrams}
	Diagrams contributing to the reaction mechanism of reaction 
	(\protect\ref{reac}).  The intermediate baryon states are 
	denoted as $N'$ for the nucleon and $\Delta$ resonances, 
	$Y,Y'$ for the $\Lambda$ and $\Sigma$ resonances, and 
	$\Xi'$ for $\Xi(1318)$ and $\Xi(1530)$. The intermediate 
	mesons in the $t$-channel are $K$ [(a) and (b)] and
	$K^\ast$ [(h) and (i)]. The diagrams (f) and (g) contain 
	the generalized contact currents that maintain gauge 
	invariance of the total amplitude. Diagrams corresponding 
	to (a)--(i) with $K(q_1) \leftrightarrow K(q_2)$ are also 
	understood. The details of the formalism, including the 
	contact amplitude, $M_c$, are discussed in 
	Refs.~\protect\cite{NOH06K,MON11K}.} } }
\end{figure}

\item \textbf{Results}

We now turn to a selected set of results of the present work which 
treats the reactions (\ref{reac1}) and (\ref{reac}) consistent with 
each other. It should be mentioned that, although similar, the 
results we show here differ from those shown in Refs.~\cite{JOHN15K,
NOH06K,MON11K}, for the model parameters have been readjusted to 
reproduce the available data for both reactions considered 
simultaneously. As far as the $S=-1$ hyperon contributions are 
concerned, our analysis reveals that a minimum of three 
above-the-threshold resonances, namely the $\Sigma(2030)7/2^+$, 
$\Sigma(2250)5/2^-$ and $\Lambda(1890)3/2^+$ resonances, in addition 
to the $\Sigma(1385)3/2^+$ and the ground states $\Lambda(1116)$ and 
$\Sigma(1193)$, suffice to reproduce all the available data in both 
the $\bar{K} + N \to K + \Xi$ and $\gamma + N \to K + K + \Xi$
reactions.

\begin{enumerate}
\item \textbf{\boldmath $\bar{K} + N \to K + \Xi$ Reaction}

\begin{figure*}[ht!]
\centering
\includegraphics[width=0.46\textwidth,clip=]{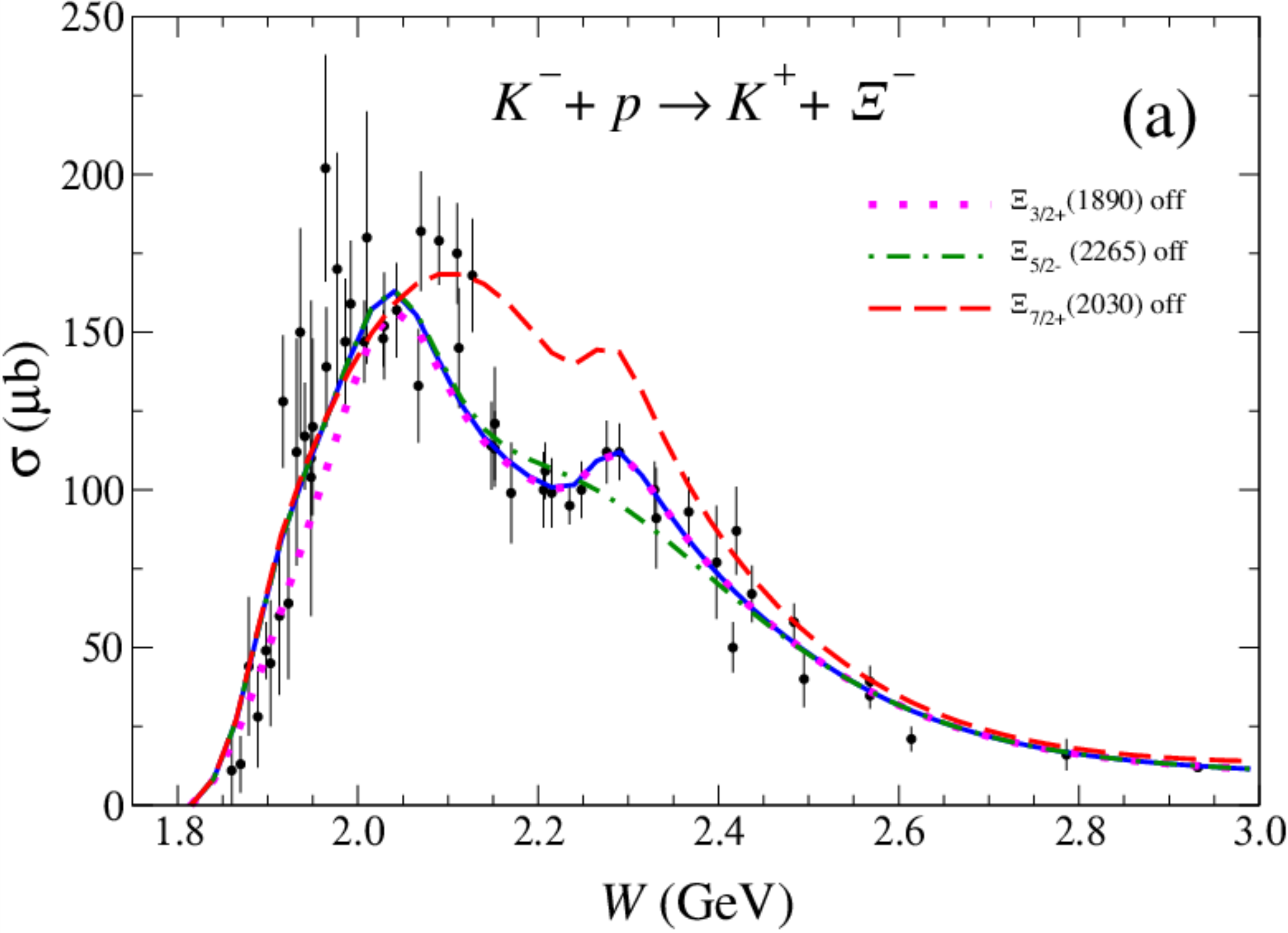} \qquad
\includegraphics[width=0.46\textwidth,clip=]{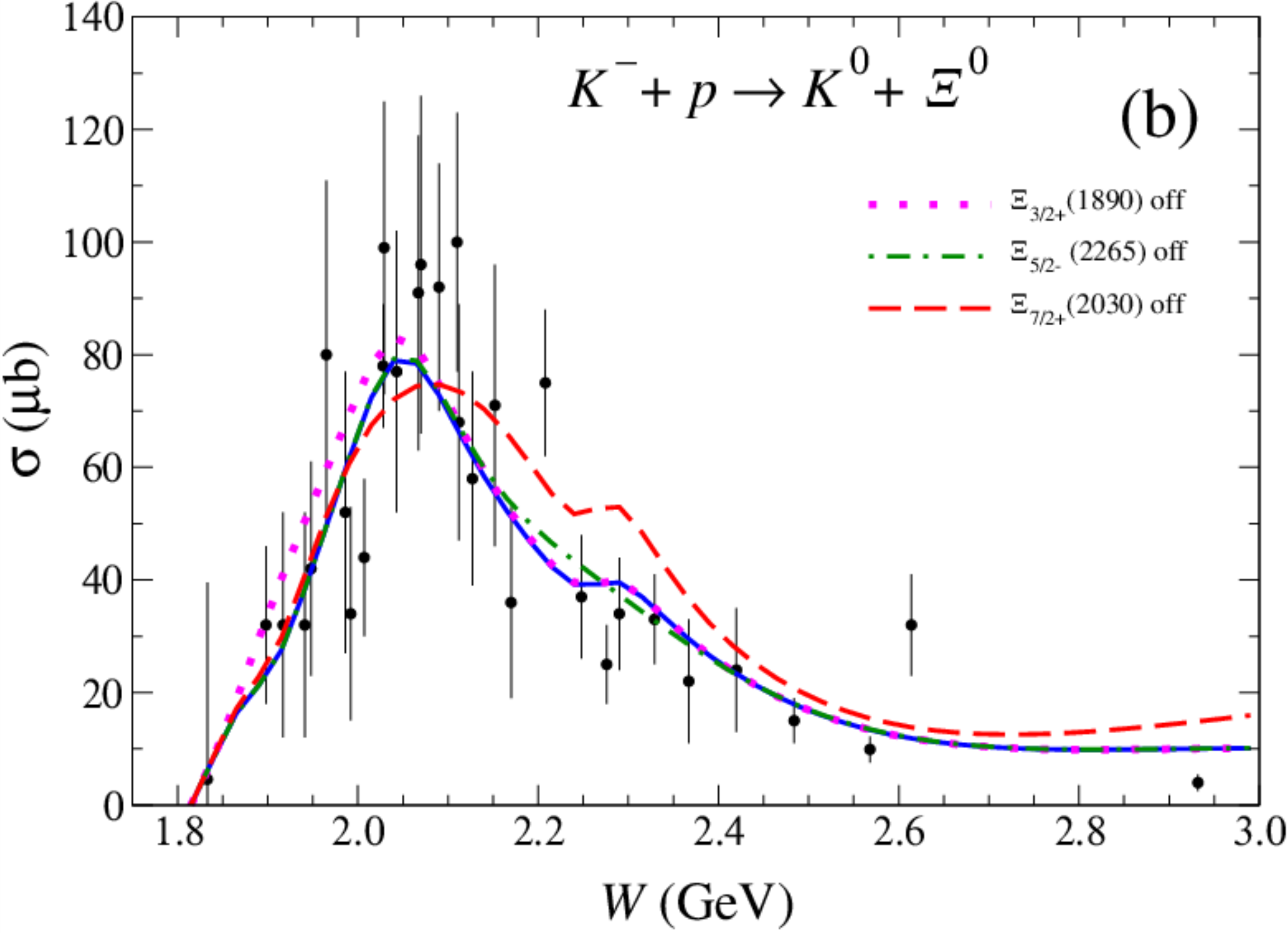}
\centerline{\parbox{0.80\textwidth}{
 \caption{Total cross section results with individual 
	resonances switched off
	(a) for $K^-+p\rightarrow K^++\Xi^-$ and 
	(b) for $K^-+p\rightarrow K^0+\Xi^0$. The blue 
	solid lines represent the full result. The red 
	dashed lines, which almost coincide with the 
	blue lines represent the result with 
	$\Lambda(1890)$ switched off. The green 
	dash-dotted lines represent the result with 
	$\Sigma(2030)$ switched off and the magenta 
	dash-dash-dotted lines represent the result 
	with $\Sigma(2250) 5/2^-$ switched off.
	The data are the digitized version from 
	Ref.~\protect\cite{SKL11K}.} \label{txsc_RR} } }
\end{figure*}

In Fig.~\ref{txsc_RR}, we illustrate the amount of the above-threshold 
resonance contributions of the present model to the total cross 
sections in reaction (\ref{reac1}). We do this by comparing the full 
results (blue solid curves) to the result found by switching off one 
resonance at a time. We see in Fig.~\ref{txsc_RR}(a) that the largest 
effect of $\Sigma(2030)$ on the cross sections is in the range of $W 
\sim 2.0$ to 2.4~GeV. This resonance is clearly needed in our model 
to reproduce the data. It also affects the recoil polarization as 
will be discussed later. We note that the present model yields the 
product of the branching ratios $\mbox{Br}\big(\Sigma(2030) \to
KN\big)\times \mbox{Br}\big(\Sigma(2030)\to K\Xi\big) \approx 15.6\%$ 
which may be contrasted with the corresponding values of $\approx 
16.1\%$ (model A) and $\approx 20.4\%$ (model B) extracted in 
Ref.~\cite{KNLS15K} within a DCC approach.
\footnote{Note that only the product of the $KYN$ and $KY\Xi$ 
	coupling constants ($Y= \Lambda,\Sigma$ resonances) is 
	sensitive to the data in the present model.}
The $\Lambda(1890)$ affects the total cross section in the range of 
$W \sim 1.9$ to 2.1~GeV, and the $\Sigma(2250)5/2^-$ contributes 
around $W \sim 2.2$~GeV, where it is needed to reproduce the observed 
bump structure. A more accurate data set is clearly needed for a more 
definitive answer about the roles of the $\Lambda(1890)$ and 
$\Sigma(2250)$ resonances. Figure~\ref{txsc_RR}(b), for the neutral 
$\Xi^0$ production, also shows a similar feature observed in the 
$\Xi^-$ case for the $\Sigma(2030)$ resonance. Here, the influence 
of the $\Sigma(2250)5/2^-$ is smaller and  that of the $\Lambda(1890)$ 
is hardly seen. Recall that there is no $u$-channel $\Lambda$ 
contribution in the neutral $\Xi^0$ production.

A peculiar feature of the $K^- + p \to K^+ + \Xi^-$ process is that 
it is dominated by the $P$ and $D$ partial-waves (not shown here). 
In particular, the $P$-wave dominates the total cross section even 
down to energies very close to threshold. This is also corroborated
by the DCC calculation of Ref.~\cite{KNLS14K}. The experimental total 
cross section data ($\sigma$) divided by the magnitude of the relative 
three-momentum in the final state ($p'$), $\sigma / p'$, as a function 
of $p'^2$, reveal essentially a linear dependence near threshold, a 
model-independent indication  of the $P$-wave contribution.
\begin{figure}[ht!]
\centering
\includegraphics[width=0.6\textwidth,clip=]{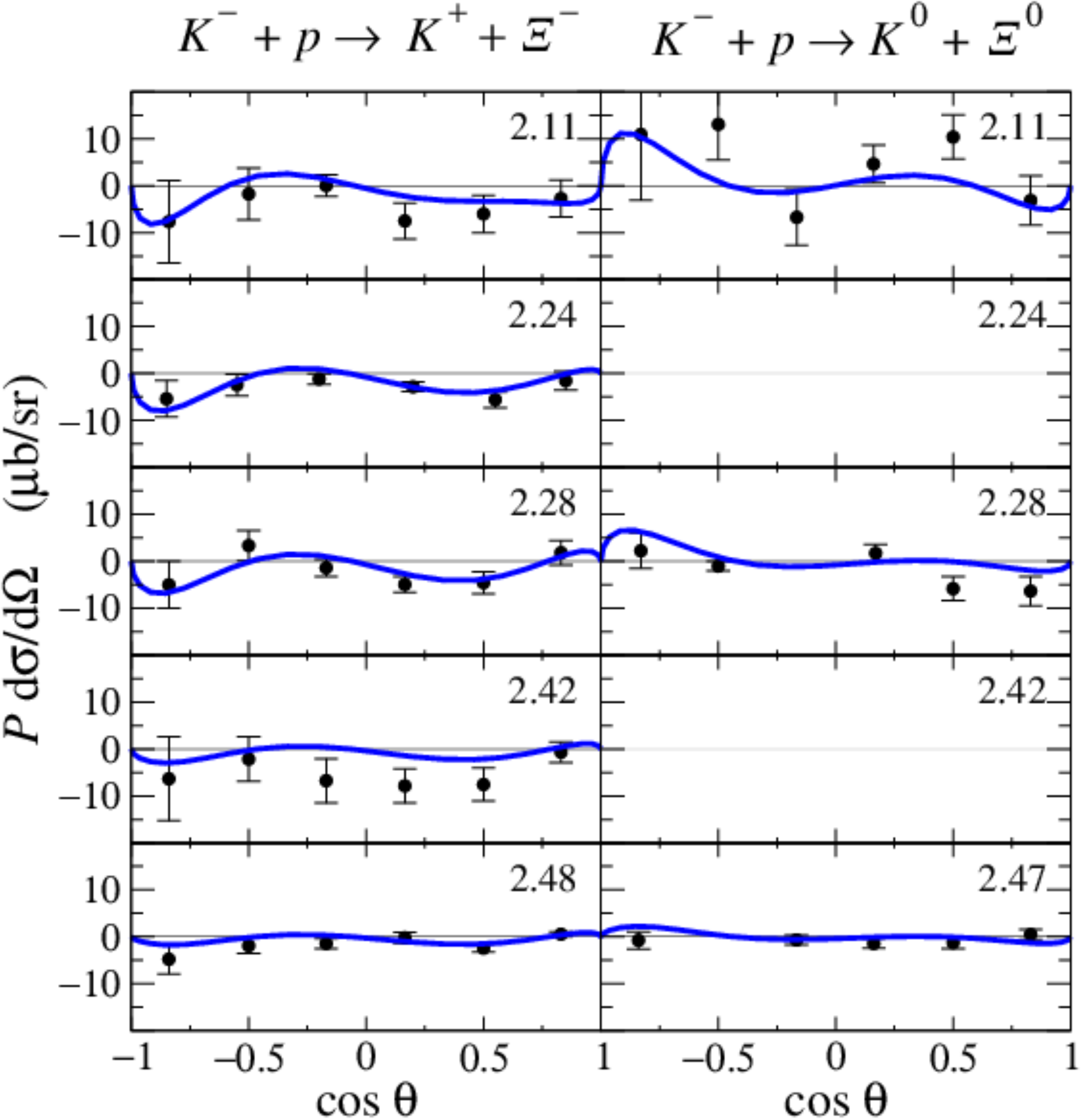}
\centerline{\parbox{0.80\textwidth}{
 \caption{The recoil asymmetry multiplied by the cross section,
	$P \frac{d\sigma}{d\Omega}$, for both the $K^-+p
	\rightarrow K^++\Xi^-$ and $K^-+p\rightarrow 
	K^0+\Xi^0$ reactions.  The blue solid lines represent 
	the full results of the current model. Data are the 
	digitized version from Ref.~\protect\cite{SKL11K}.}
	\label{P_y_12_phase} } }
\end{figure}

The results for the recoil polarization asymmetry multiplied by the 
cross section are shown in Fig.~\ref{P_y_12_phase}. Overall, we 
reproduce the data reasonably well. We also find that the results 
shown at W = 2.11~GeV are still significantly affected by the 
$\Sigma(2030)$. This corroborates the findings of Ref.~\cite{SKL11K}. 
An interesting observation here is that, although small, the
measured recoil polarization asymmetry is finite and non-vanishing. 
This offers an opportunity to measure the parity of the ground 
state $\Xi$ which has never been measured --- its positive parity 
as assigned by the Particle Data Group stems from quark-model 
predictions~\cite{PDG14K}. The reflection symmetry in the reaction 
plane implies that the target and recoil polarization asymmetries, 
$T$ and $P$, respectively, in reaction (\ref{reac1}) are related 
to each other by~\cite{NOH12K}
\begin{equation}
	T = \pi_\Xi^{} P \ ,
	\label{eq:TP}
\end{equation}
where $\pi_\Xi^{}$ stands for the parity of the $\Xi$ hyperon.

We also show in Fig.~\ref{fig:Klong} our prediction for the total
cross section in the $K_L + p \to K^+ + \Xi^0$ reaction. Unlike 
the other reaction channels considered above, this channel serves 
as a total isospin $I=1$ filter, for no contribution of $I=0$ is 
present.  However, we note that the isoscalar $\Lambda$ hyperons 
still contribute to this reaction via the $u$-channel process.
\begin{figure}[ht!] 
\centering
\includegraphics[width=0.45\textwidth,clip=]{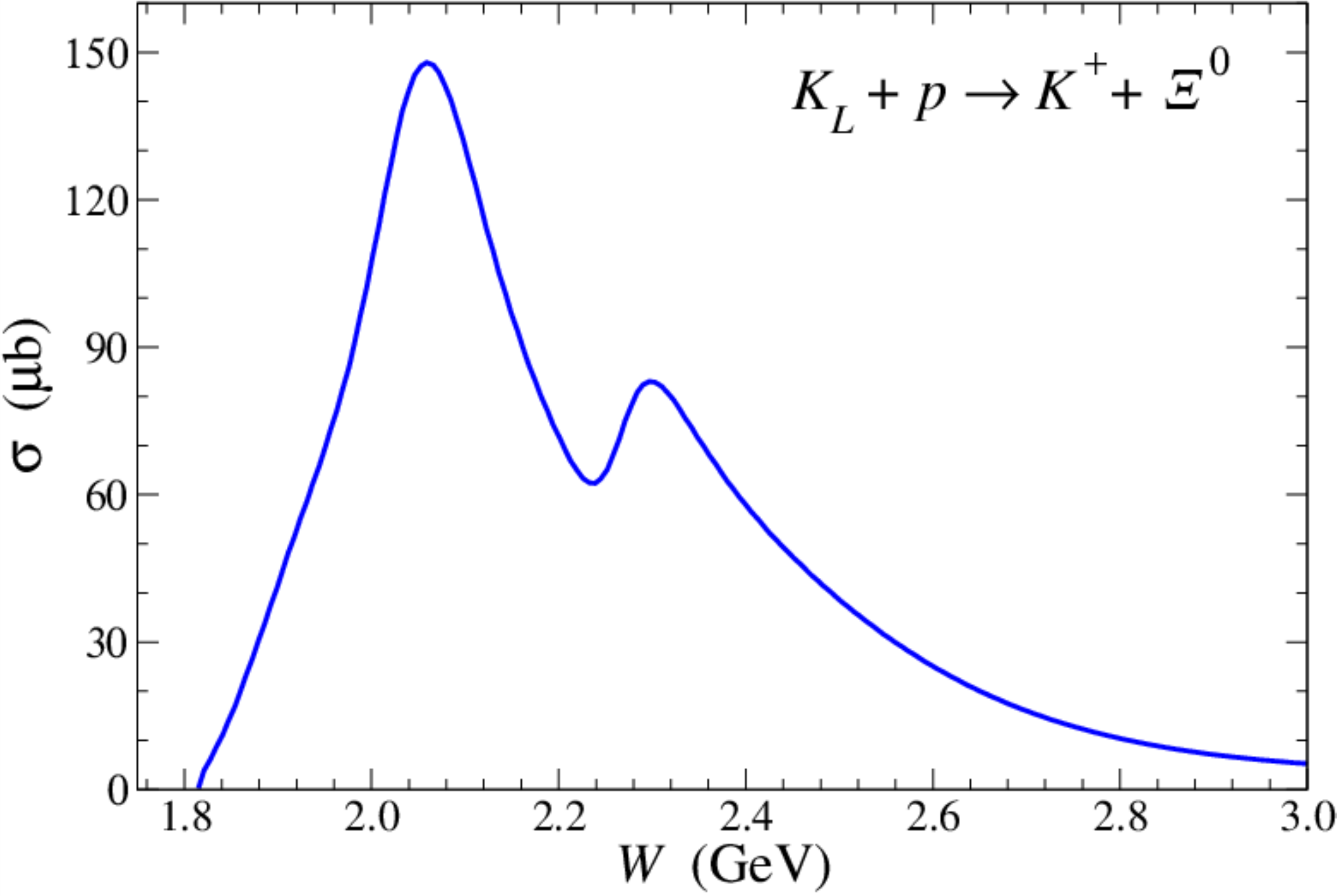}
\centerline{\parbox{0.80\textwidth}{
\caption{Prediction for the total cross section in the 
	$K_L+p\rightarrow K^++\Xi^0$ reaction.} 
	\label{fig:Klong} } }
\end{figure}

\newpage
\item \textbf{\boldmath $\gamma + N \to K + K + \Xi$ Reaction}

Figures~\ref{fig:photo_dxsc}(a),(b) display the $K^+$ and $\Xi^-$ 
angular distributions, respectively, in the center-of-mass frame 
for the reaction $\gamma + p  \to K^+ + K^+ + \Xi^-$. Overall, 
the data are reproduced very well. The same figures also show the 
results when the $t$-channel $K$-exchange current diagrams [cf.\ 
Fig.~\ref{fig:Photodiagrams}(a),(b)] involving the $S=-1$ hyperon 
resonances are switched off.  We see that they are crucial in
providing the observed behavior of the measured angular 
distributions.
\begin{figure*}[ht!]
\centering
\includegraphics[width=0.46\textwidth,clip=]{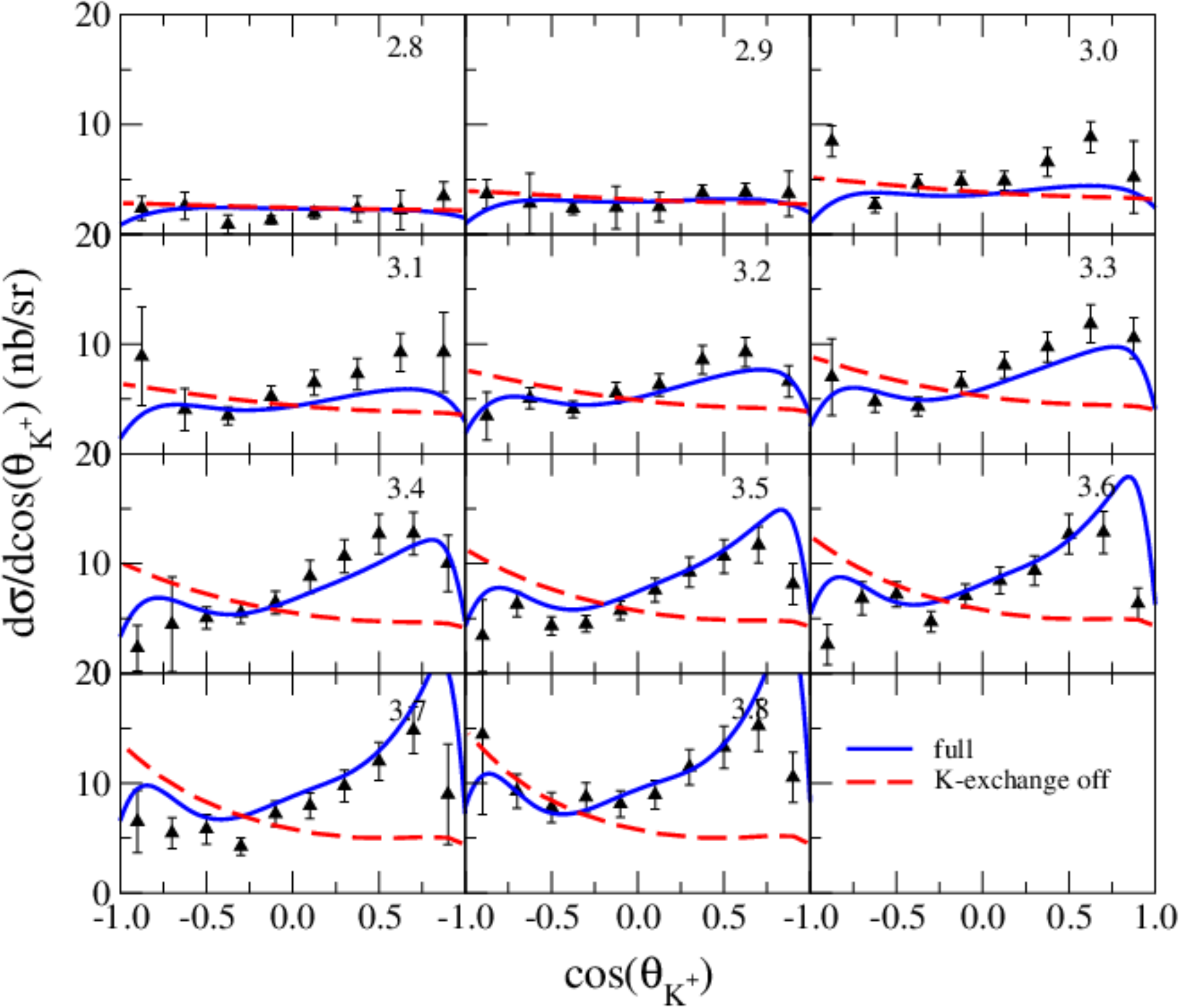} \qquad 
\includegraphics[width=0.46\textwidth,clip=]{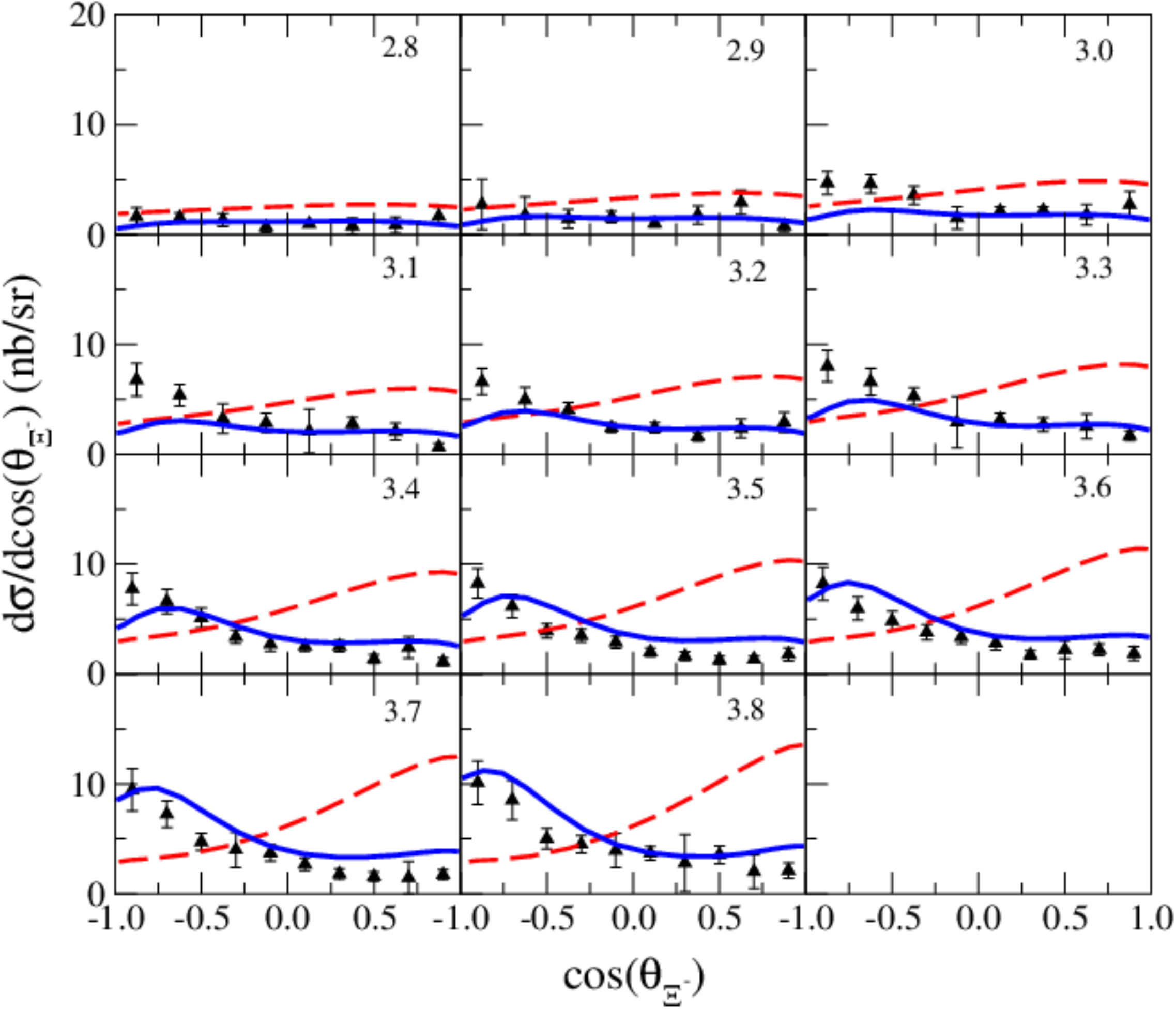}
\centerline{\parbox{0.80\textwidth}{
\caption{Differential cross sections for the reaction $\gamma 
	+ p\to K^+ + K^+ + \Xi^-$ in the center-of-mass
	frame of the system. Left panel:  $K^+$ angular 
	distribution. Right panel: $\Xi^-$ angular 
	distribution. The blue solid lines represent 
	the full result. The red dashed lines represent 
	the result with the $t$-channel $K$-exchange 
	currents [cf. Fig.~\protect\ref{fig:Photodiagrams}(a) 
	and (b)] switched off.  The number in the upper 
	right corner in each graph denotes the incident 
	photon energy in units of GeV in the laboratory 
	frame. The data are from Ref.~\protect\cite{CLAS07bK}.}
	\label{fig:photo_dxsc} } }
\end{figure*}

The results for the $K^+\Xi^-$ invariant mass distributions 
are shown in Fig.~\ref{fig:KXi-IMD}. It reveals the important 
role of the $\Sigma(2030)$ and $\Lambda(1890)$ resonances in 
reproducing the experimental data. We found that the 
$\Sigma(2250)$ resonance has a minor effect here.
\begin{figure}[ht!]
\centering
\includegraphics[width=0.7\textwidth,clip=]{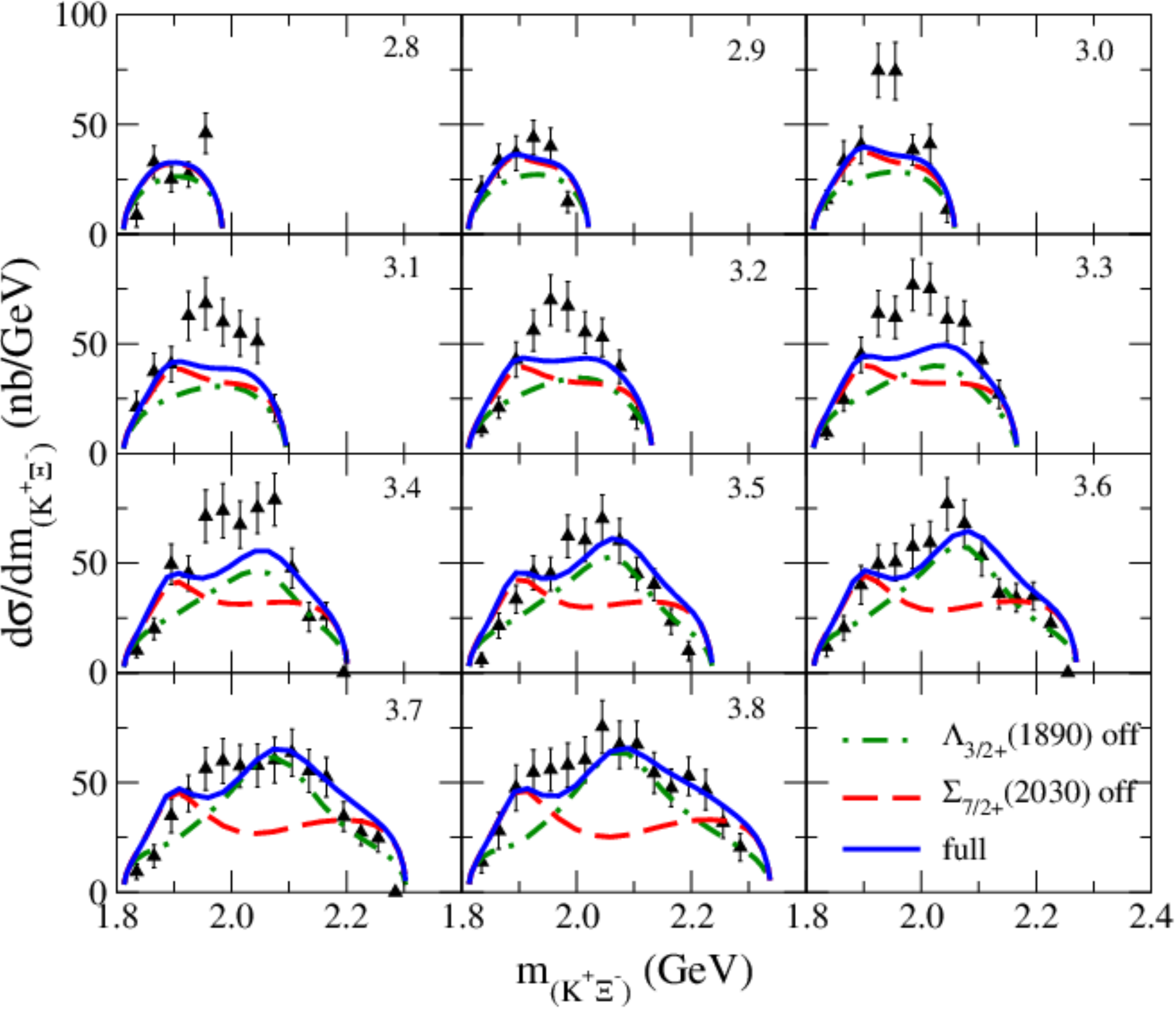}
\centerline{\parbox{0.80\textwidth}{
\caption{$K^+\Xi^-$ invariant mass distribution for the 
	reaction $\gamma + p \to K^+ + K^+ + \Xi^-$.
	The blue solid lines represent the full result.
	The red dashed lines represent the result with 
	$\Sigma(2030)$ switched off. The green 
	dash-dotted lines represent the result with 
	$\Lambda(1890)$ switched off. The number in 
	the upper right corner in each graph denotes 
	the incident photon energy in units of GeV in 
	the laboratory frame. The data are from 
	Ref.~\protect\cite{CLAS07bK}.}
	\label{fig:KXi-IMD} } }
\end{figure}
\end{enumerate}

\item \textbf{Conclusion}

In this work we have presented a combined analysis of the $\bar{K} + 
N\to K + \Xi$ and $\gamma + N\to K + K + \Xi$ reactions within an 
effective Lagrangian approach.  All the currently available data, 
in both the $K^- + p \to K^+ + \Xi^-$ and $K^- + p \to K^0 + \Xi^0$ 
processes, are well reproduced by the present model overall, and 
some of the basic features of the ground state $\Xi$ production in 
these reaction processes have been understood.

The above-threshold resonances $\Lambda(1890)$, $\Sigma(2030)$, 
and $\Sigma(2250)$ are required to achieve a good fit quality of 
the $K^-$-induced reaction data. Among them, the $\Sigma(2030)$ 
resonance is the most critical one. This resonance affects not 
only the cross sections but also the recoil asymmetry. More 
accurate data are required before a more definitive answer can
be provided for the role of the $\Lambda(1890)$ and $\Sigma(2250)$ 
resonances. In this regard, the multi-strangeness hyperon production 
programs using an intense anti-Kaon beam at J-PARC and JLab are of 
particular relevance in providing the much needed higher-precision 
data for the $\bar{K}$-induced reaction. While it may perhaps not 
be entirely clear which role any particular resonance plays for the 
$K^- + N \to K + \Xi$ reaction, the present and other calculations 
based on different approaches~\cite{SKL11K,SST11K,MFR14K,KNLS14K,KNLS15K} 
seem to agree that some $S=-1$ hyperon resonances are required to 
reproduce the existing data. To pin down the role of a particular 
resonance among them requires more precise and complete data, in
addition to more detailed theoretical models such as that of
Refs.~\cite{KNLS14K,KNLS15K}. In any case, both the $\bar{K}$- and 
photon-induced reactions studied in the present work are very well 
suited for studying $S=-1$ hyperon resonances in the $\sim 2$~GeV 
region.

We also found that the $\Lambda(1890)$ and $\Sigma(2030)$ 
resonances play an important role in the photon-induced reaction. 
In particular, they are required to bring the calculated $K^+\Xi^-$ 
invariant mass distributions in agreement with the corresponding
measurements.

Finally, the present work is our first step toward building a more 
complete reaction theory to help analyze the data and extract the 
properties of $\Xi$ resonances in future experimental efforts in 
$\Xi$ baryon spectroscopy. This is a complementary work to that of 
a model-independent analysis performed recently in Ref.~\cite{JOHN14K} 
and will also help in analyzing the data to understand the production 
mechanisms of $\Xi$ baryons.

\item \textbf{Acknowledgments}

This work was partially supported by the National Research 
Foundation of Korea (Grant No. NRF--2011--220--C00011) and 
the FFE-COSY (Grant No. 41788390).
\end{enumerate}


\newpage
\subsection{Predictions for Excited Strange Baryons}
\addtocontents{toc}{\hspace{2cm}{\sl I.P.~Fernando and J.L.~Goity}\par}
\setcounter{figure}{0}
\setcounter{table}{0}
\setcounter{footnote}{0}
\setcounter{equation}{0}
\halign{#\hfil&\quad#\hfil\cr
\large{Ishara P.~Fernando and Jos\'e~L.~Goity}\cr
\textit{Department of Physics}\cr
\textit{Hampton University}\cr
\textit{Hampton, VA 23668, U.S.A. \&}\cr
\textit{Thomas Jefferson National Accelerator Facility}\cr
\textit{Newport News, VA 23606, U.S.A.}\cr}

\begin{abstract}
An assessment is made of predictions for excited hyperon masses 
which follow from flavor symmetry and consistency with a $1/N_c$ 
expansion of QCD. Such predictions are based on presently 
established baryonic resonances. Low lying hyperon resonances 
which do not seem to fit into the proposed scheme are discussed.
\end{abstract}

\begin{enumerate}
\item \textbf{Introduction}

The present status of excited hadrons is apparently incomplete. 
This is so for mesons and especially for baryons.  There is an 
incompleteness problem known as \lq\lq missing baryon resonances
\rq\rq, which is defined by the smaller multiplicity of the 
experimentally extracted resonances vis-\'a-vis the predictions 
of quark models and emphasized by recent studies of the baryon 
spectrum in lattice QCD (LQCD). While this may turn out to be a 
lesser problem, because quark models are after all not QCD and 
LQCD calculations have been done at relatively large quark 
masses and do not fully include the coupled channels affecting 
the baryon resonances, there is another  missing resonance 
problem associated with flavor: the missing excited hyperons. 
According to approximate $SU(3)$ symmetry we would have the 
following relations between numbers of states (if excited 
baryons would only fill $\bf 1$, $\bf 8$ and $\bf {10}$ of  
$SU(3)$, and ignoring isospin): ~$\#\Sigma=\#\Xi=\#N+\#\Delta$ 
(PDG~\cite{Agashe:2014kdaJ1}:~ $26:12:49$), $\#\Omega=\#\Delta$  
(PDG:~4:22), and $\#\Lambda=\#N+\#{\rm singlets}$ (PDG:~18:29).  
An obvious question, is whether missing hyperons are only an 
experimental issue related to limited data, and thus improvable 
with future experimental efforts, or, in some cases,  due to the 
breaking of $SU(3)$ symmetry which in those cases may   be too 
large.  The latter situation  would be most likely the case for 
dynamically generated excited baryons with more non-strange 
resonances than strange ones   generated in that way, a situation 
which however does not seem to occur at least for the lower excited 
baryons, where,  as mentioned later, there are more hyperons than 
non-strange baryons identifiable with such possible dynamically 
generated states.  Here we critically analyze 
that question  anchoring the discussion in two expansions of QCD, 
namely the expansions in the light quark masses and in $1/N_c$.  
They connect QCD to the baryons  by the approximate symmetries 
they imply, namely $SU(3)\times SU(3)$ and spin-flavor 
 $SU(6)$~\cite{Dashen:1995J1}, respectively.  In both cases, the 
breaking of the symmetries is implemented by an expansion in 
quark masses and in $1/N_c$. Having a well defined framework  
leads in particular to relations that allow for predictions, 
\textit{e.g.}, among the masses of the resonances. Testing 
those relations, when possible, represents an important 
insight into the actual validity of the framework, and 
ultimately into QCD itself.  

The case of the ground state $\bf 8$ and $\bf{10}$ baryons 
is a powerful indicator. Looking at the masses, one has the 
G\"ursey-Radicati mass formula with the explicit expansion 
in $1/N_c$:
\begin{equation}
	M_B=N_c \, m_0+\frac{C_{HF}}{N_c} \, (\vec{S}^2-\frac 
	34 N_c) -c_8\, m_8 \,{\cal{S}}- c'_8\, m_8 \, S^i G^{i8}+ 
	\ord{1/N_c^2;~m_8/N_c},
\end{equation}
where $\vec S$ is the spin operator, ${\cal{S}}$ is the 
strangeness and $G^{ia}$ are the spin-flavor generators of $SU(6)$ 
(identified at leading order in $1/N_c$ with the axial vector 
currents), and $m_8=m_s-m_{u,d}$ is the octet component of the 
quark masses. The Gell-Mann-Okubo (GMO) and equal spacing relations 
(ESR) are satisfied up to deviations  $\ord{m_8^2/N_c}$ plus terms 
non-analytic in the quark masses.  In addition there is one relation  
involving simultaneously $\bf 8$ and $\bf{10}$ baryons that tests 
 $SU(6)$, namely (mass of baryon indicated by its name):
\begin{equation}
	\Sigma^\ast-\Sigma=\Xi^\ast-\Xi+\ord{1/N_c^2}~~~~~~~~~~212 \; 
	{\rm MeV} ~{\rm vs}~ 195\; {\rm MeV},
\end{equation}
which is satisfied within the expected level of accuracy.

Another test of $SU(6)$ symmetry is provided by the axial couplings, 
namely~\cite{Cordon:2013eraJ1}:
\begin{equation}
	~~~~~~~~~~~~~~~~~~~~g_A^{NN}=g_A^{N\Delta}=g_A^{\Delta\Delta}\nonumber\\
\end{equation}
\begin{equation}
	Exp: ~~1.27 : 1.24 : ~-\nonumber\\
\end{equation}
\begin{equation}
	LQCD: 1.17 : 1.07 : 0.98 , \nonumber\\
\end{equation}
where the deviations are $\ord{1/N_c^2}$ or 10\%.
 
If we look at the excited baryons, we find  that there is only one 
 $SU(3)$ multiplet that can be empirically identified (disregarding 
the $SU(3)$ singlet $\Lambda$s), namely one of the $J^P=1/2^-$ 
$\bf 8$:  $N(1532)$, $\Lambda(1676)$, $\Sigma(1667)$, $\Xi(1815)$, 
which satisfies remarkably well the GMO relation, namely $-19\pm 
26$~MeV. As shown later, a number of other relations implied by 
broken $SU(6)$ symmetry can be derived,  some of which can be tested 
with the listed PDG states. 

\item \textbf{Excited Baryons and $SU(6)\times O(3)$}

In principle, the S-matrix in the $SU(3)$ and large $N_c$ limits 
should display the exact $SU(6)$ symmetry of baryons, and one 
should be able to study all the corresponding observables via 
an expansion in quark masses and $1/N_c$ if the breaking of $SU(6)$ 
is sufficiently small. In particular, it should be possible to 
expand the resonance parameters such as pole mass and width, or 
the Breit-Wigner resonance mass, as well as partial decay widths, 
\textit{etc}. The framework for implementing such an expansion 
for excited baryons is based on expanding around an $SU(6)\times 
O(3)$ symmetry limit~\cite{Goity:1996hkJ1}. That framework allows 
for the derivation of  mass formulas for the baryons belonging 
to a given $SU(6)\times O(3)$ multiplet--(for issues of mixing 
of different multiplets see~\cite{Goity:2004pwJ1}). The importance 
of these formulas is that they lead to mass relations, and thus 
to possible predictions.  Mass formulas accurate to first order 
in the quark masses and $\ord{1/N_c}$ have been derived for the 
$[\bf{56},\ell^P=0^+,\;2^+]$~\cite{Carlson:2000zrJ1,Goity:2003abJ1} 
and for the $[\bf{70},\ell^P=1^-,\;2^+]$~\cite{Schat:2001xrJ1,
Goity:2002puJ1,Matagne:2013ccaJ1}. We briefly discuss those results, 
as they represent the best illustration of the predictivity 
issues we are discussing. Identifying the baryons in a given 
irrep  of $SU(6)\times O(3)$ by  $(R,\ell; J ,R_3 Y I)$, the 
mass formula has the general form:
\begin{equation}
	M_B(R,\ell; J , R_3 Y I) =N_c\, m_0(R,\ell)+\delta 
	M( R,\ell; J, R_3 Y I) ,
\end{equation}
where $\delta M$ provides all the symmetry breaking effects 
expanded to the given order, and  it is constructed in terms of 
a basis of composite operators built with products of the 
generators of the symmetry group~\cite{Goity:1996hkJ1} and ordered 
in powers of $m_8$ and $1/N_c$. The key observation is that we 
know how to build such a basis in a systematic way thanks to 
the fact that we know how to count in $1/N_c$ at the baryon 
level~\cite{Dashen:1995J1,Goity:1996hkJ1}.

\begin{enumerate}
\item \textbf{The $[\bf{56},2^+]$ Baryons}

As the  first illustration, we discuss the $[\bf{56},2^+]$ 
baryons, where at $\ord{1/N_c}$ and $\ord{m_8}$ there are only 
three $SU(3)$ singlet and three $SU(3)$ breaking mass operators 
leaving a total of 14 mass relations~\cite{Goity:2003abJ1}, and 
12 unknown hyperon masses can be predicted from the presently 
known states in the multiplet~\cite{Agashe:2014kdaJ1}. 
Table~\ref{masrel} shows the mass relations, where some can be 
tested by the known PDG Breit-Wigner masses. In all cases in 
this report the fits have been carried out by including PDG 
states rated with at least three stars. 
\begin{table}[htb!]
\centering
\centerline{\parbox{0.80\textwidth}{
 \caption{$[\bf{56},2^+]$  mass relations, include the  GMO 
	relations for the two octets and the two ESR for each 
	of the four decuplets.} } }
\vspace{2mm}
\begin{tabular}{|cc|}
\hline
\hline
Relation      & Test (MeV)\\
\hline
$\Delta_{5/2} - \Delta_{3/2} =  N_{5/2} - N_{3/2}$           & $-40\pm 43 \;{\rm vs}\; -17\pm 51$\\
$\frac 57 (\Delta_{7/2} - \Delta_{5/2}) = N_{5/2} - N_{3/2}$ &  $39\pm 19 \;{\rm vs}\; -17\pm 51$\\
$\frac 13 (\Delta_{7/2} - \Delta_{1/2}) = N_{5/2} - N_{3/2}$ &  $18\pm 9  \;{\rm vs}\; -17\pm 51$\\
\hline
$\frac{4}{15} (\Lambda_{3/2} - N_{3/2}) + \frac{11}{15} (\Lambda_{5/2} - N_{5/2}) = 
 \frac 12 (\Sigma_{5/2}- \Lambda_{5/2}) +  \Sigma_{7/2}-\Delta_{7/2}$      & $148\pm 17 \;{\rm vs}\; 132\pm 16$ \\
$\Lambda_{5/2} - \Lambda_{3/2} + 3(\Sigma_{5/2} - \Sigma_{3/2}) = 4 (N_{5/2} - N_{3/2})$          & $-$~~~~~~~~~~~$-$ \\
$\Lambda_{5/2} - \Lambda_{3/2} + \Sigma_{5/2} - \Sigma_{3/2} = 2 (\Sigma'_{5/2} - \Sigma'_{3/2})$ & $-$~~~~~~~~~~~$-$ \\
$ 7 \; \Sigma'_{3/2} + 5 \; \Sigma_{7/2} = 12 \; \Sigma'_{5/2}$  & $-$~~~~~~~~~~~~~$-$  \\
$ 4 \; \Sigma_{1/2} + \Sigma_{7/2} =5 \; \Sigma'_{3/2}$          & $-$~~~~~~~~~~~~~$-$ \\
\hline
{\bf 8s}:~~~~~~~~~~ $2 (N_J + \Xi_J) = 3 \; \Lambda_J  + \Sigma_J$          & $-$~~~~~~~~~~~~~~~$-$\\
{\bf {10s}}:~~~ $\Sigma_J - \Delta_J = \Xi_J - \Sigma_J = \Omega_J - \Xi_J$ & $-$~~~~~~~~~~~~~~~$-$\\
\hline
\hline
\end{tabular} \label{masrel}
\end{table}

The masses of all the missing states can be predicted, and they 
are shown in the case of $\Lambda$, $\Sigma$ and $\Xi$ hyperons 
in Fig.~\ref{fig:EvolutiontestPic1}.

\item \textbf{The $[\bf{70},1^-]$ Baryons}

In a similar way one can analyze the 70-plet baryons, in 
particular the lightest $[\bf{70},1^-]$~\cite{Schat:2001xrJ1,
Goity:2002puJ1}. Here also mass relations can be derived: there 
are 32 isospin multiplets in the  $[\bf{70},1^-]$, and the 
basis of mass operators involves a total of 15 operators, 
leaving 17 mass relations, of which 11 are GMO and ESR, and 
the rest are relations that test spin-flavor symmetry. As 
mentioned earlier, one GMO relation can be tested with the 
known PDG states. The masses of all missing states can be 
predicted, as shown in Fig.~\ref{fig:EvolutiontestPic1} for 
the hyperons. The $[\bf{70},1^-]$ is interesting because 
there is state mixing, primarily driven by the breaking of 
the spin-symmetry, but also by $SU(3)$ breaking. The two 
mixings involving the pairs of nucleons with J=1/2 and 3/2 
can be unambiguously determined by including the analysis 
in the same framework of the decays as mentioned later.
\end{enumerate}

\newpage
\item \textbf{Mass Predictions and Observations}

The predictions for yet unobserved or unidentified hyperons 
can be seen in Fig.~\ref{fig:EvolutiontestPic1}. Starting 
with the $\Lambda$s, there is a prediction for a $3/2^-$ 
with a mass of 1865~MeV which does not match any PDG listed 
state: this should therefore be a significant prediction. 
It should be noted that the $\Lambda(1405)$, which being a 
singlet of $SU(3)$ can only be constrained in the present 
framework by the spin-flavor symmetry, can be described in 
the framework without having to introduce unnatural size 
coefficients in the mass formula~\cite{Schat:2001xrJ1,
Goity:2002puJ1}. This of course does not exclude that there 
is an important role of the $\Sigma\pi$ and $NK$ coupled 
channels in its structure~\cite{Fernandez-Ramirez:2015fbqJ1,
Molina:2015uqpJ1}.  The $1/2^+$ $\Lambda(1810)$ is interesting: 
it is too light to be described within any spin-flavor 
multiplet, but it sits close to the $K\Xi$ threshold: is it 
a $K\Xi$ resonance or a threshold effect ?  The higher mass 
$\Lambda$s should belong to other multiplets not considered 
here. 

For the $\Sigma$s, the situation becomes  quite interesting. 
There are seven PDG states with undetermined $J^P$. Of those, 
four have masses between 1450 and 1750~MeV. The two lowest 
lying ones are very difficult to explain: they do not fit 
into any  flavor multiplet as they are way too light. A 
similar issue but less definite occurs with the next two 
states. The one star $3/2^-$ state at ~1570~MeV does not 
seem to fit into a multiplet, and it is about 50~MeV above 
the $\pi \Sigma^\ast$ threshold: it seems therefore to be a 
difficult state to explain.  Among the $1/2^+$ states, the 
heavier mass corresponds to the $\Sigma$ in the decuplet, 
and the PDG shows two states below it which cannot be 
matched. The lower one may be related to the $\eta\Sigma$ 
threshold, which could explain it, but the upper one is 
puzzling. In the case of the negative parity states there 
are three clear predictions as shown in the 
Fig.~\ref{fig:EvolutiontestPic1}.

Finally, for the $\Xi$s, in which the $K_L$ beam could 
play a very important role as a discovery tool, we have a 
large number of predictions and also possible 
identifications with PDG states with yet undetermined 
$J^P$. As in the case of the $\Sigma$s, there are several 
PDG listed states which are too light to belong onto 
identifiable multiplets. For further discussion on the 
$\Xi(1620)$ and $\Xi(1690)$ see~\cite{OhJ1,ZieglerJ1}.

The mentioned lower excited $\Sigma$s and $\Xi$s quoted 
in the PDG cannot be assigned to any $SU(3)$ multiplet 
because no non-strange partners sufficiently light to fit 
into such multiplets exist. It is interesting to observe 
that they are close to thresholds (50~MeV or less), and 
thus they are likely closely related to those thresholds. 
While there is one three star state, ($\Xi(1690)$), the 
rest are one or two star states. Neither quark models nor 
the LQCD calculations show those states. This can be 
understood if meson-baryon dynamics is the relevant 
physics (dynamically generated resonances or threshold  
effects), which is not captured in either of those 
approaches. More sophisticated LQCD calculations than the 
ones performed so far will be needed to find such effects. 
Those calculations are based on the L\"uscher approach to 
extract hadronic interactions from finite volume effects 
on energy levels, which in the case of  baryons are still 
in their early stages~\cite{Lang:2012dbJ1}. 

For dynamically generated states, which turn out to lie 
close to thresholds, one expects the effects of $SU(3)$ 
symmetry breaking to be magnified, and thus $SU(3)$ may 
cease to be a useful predictive tool. This is well 
known in the case of the scalar mesons, where the 
lightest $0^+$ states  cannot be accommodated into an 
octet with  a GMO relation being approximately satisfied 
\footnote{Let us remind the reader that for the vector 
	mesons, once one factors in the ideal $\omega-\phi$ 
	mixing, the GMO relation is satisfied to a very good 
	approximation.}. 
It should be emphasized that even in the large $N_c$ limit 
one should expect non-trivial meson-baryon dynamics and 
therefore  possible dynamically generated states. This 
contrasts with mesons, where the meson-meson interaction 
tends to vanish at large $N_c$, and thus the formation of 
tetra-quark states should eventually become  impossible. 
Thus in baryons, dynamically generated states which cannot 
be assigned to an $SU(3)$ multiplet are  not excluded by the 
$1/N_c$ expansion.
\begin{figure}[htbp!]
\begin{center}
\includegraphics*[width=15cm]{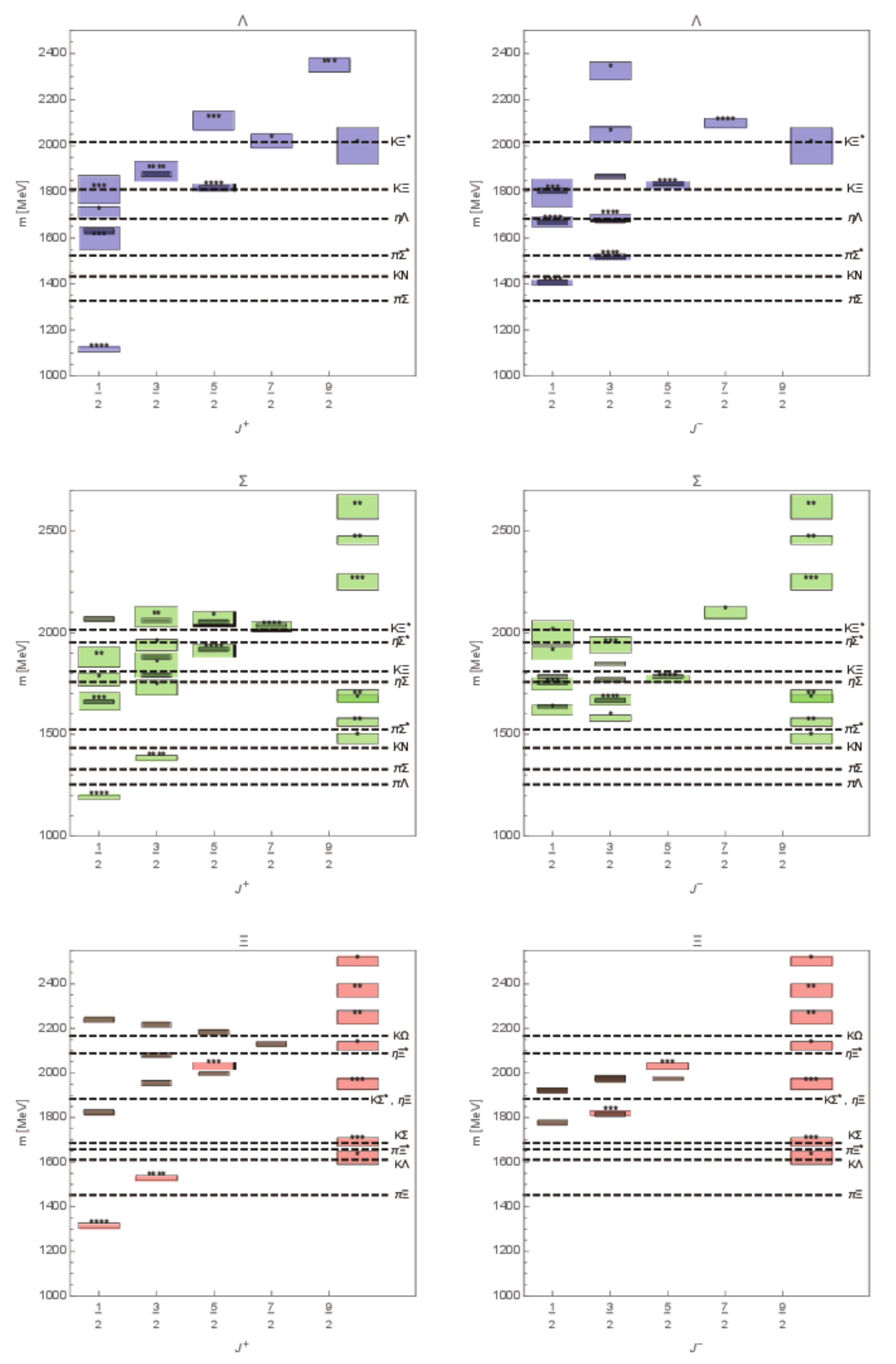}
\end{center}
\vspace{-1.0cm}
\centerline{\parbox{0.80\textwidth}{
 \caption{Excited hyperon masses. In color the PDG masses 
	with their star ratings. In dark the results of 
	the fits to the PDG states and the predictions.
        The dashed lines indicate the relevant thresholds
        corresponding to a pseudoscalar meson and a low 
	lying $\bf 8$ or $\bf{10}$ baryon. The masses on 
	the extreme right of the figures are those whose 
	$J^P$ quantum numbers are not established by the 
	PDG.}\label{fig:EvolutiontestPic1} } }
\end{figure}

\item \textbf{Other Possible Predictions: Decays}

For states which can be assigned to spin-flavor multiplets, one can 
expect that other observables beyond the masses will be constrained 
by symmetry. One case in point are the two-body strong decay partial 
widths.  It is possible to derive relations between the partial decay 
widths, and in principle also derive predictions. As illustration, we 
mention here the $[\bf{70},1^-]$ decays~\cite{Jayalath:2011ucJ1}. The 
partial decay widths play an important role in determining the mixing 
angles between the two pairs of nucleon states with $J=1/2$ and 3/2. 
Various relations, written in terms of reduced partial decay widths 
(\textit{i.e.}, by conveniently removing phase space and centrifugal 
barrier factors~\cite{Jayalath:2011ucJ1}) can be tested.  Two relations 
valid at LO in the $1/N_c$ expansion permit for a determination of the 
mixing $\theta_{J=1/2}$ via S-wave decays:
\begin{equation}
	\frac{\tilde{\Gamma}(N(1535)\to N\pi)-\tilde{\Gamma}(N(1650)\to 
	N\pi)}{\tilde{\Gamma}(N(1535)\to N\pi)+\tilde{\Gamma}(N(1650)\to 
	N\pi)}
	=
	\frac{1}{5}(3 \cos(2\theta_{1/2})-4 \sin(2\theta_{1/2}),
	\nonumber\\
\end{equation}
\begin{equation}
	\frac{\tilde{\Gamma}(N(1535)\to N\eta)-\tilde{\Gamma}(N(1650)\to 
	N\eta)}{\tilde{\Gamma}(N(1535)\to N\eta)+\tilde{\Gamma}(N(1650)\to 
	N\eta)}
	=
	2 \sin(2\theta_{1/2}),
\end{equation}
which give respectively: $\theta_{1/2}=0.46\pm 0.10~ {\rm or}~1.76\pm 0.10$ 
and $\theta_{1/2}=0.51\pm 0. 27$. The global fit in 
Ref.~\cite{deUrreta:2013koaJ1} gives $\theta_{1/2}\sim 0.40\pm0.20$.
Other relations that can be tested are the following ones:
\begin{equation}
	\text{S-wave}: 
	\frac{\tilde{\Gamma}(N(1535)\to N\pi)+\tilde{\Gamma}(N(1650)\to 
	N\pi)}{\tilde{\Gamma}(\Delta(1620)}=1 ;
        ~~~\text{PDG:}~1.64\pm 0.85, \nonumber\\
\end{equation}
\begin{equation}
	\frac{\tilde{\Gamma}(\Delta(1620)\to N\pi)}{\tilde{\Gamma}
	(\Delta(1700)\to\Delta \pi)}=0.1;
	~~~\text{PDG:} ~0.29\pm 0.15, \nonumber\\
\end{equation}
\begin{equation}
	\text{D-wave}: 
	\frac{2\tilde{\Gamma}(\Delta(1620)\to\Delta\pi)+\tilde{\Gamma}
	(\Delta(1700)\to \Delta\pi)}{15\tilde{\Gamma}(\Delta(1620)\to 
	N\pi)+32 \tilde{\Gamma}(\Delta(1700)\to N\pi)}=1;
	~~~\text{PDG:}~0.71\pm 0.30, \nonumber\\
\end{equation}
\begin{equation}
	\frac{\tilde{\Gamma}(N(1535)\to\Delta\pi)+\tilde{\Gamma}
	(N(1650)\to \Delta\pi)+11\tilde{\Gamma}(\Delta(1700)\to\Delta\pi}
	{132\tilde{\Gamma}(\Delta(1700)\to N\pi)+90 \tilde{\Gamma}(N(1675)
	\to N\pi)}=1;
	~~\text{PDG:}~0.78\pm 0.33,\nonumber\\
\end{equation}
which are to first approximation remarkably well satisfied.

A complete analysis of the $[\bf{70},1^-]$ decays involving the hyperons 
can be found in Ref.~\cite{Jayalath:2011ucJ1}. The results there could  
be of course extended to predict the partial decay widths of the yet 
unseen hyperons in the $[\bf{70},1^-]$. 

\item \textbf{Tests with Lattice QCD Results}

Calculations of the baryon spectrum in LQCD represent an important new 
source of information which can be used to test models as well as the 
approach discussed here. The advantage is that complete multiplets have 
been extracted from the calculations, and thus one can test the 
different mass relations. The disadvantage is that the calculations so 
far have been carried out at un-physically large quark masses, where 
apparently QCD dynamics shows resemblance with quark models, and where 
the framework does not allow to fully describe meson-baryon dynamics.  
Thanks to the large pions masses the lower lying excited states become 
relatively narrow, and one expects that the latter issue will not be 
so important. The LQCD results can be studied in the $SU(3)$ symmetry 
breaking and $1/N_c$ expansions, providing new tests.  This was done 
in~Ref.~\cite{Fernando:2014dnaJ1}, where states obtained in 
LQCD~\cite{Edwards:2011jjJ1,Edwards:2012fxJ1} corresponding to the 
multiplets discussed above were analyzed. Throughout it is observed 
that the mass relations are satisfied as expected. The LQCD reported 
errors on the baryon masses are in the range of $10$ to $50\;{\rm MeV}$, 
which is slightly larger than the expected NLO corrections in the mass 
formulas, and thus give a weaker test of the relations than one would 
wish. Further study is still necessary to confront our approach with 
the higher excited states determined  in the LQCD calculations. 

An open issue in the LQCD calculations is the study of possible 
dynamically generated states mentioned earlier, in particular the 
discussed hyperons. This would provide a powerful means for 
determining the existence of such states and also for answering the 
questions on symmetry breaking that arise in those cases.

\item \textbf{Comments}

A $K_L$ beam in Hall~D at Jefferson Lab would open unique 
opportunities for the study of excited hyperons, in 
particular $\Xi$s, of which much is still to be learned.   
Discerning the  multiplet structures of excited baryons 
remains an open experimental issue which would be impacted 
by that development. The coincidence and interplay with 
the progress of LQCD studies of excited baryons seems to 
be particularly auspicious for gaining the much needed 
understanding of excited baryons.

\item \textbf{Acknowledgments}

This work was supported in part by DOE Contract No. 
DE--AC05--06OR23177 under which JSA operates the Thomas
Jefferson National Accelerator Facility and by the National
Science Foundation through grant PHY--1307413.
\end{enumerate}


\newpage
\subsection{The $\bar{K} N \rightarrow K \Xi$ Reaction in a Chiral NLO 
	Model}
\addtocontents{toc}{\hspace{2cm}{\sl A.~Ramos, A.~Feijoo, and 
	V.K.~Magas}\par}
\setcounter{figure}{0}
\setcounter{equation}{0}
\halign{#\hfil&\quad#\hfil\cr
\large{Angels Ramos, A.~Feijoo and V.K.~Magas}\cr
\textit{Departament d'Estructura i Constituents de la Mat\`eria and 
	Institut de Ci\`encies del Cosmos}\cr
\textit{Universitat de Barcelona}\cr
\textit{Mart\'i Franqu\`es 1}\cr
\textit{Barcelona E08028, Spain}\cr}

\begin{abstract}
We present a model for the meson-baryon interaction in s-wave in 
the strangeness S=-1 sector, based on a chiral SU(3) Lagrangian 
up to next-to-leading order (NLO) and implementing unitarization 
in coupled channels.   A particular attention has been payed to 
fitting our model to the $K^-p\rightarrow K^+\Xi^- , K^0\Xi^0$ 
cross section data, since these processes are particularly 
sensitive to the NLO terms. Our model also includes the additional 
effect of high spin resonances believed to be important in the 
$\sim 2$~GeV energy region under study. We present predictions for 
the cross section of the $K^0_Lp\to K^+ \Xi^0$ reaction that could 
be measured with the proposed Secondary $K^0_L$ beam at Jlab. This 
process is particularly helpful for determining the properties of 
the meson-baryon interaction in $S=-1$, due to the its isospin 
$I=1$ filter character.
\end{abstract}

\begin{enumerate}
\item \textbf{Introduction}

The description of low energy hadron reactions employing SU(3) 
Chiral Perturbation Theory ($\chi$PT), which is based on an 
effective Lagrangian which respects the symmetries of QCD, has 
ben very succesful but the theory fails to describe hadron 
dynamics in the vicinity of resonances. Unitarized Chiral 
Perturbation Theory (U$\chi$PT), which combines chiral dynamics 
with unitarization techniques in coupled channels, has shown to 
be a very powerful tool that permits extending the validity of 
$\chi$PT to higher energies and to describe the physics around 
the so called dynamically generated resonances 
(see~\cite{ollerreportR} and references therein). A clear example 
of the success of U$\chi$PT is  the description of the 
$\Lambda(1405)$ resonance, located only $27$~MeV below the 
$\bar{K} N$ threshold, that emerges from coupled-channel 
meson-baryon re-scattering in the $S=-1$ sector. In fact, the 
dynamical origin of the $\Lambda(1405)$ resonance was already 
hindered more than 50 years ago~\cite{L1405R}, an idea that was 
reformulated later in terms of the chiral unitary theory in 
coupled channels~\cite{KSWR}. This success stimulated a lot of 
activity in the community, which analyzed the effects of 
including a complete basis of meson-baryon channels, differences 
in the regularization of the equations, s- and u-channel Born 
terms in the Lagrangian, next-to-leading (NLO) contributions, 
\textit{etc.} \dots~\cite{KWWR,ORR,OMR,LKR,BMWR,2poleR,BFMSR,BNWR,BMNR}. 
The various developed models could reproduce the $\bar{K} N$ 
scattering data very satisfactorily and all these efforts 
culminated in establishing the $\Lambda(1405)$ as a 
superposition of two poles of the scattering amplitude~\cite{OMR,
2poleR,PRLR}.

This topic experienced a renewed interest in the last few years, 
after the availability of a more precise measurement of the 
energy shift and width of the 1s state in Kaonic hydrogen by 
the SIDDHARTA Collaboration~\cite{SIDDR} at DA$\Phi$NE. The CLAS 
Collaboration at JLab has also recently provided mass 
distributions of $\Sigma^+\pi^-$, $\Sigma^-\pi^+$, and $\Sigma^0
\pi^0$ states in the region of the 
$\Lambda(1405)$~\cite{Moriya:2013ebAR}, as well as differential 
cross sections~\cite{Moriya:2013hwgR} and a direct determination 
of the expected spin-parity $J^\pi=1/2^-$ of the 
$\Lambda(1405)$~\cite{Moriya:2014kpvR}. Invariant $\pi\Sigma$ 
mass distributions from $pp$ scattering experiments have 
recently been  measured by the COSY Collaboration at 
J\"ulich~\cite{Zychor:2007gfR} and by the HADES Collaboration at 
GSI~\cite{Agakishiev:2012xkR}. In parallel with the increased 
experimental activity, the theoretical models have been 
revisited~\cite{IHWR,HJ_revR,GOR,MMR,MFSTR,Feijoo:2015yjaR} and 
analyses of the new reactions, aiming at pinning  down the 
properties of the $\Lambda(1405)$ better, have been 
performed~\cite{Roca:2013avAR,Roca:2013ccaAR,Mai:2014xnaAR}. 

In this contribution, we present a study of the $S=-1$ 
meson-baryon interaction focused on providing well constrained 
values of the low-energy constants of the NLO chiral 
Lagrangian~\cite{Feijoo:2015yjaR}. We employ data in the strong 
sector, including elastic and inelastic cross section data 
($K^-p\to K^-p$, ${\bar K}^0n$, $\pi^{\pm}\Sigma^{\mp}$, 
$\pi^0\Sigma^0$, $\pi^0\Lambda$) and the precise SIDDHARTA 
value of the energy shift and width of Kaonic hidrogen, as 
done by the recent works, but, in addition, we also constrain 
the parameters of our model to reproduce the $K\Xi$ production 
data via the reactions $K^-p\to K^+\Xi^-, K^0\Xi^0$. The 
motivation is that the lowest-order Lagrangian does not 
contribute directly to these reactions, which then become 
especially sensitive to the NLO terms. The model is  also 
supplemented by explicit resonant terms, which are unavoidable 
at CM energies of around 2~GeV characteristic of $K\Xi$ 
production, as proposed by several resonance-based models that 
have investigated the photoproduction of $\Xi$ particles off 
the proton~\cite{Nakayama:2006tyR,Man:2011npR} or via the strong 
reactions $K^-p\to K^+\Xi^-, K^0\Xi^0$~\cite{Sharov:2011xqR,
Shyam:2011ysR,Jackson:2015dvaAR}, as in this contribution.

\item \textbf{Formalism}

The lagrangian implementing the interactions between mesons 
and baryons at lowest order reads
\begin{eqnarray} \label{LagrphiB1}
	\Lagr_{\phi B}^{(1)} 
	= i \langle \bar{B} \gamma_{\mu} 
	[D^{\mu},B] \rangle
	- M_0 \langle \bar{B}B \rangle 
	- \frac{1}{2} D \langle \bar{B} \gamma_{\mu}
	\gamma_5 \{u^{\mu},B\} \rangle \no \\
	- \frac{1}{2} F \langle \bar{B} \gamma_{\mu} 
        \gamma_5 [u^{\mu},B] \rangle \ ,
\end{eqnarray}
where $u_\mu = i u^\dagger \partial_\mu U u^\dagger$, with
$U(\phi) = u^2(\phi) = \exp{\left( \sqrt{2} i \phi/ f \right)}$
containing the field $\phi$ of the pseudoscalar octet, $B$ 
stands for the $J^P=1/2^+$ baryon octet, $f$ is the pseudoscalar 
decay constant, $M_0$ the common baryon octet mass in the 
chiral limit, the constants $D$, $F$ denote the axial vector 
couplings of the baryons to the mesons, and the symbol $\langle 
\dots\rangle$ stands for the trace in flavor space. Finally, 
$[D_\mu, B]$ stands for the covariant derivative $[D_\mu, B] = 
\partial_\mu B + [ \Gamma_\mu, B] $, with $\Gamma_\mu =  
[ u^\dagger,  \partial_\mu u]/2 $.

At next-to-leading order, the contributions of $\Lagr_{\phi B}$ 
to meson-baryon scattering are:
\begin{eqnarray} \label{LagrphiB2}
	\Lagr_{\phi B}^{(2)}
	= b_D \langle \bar{B} \{\chi_+,B\} \rangle
	+ b_F \langle \bar{B} [\chi_+,B] \rangle
	+ b_0 \langle \bar{B} B \rangle \langle \chi_+ \rangle \no \\ 
	+ d_1 \langle \bar{B} \{u_{\mu},[u^{\mu},B]\} \rangle
	+ d_2 \langle \bar{B} [u_{\mu},[u^{\mu},B]] \rangle    \no \\ 
	+ d_3 \langle \bar{B} u_{\mu} \rangle \langle u^{\mu} B \rangle
	+ d_4 \langle \bar{B} B \rangle \langle u^{\mu} u_{\mu} \rangle \ ,
\end{eqnarray}
where $\chi_+ = 2 B_0 (u^\dagger \mathcal{M} u^\dagger + u 
\mathcal{M} u)$ breaks chiral symmetry explicitly via the quark 
mass matrix  $\mathcal{M} = {\rm diag}(m_u, m_d, m_s)$ and $B_0 
= - \bra{0} \bar{q} q \ket{0} / f^2$ relates to the order 
parameter of spontaneously broken chiral symmetry.

From these lagrangians one can derive the interaction kernel up to 
NLO in the non-relativistic limit 
$$
	V_{ij}=V^{\scriptscriptstyle WT}_{ij}+V^{\scriptscriptstyle 
	NLO}_{ij}=
$$
\begin{equation}\label{V_TOT}
	- \frac{C_{i j}(2\sqrt{s} - M_{i}-M_{j})}{4 f^2}\! N_{i} N_{j}
	+ \frac{D_{ij}-2(k_\mu k^{\prime\,\mu})L_{ij}}{f^2}\! N_{i} N_{j}\,,
\end{equation}
where
\begin{equation}
	N_{i}=\sqrt{\frac{M_i+E_i}{2M_i}},\,\, N_{j}=\sqrt{\frac{M_j
	+E_j}{2M_j}} \no 
\end{equation}
with $M_i,M_j$ and $E_i,E_j$ the masses and energies, respectively, of 
the baryons involved in the transition. The indices $(i,j)$ cover all 
the initial and final channels, which, in the case of strangeness 
$S=-1$ and charge $Q=0$ explored here, amount to ten: $K^-p$, 
$\bar{K}^0 n$, $\pi^0\Lambda$, $\pi^0\Sigma^0$, $\pi^-\Sigma^+$, 
$\pi^+\Sigma^-$, $\eta\Lambda$, $\eta\Sigma^0$, $K^+\Xi^-$, and 
$K^0\Xi^0$. The matrices of coefficients $C_{ij}$, $D_{ij}$ and 
$L_{ij}$ are shown in the appendix of~\cite{Feijoo:2015yjaR}. They 
depend on the pion decay constant $f$ and  the parameters $b_0$, 
$b_D$, $b_F$, $d_1$, $d_2$, $d_3$ and $d_4$, which will be 
determined in our fits.

The U$\chi$PT method consists in solving the Bethe-Salpether 
equation in coupled channels
\begin{equation}
	T_{ij} =V_{ij}+V_{il} G_l T_{lj}  \ ,
\label{LS}
\end{equation} 
where the loop function $G_i$ stands for the propagator of the 
$i^{\rm th}$ meson-baryon state, which is regularized employing 
dimensional regularization
\begin{eqnarray} \label{Loop_integral}
	G_l = {\rm i}\int \frac{d^4q_l}{{(2\pi)}^4}\frac{2M_l}{{(P
	-q_l)}^2-M_l^2+{\rm i}\epsilon}\frac{1}{q_l^2-m_l^2+{\rm i}
	\epsilon} \nonumber \\
	= \frac{2M_l}{(4\pi)^2} \Bigg \lbrace a_l+\ln\frac{M_l^2}{\mu^2}
	+\frac{m_l^2-M_l^2+s}{2s}\ln\frac{m_l^2}{M_l^2} + \nonumber \\ 
	\frac{q_{\rm cm}}{\sqrt{s}}\ln\left[\frac{(s+2\sqrt{s}
	q_{\rm cm})^2-(M_l^2-m_l^2)^2}{(s-2\sqrt{s}q_{\rm cm})^2
	-(M_l^2-m_l^2)^2}\right]\Bigg \rbrace ,  
\end{eqnarray}
where $\mu=1$~GeV is the regularization scale and $a_l$ are the 
so called subtraction constants, which are taken as free 
parameters to be fitted to data. There are only 6 independent 
subtraction constants in $S=-1$ meson-baryon scattering due to 
isospin symmetry.

This model will be supplemented by adding, to the chiral  $\bar{K} 
N\rightarrow K^+ \Xi^-$,  $K^0 \Xi^0$ amplitudes, some of the 
known resonances in the $1.89<M<2.35$~GeV energy range. From the 
eight resonances rated with three- and four-stars in~\cite{PDGrR}, 
we take the $\Sigma(2030)$ and $\Sigma(2250)$ as good candidates, 
coinciding with the findings of Ref.~\cite{Sharov:2011xqR}, which 
examined various combinations of several resonances. The spin and 
parity $J^\pi =7/2^+$ of the $\Sigma(2030)$ are well established. 
Those of the $\Sigma(2250)$ are not known,  but the most probable 
assignments are $5/2^-$ or $9/2^-$~\cite{PDGrR}. We choose $J^\pi 
=5/2^-$ to simplify the calculations. An ad-hoc exponential 
form-factor function is also introduced to modify the energy 
dependence of the resonance contributions. Details on how the 
resonant terms are implemented can be found in 
Ref.~\cite{Feijoo:2015yjaR}. The fit to the data will determine 
the masses, widths, form-factor cut-off values, and the product 
of the resonance couplings to ${\bar K}N$ and $K \Xi$ states.

Once the $T$-matrix is known, one can obtain the differential 
and total cross sections, the $K^-p$ scattering length, the 
related energy shift and width of Kaonic hidrogen via the 
second order corrected Deser-type formula~\cite{Meissner:2004jrAR}, 
the branching ratios at threshold, \textit{etc}.
\begin{figure*}[htb!]
\centering
\includegraphics[width=4.3in]{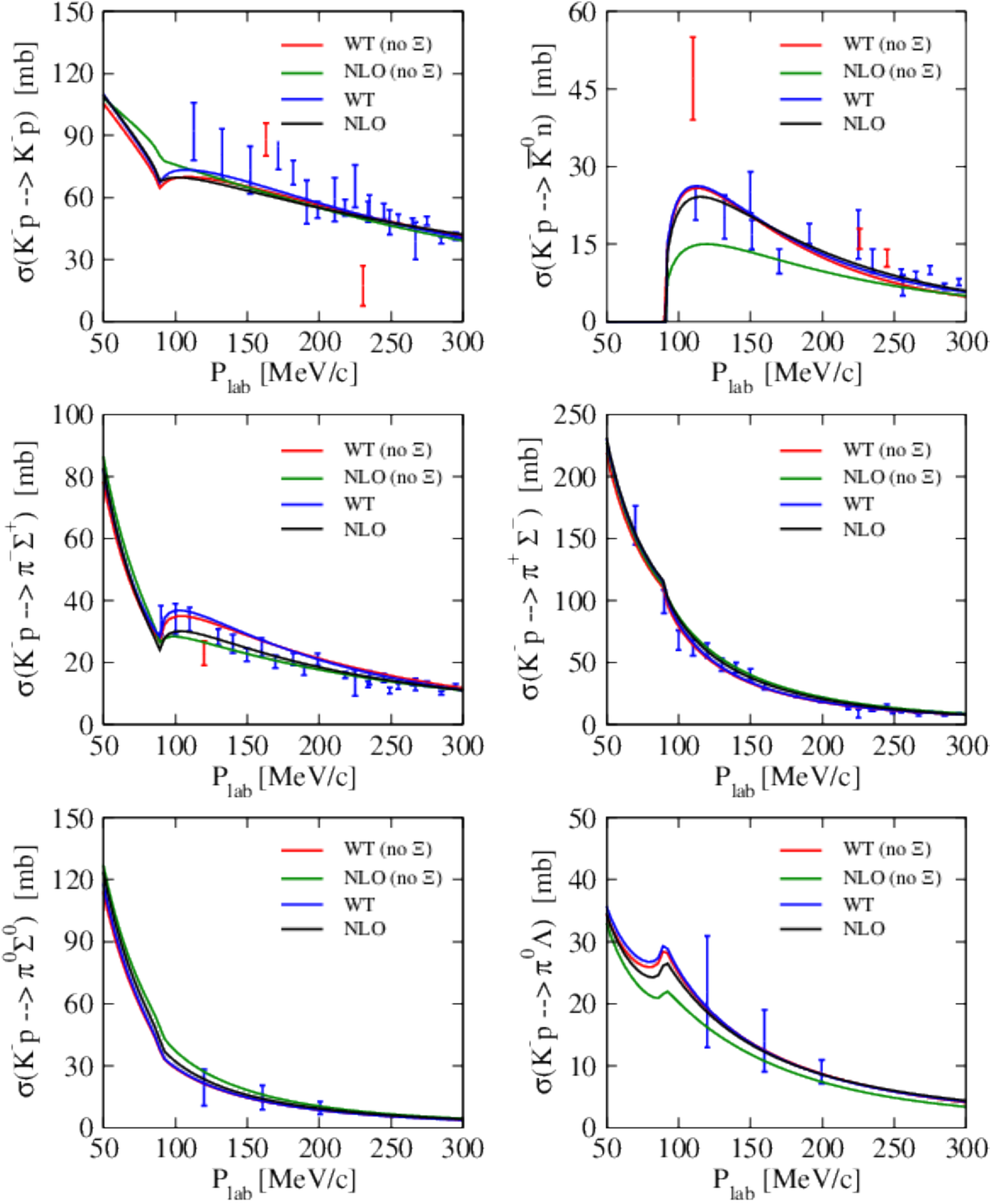}
\centerline{\parbox{0.80\textwidth}{
 \caption{Total cross sections for the $K^-p\to K^-p$, 
	$\bar{K}^0n$, $\pi^-\Sigma^+$, $\pi^+\Sigma^-$, 
	$\pi^0\Sigma^0$, and $\pi^0\Lambda$ reactions 
	obtained from the WT (no $K\Xi$) fit (red line), 
	the NLO (no $K\Xi$) fit (green line), the WT fit 
	(blue line) and the NLO fit (black line), where 
	the last two cases take into account the 
	experimental data of the $K\Xi$ channels, see 
	text for more details. Experimental data are 
	from~\protect\cite{exp_1R,exp_2R,exp_3R,exp_4R}. 
	The points in red have not been included in the 
	fitting procedure.} \label{fig1} } }
\end{figure*}

\item \textbf{Results}

The elastic and inelastic cross sections are shown in 
Fig.~\ref{fig1} for several fits: 'WT (no $\Xi$)' and 'NLO 
(no $\Xi$)' correspond to cases employing the chiral kernel 
up to lowest order or NLO, respectively, without considering 
the $K\Xi$ production data in the fits, as customary done by 
the chiral unitary models existing in the literature. The 
other two models, 'WT'  and 'NLO', do fit in addition the 
$K^-p\to K^+\Xi^-, K^0\Xi^0$ data. It is evident from 
Fig.~\ref{fig1} that, for the observables represented there, 
the four fits are similarly good. The situation changes 
drastically for the cross sections of the $\Xi$ production 
reactions represented in Fig.~\ref{fig2} for the same four 
models. The WT (no $K\Xi$) fit, represented by red lines, 
cannot even reproduce the size of the cross section in 
either reaction, which is not a surprising result because 
there is no direct contribution from the reactions $K^-p\to 
K^0\Xi^0, K^+\Xi^-$ at lowest order. This is why these 
reactions are very sensitive to the NLO corrections, which 
is confirmed by the NLO (no $K\Xi$) results  represented by 
green lines in Fig.~\ref{fig2}. Even if the experimental 
data for the $K^-p\to K^0\Xi^0, K^+\Xi^-$ reactions 
have not been employed in this fit, the NLO (no $K\Xi$)  
result gives a larger amount of strength for this channels, 
especially in the case of the $K^+\Xi^-$ production reaction, 
where the prediction even overshoots the data considerably. 
When the $K\Xi$ data is included  in the fitting procedure, 
the NLO results, represented by the black lines, reproduce 
quite satisfactorily the $K^-p\to K^0\Xi^0, K^+\Xi^-$ cross 
sections. For completeness, we have also attempted to 
reproduce these reactions employing only the lowest order 
Lagrangian. The corresponding WT results, represented by the 
blue lines, improve considerably over those of the WT (no 
$K\Xi$) fit, but at the expense of unphysical values for the 
fitted subtraction constants since the strength in these 
channels is built mainly through unitarization.
\begin{figure}[htb!]
\centering
\includegraphics[width=3in]{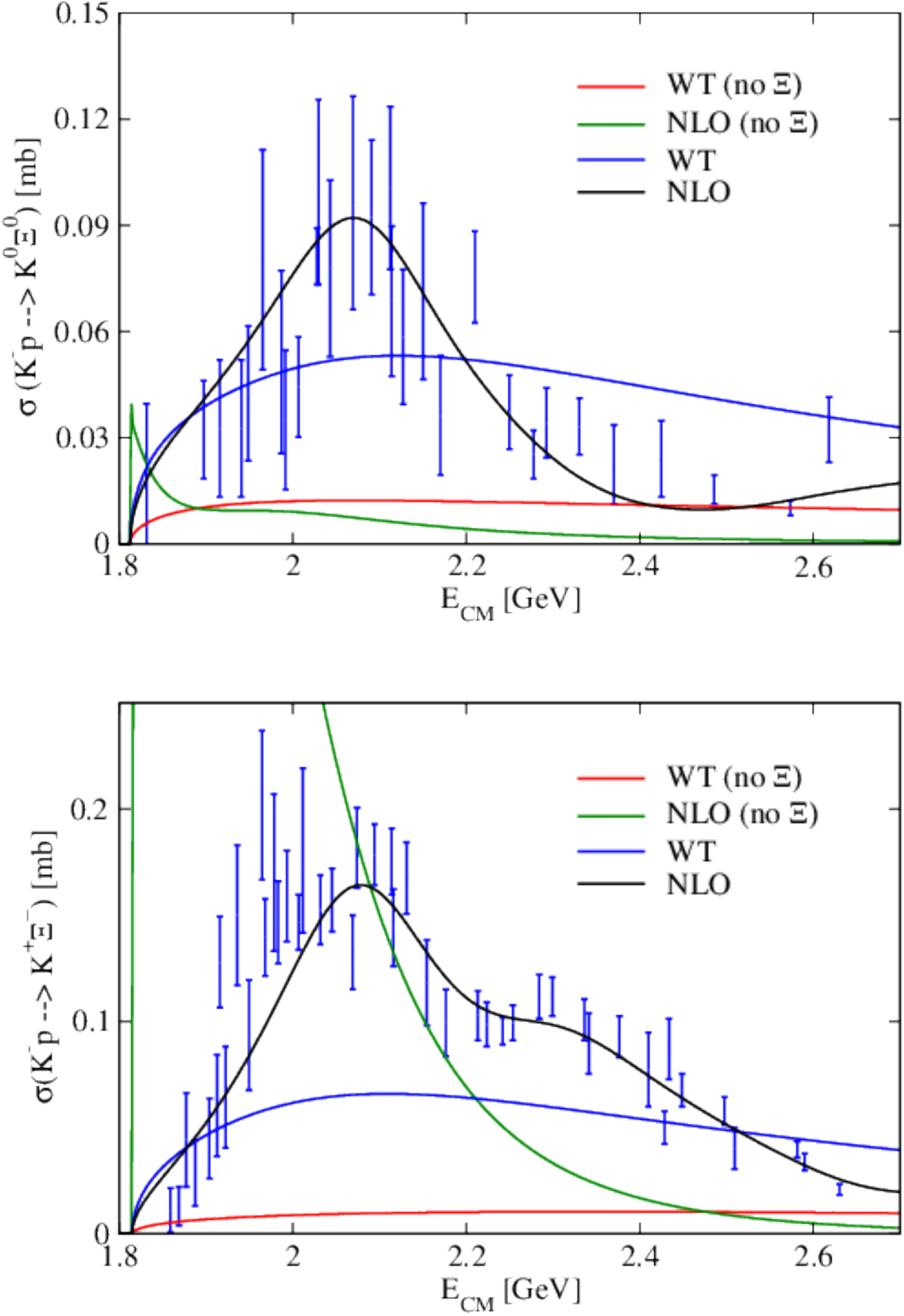}
\centerline{\parbox{0.80\textwidth}{
 \caption{The total cross sections of the $K^-p\to
        K^0\Xi^0, K^+\Xi^-$ reactions obtained from
        the WT (no $K\Xi$) fit (red line), the NLO (no
        $K\Xi$) fit (green line), the WT fit (blue line)
        and the NLO fit (black line). Experimental data
        are from~\protect\cite{exp_5R,exp_6R,exp_7R,exp_8R,
        exp_9R,exp_10R,exp_11R}.} \label{fig2} } }
\end{figure}
\begin{figure}[ht!]
\centering
\includegraphics[width=3in]{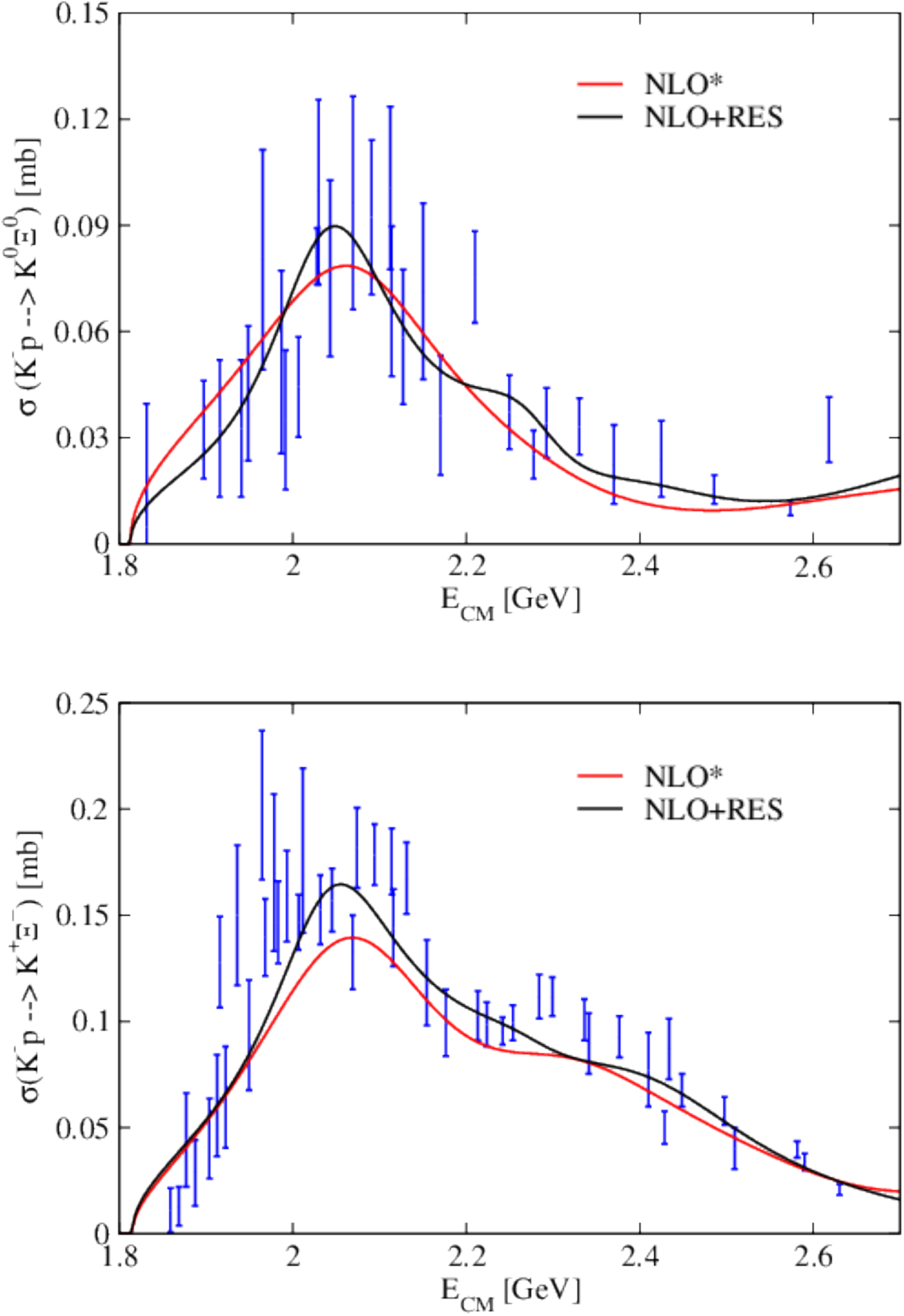}
\centerline{\parbox{0.80\textwidth}{
 \caption{Total cross sections of the $K^-p\to K^0\Xi^0, 
	K^+\Xi^-$ reactions for the NLO$^\ast$ fit (red 
	line) and the NLO+RES fit (black line). 
	Experimental data are from~\protect\cite{exp_5R,
	exp_6R,exp_7R,exp_8R,exp_9R,exp_10R,exp_11R}.} 
	\label{fig3} } }
\end{figure}

The discrepancies between the NLO model and the data, which 
are larger in the vicinity of $2$~GeV and around $2.2$~GeV, 
can be improved by the explicit implementation of resonance 
terms coupling to the $K \Xi$ channels. Since the resonant 
terms produce angular dependent scattering amplitudes, we 
can now consider, in addition to the total cross sections 
and threshold observables already employed in the earlier 
fits, the differential cross section data of the $K^-p\to 
K\Xi$ reactions. In Figs.~\ref{fig3}, \ref{fig4}, and 
\ref{fig5}, we present total and differential cross section 
data for two different fits: 'NLO$^ast$' stands for a fit 
that considers the chiral lagrangian up to NLO and includes 
all the data than the previous NLO fit plus the differential 
cross section data in the $\Xi$ production channels. 
'NLO+RES' stands for the fit employing a model that 
incorporates the additional effect of the two high spin 
resonances. 
\begin{figure}[htb!]
\begin{center}
\includegraphics[width=4in]{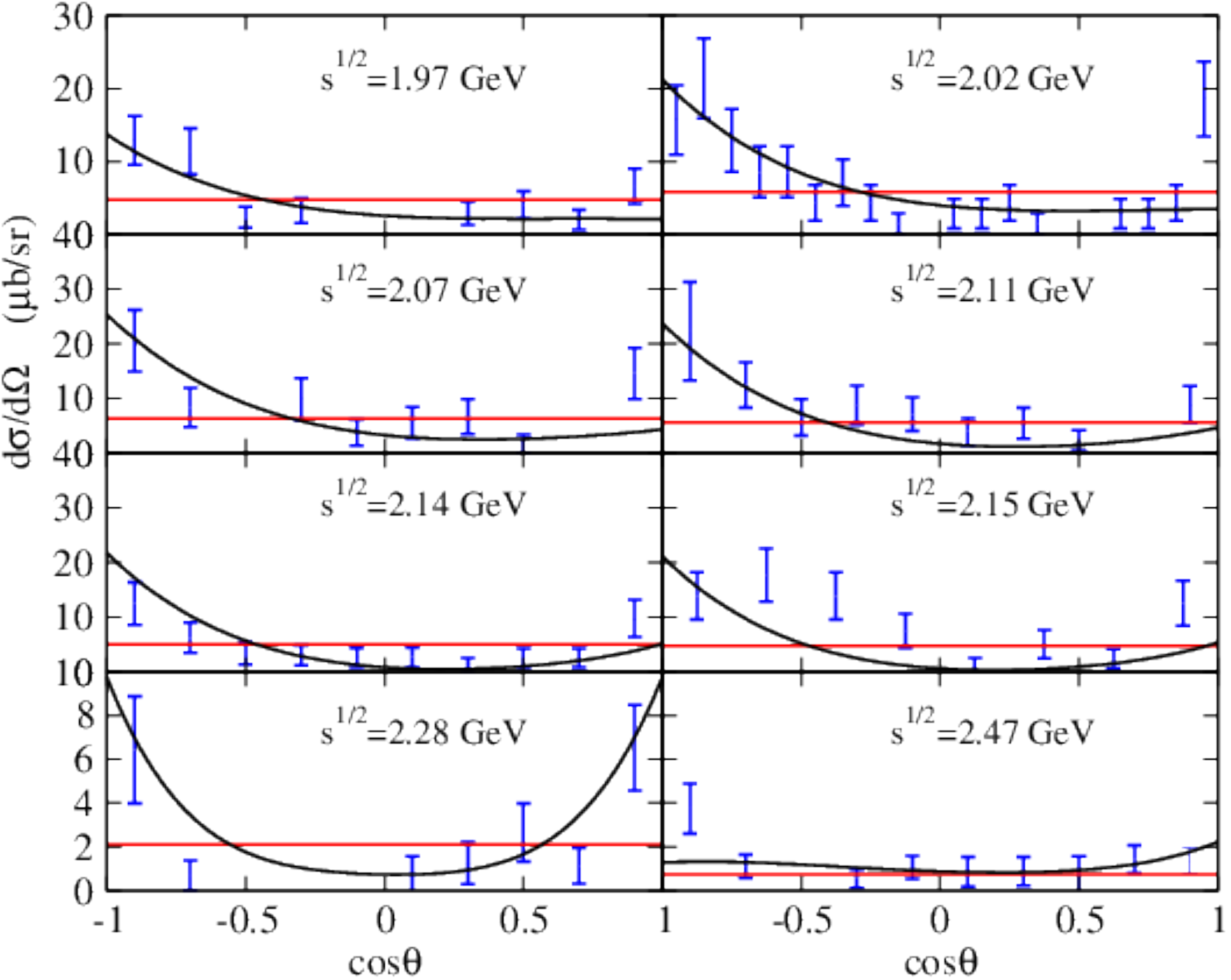}
\end{center}
\centerline{\parbox{0.80\textwidth}{
 \caption{Differential cross section of the $K^-p\to K^0
	\Xi^0$ reaction for the NLO$^\ast$ fit (red line) 
	and the NLO+RES fit (black line), see the text for 
	more details. Experimental data are 
	from~\protect\cite{exp_5R,exp_6R,exp_7R,exp_8R,exp_9R,
	exp_10R,exp_11R}.} \label{fig4} } }
\end{figure}
\begin{figure}[ht!]
\centering
\includegraphics[width=4in]{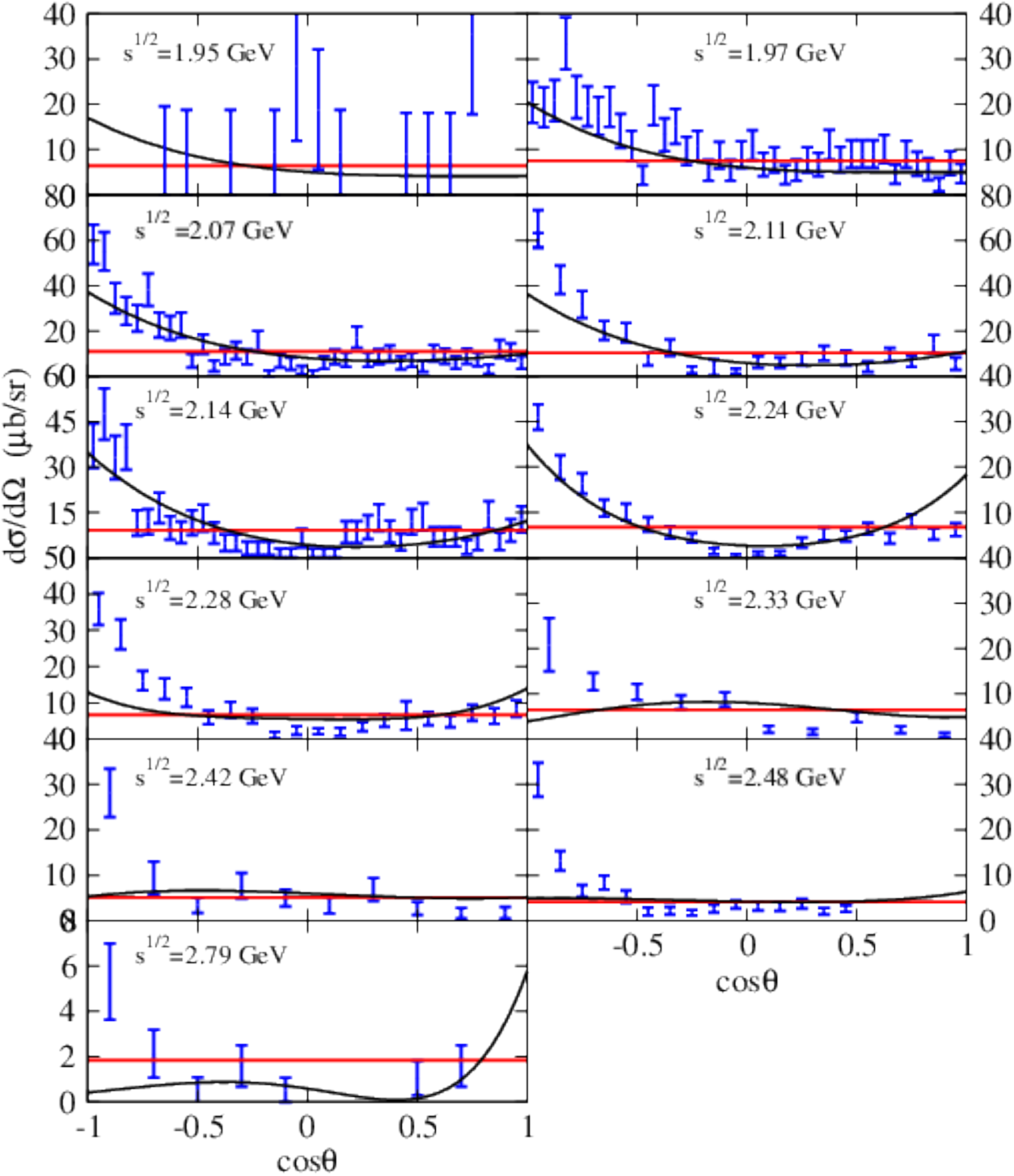}
\centerline{\parbox{0.80\textwidth}{
 \caption{Differential cross section of the $K^-p\to 
	K^+ \Xi^-$ reaction for the NLO$^\ast$ fit (red 
	line) and the NLO+RES fit (black line), see the 
	text for more details.  Experimental data are 
	from~\protect\cite{exp_5R,exp_6R,exp_7R,exp_8R,exp_9R,
	exp_10R,exp_11R}.} \label{fig5} } }
\end{figure}

The total cross sections for $K\Xi$ production obtained 
from the NLO$^\ast$ fit (red lines in Fig.~\ref{fig3}) are 
in reasonable agreement with the data, even if the resonant 
terms are not included. This NLO$^\ast$ fit is in fact very 
similar to the NLO one but it also tries to accommodate 
the differential $K\Xi$ production cross section data,  
which can only be adjusted on average, as shown by the 
red lines in Figs.~\ref{fig4} and \ref{fig5}, because of 
the flat distribution characteristic of $s$-wave models. 
In order to account for some structure in the differential 
$K\Xi$ production cross sections we need to implement the 
resonant terms. On inspecting the black lines in 
Figs.~\ref{fig3}, \ref{fig4}, and \ref{fig5} one clearly 
sees that the NLO+RES fit reproduces satisfactorily the 
$K\Xi$ total cross sections, while accounting  quite 
reasonably for the differential ones.
\begin{figure}[ht!]
\centering
\includegraphics[width=4in]{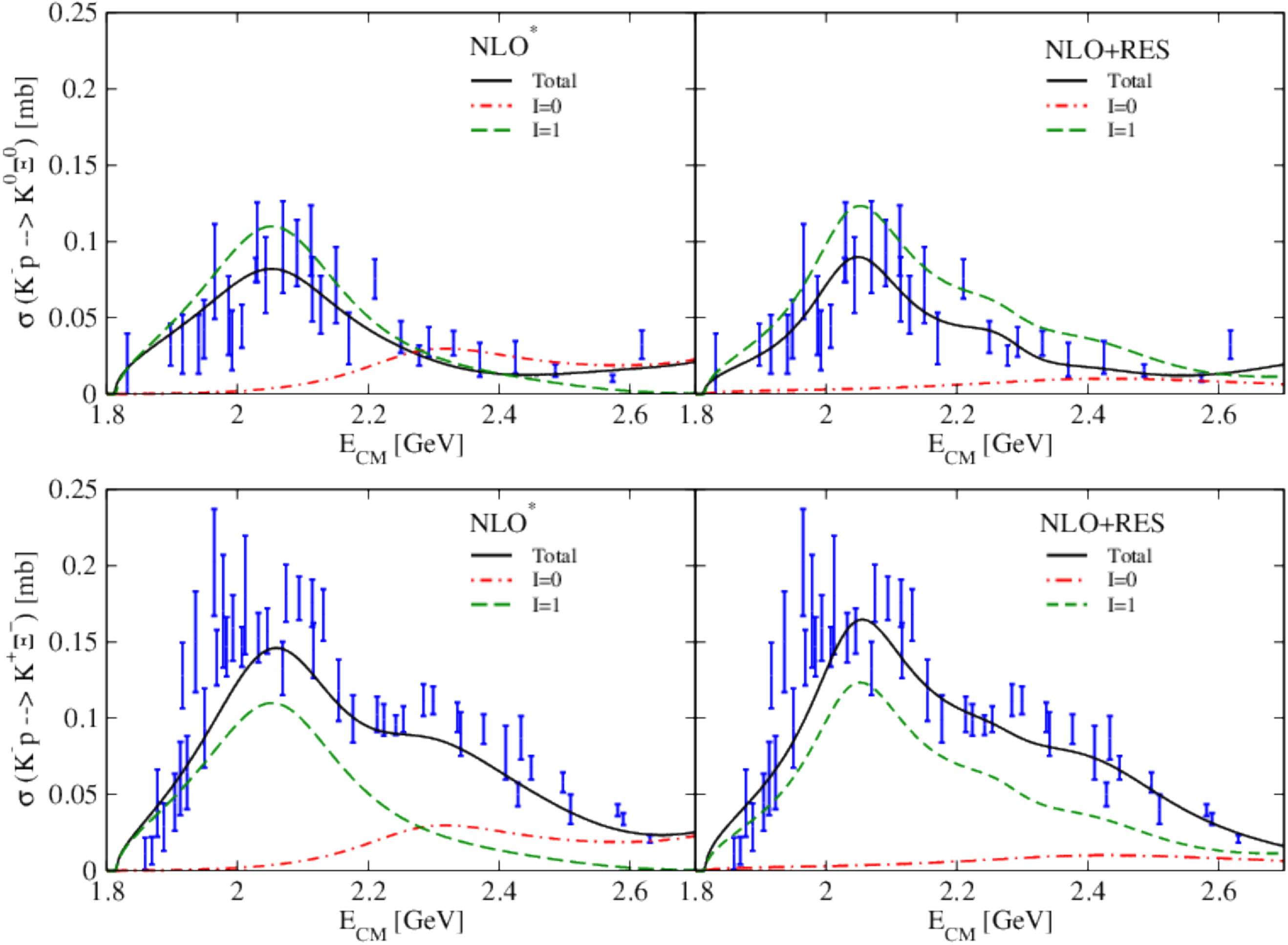}
\centerline{\parbox{0.80\textwidth}{
 \caption{Total cross sections of the $K^-p\to K^0\Xi^0, K^+
	\Xi^-$ reactions for the NLO$^\ast$ fit (left panels) 
	and the NLO+RES fit (right panels), splitted into 
	their $I=0$ (red dot-dashed lines) and $I=1$ (green 
	dashed lines) components. Experimental data are 
	from~\protect\cite{exp_5R,exp_6R,exp_7R,exp_8R,exp_9R,
	exp_10R,exp_11R}.} \label{fig6} } }
\end{figure}

In Fig.~\ref{fig6} we show the isopin $I=0$ and $I=1$ 
contributions to the $K^-p \to K^0 \Xi^0, K^+ \Xi^-$ 
cross sections, which must be summed coherently. We see 
that the chiral NLO$^\ast$ model produces stronger 
$I=1$ amplitudes than $I=0$, which is the reversed 
situation than one finds in the absence of unitarization. 
The dominance of $I=1$ contributions is obviously 
enhanced in the NLO+RES model that includes two $I=1$ 
resonances explicitly.  However, other models in the 
literature, as that of Ref.~\cite{Jackson:2015dvaAR} which 
is fitted to the same data, find a different distribution 
over isospin channels. For this reason, in order to pin 
down the details of the meson-baryon interaction in the 
$S=-1$ sector, it would be very valuable to have data in 
this energy range with a definite value of isospin. 

A recent possibility has emerged with the measurement at 
LHCb of the decay of the $\Lambda_b$ into $J/\Psi$ and a 
meson-baryon pair in $S=-1$. From the three-body final 
state, the reconstruction of $J/\Psi ~ p$ pairs permitted 
to find the signal of the $P_c(4450)$ pentaquark 
state~\cite{Aaij:2015tgaAR}. On the other hand, the 
invariant mass spectrum of the $K^-p$ pairs gives access 
to study their interaction, and actually, that of any of 
its related coupled-channel meson-baryon ($MB$) pairs, in 
$I=0$, since it can be shown that the decay $\Lambda_b 
\to J/\Psi M B$ acts as an $I=0$ filter~\cite{rocamaiR,
feijoomagasR}. The invariant masses of  $I=0$ $K \Xi$ 
states calculated in~\cite{feijoomagasR} show indeed the 
different predictions for the invariant mass distributions 
of $K\Xi$ pairs coming from the decay  of the $\Lambda_b$, 
obtained using the NLO$^\ast$ and  the NLO+RES models.
\begin{figure}[ht!]
\centering
\includegraphics[width=3.5in]{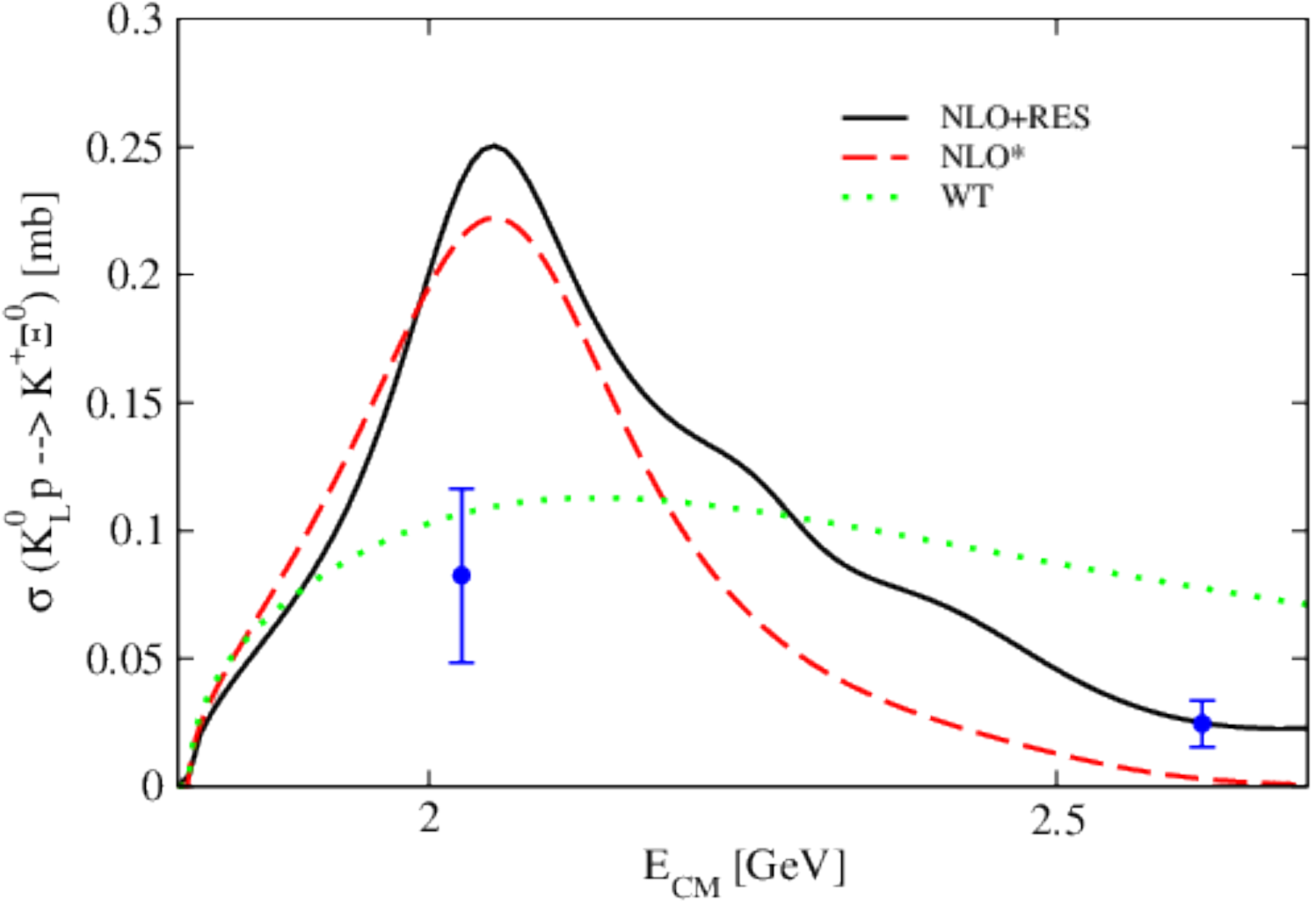}
\centerline{\parbox{0.80\textwidth}{
 \caption{Total cross sections of the $K^0_L\to K^+ 
	\Xi^0$ reactions for the for the WT (green 
	dotted line), NLO$^\ast$ (red dashed line) and 
	NLO+RES (black solid line) models, together 
	with the experimental points of the $I=1$ 
	$K^-n\to  K^0\Xi^-$ reaction (divided by two). 
	Experimental data are from \cite{iso1exp1R,
	iso1exp2R}.} \label{fig7} } }
\end{figure}

It would also be interesting to obtain information on the 
${\bar K}N \to K \Xi$ interacion in $I=1$, of which only 
two points, obtained from $K^-$ deuteron reactions in 
bubble chamber experiments, are known~\cite{iso1exp1R,
iso1exp2R}. The recent proposal of creating a secondary 
$K^0_L$ beam at JLab offers a great opportunity for 
measuring the $K^0_Lp\to K^+\Xi^0$ reaction, which would 
proceed through the ${\bar K}^0$ component of the $K^0_L$ 
and is of pure $I=1$ character. Our predictions for this 
reaction are shown in Fig.~\ref{fig7} for the WT (green 
dotted line), NLO$^\ast$ (red dashed line) and NLO+RES 
(black solid line) models, together with the experimental 
points of the $I=1$ $K^-n\to K^0\Xi^-$ reaction, which 
have been divided by two to properly account for the size 
of the strangeness $-1$ component of the $K^0_L$. Since 
these two data points have not been used in the fit, our 
predictions for the most complete NLO$^\ast$ or NLO+RES 
models are not good, especially for the data point around 
2~GeV. New data from modern experiments, as the proposed 
measurement of the $K^0_Lp\to K^+\Xi^0$ reaction with a 
secondary $K^0_L$ beam at Jlab, would certainly be very 
helpful in constraining  the models describing the 
meson-baryon interaction in the $S=-1$ sector around 
2~GeV tighter. 

\item \textbf{Conclusions}

We have presented a study of the $S=-1$ meson-baryon 
interaction, employing a chiral SU(3) Lagrangian up to 
next-to-leading order and implementing unitarization in 
coupled channels. The model has been supplemented by the 
explicit consideration of two resonances. The parameters 
of the Lagrangian and of the resonances have been fitted 
to a large set of experimental scattering data in 
different two-body channels, to threshold branching 
ratios, and to the precise SIDDHARTA value of the energy 
shift and width of Kaonic hidrogen. In contrast to other 
works, we have also constrained our model to also 
reproduce the $K^-p\to K^+\Xi^-, K^0\Xi^0$ reactions, 
since they become especially sensitive to the NLO terms.

While a good account of the $K^-p\to K^+\Xi^-, K^0\Xi^0$ 
total and differencial cross sections is achieved, the 
isospin decompositions of the models studied here show 
some differences, and they also differ from other works 
in the literature. We find a disagreement of our 
predictions with the scarce available data in the $I=1$ 
sector. Measuring the $K^0_L p \to K^+ \Xi^0$ reactions 
with a secondary $K^0_L$ beam at Jlab would be extremely 
valuable to further constrain the parameters of the chiral 
lagrangian describing the meson-baryon interaction in the 
$S=-1$ sector.

\item \textbf{Acknowledgments}

This work is partly supported by the Spanish Ministerio 
de Economia y Competitividad (MINECO), under the project 
MDM--2014--0369 of ICCUB (Unidad de Excelencia 'Mar\'\i a 
de Maeztu') and  under the contract FIS2014--54762--P 
(with additional European FEDER funds), and by the 
Ge\-ne\-ra\-li\-tat de Catalunya under the contract 
2014SGR--401.
\end{enumerate}


\newpage
\subsection{Hyperon Studies at JPAC}
\addtocontents{toc}{\hspace{2cm}{\sl C.~Fern\'andez-Ram\'{\i}rez and
	A.~Szczepaniak}\par}
\setcounter{figure}{0}
\setcounter{equation}{0}
\halign{#\hfil&\quad#\hfil\cr
\large{C\'esar Fern\'andez-Ram\'{\i}rez}\cr
\textit{Instituto de Ciencias Nucleares}\cr
\textit{Universidad Nacional Aut\'onoma de M\'exico}\cr 
\textit{A.P.~70-543, Ciudad de M\'exico 04510, Mexico}\cr\cr
\large{Adam Szczepaniak}\cr
\textit{Center for Exploration of Energy and Matter}\cr
\textit{Indiana University}\cr
\textit{Bloomington, IN 47403, U.S.A. \&}\cr
\textit{Theory Center}\cr
\textit{Thomas Jefferson National Accelerator Facility}\cr
\textit{Newport News, VA 23606, U.S.A.}\cr}

\begin{abstract}
We provide an overview of the recent work developed at the Joint 
Physics Analysis Center (JPAC) regarding $\bar{K}N$ scattering 
and the hyperon spectrum. We emphasize our findings on the nature 
of the two $1/2^+$ poles present in the $\Lambda(1405)$ region.
\end{abstract}

\begin{enumerate}
\item \textbf{Hadron Reactions and Resonances as Probes of 
	Strong QCD}

Recently there have been dramatic advancements in accelerator 
technologies, detection techniques and on the theoretical side,  
algorithms for first principle  QCD analyses~\cite{ATHOSC}. 
These have led to several candidates for possible 
\textquotedblleft exotic \textquotedblright ~hadrons, {\it 
i.e.}, quark-gluon hybrids or quark-hadron molecular states. 
It thus appears that interpretation of the entire hadron 
spectrum in terms of the valence constituents of the quark 
model is no longer possible. If confirmed, such 
\textquotedblleft exotic \textquotedblright ~hadrons could 
drastically alter our understanding of strong QCD and shed 
new light on the confinement of quarks. 
 
Given the wide interest in hadron spectroscopy, the Joint 
Physics Analysis Center (JPAC)~\cite{JPACwebpageC} has been 
dedicated to the development of theoretical and phenomenological 
analysis methods for analysis of hadron reactions. To achieve 
these goals researchers affiliated with JPAC are developing 
amplitude models based on principles of S-matrix theory to 
formulate scattering amplitudes for various reactions of 
interest to the hadron physics community and QCD practitioners. 
JPAC members work in close collaboration with experimentalists 
on implementing theoretical innovations into the existing data 
analysis streams, preserving knowledge for future use and 
disseminating the methodology across various experiments. 

Resonances are characterized by their mass and spin. Near the 
resonance mass a two-body cross section  vary as a function of 
the center of mass energy while angular variations of the 
differential cross section reflect the resonance spin. 
Variations in the cross section, albeit smooth, are a 
manifestation of a singularity in a partial wave amplitude, 
which is seen when the amplitude is continued beyond the real 
energy axis and/or beyond the integer (or half-integer) values 
of spin to complex domains of these variables. Resonance 
parameters are therefore determined by analytical continuation 
of reaction amplitude models. 

Analytical behavior of reaction amplitudes as functions of 
Mandelstam variables follows from principles of the S-matrix 
theory~\cite{EdenC}. For example, the absence of singularities 
in the complex energy plane, {\it a.k.a.}, the physical sheet,  
follows from causality and crossing symmetry. On the physical 
sheet the only allowed singularities are bound state poles, 
{\it e.g.}, the nucleon pole, and cuts induced by opening of 
particle production thresholds. The latter  lie on the real 
axis and discontinuity of the amplitude across cuts is 
constrained by unitarity. Resonance poles appear as 
singularities on complex planes (unphysical sheets) that are 
connected to the physical sheet along these discontinuities. 
These unphysical sheets are connected in such a way that the 
amplitudes change smoothly when passing from the physical to 
an unphysical sheet. Therefore, a resonance pole located on 
an unphysical sheet close to the real axis has a strong 
influence on the amplitude in the physical region of scattering. 
   
S-matrix analyticity does not predict whether a resonance 
exists or it does not. It is the underlying dynamics, {\it i.e.}, 
QCD, that determines that. Given a model that specifies the 
number of expected resonances, S-matrix principles, however, 
enable to write amplitudes that properly continue amplitudes 
from poles to the real axis where they can be compared with 
experimental data.  Resonances of different spins are not 
independent. This follows from crossing relations and unitarity 
implying analyticity of partial waves in the complex angular 
momentum plane~\cite{GribovC}. This is the  essence of the Regge 
theory~\cite{PDBCollinsC}. Therefore, partial waves should be 
analyzed simultaneously as function of mass and spin. This, 
however, is almost never done. Typically mass dependent partial 
wave analyses deal with each partial wave independently. 
Without imposing relations between resonances of different 
spins, the various methods for implementing resonances in a 
single partial wave are closely related. These methods include, 
for example, the K-matrix parametrization~\cite{kmatrixC}, N/D 
parametrizations~\cite{noverdC}, or solutions of  Bethe-Salpeter 
motivated equations with contact interactions~\cite{Oset1C}.
When appropriate, all these parameterizations can be supplemented  
with additional constraints, {\it e.g.}, from chiral symmetry at 
low-energies or Regge asymptotics.  It is worth noting that the 
extraction of resonance properties based on a first principle 
QCD analysis, {\it i.e.}, using  lattice gauge techniques, 
requires the same amplitude parametrizations as data 
analysis~\cite{latticeC}. 

\item \textbf{Hyperons: Terra Incognita in the Baryon 
	Landscape}

Even though the spectrum of baryons has been investigated for 
several decades, only a handful of hyperon resonances has been 
well established~\cite{PDG2014C}. One of the key aspects of 
confinement is the (approximate) linearity of Regge trajectories. 
In the isoscalar sector [$\Lambda$, Fig.~\ref{fig:polesreg}] 
the leading natural parity, even ($1/2^+,5/2^+,\dots$) and odd 
($3/2^-,7/2^-, \dots$) trajectories have two well-established 
states each. If one assumes exchange degeneracy, then the four 
states do appear to lie on a straight line. Unfortunately,  
only a few more $\Lambda$'s  are reasonably well established 
and the identification of other trajectories remains ambiguous.  
For example, it is unclear whether the first excited $1/2^+$ 
state, the $\Lambda(1600)$, belongs to the same trajectory as 
the $5/2^+$ $\Lambda(2110)$ resonance or to a different one. 
Similar ambiguity appears in the leading unnatural parity ($1/2^-, 
5/2^-, \dots$) trajectory where, given that there is a strong 
indication that what is known as the $\Lambda(1405)$ could 
actually correspond to two resonances, it is unclear which 
$1/2^-$ pole should be connected with the $5/2^-$ $\Lambda(1830)$. 
Similar ambiguities exist in the isovector sector [$\Sigma$, 
Fig.~\ref{fig:polesreg}]. Except for the leading even natural 
parity trajectory, other trajectories have, at best, one 
well-established resonance, thus their shape cannot be 
established. The $3/2^-$ $\Sigma(1940)$ should be the first 
resonance on the leading odd natural-parity trajectory, 
however there is weak evidence of such a pole in mass dependent 
partial-wave analysis of $\bar{K}N$ scattering. The situation 
with the $\Xi$ and $\Omega$ baryons is even worse~\cite{PDG2014C}.
\begin{figure}[ht!]
\begin{center}
\includegraphics[angle=0, width=0.45\textwidth]{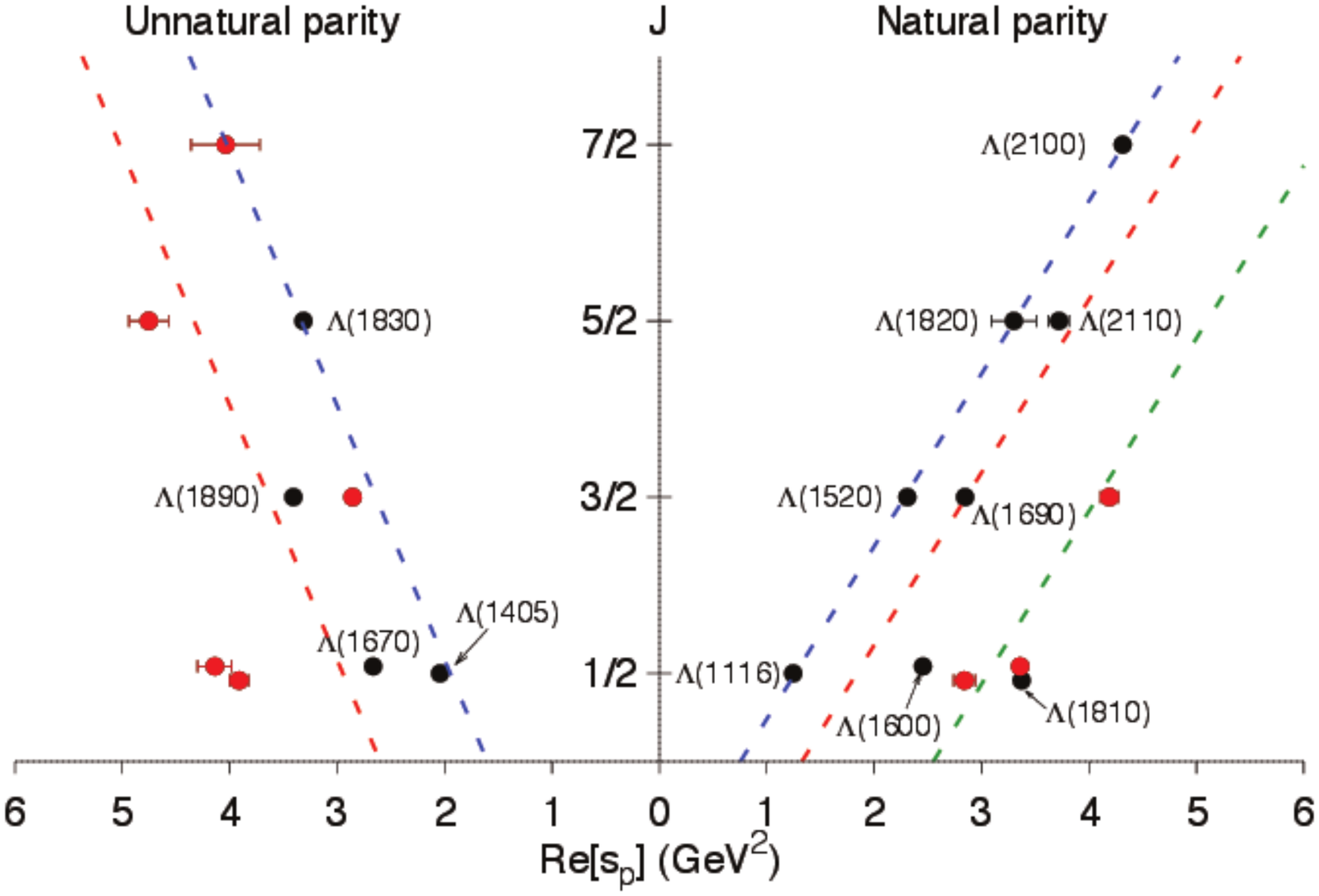}
\includegraphics[angle=0, width=0.45\textwidth]{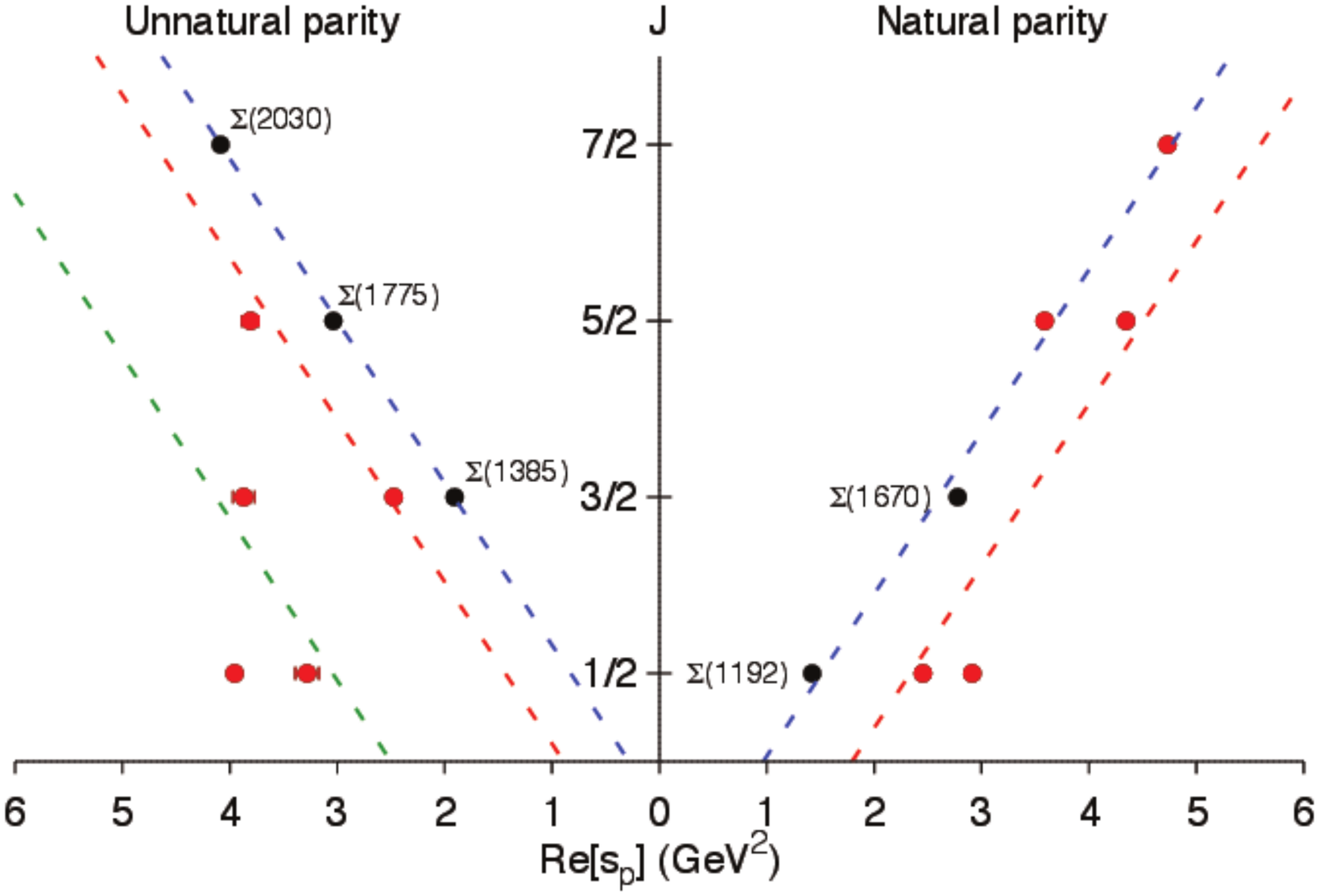} \\
\end{center}
\vspace{-1.0cm}
\centerline{\parbox{0.80\textwidth}{
 \caption[] {\protect\small Chew--Frautschi plot for the $\Lambda$ and $\Sigma$
        Regge trajectories. Black dots represent resonances with
        a four-star status in the {\it Review of Particle
        Properties}~\protect\cite{PDG2014C}. All resonances are
        taken from JPAC analysis in~\protect\cite{FR2015C}.
        Dashed lines are displayed to guide the eye.}
        \label{fig:polesreg} } }
\end{figure}

The most recent (and advanced)  mass-dependent partial-wave 
analysis of the single-energy partial waves of the KSU data 
analysis of $\bar{K}N$ scattering in the resonant 
region~\cite{Manley13aC} was performed by JPAC in~\cite{FR2015C}.
The partial-wave model is based on a coupled-channel K-matrix 
approach.  It incorporates up to 13 channels per partial wave, 
analyticity, unitarity, and the appropriate threshold (angular 
momentum barrier) factors  for the partial waves. Here we sketch 
the building blocks of the model. For further details, the 
analysis, the fitting, and the obtained spectrum we refer the 
reader to~\cite{FR2015C}.

The partial-wave expanded scattering matrix $S_\ell$ can be 
written in terms of the amplitude $T_\ell$ as follows
\begin{equation}
	S_\ell=\mathbb{I}+2iR_\ell(s)=\mathbb{I}+2i \left[C_\ell 
	(s) \right]^{1/2} T_\ell(s) \left[C_\ell (s) \right]^{1/2}, 
	\label{eq:smatrix}
\end{equation}
where $\mathbb{I}$ is the identity matrix and the diagonal matrix
\begin{equation}
	C_\ell (s) = \frac{q_k (s)}{q_0}\left[ \frac{r^2q^2_k(s)}{1
	+ r^2q^2_k(s) }\right]^{\ell}  \label{C}
\end{equation}
takes into account the phase space. We define $q_k(s)  = 
\sqrt{( m_a m_b) (s-s_k )}/ (m_a+m_b)$, where $m_a$ and $m_b$ are 
the masses of the outgoing particles, $s_k$ is the threshold 
center-of-mass energy squared of channel $k$, and $q_0=2$~GeV 
and $r=1$~fm are normalization factors.

We employ the $K$ matrix approach to guarantee unitarity through
\begin{equation}
	T_\ell (s)= \left[ K(s)^{-1} -i \rho(s,\ell) \:  
	\right]^{-1}, \label{eq:kmatrix}
\end{equation}
where $\rho(s,\ell )$  is obtained from $C_\ell (s)$ by means of 
a dispersive integral
\begin{equation}
	i \rho (s,\ell)  =   \frac{s-s_k}{\pi} \int_{s_k}^\infty\frac{ 
	C_\ell (s')  }{s'-s} \frac{ds'}{s'-s_k}. \label{eq:rho}
\end{equation}
In this way, $T_\ell (s)$ can be analytically continued to both $s$ 
and $\ell$ complex planes. To build the $K(s)$ matrix in equation 
(\ref{eq:kmatrix}) we add up to six $K$ matrices
\begin{equation}
	\left[ K(s) \right]_{kj} =  \sum_a x^a_k\:K_a(s)\: x^a_j  \:.
\end{equation}
Each $K_a(s)$ matrix can either represent a {\it pole},
\begin{equation}
	\left[ K_P (s)\right]_{kj} = x^P_k \: \frac{M_P}{M_P^2-s} \:  
	x^P_j \:, \label{eq:kr}
\end{equation}
or a {\it background} term, 
\begin{equation}
	\left[ K_B (s)\right]_{kj} = x^B_k \: \frac{M_B}{M_B^2+s} \:  
	x^B_j \:. \label{eq:kb}
\end{equation}
The relative contribution of {\it pole} vs. {\it background} terms 
depends on the individual partial wave.  The parameters $M_P$, 
$x^P_k$, $M_B$ and $x^B_k$ are fixed by fitting the single-energy 
partial waves from the KSU analysis~\cite{Manley13aC} 
using~\textsc{MINUIT}~\cite{MINUITC} and a genetic 
algorithm~\cite{geneticC}. Once these parameters have been fitted 
to the data we can analytically continue the amplitudes to the 
unphysical Riemann sheets and search for poles (hyperon resonances).
In Fig.~\ref{fig:polesreg} we show the Chew-Frautschi~\cite{CFplotC} 
plot of the obtained hyperon resonances.

The codes to compute the partial waves and the observables (cross 
sections and asymmetries) can be downloaded from or run online on  
the JPAC web page~\cite{JPACwebpageC,Mathieu16C}.

\item \textbf{On the Nature of the $\Lambda(1405)$}

The nature of $\Lambda$(1405) hyperon resonance has been an open 
question for more than half a century~\cite{lambdaC}. The interest 
in this state has been renewed recently because of the new 
precision data from CLAS that enabled to resolve its spin and 
parity and confirm the $J^P=1/2^-$ assignment~\cite{Moriya2014C}.
There have also been new developments in chiral unitary 
models~\cite{OsetC,Mai2015C,twopolesC}, large $N_c$ QCD 
calculations~\cite{LargeNcC}, lattice QCD~\cite{Engel13C,Hall2015C}, 
quark-diquark models~\cite{Santopinto14C,Faustov15C}, and Regge 
phenomenology~\cite{JPACLambda1405C}.

Chiral unitary models applied to $\bar{K}N$ scattering and
$\pi\Sigma K^+$ photoproduction have been able to establish that 
the $\Lambda(1405)$ is not a single state but it corresponds to 
two resonances~\cite{OsetC,Mai2015C,twopolesC,moleculeC} located at 
$1429^{+8}_{-7}-i \: 12 ^{+2}_{-3}$~MeV  and at $1325^{+15}_{-15}
-i\: 90 ^{+12}_{-18}$~MeV~\cite{Mai2015C}.  In these papers the 
$\Lambda$(1405) poles are interpreted as of molecular nature. 
However, the poles are generated by effective interactions and
there is no reference to the fundamental, {\it i.e.}, QCD, 
composition of the resonances. Hence, the answer to the question 
of the nature of the poles remains open.  Especially if we 
consider that large $N_c$ QCD calculations of the baryon 
spectrum indicate that a (mostly) three-quark state should 
appear in the $\Lambda$(1405) region~\cite{LargeNcC}. Furthermore, 
recent quark-diquark models obtain a state in the $\Lambda$(1405) 
region at 1431~MeV in~\cite{Santopinto14C} and at 1406~MeV 
in~\cite{Faustov15C}. 

Regarding lattice QCD, the available simulations lead to 
inconclusive results. In~\cite{Engel13C} $\Lambda(1405)$ appears 
to be a three-quark state while in~\cite{Hall2015C} it seems to 
be more like a $\bar{K}N$ molecule.  However, it has to be taken 
into account that the resonant nature of the $\Lambda(1405)$ has 
been ignored in these calculations~\cite{FRcocoyoc2016C}.

In~\cite{JPACLambda1405C}, we use Regge phenomenology and connect 
the $\Lambda(1405)$ to the rest of the  hyperon spectrum.  In 
Fig.~\ref{fig:polesreg1} we show the leading Regge trajectories 
for $\Lambda$ and $\Sigma$ hyperons.  It is apparent how one of 
the $\Lambda$(1405) poles ($1429^{+8}_{-7}-i \: 12 ^{+2}_{-3}$~MeV) 
fits within the leading natural-parity $\Lambda$ Regge trajectory 
while the other ($1325^{+15}_{-15}-i\: 90 ^{+12}_{-18}$~MeV) 
does not. The presence of a new narrow-width $3/2^+$ state in the 
leading natural-parity $\Lambda$ Regge trajectory is essential to 
reach this conclusion.  The first evidence of such state was 
provided in~\cite{lambda1680C} and latter confirmed by JPAC 
analysis~\cite{FR2015C}.
\begin{figure}[ht!]
\begin{center}
\includegraphics[angle=0, width=0.45\textwidth]{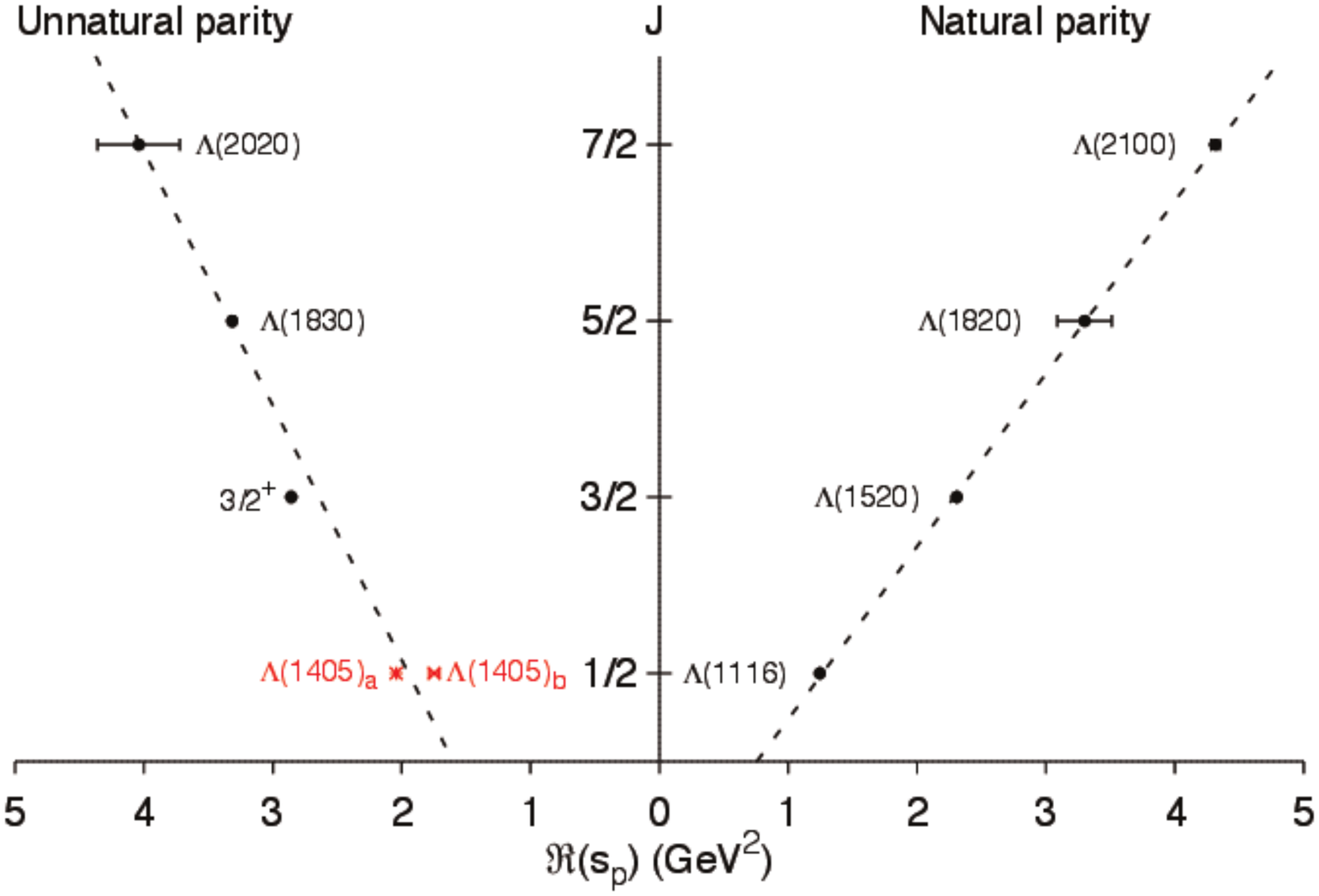}
\includegraphics[angle=0, width=0.45\textwidth]{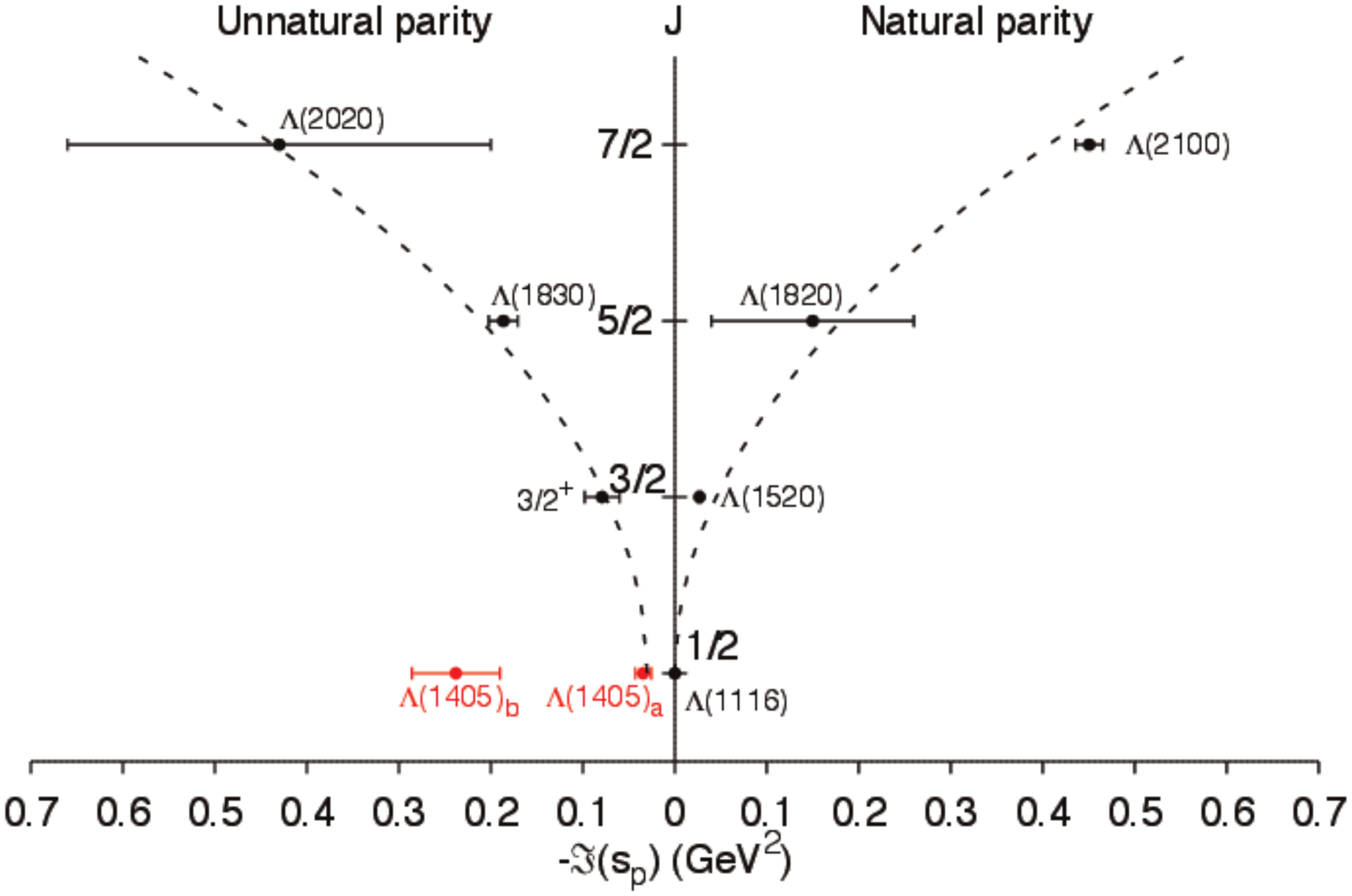} \\
\end{center}
\vspace{-1.0cm}
\centerline{\parbox{0.80\textwidth}{
 \caption[] {\protect\small Leading Regge trajectories for the $\Lambda$
        resonances. Dashed lines are displayed to guide
        the eye.}
        \label{fig:polesreg1} } }
\end{figure}

In~\cite{JPACLambda1405C}, we confirm what is apparent from 
inspecting Fig.~\ref{fig:polesreg1} through extensive numerical 
calculations by performing fits to various  physically 
motivated parametrizations of the Regge trajectories. We used 
$\Sigma$ Regge trajectories and the natural-parity $\Lambda$ 
trajectory to benchmark our approach. This analysis lead to 
the conclusion that the higher-mass pole belongs to the leading 
Regge trajectory and is compatible with a three-quark structure, 
while the lower-mass one does not belong either to the leading 
or to a nearby daughter Regge trajectory.

\item \textbf{Conclusions}

JPAC encourages close collaboration between experimentalists and 
theorists on implementing theoretical innovations  into the 
existing data analysis stream to order to obtain robust 
information on the hadron spectrum and hadron structure. 

Recently we have developed in~\cite{FR2015C} a $\bar{K}N$ model 
in the resonance region guided by S-matrix unitarity and 
analyticity (both in the $s$ and the $\ell$ complex planes).
We have obtained the most comprehensive picture of the 
$\Lambda$ and $\Sigma$ spectra to the date. Both spectra show 
a remarkable alignment of hyperons in Regge trajectories. The 
codes to compute the partial waves and the observables (cross 
sections and asymmetries) can be run online and downloaded 
from~\cite{JPACwebpageC}.

Regge phenomenology seems to indicate that the higher-mass 
$\Lambda$(1405) pole belongs to the leading Regge trajectory 
and that it is compatible with a (mostly) three-quark structure,
while the lower-mass one is either a molecule or a pentaquark.
If confirmed, by for example new data with $K_L$ beams, this 
finding becomes of prominent importance in identifying a 
two-component hybrid resonance.  It also sheds a valuable 
light on the success/failure of approaches based on $\bar{K}N$, 
three-quark, or five-quark states embodying genuine intrinsic 
light quark-antiquark pairs.  The work presented 
in~\cite{JPACLambda1405C} gives a new direction for investigation 
of the origin of poles associated  with the $\Lambda$(1405) 
that can also be extended to other sectors of the baryon 
spectrum. 

\item \textbf{Acknowledgments}

This material is based upon work supported in part by the 
U.S.~Department of Energy, Office of Science, Office of 
Nuclear Physics under contract DE--AC05--06OR23177. This 
work was also supported in part by the U.S.~Department of 
Energy under Grant DE--FG0287ER40365, National Science 
Foundation under Grants PHY--1415459 and PHY-1205019, and 
IU Collaborative Research Grant.
\end{enumerate}


\newpage
\subsection{Spectrum and Quantum Numbers of $\Xi$ Resonances}
\addtocontents{toc}{\hspace{2cm}{\sl Y.~Oh}\par}
\setcounter{figure}{0}
\setcounter{table}{0}
\setcounter{equation}{0}
\halign{#\hfil&\quad#\hfil\cr
\large{Yongseok Oh}\cr
\textit{Department of Physics}\cr
\textit{Kyungpook National University}\cr
\textit{Daegu 41566, Korea \&}\cr
\textit{Institute for Nuclear Studies and Department of Physics}\cr
\textit{The George Washington University}\cr
\textit{Washington, DC 20052, U.S.A.}\cr}

\begin{abstract}
Hyperons with strangeness $-2$ and $-3$ provide a useful tool 
to investigate the structure of baryons and the underlying 
dynamics.  They are expected to give information which is hard 
to be seen in nonstrange or strangeness $-1$ baryons. In this 
presentation, we analyze the spectrum of $\Xi$ resonances in 
various models and make comments on the problems and puzzles.
In particular, the hyperon spectrum in the bound state 
approach of the Skyrme model is discussed to identify the 
analog states of $\Lambda(1405)$ in strangeness $-2$ and $-3$ 
sector. In addition, sum rules in hyperons masses and magnetic 
moments are presented, which can be used to predict unmeasured 
masses and magnetic moments of hyperons. The planned Kaon beam 
facility will have a crucial role to resolve these issues and 
open a new way to understand baryon structure.
\end{abstract}

\begin{enumerate}
\item \textbf{Introduction}

Understanding the structure of baryons is essential to investigate 
strong interactions. In particular, establishing baryon spectrum 
requires rigorous studies both in theoretical and experimental 
investigations and identifying quantum numbers and various physical 
quantities of baryons is crucial to resolve questions and puzzles 
in baryon structure and strong interactions. In theoretical side, 
there many phenomenological models have been developed to explain 
the observed baryon spectrum and to predict unobserved states.
Since Quantum Chromodynamics, the underlying theory of strong 
interactions, cannot be directly used to explain baryon spectrum, 
those models inevitably introduce several model parameters. These 
parameters are fitted to observed physical properties of baryons 
and then the model can be used to make predictions. Since such 
parameters are mostly determined in the nonstrange baryon sector 
or strangeness $-1$ sector, there is little, if any, freedom to 
introduce more parameters in the multi-strangeness sector. In 
this respect, the importance of studying multistrangeness sector 
cannot be overemphasized as it can provide a very useful tool to 
test various models.

In spite of the early efforts for studying $\Xi$ and $\Omega$ 
spectra, our understanding on the multistrangeness sector is 
still far from complete and opens many questions. As mentioned 
by the Particle Data Group, there are several serious difficulties 
in studying $\Xi$ and $\Omega$ resonances experimentally. First of 
all, the cross sections of producing multistrangeness from 
nonstrange initial state are very small and precise studies on the 
spectrum and properties of $\Xi$ and $\Omega$ hyperons are 
extremely difficult.  An example can be found in the case of the 
$\Omega^-$ hyperon that has strangeness $-3$.  Although the ground 
state of $\Omega^-$ baryon was discovered at BNL in mid 
60s~\cite{BCCC64H}, which confirmed the prediction of 
Gell-Mann~\cite{Gell-Mann64H}, it took about 40 years to confirm 
that it has spin-3/2~\cite{BABAR-06aH}.  Furthermore, the parity 
of the ground state of $\Xi$ hyperon has not been 
measured~\cite{PDG14yH}.  The lack of Kaon beam facility, therefore, 
lead to the conclusion that any new significant information on 
the multi-strangeness baryons has not been accumulated during the 
last two decades~\cite{PDG14yH}.  Nevertheless, the continuous 
efforts to study multistrangeness system cause, albeit slow, 
progress in our knowledge in this system. This includes the 
measurement of the magnetic moment of 
$\Omega^-(1672)$~\cite{WBCG95H}, experimental studies on $\Xi$ 
resonances~\cite{WA89-99H}, measuring weak decays of $\Xi^0$ 
hyperon~\cite{KTeV-05H,NA48/1-07H}, and the production of $\Xi$ 
resonances in relativistic heavy-ion collisions~\cite{STAR07H}.
In addition, BABAR Collaboration claimed that the spin-parity 
quantum number of the $\Xi(1690)$ would be 
$\frac12^-$~\cite{BABAR08H}.

Recently interests in multistrangeness systems are renewed by 
the advent of new facilities.  Indeed, the cascade physics 
program of the CLAS Collaboration at the Thomas Jefferson 
National Accelerator Facility (JLab) has been launched and 
some preliminary results were already reported~\cite{PDGN05H,
CLAS04dH,CLAS06dH,CLAS07bH}.  This activity continues to 
initiate a $\Xi$ spectroscopy program using photoproduction 
reactions at the upgraded 12-GeV machine, which also includes 
a plan to measure exclusive $\Omega$ 
photoproduction~\cite{VSC12H}.  J-PARC has a Kaon beam facility 
and is going to study the $\Xi$ baryons via the $\bar{K} N$ 
scattering although the energy is not large enough to produce
most $\Xi$ resonances. It also plans to study the $\pi N$ 
reaction to produce $\Xi$ and $\Omega$ hyperons. These 
reactions can also be used to identify the spin-parity 
quantum numbers of produced hyperons~\cite{NOH12H,JOHN14H,JOHN15H}.
The plan for having a $K_L$ beam at JLab will be complementary 
to J-PARC facility and unique place to produce $\Xi$ and 
$\Omega$ resonances by offering higher energy Kaon beams.

Table~\ref{tab:pdg} lists the multistrangeness baryons compiled 
in the review of the Particle Data Group (PDG)~\cite{PDG14yH}, 
which includes eleven $\Xi$ baryons and four $\Omega$ baryons.
Among them only the ground states, $\Xi(1318)$, $\Xi(1530)$, 
and $\Omega(1672)$, have four-star ratings with definite 
spin-parity, and there are four $\Xi$ baryons and one $\Omega$ 
baryon with three-star ratings.  Among the three-star-rated 
baryons, the $\Xi(1820)$ is the only state whose spin-parity 
quantum numbers are reported.
\begin{table}[t]
\centerline{\parbox{0.80\textwidth}{
\caption{$\Xi$ and $\Omega$ baryons compiled by the
        Particle Data Group~\protect\cite{PDG14yH}.} } }
\centering
\vspace{0.5cm}
\begin{tabular}{ccc|ccc} \hline\hline
Particle & $I(J^P)$ & rating & Particle & $I(J^P)$ & rating \\
\hline\hline
$\Xi(1318)$    & $\frac12(\frac12^+)$    & **** &
$\Omega(1672)$ & $0(\frac32^+)$          & **** \\
$\Xi(1530)$    & $\frac12(\frac32^+)$    & **** &
$\Omega(2250)$ & $0(?^?)$                & *** \\
$\Xi(1620)$    & $\frac12(?^?)$          & *    &
$\Omega(2380)$ & $?(?^?)$                & ** \\
$\Xi(1690)$    & $\frac12(\frac12^-?)$   & ***  &
$\Omega(2470)$ & $?(?^?)$                & ** \\
$\Xi(1820)$    & $\frac12(\frac32^-)$    & ***  & \\
$\Xi(1950)$    & $\frac12(?^?)$          & ***  & \\
$\Xi(2030)$    & $\frac12(\ge\frac52^?)$ & ***  & \\
$\Xi(2120)$    & $\frac12(?^?)$          & *    & \\
$\Xi(2250)$    & $\frac12(?^?)$          & **   & \\
$\Xi(2370)$    & $\frac12(?^?)$          & **   & \\
$\Xi(2500)$    & $\frac12(?^?)$          & *    & \\
\hline\hline
\end{tabular}
\label{tab:pdg}
\end{table}

In this presentation, we discuss the predictions of various 
models on hyperon spectrum. We will see that these 
predictions are not consistent with experimental observations
and are even contradictory to each other.  This shows the 
importance of high-quality and high-energy Kaon beam 
facilities to understand the underlying dynamics of hyperon 
structure.

\item \textbf{Model Dependence of Hyperon Spectrum}

There have been various theoretical studies on the excited 
states of multi-strangeness baryons based on phenomenological 
models. Although those models could reproduce the masses of 
the ground states of octet and decuplet, it is mainly due to 
the SU(3) group structure. Because of this, most models have 
the same mass sum rules for the ground state baryons. However 
they have very different and even contradictory predictions 
on the spectrum of excited states, in particular, for 
multistrangeness baryons. The most evident example is the 
third state of $\Xi$ baryons as will be explained below. (See 
also Ref.~\cite{Oh07H}.)

The most straightforward application of the SU(3) group 
structure is finding the SU(3) multiplets and their members.
Early efforts in this direction were summarized, for example, 
in Refs.~\cite{PBBF70H,SGM74H,Horgan74H}. However, such 
approaches ignore the dynamics of the constituents of baryons. 
A more detailed study on the excited states of $\Xi$ and 
$\Omega$ baryons was done by Chao, Isgur, and Karl~\cite{CIK81H} 
employing a non-relativistic quark model as the quark dynamics.
In this model, the $\Xi(1820)\frac32^-$ is well explained and 
the third lowest state following $\Xi(1318)$ and $\Xi(1530)$ 
is predicted to be at a mass of 1695 MeV having $J^P = 
\frac12^+$.  Although this model predicts the almost correct 
mass for $\Xi(1690)$, its prediction on the spin-parity
quantum numbers is not consistent with the experimental 
observation of Ref.~\cite{BABAR08H}.

The relativized quark model of Capstick and Isgur, however, 
gives a very different predictions on $\Xi$ 
resonances~\cite{CI86H}.  In this model, the first excited 
state of $\Xi(\frac12^+)$ has a higher mass of around 
$1840$~MeV. Furthermore, the third lowest $\Xi$ state would 
have a mass of $1755$~MeV with $J^P = \frac12^-$. Although 
the spin-parity quantum numbers are consistent with 
$\Xi(1690)$, its mass is much larger than the mass of 
$\Xi(1690)$. This pattern is also confirmed by a more recent 
relativistic quark model of Ref.~\cite{PR07H}.

In the one-boson exchange model, Glozman and Riska predicted 
that the third lowest state would have odd parity at a mass 
of 1758~MeV with $J=1/2$ or $3/2$~\cite{GR96bH}. This mass is 
in the middle of the masses of $\Xi(1820)$ and $\Xi(1690)$.
As a result, this model overestimates the mass of $\Xi(1690)$ 
and underestimates that of $\Xi(1820)$~\cite{GPVW97H,VGV05H}.

One may construct a mass operator based on large $N_c$ 
approximation of QCD, where $N_c$ is the number of colors.
Then the coefficients of the mass operator can be fitted by 
some known masses and then it can predict the masses of other 
baryons.  The results can be found in Refs.~\cite{CC00H,SGS02H,
GSS03H,MS04bH,MS06bH}, where a quite different $\Xi$ spectrum can 
be found.  In this approach, the third lowest state would have 
$J^P = \frac12^-$ at a mass of 1780~MeV. Therefore, this model 
gives a prediction on the third lowest $\Xi$ resonance mass 
similar to that of relativistic quark models. (See 
Ref.~\cite{SBMS07H} for a connection between the quark models 
and the large $N_c$ expansion.)

In the algebraic model of Bijker~{\it et al.\/}~\cite{BIL00H}, 
the third lowest state has $J^P = \frac12^+$ at a mass around 
$1730$~MeV. Thus it is not consistent with the BABAR result.
This model predicts two $\Xi(\frac32^-)$ states which lie 
close to the observed $\Xi(1820)$. As a result, it predicts 
richer hyperon spectrum than quark models and, in particular, 
it makes very different predictions for the $J^P = \frac12^-$ 
states.

The QCD sum rule approaches were also used to identify the 
lowest states of each spin-parity quantum numbers~\cite{LL02H,
JO96H}.  All results are summarized in Table~\ref{tab:xi-omega} 
for low-lying $\Xi$ and $\Omega$ resonances. It shows that the 
predictions on $\Xi$ and $\Omega$ spectrum are highly 
model-dependent and having precise information on these 
resonances is crucial to distinguish the underlying dynamics 
and baryon structure.
\begin{table*}[t]
\centerline{\parbox{0.80\textwidth}{
\caption{Low-lying $\Xi$ and $\Omega$ baryon spectrum of spin 
	$1/2$ and $3/2$ predicted by the non-relativistic quark 
	model of 
	Chao {\it et al.\/}~\protect\cite{CIK81H} (CIK), 
	relativized quark model of Capstick and 
	Isgur~\protect\cite{CI86H} (CI), 
	Glozman-Riska model~\protect\cite{GR96bH} (GR), 
	large $N_c$ analysis~\protect\cite{CC00H,SGS02H,GSS03H,
	MS04bH,MS06bH}, 
	algebraic model~\protect\cite{BIL00H} (BIL), and 
	QCD sum rules~\protect\cite{LL02H,JO96H} (SR). 
	The recent quark model prediction~\protect\cite{PR07H} 
	(QM), 
	and the Skyrme model results~\protect\cite{Oh07H}
	(SK) are given as well. 
	The mass is given in the unit of MeV.} } }
\label{tab:xi-omega}
\vspace{0.5cm}
\centering
\begin{tabular}{c|cccccccc} \hline\hline
State & CIK & CI & GR & Large-$N_c$ & BIL & SR & QM & SK \\ 
\hline
$\Xi(\frac12^+)$ & $1325$ & $1305$ & $1320$ &        & $1334$ &
$1320$ ($1320)$  & $1325$ & $1318$  \\
                 & $1695$ & $1840$ & $1798$ & $1825$ & $1727$ &  & $1891$ & $1932$   \\
                 & $1950$ & $2040$ & $1947$ & $1839$ & $1932$ &  & $2014$ &  \\ 
\hline
$\Xi(\frac32^+)$ & $1530$ & $1505$ & $1516$ &        & $1524$ &  & $1520$ & $1539$ \\
                 & $1930$ & $2045$ & $1886$ & $1854$ & $1878$ &  & $1934$ & $2120$ \\
                 & $1965$ & $2065$ & $1947$ & $1859$ & $1979$ &  & $2020$ & \\ 
\hline
$\Xi(\frac12^-)$ & $1785$ & $1755$ & $1758$ & $1780$ & $1869$ &
$1550$ ($1630)$  & $1725$ & $1614$ \\
                 & $1890$ & $1810$ & $1849$ & $1922$ & $1932$ &  & $1811$ & $1660$  \\
                 & $1925$ & $1835$ & $1889$ & $1927$ & $2076$ &    \\ 
\hline
$\Xi(\frac32^-)$ & $1800$ & $1785$ & $1758$ & $1815$ & $1828$ &
$1840$ & $1759$  & $1820$ \\
                 & $1910$ & $1880$ & $1849$ & $1973$ & $1869$ &  & $1826$ &  \\
                 & $1970$ & $1895$ & $1889$ & $1980$ & $1932$ &  & &  \\
\hline
$\Omega(\frac12^+)$ & $2190$ & $2220$ & $2068$ & $2408$ & $2085$ &  & $2175$ & $2140$ \\
                    & $2210$ & $2255$ & $2166$ &        & $2219$ &  & $2191$ & \\ 
\hline
$\Omega(\frac32^+)$ & $1675$ & $1635$ & $1651$ &        & $1670$ &  & $1656$ & $1694$ \\
                    & $2065$ & $2165$ & $2020$ & $1922$ & $1998$ &  & $2170$ & $2282$ \\
                    & $2215$ & $2280$ & $2068$ & $2120$ & $2219$ &  & $2182$ & \\ 
\hline
$\Omega(\frac12^-)$ & $2020$ & $1950$ & $1991$ & $2061$ & $1989$ &  & $1923$ & $1837$ \\ 
\hline
$\Omega(\frac32^-)$ & $2020$ & $2000$ & $1991$ & $2100$ & $1989$ &  & $1953$ & $1978$ \\
\hline\hline
\end{tabular}
\end{table*}

\item \textbf{The Skyrme Model}

As can be seen in Table~\ref{tab:xi-omega}, the quark models have a 
difficulty in explaining the mass and spin-parity quantum numbers of 
$\Xi(1690)$. Furthermore, the presence of $\Xi(1620)$, if confirmed, 
raises a serious question on the prediction of quark models. This is 
very similar to the puzzle of $\Lambda(1405)$, where the low mass of 
the $\Lambda(1405)$ hyperon makes it hard to be described as a 
$P$-wave three-quark state~\cite{IK78aH,AMS94H}. Instead, interpreting 
the $\Lambda(1405)$ as a $\bar{K}N$ bound state has been successful 
to understand its physical properties~\cite{DT59H,AS62H,VJTB85H}. It is 
then natural to search for other hyperons that have similar structure 
as the $\Lambda(1405)$ and we claim that $\Xi(1620)$ and $\Xi(1690)$ 
are such states through investigating the hyperon spectrum in the 
bound state approach in the Skyrme model.

In the bound state approach to the Skyrme model~\cite{CK85H}, hyperons 
are described as bound states of the SU(2) soliton and the Kaon. (The 
$K^\ast$ vector meson can also be included.) The underlying dynamics 
between the soliton and Kaon is described by a chiral Lagrangian of 
mesons. As shown in Ref.~\cite{CK85H}, the Wess-Zumino term in an SU(3) 
chiral Lagrangian has a very crucial role in hyperon spectrum by 
pushing up the $S=+1$ state while pulling down the $S=-1$ state. As a 
result,  the $S=+1$ pentaquark $\Theta^+$ cannot be a bound state, 
and the bound states of $S=-1$ orrespond to the normal hyperons. 
Furthermore, this model renders two bound states, a positive parity 
state in $P$-wave and a negative parity state in $S$-wave. The 
$P$-wave state is strongly bound and, when quantized, it gives the 
ground states of hyperons with $j^P = 1/2^+$ and $3/2^+$. On the 
other hand, the $S$-wave state is an excited state and, when quantized, 
it corresponds to the $\Lambda(1405)$ with $j^P = 1/2^-$. Therefore, 
this model gives a natural way to describe both the 
$\Lambda(1116,1/2^+)$ and the $\Lambda(1405,1/2^-)$ on the same 
footing~\cite{SSG95H}.

In this model, the mass of a hyperon with isospin $i$ and spin $j$ is 
written as~\cite{Oh07H} 
\begin{eqnarray}
	M(i,j,j_m^{}) 
	M_{\rm sol} + n_1^{} \omega_1^{} + n_2^{} \omega_2^{}\nonumber \\ 
	+ \frac{1}{2\mathcal{I}} \Biggl\{i(i+1) + c_1^{} c_2^{} j_m^{}(j_m^{}+1)
	+ (\bar{c}_1^{} - c_1^{} c_2^{}) j_1^{}(j_1^{}+1)
	+ (\bar{c}_2^{} - c_1^{} c_2^{}) j_2^{}(j_2^{}+1)\nonumber \\ 
	\mbox{} \qquad +\frac{c_1^{}+c_2^{}}{2} [j(j+1) - j_m^{}(j_m^{}+1) 
	- i(i+1)] + \frac{c_1^{} - c_2^{}}{2} \, \vec{R}\cdot (\vec{J}_1 - 
	\vec{J}_2)\Biggr\}, 
	\label{eq:mass0}
\end{eqnarray}
where $\vec{J}_1$ and $\vec{J}_2$ are the grand spins of the 
$P$-wave and $S$-wave Kaon, respectively, and $\vec{J}_m = 
\vec{J}_1 + \vec{J}_2$. The total spin of the system is then 
given by $\vec{J} = \vec{J}_{\rm sol} + \vec{J}_m$, where 
$\vec{J}_{\rm sol}$ is the soliton spin. The number and energy 
of the bound Kaons are $n_i$ and $\omega_i$, respectively, and 
$c_i$ are the hyperfine splitting constants of the bound states.  
We refer the details to Ref.~\cite{Oh07H}, but we emphasize that 
this mass formula has three parts in large $N_c$ expansion. 
First, the soliton mass $M_{\rm sol}$ is of $O(N_c)$ and the 
energies of the bound Kaons are of $O(N_c^0)$. The hyperfine 
term, which contains $1/2\mathcal{I}$ with $\mathcal{I}$ being 
the moment of inertia, is of $O(1/N_c)$. This shows that the 
mass splitting between the $\Lambda(1405)$ and the 
$\Lambda(1116)$ mainly comes from the energy difference between 
the $P$-wave Kaon and the $S$-wave Kaon. In fact, its empirical 
value is about 300~MeV and this pattern repeats in the $\Xi$ 
baryon spectrum and in the $\Omega$ baryon spectrum.

In principle, the mass parameters in Eq.~(\ref{eq:mass0}) can 
be calculated for a given dynamics of the meson-soliton system, 
for example, by extending the work of Refs.~\cite{MOYHLPR12H,
MYOH12H}. However, this is a highly nontrivial and complicated 
calculation. Therefore, instead of calculating the mass 
parameters, we fit them to some known hyperon masses and 
predict the masses of other hyperons. The obtained results are 
illustrated in Table~\ref{tab:mass}.

In this model, the parity of a hyperon changes if the $P$-wave 
Kaon is replaced by the $S$-wave state.  Since the energy 
difference between the two Kaons is about 300~MeV, there always 
exist pairs of hyperons of having same spin and the opposite 
parity with a mass difference of about 300~MeV.  Since the mass 
of the ground state of the $\Xi(1/2^+)$ is 1318~MeV, a 
$\Xi(1/2^-)$ state is expected at a mass of about 1620~MeV. 
Furthermore, this model requires two $\Xi$ states of this mass 
scale.  This is because the two Kaons, one in $P$-wave and one 
in $S$-wave, can make either $j_m = 0$ or $j_m = 1$ states. 
Considering the soliton spin $j_{\rm sol} = 1/2$, these states 
can give two $j=1/2$ states and one $j=3/2$ state. This explains 
naturally the existence of two $\Xi$ baryons with $j^P = 1/2^-$ 
at similar masses, namely, the one-star rated $\Xi(1620)$ and 
the three-star rated $\Xi(1690)$. However, since the observation 
of the $\Xi(1620)$ at early 1980s~\cite{HACN81H}, there is no 
other experimental confirmation of this state. Therefore, it is 
strongly required to resolve this issue urgently at current 
experimental facilities.
\begin{table}[t]
\centerline{\parbox{0.80\textwidth}{
\caption{Mass spectrum of the Skyrme model.
        The underlined values are used to determine the
        mass parameters. The values within the parenthesis
        are obtained by considering the mixing
        effect~\protect\cite{Oh07H}. The question marks
        after the particle name mean that the spin-parity
        quantum numbers are not identified yet.} } }
\vspace{0.5cm}
\centering
\begin{tabular}{ccc} \hline\hline
Particle Name & Mass (MeV) & Assigned State \\ 
\hline
$N$ & $\underline{939}$ \\
$\Delta$ & $\underline{1232}$ \\ 
\hline
$\Lambda_{1/2^+,0}$ & $\underline{1116}$ & $\Lambda(1116)$ \\
$\Lambda_{1/2^-,1}$ & $\underline{1405}$ & $\Lambda(1405)$ \\
$\Sigma_{1/2^+,0}$  & $1164$           & $\Sigma(1193)$ \\
$\Sigma_{3/2^+,0}$  & $\underline{1385}$ & $\Sigma(1385)$ \\
$\Sigma_{1/2^-,1}$  & $1475$             & $\Sigma(1480)?$ \\
$\Sigma_{3/2^-,1}$  & $1663$             & $\Sigma(1670)$ \\ 
\hline
$\Xi_{1/2^+,0}$     & $\underline{1318}$ & $\Xi(1318)$ \\
$\Xi_{3/2^+,0}$     & $1539$             & $\Xi(1530)$ \\
$\Xi_{1/2^-,1}$     & $1658(1660)^\ast$  & $\Xi(1690)?$ \\
$\Xi_{1/2^-,2}$     & $1616(1614)^\ast$  & $\Xi(1620)?$ \\
$\Xi_{3/2^-,1}$     & $\underline{1820}$ & $\Xi(1820)$ \\
$\Xi_{1/2^+,1}$     & $1932$             & $\Xi(1950)?$ \\
$\Xi_{3/2^+,1}$     & $\underline{2120}$ & $\Xi(2120)?$ \\ 
\hline
$\Omega_{3/2^+,0}$  & $1694$             & $\Omega(1672)$ \\
$\Omega_{1/2^-,1}$  & $1837$ \\
$\Omega_{3/2^-,1}$  & $1978$ \\
$\Omega_{1/2^+,1}$  & $2140$ \\
$\Omega_{3/2^+,1}$  & $2282$             & $\Omega(2250)?$ \\
$\Omega_{3/2^-,2}$  & $2604$ \\ 
\hline\hline
\end{tabular}
\label{tab:mass}
\end{table}

This analysis reveals that the $\Xi(1620)$ and the $\Xi(1690)$ 
are analogue states of the $\Lambda(1405)$. In addition, by 
replacing two $P$-wave Kaons in the $\Xi(1382)$ and in the 
$\Xi(1530)$, we predict that the $\Xi(1950)$ has $j^P=1/2^+$ 
and the $\Xi(2120)$ has $j^P=3/2^+$.  Their spin-parity 
quantum numbers are not known yet and should be identified by 
future experiments.

Comparing the predictions presented in Tables~\ref{tab:xi-omega} 
and \ref{tab:mass} shows that the predictions on the $\Omega$ 
hyperon spectrum are drastically different from the quark model
predictions. In quark models, the second lowest $\Omega$ hyperon 
has a mass of around 2~GeV, while the second state has a mass of 
around 1840~MeV with $j^P = 1/2^-$. Again, this low mass of the 
$\Omega$ excited state can hardly be explained by quark models. 
Thus, it is very interesting and crucial to see whether such low 
mass $\Omega$ hyperon really exists. Furthermore, most quark 
models predict that the lowest $\Omega$ baryon with $j^P=1/2^-$ 
is degenerate or almost degenerate in mass with the lowest  
$\Omega$ baryon with $j^P = 3/2^-$, which is in contradiction 
to our predictions.  These inconsistency with quark model 
predictions can be tested by future experiments.

If we extend our model to heavy quark baryons with a charm or 
a bottom quark~\cite{RRS92H}, we can also find a similar pattern 
in heavy baryon spectra. In Ref.~\cite{OP97H}, the binding 
energies of the soliton--heavy-meson system were calculated in 
the rest frame of the heavy meson, which shows that the energy 
difference between the positive parity state and the negative 
parity state is again close to 300~MeV, which can explain the 
observed masses of $\Lambda_c(2286)$ of $j^P = 1/2^+$ and the 
$\Lambda_c(2595)$ of $j^P = 1/2^-$. In quark models, the mass 
difference between the two states are estimated to be $250 \sim 
350$~MeV depending on the details of the quark 
dynamics~\cite{CI86H,RP08H}. Therefore, more detailed studies are
needed to clarify the structure of the $\Lambda_c(2595)$.

\item \textbf{Mass and Magnetic Moment Sum Rules}

The mass formula of Eq.~(\ref{eq:mass0}) can be used to derive 
mass sum rules.  Since it contains the second order of 
strangeness, it satisfies the modified Gell-Mann--Okubo mass 
relation and the modified decuplet equal spacing rule~\cite{OW99H},
\begin{eqnarray}
	3\Lambda + \Sigma - 2(N+\Xi) = \Sigma^\ast - \Delta - 
	(\Omega - \Xi^\ast),\nonumber \\
	(\Omega - \Xi^\ast) - (\Xi^\ast - \Sigma^\ast) = (\Xi^\ast 
	- \Sigma^\ast) - (\Sigma^\ast - \Delta),
	\label{eq:mass-rel-1}
\end{eqnarray}
where the symbols denote the masses of the corresponding 
octet and decuplet ground states. On the other hand, the 
hyperfine relation holds even with the second order of 
strangeness, and, therefore, the mass formula of 
Eq.~(\ref{eq:mass0}) satisfies
\begin{equation}
	\Sigma^\ast - \Sigma + \frac32 ( \Sigma-\Lambda) = \Delta - N.
	\label{eq:hyp}
\end{equation}

Since the mass relations (\ref{eq:mass-rel-1}) and (\ref{eq:hyp}) 
are obtained for the hyperons with the $P$-wave Kaons, the same 
relations should be true for the hyperons containing the $S$-wave 
Kaons only.  Therefore, those relations are expected to be valid 
by replacing $\Lambda$, $\Sigma$, $\Sigma^\ast$, $\Xi$, $\Xi^\ast$, 
and $\Omega$ by $\Lambda_{1/2^-,1}$, $\Sigma_{1/2^-,1}$, 
$\Sigma_{3/2^-,1}$, $\Xi_{1/2^+,1}$, $\Xi_{3/2^+,1}$, and 
$\Omega_{3/2^-,2}$, respectively. Note that those mass sum rules 
relate the mixed parity states of hyperons, \textit{i.e.}, 
odd-parity $\Lambda$ and $\Sigma$, even-parity $\Xi$, and 
odd-parity $\Omega$, and, therefore, should be distinguished by 
the quark model predictions.

We also derive a mass sum rule of
\begin{equation}
	\Omega_{3/2^+,1} - \Omega_{3/2^-,1} =
	\Omega_{1/2^+,1} - \Omega_{1/2^-,1}
\end{equation}
for $\Omega$ resonances.

This reveals the character of the bound state model, namely, the 
mass differences between the baryons of the same spin but of 
opposite parity, which we call ``parity partners", are the same.
Although the mass splitting of other parity partner hyperons are 
not exactly equal to the above formula, we observe that their 
mass differences are always close to $\sim 290$~MeV and this 
pattern can be actually observed in some experimental data.

The magnetic moment operator in this approach can be written 
as~\cite{OMRS91H}
\begin{equation}
	\hat\mu = \hat\mu_s + \hat\mu_v,
\end{equation}
where
\begin{eqnarray}
	\hat\mu_s &=& \mu_{s,0} R^z + \mu_{s,1} J^z_1 + \mu_{s,2} 
	J^z_2,\nonumber \\
	\hat\mu_v &=& -2 (\mu_{v,0} + \mu_{v,1} n_1 + \mu_{v,2} 
	n_2) D^{33},
\end{eqnarray}
with $D^{33} = -I^z R^z / \bm{I}^2$.  Here, $\mu_{s,0}$ and 
$\mu_{v,0}$ are the magnetic moment parameters of the SU(2) 
sector while $\mu_{s,1}$ and $\mu_{v,1}$ ($\mu_{s,2}$ and
$\mu_{v,2}$) are the parameters for the $P$-wave ($S$-wave) 
Kaon.

Instead of making predictions for the magnetic moment of each 
hyperon, we develop sum rules for magnetic moments. For the 
ground state baryons, we have the well-known results,
\begin{eqnarray}
	\mu(\Sigma^{\ast +}) - \mu(\Sigma^{\ast -}) &=& \frac32 
	\left\{ \mu(\Sigma^+) - \mu(\Sigma^-) \right\},\nonumber \\ 
	\mu(\Sigma^{+}) + \mu(\Sigma^{-}) &=& \frac43 \left\{ \mu(p) 
	+ \mu(n) \right\}- \frac23 \mu(\Lambda),\nonumber \\ 
	\mu(\Sigma^{\ast +}) + \mu(\Sigma^{\ast -}) &=& 2 \left\{ 
	\mu(p) + \mu(n) \right\}+ 2 \mu(\Lambda),\nonumber \\
	\mu(\Xi^0) + \mu(\Xi^-) &=& - \frac13 \left\{ \mu(p) 
	+ \mu(n) \right\} + \frac83 \mu(\Lambda),\nonumber \\
	\mu(\Xi^{\ast 0}) + \mu(\Xi^{\ast -}) &=& \mu(p) + \mu(n) 
	+ 4 \mu(\Lambda),\nonumber \\
	\mu(\Xi^{\ast 0}) - \mu(\Xi^{\ast -}) &=& -3 \left\{ 
	\mu(\Xi^{0}) - \mu(\Xi^{-})\right\},\nonumber \\
	\mu(\Omega) &=& 3 \mu(\Lambda)
\end{eqnarray}
for the octet and decuplet ground state baryons. It should 
also be mentioned that these relations are valid by replacing 
$\Sigma$, $\Sigma^\ast$, $\Xi$, $\Xi^\ast$, $\Omega$ by 
$\Sigma_{1/2^-,1}$, $\Sigma_{3/2^-,1}$, $\Xi_{1/2^+,1}$, 
$\Xi_{3/2^+,1}$, $\Omega_{3/2^-,2}$, respectively.

Other interesting sum rules include
\begin{eqnarray}
	\mu(\Xi_{3/2^-,1}^0) - \mu(\Lambda_{1/2^-,1})
	- \frac12 \left\{ \mu(\Sigma_{1/2^-,1}^+)
	- \mu(\Sigma_{1/2^-,1}^-) \right\}\nonumber \\
	= \mu(\Xi_{3/2^+,0}^0) - \mu(\Lambda_{1/2^+,0})
	- \frac12 \left\{ \mu(\Sigma_{1/2^+,0}^+)
	- \mu(\Sigma_{1/2^+,0}^-) \right\},
\end{eqnarray}
\begin{eqnarray}
	\mu(\Xi_{3/2^-,1}^-) - \mu(\Lambda_{1/2^-,1}) 
	+ \frac12 \left\{\mu(\Sigma_{1/2^-,1}^+) - 
	\mu(\Sigma_{1/2^-,1}^-) \right\}\nonumber \\
	= \mu(\Xi_{3/2^+,0}^-) - \mu(\Lambda_{1/2^+,0}) 
	+ \frac12 \left\{\mu(\Sigma_{1/2^+,0}^+) - 
	\mu(\Sigma_{1/2^+,0}^-) \right\},
\end{eqnarray} 
\begin{eqnarray}
	\mu(\Xi_{1/2^-,1}^0) + 3 \mu(\Xi_{1/2^-,2}^0) = 
	\mu(\Xi_{1/2^-,1}^-) + 3 \mu(\Xi_{1/2^-,2}^-)\nonumber \\
	= 2 \left\{\mu(\Lambda_{1/2^+,0}) + \mu(\Lambda_{1/2^-,1}) 
	\right\},
\end{eqnarray}
and
\begin{eqnarray}
	\mu(\Omega_{1/2^-,1}) &=& 
	\frac43 \mu(\Lambda_{1116}) - \frac13 
	\mu(\Lambda_{1405}),\nonumber \\
	\mu(\Omega_{3/2^-,1}) &=& 2 \mu(\Lambda_{1116}) 
	+ \mu(\Lambda_{1405}),\nonumber \\
	\mu(\Omega_{1/2^+,1}) &=& -\frac13 \mu(\Lambda_{1116}) 
	+ \frac43\mu(\Lambda_{1405}),\nonumber \\
	\mu(\Omega_{3/2^+,1}) &=& \frac13 \mu(\Lambda_{1116}) 
	+ 2\mu(\Lambda_{1405}).
\end{eqnarray}
These sum rules relate the magnetic moments of positive and 
negative parity hyperons.

\item \textbf{Summary}

The predictions on multistrangeness baryon spectrum are highly 
model-dependent and new precise experimental data are strongly 
called for to distinguish the models on baryon structure. This 
shows that multistrangeness baryons provide a unique tool for 
investigating the underlying dynamics and the role of strange 
quarks.  One issue is the low masses of the $\Xi(1620)$ and 
the $\Xi(1690)$, which are hard to be explained by quark 
models but are regarded as the analogous states of the 
$\Lambda(1405)$ in the Skyrme model. This model also leads to 
mass sum rules and magnetic moment sum rules which can 
distinguish the model from quark model predictions. Therefore, 
the suggested $K_L$ beam facility will shed light on our 
understanding of strong interactions through the excited 
states of $\Xi$ and $\Omega$ baryons.

\item \textbf{Acknowledgments}

This research was supported by Basic Science Research Program 
through the National Research Foundation of Korea (NRF) under 
Grant No. NRF--2015R1D1A1A01059603.
\end{enumerate}


\newpage
\subsection{Hyperon Resonance Studies from Charm Baryon Decays
	at BaBar}
\addtocontents{toc}{\hspace{2cm}{\sl V.~Ziegler}\par}
\setcounter{figure}{0}
\setcounter{equation}{0}
\halign{#\hfil&\quad#\hfil\cr
\large{Veronique Ziegler}\cr
\textit{Thomas Jefferson National Accelerator Facility}\cr
\textit{Newport News, VA 23606, U.S.A.}\cr}

\begin{abstract}
We present studies of hyperon and hyperon resonance production 
in charm baryon decays at BaBar. The helicity formalism employed 
to measure the spin of $\Omega^-$ was extended to three-body 
final states whereby the properties of the $\Xi(1690)^0$ and 
$\Xi(1690)^0$ produced in $\Lambda_c^+$ decay where obtained.
\end{abstract}

\begin{enumerate}
\item \textbf{Introduction}

The data samples used for the analyses described in this note 
were collected with the BaBar detector at the PEP-II 
asymmetric-energy $e^+e^-$ collider.  In these studies, the 
charm baryons are inclusively produced in $e^+e^-$ collisions 
at center-of-mass energies 10.58 and 10.54~GeV. The BaBar 
detector and reconstruction software are described 
elsewhere~\cite{ref:babar}.

\item \textbf{General Procedure for Charm Baryon Selection}

The selection of charm baryon candidates requires the 
sequential reconstruction of initial and intermediate
state candidates using four-momentum addition of tracks.
Particle identification selectors based on specific energy 
loss (${\rm d}E/{\rm d}x$) and Cherenkov angle measurements 
have been used to identify proton, pion and kaon final 
tracks.  Each intermediate state candidate is required to 
have its invariant mass within $\pm 3\sigma$ of the fitted 
peak position of the relevant distribution, where $\sigma$ 
is the mass resolution.  In all cases, the fitted peak mass 
is consistent with the expected value, and the intermediate 
state invariant mass is then constrained to this value. Due 
to the fact that each weakly-decaying intermediate state 
(\textit{i.e.}, the $K_S$ and hyperons) is long-lived, the 
signal-to-background ratio is improved by imposing a vertex 
displacement criterion (in the direction of the momentum 
vector).  In order to further enhance signal-to-background 
ratio, a selection criterion is imposed on the 
center-of-mass momentum $p^*$ of the parent charm baryon.  
The use of charge conjugate states is implied throughout 
in this note.

\item \textbf{Formalism}

Measurements of the $\Omega^-$ spin are obtained using
a primary sample obtained from the decay sequence 
$\Xi_{c}^{0} \rightarrow \Omega^{-} K^+$, with
$\Omega^{-} \rightarrow \Lambda K^-$~\cite{ref:omesp}.
It is assumed that each charm baryon type has spin 1/2 
and, as a result of its inclusive production, that it is 
described by a diagonal spin projection density matrix.  
The analysis does not require that the diagonal matrix 
elements be equal.

By choosing the quantization axis along the direction of 
the $\Omega^-$ in the charm baryon rest-frame, the 
$\Omega^-$ inherits the spin projection of the charm 
baryon~\cite{ref:omesp}.  It follows that, regardless of 
the spin $J$ of the $\Omega^-$, the density matrix 
describing the $\Omega^-$ sample is diagonal, with 
non-zero values only for the $\pm 1/2$ spin projection 
elements, \textit{i.e.}, the helicity $\lambda_i$ of the 
$\Omega^-$ can take only the values  $\pm 1/2$.  Since 
the final state $\Lambda$ and $K^-$ have spin values 1/2 
and 0, respectively, the net final state helicity 
$\lambda_f$ also can take only the values $\pm 1/2$.  

Defining the helicity angle $\theta_h$ as the angle 
between the direction of the $\Lambda$ in the rest-frame 
of the $\Omega^-$ and the quantization axis, the 
probability for the $\Lambda$ to be produced with Euler 
angles $(\phi, \theta_h, 0)$ with respect to the 
quantization axis is given by the square of the amplitude 
$\psi$,  characterizing the decay of an $\Omega^-$ with
spin $J$ and helicity $\lambda_{i}$ to a 2-body system 
with net helicity $\lambda_{f}$, where
$
	\psi = A^J_{\lambda_f} D^{J\, *}_{\lambda_{i} 
	\lambda_{f}}(\phi, \theta_{h}, 0),
$
and the transition matrix element $A^J_{\lambda_f}$
represents the coupling of the $\Omega^-$ to the final 
state. The angular distribution of the $\Lambda$ is then 
given by 
\begin{eqnarray}
	I\propto \sum_{\lambda_{i}, \lambda_{f}} \rho_{i\, 
	i}\left |A^J_{\lambda_f}D^{J\, *}_{\lambda_{i} 
	\lambda_{f}}(\phi, \theta_h, 0)\right |^2,
\end{eqnarray}
\noindent where the $\rho_{i\, i}$ ($i= \pm 1/2$) are the 
diagonal density matrix elements inherited from the charm 
baryon, and the sum is over all initial and final helicity 
states.  

The $\Lambda$ angular distribution integrated over $\phi$ 
is then obtained for spin hypotheses $J_{\Omega}=1/2$, 
$3/2,$ and $5/2$, respectively as follows:
\begin{eqnarray}
	{dN}/{d\rm{cos} \theta_{\it h}}&\propto& 1+\beta\, 
	\rm {cos}\theta_{\it h},\\
	{dN}/{d\rm{cos} \theta_{\it h}}&\propto& 1 + 3\,{\rm 
	cos}^2\theta_{\it h}+\beta\, \rm {cos}\theta_{\it h}
	(5- 9\,\rm cos^2\theta_{\it h}), \;\;\;\;\;\;\\ \nonumber 
	{dN}/{d\rm{cos} \theta_{\it h}}&\propto&  1-2\,{\rm 
	cos}^2\theta_{h}+5\,{\rm cos}^4\theta_{\it h} \\
	&&+\beta\, \rm {cos}\theta_{h}(5-26\,{\rm cos}^2
	\theta_{\it h}+25\,{\rm cos}^4\theta_{\it h}),
\end{eqnarray}
where the coefficient of the asymmetric term, 
$\beta$~\cite{ref:omesp}, may be non-zero as a consequence 
of parity violation in charm baryon and $\Omega^-$ weak 
decay.

The angular distributions of the decay products of the 
$\Omega^-$ baryon resulting from a spin 1/2 charm baryon 
decay are well-described by a function $ \propto 1 + 3\,{\rm 
cos}^2\theta_{\it h} $. These observations are consistent 
with spin assignments 3/2 for the $\Omega^-$. Values of 1/2 
and greater than 3/2 for the spin of the $\Omega^-$ yield 
C.L. values significantly less than 1$\%$ when spin 1/2 is 
assumed for the parent charm baryon.

\begin{enumerate}
\item \textbf{The Use of Legendre Polynomial Moments in 
	Spin Determination}

For spin $J$, the corrected angular distributions can be 
written  
\begin{eqnarray*}
	\frac{dN}{d\rm{cos} \theta_{\it h}}=N\left [ 
	\sum_{{\it l}=0}^{{\it l}_{max}} \langle P_{\it 
	l}\rangle P_{\it l}\left (\rm{cos} \theta_{\it h}
	\right )\right],
\end{eqnarray*}
where $P_{\it l}\left (\rm{cos} \theta_{\it h}\right )$ are 
normalized Legendre Polynomial functions such that ${\it l}_{max}
=2J-1$, and if ${\it l}$ is odd $\langle P_{\it l}\rangle=0.$  
Each assumed $J$ defines ${\it l}_{max}$, so that $\langle P_{\it 
l}\rangle=0$ for ${\it l}>{\it l}_{max}$ and $\langle P_{\it l}
\rangle$ is calculable. The number of $\Omega^-$ signal events 
in a given mass bin  is obtained by giving each event, $j$, in 
that bin, a weight $w_j = \frac{P_{\it l_{max}}\left (\rm{cos} 
\theta_{\it h_j}\right )}{\langle P_{\it l_{max}}\rangle }.$
     
In particular, for $J=3/2$, giving each event a weight $w_j = 
\sqrt{10}  P_2(\rm{cos}\theta_{\it h_j})$ projects the complete 
$\Omega^-$ signal. In order to test the $J=5/2$ hypothesis, each 
event is given a weight $w_j = \frac{7}{\sqrt{2}}  P_4(\rm{cos}
\theta_{\it h_j})$.

As expected, the $\sqrt{10} P_2(\rm{cos}\theta_{\it h})$ moment 
projects out the signal of a spin 3/2 hyperon, whereas the
$7/\sqrt{2} P_4(\rm{cos}\theta_{\it h})$ moment does not.  
These moments are used in the analysis of the $\Xi(1690)$ and 
$\Xi(1530)$ resonances described in the next section.
\end{enumerate}
 
\item \textbf{Study of Cascade Resonances Using Three-body 
	Charm Baryon Decays}

Although considerable advances have been made in baryon 
spectroscopy over the past decade, there has been very little 
improvement in our knowledge of hyperon resonances since 1988. 
The $\Xi(1690)$ has been observed in the $\Lambda \bar{K}$, 
$\Sigma \bar{K}$ and $\Xi \pi$ final states with various 
degrees of certainty.   

\begin{enumerate}
\item \textbf{The $\Xi(1530)^0$ from $\Lambda_c^+\rightarrow 
	\Xi^-\pi^+K^+$ Decay}

The properties of the Xi(1530) resonance are investigated in 
the $\Lambda_c^+\rightarrow \Xi^- \pi^+ K^+$  decay 
process~\cite{ref:xi1530}. The Dalitz plot for $\Lambda_c^+
\rightarrow \Xi^- \pi^+ K^+$ is dominated by the contribution 
from $\Lambda_c^+\rightarrow \Xi(1530)^0 K^+$, where 
$\Xi(1530)^0\rightarrow \Xi^- \pi^+$ by strong decay. The 
Dalitz plot (Fig.~\ref{fig:fig1}) shows evidence for only one 
resonant structure.  A clear band can be seen at the nominal 
mass squared of the $\Xi(1530)^0 \rightarrow \Xi^- \pi^{+}$.
The analysis of the Legendre polynomial moments of the 
$\Xi(1530)^0\rightarrow \Xi^- \pi^+$ system established quite 
clearly, on the basis of Figs.~\ref{fig:fig2}(b) 
and~\ref{fig:fig2}(c), that the $\Xi(1530)^0$ hyperon 
resonance has spin 3/2. In conjunction with previous 
analyses~\cite{ref:schlein} this also definitively 
establishes positive parity. However, comparison of the 
$P_2(\rm{cos}\theta_{\Xi})$ moment to the $\Xi^- \pi^+$ 
mass distribution and fits to the angular decay distribution 
in the $\Xi(1530)^0$ region indicate that it is necessary 
to include other $\Xi^- \pi^+$ amplitudes in order to obtain 
a complete description of the data. In particular, the 
observation of a $P_1(\rm{cos}\theta_{\Xi})$ moment exhibiting 
oscillatory behavior in the $\Xi(1530)^0$ region indicates the 
need for an $S_{1/2}$ amplitude, while providing first evidence 
for the expected rapid BW phase motion of the $P_{3/2}$ 
$\Xi(1530)^0$ amplitude. However, a simple model incorporating 
only these amplitudes and a $D_{5/2}$ amplitude is ruled out 
because of the failure to describe the $\Xi(1530)^0$ line 
shape. The presence of the $S_{1/2}$ amplitude at high mass 
and the behavior of the mass distribution near 1.7~GeV/$c^2$ 
suggest that a resonant $\Xi(1690)^0$ amplitude may be adding 
coherently to this amplitude, thus leading to the inference 
of spin-parity 1/2$^-$ for the $\Xi(1690)^0$. It appears that 
a quantitative description of the $\Xi(1530)^0$ line shape, 
and indeed of the entire Dalitz plot, must incorporate these 
features together with amplitude contributions associated 
with the $K^+\pi^+$ and/or the $\Xi^-K^+$ systems.  However 
such an analysis requires a higher statistics data sample.
\begin{figure}[ht!]
  \centering\small
  \includegraphics[width=0.46\textwidth]{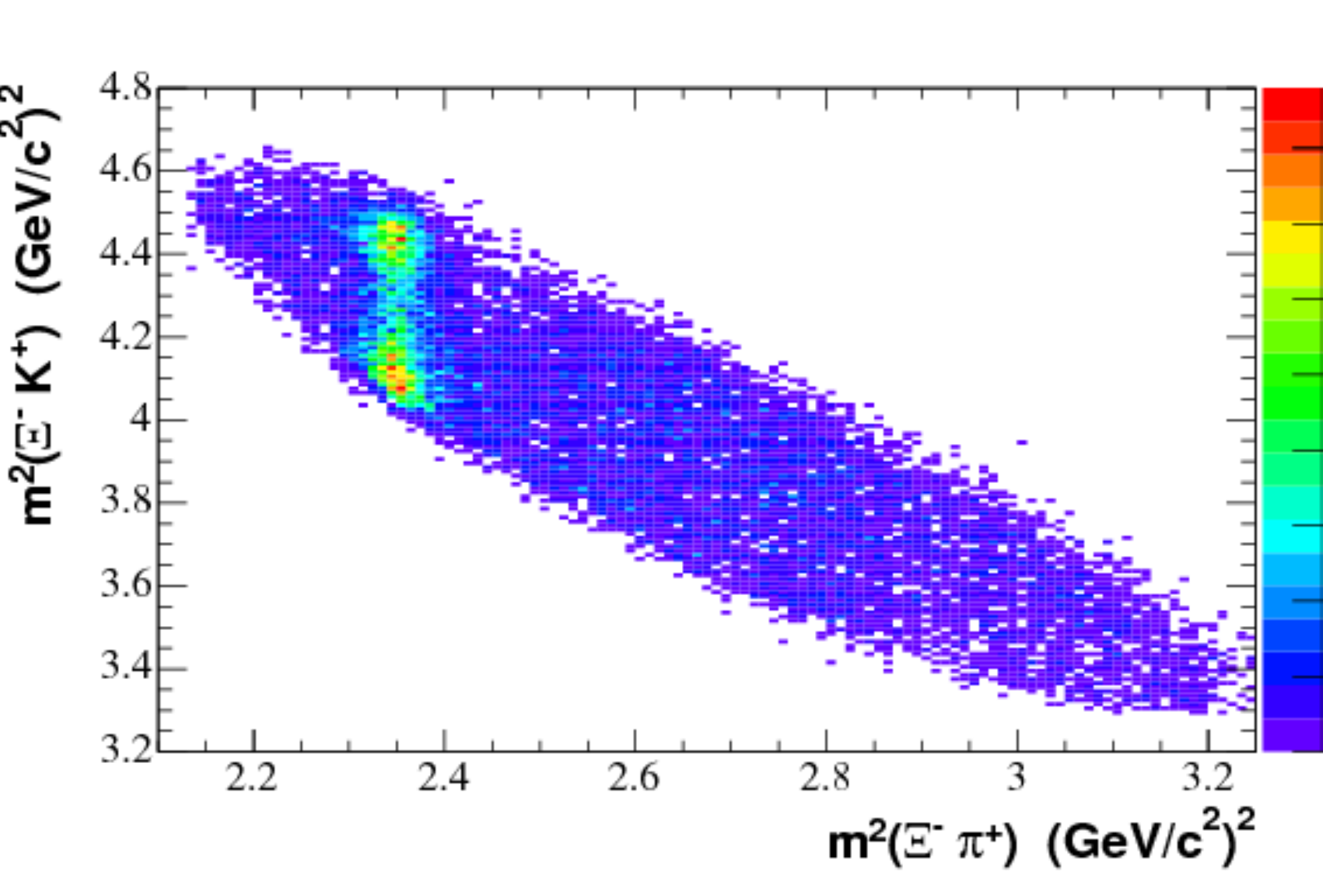}
  \begin{picture}(0.,0.)
        \end{picture}
\centerline{\parbox{0.80\textwidth}{
  \caption{The Dalitz plot for $\Lambda_c^+\rightarrow 
	\Xi^-\pi^-K^{+}$ corresponding to the 
	$\Lambda_c^+$ signal region.}  \label{fig:fig1} } }
\end{figure}
\begin{figure}[ht!]
  \centering\small
  \includegraphics[width=0.46\textwidth]{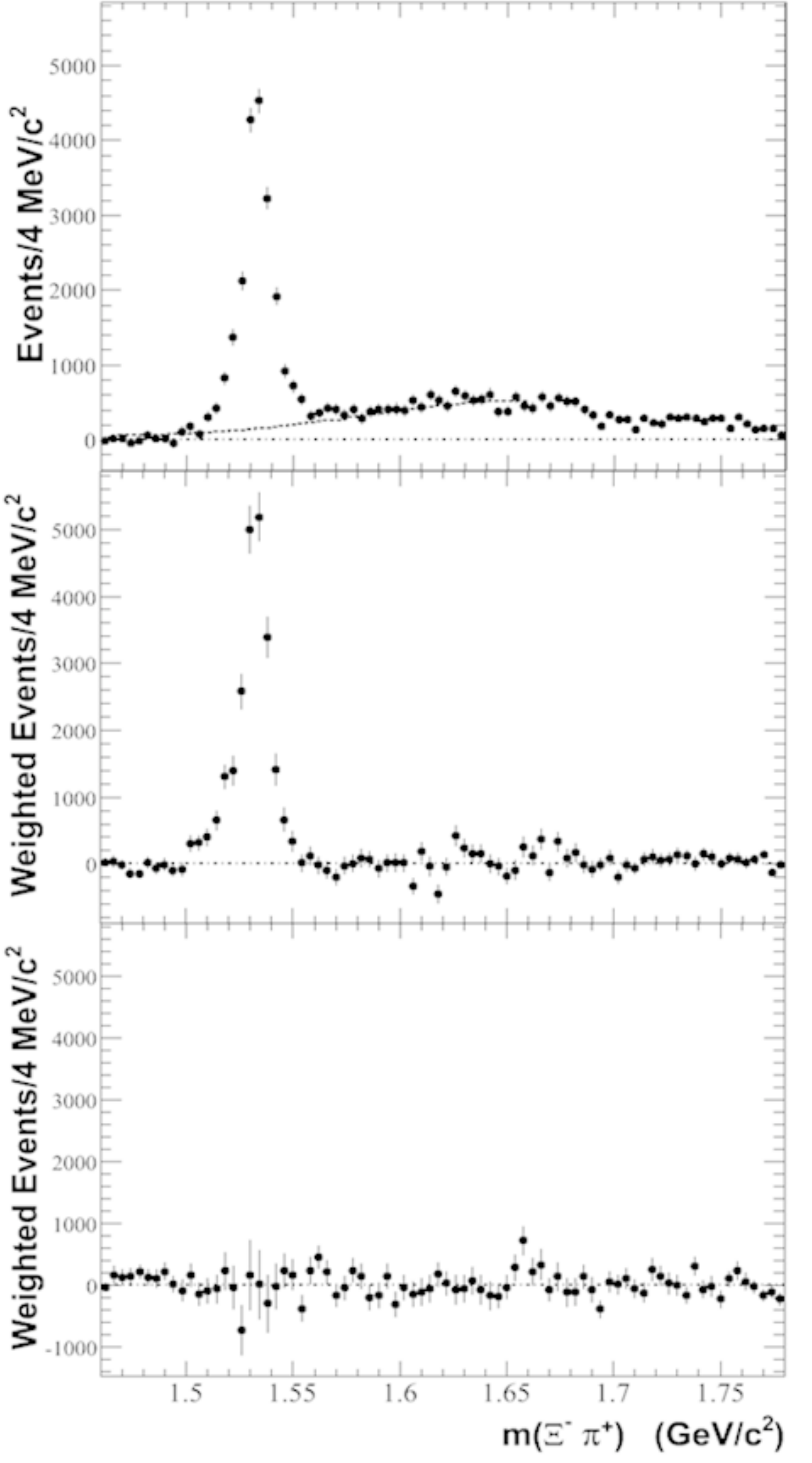}
\begin{picture}(0.,0.)
    \end{picture}
\centerline{\parbox{0.80\textwidth}{
  \caption{The efficiency-corrected $P0,2,4$ moments 
	of the $\Xi^-\pi^+$ system invariant mass 
	distribution for the  $\Lambda_c^+$ signal 
	region. In (a) the dashed curve represents the 
	estimated background contribution in the  
	$\Lambda_c^+$ region.} \label{fig:fig2} } }
\end{figure}

\item \textbf{The $\Xi(1690)^0$ from $\Lambda_c^+ 
	\rightarrow (\Lambda \bar K^{0}) K^{+}$ Decay}

The $\Xi(1690)^0$ is observed in the $\Lambda \bar K^0$
system produced in the decay $\Lambda_c^+ \rightarrow 
(\Lambda \bar K^{0}) K^{+}$, where the $\bar K^{0}$ is 
reconstructed via $K_S\rightarrow \pi^+ \pi^-$.
 
The selection of $\Lambda_{c}^{+}$ candidates requires
the intermediate reconstruction of oppositely-charged 
track pairs consistent with
$\Lambda \rightarrow p \: \pi^-$ and
$K_S \rightarrow \pi^+ \: \pi^-$ decays.
A clear peak, significant skewed toward high mass, is 
seen in the vicinity of the $\Xi(1690)^0$.

The second and fourth order Legendre polynomial moments 
as a function of the mass of the $(\Lambda K_S)$ system
display no peaking structure at the position of the 
$\Xi(1690)^0$, which suggests that the $\Xi(1690)^0$ 
spin is probably 1/2. However, the $\Lambda$ helicity 
cosine (cos$\theta_{\Lambda}$) distribution is not flat 
in contrast to the expectation for a spin 1/2 to 1/2 
transition.  The Dalitz plot of $\Lambda_c^+\rightarrow 
\Lambda \bar K^{0} K^{+}$ signal candidates is shown, 
without efficiency-correction, in Fig.~\ref{fig:fig3}(a).  
A clear band is observed in the mass-squared region of 
the $\Xi(1690)^0$, together with an accumulation of 
events in the $\bar K^0 K^+$ threshold region; the 
latter is consistent with a contribution to the Dalitz 
plot due to the $a_0(980)^+$ resonance. In contrast, the 
Dalitz plots corresponding to the $\Lambda_c^+$ 
mass-sideband regions exhibit no structure.
\begin{figure}[ht!]
  \centering\small
  \includegraphics[width=0.46\textwidth]{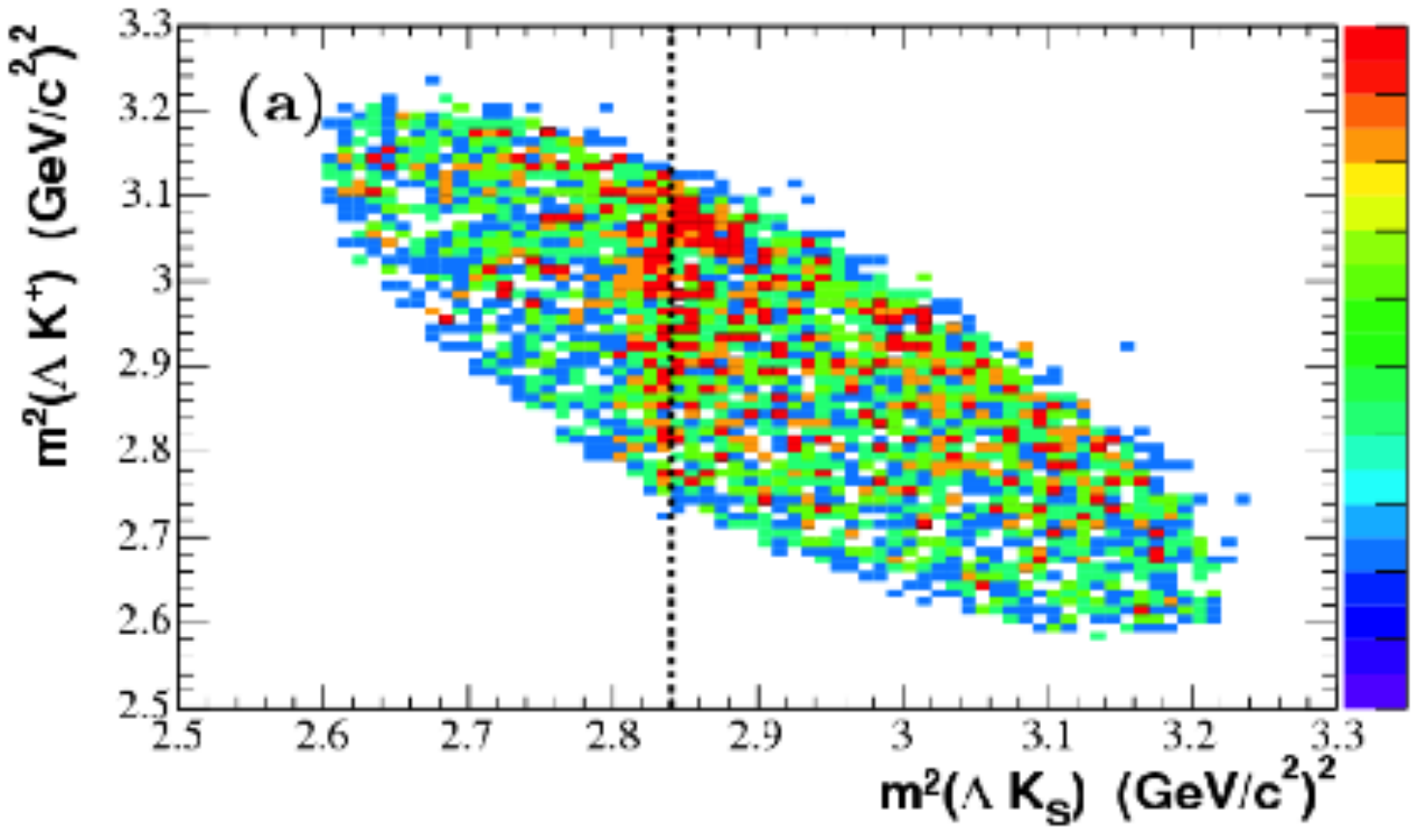}
  \includegraphics[width=0.46\textwidth]{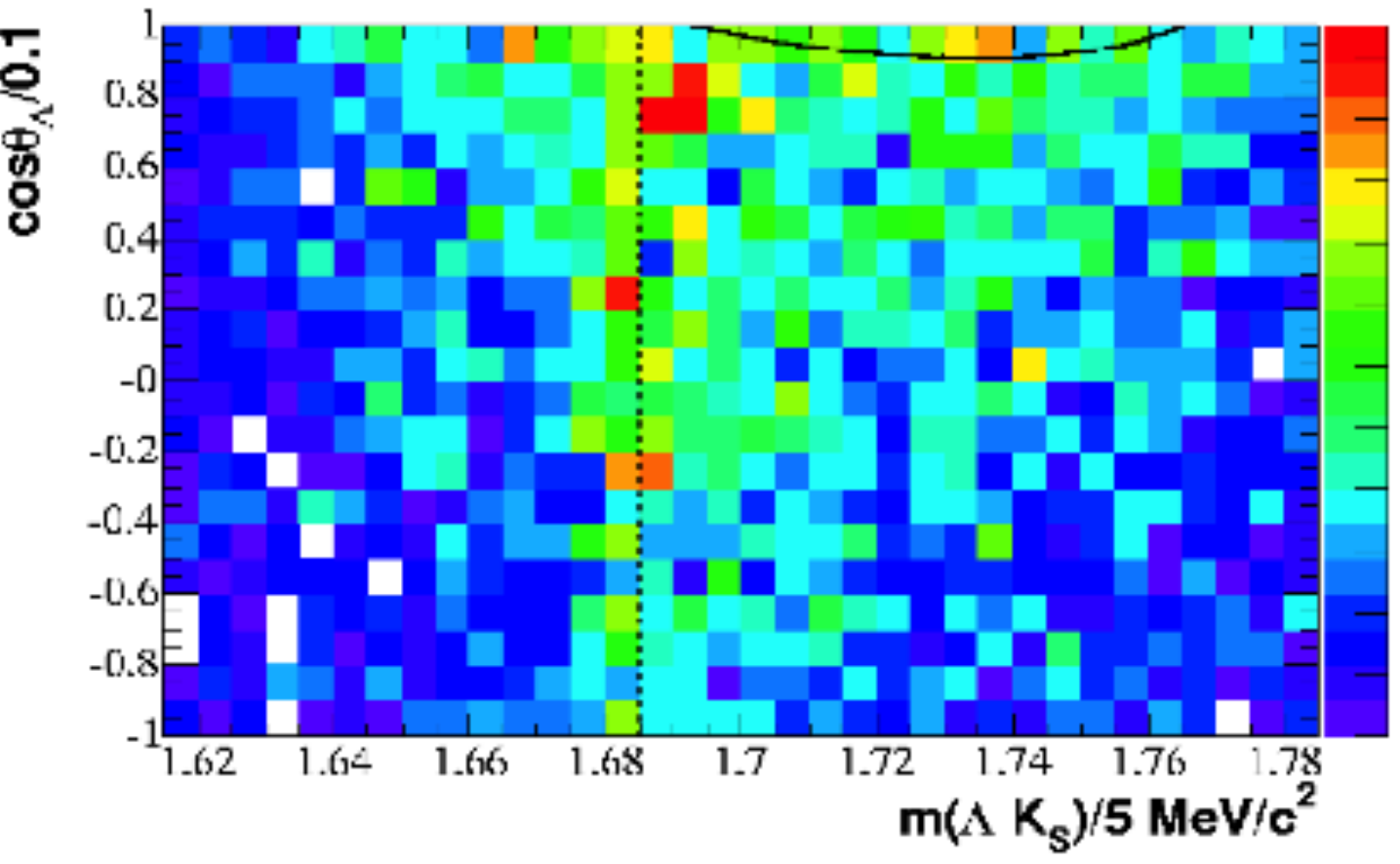}
\begin{picture}(0.,0.)
    \put(-155,96){\bf{(b)}}
    \put(-70,200){Preliminary}
    \end{picture}
\centerline{\parbox{0.80\textwidth}{
  \caption{(a) The Dalitz plot for $\Lambda_c^+
	\rightarrow\Lambda\bar K^0 K^{+}$ corresponding 
	to the $\Lambda_c^+$ signal region. The dashed 
	line indicates the nominal mass-squared region 
	of the $\Xi(1690)^0$. (b) The rectangular Dalitz 
	plot for $\Lambda_c^+\rightarrow\Lambda\bar K^0 
	K^{+}$ corresponding to the $\Lambda_c^+$ signal 
	region. The black curve corresponds to the 
	$a_0(980)^+$ pole position.}  \label{fig:fig3} } }
\end{figure}

We describe the event distribution in the Dalitz plot of 
Fig.~\ref{fig:fig3}(b) in terms of an isobar model 
consisting of the coherent superposition of amplitudes 
characterizing $(\Lambda a_0(980)^+)$ and $(\Xi(1690)^{0} 
K^+)$ decay of the $\Lambda_c^+$.  The $a_0(980)$ is known 
to couple to both $\eta \pi$ and $\bar K K$ and is 
characterized by the Flatt\'{e} 
parametrization~\cite{ref:Flatte}, while a Breit-Wigner 
function is used to describe the amplitude for the 
$\Xi(1690)^0$.

This model is used to describe the intensity distribution 
at a point on the Dalitz plot by means of the squared 
modulus of a coherent superposition of these two amplitudes, 
under the assumption that the $\Xi(1690)^0$ has spin 1/2, 
since the moment projections favor this choice. Fits to the 
Dalitz plot under the assumptions of spin 3/2 and 5/2 are 
ongoing. We find that no additional isobars are needed in 
order to accurately model the data.  In order to extract 
the mass and width parameters of the $\Xi(1690)^0$, we 
perform a binned maximum Likelihood fit to the rectangular 
Dalitz plot of Fig.~\ref{fig:fig3}(b) (incorporating 
resolution smearing in mass, and a background parametrization 
obtained from the $\Lambda_c^+$ mass-sidebands).

The fit reproduces accurately the skewed lineshape of the 
$\Lambda K_S$ invariant mass projection (Fig.~\ref{fig:fig4}).
The skewing results from  the interference between the 
$a_0(980)^+$ and the $\Xi(1690)^0$.  The actual $\Xi(1690)^0$ 
signal is symmetric and significantly smaller than the apparent 
signal, which is dominated by this interference effect.
The fit also provides an excellent representation of the 
other invariant mass projections.   
\begin{figure}[ht!]
  \centering\small
  \includegraphics[width=.45\textwidth]{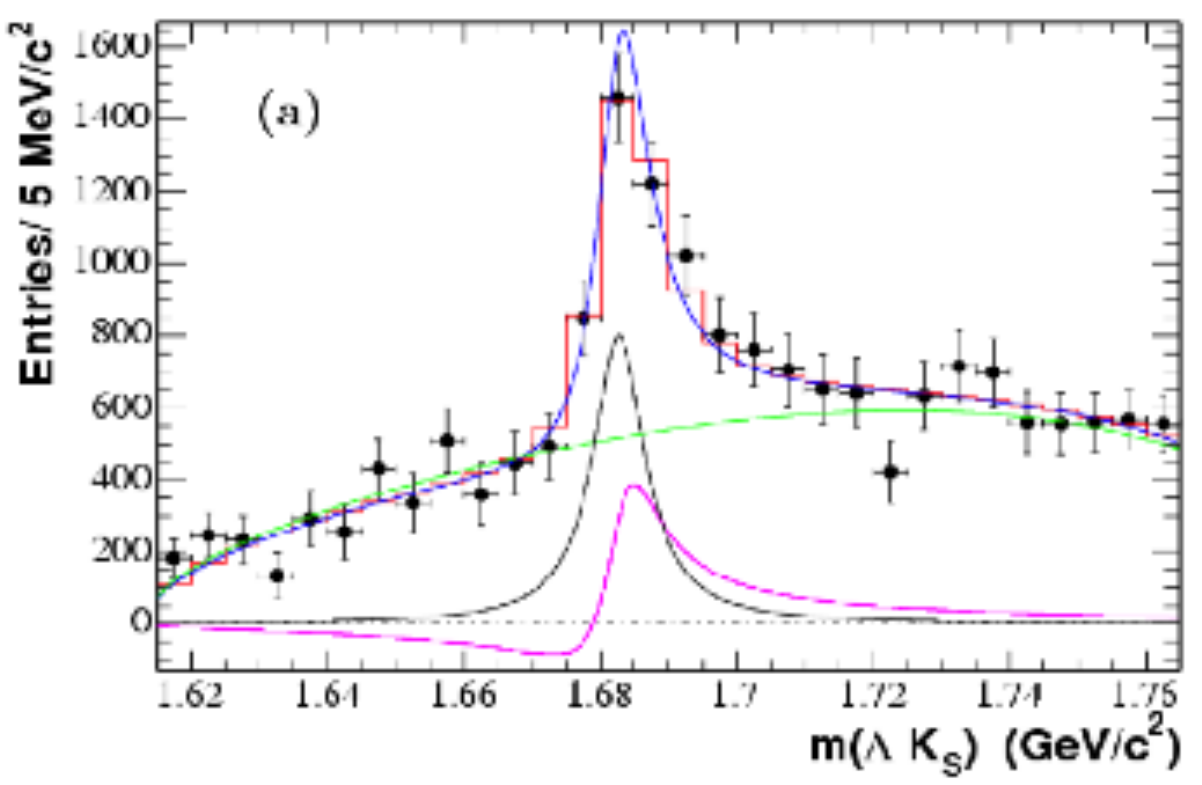}
  \includegraphics[width=.45\textwidth]{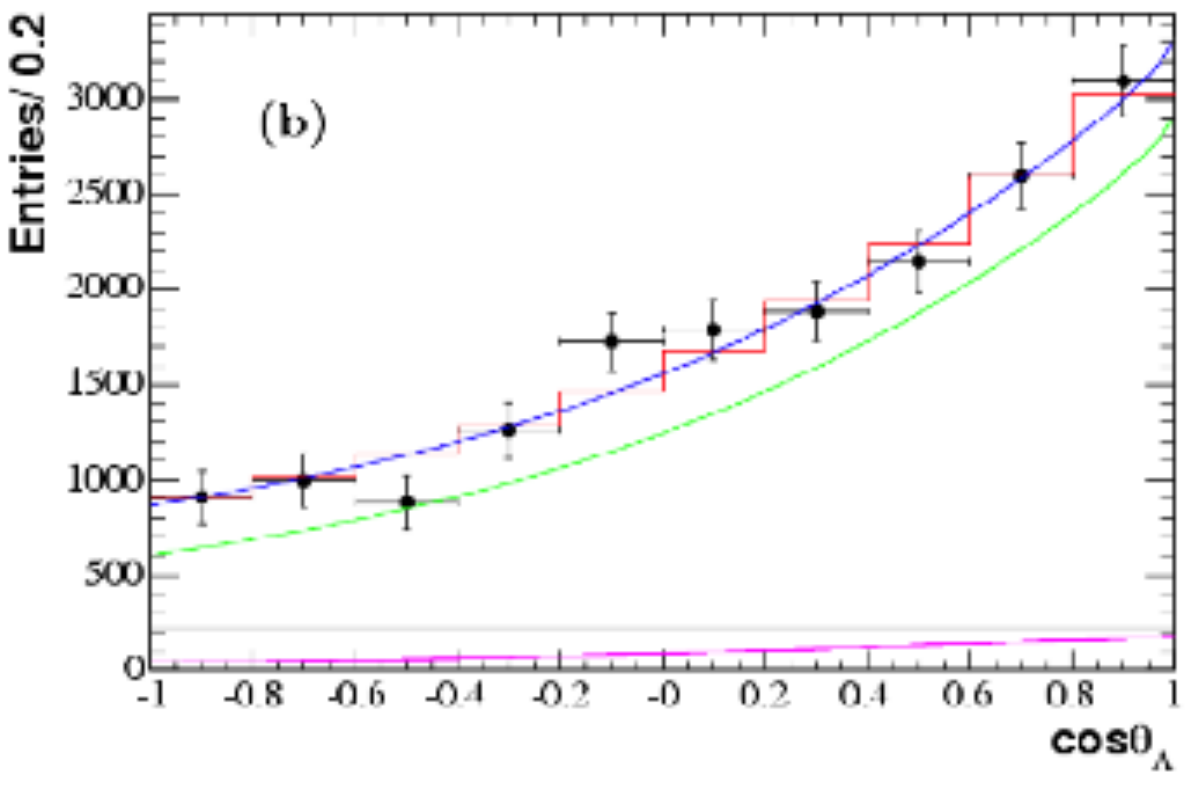}
  \begin{picture}(0.,0.)
  \put(-60,210){Preliminary}
    \end{picture}
\centerline{\parbox{0.80\textwidth}{
  \caption{(a) $\Lambda_c$-mass-sideband-subtracted 
	efficiency-corrected $\Lambda K_S$ invariant 
	mass projection. (b) 
	$\Lambda_c$-mass-sideband-subtracted 
	efficiency-corrected $\rm {cos} 
	\theta_{\Lambda}$ spectrum.  }\label{fig:fig4} } }
\end{figure}
\end{enumerate}

\item \textbf{Conclusions}

Mass and width measurements for the $\Xi(1690)^0$ have 
been obtained from fits to the $\Lambda_c^+\rightarrow 
\Lambda K_S K^+$ Dalitz plot. The results indicate that 
the spin of the $\Xi(1690)$ is consistent with 1/2. The 
properties of the $\Xi(1530)^0$ are studied using the 
decay $\Lambda_c^+\rightarrow \Xi^- \pi^+ K^+$.  The 
spin of the $\Xi(1530)$ is consistent with 3/2. 

Similar studies for cascade resonance production and 
associated spectra done at BaBar using charm baryon 
production can be done at GlueX with a $K_L$ beam.
Three-body systems involving two-body Cascade resonance 
decays require the analysis of the entire Dalitz plot 
when the statistical level is such that the shortcomings 
of a quasi-two-body approach become apparent.  Therefore 
it is essential to have high statistics to allow for a 
proper to fit to the entire Dalitz plot.

\newpage
\item \textbf{Acknowledgments}

This work was supported by DOE and NSF (USA), NSERC (Canada),
IHEP (China), CEA and CNRS-IN2P3 (France), BMBF and DFG
(Germany), INFN (Italy), FOM (The Netherlands), NFR (Norway),
MIST (Russia), and PPARC (United Kingdom).

\end{enumerate}


\newpage
\subsection{Evidence of Some New Hyperon Resonances $-$ to be 
	Checked by KL Beam}
\addtocontents{toc}{\hspace{2cm}{\sl B.~Zou}\par}
\setcounter{figure}{0}
\setcounter{equation}{0}
\halign{#\hfil&\quad#\hfil\cr
\large{Bingsong Zou}\cr
\textit{State Key Laboratory of Theoretical Physics}\cr
\textit{Institute of Theoretical Physics}\cr
\textit{Chinese Academy of Sciences}\cr
\textit{Beijing 100190, People's Republic of China}\cr}

\begin{abstract}
Quenched and unquenched quark models predict very different 
patterns for the spectrum of the low excited hyperon states. 
Evidence is accumulating for the existence of some new hyperon 
resonances, such as a $\Sigma^\ast$ of spin-parity $J^P=1/2^-$ 
around 1400~MeV instead of 1620~MeV as listed in PDG, a new 
$\Sigma(1540)3/2^-$ resonance, a new narrow $\Lambda(1670)3/2^-$ 
resonance and a new $\Lambda(1680)3/2^+$ resonance. All these 
new hyperon resonances fit in the predicted pattern of the 
unquenched quark models very well.  It is extremely important 
to check and establish the spectrum of these low excited 
hyperon states by the proposed $K_L$ beam experiments at JLab.
\end{abstract}

\begin{enumerate}
\item \textbf{Why hyperon resonances ?}

Creation of quark-anti-quark pairs from gluon field plays a 
crucial role for understanding quark confinement and hadron 
spectroscopy. In the classical quenched quark model for a 
$q_1\bar q_1$ meson, the $q_1$ quark cannot be separated 
from the $\bar q_1$ anti-quark due to a infinitely large 
confinement potential. But in realty, we know the $q_1$ and 
$\bar q_1$ can be easily separated from each other by
creation of another quark-anti-quark pair $q_2\bar q_2$ to 
decay to two mesons,  $q_1\bar q_2$ and $q_2\bar q_1$. With 
the creation of the $q_2\bar q_2$, instead of forming two 
colorless mesons, the system could also exist in the form 
of a tetra-quark state $[q_1q_2][\bar q_1\bar q_2]$. 
Therefore both lattice QCD and quark models should go 
beyond the quenched approximation which ignore the creation 
of quark-anti-quark pairs.

Quenched $qqq$ quark models and unquenched $qqq\leftrightarrow
qqqq\bar q$ quark models give very different predictions for 
the hyperon spectroscopy. For example, for the $J^P=
{1\over 2}^-$ SU(3) nonet partners of the $N(1535)$ and 
$\Lambda(1405)$. While quenched quark models~\cite{Capstick0Z,
capstickZ,glozZ,loringZ} predict the $J^P={1\over 2}^-$ $\Sigma$ 
and $\Xi$ resonances to be around 1650~MeV and 1760~MeV, 
respectively, the unquenched quark models~\cite{pentq1Z,pentq2Z,
zoupentZ} expect them to be around 1400~MeV and 1550~MeV, 
respectively, a meson-soliton bound-state approach of the 
Skyrme model~\cite{OhZ} and other meson-baryon dynamical
models~\cite{Kanchan11Z,RamosZ} predict them to be around 
1450~MeV and 1620~MeV, respectively. In Fig.~\ref{fig:graph}, 
we show prediction of the lowest penta-quark states with
$J^P=1/2^\pm,3/2^\pm$~\cite{pentq1Z,pentq2Z} (red solid) 
compared with those from the classical quenched $qqq$ 
model~\cite{Capstick0Z} (black solid). The major differences 
are that the lowest penta-quark hyperon states with 
$J^P=1/2^-$ and $3/2^+$ are about 200~MeV lower those from the 
classical quenched $qqq$ models~\cite{Capstick0Z}.
\begin{figure}[htbp]
\begin{center}
\includegraphics*[width=15cm]{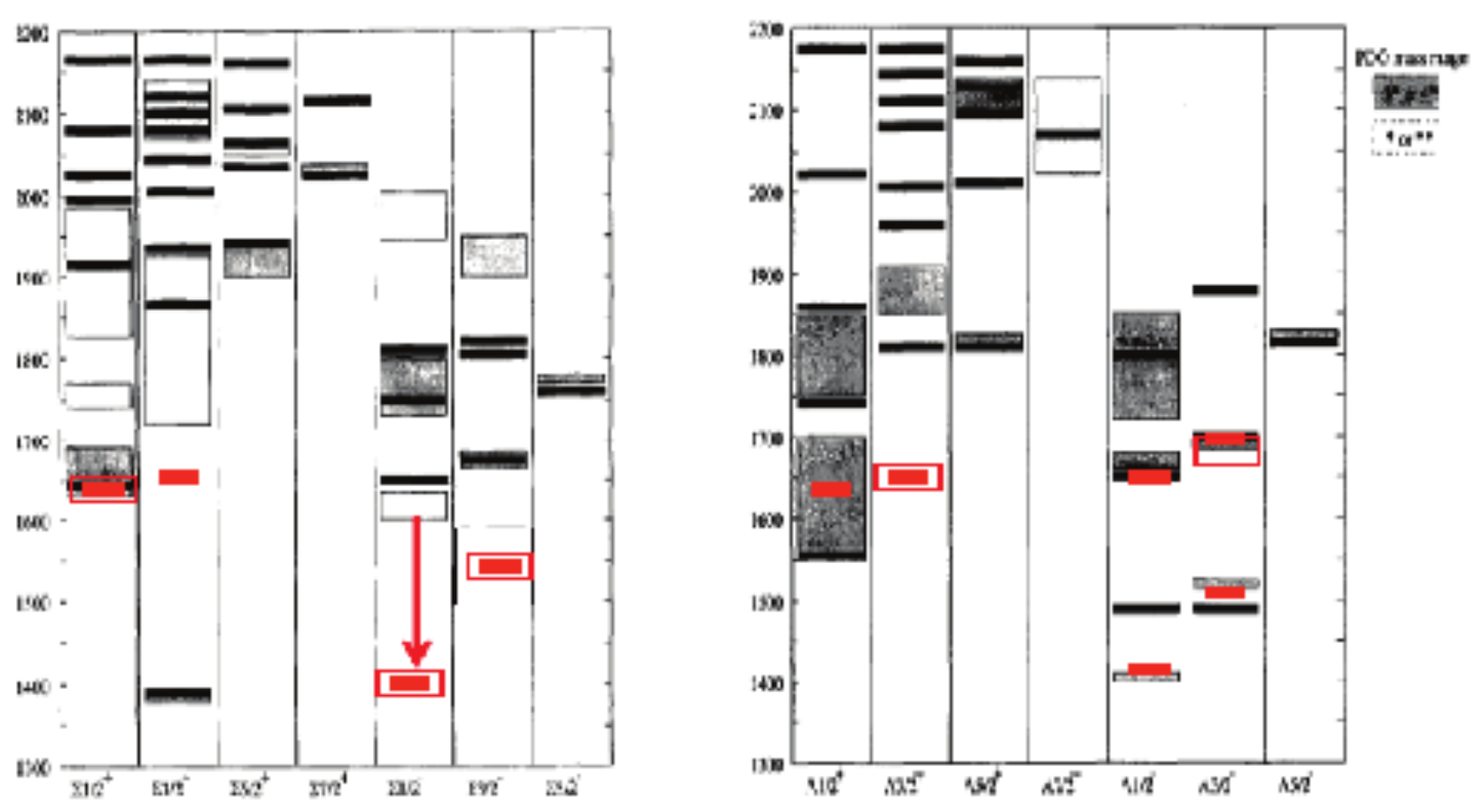}
\end{center}
\centerline{\parbox{0.80\textwidth}{
 \caption{Prediction of the lowest penta-quark states with
	$J^P=1/2^\pm,3/2^\pm$~\cite{pentq1Z,pentq2Z} (red solid) 
	compared with those from the classical quenched $qqq$ 
	model~\protect\cite{Capstick0Z} (black solid). The black 
	boxes are experimental results from PDG while the red 
	box are from recent new analyses. }\label{fig:graph} } }
\end{figure}

Although various phenomenological models give distinguishable
predictions for the lowest excited hyperon states, most of them 
are not experimentally established or even listed in 
PDG~\cite{pdgZ}. Most of our knowledge for the hyperon 
resonances came from analyses of old ${\overline K}N$ 
experiments in the 1970s~\cite{pdgZ}. In the new century, some 
new measurements from Crystal Ball (CB)~\cite{CBdata0Z,CBdata1Z,
CBdata2Z}, LEPS~\cite{LEPSZ} and CLAS~\cite{CLASZ} have started 
to provide us new information on $\Sigma^\ast$ and 
$\Lambda^\ast$ resonances. It is crucial to use them to clarify 
the spectrum of low-lying hyperon resonances to pin down the
underlying dynamics for baryon spectrum and structure. Recent
analyses of these new data together with old data reveal some
interesting new features of the low-lying excited hyperon 
states. Here I will give a brief review of these new results 
and discuss about their further confirmation from the proposed 
$K_L$ beam and other experiments.

\item \textbf{New Results on $\Sigma^\ast$ and $\Lambda^\ast$ 
	Resonances}

\begin{enumerate}
\item \textbf{On the Lowest $\Sigma^\ast$ Resonances with 
	Negative Parity}

The lowest $\Sigma^\ast$ resonances with $J^P=1/2^-$ or $3/2^-$ 
are still far from established. There is a $\Sigma(1620){1\over 
2}^-$ listed as a 2-star resonance in the previous versions of 
PDG tables and downgraded to 1-star in the newest 
version~\cite{pdgZ}. There is also a $\Sigma(1580){3\over 2}^-$ 
listed as 1-star resonance~\cite{pdgZ}.

The $\Sigma(1620){1\over 2}^-$ seems supporting the prediction 
of quenched quark models. However, for the 2-star $\Sigma(1620)
\frac{1}{2}^-$ resonance, only four references~\cite{16208Z,16207Z,
16202Z,16201Z} are listed in PDG tables with weak evidence for its 
existence. Among them, Ref.~\cite{16208Z} and Ref.~\cite{16207Z} 
are based on multi-channel analysis of the $\overline{K}N$ 
reactions.  Both claim evidence for a $\Sigma(\frac{1}{2}^-)$ 
resonance with mass around 1620~MeV, but give totally different 
branching ratios for this resonance. Ref.~\cite{16208Z} claims 
that it couples only to $\pi\Lambda$ and not to $\pi\Sigma$ 
while Ref.~\cite{16207Z} claims the opposite way. Both analyses 
do not have $\Sigma(1660)\frac{1}{2}^+$ in their solutions. 
However, Ref.~\cite{NPB94Z} shows no sign of $\Sigma(\frac{1}
{2}^-)$ resonance between 1600 and 1650~MeV through analysis 
of the reaction $\overline{K}N\rightarrow\Lambda\pi$ with the 
c.m. energy in the range of 1540-2150~MeV, instead it suggests
the existence of $\Sigma(1660)\frac{1}{2}^+$. Later 
multi-channel analyses of the $\overline{K}N$ reactions 
support the existence of the $\Sigma(1660)\frac{1}{2}^+$ 
instead of $\Sigma(1620)\frac{1}{2}^-$~\cite{pdgZ}. In 
Ref.~\cite{16202Z}, the total cross sections for $K^-p$ and 
$K^-n$ with all proper final states are analyzed and indicate 
some $\Sigma$ resonances near 1600~MeV without clear quantum 
numbers. Ref.~\cite{16201Z} analyzes the reaction $K^-n
\rightarrow\pi^-\Lambda$ and gets two possible solutions, with 
one solution indicating a $\Sigma({1\over 2}^-)$ near 1600~MeV, 
and the other showing no resonant structure below the
$\Sigma(1670)$. So all these claims of evidence for the
$\Sigma(1620){1\over 2}^-$ are very shaky. Instead, some 
re-analyses of the $\pi\Lambda$ relevant data suggest that 
there may exist a $\Sigma({1\over 2}^-)$ resonance around 
1380~MeV~\cite{WujjZ}, which supports the prediction of 
unquenched quark models~\cite{pentq1Z,pentq2Z}. This is supported 
by the new CLAS data on $\gamma p\to K\Sigma\pi$~\cite{CLASZ}, 
although a more delicate analysis~\cite{Roca:2013ccaZ} of the 
data suggests the resonant peak to be at a higher mass around 
1430~MeV.

For the study of $\Sigma$ resonances, the ${\bar K} N\to\pi
\Lambda$ reaction is the best available channel, where the 
s-channel intermediate states are purely hyperons with 
strangeness $S=-1$ and isospin $I=1$. Recently, high statistic 
new data for the reaction $K^-p\rightarrow\pi^0\Lambda$ are 
presented by the Crystal Ball Collaboration with the c.m. 
energy of 1560 -- 1676~MeV for both differential cross sections 
and $\Lambda$ polarizations~\cite{CBdata1Z}. In order to clarify 
the status of the $\Sigma(1620)\frac{1}{2}^-$ and the 
$\Sigma(1660)\frac{1}{2}^+$, we analyzed the differential cross 
sections and $\Lambda$ polarizations for both $K^-p\to\pi^0
\Lambda$ and $K^-n\to\pi^-\Lambda$ reactions with an effective 
Lagrangian approach, using the new Crystal Ball data on $K^-p
\to\pi^0\Lambda$ with the c.m. energy of 1560 --
1676~MeV~\cite{CBdata1Z}, and the $K^-n\to\pi^-\Lambda$ data of
Ref.~\cite{16201Z} with the c.m. energy of 1550 -- 1650~MeV, 
where the evidence of the $\Sigma(1620)\frac{1}{2}^-$ was 
claimed. The new Crystal Ball data clearly shows that the 
Crystal Ball $\Lambda$ polarization data demand the existence 
of a $\Sigma$ resonance with $J^P=\frac{1}{2}^+$ and mass near 
1635~MeV~\cite{pzgaoZ}, compatible with $\Sigma(1660)\frac{1}
{2}^+$ listed in PDG, while the $\Sigma(1620){1\over 2}^-$ is 
not needed by the data. The differential cross sections alone 
cannot distinguish the two solutions with either $\Sigma(1660)
\frac{1}{2}^+$ or $\Sigma(1620)\frac{1}{2}^-$.

This analysis also suggests a possible $\Sigma(\frac{3}{2}^-)$
resonance with mass around 1542~MeV and width about 25.6~MeV. 
This seems consistent with the resonance structure $\Sigma(1560)$ 
or\\ 
$\Sigma(1580){3\over 2}^-$ in PDG and compatible with 
expectation from penta-quark model~\cite{pentq1Z}. Ref.~\cite{D13Z} 
also proposes a $\Sigma({3\over 2}^-)$ resonance with mass around 
1570~MeV and width about 60~MeV from $\overline{K}N\pi$ system.

After our analysis, there were three groups~\cite{Zhang:2013svaZ,
Kamano:2015hxaZ,Fernandez-Ramirez:2015tfaZ} having made more 
sophisticated coupled channel analysis of the $\bar KN$ 
scattering data including those from the Crystal Ball experiment. 
The newest analysis~\cite{Fernandez-Ramirez:2015tfaZ} gives 
roughly consistent results for the lowest $\Sigma^\ast(1/2^\pm)$
resonances as ours. In both analyses, there is no $\Sigma(1620)
1/2^-$. While in our analysis, the $\Sigma(1635)1/2^+$ is 
definitely needed, in Ref.~\cite{Fernandez-Ramirez:2015tfaZ}, the
$\Sigma(1635)1/2^+$ is split to two $1/2^+$ resonances:
$\Sigma(1567)$ and $\Sigma(1708)$. The other two analyses claim 
the need of the $\Sigma(1620)1/2^-$, but with much lower energy 
at 1501~MeV~\cite{Zhang:2013svaZ} and 1551~MeV~\cite{Kamano:2015hxaZ},
respectively.

For the lowest $\Sigma^\ast(3/2^-)$, Ref.~\cite{Kamano:2015hxaZ} 
gives a similar result as ours with mass around 1550~MeV.
Refs.~\cite{Zhang:2013svaZ,Fernandez-Ramirez:2015tfaZ} give a 
higher mass around 1670~MeV.

So there are strong evidences for the lowest $\Sigma^\ast(1/2^-)$ 
to be in the range of $1380\sim 1500$~MeV and the lowest 
$\Sigma^\ast(3/2^-)$ to be around 1550~MeV. But this is not 
conclusive.

\item \textbf{On the Lowest $\Lambda^\ast(3/2^\pm)$ Resonances}

Many studies have been carried out to investigate the $\Lambda$
resonances. Oset \textit{et al.}~\cite{Oset1Z,Oset2Z} used a chiral 
unitary approach for the meson-baryon interactions and got two 
$J^P={1\over 2}^-$ resonances with one mass near 1390~MeV and the 
other around 1420~MeV. They believe the well established 
$\Lambda(1405){1\over 2}^-$ resonance listed in PDG~\cite{pdgZ} 
is actually a superposition of these two ${1\over 2}^-$ 
resonances. Manley \textit{et al.}~\cite{Zhang:2013svaZ} and 
Kamano \textit{et al.}~\cite{Kamano:2015hxaZ} made multichannel 
partial-wave analysis of $\overline{K}N$ reactions and got 
results with some significant differences. Zhong \textit{et 
al.}~\cite{xhZhongZ} analyzed the $K^-p\rightarrow\pi^0\Sigma^0$
reaction with the chiral-quark model and discussed 
characteristics of the well established $\Lambda$ resonances. 
Liu \textit{et al.}~\cite{Xie_etaZ} analyzed the $K^-p\rightarrow
\eta\Lambda$ reaction~\cite{CBdata0Z} with an effective Lagrangian 
approach and implied a $D_{03}$-resonance with mass about 
1670~MeV but much smaller width compared with the well 
established $\Lambda(1690){3\over 2}^-$. So there are still some 
ambiguities of the $\Lambda$ resonant structures needing to be 
clarified.

Recently, the most precise data on the differential cross 
sections for the $K^-p\to\pi^0\Sigma^0$ reaction have been 
provided by the Crystal Ball experiment at AGS/BNL~\cite{CBdata1Z,
CBdata2Z}. The $\Sigma^0$ polarization data were presented for 
the first time. However, with different data selection cuts and 
reconstructions, two groups in the same collaboration, {\it i.e.}, 
VA group~\cite{CBdata2Z} and UCLA group~\cite{CBdata1Z}, got 
inconsistent results for the $\Sigma^0$ polarizations. Previous 
multi-channel analysis~\cite{Zhang:2013svaZ,Kamano:2015hxaZ,
xhZhongZ} of the $\overline{K}N$ reactions failed to reproduce 
either set of the polarization data.

In our recent work~\cite{Shi:2014vhaZ}, we concentrate on the 
most precise data by the Crystal Ball Collaboration on the pure 
isospin scalar channel of $\overline{K}N$ reaction to see what 
are the $\Lambda$ resonances the data demand and how the two 
groups' distinct polarization data~\cite{CBdata1Z,CBdata2Z} 
influence the spectroscopy of $\Lambda$ resonances. Consistent 
differential cross sections of earlier work by Armenteros 
\textit{et al.}~\cite{LowEnergyDataZ} at lower energies are also 
used. It is found that the 4-star $\Lambda(1670){1\over 2}^-$ 
and 3-star $\Lambda(1600){1\over 2}^+$ resonances listed in 
PDG~\cite{pdgZ} are definitely needed no matter which set of CB 
data is used. In addition, there is strong evidence for the 
existence of a new $\Lambda({3\over 2}^+)$ resonance around
1680~MeV no matter which set of data is used. It gives large
contribution to this reaction, replacing the contribution from 
the 4-star $\Lambda(1690){3\over 2}^-$ resonance included by 
previous fits to this reaction.

Replacing the PDG $\Lambda(1690){3\over 2}^-$ resonance by a 
new $\Lambda(1680){3\over 2}^+$ resonance has important 
implications on hyperon spectroscopy and its underlying 
dynamics. While the classical qqq constituent quark 
model~\cite{capstickZ} predicts the lowest $\Lambda({3\over 
2}^+)$ resonance to be around 1900~MeV in consistent with the 
$\Lambda(1890){3\over 2}^+$ listed in PDG, the penta-quark 
dynamics~\cite{pentq1Z} predicts to be below 1700~MeV in
consistent with $\Lambda(1680){3\over 2}^+$ claimed in this 
work.

A recent analysis~\cite{Xie_etaZ} of CB data on the
$K^-p\to\eta\Lambda$ reaction requires a $\Lambda({3\over 
2}^-)$ resonance with mass about 1670~MeV and width about 
1.5~MeV instead of the well established $\Lambda(1690){3\over 
2}^-$ resonance with width around 60~MeV. Together with 
$N^\ast(1520){3\over 2}^-$, $\Sigma(1542){3\over 2}^-$ 
suggested in Ref.~\cite{pzgaoZ} and either $\Xi(1620)$ or 
$\Xi(1690)$, they fit in a nice $3/2^-$ baryon nonet with 
large penta-quark configuration, {\it i.e.}, $N^\ast(1520)$ 
as $|[ud]\{uq\}\bar q>$ state, $\Lambda(1520)$ as 
$|[ud]\{sq\}\bar q>$ state, $\Lambda(1670)$ as $|[ud]\{ss\}
\bar s>$ state, and $\Xi(16xx)$ as $|[ud]\{ss\}\bar q>$ 
state. Here $\{q_1q_2\}$ means a diquark with configuration 
of flavor representation ${\bf 6}$, spin 1 and color $\bar 
3$. The $\Lambda(1670)$ as $|[ud]\{ss\}\bar s>$ state gives 
a natural explanation for its dominant $\eta\Lambda$ decay 
mode with a very narrow width due to its very small phase
space meanwhile a D-wave decay~\cite{ZouHypZ}.

Recent analyses~\cite{Kamano:2015hxaZ,Fernandez-Ramirez:2015tfaZ} 
also support possible existence of the $\Lambda(1680){3\over 
2}^+$, but with a narrower width.
\end{enumerate}

\item \textbf{Summary and Prospects}

Taking into account new data from Crystal Ball
(CB)~\cite{CBdata0Z,CBdata1Z,CBdata2Z}, LEPS~\cite{LEPSZ} and
CLAS~\cite{CLASZ}, new analyses show strong evidences for the 
lowest $\Sigma^\ast(1/2^-)$ to be in the range of $1380\sim 
1500$~MeV, the lowest $\Sigma^\ast(3/2^-)$ to be around 
1550~MeV and the lowest $\Lambda^\ast(3/2^+)$ to be around 
1680~MeV. There is also evidence for a very narrow 
$\Lambda^\ast(3/2^-)$ around 1670~MeV decaying to 
$\Lambda\eta$. All these new hyperon resonances fit in the 
expected pattern of unquenched quark models very well. It is 
very important to pin down the existence of these new 
resonances.

Various processes could be used to study these hyperon 
resonances. The neutrino induced hyperon production processes 
$\bar{\nu}_{e/\mu} + p \to e^+/\mu^+ + \pi + \Lambda/\Sigma$ 
may provide a unique clean place for studying low energy 
$\pi\Lambda/\Sigma$ interaction and hyperon resonances below 
$KN$ threshold~\cite{Wu:2013klaZ}. With plenty production of 
$\Lambda_c$ at BESIII, J-PARC, BelleII, $\Lambda_c^+\to\pi^+
\pi^0\Lambda$ could also be used to study $\Sigma^\ast$. The 
$K^-$, $K_L$ beam experiments at JPARC and Jlab could provide 
an elegant new source for $\Lambda^\ast$, $\Sigma^\ast$, and
$\Xi^\ast$ hyperon spectroscopy.  $K_Lp\to\Lambda\pi^+$,
$\Sigma^0\pi^+$, $\Sigma^+\pi^0$, $\Sigma^{\ast 0}\pi^+$, and
$\Sigma^{\ast +}\pi^0$ could pin down the $\Sigma^\ast(1540) 
3/2^-$; $K_Lp\to\Sigma^0\pi^0\pi^+$, and $\Lambda\pi^0\pi^+$ 
could shed light on the $\Sigma^\ast(1380\sim 1500)1/2^-$,
$\Sigma^\ast(1540)3/2^-$, $\Lambda^\ast(1680)3/2^+$; 
$K_Lp\to\Sigma^0\eta\pi^+$, and $\Lambda\eta\pi^+$ may check 
$\Sigma^\ast(1380\sim 1500)1/2^-$, \\
$\Sigma^\ast(1540)3/2^-$, and $\Lambda^\ast(1670)3/2^-$. We 
believe the proposed $K_L$ beam experiments at JLab could 
settle down the spectrum of the low excited hyperon states 
which provide complimentary information to the study of 
penta-quark states with hidden charm~\cite{Wu:2010jyZ,
Aaij:2015tgaZ} and play a crucial role for understanding the 
hadron dynamics and hadron structure.

\item \textbf{Acknowledgments}

I thank S.~Dulat, Puze~Gao, Jun~Shi, J.J.~Wu, and J.J.~Xie for
collaboration works reviewed here. This work is supported by 
the National Natural Science Foundation of China under Grant 
11261130311 (CRC110 by DFG and NSFC).
\end{enumerate}


\newpage
\subsection{Can Spectroscopy with Kaon Beams at JLab Discriminate 
	between Quark Diquark and Three Quark Models ?}
\addtocontents{toc}{\hspace{2cm}{\sl E.~Santopinto}\par}
\setcounter{figure}{0}
\setcounter{table}{0}
\setcounter{equation}{0}
\halign{#\hfil&\quad#\hfil\cr
\large{Elena Santopinto}\cr
\textit{I.N.F.N., Sezione di Genova}\cr
\textit{via Dodecaneso 33}\cr
\textit{16146 Genova, Italy}\cr}

\begin{abstract}
Different three quark models exhibit different missing states 
but also quark diquark models still exhibit missing states, 
even if they have a reduced  space states. Moreover even quark 
diquark models show some differences in their missing states.  
After many years still we are not able to answer the question 
if  nature is completely described by three quark models or if 
diquark correlations in quark  diquark models have to be 
dismissed, or even if one of  the two pictures is the dominant 
one at different scales, as suggested 
in~\cite{Santopinto:2004V,Galata:2012xtV}. A new experiment based 
on Kaon beam and with polarization techniques, just as can be 
planned at Jlab will be able to answer to that fundamental 
open question.  The most recent LQCD effort show a threee quark  
clustering of their states at least at lower energy, but still 
they  are not at he  pion mass physical point, thus they are 
still not able to encode the complexity of the chiral symmetry 
breaking that as shown on he other side by eroic efforts in 
Dyson Swinger approach to QCD, underline the emerging of the 
importance of  diquark correlations. The quark diquark model 
corresponds in first approximation to the leading Regge 
trajectories and still all the resonances belonging to those 
trajectories are waiting for, to be discovered, but we expect 
that at least those that correspond to the leading Regge 
trajectories should be there, so considering that each piece 
of knowledge is closely interlocked and interconnected, the 
poor knowledge of some of the Lambda excited states, can be 
reflected  also in a early stage  in the Pentaquark analysis.
Finally, a review of the underlying ideas of the Interacting 
Quark Diquark Model (IQDM) that asses the  baryon spectroscopy 
and structure  in terms of quark diquark  degrees of freedom 
is  given, together with a discussion of the missing resonance 
problem. In respect to the early quark diquark models, we found 
that the IQDM is able to the describe the three star. 
$N^{3/2+}$(1930), that is missing in the old quark diquark 
models. 
\end{abstract}

\begin{enumerate}
\item \textbf{Introduction: Missing States and Kaon Beams}

Different three quark models exhibit different missing states  
(as confirmed~\cite{BSV} also in the study of strong decays 
with different quark models) but also quark diquark models 
still exhibit missing states~\cite{JPSV,Ferretti:2011V,
Santopinto:2015V}, even if they have a reduced space state. 
Moreover even quark diquark models show some differences in 
their missing states( let's compare the old~\cite{idaV,lichV} 
with the new~\cite{JPSV,Ferretti:2011V,Santopinto:2015V,
desanctisV}). After many years still we are 
not able to answer the question if  nature is completely 
described by three quark models or if diquark correlations 
in quark diquark models have to be dismissed, or even if 
one of the two pictures is the dominant one at different 
scales, as suggested in Ref.~\cite{Santopinto:2004V,
Galata:2012xtV}.  A new experiment based on Kaon beam and 
with polarization techniques, just as can be planned at 
Jlab will  be able  to  answer to that  fundamental open 
question. 

In parallel, recently, theoretical approaches based on QCD 
have been strongly developed. Lattice QCD performs ab initio 
calculation for hadron spectroscopy, even if  it is not easy 
to approach hadron states at the physical pion mass or with 
heavy flavor. Nevertheless, the  recent progress of the 
Lattice simulations  are really impressive and hadron 
structures and interactions have been discussed extensively 
in Refs.~\cite{Dudek:2009qfV,Aoki:2012omaV}.

The most recent LQCD effort show a threee quark like 
clustering at least at lower energy, but still LQCD results 
are not at the pion mass physical point, thus they are still 
not able to encode the complexity and richness of the chiral 
symmetry breaking. A bit of that, it is on the contrary kept 
by the efforts in Dyson Swinger approach to 
QCD~\cite{RobertsV}, that on the contrary is able to show that 
any interaction that binds $\pi$ mesons in the rainbow-ladder 
approximation of the DSE will produce also diquarks as can be 
seen in Ref.~\cite{RobertsV}. Nevertheless even if starting 
from the QCD Lagrangian with a DSE equation, due to the many 
approximations that are  necessary to be able to do 
calculations, we still turn out dealing with a model even if 
rooted in QCD.
 
On the contrary, quark diquark models are by definition only  
phenomenological models, and they corresponds in first 
approximation to the leading Regge trajectories, but many of 
those resonances belonging to those trajectories are still 
waiting to be discovered. It is reasonable to expect that at 
least those resonances that correspond to the leading Regge 
trajectories should exist. 
   
Considering up to only 2~GeV the Interacting Quark Diquark 
model has 8 missing $\Lambda$s in the octet and 6 in the 
singlet, so that many more  can be expected  up to 10~GeV.
It seems reasonable to expect that at least the quark 
diquark subset of states will be found by the experiments if 
we believe in a string like Regge behavior at higher energies  
where the quark diquark picture should be the dominant one, 
but also those resonances are still waiting to be discovered.  
In this respect, the study of the higher energy  part of the 
spectrum will shed light on the confinement 
mechanism~\cite{Ostrander:2012kzV,Roberts:2016dnbV} and the 
generation of the strange baryon and meson masses, as due to 
the breaking of chiral symmetry, and this will be one of the 
main task for a JLab Kaon beam experiment.  

Considering the same problem but as a three quark follower, 
we can argue in another way, but stiil the conclusions will 
be the same:  the number of $\Lambda$'s states (but the 
same can be said for $\Sigma$ or $\Omega$'s states) should 
be expected in nature  at least in equal number than the 
N$^\ast$ or $\Delta^\ast$ states (around 26), if we believe 
in three quark $SU(3)$ flavor symmetry (or at least only a  
subset of those if we on the contrary believe in a quark 
diquark like clustering of states). Considering that up to  
now only few strange states are  experimentally known, and 
very few also with their quantum numbers \textit{etc.}, 
thus for sure a 10~GeV Kaon beam experiment, as it can be  
planned at JLab, should  be rated to have a sure important 
result, also considering that there will be not only 
expertise in the hardware, but also in the analysis 
tecniques.

Considering that each piece of knowledge is closely 
interlocked and interconnected, for example the poor 
knowledge of some of the $\Lambda$'s  excited states, can 
be reflected  also in a early stage of the charmonium like 
Pentaquark analysis~\cite{Aaij:2015tgaV}.  Comparing the 
number of $\Lambda$'s states predicted by the relativistic 
Interacting Quark Diquark models (8 for the octet and 6 
for the singlet under 2~GeV) that are only a subset of 
those predicted by three-quark models, we can try to 
suggest a next generation Pentaquark analysis that 
evaluates the systematic error on the background due to 
the missing $\Lambda$'s states (see Ref.~\cite{JPSV}). The 
future discovering of missing $\Lambda$ resonances by a 
new JLab Kaon beam experiment maybe will not change the 
structures seen in the Dalitz Plot by the LHCb analysis, 
but eventually modify some parameters.  In a similar way, 
the poor knowledge of strange hadrons can be reflected 
into an early stage of strangeness physics and beyond 
the standard model analysis, if (as very often happen) 
hadron physics pieces are involved in the analysis too.

Various aspects of the hadron structures have been 
investigated by many experimental and theoretical 
approaches in the last years. The observations of the 
hadron states with an exotic structure have attracted a 
lot of interest. In particular, regarding the light 
flavor region, we can remind the exotic states found in 
the accelerator facilities such as the scalar mesons 
$a_0(980)$ and $f_0(980)$,or the $\Lambda(1405)$ which 
are expected to have an exotic structure as multiquarks, 
hadronic molecules, but also hybrid states  and so 
forth~\cite{Klempt:2007cpV,Brambilla:2010csV}, that with 
Kaon beam could be better studied. On the other hand in 
the heavy counterpart, there are now accumulating 
evidences of exotic heavy hadrons, we can cite states 
such as the $Z_c$~\cite{Ablikim:2013mioV,Liu:2013dauV} 
and $Z^{(\prime)}_b$~\cite{Belle:2011aaV} which can not 
be explained by the simple quark model picture. 

The chiral effective field theory respecting  the chiral 
symmetry provides the hadron-hadron scatterings at low 
energy with the Nambu-Goldstone bosons exchange. This is a 
powerful tool to investigate hadronic molecules as the 
meson-meson~\cite{Oller:1997tiV,Wang:2013kvaV,Baru:2015neaV}, 
meson-baryon~\cite{Hyodo:2011urV,Yamaguchi:2011xbV}, and 
baryon-baryon~\cite{Machleidt:2011zzV,Haidenbauer:2011zaV} 
states appearing near thresholds, but they need a fine 
tuning of their parameters that can only be obtained with 
high precision Kaon beam experiments. 
  
Finally , in the last part of this article, we will discuss  
briefly some new results obtained within the  formalism of 
the Unquenched Quark Model (UQM):  when LQCD or Chiral 
effective models can not be applied, it can provide anyway 
predictions, making up  with the three quark model defects, 
but again also the UQM like chiral effective field theory 
needs a good knowledge of the  strange couplings that can 
be a sub-product of a Kaon beam experiment. 

\item \textbf{Phenomenological Motivation for Quark Diquark 
	Model}

The  notion of diquark is as old as the quark model itself. 
Gell-Mann~\cite{gellV} mentioned the possibility of diquarks 
in his original paper on quarks, just as the possibility of 
tetra and pentaquark. Soon afterwards, Ida and 
Kobayashi~\cite{idaV} and Lichtenberg and Tassie~\cite{lichV} 
introduced effective degrees of freedom of diquarks in order 
to describe baryons as composed of a constituent diquark and 
quark. Since its introduction, many articles have been 
written on this subject~\cite{ansV,Jakob:1997V,Brodsky:2002V,
Gamberg:2003V,Jaffe:2003V,WilczekV,Jaffe:2004phV,Santopinto:2004V,
Selem:2006ndV,DeGrand:2007vuV,Forkel:2008unV,Anisovich:2010wxV} 
up to the most recent ones~\cite{Ferretti:2011V,Santopinto:2015V,
desanctisV}, and, more recently, also in tetraquark 
spectroscopy.
Moreover different phenomenological indications for diquark 
correlations have been collected during the years, such as 
some regularities 
in hadron spectroscopy, the $\Delta I = \frac{1}{2}$ rule in 
weak nonleptonic decays~\cite{NeubertV}, some regularities in 
parton distribution functions and in spin-dependent structure 
functions~\cite{CloseV} and in the $\Lambda(1116)$ and 
$\Lambda(1520)$  fragmentation functions. Although the 
phenomenon of color superconductivity~\cite{bailingwilczekV} 
in quark dense matter cannot be considered an argument in 
support of diquarks in the vacuum, it is nevertheless of 
interest since it stresses the important role of Cooper 
pairs of color superconductivity, which are color antitriplet, 
flavor antisymmetric, scalar diquarks. The concept of diquarks 
in hadronic physics has some similarities to that of correlated 
pairs in condensed matter physics (superconductivity~\cite{bcsV}) 
and in nuclear physics (interacting boson model~\cite{ibmV}), 
where effective bosons emerge from pairs of electrons~\cite{coopV} 
and nucleons~\cite{ibm2V}, respectively. Any interaction that
binds $\pi$ and $\rho$  mesons in the rainbow-ladder 
approximation of the DSE will produce diquarks as can be seen 
in Ref.~\cite{RobertsV}, and finally  there are even some 
indication of diquark confinement.
The quark-diquark effective degrees of freedom have shown their 
usefulness also in the study of transversity problems and 
fragmentation functions (see  Ref.~\cite{BacchettaV}), even 
in an oversimplified form, \textit{i.e.} with the spatial 
part of the quark-diquark ground state wave function 
parametrized by means of a gaussian.  The microscopic origin 
of the diquark as an effective degrees of freedom, it is not 
completely clear, nevertheless, as in nuclear physics, one 
may attempt to correlate the data in terms of a 
phenomenological model, and in many cases it has already 
shown it usefulness. In this short contribution, we will 
review the Interacting Quark Diquark model in its original 
formulation~\cite{Santopinto:2004V}, discussing also the 
Point Form relativistic reformulation~\cite{Santopinto:2004V,
Santopinto:2015V,Ferretti:2011V}. We shall focus on its 
differences and extension to the strange 
spectra~\cite{Santopinto:2015V}. We will point out some 
important consequences  on the ratio of the electric and 
magnetic form factor of the proton, that is a  presence of 
a zero at $Q^2= 8\ GeV^2$, while impossible with three quark 
models. The new 12~GeV$^2$ experiment  planned at JLab 
will eventually shed light on the three quark versus 
diquark structure  of the nucleon. 

\item \textbf{The Interacting Quark Diquark Model}

The  model is an attempt to arrive to a systematic 
description  and correlation of data in term of q-diquark 
effective degrees of freedom.  By formulating a quark- 
diquark model with explicit interactions, in particular 
with a direct and an exchange interaction, we will show 
the spectrum which emerges from this model.  In respect 
to the  prediction shown in Ref.~\cite{Santopinto:2004V} 
we have  extended our calculation up to 2.4~GeV, and so 
we have predicted more states. Up to an energy of about 
2~GeV, the diquark can be described as two correlated 
quarks with no internal spatial 
excitations~\cite{Santopinto:2004V,Ferretti:2011V}, thus 
its color-spin-flavor wave function must be antisymmetric. 
Moreover, as we consider only light baryons, made up of 
$u$, $d$, $s$ quarks, the internal group is restricted to 
SU$_{\mbox{sf}}$(6). If we denote spin by its value, 
flavor and color by the dimension of the representation, 
the quark has spin $s_2 = \frac{1}{2}$, flavor $F_2={\bf 
{3}}$, and color $C_2 = {\bf {3}}$. The diquark must 
transform as ${\bf {\overline{3}}}$ under 
SU$_{\mbox{c}}$(3), hadrons being color singlets. Then, 
one only has the symmetric SU$_{\mbox{sf}}$(6) 
representation $\mbox{{\boldmath{$21$}}}_{\mbox{sf}}$(S), 
containing $s_1=0$, $F_1={\bf {\overline{3}}}$, and $s_1
=1$, $F_1={\bf {6}}$, i.e. the scalar and axial-vector 
diquarks, respectively. This is because we think of the 
diquark as two correlated quarks in an antisymmetric 
nonexcited state. We assume that the baryons are composed 
of a valence quark and a valence diquark.

The relative configurations of two body can be described 
by the relative coordinate $\vec{r}$ and its conjugate 
momenta $\vec{p}$. The Hamiltonian  contains a direct and 
an exchange interaction. The direct interaction is Coulomb 
plus linear interaction, while the exchange one is of the 
type spin-spin, isospin-isospin etc.  A contact term has 
to be present to describe the splitting between the nucleon 
and the $\Delta$:
\begin{equation}
	H =E_{0}+\frac{p^{2}}{2m}-\frac{\tau }{r}+{\beta }r
	+(B+C\delta_{0})\delta _{S_{12},1}~~~~~~~~~\nonumber \\
\end{equation}
\begin{equation}
	+(-1)^{l+1}2Ae^{-\alpha r}[\vec{s_{12}}\cdot\vec{s_{3}}
	+\vec{t_{12}} 
	\cdot \vec{t_{3}}+2\vec{s_{12}}\cdot\vec{s_{3}}~\vec{t_{12}}
	\cdot\vec{t_{3}}.
\end{equation}
For a purely Coulomb-like interaction the problem is
analytically solvable. The solution is trivial, with 
eigenvalues 
\begin{equation}
	E_{n,l}~=~-\frac{\tau ^{2}m}{2~n^{2}}~~~~,~n~=~1,~2~...
\end{equation}
Here $m$ is the reduced mass of the diquark-quark configuration 
and $n$ the principal quantum number. The eigenfunctions are the 
usual Coulomb functions 
\begin{equation}
	R_{n,l}(r)=\sqrt{\frac{(n-l-1)!(2g)^{3}}{2n[(n+l)!]^{3}}}
	(2gr)^{l}~e^{-gr}L_{n-l-1}^{2l+1}(2gr),
\end{equation}
where for the associated Laguerre polynomials 
$g=\frac{\tau m}{n}$. We treat all the 
other interactions as perturbations, so the  model is completely 
analytical.  The matrix elements of $\beta r$ can be evaluated  
in closed form as 
\begin{equation}
	\Delta E_{n,l}=\int_{0}^{\infty }\beta r[R_{n,l}(r)]^{2}
	r^2dr=\frac{\beta }{2m\tau }[3n^{2}-l(l+1)].
\end{equation}
Next comes the exchange interaction of Eq. (5). The spin-isospin 
part is obviously diagonal in the basis of Eq. (7)
\begin{eqnarray}
	\langle \vec{s}_{12}\cdot \vec{s}_{3}\rangle =\frac{1}{2}
	\left[S(S+1)-s_{12}(s_{12}+1)-s_{3}(s_{3}+1)\right], \nonumber \\
	\langle \vec{t}_{12}\cdot \vec{t}_{3}\rangle =\frac{1}{2}\left[
	T(T+1)-t_{12}(t_{12}+1)-t_{3}(t_{3}+1)\right] .
\end{eqnarray}
To complete the evaluation, we need the matrix elements of the 
exponential. These can be obtained in analytic form 
\begin{equation}
	I_{n,l}(\alpha )~=~\int_{0}^{\infty }~e^{-\alpha ~r}~
	[R_{n,l}(r)]^{2}r^2dr~~.
\end{equation}
The results are straightforward. Here, by way of example, we 
quote the result for $l =n-1$ 
\begin{equation}
	I_{n,l=n-1}(\alpha )~=~(\frac{1}{1+\frac{n~\alpha }
	{2\tau ~m}})^{2n+1}~~~.
\end{equation}
Our results are in present in Tables~\ref{tab:nuc-spectrum} and 
\ref{tab:del-spectrum}.
\begin{table}
\centerline{\parbox{0.80\textwidth}{
 \caption{Mass spectrum of $N$-type resonances (up to 2.1~GeV)
        in the interacting quark diquark
        model~\protect\cite{Santopinto:2004V}. The value of the
        parameters are those obtained and reported in Ref.~[10]
        based on the fit of the 3 and 4 star resonances known
        at the time. The table reports also the prediction for
        the remaining resonances, including the recent upgraded
        3$^\ast$ $P13(1900)$. The experimental values are taken
        from Ref.~\protect\cite{Olive:2014V}.}
	\label{tab:nuc-spectrum} } }
\vspace{0.5cm}
\centering 
\vspace{15pt} 
\begin{tabular}{ccccc}
\hline
\hline
& & & &  \\
Baryon $L_{2I,2J}$ & Status & Mass & $J^p$ & $M_{{\rm cal}}$ \\
& & (MeV) & & (MeV)  \\
\hline
$N( 939)P_{11}$   & **** & 939       & $1/2^+$ &  940 \\
$N(1440)P_{11}$   & **** & 1410-1450 & $1/2^+$ & 1538 \\
$N(1520)D_{13}$   & **** & 1510-1520 & $3/2^-$ & 1543 \\
$N(1535)S_{11}$   & **** & 1525-1545 & $1/2^-$ & 1538 \\
$N(1650)S_{11}$   & **** & 1645-1670 & $1/2^-$ & 1673 \\
$N(1675)D_{15}$   & **** & 1670-1680 & $5/2^-$ & 1673 \\
$N(1680)F_{15}$   & **** & 1680-1690 & $5/2^+$ & 1675 \\
$N(1700)D_{13}$   &  *** & 1650-1750 & $3/2^-$ & 1673 \\
$N(1710)P_{11}$   &  *** & 1680-1740 & $1/2^+$ & 1640 \\
$N(1720)P_{13}$   & **** & 1700-1750 & $3/2^+$ & 1675 \\
$N(1860)F_{15}$   &   ** & 1820-1960 & $5/2^+$ & 1975 \\
$N(1875)D_{13}$   &  *** & 1820-1920 & $3/2^-$ & 1838 \\
$N(1880)P_{11}$   &   ** & 1835-1905 & $1/2^+$ & 1838 \\
$N(1895)S_{11}$   &   ** & 1880-1910 & $1/2^-$ & 1838 \\
$N(1900)P_{13}$   &  *** & 1875-1935 & $3/2^+$ & 1967 \\
$N(1990)F_{17}$   &   ** & 1995-2125 & $7/2^+$ & 2015 \\
$N(2000)F_{15}$   &   ** & 1950-2150 & $5/2^+$ & 2015 \\
$N(2040)P_{13}$   &    * & 2031-2065 & $3/2^+$ & 2015 \\
$N(2060)D_{15}$   &   ** & 2045-2075 & $5/2^-$ & 2078 \\
$N(2100)P_{11}$   &   ** & 2050-2200 & $1/2^+$ & 2015\\
$N(2120)D_{13}$   &   ** & 2090-2210 & $3/2^-$ & 2069 \\
\hline
\end{tabular}
\end{table}
\begin{table}
\centerline{\parbox{0.80\textwidth}{
 \caption{As Table~\protect\ref{tab:nuc-spectrum}, but for
        $\Delta$-type resonances.} \label{tab:del-spectrum} } }
\vspace{0.5cm}
\centering 
\begin{tabular}{ccccc}
\hline
\hline
Baryon $L_{2I,2J}$ & Status & Mass & State & $M_{{\rm cal}}$ \\
 & & (MeV) & & (MeV)  \\
\hline
$\Delta(1232)P_{33}$ & **** & 1230-1234    & $3/2^+$ & 1235 \\
$\Delta(1600)P_{33}$ &  *** & 1500-1700    & $3/2^+$ & 1709 \\
$\Delta(1620)S_{31}$ & **** & 1600-1660    & $1/2^-$ & 1673 \\
$\Delta(1700)D_{33}$ & **** & 1670-1750    & $3/2^-$ & 1673 \\
$\Delta(1900)S_{31}$ & **   & 1840-1920    & $1/2^-$ & 2003 \\
$\Delta(1905)F_{35}$ & **** & 1855-1910    & $5/2^+$ & 1930 \\ 
$\Delta(1910)P_{31}$ & **** & 1860-1910    & $1/2^+$ & 1967 \\ 
$\Delta(1920)P_{33}$ &  *** & 1900-1970    & $3/2^+$ & 1930 \\ 
$\Delta(1930)D_{35}$ &  *** & 1900-2000    & $5/2^-$ & 2003 \\ 
$\Delta(1940)D_{33}$ & **   & 1940-2060    & $3/2^-$ & 2003 \\
$\Delta(1950)F_{37}$ & **** & 1915-1950    & $7/2^+$ & 1930 \\ 
$\Delta(2000)F_{35}$ & **   & $\approx$ 2000 & $5/2^+$ & 2015 \\ 
\hline
\hline
\end{tabular}
\end{table}

\item \textbf{The Relativistic Interacting Quark Diquark Model}

The exstention of the Interacting quark diquark 
model~\cite{Santopinto:2004V} in Point Form can be easily 
done~\cite{Ferretti:2011V,Santopinto:2015V}. This is a 
potential model, constructed within the point form 
formalism~\cite{Klink:1998zzV}, where baryon resonances are 
described as two-body quark-diquark bound states; thus, the 
relative motion between the two constituents and the Hamiltonian 
of the model are functions of the relative coordinate $\vec r$ 
and its conjugate momentum $\vec q$. The Hamiltonian contains  
just as in the 2005 paper~\cite{Santopinto:2004V}, the two basic 
ingredients: a Coulomb-like  plus linear confining interaction 
and an exchange one, depending on the spin and isospin of the 
quark and the diquark. The mass operator is given by
\begin{equation}
\begin{array}{rcl}
	M & = & E_0 + \sqrt{\vec q\hspace{0.08cm}^2 + m_1^2} 
	+ \sqrt{\vec q\hspace{0.08cm}^2 + m_2^2} 
	+ M_{\mbox{dir}}(r)  + M_{\mbox{ex}}(r)  
\end{array}  \mbox{ }, \label{eqn:H0}
\end{equation}
where $E_0$ is a constant, $M_{\mbox{dir}}(r)$ and $M_{\mbox{ex}}(r)$ 
the direct and the exchange diquark-quark interaction, respectively, 
$m_1$ and $m_2$ stand for diquark and quark masses. The direct term, 
we consider, 
\begin{equation} \label{eq:Vdir}
	M_{\mbox{dir}}(r)=-\frac{\tau}{r} \left(1 - e^{-\mu 
	r}\right)+ \beta r 
\end{equation}
is the sum of a Coulomb-like interaction with a cut off plus a 
linear confinement term. We also have an exchange interaction, 
since this is the crucial ingredient of a quark-diquark 
description of baryons that has to be extended to contain 
flavor $\lambda$ matrices in such a way to be able to describe 
in a simultaneous way  both the non strange and the strange 
sector~\cite{Santopinto:2004V,Santopinto:2015V}. We have also 
generalized the exchange interaction in such a way to be able 
to describe strange baryons, simply considering 
\begin{equation}
	\begin{array}{rcl}
	M_{\mbox{ex}}(r) & = & \left(-1 \right)^{L + 1} \mbox{ } 
	e^{-\sigma r} \left[ A_S \mbox{ } \vec{s}_1 
	\cdot \vec{s}_2  + A_F \mbox{ } \vec{\lambda}_1^f \cdot 
	\vec{\lambda}_2^f \mbox{ } 
	+ A_I \mbox{ } \vec{t}_1 \cdot \vec{t}_2  \right]  
	\end{array}  \mbox{ },
	\label{eqn:Vexch-strange}
\end{equation}
where $\vec{\lambda}^f$ are the SU$_{\mbox{f}}$(3) Gell-Mann 
matrices. In a certain sense, we can consider it as a 
G\"ursey-Radicati inspired interaction~\cite{desanctisV,
Gursey:1992dcV}.  In the nonstrange sector, we also have to 
keep a contact interaction~\cite{Ferretti:2011V} in the mass 
operator
\begin{equation}
\begin{array}{rcl} \label{eqn:Vcont}	
	M_{\mbox{cont}} & = & \left(\frac{m_1 m_2}{E_1 E_2}\right)^{1/2
	+\epsilon} \frac{\eta^3 D}{\pi^{3/2}} 
	e^{-\eta^2 r^2} \mbox{ } \delta_{L,0} \delta_{s_1,1} 
	\left(\frac{m_1 m_2}{E_1 E_2}\right)^{1/2+\epsilon}
\end{array}  \mbox{ }
\end{equation}
as necessary to reproduce the $\Delta-N$ mass splitting. 
\begin{table}
 \caption{Mass predictions~\protect\cite{Santopinto:2015V} for $\Lambda$-type
        resonances compared with PDG data; APS copyright.}
        \label{tab:lam-spectrum}
\vspace{0.5cm}
\begin{tabular}{cccccccccccccc}
\hline
\hline
Resonance & Status & $M^{\mbox{exp.}}$ & $J^P$ & $L^P$ & $S$ & $s_1$ & $Q^2q$ & ${\bf F}$ & ${\bf {F_1}}$ & $I$ & $t_1$ & $n_r$ & $M^{\mbox{calc.}}$ \\
	&  & (MeV) &  &  &  &  &  &  & & & & & (MeV) \\ 
\hline
$\Lambda(1116)$ $P_{01}$ & **** & 1116        & $\frac{1}{2}^+$ & $0^+$ & $\frac{1}{2}$ & 0   & $[n,n]s$   & ${\bf 8}$ & ${\bf {\bar 3}}$ & 0 & 0             & 0 & 1116  \\
$\Lambda(1600)$ $P_{01}$ & ***  & 1560 - 1700 & $\frac{1}{2}^+$ & $0^+$ & $\frac{1}{2}$ & 0   & $[n,s]n$   & ${\bf 8}$ & ${\bf {\bar 3}}$ & 0 & $\frac{1}{2}$ & 0 & 1518  \\
$\Lambda(1670)$ $S_{01}$ & **** & 1660 - 1680 & $\frac{1}{2}^-$ & $1^-$ & $\frac{1}{2}$ & 0   & $[n,n]s$   & ${\bf 8}$ & ${\bf {\bar 3}}$ & 0 & 0             & 0 & 1650  \\
$\Lambda(1690)$ $D_{03}$ & **** & 1685 - 1695 & $\frac{3}{2}^-$ & $1^-$ & $\frac{1}{2}$ & 0   & $[n,n]s$   & ${\bf 8}$ & ${\bf {\bar 3}}$ & 0 & 0             & 0 & 1650  \\
$\Lambda(1800)$ $S_{01}$ & ***  & 1720 - 1850 & $\frac{1}{2}^-$ & $1^-$ & $\frac{1}{2}$ & 0   & $[n,s]n$   & ${\bf 8}$ & ${\bf {\bar 3}}$ & 0 & $\frac{1}{2}$ & 0 & 1732  \\
$\Lambda(1810)$ $P_{01}$ & ***  & 1750 - 1850 & $\frac{1}{2}^+$ & $0^+$ & $\frac{1}{2}$ & 0   & $[n,n]s$   & ${\bf 8}$ & ${\bf {\bar 3}}$ & 0 & 0             & 1 & 1666  \\
$\Lambda(1820)$ $F_{05}$ & **** & 1815 - 1825 & $\frac{5}{2}^+$ & $2^+$ & $\frac{1}{2}$ & 0   & $[n,n]s$   & ${\bf 8}$ & ${\bf {\bar 3}}$ & 0 & 0             & 0 & 1896  \\
$\Lambda(1830)$ $D_{05}$ & **** & 1810 - 1830 & $\frac{5}{2}^-$ & $1^-$ & $\frac{3}{2}$ & 1   & $\{n,s\}n$ & ${\bf 8}$ & ${\bf {6}}$      & 0 & $\frac{1}{2}$ & 0 & 1785  \\
$\Lambda(1890)$ $P_{03}$ & **** & 1850 - 1910 & $\frac{3}{2}^+$ & $0^+$ & $\frac{3}{2}$ & 1   & $\{n,s\}n$ & ${\bf 8}$ & ${\bf {6}}$      & 0 & $\frac{1}{2}$ & 0 & 1896  \\ 
\hline
missing                 & --   & --          & $\frac{3}{2}^-$ & $1^-$ & $\frac{1}{2}$ & 0   & $[n,s]n$   & ${\bf 8}$ & ${\bf {\bar 3}}$ & 0 & $\frac{1}{2}$ & 0 & 1732  \\
missing                 & --   & --          & $\frac{1}{2}^-$ & $1^-$ & $\frac{3}{2}$ & 1   & $\{n,s\}n$ & ${\bf 8}$ & ${\bf {6}}$      & 0 & $\frac{1}{2}$ & 0 & 1785  \\
missing                 & --   & --          & $\frac{3}{2}^-$ & $1^-$ & $\frac{1}{2}$ & 0   & $[n,n]s$   & ${\bf 8}$ & ${\bf {\bar 3}}$ & 0 & 0             & 1 & 1785  \\
missing                 & --   & --          & $\frac{1}{2}^+$ & $0^+$ & $\frac{1}{2}$ & 1   & $\{n,s\}n$ & ${\bf 8}$ & ${\bf {6}}$      & 0 & $\frac{1}{2}$ & 0 & 1955  \\ 
missing                 & --   & --          & $\frac{1}{2}^+$ & $0^+$ & $\frac{1}{2}$ & 0   & $[n,s]n$   & ${\bf 8}$ & ${\bf {\bar 3}}$ & 0 & $\frac{1}{2}$ & 1 & 1960  \\
missing                 & --   & --          & $\frac{1}{2}^-$ & $1^-$ & $\frac{1}{2}$ & 1   & $\{n,s\}n$ & ${\bf 8}$ & ${\bf {6}}$      & 0 & $\frac{1}{2}$ & 0 & 1969  \\
missing                 & --   & --          & $\frac{3}{2}^-$ & $1^-$ & $\frac{1}{2}$ & 1   & $\{n,s\}n$ & ${\bf 8}$ & ${\bf {6}}$      & 0 & $\frac{1}{2}$ & 0 & 1969  \\ 
\hline
$\Lambda^*(1405)$ $S_{01}$ & **** & 1402 - 1410 & $\frac{1}{2}^-$ & $1^-$ & $\frac{1}{2}$ & 0 & $[n,n]s$   & ${\bf 1}$ & ${\bf {\bar 3}}$ & 0 & 0             & 0 & 1431  \\
$\Lambda^*(1520)$ $D_{03}$ & **** & 1519 - 1521 & $\frac{3}{2}^-$ & $1^-$ & $\frac{1}{2}$ & 0 & $[n,n]s$   & ${\bf 1}$ & ${\bf {\bar 3}}$ & 0 & 0             & 0 & 1431\\ 
missing                 & --     & --         & $\frac{1}{2}^-$ & $1^-$ & $\frac{1}{2}$ & 0 & $[n,s]n$   & ${\bf 1}$ & ${\bf {\bar 3}}$ & 0 & $\frac{1}{2}$ & 0 & 1443  \\
missing                 & --     & --         & $\frac{3}{2}^-$ & $1^-$ & $\frac{1}{2}$ & 0 & $[n,s]n$   & ${\bf 1}$ & ${\bf {\bar 3}}$ & 0 & $\frac{1}{2}$ & 0 & 1443  \\
missing                 & --     & --         & $\frac{1}{2}^-$ & $1^-$ & $\frac{1}{2}$ & 0 & $[n,n]s$   & ${\bf 1}$ & ${\bf {\bar 3}}$ & 0 & 0               & 1 & 1854  \\
missing                 & --     & --         & $\frac{3}{2}^-$ & $1^-$ & $\frac{1}{2}$ & 0 & $[n,n]s$   & ${\bf 1}$ & ${\bf {\bar 3}}$ & 0 & 0               & 1 & 1854  \\
missing                 & --     & --         & $\frac{1}{2}^-$ & $1^-$ & $\frac{1}{2}$ & 0 & $[n,s]n$   & ${\bf 1}$ & ${\bf {\bar 3}}$ & 0 & $\frac{1}{2}$   & 1 & 1928  \\ 
missing                 & --     & --         & $\frac{3}{2}^-$ & $1^-$ & $\frac{1}{2}$ & 0 & $[n,s]n$   & ${\bf 1}$ & ${\bf {\bar 3}}$ & 0 & $\frac{1}{2}$   & 1 & 1928  \\
\hline
\hline
\end{tabular}
\end{table}

The results  for the strange and non-strange baryon spectra 
from Ref.~\cite{Santopinto:2015V,Santopinto:2004V} (See 
Tables~\ref{tab:nuc-spectrum}, \ref{tab:del-spectrum}, and 
\ref{tab:lam-spectrum}) were obtained by diagonalizing the 
mass operator of Eq.(\ref{eqn:H0}) by means of a numerical 
variational procedure, based on harmonic oscillator trial 
wave functions. With a basis of 150 harmonic oscillator 
shells, the results converge very well.

It is interesting to compare our results~\cite{Santopinto:2015V} 
to those of three-quark quark models (see 
Refs.~\cite{Isgur:1979beV,Godfrey:1985xjV,Capstick:1986bmV,
Giannini:2001kbV,Glozman-RiskaV,Loring:2001kxV,Ferretti:2011V,
Galata:2012xtV,BILV}).
It is clear that a larger number of experiments 
and analysis, looking for missing resonances, are necessary 
because many aspects of hadron spectroscopy are still unclear. 
In particular the number of $\Lambda $ states   reported by 
the PDG are still very few in respect to the predictions both 
of the Lattice QCD and by the models. In particular the  
relativistic version of the  interacting quark diquark model  
predict seven $\Lambda$  missing  states belonging to the 
octet and other six missing states  belonging to the singlet 
(considering only states under 2.0~GeV), otherwise much more 
states should be considered, and looked for, by a 10~GeV 
secondary Kaon beam experiment at Jlab. Typical three quark 
models  will predict much more $\Lambda $  missing states,  
and in short they should be in the same number then the 
already known  $N$ or $\Delta$ states,so  at least so 24 
for the octet and the same for the singlet. New experiments 
should be dedicated to the hunting of those elusive missing 
$\Lambda$ states.

Without relying on models only considering the *** and ***  
Nstar and using $SU(3)$ symmetry, for each Nstar belonging to 
an octet, one can expect to complete with the corresponding 
$\Lambda$ state belonging to the same octet. That  will give 
us an expectation  for its mass  by means of  an evaluation 
via a Guersey and Radicati  mass formula (see table  ).    
The $\Lambda$'s states that are partners  of the same octet  
multiplet for which at least an $N$ star state has been 
already seen or viceversa will be denoted with the same 
colors.

It is also worthwhile noting that in our model~\cite{Santopinto:2015V}  
$\Lambda(1116)$ and $\Lambda^*(1520)$ are described as bound states 
of a scalar diquark $[n,n]$ and a quark $s$, where the quark-diquark 
system is in $S$ or $P$-wave, respectively~\cite{Santopinto:2015V}. 
This is in accordance with the observations of 
Refs.~\cite{Jaffe:2004phV,Selem:2006ndV} on $\Lambda$'s fragmentation 
functions, that the two resonances can be described as $[n,n]-s$ 
systems.

The present work can be expanded to include charmed and/or bottomed 
baryons, which can be quite interesting in light of the recent 
experimental effort to study the properties of heavy hadrons. 

We should also underline that the interacting quark-diquark model  
gives origin to wave functions that can describe in a  reasonable 
way the elastic electromagnetic form factors of the nucleon. In 
particular they give origin to  a reproduction of the existing 
data for the ratio of the electric and magnetic form factor of 
the proton that predict a zero at  $Q^2= 8~GeV^2$ (see 
Fig.~\ref{ratiomp1}) like in vector meson parametrizations. On 
the contrary, we have found impossible to get this zero with a 
three quark model~\cite{Santopinto:VasalloV} (see 
Fig.~\ref{ratiomp2}). The new experiment planned at JLab will be 
able to distinguish between the two scenarios ruling out one of 
the two models.  
\begin{figure}[ht!]
\begin{center}
\includegraphics[width=10cm]{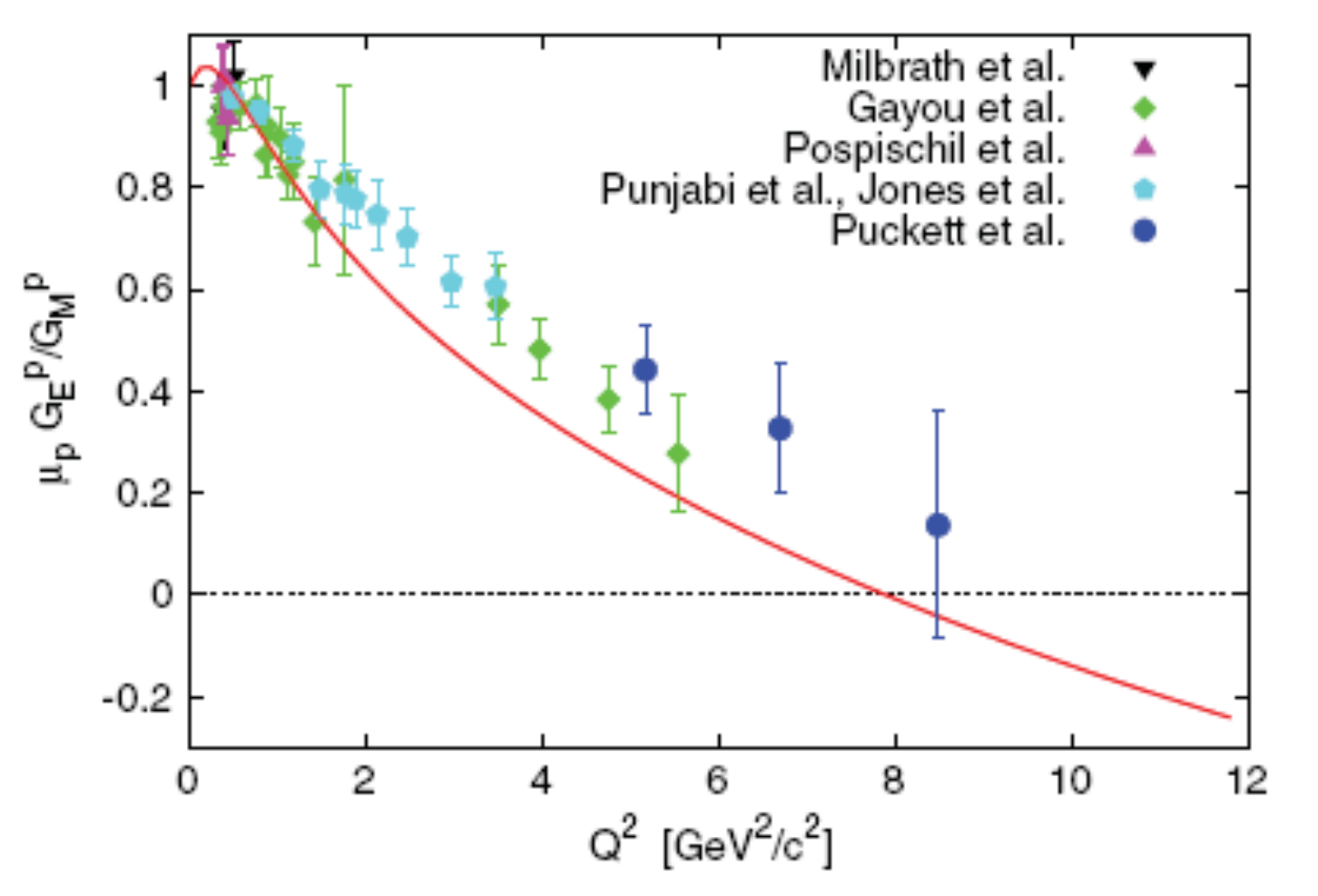}
\end{center}
\centerline{\parbox{0.80\textwidth}{
 \caption{\label{ratiomp1}Ratio $\mu_pG^p_E(Q^2)/G^p_M(Q^2)$, the 
	solid line correspond to the relativistic quark-diquark 
	calculation, figure taken from 
	Ref.\protect\cite{Ferretti_ff:2011V}; APS copyright.} } }
\end{figure}
\begin{figure}[ht!]
\begin{center}
\includegraphics[width=10cm]{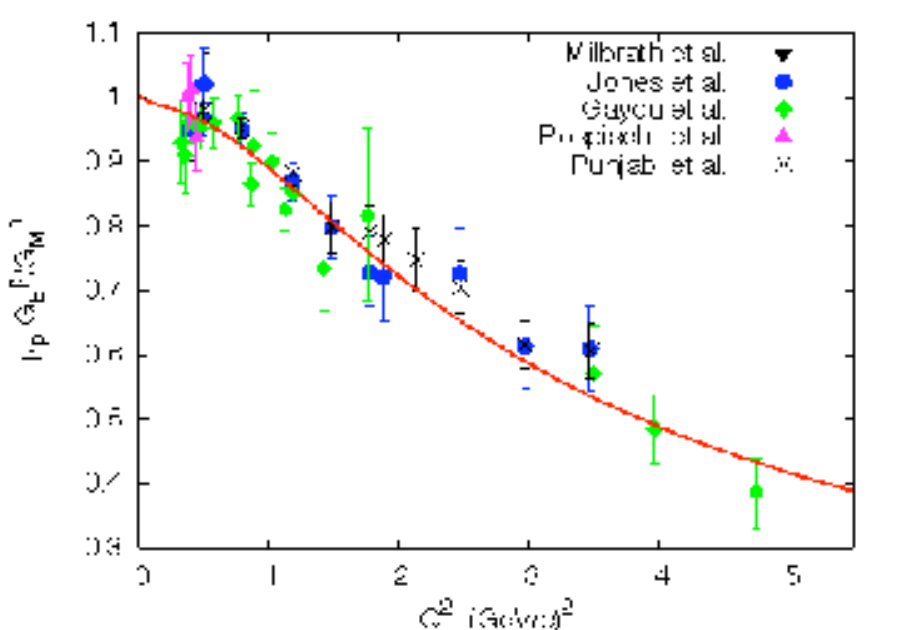}
\end{center}
\centerline{\parbox{0.80\textwidth}{
 \caption{\label{ratiomp2} Ratio $\mu_pG^p_E(Q^2)/G^p_M(Q^2)$, the 
	solid line correspond to the relativistic Hypercetral quark 
	model, Figure taken from 
	Ref.~\protect\cite{Santopinto:VasalloV}; APS copyright.} } }
\end{figure}

\item \textbf{The Unquenched Quark Model}

The behavior of observables such as the spectrum and the magnetic 
moments of hadrons are well reproduced by the constituent quark 
model (CQM)~\cite{Eichten:1974afV,Isgur:1979beV,Godfrey:1985xjV,
Capstick:1986bmV,Giannini:2001kbV,Glozman-RiskaV,Loring:2001kxV,
Ferretti:2011V,Galata:2012xtV,BILV}, but it neglects quark-antiquark 
pair-creation (or continuum-coupling) effects. The unquenching of 
the quark model for hadrons is a way to take these components 
into account.

The unquenching of CQM were initially done by T\"ornqvist and 
collaborators, who used an unitarized quark model~\cite{Ono:1983rdV,
TornqvistV}, while Van Beveren and Rupp used an t-matrix 
approach~\cite{vanBeveren:1979bdV,vanBeveren:1986eaV}.  These 
techniques were applied to study of scalar meson nonet ($a_0$, 
$f_0$, \textit{etc.}) of Ref.~\cite{vanBeveren:1986eaV,
Tornqvist:1995krV} in which the loop contributions are given by 
the hadronic intermediate states that each meson can access. It 
is via these hadronic loops that the bare states become ``dressed" 
and  the hadronic loop contributions totally dominate the dynamics 
of the process.  A similar approach was developed by Pennington in 
Ref.~\cite{Pennington:2002V}, where they have  investigated the 
dynamical generation of the scalar mesons by initially inserting 
only one ``bare seed".  Also, the strangeness content of the 
nucleon and electromagnetic form factors were  investigated 
in~\cite{Bijker:2012zzaV}, whereas  Capstick and Morel in 
Ref.~\cite{CapstickV} analyzed baryon meson loop effects on the 
spectrum of nonstrange baryons. In the meson sector, Eichten 
{\it et al.} explored the influence of the open-charm channels on 
the charmonium properties using the Cornell coupled-channel 
model~\cite{Eichten:1974afV} to assess departures from the 
single-channel potential-model expectations.

In this work we present  the latest applications of the UQM to 
study the orbital angular momenta contribution to the spin of 
the proton in which the effects of the sea quarks were introduced  
into the CQM in a systematic way and the wave functions given 
explicitly. In another contribution of the same workshop are on 
the contrary discussed the flavor asymmetry and strangeness of 
the proton. Finally, the UQM is applied to describe meson 
observables and the spectroscopy of the charmonium and 
bottomonium, developing the formalism to take into account in a 
systematic way, the continuum components.  

\item \textbf{The UQM Formalism}

In the UQM for baryons~\cite{Bijker:2012zzaV,Santopinto:2010zzaV,
Bijker:2009upV,Bijker:210V} and mesons \cite{bottomoniumV,
charmoniumV,Ferretti:2013vuaV,Ferretti:2014xqaV}, the hadron wave 
function is made up of a zeroth order $qqq$ ($q \bar q$) 
configuration plus a sum over the possible higher Fock 
components, due to the creation of $^{3}P_0$ $q \bar q$ pairs. 
Thus, we have 
\begin{eqnarray} 
	\label{eqn:Psi-A}
	\mid \psi_A \rangle
	 ={\cal N} \left[ \mid A \rangle 
	+ \sum_{BC \ell J} \int d \vec{K} \, k^2 dk \, \mid BC \ell 
	J;\vec{K} k \rangle \right.
	\left.  \frac{ \langle BC \ell J;\vec{K} k \mid T^{\dagger} 
	\mid A \rangle } {E_a - E_b - E_c} \right] ~, 
\end{eqnarray}
where $T^{\dagger}$ stands for the $^{3}P_0$ quark-antiquark 
pair-creation operator~\cite{bottomoniumV,charmoniumV,Ferretti:2013vuaV,
Ferretti:2014xqaV}, $A$ is the\\ 
baryon/meson, $B$ and $C$ represent 
the intermediate state hadrons. $E_a$, $E_b$ and $E_c$ are the 
corresponding energies, $k$ and $\ell$ the relative radial momentum 
and orbital angular momentum between $B$ and $C$ and $\vec{J} 
= \vec{J}_b + \vec{J}_c + \vec{\ell}$ is the total angular momentum. 
It is worthwhile noting that in Refs.~\cite{bottomoniumV,charmoniumV,
Ferretti:2013vuaV,Ferretti:2014xqaV}, the constant pair-creation 
strength in the operator (\ref{eqn:Psi-A}) was substituted with an 
effective one, to suppress unphysical heavy quark pair-creation. 

The introduction of continuum effects in the CQM can thus be essential 
to study observables that only depend on $q \bar q$ sea pairs, like 
the strangeness content of the nucleon electromagnetic form 
factors~\cite{Bijker:2012zzaV} or the flavor asymmetry of the nucleon 
sea~\cite{Santopinto:2010zzaV} it has been discussed in another 
contribution of the same conference (see Garc\'ia-Tecocoatzi \textit{et 
al.}) The continuum effects can give important corrections to 
baryon/meson observables, like the self-energy corrections to meson 
masses~\cite{bottomoniumV,charmoniumV,Ferretti:2013vuaV,Ferretti:2014xqaV} 
or the importance of the orbital angular momentum in the spin of the 
proton~\cite{Bijker:2009upV}. 

\item \textbf{Orbital Angular Momenta Contribution to Proton Spin 
	in the UQM Formalism}

The inclusion of the continuum higher Fock components has a 
dramatic effect on the spin content of the proton~\cite{Bijker:210V}. 
Whereas in the CQM the proton spin is carried entirely by the 
(valence) quarks, while in the unquenched calculation 67.6\% 
is carried by the quark and antiquark spins and the remaining 
32.4\% by orbital angular momentum. The orbital angular momentum 
due to the relative motion of the baryon with respect to the meson 
accounts for 31.7\% of the proton spin, whereas the orbitally 
excited baryons and mesons in the intermediate state only 
contribute 0.7\%. Finally we note, that the orbital angular 
momentum arises almost entirely from the relative motion of the 
nucleon and $\Delta$ resonance with respect to the $\pi$-meson in 
the intermediate states. 

\item \textbf{Self-Energy Corrections in the UQM}

The formalism  was used to compute the charmonium ($c \bar c$) 
and bottomonium ($b \bar b$) spectra with self-energy corrections, 
due to continuum coupling effects~\cite{bottomoniumV,charmoniumV,
Ferretti:2013vuaV,Ferretti:2014xqaV}. In the UQM, the physical mass 
of a meson
\begin{equation}\label{eqn:self-trascendental}
	M_a = E_a + \Sigma(E_a)  \mbox{ }
\end{equation}
is given by the sum of two terms: a bare energy, $E_a$, calculated 
within a potential model~\cite{Godfrey:1985xjV}, and a self energy 
correction
\begin{equation}\label{eqn:self-a}
	\Sigma(E_a) = \sum_{BC\ell J} \int_0^{\infty} k^2 dk 
	\mbox{ } \frac{\left| M_{A \rightarrow BC}(k) \right|^2}
	{E_a - E_b - E_c}  \mbox{ },
\end{equation}
computed within the UQM formalism. 
\begin{figure}[ht!]
\begin{center}
\includegraphics[width=25pc]{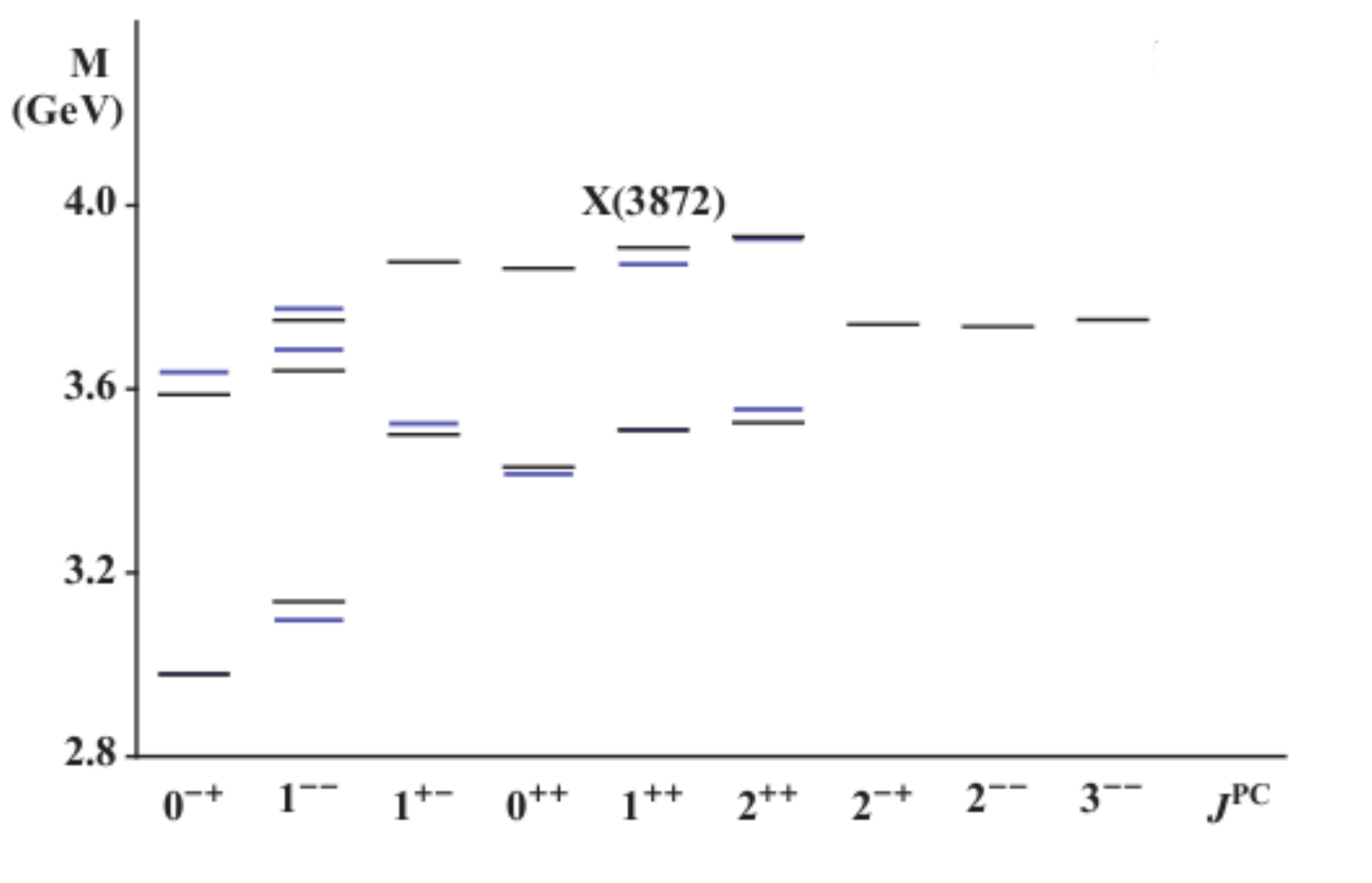}
\end{center}
\centerline{\parbox{0.80\textwidth}{
 \caption{\label{charm} Charmonium spectrum with self energies 
	corrections. Black lines are theoretical predictions and 
	blue lines are experimental data available. Figure taken 
	from Ref.~\protect\cite{charmoniumV}; APS copyright.} } }
\end{figure}
\begin{figure}[ht!]
\begin{center}
\includegraphics[width=25pc]{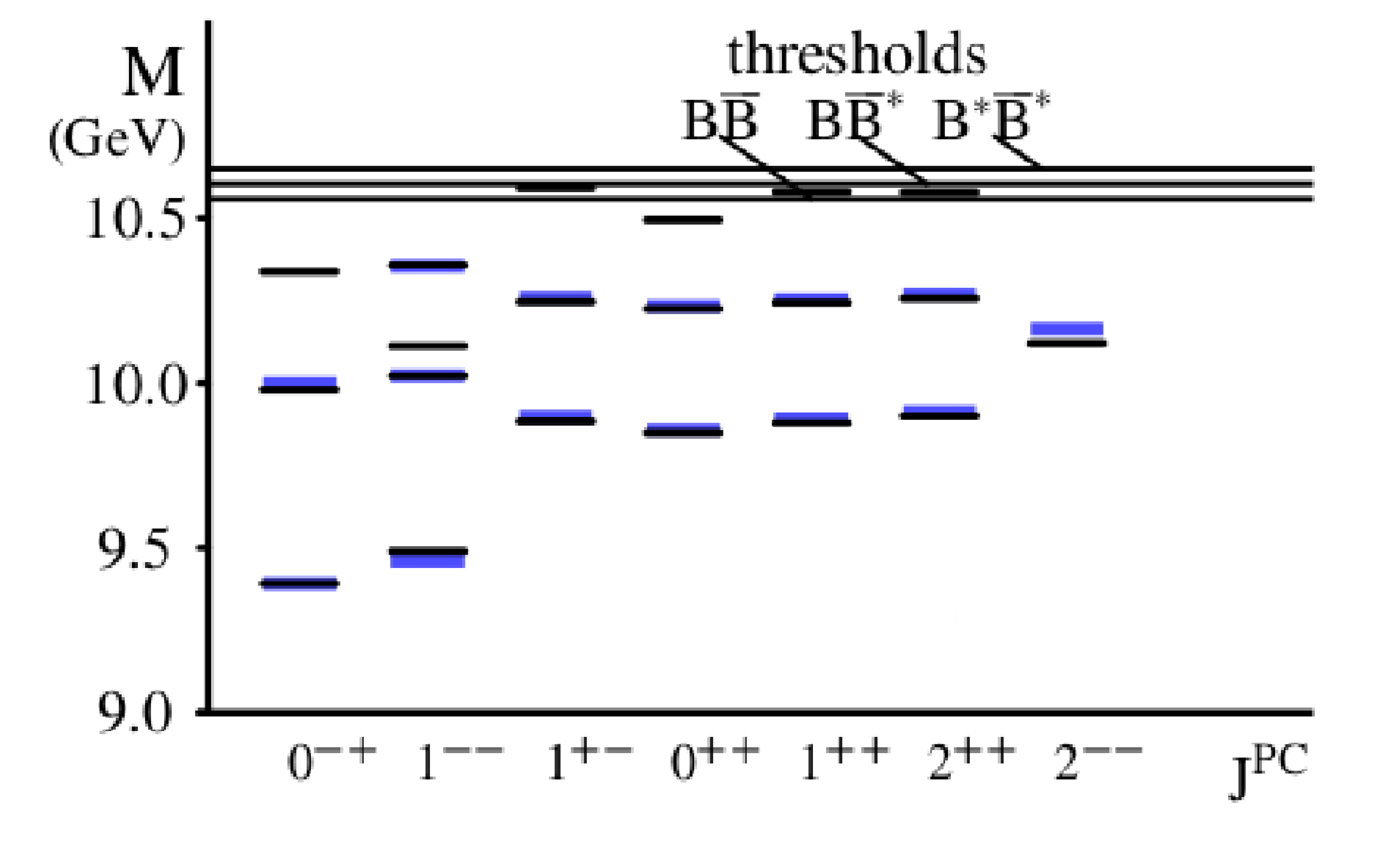}
\end{center}
\centerline{\parbox{0.80\textwidth}{
 \caption{\label{botton}Bottomonium spectrum with self energies 
	corrections. Black lines are theoretical predictions and 
	blue lines are experimental data available. Figure taken 
	from Ref.~\protect\cite{Ferretti:2013vuaV}; APS copyright.} } }
\end{figure}

Our results for the self energies corrections of 
charmonia~\cite{charmoniumV,Ferretti:2014xqaV} and 
bottomonia~\cite{bottomoniumV,Ferretti:2013vuaV,Ferretti:2014xqaV} 
spectrums, are shown in Figures~\ref{charm} and \ref{botton}. 

In our framework the $X(3872)$ can be interpreted as a $c \bar c$ 
core [the $\chi_{c1}(2^3P_1)$], plus higher Fock components 
due to the coupling to the meson-meson continuum. In 
Ref.~\cite{Ferretti:2014xqaV}, we  were the first to predict  
analogous states (as $X(3872)$) with strong continuum components  
in the bottomonium sector but in the $\chi_{b1}(3^3P_1)$ sector, 
due to opening of threshold of $B\bar{B}$, $B\bar{B}^\ast$ and 
$B^\ast\bar{B}^\ast$. We expect similar interesting effects near 
threshold also in the $N^\ast$ sector.

It is interesting to compare the present results to those of the 
main three-quark quark models~\cite{Isgur:1979beV,Godfrey:1985xjV,
Capstick:1986bmV,Giannini:2001kbV,Glozman-RiskaV,Loring:2001kxV,
Ferretti:2011V,Galata:2012xtV,BILV}.
It is clear that a larger number of experiments and analyses, 
looking for missing resonances, are necessary because many aspects 
of hadron spectroscopy are still unclear.
\end{enumerate}


\newpage
\subsection{Reducing the Ambiguity of the AntiKaon-Nucleon Amplitude
        Using Modern Experimental Data}
\addtocontents{toc}{\hspace{2cm}{\sl M.~Mai}\par}
\setcounter{figure}{0}
\setcounter{table}{0}
\setcounter{equation}{0}
\halign{#\hfil&\quad#\hfil\cr
\large{Maxim Mai}\cr
\textit{Helmholtz--Institut f\"ur Strahlen- und Kernphysik (Theorie)}\cr
\textit{and Bethe Center for Theoretical Physics}\cr
\textit{Universit\"at Bonn}\cr
\textit{D-53115 Bonn, Germany}\cr}

\begin{abstract}
AntiKaon-nucleon scattering is studied utilizing an analytic solution
of the Bethe-Salpeter equation with the interaction kernel derived
from the leading and next-to leading order chiral Lagrangian. In the
on-shell factorization of this solution multiple sets of parameters
are found, which all allow for a good description of the existing
hadronic data. We confirm the two-pole structure of the
$\Lambda(1405)$. The narrow $\Lambda(1405)$ pole appears at 
comparable positions in the complex energy plane, whereas the 
location of the broad pole suffers from a large uncertainty. It is 
demonstrated how experimental data on the photoproduction of $K^+
\pi\Sigma$ off the proton measured by the CLAS Collaboration can be 
used to reduce this ambiguity. Finally, an estimation is made on the 
desired quality of the new scattering data to constrain the parameter 
space of the presented model.
\end{abstract}

\begin{enumerate}
\item \textbf{Introduction}

The strangeness $S=-1$ resonance $\Lambda(1405)$ is believed to be
dynamically generated through coupled-channel effects in the
antiKaon-nucleon interaction. A further intricate feature is its
two-pole structure. Within chiral unitary approaches, which are
considered to be the best tool to address the chiral SU(3) dynamics
in such type of system, the investigation of the two-pole structure
was initiated in Ref.~\cite{Oller:2000fjMM} and thoroughly analyzed
in many publications, for a (PDG) review see Ref.~\cite{PDGreviewMM}.
However, the available scattering data alone do not allow to pin
down the poles with good precision, as it is known since long, see, 
{\it e.g.}, Ref.~\cite{Borasoy:2006srMM}.

Recently, very sophisticated measurements of the reaction $\gamma
p\to K^+\Sigma\pi$ were performed by the CLAS Collaboration at JLab,
see Ref.~\cite{Moriya:2013ebMM}. There, the invariant mass distribution
of all three $\pi\Sigma$ channels was determined in a broad energy
range and with high resolution. First theoretical analyses of this
data have already been performed on the basis of a chiral unitary
approach in Refs.~\cite{Roca:2013avMM,Nakamura:2013boaMM}. In this 
work, we take up the challenge to combine our next-to-leading order
approach of antiKaon-nucleon scattering~\cite{Mai:2012dtMM} in an
on-shell approximation with the CLAS data.

First, we construct a family of solutions that lead to a good
description of the scattering and the SIDDHARTA data. This
reconfirms the two-pole structure of the $\Lambda(1405)$. As before,
we find that the location of the second pole in the complex energy
plane is not well determined from these data alone. Then, we address
the issue how this ambiguity can be constrained from the CLAS data.
Similar to Ref.~\cite{Roca:2013avMM}, we use a simple
semi-phenomenological model for the photoproduction process that
combines the description of the hadronic scattering with a simple
polynomial and energy-dependent ansatz for the photoproduction of
$K^+$ and a meson-baryon pair of strangeness $S=-1$ off the proton.
The corresponding energy- and channel-dependent constants are
fitted to the CLAS data. However, it appears that not all solutions,
consistent with the scattering data, lead to a decent fit to the
photoproduction data. Moreover, we find that the solutions,
consistent with photoproduction and scattering data lead to similar
positions of both poles of $\Lambda(1405)$.

\item \textbf{AntiKaon-nucleon Scattering}

The starting point of the present analysis is the meson-baryon
scattering amplitude - a simplified version of the amplitude
constructed and described in detail in the original
publication~\cite{Mai:2014xnaMM} as well as in 
Refs.~\cite{Mai:2012wyMM,Bruns:2010svMM}, to which we refer 
the reader for conceptual details. We start from the chiral 
Lagrangian of leading (LO) and next-to-leading (NLO) order. 
For the reasons given in Refs.~\cite{Mai:2014xnaMM,
Mai:2012wyMM,Bruns:2010svMM}, the $s$- and $u$-channel 
one-baryon exchange diagrams are neglected, leaving us with 
the following chiral potential
\begin{equation}\label{eqn:potential}
         V(\slashed{q}_2, \slashed{q}_1; p)=A_{WT}(\slashed{q_1}
        +\slashed{q_2}) +A_{14}(q_1\cdot q_2)+A_{57}
        [\slashed{q_1},\slashed{q_2}] +A_{M} +A_{811}\Big(\slashed{q_2}
        (q_1\cdot p)+\slashed{q_1}(q_2\cdot p)\Big)\,,
\end{equation}
where the incoming- and outgoing-meson four-momenta are denoted by
$q_1$ and $q_2$, respectively, whereas the overall four-momentum of
the meson-baryon system is denoted by $p$. The $A_{WT}$, $A_{14}$,
$A_{57}$, $A_{M}$ and $A_{811}$ are 10-dimensional  matrices which
encode the coupling strengths between all 10 channels of the
meson-baryon system for strangeness $S=-1$, {\it i.e.}, $\{K^-p$, 
$\bar K^0 n$, $\pi^0\Lambda$, $\pi^0\Sigma^0$, $\pi^+\Sigma^-$, 
$\pi^- \Sigma^+$, $\eta\Lambda$, $\eta \Sigma^0$, $K^+\Xi^-$, 
$K^0\Xi^0\}$.  These matrices depend on the meson decay constants, 
the baryon and meson masses as well as 14 low-energy constants 
(LECs) as specified in the original publication~\cite{Mai:2014xnaMM}.

Due to the appearance of the $\Lambda(1405)$ resonance just below
the $\bar KN$ threshold and large momentum transfer, the strict
chiral expansion is not applicable for the present system. Instead,
the above potential is used as a driving term of the coupled-channel
Bethe-Salpeter equation (BSE), for NLO approaches see, {\it e.g.},
Ref.~\cite{Borasoy:2006srMM,Mai:2012dtMM,Ikeda:2012auMM,Guo:2012vvMM}. 
For the meson-baryon scattering amplitude $T(\slashed{q}_2,
\slashed{q}_1; p)$ the integral equation to be solved reads
\begin{equation}\label{eqn:BSE}
        T(\slashed{q}_2, \slashed{q}_1; p)=V(\slashed{q}_2,
        \slashed{q}_1;p)  +i\int\frac{d^d l}{(2\pi)^d}
        V(\slashed{q}_2, \slashed{l}; p)  S(\slashed{p}
        -\slashed{l})\Delta(l)T(\slashed{l}, \slashed{q}_1; p)\,,
\end{equation}
where $S$ and $\Delta$ represent the baryon (of mass $m$) and the
meson (of mass $M$) propagator, respectively, and are given by
$iS(\slashed{p}) = {i}/({\slashed{p}-m+i\epsilon})$ and $i\Delta(k)
={i}/({k^2-M^2+i\epsilon})$. Moreover, $T$, $V$, $S$ and $\Delta$
in the last expression are matrices in the channel space. The loop
diagrams appearing above are treated using dimensional 
regularization and applying the usual $\overline{\rm MS}$ 
subtraction scheme in the spirit of our previous 
work~\cite{Bruns:2010svMM}. Note that the modified loop integrals 
are still scale-dependent. This scale $\mu$ reflects the influence 
of the higher-order terms not included in our potential and is 
used as a fit parameter. To be precise, we have 6 such parameters 
in the isospin basis.
\begin{figure}[ht!]
\includegraphics[width=\linewidth]{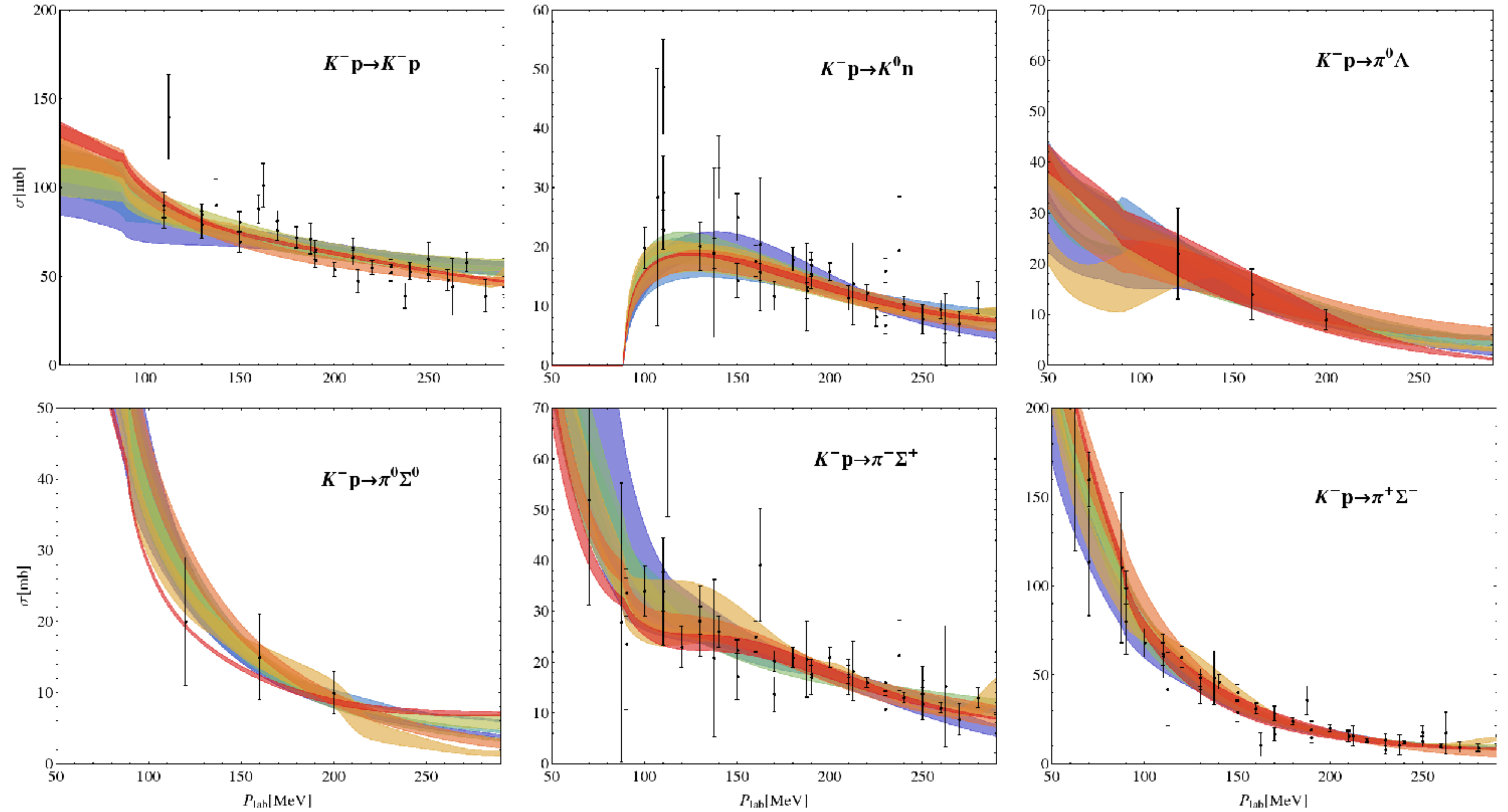}
\centerline{\parbox{0.80\textwidth}{
\caption{Fit results compared to the experimental data from
        Refs.~\protect\cite{Ciborowski:1982etMM,Humphrey:1962zzMM,
        Sakitt:1965khMM,Watson:1963zzMM}. Different colors
        correspond to the eight best solutions, while the
        bands represent the $1\sigma$ uncertainty due to
        errors of the fit parameters. The color coding is
        specified in Fig.~\protect\ref{fig:poles1}.}
        \label{fig:cs} } }
\end{figure}

The above equation can be solved analytically if the kernel
contains contact terms only, see Ref.~\cite{Mai:2012wyMM} for the
corresponding solution. Using this solution for the strangeness
$S=-1$ system, we have shown in Ref.~\cite{Mai:2012dtMM} that once
the full off-shell amplitude is constructed, one can easily
reduce it to the on-shell solution, {\it i.e.}, setting all 
tadpole integrals to zero. It appears that the double pole 
structure of the $\Lambda(1405)$ is preserved by this reduction 
and that the positions of the two poles are changing only by 
about $20$~MeV in imaginary part. On the other hand, the use 
of the on-shell approximation of the Eq.~\eqref{eqn:BSE} 
reduces the computational time roughly by a factor of 30. 
Therefore, in order to explore the parameter space in more 
detail, it seems to be safe and also quite meaningful to start 
from the solution of the BSE~\eqref{eqn:BSE} with the chiral 
potential~\eqref{eqn:potential} on the mass-shell. Once the 
parameter space is explored well enough we can slowly turn on 
the tadpole integrals obtaining the full off-shell solution. 
Such a solution will become a part of a more sophisticated 
two-meson photoproduction amplitude in a future work.
\renewcommand{\baselinestretch}{1.25}
\begin{table}
\centerline{\parbox{0.80\textwidth}{
\caption{Quality of the various fits in the description of the
        hadronic and the photoproduction data from CLAS. For
        the definition of $\chi_{\rm p.p.}^2$, see the text.}
        \label{tab:photo} } }
\begin{center}
\begin{tabular}{ccccccccc}
\hline
Fit \# & 1&2&3&4&5&6&7&8\\
\hline
$\chi_{\rm d.o.f.}^2$ (hadronic data)~~~ &1.35 &1.14&0.99 &0.96 &1.06 &1.02&1.15 &0.90\\
\hline
$\chi_{\rm p.p.}^2$   ~~(CLAS data)~~~~~ &3.18&1.94&2.56&1.77&1.90&6.11&2.93&3.14\\
\hline
\end{tabular}
\end{center}
\end{table}

The free parameters of the present model, the low-energy constants
as well as the regularization scales $\mu$ are adjusted to reproduce
all known experimental data in the meson-baryon sector of strangeness
$S=-1$. The main bulk of this data consists of the cross sections
for the processes $K^-p\to MB$, where $MB\in \{K^-p, \bar K^0n,
\pi^0\Lambda, \pi^+\Sigma^-, \pi^0\Sigma^0, \pi^-\Sigma^+\}$, and
laboratory momentum $P_{\rm lab}<300$~MeV, from
Refs.~\cite{Ciborowski:1982etMM,Humphrey:1962zzMM,Sakitt:1965khMM,
Watson:1963zzMM}. Electromagnetic effects are not included in the
analysis and assumed to be negligible at the measured values of
$P_{\rm lab}$. Additionally, at the antiKaon-nucleon threshold, we
consider the decay ratios from Refs.~\cite{Tovee:1971gaMM,
Nowak:1978auMM} as well as the energy shift and width of Kaonic
hydrogen in the 1s state from the SIDDHARTA experiment at
DA$\Phi$NE~\cite{Bazzi:2011zjMM} related to the $K^-p$ scattering
length via the modified Deser-type formula~\cite{Meissner:2004jrMMM}.
Due to the precision of the experiment, the latter two values have
already become the most important input in this sector. In
principle, both $\bar KN$ scattering lengths can be determined
directly, performing a complementary measurement on the Kaonic
deuterium, see Refs.~\cite{LNFMM,JPARKMM} for the proposed 
experiments.  The strong energy shift and width of the latter can 
again be related to the antiKaon-deuteron scattering length, using 
the the modified Deser-type formula~\cite{Meissner:2004jrMMM} and 
finally to the antiKaon-nucleon scattering lengths as described in
Ref.~\cite{Mai:2014umaMM}.

The fit to the above data was performed minimizing $\chi_{\rm
d.o.f.}^2$ using several thousands randomly distributed sets of
starting values of the free parameters. The latter were assumed
to be of natural size, while the unphysical solutions, {\it e.g.},
poles on the first Riemann sheet for ${\rm Im}(W)<200$~MeV
($W:=\sqrt{p^2}$), were sorted out. For more details on the
fitting procedure and results, we refer the reader to the original
publication~\cite{Mai:2014xnaMM}. Eight best solutions were obtained
by this, see second row of Tab.~\ref{tab:photo}, whereas the next
best $\chi_{\rm d.o.f.}^2$ are at least one order of magnitude
larger. The results of the fits compared to experimental data are
presented in Fig.~\ref{fig:cs}, where every solution is 
represented by a distinct color.
\begin{figure}[ht!]
\begin{center}
\includegraphics[width=\linewidth]{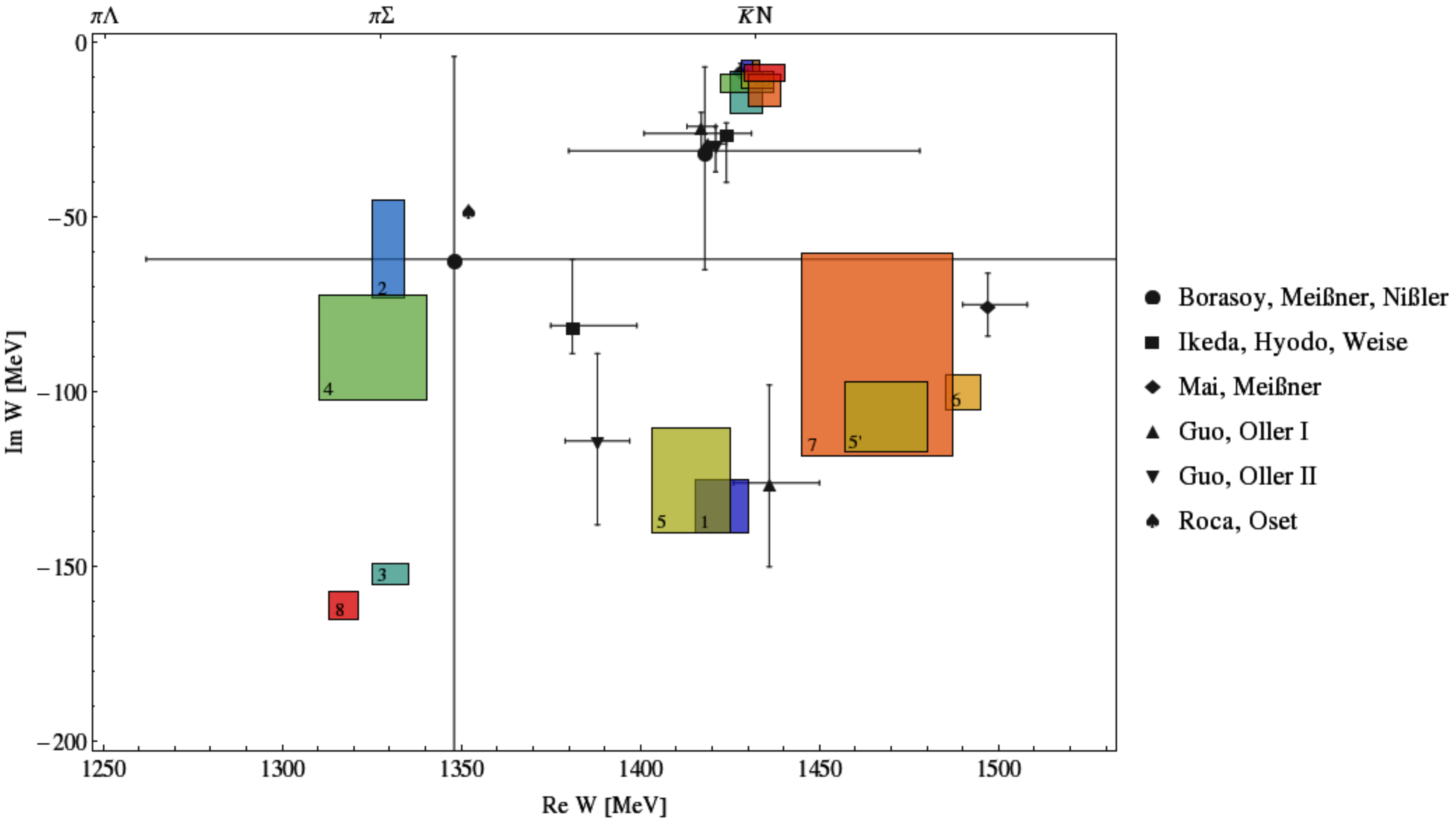}
\centerline{\parbox{0.80\textwidth}{
\caption{Double pole structure of the $\Lambda(1405)$ in the
        complex energy plane for the eight solutions that
        describe the scattering and the SIDDHARTA data. For
        easier reading, we have labeled the second pole of
        these solutions by the corresponding fit \#, where
        $5$ and $5'$ denote the second pole on the second
        Riemann sheet, connected to the real axis between the
        $\pi\Sigma-\bar KN$ and $\bar KN-\eta\Lambda$
        thresholds, respectively. For comparison, various
        results from the literature are also shown, see
        Refs.~\protect\cite{Borasoy:2006srMM,Roca:2013avMM,
        Mai:2012dtMM,Ikeda:2012auMM,Guo:2012vvMM}. }
        \label{fig:poles1} } }
\end{center}
\end{figure}

The data are described equally well by all eight solutions,
showing, however, different functional behaviour of the cross
sections as a function of $P_{\rm lab}$. When continued
analytically to the complex $W$ plane, all eight solutions
confirm the double pole structure of the $\Lambda(1405)$,
see Fig.~\ref{fig:poles1}. There, the narrow pole lies on the
Riemann sheet, connected to the real axis between the
$\pi\Sigma-\bar KN$ thresholds for every solution. The second
poles lie on the Riemann sheets, connected to the real axis
between the following thresholds: $\pi\Sigma-\bar KN$ for
solution 1, 2, 4, 5  and 8; $\pi\Lambda-\pi\Sigma$ for solution
3; $\bar KN-\eta\Lambda$ for solutions 6 and 7. Please note
that the second pole of the solution 5 has a shadow pole (5'
in Fig.~\ref{fig:poles1}) on the Riemann sheet, connected to
the real axis between $\bar KN-\eta\Lambda$ thresholds. The
scattering amplitude is restricted around the $\bar KN$
threshold by the SIDDHARTA measurement quite strongly.
Therefore, in the complex $W$ plane we observe a very stable
behaviour of the amplitude at this energy, {\it i.e.}, the 
position of the narrow pole agrees among all solutions within 
the $1\sigma$ parameter errors, see Fig.~\ref{fig:poles1}. 
This is in line with the findings of other
groups~\cite{Borasoy:2006srMM,Ikeda:2012auMM,Guo:2012vvMM}, {\it 
i.e.}, one observes stability of the position of the narrow 
pole. Quantitatively, the first pole found in these models 
is located at somewhat lower energies and is slightly broader
than those of our model. In view of the stability of the pole
position, we trace this shift to the different treatment of
the Born term contributions to the chiral potential utilized
in Refs.~\cite{Borasoy:2006srMM,Ikeda:2012auMM,Guo:2012vvMM}.

The position of the second pole is, on the other hand, less
restricted. To be more precise, for the real part we find
three clusters of these poles: around the $\pi\Sigma$
threshold, around the $\bar K N$ threshold as well as around
$1470$~MeV. For several solutions, there is some agreement in
the positions of the second pole between the present analysis
and the one of Ref.~\cite{Guo:2012vvMM} and of our previous
work~\cite{Mai:2012dtMM}. However, as the experimental data is
described similarly well by all fit solutions, one can not
reject any of them. Thus, the distribution of poles represents
the systematic uncertainty of the present approach.  It
appears to be quite large, but is still significantly smaller
than the older analysis of Ref.~\cite{Borasoy:2006srMM}. Recall
that no restrictions were put on the  parameters of the model,
except for naturalness.

\item \textbf{Photoproduction Amplitude}

We have demonstrated above that the present model for the
meson-baryon interaction possesses at least eight different
solutions, which all describe the hadronic data similarly
well. In this Section, we wish to see whether these solutions
are compatible with the photoproduction data, if they are
considered as a final-state interaction of the reaction
$\gamma p \to K^+\Sigma \pi$. For this purpose it is
sufficient to consider the simple ansatz
\begin{equation}\label{eq:photo}
        \mathcal M^j(\tilde W,M_{\rm inv}) = \sum_{i=1}^{10}
        C^i(\tilde W)\,G^i(M_{\rm inv})\,f_{0+}^{i,j}(M_{\rm inv})\,,
\end{equation}
where $\tilde W$ and $M_{\rm inv}$ denote the total energy of
the system and the invariant mass of the $\pi\Sigma$ subsystem,
respectively. For a specific meson-baryon channel $i$, the
energy-dependent (and in general complex valued) constants
$C^i(\tilde W)$ describe the reaction mechanism of $\gamma
p\to K^+M_iB_i$, where\-as the final-state interaction is
captured by the standard H\"oh\-ler partial waves $f_{0+}$.
For a specific meson-baryon channel $i$, the Greens function
is denoted by $G^i(M_{\rm inv})$ and is given by the one-loop
meson baryon function in dimensional regularization.

The regularization scales appearing in the Eq.~(\ref{eq:photo})
via the $G^i(M_{\rm inv})$ have already been fixed in the fit
to the hadronic cross sections and the SIDDHARTA data.  Thus,
the only new parameters of the photoproduction amplitude are
the constants $C^i(\tilde W)$ which, however, are quite numerous
(10 for each $\tilde W$). These parameters are adjusted to
reproduce the invariant mass distribution $d\sigma/dM_{\rm
inv}(M_{\rm inv})$ for the final $\pi^+\Sigma^-$, $\pi^0
\Sigma^0$ and $\pi^-\Sigma^+$ states and for all 9 measured
total energy values $\tilde W=2.0,\,2.1,\,\ldots,\,2.8$~GeV.
The achieved  quality  of the photoproduction fits is listed
in the third row of Tab.~\ref{tab:photo}, whereas the $\chi_{\rm
d.o.f.}^2$ of the hadronic part are stated in the second row.
Note that for the comparison of the photoproduction fits the
quantity $\chi_{\rm d.o.f.}^2$ is not meaningful due to the
large number of generic parameters $C_i(\tilde W)$. Therefore,
we compare the total $\chi^2$ divided by the total number of
data points for all three $\pi\Sigma$ final states, denoted
by $\chi_{\rm p.p.}^2$. For the same reason it is not
meaningful to perform a global fit, minimizing the total
$\chi_{\rm d.o.f.}^2$.

It turns out that even within such a simple and flexible
photoproduction amplitude, only the solutions~\#2, \#4 and
\#5 of the eight hadronic solutions allow for a decent
description of the CLAS data. While the total $\chi^2$ per
data point of these solutions is very close to each other,
the next best solution has a 40\% larger total $\chi^2_{\rm
p.p.}$ than the best one.

We have checked this statement for a large number of hadronic
solutions randomly distributed within $1\sigma$ band around
the central ones. For every such solution a fit to the CLAS
data was performed independently and no significantly better
fit was found to those of the central solution. Therefore, we
consider the above exclusion principle of the hadronic solutions
as statistically stable. For further discussion on this aspect
see Ref.~\cite{Mai:2014xnaMM}.

The best solution is indeed \#4, which, incidentally, has also
the lowest $\chi_{\rm d.o.f.}^2$ for the hadronic part. This
solution also gives an excellent description of the $\Sigma
\pi\pi$ mass distribution from Ref.~\cite{Hemingway:1984pzMM},
calculated using the method developed in Ref.~\cite{Oller:2000fjMM}.
With respect to these data, solution \#2 is also satisfactory
but \#5 is not. Therefore, the photoproduction data combined
with the scattering and the SIDDHARTA data lead to a sizable
reduction in the ambiguity of the second pole of the
$\Lambda(1405)$. The locations of the two poles in these
surviving solutions are $(1434^{+2}_{-2} - i \,
10^{+2}_{-1})$~MeV  ($(1330^{+4~}_{-5~} - i \,
56^{+17}_{-11})$~MeV) and $(1429^{+8}_{-7} - i \,
12^{+2}_{-3})$~MeV ($(1325^{+15}_{-15} - i \,
90^{+12}_{-18})$~MeV) for the first (second) pole of the
solution \#2 and \#4, respectively. In fact, the second pole
of the surviving solutions is close to the value found in
Ref.~\cite{Roca:2013avMM}, see Fig.~\ref{fig:poles1}, and
also close to the central value of the analysis based on
scattering data only~\cite{Borasoy:2006srMM}.

We conclude that the inclusion of the CLAS data as experimental
input can serve as a new important constraint on the
antiKaon-nucleon scattering amplitude. However, for future
studies a theoretically more robust model for the two-meson
photoproduction amplitude is required. We propose that a
generalization of the one-meson photoproduction model,
presented in Ref.~\cite{Borasoy:2007kuMM,Mai:2012wyMM}, may be
the next logical step for this endeavor.

\item \textbf{New Scattering (Pseudo-) Data}

\begin{figure}[ht!]
\begin{center}
\includegraphics[width=\linewidth]{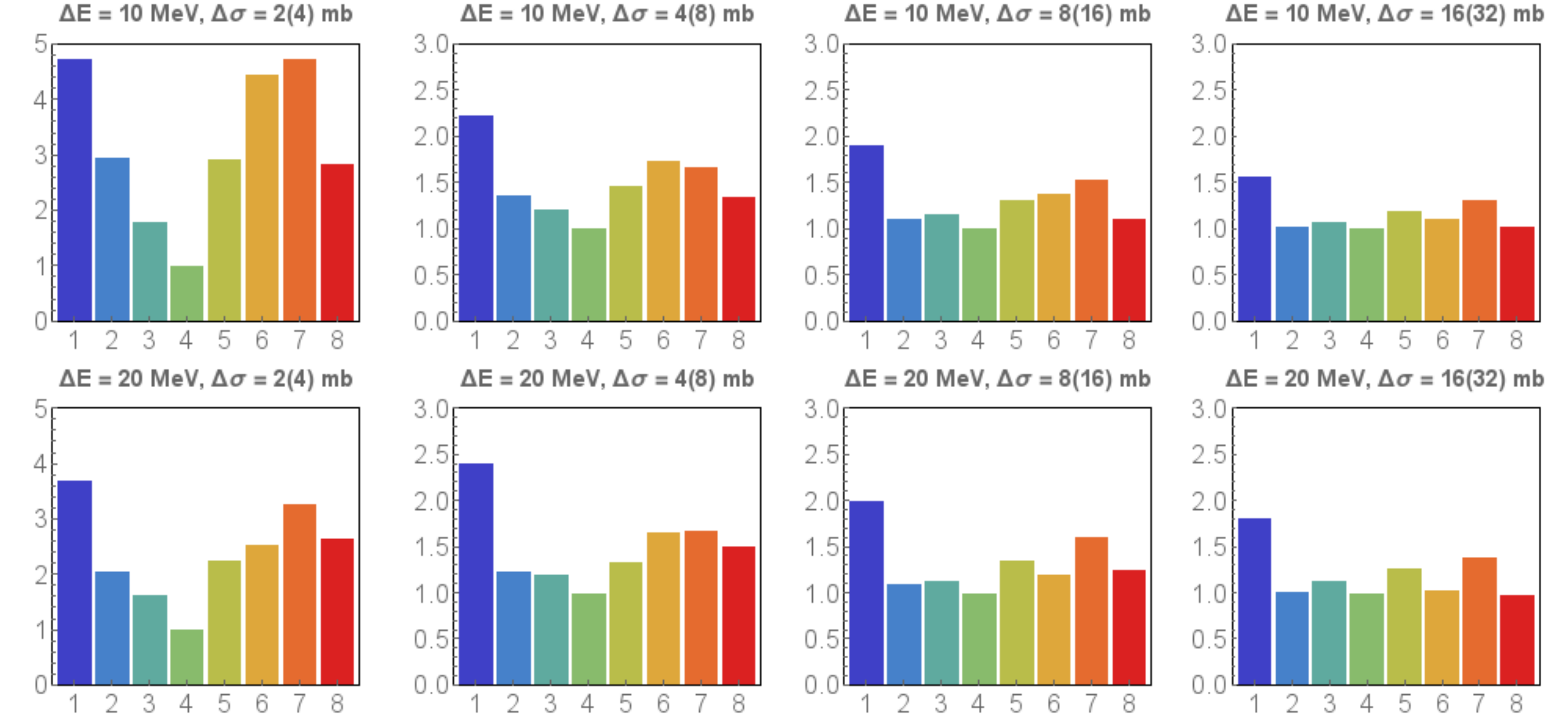}
\centerline{\parbox{0.80\textwidth}{
        \caption{Comparison of $\chi^2_{\rm d.o.f.}$ for all
        8 solutions and all available as well as pseudo
        scattering data. The values are normalized by the
        $\chi^2_{\rm d.o.f.}$ of solution \#4, which is used
        to generate the pseudo-data. Each plot label denotes
        the assumed energy binning $\Delta E$ and measurement
        uncertainty $\Delta\sigma$ for the charged (neutral)
        final states.} \label{pic:CROSSCHECK} } }
\end{center}
\end{figure}

In the last Section, we have demonstrated that modern data indeed
allows to put additional constraints on the antiKaon-nucleon
scattering amplitude. To use these data we have used a very
simple ansatz for the two-meson photoproduction amplitude.
Another and actually more direct way to put new constraints on
the scattering amplitude is to improve the (currently very old)
cross section data on $K^-p\to ...$ as proposed in, {\it e.g.},
Ref.~\cite{Briscoe:2015qiaMM} using Kaon beams at JLab. In the
following we study the impact of such, in the future available
data if used complementary to the already available data. In
particular, we will estimate the minimal resolution required
for such new data to be capable to rule out some of our 8
solutions.

We start from generating realistic pseudo-data, without
discussing further the details of such measurements. For this
we assume our best solution (\#4) to be a realistic one and
calculate total cross sections in all six channels $K^-p\to
MB$, where $MB\in \{K^-p, \bar K^0n, \pi^0\Lambda, 
\pi^+\Sigma^-$, $\pi^0\Sigma^0, \pi^-\Sigma^+\}$ in the energy 
range $P_{lab}= 100\ldots300$~MeV for various values of the 
energy resolution $\Delta E$. To account for the uncertainty 
of the new measurement, we assume several values $\Delta\sigma$. 
Since neutral channels are usually more complicated to measure 
the uncertainty $\Delta\sigma$ is assumed to be twice as large 
as in the charged channels. Finally, realistic pseudo-data is 
obtained as a random value around the central one (predicted 
by the solution \#4) normally distributed with the standard 
deviation of $\Delta\sigma$.

For the fixed parameter sets of the model we calculate the new
$\chi^2_{\rm d.o.f.}$ using all available data together with
the new pseudo-data. The results of such a test for different
values of $\Delta E$ and $\Delta \sigma$ are depicted in
Fig.~\ref{pic:CROSSCHECK}. There, the individual values
$\chi^2_{\rm d.o.f.}$ are normalized to the one of the solution
\#4, which is used to generate  pseudo-data. It is seen that
even at quite low energy resolution of $\Delta E=20$~MeV already
3 solutions have twice as large $\chi^2_{\rm d.o.f.}$ as the one
of solution \#4. Thus such solutions could presumably be ruled
out by the new data. On the other hand, it looks like given a
too large measurement uncertainty $\Delta\sigma$ none of the
above solutions can be ruled out that easily. Therefore, we
conclude from this very preliminary and qualitative study that
for the new data to be restrictive enough the desired measurement
precision should be $\Delta\sigma\lesssim4(8)$~mb for the
charged (neutral) final states. The energy resolutions seems to
play a minor role.

\item \textbf{Acknowledgments}

The speaker thanks the organizers and in particular Igor
Strakovsky for the invitation to this very inspiring workshop.
The work of the speaker was performed with Ulf-G. Mei{\ss}ner
and supported in part by the DFG and the NSFC through funds
provided to the Sino-German CRC~110 ``Symmetries and the
Emergence of Structure in QCD" (NSFC Grant No.~11261130311).
\end{enumerate}


\newpage
\subsection{Opportunities  in the Hyperon Spectrum with Neutral 
	Kaon Beam}
\addtocontents{toc}{\hspace{2cm}{\sl V.~Mathieu}\par}
\setcounter{figure}{0}
\setcounter{footnote}{0}
\setcounter{equation}{0}
\halign{#\hfil&\quad#\hfil\cr
\large{Vincent Mathieu}\cr
\textit{Center for Exploration of Energy and Matter}\cr
\textit{Indiana University}\cr
\textit{Bloomington, IN 47403, U.S.A.}\cr}

\begin{abstract}
In this talk, I presented the features of the webpage of 
our model for $\bar K N$ scattering. The future directions 
concering these reactions were also discussed. 
\end{abstract}

\begin{enumerate}
\item 

As explained in these Proceedings,
\footnote{See the contribution, in these Proceedings, by 
C\'esar Fern\'andez-Ram\'irez and Adam Szczepaniak.} 
the hyperon spectrum is of importance for the understanding of 
the strong interaction. In Ref.~\cite{Fernandez-Ramirez:2015tfaT}, 
we presented a unitary multichannel model for $\bar KN$ 
scattering in the resonance region that fulfills unitarity. 
This project has led to deliverables (such as the partial waves 
or codes for the various observables). We decided to create an 
interactive webpage~\cite{webpageT}, where the practitioners can 
download and simulate online our models.
\footnote{Other projects perfomed by the Joint Physics Analysis 
Center are also availble online. A short description of these 
projects are presented in Ref.~\cite{Mathieu:2016mcyT}.} 

Several coupled channels, indicated in the publication, were 
considered in the fitting procedure. In the JPAC webpage, 
the observables and partial waves for the following channels 
can be computed 
\begin{eqnarray}\nonumber
	K^-p &\to& K^-p,\ \bar K^0n, \quad
	\pi^-\Sigma^+,\ \pi^+\Sigma^-,\ \pi^0\Sigma^0,
	\quad  \pi^0 \Lambda.
\end{eqnarray}
All observables, differential cross section $d\sigma/dz_s$, 
polarization observable $P$ and total cross section $\sigma$, 
are expressed in terms of the spin-non-flip $f(s,z_s)$ and 
the spin-flip $g(s,z_s)$ amplitudes with the relations
\begin{equation}
	\frac{d\sigma}{dz_s}(s,z_s)  = \frac{1}{q^2} \left[ 
	|f(s,z_s)|^2 + |g(s,z_s)|^2 \right], \quad\nonumber\\
\end{equation}
\begin{equation}
	P\frac{d\sigma}{dz_s}(s,z_s)  = \frac{2}{q^2}  
	{Im}\left[ f(s,z_s)g^*(s,z_s) \right],  \quad
	\sigma(s)  = \int_{-1}^1 \frac{d\sigma}{dz_s}(s,z_s)\ d z_s .
\end{equation}
For a given channel (the channel index is omitted), the 
amplitudes admit a partial wave expansion
\begin{eqnarray}
	f(s,z_s)  &=& \sum_{\ell=0}^{\infty} \left[ (\ell+1) 
	R_{\ell+}(s) + \ell R_{\ell-}(s)\right] P_\ell(z_s), \\
	g(s,z_s)  &=& \sum_{\ell=1}^{\infty} \left[ R_{\ell+}(s) 
	- R_{\ell-}(s)\right] \sqrt{1-z_s^2}P'_\ell(z_s).
\end{eqnarray}
In a given meson-baryon channel $\ell$ labels the relative 
orbital angular momentum and the total angular momentum is 
given by $J = \ell \pm 1/2$. For a detailed relation, in all 
channels, between the orbital momentum and the partial waves 
we refer the reader to Ref.~\cite{Fernandez-Ramirez:2015tfaT}.  
\begin{figure}[htb!]
\begin{center}
\includegraphics[width=0.49\linewidth]{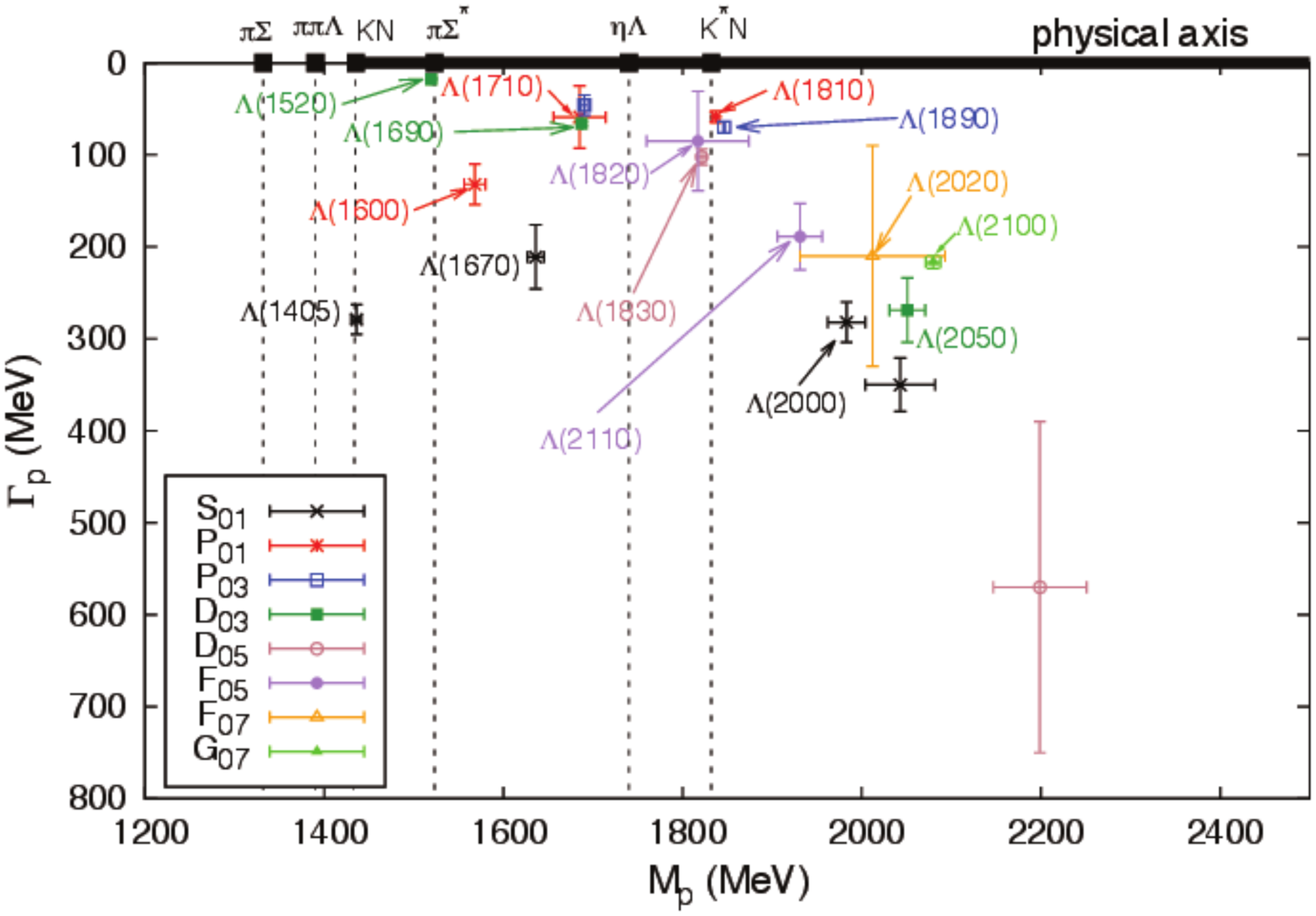}
\includegraphics[width=0.49\linewidth]{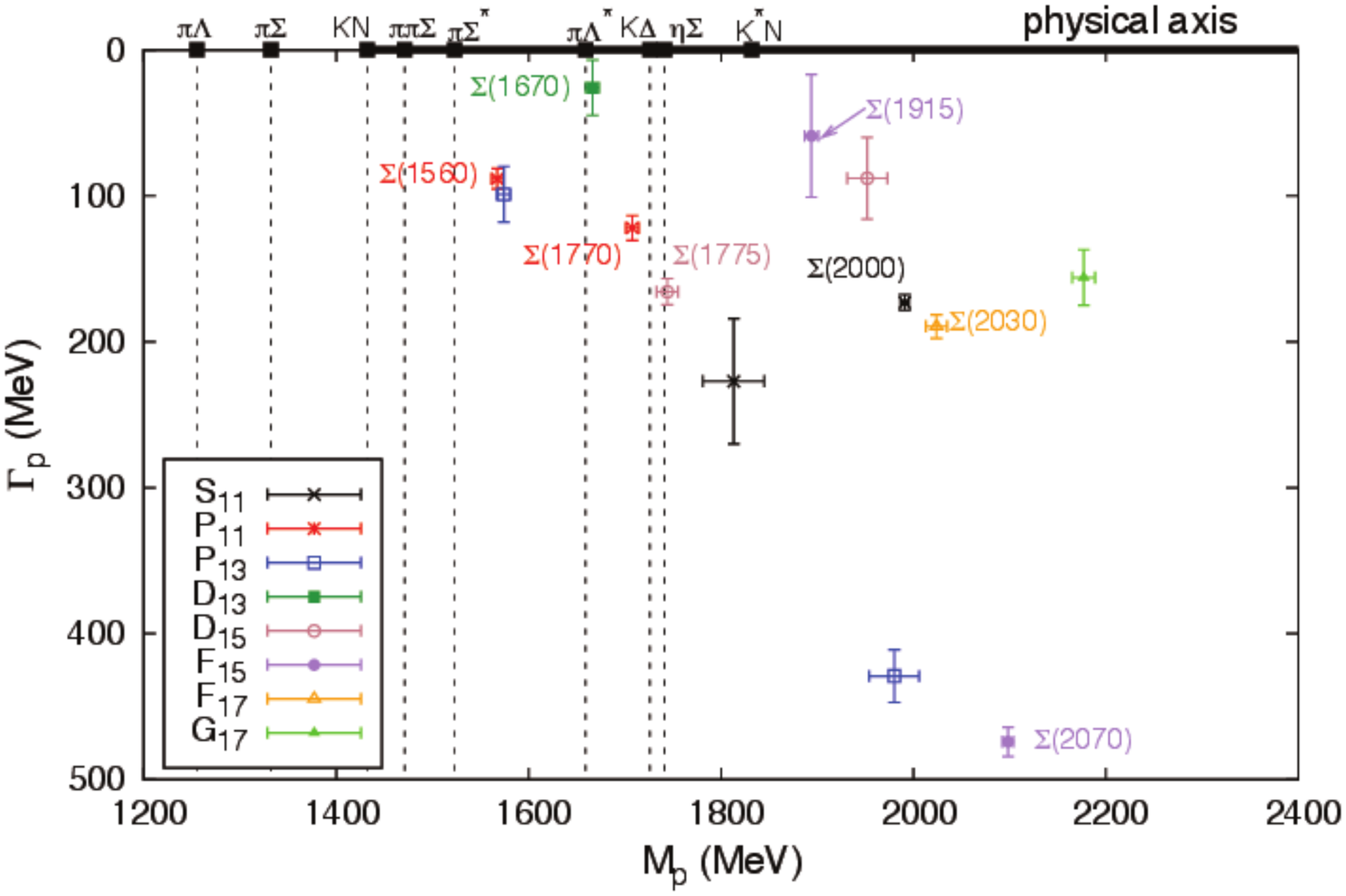}
\end{center}
\vspace{-0.5cm}
\centerline{\parbox{0.80\textwidth}{
 \caption{Spectrum of the $\Lambda$ (I=0) and $\Sigma$ 
	(I=1) baryons from 
	Ref.~\protect\cite{Fernandez-Ramirez:2015tfaT}.}       
	\label{fig:KN} } }
\end{figure}

After removing the barrier factor, the partial waves are 
parametrized with a $K-$matrix to ensure proper 2-body 
unitarity in the resonance region. The $K-$matrix is the sum 
of the resonance contributions and a empirical background 
term. Each wave is parametrized and fitted independently. 
The detailed procedure is described in 
Ref.~\cite{Fernandez-Ramirez:2015tfaT}. Finally the partial 
waves are analytically continue on the unphysical sheet and 
the pole positions are extracted. The resulting spectrum 
for $\Lambda$ and $\Sigma$ baryons is displayed on 
Figure~\ref{fig:KN}. 

The partial waves, binned in energy supplied by the user, 
can be dowloaded online . The Fortran code yielding the 
partial is also available. The differential cross section 
(together with the polarization) and the total cross section 
have also their dedicated pages. The codes for producing the 
observables can be both simulated online and dowloaded. 
\begin{figure}[htb!]
\begin{center}
\includegraphics[width=\linewidth]{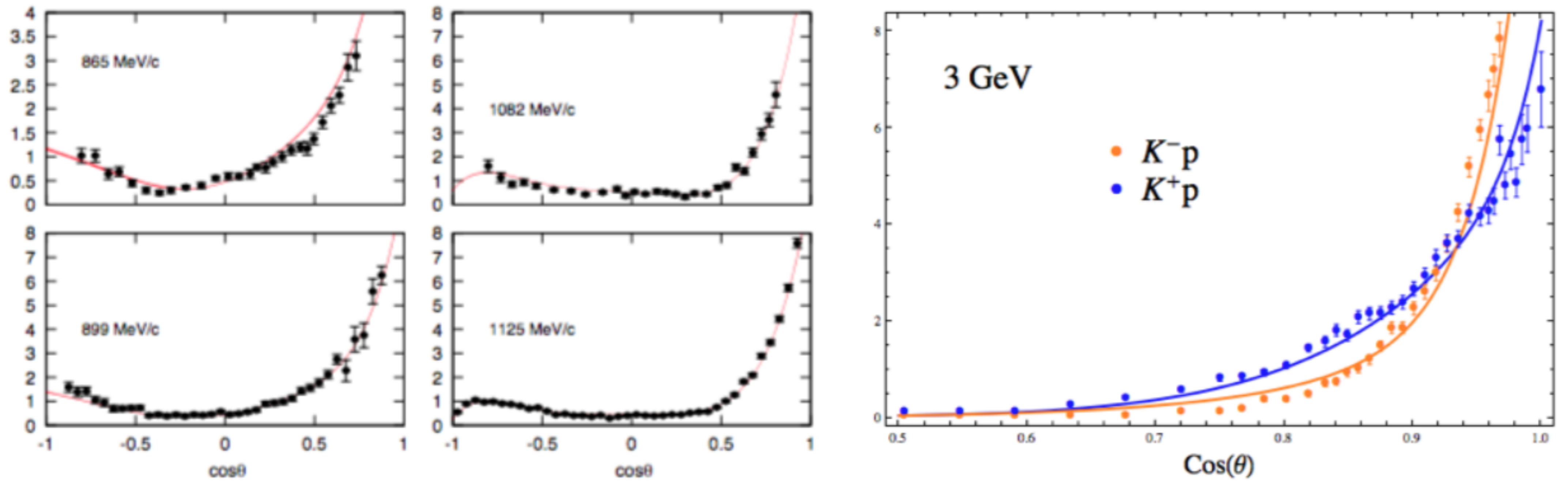}
\end{center}
\vspace{-0.5cm}
\centerline{\parbox{0.80\textwidth}{
 \caption {$K^-p\to K^-p$ differential cross section 
	from Ref.~\protect\cite{Fernandez-Ramirez:2015tfaT} 
	(left) and from Ref.~\protect\cite{moiT} 
	(right).} \label{fig:KN2} } }
\end{figure}

The differential cross section peaks in the forward 
direction as the energy increases as can be seen on 
Fig.~\ref{fig:KN2}. This is characteristic of the Regge 
poles. Indeed at high energy, the reaction is driven by 
singularities in the complex angular momentum plane. Those 
Regge poles display an exponential suppression in the 
momentum transferred squared $t= - 2q^2(1-\cos\theta)$. 
The smooth continuation from the resonance region to the 
Regge region is suggested on Fig.~\ref{fig:KN2}. 

We can formalize this phenomenon by a mathematical relation 
between the resonances and the Regge poles, the so-called 
finite energy sum rules.  An example of these sum rules read 
(for the notation, their derivation and their applications 
to pion-nucleon scattering, see Ref.~\cite{Mathieu:2015gxaT}):
\begin{equation}
	\frac{1}{\Lambda} \int_0^{\Lambda} \text{Im}\, 
	A(\nu,t) d\nu = \frac{\beta(t)\Lambda^{\alpha(t)}}
	{\alpha(t)+1} .
\end{equation}
The left-hand side consists in an integration over the 
resonance region. The right-hand side is determined by the 
residue $\beta(t)$ and the trajectory $\alpha(t)$ of the 
Regge pole(s) contributing to the scalar amplitudes 
$A(\nu,t)$ ($\nu = (s-u)/2$ being the crossing variable).
In practice, that means that the high energy data can be 
used to constrain the parameters of the fit in the 
resonances region. Since the energy range for the kaon 
long beam in this proposal extend above the resonance 
region, one could fully exploit the data and the finite 
energy sum rules to better constrain the hyperon spectrum. 

\item \textbf{Acknowledgments}

This material is based upon work supported in part by the 
U.S. Department of Energy, Office of Science, Office of 
Nuclear Physics under contract DE--AC05--06OR23177. This 
work was also supported in part by the U.S. Department of 
Energy under Grant No. DE--FG0287ER40365 and National 
Science Foundation under Grants PHY-1415459 and 
NSF--PHY--1205019.
\end{enumerate}


\newpage
\subsection{Establishing S = -1 Hyperon Resonances Using Kaon-Induced 
	Meson Productions within Dynamical Coupled-Channels Approach}
\addtocontents{toc}{\hspace{2cm}{\sl H.~Kamano}\par}
\setcounter{figure}{0}
\halign{#\hfil&\quad#\hfil\cr
\large{Hiroyuki Kamano}\cr
\setcounter{equation}{0}
\textit{Research Center for Nuclear Physics}\cr
\textit{Osaka University}\cr
\textit{Ibaraki, Osaka 567-0047, Japan}\cr}

\begin{abstract}
We give an overview of our recent effort for the spectroscopy of 
strangeness $S=-1$ hyperon resonances ($Y^\ast$), which has been 
made through a comprehensive partial-wave analysis of the $K^-p\to 
\bar KN, \pi\Sigma, \pi\Lambda, \eta\Lambda,K\Xi$ reactions within 
a dynamical coupled-channels (DCC) approach. It is found that the 
existing $K^-p$ reaction data are not sufficient to unambiguously 
determine partial-wave amplitudes and properties (pole masses and 
residues, etc.) of $Y^\ast$ resonances.  We then discuss what new 
data are actually needed for further establishing $Y^\ast$ 
resonances.
\end{abstract}

\begin{enumerate}
\item \textbf{Introduction}

So far, a number of $\Lambda^\ast$ and $\Sigma^\ast$ resonances 
with strangeness $S=-1$ (collectively referred to as $Y^\ast$)
have been reported as listed by Particle Data Group 
(PDG)~\cite{pdg2014B}.  However, those are much less understood 
than the nonstrange $N^\ast$ and $\Delta^\ast$ resonances. For 
example, most of the $\Sigma^\ast$ resonances are poorly 
established.  In fact, just 6 out of 26 reported $\Sigma^\ast$ 
resonances are rated as ``four-star" by PDG, and unlike 
the $N^\ast$ and $\Delta^\ast$ resonances, even the existence 
of low-lying resonances is still uncertain.  Furthermore, the 
spin and parity quantum numbers have not been determined for 
quite a few $Y^\ast$ resonances~\cite{pdg2014B}. 

Another issue that should be noted for $Y^\ast$ resonances is 
that until very recently, only the so-called Breit-Wigner mass 
and width were listed by PDG with a few exception (see, {\it e.g.},
2012 edition of PDG~\cite{pdg2012B}).  This is also in contrast 
to the $N^\ast$ and $\Delta^\ast$ cases, where resonances 
defined by poles of scattering amplitudes have also been 
extensively studied, and both of the pole and Breit-Wigner 
results have been listed by PDG for a long time.  It is known 
(see, {\it e.g.}, Ref.~\cite{resoB}) that a resonance mass 
defined by the pole of scattering amplitudes is equivalent to 
an exact (complex-)energy eigenvalue of the {\it full} 
Hamiltonian of the underlying theory, namely the Quantum 
Chromodynamics (QCD) in this case, under the purely outgoing 
boundary condition. Thus extracting resonances defined by poles 
from reaction data is essential to testing QCD in the confinement 
domain, and such a test is now becoming reality with the help of 
the Lattice QCD simulations (see, {\it e.g.}, Refs.~\cite{dudekB,
wuB,molinaB,leeB}). 

In this situation, a first attempt of a comprehensive and 
systematic partial-wave analysis of $K^-p$ reactions to extract 
$Y^\ast$ resonances defined by poles was accomplished by the 
Kent State University (KSU) group in 2013~\cite{ksu1B,ksu2B}, and 
then by our group using the dynamical coupled-channels (DCC) 
approach~\cite{knlskp1B,knlskp2B}. (Recently, a reanalysis of the 
KSU single-energy solution~\cite{ksu1B} using an on-shell 
$K$-matrix approach has been done in Ref.~\cite{ksu3B}.)  Here 
it is emphasized again that it is only in recent years that 
this kind of comprehensive study to extract pole information 
from reaction data began for the $Y^\ast$ resonances.

The basic formula of our DCC approach~\cite{msl07B,knls13B,knlskp1B} 
is the coupled-channels integral equation for the partial-wave 
amplitudes: 
\begin{equation}
	T^{(J^P I)}_{b,a} (p_b,p_a;W) = 
	V^{(J^PI)}_{b,a} (p_b,p_a;W)
	+\sum_c \int dp_c\,p_c^2 V^{(J^PI)}_{b,c} (p_b,p_c;W) 
	G_c(p_c;W) T^{(J^PI)}_{c,a} (p_c,p_a;W) \,.
	\label{lseq}
\end{equation}
Here, the subscripts ($a,b,c$) represent the meson-baryon 
channels we have considered, {\it i.e.}, $\bar KN$, $\pi\Sigma$, 
$\pi\Lambda$, $\eta\Lambda$, $K\Xi$, $\pi\Sigma^\ast$, and $\bar 
K^\ast N$, where the last two are the quasi-two-body channels that 
subsequently decay into the three-body $\pi\pi\Lambda$ and $\pi\bar 
KN$ channels, respectively; $V^{(J^PI)}_{b,a} (p_b,p_a;W)$ denotes 
the potential driving the transition from the channel $a$ to the 
channel $b$; and $G_c(p_c;W)$ denotes the Green's function for the 
channel $c$. In our approach, the transition potential consists of 
hadron-exchange diagrams derived from effective Lagrangians. By 
solving Eq.~(\ref{lseq}), one can sum up all the possible 
transition processes between reaction channels considered, and 
this ensures the multichannel two-body as well as three-body 
unitarity for the resulting amplitudes.  Furthermore, off-shell 
rescattering effects, which are usually neglected in on-shell 
approaches, are also taken into account properly through the 
momentum integral appearing in the right hand side of 
Eq.~(\ref{lseq}). Actually, these features make our model quite 
unique among existing models of meson production reactions.

In this contribution, we first give an overview of our recent 
efforts for the $Y^\ast$ spectroscopy through the comprehensive 
partial-wave analysis of the $K^-p$ reactions within our DCC 
approach in Sec.~2. Then, in Sec.~3, we discuss and give 
prospects for what new data for anti-Kaon induced reactions are 
needed for further establishing $Y^\ast$ resonances.

\item \textbf{Results of DCC Analysis for $K^-p$ Reactions}

In Ref.~\cite{knlskp1B}, we have constructed models for the $S=-1$ 
sector within the DCC approach by analyzing the data of $K^-p\to 
\bar KN, \pi\Sigma, \pi\Lambda, \eta\Lambda$, and $K\Xi$ reactions 
up to $W =2.1$~GeV.  The analysis takes into account all available 
data for both unpolarized and polarized observables as far as we 
found in the considered energy region, and it results in fitting 
to more than 17,000 data points.  From this analysis, we have 
determined the partial-wave amplitudes for the $K^-p\to\bar KN, 
\pi\Sigma, \pi\Lambda, \eta\Lambda$, and $K\Xi$ reactions not only 
for $S$ wave but also higher partial waves including $P$, $D$, and 
$F$ waves. Furthermore, the threshold parameters such as scattering 
lengths and effective ranges have also been determined for the 
$\bar KN$, $\eta\Lambda$, and $K\Xi$ scatterings. We then extracted 
in Ref.~\cite{knlskp2B} the parameters associated with $Y^\ast$ 
resonances such as mass, width, and coupling constants defined by 
poles of scattering amplitudes within our constructed models.
\begin{figure}[ht!]
\begin{minipage}{0.3\textwidth}
\includegraphics[clip,width=\textwidth]{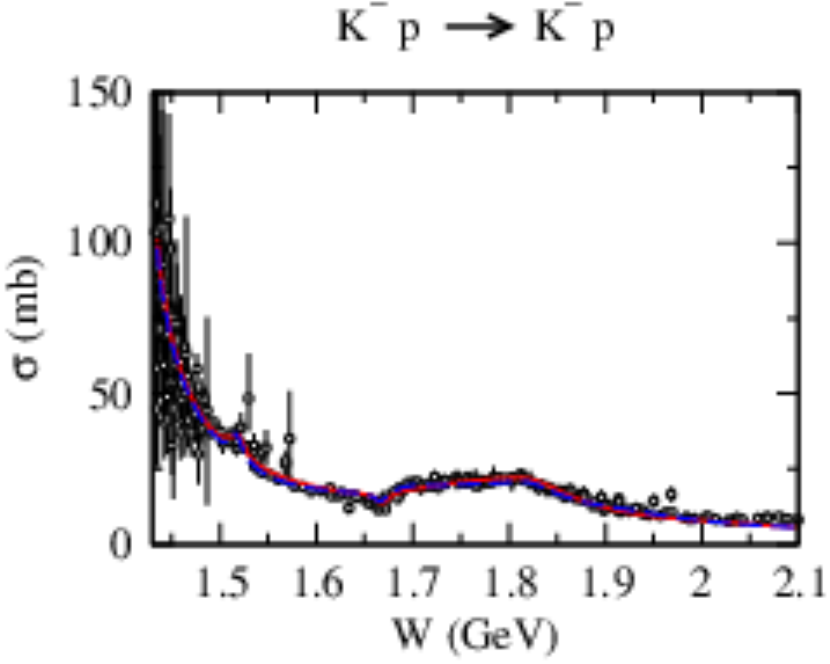}
\end{minipage}
\qquad
\begin{minipage}{0.33\textwidth}
\includegraphics[clip,width=\textwidth]{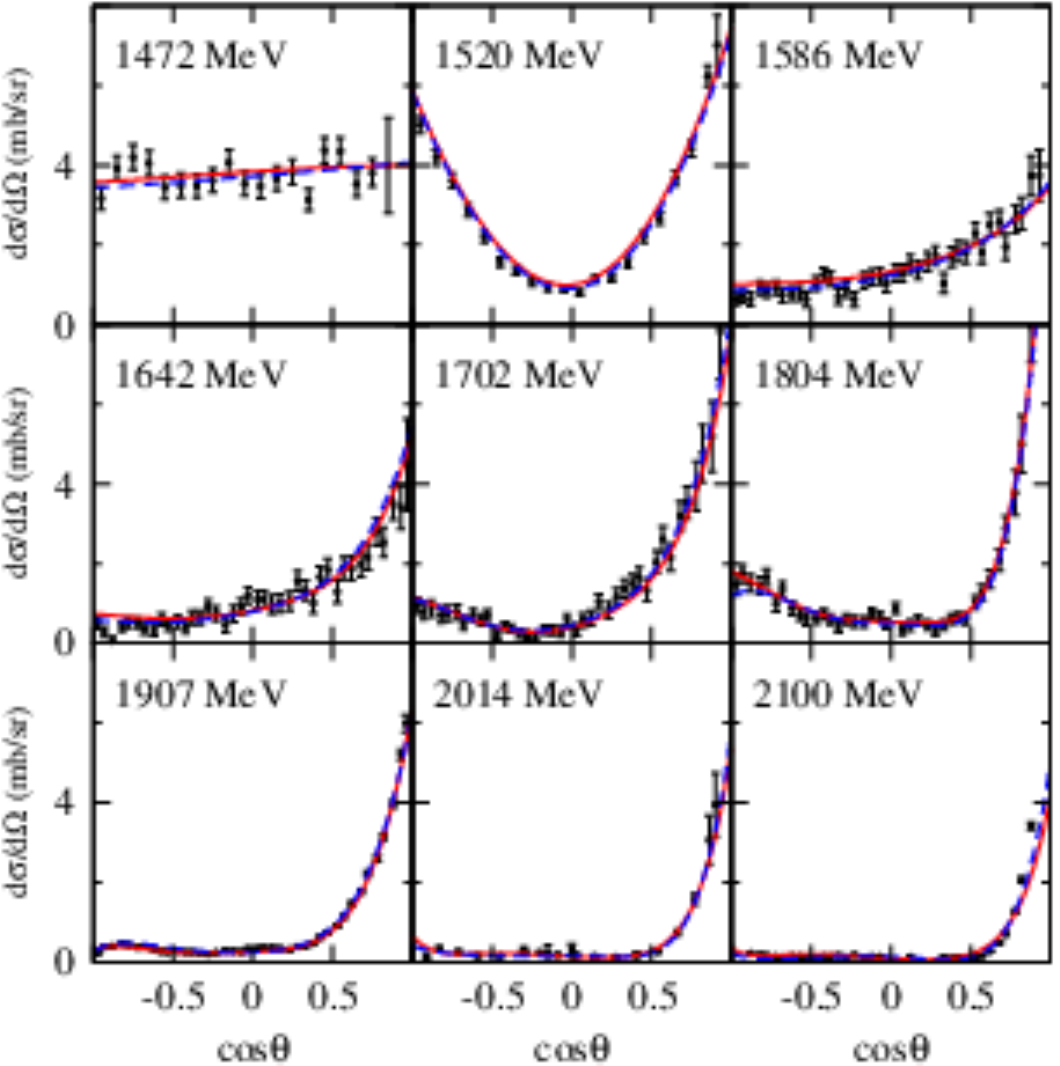}
\end{minipage}
\qquad
\begin{minipage}{0.22\textwidth}
\includegraphics[clip,width=\textwidth]{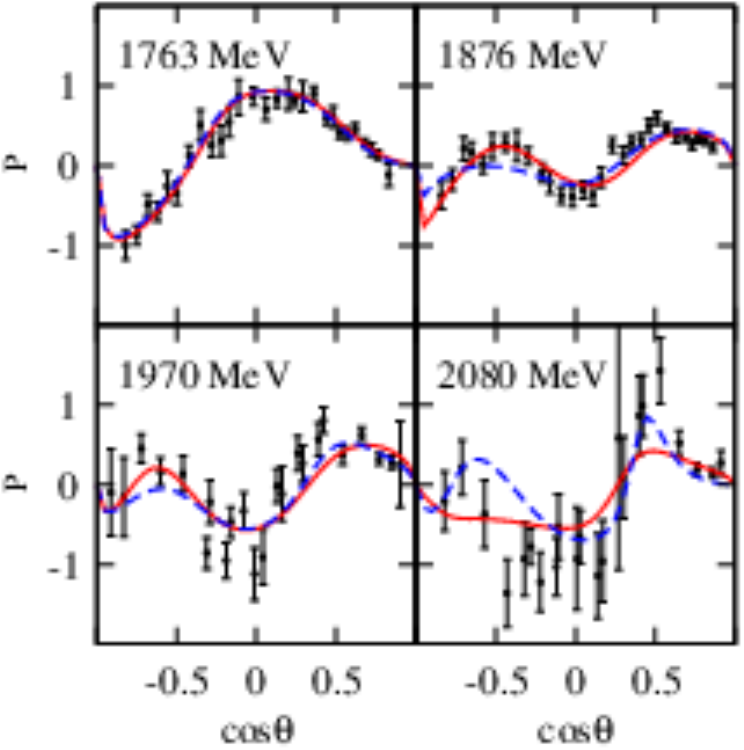}
\end{minipage}
\centerline{\parbox{0.80\textwidth}{
\caption{Several results of our fits to the data for $K^-p\to K^- 
	p$ scattering from Ref.~\protect\cite{knlskp1B}.
	Total cross section $\sigma$ (the leftest panel), 
	differential cross section $d\sigma/d\Omega$ (middle 
	panels), and recoil polarization $P$ (right panels) 
	are presented. Red solid and blue dashed curves 
	represent the results of our two analyses, Model~A 
	and Model~B, respectively. The references for the 
	data can be found in Ref.~\protect\cite{knlskp1B}.}
	\label{fig:kmpkmp} } }
\end{figure}

In Fig.~\ref{fig:kmpkmp}, we present several results of our fits 
to the data for $K^-p\to K^-p$ (see Refs.~\cite{knlskp1B,knlskp2B} 
for the full details of our analysis). Here it is found that two 
curves (red solid and blue dashed curves) are plotted in each 
panel. As will be discussed later, this is because the available 
$K^-p$ reaction data are not sufficient to constrain our reaction 
model unambiguously, but it allows us to have two distinct sets 
of our model parameters, yet both give almost the same $\chi^2$ 
value.  Thus these two curves may be viewed as a measure of 
ambiguity in our analysis originating from the insufficient 
amount and accuracy of the current existing data. Hereafter we 
call them Model~A and Model~B, respectively.  Our two models 
reproduce not only the total cross sections but also the 
``angle-dependent quantities" such as the differential 
cross section ($d\sigma/d\Omega$) and recoil polarization ($P$)
very well over the entire kinematical region where the data are 
available.  We have confirmed that the other reaction data are 
also well reproduced.

Figure~\ref{fig:spectrum} shows a comparison of extracted $Y^\ast$ 
mass spectra between our two models~\cite{knlskp1B,knlskp2B} and 
the KSU analysis~\cite{ksu1B,ksu2B}.  The extracted masses show an 
excellent agreement for several resonances, but in overall, they 
are still fluctuating between our two models and the KSU analysis.
Again, this is because the existing $K^-p$ reaction data are not 
sufficient to determine the $Y^\ast$ mass spectrum, and without 
new data this level of analysis dependence will not be avoidable.
Although the extracted spectrum is still analysis dependent, we 
found a couple of new $Y^\ast$ resonances that are quite interesting.
One is a new $J^P = 3/2^-$ $\Lambda$ resonance located near the 
$\eta\Lambda$ threshold found in Model~B. The width of this new 
resonance is $\sim 10$~MeV, which is much narrower than usually
expected for light-quark baryon resonances.  As shown in 
Ref.~\cite{knlskp1B}, the contribution of this new resonance is 
hardly seen in most of the reaction observables considered in our 
analysis, but it is turned out that the new resonance is 
responsible for reproducing the concave-up behavior of differential 
cross section for $K^-p\to\eta\Lambda$ near the threshold. Thus 
the angular dependence of the $K^-p\to\eta\Lambda$ differential 
cross section data seems to favor the existence of this narrow 
resonance. Another interesting finding is that Model~B further 
presents a new $J^P = 1/2^-$ $\Lambda$ resonance with the mass 
close to $\Lambda(1520)3/2^-$.  It is often discussed in quark 
models that the $\Lambda(1405)1/2^-$ is the spin partner of 
$\Lambda(1520)3/2^-$. However, from this result, this new 
$S$-wave resonance might be the true spin partner.  It would also 
be worthwhile to mention that a number of low-lying $\Sigma^\ast$ 
resonances located just above the $\bar KN$ threshold are found, 
and those may correspond to the one-star and two-star resonances 
assigned by PDG.  For further confirmation of these interesting 
$Y^\ast$ resonances, however, more extensive and accurate data of 
anti-Kaon induced reactions are definitely needed.
\begin{figure}[ht!]
\begin{center}
\includegraphics[clip,width=0.9\textwidth]{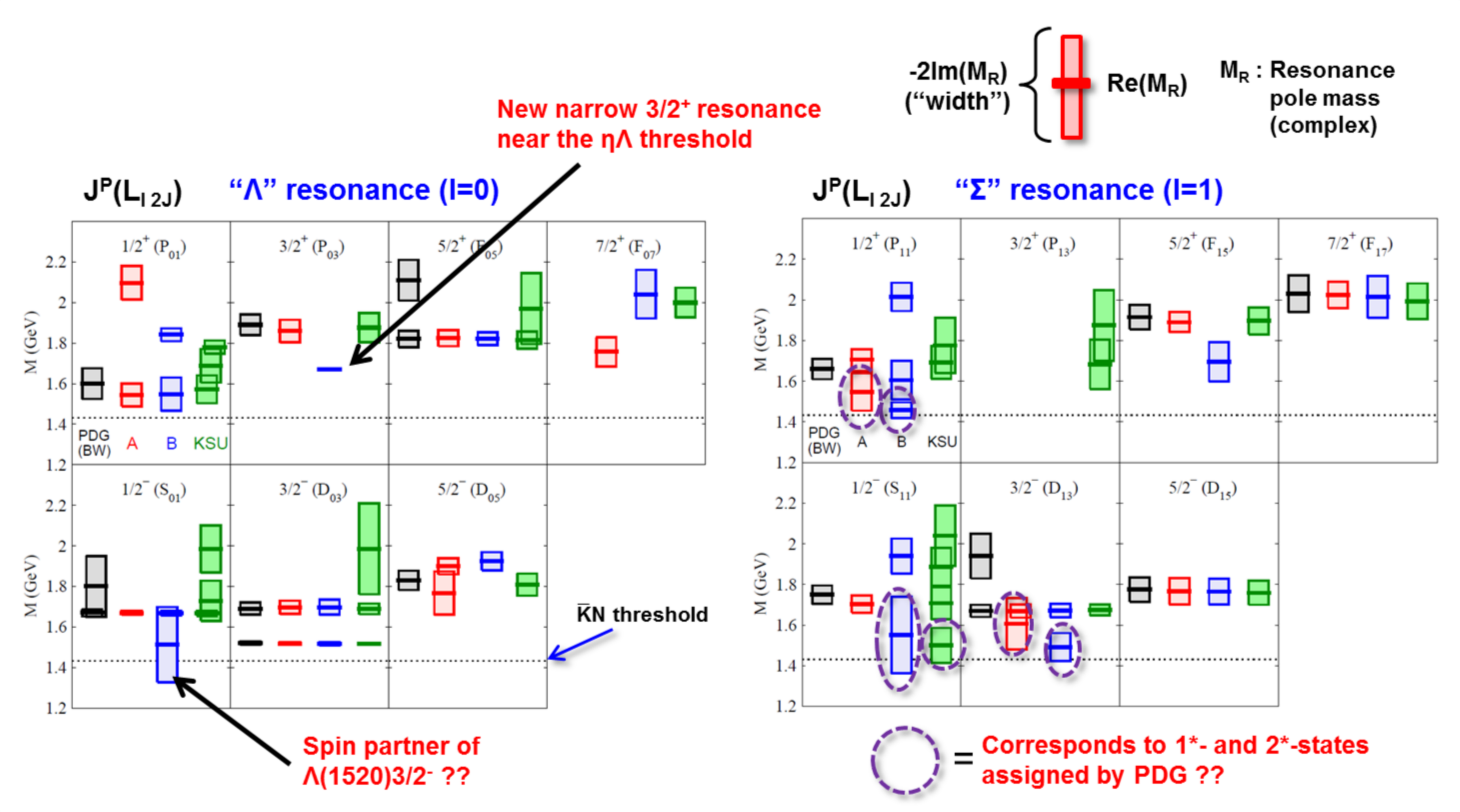}
\end{center}
\centerline{\parbox{0.80\textwidth}{
\caption{Comparison of mass spectra for $Y^\ast$ resonances 
	defined by poles of scattering 
	amplitudes~\protect\cite{knlskp2B}. Spectra in red 
	and blue are the results from Model~A and 
	Model~B~\protect\cite{knlskp2B}, respectively, while 
	the spectrum in green is obtained by the KSU 
	analysis~\protect\cite{ksu2B}. As a reference, the 
	Breit-Wigner masses and widths for the four- and 
	three-star resonances assigned by 
	PDG~\protect\cite{pdg2014B} are also presented in 
	black.} \label{fig:spectrum} } }
\end{figure}
\begin{figure}[ht!]
\begin{center}
\includegraphics[clip,width=0.9\textwidth]{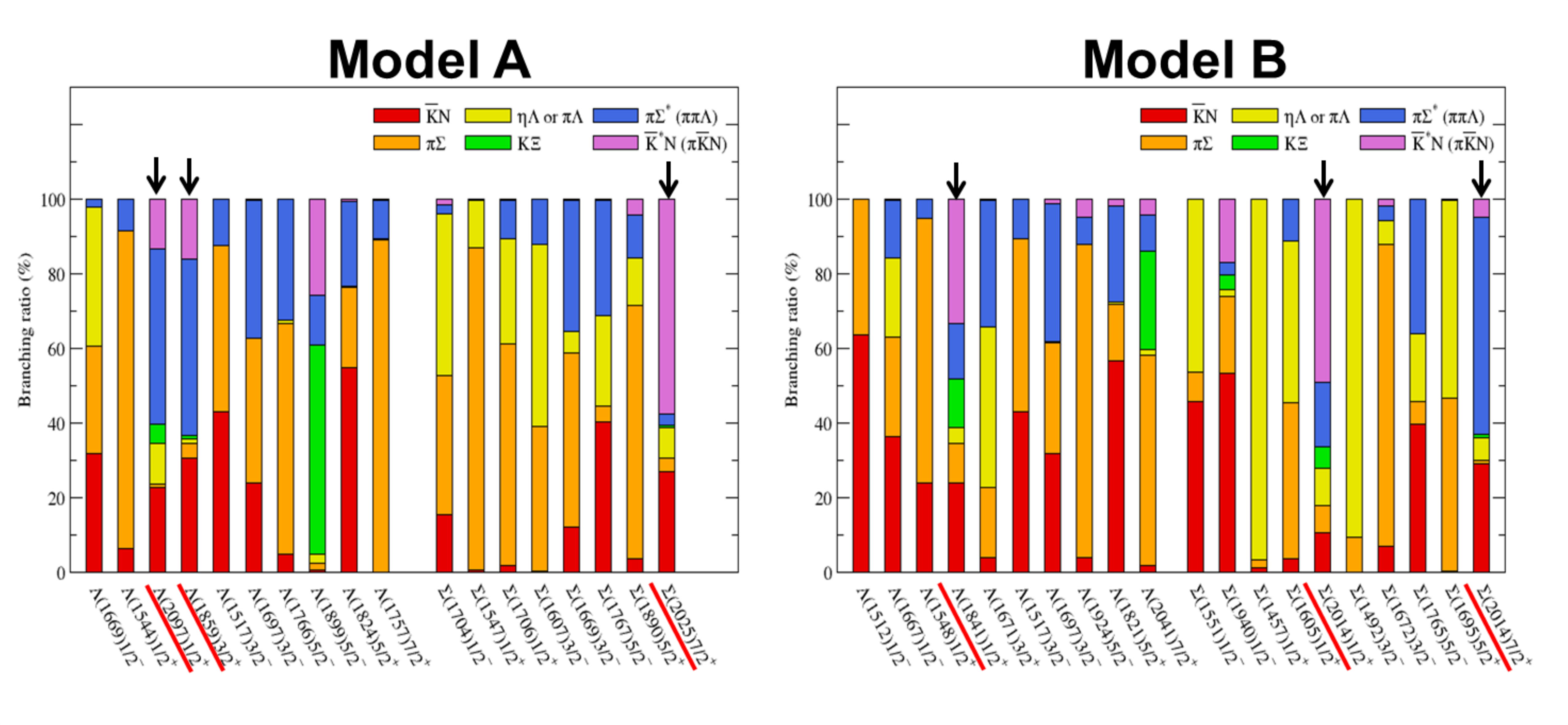}
\end{center}
\centerline{\parbox{0.80\textwidth}{
\caption{Branching ratios for $Y^\ast$ resonances extracted from 
	Model~A and Model~B~\protect\cite{knlskp2B}.}
	\label{fig:br} } }
\end{figure}

Figure~\ref{fig:br} shows the branching ratios for the $Y^\ast$ 
resonances extracted from our two models, Models~A and B. It is 
found that most resonances have large branching ratios for the 
$\bar KN$ and $\pi\Sigma$ channels, and also for the $\pi\Lambda$ 
channel for the $\Sigma^\ast$ resonances. However, as indicated 
by black arrows in Fig.~\ref{fig:br}, high-mass resonances also 
have large branching ratios for the quasi-two-body $\pi
\Sigma^\ast$ and $\bar K^\ast N$ channels, which subsequently 
decay into the three-body $\pi\pi\Lambda$ and $\pi\bar KN$ 
channels.  This suggests that the three-body production 
reactions would also play an important role for establishing 
high-mass $Y^\ast$ resonances.  This situation is quite similar 
to the $N^\ast$ and $\Delta^\ast$ resonances, where the 
high-mass resonances decay dominantly to the three-body $\pi\pi 
N$ channel~(see, {\it e.g.}, Fig.~6 of Ref.~\cite{e45B}), and the 
$\pi\pi N$ production data are expected to be the key to 
establishing high-mass $N^\ast$ and $\Delta^\ast$ resonances (see, 
{\it e.g.}, Refs.~\cite{kpi2pi13B,kjlms09-1B}). In fact, this is the 
motivation for the planned measurement of $\pi N\to\pi\pi N$ 
reactions at the J-PARC $E45$ experiment~\cite{e45B}.  It is also 
worthwhile to mention the $J^P = 7/2^+$ $\Sigma$ resonance.  This 
resonance is assigned as a four-star resonance by PDG~\cite{pdg2014B} 
[$\Sigma(2030)7/2^+$ in the PDG notation], and our two models 
actually give almost the same mass value, {\it i.e.}, ${\rm Re}(M_R) 
=2025$~MeV for Model~A and ${\rm Re}(M_R) =2014$~MeV for Model~B.
However, one can see that the component of the branching ratios 
for the quasi-two-body channels is rather different between the 
two models: In Model~A, it is dominated by $\bar K^\ast N$ 
($\pi\bar KN$) channel, while in Model~B it is dominated by 
$\pi\Sigma^\ast$ ($\pi\pi\Lambda$) channel. This indicates that 
our knowledge on the three-body channels is still poor even for 
the four-star resonance.
\begin{figure}[ht!]
\begin{center}
\includegraphics[clip,width=\textwidth]{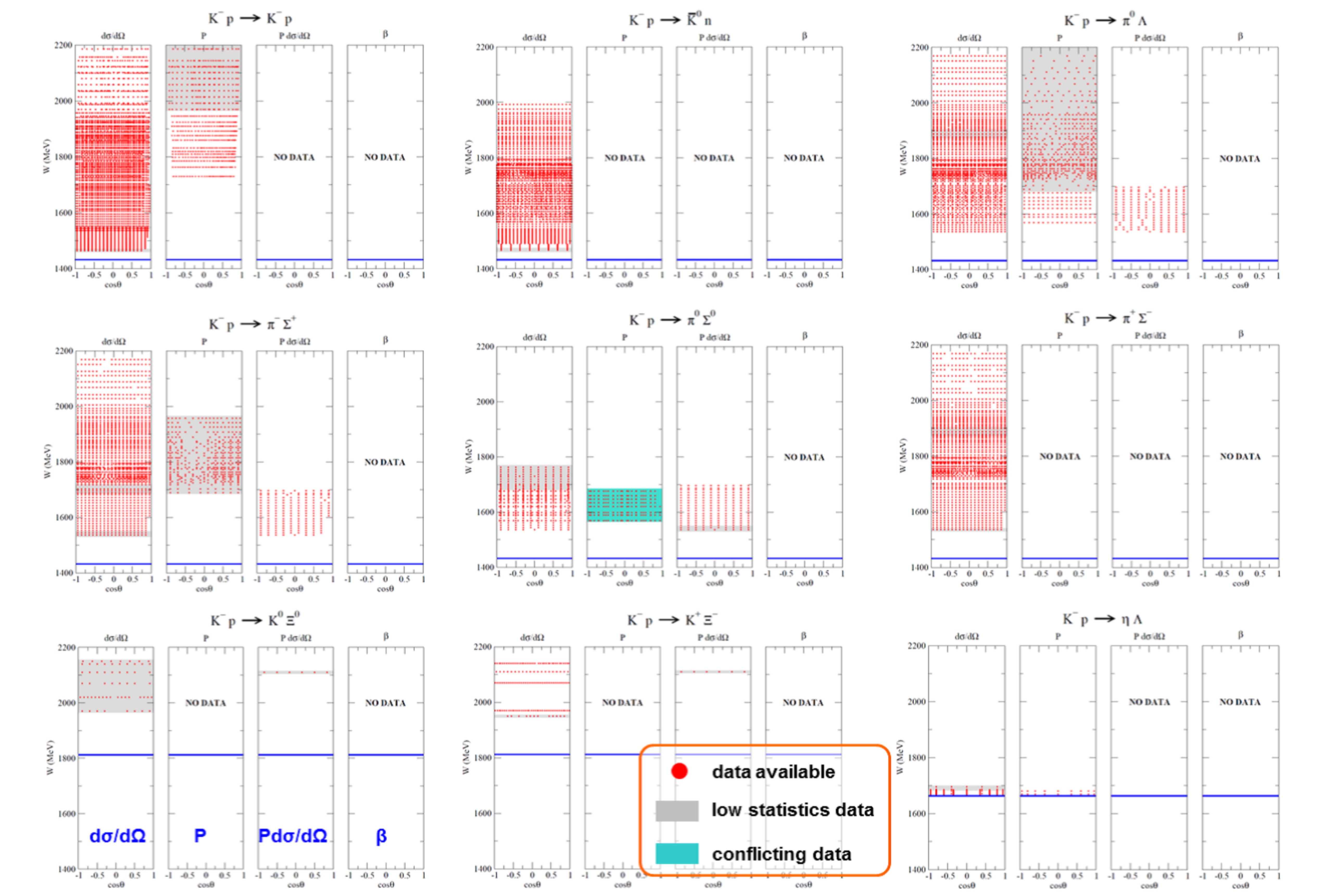}
\end{center}
\centerline{\parbox{0.80\textwidth}{
\caption{Kinematical coverage of $K^-p$ reaction data included 
	in our analysis. The horizontal axis is cosine of the 
	scattering angle in the center-of-mass frame and the 
	vertical axis is the total scattering energy. The blue 
	line in each panel represents the threshold of the 
	corresponding reaction.} \label{fig:data} } }
\end{figure}

Now we make some comments on the kinematical coverage of 
the data for $K^-p\to\bar KN, \pi\Sigma$, $\pi\Lambda$, 
$\eta\Lambda$, and $K\Xi$ included in our analysis. It is 
summarized in Fig.~\ref{fig:data}. It is well known that for 
$(0^-) + (1/2^+) \to (0^-) + (1/2^+)$ reactions the differential 
cross section ($d\sigma/d\Omega$), recoil polarization ($P$), 
and spin-rotation angle ($\beta$) form a ``complete set" of 
observables~\cite{kellyB,saxonB}. Therefore, one needs to have 
high statistics data for all the three observables in order 
to determine the scattering amplitudes accurately and accomplish 
a reliable extraction of resonance parameters. However, from 
Fig.~\ref{fig:data} one can see that the existing data are 
{\it far from complete}. The kinematical coverage of 
$d\sigma/d\Omega$, $P$, and their products $P\times 
d\sigma/d\Omega$ are still small for most reactions, and no 
data of $\beta$ are available for all reactions. Furthermore, 
even though the data exist, some data sets have large 
statistical errors and are even conflicting with each other 
(see Ref.~\cite{knlskp1B} for the details). Because of this, 
there still exist ambiguities in our constructed models, even 
though they reproduce the existing data very well. Actually, 
this can be seen, {\it e.g.}, from the spin-rotation angles 
predicted from our two models and the KSU analysis, which can 
be found in Fig.~28 of Ref.~\cite{knlskp1B}. Although the three 
analyses reproduce the existing data equally well, significant 
difference appears in the predicted value of the spin-rotation 
angles, particularly at higher $W$.  It is therefore highly 
desirable that the complete experiments of $\bar KN$ reactions 
will be performed at J-PARC using charged Kaon beam as well as 
at JLab using neutral Kaon beam.

\item \textbf{Discussions and Prospects for $Y^\ast$ 
	Spectroscopy using anti-Kaon Induced Reactions}

\begin{figure}[ht!]
\centering
\includegraphics[clip,width=0.9\textwidth]{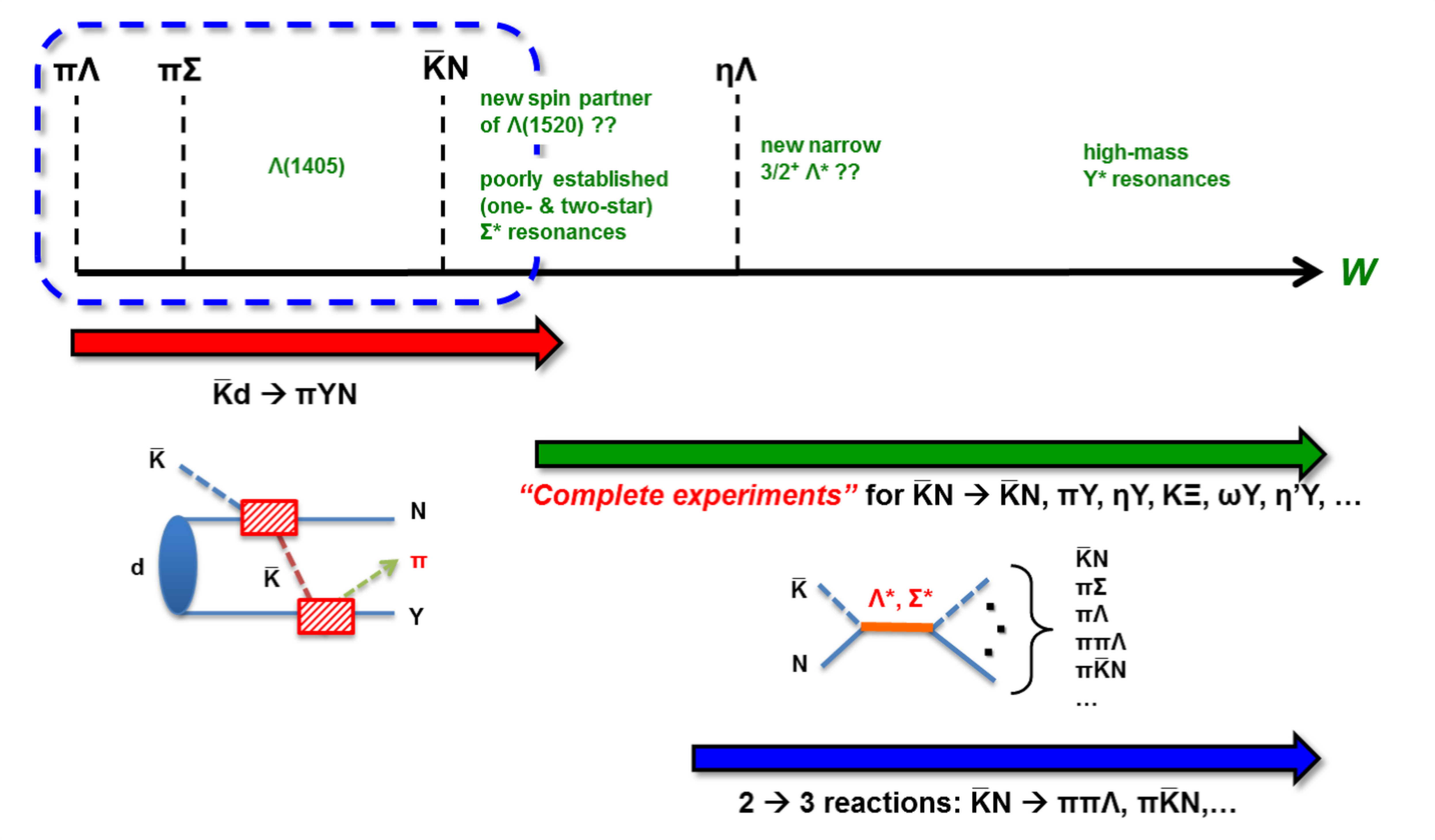}
\centerline{\parbox{0.80\textwidth}{
\caption{Strategy and necessary data for establishing $Y^\ast$ 
	resonances using anti-Kaon induced reactions.}
	\label{fig:strategy} } }
\end{figure}

As we have seen in the previous Section, the existing $\bar KN$ 
reaction data are not sufficient to eliminate analysis dependence 
in the extracted $Y^\ast$ resonance parameters, and, as depicted 
in Fig.~\ref{fig:strategy}, more extensive and accurate data for 
both two-body and three-body productions are necessary for 
further establishing $Y^\ast$ resonances.  In particular, the 
complete experiments measuring all polarization observables will 
be the key to resolving this issue.  In this regard, some 
discussions are ongoing with experimentalists to examine possible 
new experiments measuring $\bar KN$ reactions at 
J-PARC~\cite{sakoB}.  Experiments using neutral Kaon beam at JLab 
is also highly desirable.  Actually, $K^0_Lp$ reaction has a 
great advantage since it exclusively produces $\Sigma^\ast$ 
resonances in the direct $s$-channel processes due to the 
isospin selection, and thus the combined analysis of both 
charged- and neutral-Kaon induced reactions off the nucleon 
would be the best way for completing the $Y^\ast$ spectroscopy.

On the other hand, the $\bar K N$ reactions are not so suitable 
for studying the low energy region, indicated by the blue dashed 
circle in Fig.~\ref{fig:strategy}.  There are two reasons for 
this: (i) the $\bar KN$ reactions cannot directly access the 
region below the $\bar K N$ threshold, and (ii) experimentally 
it is not easy to measure the $\bar K N$ reactions in the region 
just above the $\bar K N$ threshold because of difficulty in 
producing low momentum Kaon beam. However, studying this energy 
region is very important because a number of interesting $Y^\ast$ 
resonances exist or are suggested to exist, such as 
$\Lambda(1405)$, poorly established low-lying $\Sigma^\ast$ 
resonances, and a possible new $S$-wave $\Lambda$ resonance that 
might be a ``true" spin partner of $\Lambda(1520)3/2^+$ as 
mentioned in Sec.~2. Therefore, we have recently started an 
application of our DCC approach to the deuteron-target reaction, 
$\bar Kd\to\pi YN$, which is being measured at the J-PARC $E31$ 
experiment~\cite{e31B}. This is because for this reaction the 
$\pi Y$ system in the final state can access the low energy 
region indicated by the blue dashed circle in 
Fig.~\ref{fig:strategy}, even if the momentum of the incoming 
Kaon is rather high. With this application, we aim at a combined 
analysis of both $\bar KN$ and $\bar Kd$ reactions so that we 
can cover the whole energy region relevant to the $Y^\ast$ 
spectroscopy.  The construction of the model for the 
deuteron-target reactions is underway, and it will be presented 
elsewhere.

\item \textbf{Acknowledgments}

The author thanks T.-S.~H.~Lee, S.X.~Nakamura, and T.~Sato for 
their collaboration.  This work was supported by the Japan 
Society for the Promotion of Science (JSPS) KAKENHI Grant
No.~25800149 and by the HPCI Strategic Program (Field 5 
\textit{The Origin of Matter and the Universe}) of Ministry of 
Education, Culture, Sports, Science and Technology (MEXT) of 
Japan.
\end{enumerate}


\newpage
\subsection{Strangeness Physics at CLAS in the 6 GeV Era}
\addtocontents{toc}{\hspace{2cm}{\sl R.~Schumacher}\par}
\setcounter{figure}{0}
\setcounter{equation}{0}
\halign{#\hfil&\quad#\hfil\cr
\large{Reinhard Schumacher}\cr
\textit{Department of Physics}\cr
\textit{Carnegie Mellon University}\cr
\textit{Pittsburgh, PA 15213, U.S.A.}\cr}

\begin{abstract}
A very brief overview is presented of varied strangeness-physics 
studies that have been conducted with the CLAS system in the era 
of 6~GeV beam at Jefferson Lab. A full bibliography of articles 
related to open strangeness production is given, together with 
some physics context for each work.  One natural place where 
these studies could be continued, using a $K_L$ beam and the 
GlueX detector, is in the further investigation of the 
$\Lambda(1405)$ baryon.  The line shapes and cross sections of 
this state were found, using photoproduction at CLAS, to differ 
markedly in the three possible $\Sigma\pi$ final states. The 
analogous strong-interaction reactions using a $K_L$ beam could 
further bring this phenomenon into focus.
\end{abstract}

\begin{enumerate}
\item The CLAS program ran from 1998 to 2012, during the time when 
the maximum Jefferson Lab beam energy was 6~GeV.  An important 
thrust of this program was to investigate the spectrum of 
$N^\ast$ and $\Delta^\ast$ (non-strange) baryon resonances 
using photo- and electro-production reactions.   To this end, 
final states containing strange particles ($K$ mesons and 
low-mass hyperons) played a significant role.  The reason for 
this is partly due to favorable kinematics.  When the total 
invariant energy $W (=\sqrt{s})$ of a baryonic system exceeds 
1.6~GeV it becomes possible to create the lightest 
strangeness-containing final state, $K^+\Lambda$.   This is a 
two-body final state that is straightforward to reconstruct 
in the CLAS detector system~\cite{CLAS-NIMS}, and 
theoretically it is easier to deal with two-body reaction 
amplitudes than with three- and higher-body reaction 
amplitudes.  In the mass range $W > 1.6$~GeV the decay modes 
of excited nucleons tend to not to favor two-body  
$\pi$-nucleon final states but rather multi-pion states.   
As input to partial-wave decompositions and 
resonance-extraction models, therefore, the 
strangeness-containing final states of high-mass nucleon 
excitations have had importance. Excited baryons decay 
through all possible channels simultaneously, constrained by 
unitarity of course, and channel-coupling is crucial to 
determining the spectrum of excitations.  Within this mix of 
amplitudes, however,  the $KY$ decay modes have proven useful. 
The end result has been, as summarized in the recent edition 
of the Review of Particle Properties~\cite{Agashe:2014kdaS}, 
clearer definition of the spectrum of baryonic excitations, 
with definite contributions from the strangeness sector 
channels.

To this end, strangeness photoproduction cross sections 
measurements at CLAS for the $K^+\Lambda$, $K^+\Sigma^0$ and 
$K^0\Sigma^+$ channels on a proton target were 
published~\cite{McNabbS,Bradford:2005ptS,McCracken:2009raS, 
Klein:2005mvS}.   Cross sections are not enough, in general, 
to define the reaction mechanism, including the underlying 
$N^\ast$ excitation spectrum.  Photoproduction of pseudo-scalar 
mesons is described by four complex amplitudes, leading to 
fifteen spin observables in addition to the cross section.  
Full knowledge of these spin observables would exhaust the 
information that can be gleaned experimentally about any 
given reaction channel.  Here the hyperonic channels offer 
another advantage when compared with the non-strange reaction 
channels: the polarization of most hyperons can be measured 
directly through their parity-violating weak decay 
asymmetries.  Unlike the polarization of nucleons, that 
require recoil polarimeter instrumentation and secondary 
scattering to measure, the hyperons reveal their polarization 
states directly in the angular distribution of their decay 
products.  CLAS published photoproduction measurements for a 
number of the observables that involve the recoiling hyperons 
and/or circularly polarized photons, specifically $P$, $C_x$ 
and $C_z$~\cite{McNabbS,Bradford:2006baS,McCracken:2009raS,
Dey:2010hhS,Nepali:2013bpS}.  Additional work was done on 
observables involving linearly polarized photons~\cite{PatersonS}  
and/or a polarized hydrogen target in the FROST program, both 
in combination with hyperon polarization. This included the 
observables $\Sigma$, $T$, $E$, $O_x$ and $O_z$.  Other spin 
observables are still under analysis. In principle, a complete 
set of observables can be measured for the $KY$ photoproduction 
reactions, meaning that at any given energy and production 
angle all 16 observables can be separated using a finite number 
of measurements.  The CLAS program has, in principle, enough 
data in hand to make this a reality for the $K^+\Lambda$ and 
$K^+\Sigma^0$ reactions on the proton, but is has not been 
achieved so far, mainly on account of limits in statistical 
precision.  Exploratory cross section measurements on a neutron 
(deuteron) target have also been 
published~\cite{AnefalosPereira:2009zwS}.  On the model-building 
side, all these observables have been analyzed, for example, in 
the framework of K-matrix coupled channels PWA calculations by 
the Bonn-Gatchina group and collaborators~\cite{Sarantsev:2005tgS,
Anisovich:2007bqS,Nikonov:2007brS,Anisovich:2011yeS,Anisovich:2011fcS,
Anisovich:2012ctS,Anisovich:2014yzaS}, and also the Argonne-Osaka 
Collaboration~\cite{Kamano:2013ivaS}.

Photoproduction measurements probe nucleon excitations with zero 
net 4-momentum transfer to the target ($Q^2=0$). Electroproduction 
measurements add the degree of freedom of $Q^2 >0$, which brings 
in the ``longitudinal" virtual photon degree of freedom and the 
associated interference amplitudes between the photon polarization 
components.  CLAS results on electroproduction of $K^+\Lambda$ and 
$K^+\Sigma^0$ final states off the proton have been published for  
values of $0.3 <Q^2  < 2.6$~GeV$^2$~\cite{Carman:2002seS,Raue:2004usS, 
Ambrozewicz:2006zjS,Carman:2012qjS,Nasseripour:2008aaS,Carman:2009fiS,
Gabrielyan:2014zunS}. The results include include separation of the 
cross sections into 5 structure functions, as well as measurement 
of the hyperon recoil polarization and the beam-hyperon 
polarization transfer observables.  Model-building approaches have 
generally followed in the path of photoproduction work, with the 
addition of hadronic form factors to address the $Q^2$ dependence 
of resonance contributions. 

One can inquire into hadronization properties of quarks propagating 
in the nuclear medium as a function of variable such as the hadronic 
fraction $z$ of the detected final-state meson and the hadron 
transverse momentum.  CLAS investigated this nuclear dependence 
for neutral Kaons, comparing multiplicities of Kaons in heavy 
nuclei to those in deuterium~\cite{Daniel:2011nqS}.  This was the 
only  study done at CLAS in the area of nuclear effects with 
strange particles.

Cross sections for pseudo-scalar meson photoproduction are predicted 
to scale as $s^7$ at high energies and large $t$ in perturbative 
QCD~\cite{Brodsky:1973krS} and also some other reaction models. This 
phenomenon was confirmed decades ago for pion photoproduction.  
Analysis of CLAS data confirmed this behavior for the first time 
in $K^+\Lambda$ photoproduction as well~\cite{Schumacher:2010qxS}.  
The same study also illustrated the transition in $W$ from the 
resonance region, where high-mass resonances near 1920 and 2100~MeV 
contribute to the mechanism, to  the scaling behavior that dominates 
at higher $W$.   

At the partonic level, in many high-energy reactions, phenomenology 
shows that the creation of $s\bar{s}$ quark pairs is suppressed by 
a factor of roughly 5 compared to creation of $u\bar{u}$ and 
$d\bar{d}$ pairs.  The mass difference between the quarks is 
presumably at the root of this effect.  It was shown with CLAS 
that this suppression extends to  electroproduction reactions at 
the comparatively low energies leading to exclusive $KY$ final 
states when compared to exclusive $\pi N$ final 
states~\cite{Park:2014zraS}.  Also, new limits on baryon 
non-conservation in the decay of the $\Lambda$ 
hyperon~\cite{McCracken:2015coaS} have been published.  Furthermore, 
the structure of excited baryons can be related to their radiative 
decay branching fractions. CLAS published radiative decay fractions 
for the $\Sigma^0(1385)$, the $\Sigma^+(1385)$, and the 
$\Lambda(1520)$~\cite{Taylor:2005zwS,Keller_PhysRevD.83.072004S,
Keller:2011awS}.

The possibility of an exotic pentaquark state known as the 
$\Theta^+(1520)$  caused world-wide excitement in the hadronic 
physics community over a decade ago. Initially, a hint of such 
a state was seen and reported by CLAS based on some 
already-existing data~\cite{Stepanyan:2003qrS,
Kubarovsky_PhysRevLett.92.032001S}, but subsequent dedicated 
searches with higher statistics and better background control 
found no support for this or similar states~\cite{McKinnon:2006zvS,
Battaglieri:2005erS,DeVita:2006aaqS,Kubarovsky:2006nsS,Niccolai:2006tdS}.

Differential photoproduction cross sections for the excited hyperons 
were measured simultaneously for the $\Sigma(1385)$, the 
$\Lambda(1405)$ and the $\Lambda(1520)$~\cite{Moriya:2013hwgRS}, and 
for the $\Lambda(1520)$ in electroproduction~\cite{Barrow:2001dsS}.
All three hyperons showed $t$-channel-like behavior at high $W$, 
with some evidence for high-mass $N^\ast$ contributions near their 
respective thresholds.  However, the $\Lambda(1405)$ showed an 
unexpected and unexplained charge dependence among the three 
available decay modes, $\Sigma^+\pi^-$, $\Sigma^-\pi^+$ and 
$\Sigma^0\pi^0$, near the reaction threshold.  The effect was not 
seen in the case of the $\Lambda(1520)$, indicated that something 
is special about the $\Lambda(1405)$.

The structure of the $\Lambda(1405)$ was investigated in the 
reaction $\gamma + p \to K^+ + \Sigma + \pi$  to determine the 
invariant mass distributions or ``line shapes" of the $\Sigma^+\pi^-$, 
$\Sigma^-\pi^+$ and $\Sigma^0\pi^0$ final states, from threshold at 
$1328$~MeV/$c^2$ through the mass range of the $\Lambda(1405)$ and 
the $\Lambda(1520)$, for center-of-mass energies $1.95 < W < 
2.85$~GeV~\cite{Moriya:2013ebRS,Schumacher:2013vmaS}.  The three mass 
distributions differ strongly in the vicinity of the $I=0$ 
$\Lambda(1405)$, indicating the presence of substantial $I=1$ 
strength in the reaction.  Background contributions to the data 
from the $\Sigma^0(1385)$ and from $K^\ast\Sigma$ production were 
studied and shown to have negligible influence.   The nature of 
this $I=1$ component has not been understood fully, but initial 
model calculations based on the chiral unitary ansatz have been 
made~\cite{Nacher:1998miRS,Roca:2013avRS,Roca:2013ccaRS,Zou:2010tcRS}.  
The $I=0$ nature of the $\Lambda(1405)$ is consistent with chiral 
unitary model approaches that indicate a composite structure of 
two pole.  One couples strongest to the $\bar{K}N$ channel 
(subthreshold) and the other couples strongest to the $\Sigma\pi$ 
channel (the open decay channels)~\cite{Oller:2000fjRS,Jido:2003cbRS,
Magas-Oset-RamosRS,BorasoyRS,Hyodo:2011urRS}. It was furthermore 
experimentally determined at CLAS for the first time that the 
spin and parity of the $\Lambda(1405)$ is $J^P = 
(1/2)^-$~\cite{Moriya:2014kpvRS}, as had long been assumed. A 
first-time measurement of electroproduction of the $\Lambda(1405)$ 
was published~\cite{Lu:2013nzaRS} that suggested again that the line 
shape of this hyperon is indicative of structure more complex than 
a single Breit-Wigner-type  resonance.

The next category of hyperons investigated at CLAS were the $S=-2$ 
cascade resonances.   Photoproduction of the $\Xi^-(1321)$ and its 
first excitation, the $\Xi^-(1530)$, were measured~\cite{Price:2004xmRS,
Guo:2007dwRS}.  However, none of the expected higher-mass cascades 
that are expected in the quark model were revealed in the available 
energy range, despite considerable effort.  

All the reaction channels itemized above involved hyperon production 
in association with the ground-state pseudo-scalar Kaons.  But there 
has also been some investigation of vector strange meson 
photoproduction leading to the final states $K^{\ast +}\Lambda$ and 
$K^{\ast +}\Sigma^0$~\cite{Tang:2013gsaRS} and also  
$K^0\Sigma^\ast$~\cite{Hleiqawi:2007adRS}.

Looking ahead, the CLAS12 program~\cite{Stepanyan:2010kxS} is 
scheduled to begin data taking in about 2017, and there are hyperon 
spectroscopy measurements planned for that new era of research.  A 
continued search for excited cascade hyperons is one example, and 
there are even plans to detect photoproduction of the triply-strange 
$\Omega^-$ baryon.  
\begin{figure}[htpb]
\begin{center}
\resizebox{0.65\textwidth}{!}{\includegraphics[angle=0.0]{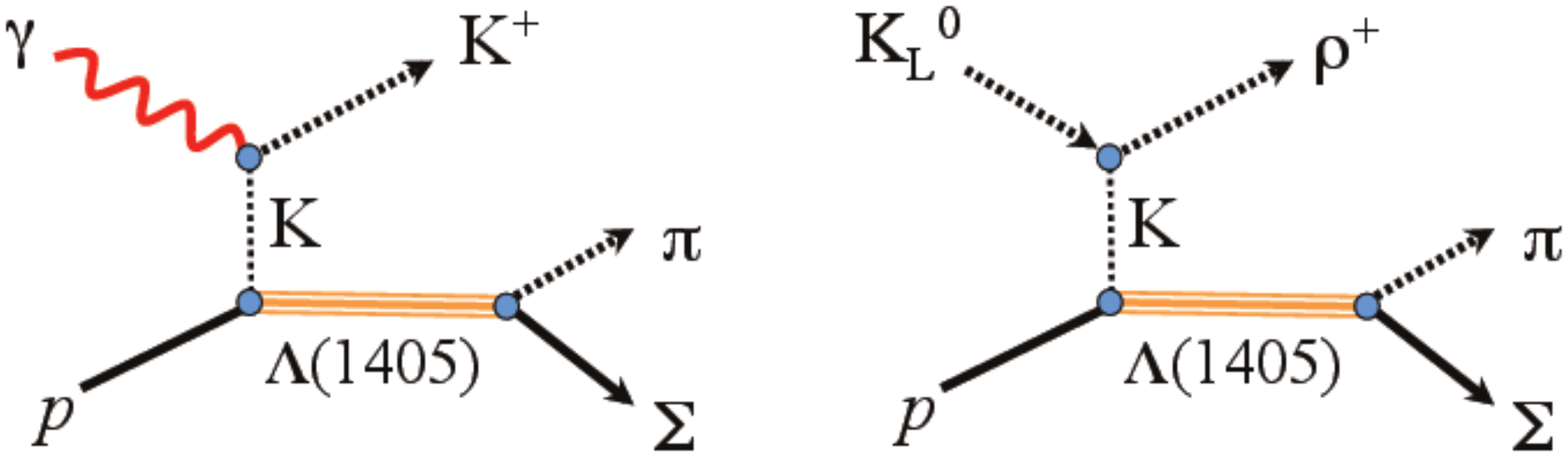}}
\end{center}
\vspace{-4cm}
\centerline{\parbox{0.80\textwidth}{
 \caption{Creation and decay of the $\Lambda(1405)$ baryon via 
	photoproduction (left) and via hadronic production using 
	a $K_L$ beam (right).  In either case the exchange of an 
	off-shell Kaon allows production of the $\Lambda(1405)$, 
	which lies below the $\bar{K}N$ threshold.}
	\label{fig:cartoon} } }
\end{figure}

Turning to the prospects of using a $K_L$ beam in conjunction 
with the GlueX detector, one can ask whether the CLAS program, 
as it has been very briefly outlined here, serves as an impetus 
for new research.  One idea to consider is to extend the cited 
studies of the still-mysterious $\Lambda(1405)$ hyperon.  Does 
its line shape depend on the manner in which it is produced, as 
currently thought, in light of the chiral unitary model 
approaches ?  Can its line shape be precisely measured using a 
$K_L$ beam and the excellent resolution and  particle 
identification capabilities of GlueX~\cite{Ghoul:2015ifwS} ?  
Figure~\ref{fig:cartoon} shows a comparison of sample 
diagrams for production of this hyperon via photon and Kaon 
beams.  What they have in common is that the exchanged Kaon is 
off shell, allowing entry into the sub-threshold $\bar{K}N$ 
regime where this hyperon exists.  The method proved fruitful 
in the photoproduction case, and so by analogy one may expect 
to create the state in the $K_L$ beam case as well.  A very 
preliminary check of existing data, which is exceedingly sparse, 
suggests that the cross section is in the range of 250 
micro-barns, which in turn could lead to a reaction rate of, 
very roughly, one event per few seconds.  One favorable aspect 
of such a measurement program would be that no Kaons need to be 
detected in the final state with GlueX, while photon detection 
capability would be very important indeed.  No acceptance 
calculations for an actual experiment have been carried out 
yet for this discussion.  

In conclusion, the CLAS program of hyperon physics has produced 
a sizable harvest of hyperon and strangeness-related results.  
These have helped define the spectrum of non-strange excited 
states, reveal the reaction mechanisms for the photo- and 
electro- production of several ground state and excited state 
hyperons, test some quark model and QCD-related phenomenology, 
and shed further light onto the nature of the $\Lambda(1405)$ 
state.  The further study of the latter state may by one place 
where future work using a $K_L$ beam in conjunction with GlueX 
would be of interest.   

\item \textbf{Acknowledgments}

This work  was supported by DOE grant DE--FG02--87ER40315.  
\end{enumerate}


\newpage
\subsection{Lattice Studies of Hyperon Spectroscopy}
\addtocontents{toc}{\hspace{2cm}{\sl D.~Richards}\par}
\setcounter{figure}{0}
\setcounter{table}{0}
\setcounter{equation}{0}
\halign{#\hfil&\quad#\hfil\cr
\large{David Richards}\cr
\textit{Thomas Jefferson National Accelerator Facility}\cr
\textit{Newport News, VA 23606, U.S.A.}\cr}

\begin{abstract}
I describe recent progress at studying the spectrum of hadrons
containing the strange quark through lattice QCD calculations.  
I emphasise in particular the richness of the spectrum revealed 
by lattice studies, with a spectrum of states at least as rich 
as that of the quark model.  I conclude by prospects for future
calculations, including in particular the determination of the 
decay amplitudes for the excited states.
\end{abstract}

\begin{enumerate}
\item \textbf{Introduction}

The calculation of the spectrum of QCD using lattice simulations 
has long been a key quest of the lattice community.  For the case 
of the lowest-lying hadrons, they form a vital benchmark of our 
ability to describe the strong interactions through lattice 
calculations: the spectrum is well established from experiment.  
A noticeably high-profile example of such a calculation is that 
of the BMW Collaboration~\cite{Durr:2008zzI}, whereby the spectrum 
of the lowest-lying states containing the $u,d$ and $s$ quarks 
exhibited remarkable agreement with experiment.  Such 
calculations require a high degree of control over the systematic 
uncertainties inherent to lattice calculations, namely those 
arising from the finite volume in which they are performed, 
finite lattice spacing, and, finally, the need until recently to 
extrapolate from unphysical $u$ and $d$ quark masses to the 
physical light quark masses.  It worth noting here that the 
quark masses are not themselves physical observables, but are
tunable parameters in the calculations that are tuned to ensure 
that certain mass ratios attain their physical values.

The calculation of the excitations of the theory, the 
excited-state spectrum, provides an important \textit{predictive} 
opportunity for lattice QCD, yet imposes still further challenges.  
Nowhere are the opportunites for predictions more apparent than 
in the spectrum of hyperons, where so few of the states 
anticipated have been seen in experiment, and where the quantum 
numbers of the states that have been observed are often poorly 
established.  The aim of this talk is to review our current 
knowledge of hyperon spectroscopy, to outline the challenges 
that are being overcome to advance that knowledge, and to 
emphasise the role that lattice calculations will play in the 
future hyperon physics program.

\item \textbf{The flavor structure of excited baryons}

Lattice QCD is formulated in Euclidean space, thereby admitting 
the use of importance sampling that is essential to numerical 
calculations of QCD.  The spectrum of the theory is determined 
through observing the temporal decay of time-sliced correlation 
functions, \textit{i.e.},
\begin{equation}
	C(t) = \sum_{\vec{x}} \langle 0 \mid {\cal O}^{J^P}
	(\vec{x},t)\bar{\cal O}^{J^P}(\vec{0},0)\mid 0 
	\rangle \longrightarrow A^n e^{- M_n t},\label{eq:corr}
\end{equation}
where ${\cal O}^{J^P}$ is an interpolating operator of specified 
spin and parity $J^P$, and $M_n$ are the masses of the states of 
those quantum numbers.  That these are real, rather than 
imaginary, exponentials is reflective of working in Euclidean 
space.  Extracting subleading terms in a sum of exponentials 
from a single correlation function is a challenging task, though 
various techniques, such as the sequential-Bayes 
method~\cite{Chen:2004gpI}, have been attempted.  A robust way of 
extracting the subleading contributions is by means of the 
variational method, whereby we compute a matrix of correlation
functions
\begin{equation}
	C_{ij}(t) = \sum_{\vec{x}} \langle 0 \mid {\cal O}^{J^P}_i
	(\vec{x},t)\bar{\cal O}^{J^P}_j(\vec{0},0)\mid 0 \rangle 
	\longrightarrow A^n_{ij} e^{- M_n t},\label{eq:matrix}
\end{equation}
where $\{ {\cal O}^{J^P}_i: i = 1,\dots,N\}$ is a basis of 
operators, each having common quantum numbers.  We now solve the 
generalized eigenvalue equation
\begin{equation}
	C(t) u(t,t_0) = \lambda(t,t_0) C(t_0) u(t,t_0)
\end{equation}
yielding a set of real eigenvalues $\{ \lambda_n(t,t_0): n =
1,\dots,N\}$ with corresponding eigenvectors $\{ u^n(t,t_0): 
n = 1,\dots,N\}$, where, for sufficiently large $t$, we have 
$\lambda_0\ge \lambda_1 \ge \dots$.  These eigenvalues have 
the property that, for sufficiently large $t$ and $t_0$
\begin{equation}
	\lambda_n(t,t_0) \rightarrow (1 - A) e^{-M_n (t - t_0)} 
	+ A e^{ - (M_n  + \Delta M) (t - t_0)}.
\end{equation}
We have thus delineated between the different subleading
exponentials. Furthermore, the eigvenvector corresponding 
to a particular state in the spectrum provides important 
information as to the structure of the state.  In particular, 
we can express the spectral decomposition of the correlation 
functions as
\[
	C_{ij}(t) = \sum_n \frac{Z^{n*}_i Z^n_j}{2 M_n} e^{-M_n t} ,
\]
where
\begin{equation}
	Z^n_i \equiv \langle n \mid {\cal O}_i^{\dagger} \mid 0 
	\rangle = \sqrt{2 M_n} e^{M_n t_0/2} u^{n*}_j C_{ji}(t_0)
	\label{eq:z}
\end{equation}
for an eigenvectors $u^{n}$ obtained at some reference time $t = 
t_{\rm ref}$.  We will exploit this feature below to determine the 
dominant operators corresponding to the varioius states in the 
spectrum.

The efficacy of the variational method relies on three features. 
Firstly, on a basis of operators that faithfully spans the
structure of the different states. Secondly, on an efficient 
means of computing the correlation functions of 
Eqn.(\ref{eq:matrix}). Thirdly, on having a sufficiently large 
signal-to-noise ratio to render the solution of the generalized 
eigenvalue equation feasible.  In the calculations of the 
\textit{Hadron Spectrum Collaboration} that I will emphasise in 
this talk, the second requirement is satisfied through the use 
of ``distillation"~\cite{Peardon:2009ghI} and its stochastic
variants~\cite{Morningstar:2011kaI}, whilst the third through the 
use of a so-called anisotropic lattice~\cite{Morningstar:1997ffI,
Lin:2008prI} with a fine temporal lattice spacing to allow the 
expontial fall-off of correlation functions to be resolved at 
small temporal separations, and through exploiting the 
translational symmetry of the lattice to make many calculations 
of the correlation function on a single gauge configuration.  
For this workshop, I will focus on thie first of these issues, 
namely the construction of a suitable operator basis.

We begin by expressing baryon interpolating operators of definite
$J^P$ as~\cite{Edwards:2011jjI,Edwards:2012fxI} 
\begin{equation}
	{\cal O}^{J^P} \sim  \left( F_{\Sigma_F} \otimes 
	(S^{P_s})_{\Sigma_S}^n \otimes D^{[d]}_{L,
	\Sigma_D} \right)^{J^P},
\end{equation}
where $F$, $S$ and $D$ are the flavor, Dirac spin and orbital 
angular momentum parts of the wave function, and the 
$\Sigma$'s express the corresponding permutation symmetry: - 
Symmetric (S), mixed-symmetric (MS), mixed anti-symmetric 
(MA), and antisymmetric (A). The construction of the flavor 
and spin components of the interpolating operators is 
straightforward.

Non-zero orbital angular momentum is introduced through the 
use of gauge-covariant derivatives, written in a circular basis 
and acting on the quark fields.  In the notation above, 
$D^{[d]}_{L, \Sigma_D}$ corresponds to an orbital wave function 
constructure from $d$ derivatives, and projected onto orbital 
angular momentum $L$. In the calculations described here, up to 
two covariant derivatives and employed, enabling orbital angular 
momentum up to $l = 2$ to be acessed.  Of particular note is the 
mixed-symmetric combination $D^{[2]}_{L = 1, M}$, the commutator 
of two covariant derivatives projected to $L = 1$, that 
corresponds to a chromo-magnetic field that would vanish for 
trvial gauge field configuration; operators with this construction 
we identify as \textit{hybrid} operators, associated with a 
manifest gluon content~\cite{Dudek:2012agI}.  Whilst the lack of 
rotational symmetry introduced through the discretization onto a 
finite space-time lattice has the consequence that angular 
momentum is no longer a good quantum number at any finite spacing, 
we find in practice that the spectrum shows a remarkable 
realization of rotational symmetry, enabling the 
``single-particle" spectrum to be classified according to the 
total angular momentum of the states, illustrated in 
Figure~\ref{fig:flavor} below.
\begin{table}
\centerline{\parbox{0.80\textwidth}{
 \caption{The table shows the different baryons that can
        be constructed from the light $u$, $d$ and $s$
        quarks, together with their isospin, strangeness
        and $SU(3)_F$ flavor content. \label{tab:flavor}} } }
\begin{center}
\begin{tabular}{cccc}
\hline
    Baryon    & $I$           & $S$  & $SU(3)_F$\\ 
\hline
    $N$       & $\frac{1}{2}$ & $0$  & $\mathbf{8_F}$\\
    $\Delta$  & $\frac{3}{2}$ & $0$  & $\mathbf{10_F}$\\
    $\Lambda$ & $0$           & $0$  & $\mathbf{1_F}$\\
              &               &      & $\mathbf{8_F}$\\
    $\Sigma$  & $1$           & $-1$ & $\mathbf{8_F}$\\
              &               &      & $\mathbf{10_F}$\\
    $\Xi$     & $\frac{1}{2}$ & $-2$ & $\mathbf{8_F}$\\
              &               &      & $\mathbf{10_F}$\\
    $\Omega$  & $0$           & $-3$ & $\mathbf{10_F}$\\
\hline
\end{tabular} 
\end{center}
\end{table}

Our published work on the flavor structure of the excited 
baryon spectrum was obtained for three difference quark 
masses, corresponding to $m_\pi = 702, 524~{\rm and}~ 391$~MeV, 
with the strange quark maintained at its physical value; the 
largest pion mass corresponds to three degenerate quark masses, 
\textit{i.e.}, the SU(3) flavor-symmetric point. We classify 
the flavor structure of the three-quark interpolating operators 
that can be constructed from the $ud$, $d$ and $s$ quarks
according to their $SU(3)_F$ flavor composition, corresponding 
to octet ($\mathbf{8_F}$), decuplet ($\mathbf{10_F}$) and 
singlet ($\mathbf{1_F}$), detailed in Table~\ref{tab:flavor}.  
We find that, even at the lightest pion mass used in our 
calculations, where $SU(3)_F$ is most severely broken, the 
states are dominated by a particular $SU(3)_F$ representation, 
illustrated in Figure~\ref{fig:flavor}.
\begin{figure}[ht!]
\begin{center}
\includegraphics[width=0.5\textwidth]{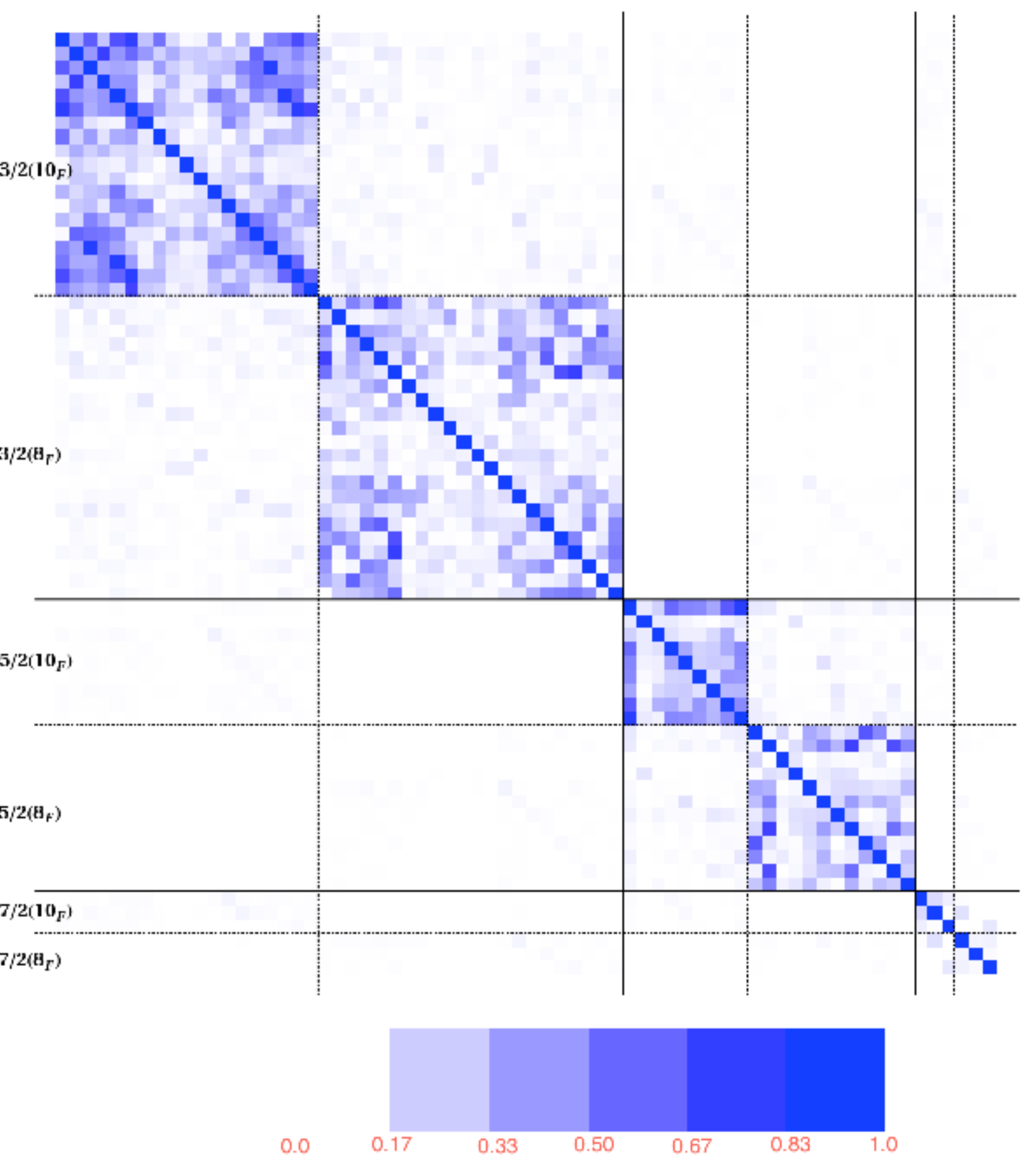}
\centerline{\parbox{0.80\textwidth}{
 \caption{The magnitudes $C_{ij}/\sqrt{C_{ii} C_{jj}}$ 
	of the elements of the correlator matrix of 
	Eqn.(\protect\ref{eq:matrix}), normalized to the 
	diagonal elements, at a separation of 5 time 
	slices and at a pion mass of 391~MeV. The plot 
	shows the correlation matrix not only to be 
	block-diagonal in spin, but also block diagonal 
	in flavor. \label{fig:flavor}} } }
\end{center}
\end{figure}

The spectrum is encapsulated in Figure~\ref{fig:lattice_lambda}, 
for the calculation at $m_\pi = 391~{\rm MeV}$, showing the 
dominant flavor structure for each state represented as blue 
($\mathbf{8_F}$), yellow ($\mathbf{10_F}$) and beige 
($\mathbf{1_F}$).  The spectrum that emerges has several 
remarkable features:
\begin{enumerate}
\item The spectrum is remarkably rich, displaying a counting 
	of states satisfying $SU(6) \otimes O(3) $ symmetry, 
	and beyond that expected in a simple quark-diquark 
	picture of a baryon.
\item There are additional positive-parity states which we 
	label ``hybrid" baryons, denoted in the figure by the 
	bold borders, with a mass around $1.2~{\rm GeV}$ above 
	their non-hybrid cousin,s that we identify through 
	their dominant coupling to the hybrid operators 
	introduced above, illustrated in 
	Figure~\ref{fig:hybrid}~\cite{Dudek:2012agI}.
\item The mixed-flavor states, the $\Lambda$, $\Sigma$ and
	$\Xi$, exhibit multiplicities expected from exact 
	$SU(3)$ flavor symmetries.  Thus the $\Xi$, for example, 
	has a spectrum corresponding to the superposition of 
	multiplicities of the $\mathbf{8_F}$ and $\mathbf{10_F}$ 
	exact $SU(3)_F$ expectations, seen clearly by comparing 
	the low-lying band for the $\Xi$ with that of the octet 
	Nucleon and decuplet $\Delta$.
\end{enumerate}
\begin{figure}[ht!]
\begin{center}
\includegraphics[width=0.8\textwidth]{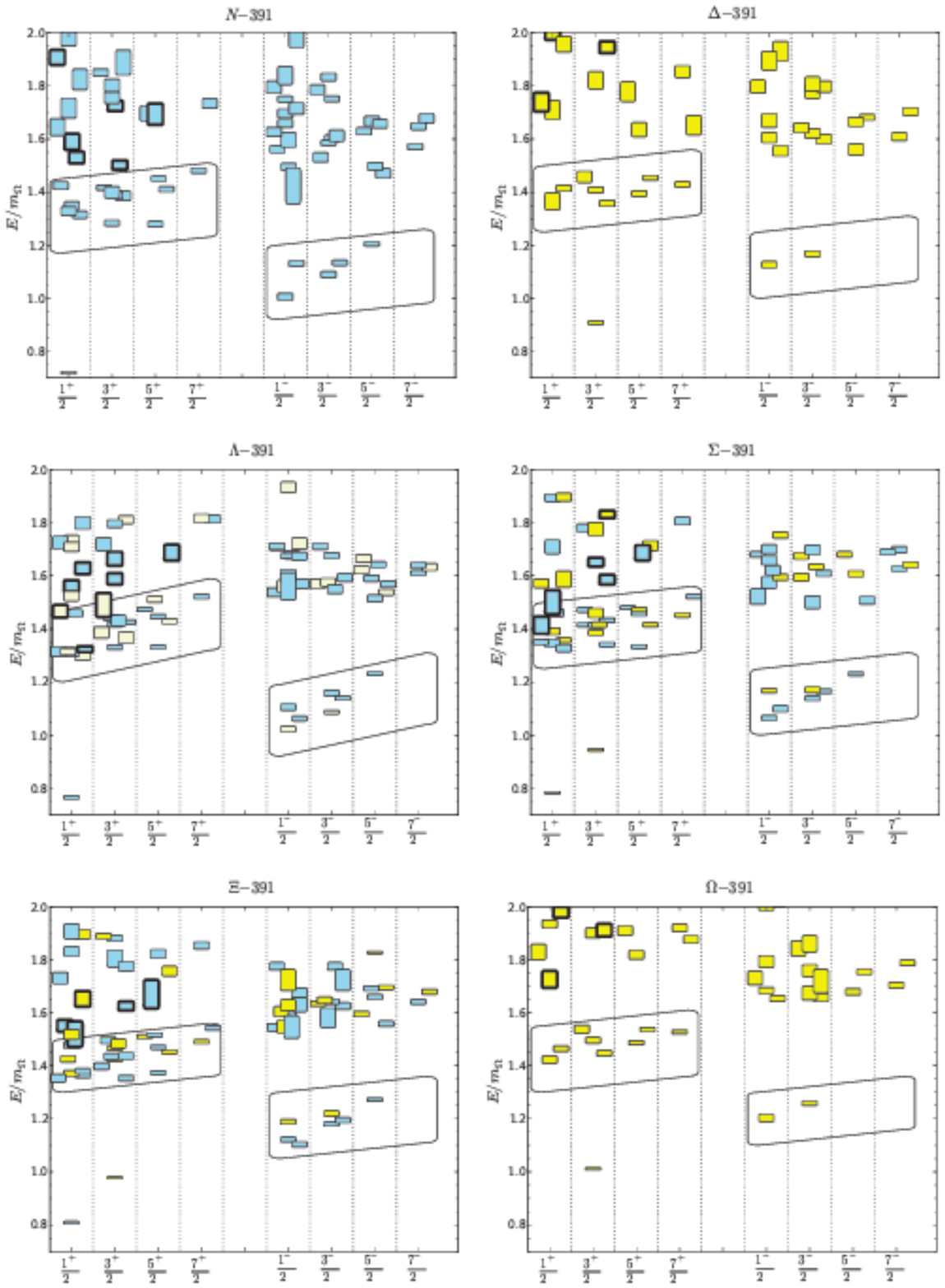}
\centerline{\parbox{0.80\textwidth}{
 \caption{The figures show the spectrum of baryon states 
	composed of $u$, $d$ and $s$ quarks, obtained on a 
	$16^3 \times 48$ anisotropic 
	lattice~\protect\cite{Edwards:2012fxI}.  The bands 
	denote the lowest-lying states identified with
	$SU(6) \otimes O(3)$ symmetry, whilst the states 
	with bold borders denote those identified as 
	\textit{hybrid} sates, as discussed in the text.
	\label{fig:lattice_lambda}} } }
\end{center}
\end{figure}
\begin{figure}[ht!]
\begin{center}
\includegraphics[width=0.7\textwidth]{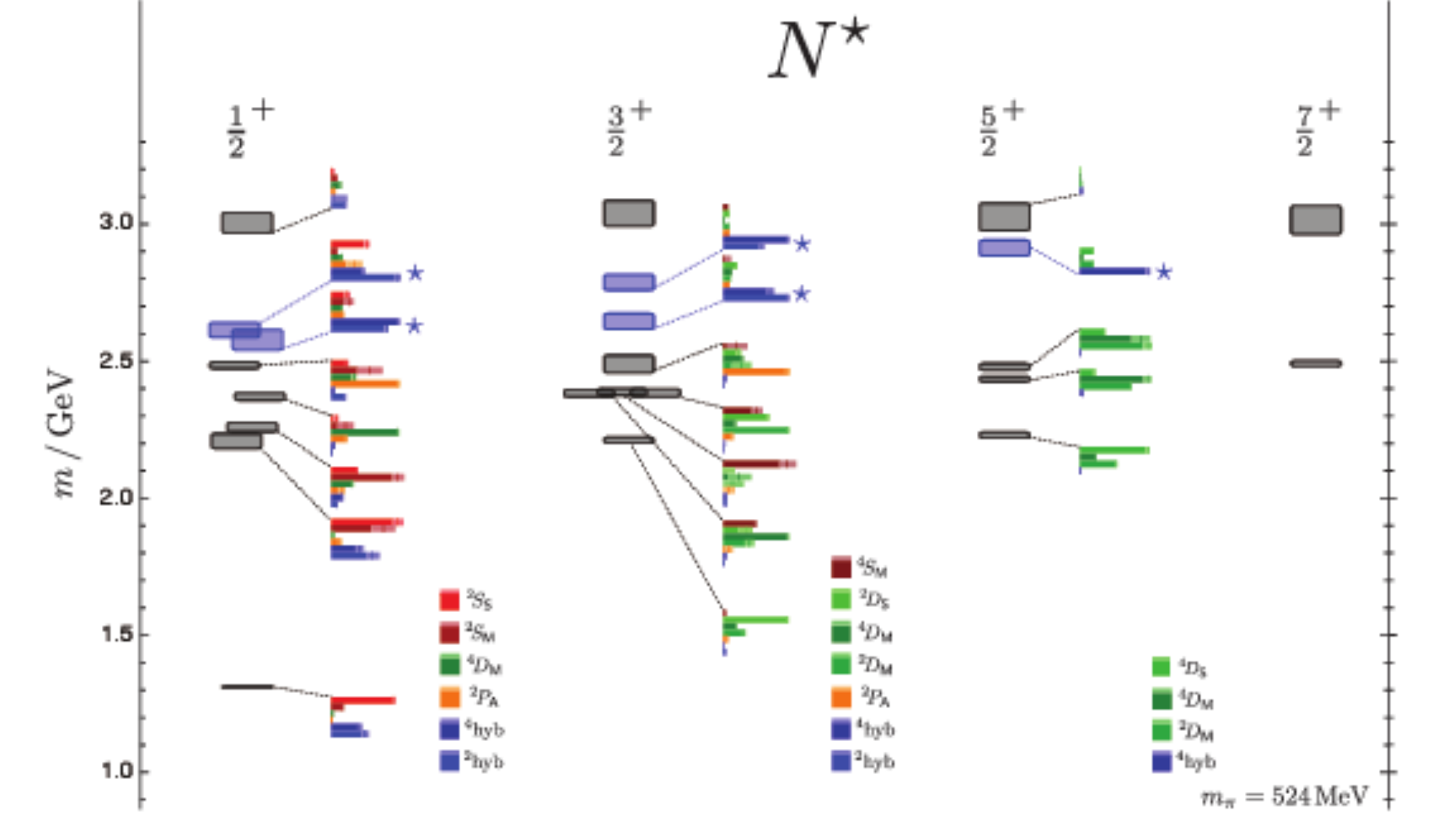}
\centerline{\parbox{0.80\textwidth}{
  \caption{The figure, taken from 
	Ref.~\protect\cite{Dudek:2012agI}, shows the relative 
	overlaps, defined through Eqn.(\protect\ref{eq:z}), of 
	the lowest-lying states in the positive parity nucleon 
	spectrum. Those operators labelled $^2$hyb and $^4$hyb 
	correspond to the commutator of two gauge-covariant 
	derivatives; the states in blue are those dominated by 
	such operators which we identify as ``hybrid" baryons.
	\label{fig:hybrid}} } }
\end{center}
\end{figure}

There are features of the calculated spectrum that are 
qualitatively different to experiment, notably the 
anomalously low masses of the Roper and of the 
$\Lambda(1405)^-$. In the case of the latter, a recent 
calculation using different source and sink smearing radii 
found a level ordering consistent with that 
observed~\cite{Menadue:2011pdI}, and there are arguments 
that the $\Lambda(1405)^-$ is a $\bar{K} N$ molecule, based 
on the calculated vanishing of its strange magnetic form 
factor~\cite{Hall:2014ucaI}; it should be noted, however, 
that the interpretation of the electromagnetic properties 
of a two-particle state from a calculation on a finite 
lattice requires considerable theoretical 
analysis~\cite{Briceno:2015tzaI}.  However, a comprehensive 
understanding of the spectrum of hyperon must reflect the 
fact that in general these states are resonances, unstable 
under the strong interactions.

\item \textbf{Resonances and Lattice QCD}

Lattice QCD is formulated in Euclidean space, and the 
energies entering into the spectral decomposition of 
Eqn.(\ref{eq:matrix}) are real.  In the finite spatial volume 
in which are calculations are performed, those energies are 
quantised and should include the two- and higher-body 
scattering states that must be present in the spectrum. For 
non-interacting particles, the energies of those multi-hadron 
states are given by the symmetries of the box in which were 
are performing our calculations, and the allowed three 
momenta of the states.  The finite spatial volume forces 
those scattering states to interact thereby shifting the 
energies from their non-interacting values. For the case 
of elastic scattering, the so-called L\"{u}scher method 
enables the shift in energies at a finite volume to be 
related to the infinite volume phase 
shift~\cite{Luscher:1986pfI,Luscher:1990uxI}; this was 
shortly thereafter extended to states with non-zero total
momentum~\cite{Rummukainen:1995vsI}, enabling a far finer 
resolution of the momentum-dependent phase shifts.

In the meson sector, and using the distillation method 
introduced above, the precise calculation of the 
momentum-dependent phase shifts for states such as the 
$\rho$ meson in $I=1 \, \pi\pi$ scattering has now been 
accomplished~\cite{Dudek:2012xnI}.  More recently, the 
formalism has been extended to the extraction of the
momentum-dependent amplitudes for inelastic 
scattering~\cite{He:2005eyI,Guo:2012hvI,Hansen:2012tfI,
Briceno:2012yiI}, and applied to the $\pi K - \eta
K$~\cite{Dudek:2014qhaI,Wilson:2014cnaI} and the coupled 
$\pi\pi - \bar{K} K$ system~\cite{Wilson:2015dqaI}.

Analogous calculations in the baryon sector are less 
advanced. Firstly, we note that the multi-hadron states 
that we would expect to observe in the baryon spectrum 
appear absent in the spectrum presented in 
Figure~\ref{fig:lattice_lambda}.  The reason is simply 
that the three-quark interpolating operators that forms 
the basis for the application of the variational method 
have a coupling to multi-hadron states in Eqn.(\ref{eq:z}) 
that is suppressed by the spatial volume compared to that 
for ``single-hadron" states.  Key to the meson results 
cited above has been the introduction into the variational 
basis of multi-meson operators, at both zero and non-zero 
total momentum, for which the coupling to two-hadron 
states is not suppressed.  The first steps have been taken 
to include multi-hadron $N\pi$ operators into the excited 
nucleon basis~\cite{Verduci:2014csaI,Kiratidis:2015vpaI}, 
with the expected spectrum $N\pi$ energy emerging, and 
meson-baryon phase shifts for several channels have been 
computed using the same anisotropic lattice formulation
employed here~\cite{Detmold:2015qwfI}.  The application of 
the variational method with as faithful a basis as that 
used in the meson sector is more limited by computational 
requirements, which are considerably more demanding, than 
by theoretical background.

\item \textbf{Summary}

There has been enormous progress, both theoretical and
computational, aimed at extracting the excited-state 
spectrum of QCD.  Precise calculations of 
momentum-dependent phase shifts have been obtained for 
the meson spectrum, and the formalism developed that can 
be applied to baryons.  The most important conclusion of 
this talk is the richness of the hyperon spectrum,
encompassing not only those states expected in the quark 
model, but additional hybrid states in which the gluons 
are manifest.  In the coming years, the tools that have 
been developed for understanding the amplitudes and decays 
of resonances from lattice QCD calculations, and applied 
to the excited meson spectrum, will be applied to the 
excited baryon sector, both guiding and interpreting an 
exciting experimental program in hyperon spectroscopy.

\item \textbf{Acknowledgments}

The author would like to thank his colleagues in the 
\textit{Hadron Spectrum Collaboration} for their 
collaboration on the work presented here.  This material 
is based upon work supported by the U.S. Department of 
Energy, Office of Science, Office of Nuclear Physics 
under contract DE--AC05--06OR23177.
\end{enumerate}


%
%
%
%
%
%
\newpage
\subsection{Formation of the $f_0(980)$ and $a_0(980)$ resonances by
        $\bar K$ Induced Reactions on Protons}
\addtocontents{toc}{\hspace{2cm}{\sl E.~Oset, Ju-Jun~Xie, Wei-Hong~Liang}\par}
\setcounter{figure}{0}
\setcounter{table}{0}
\setcounter{equation}{0}
\halign{#\hfil&\quad#\hfil\cr
\large{Eulogio Oset}\cr
\textit{Departamento de F\'{\i}sica Te\'orica and IFIC}\cr
\textit{Centro Mixto Universidad de
        Valencia-CSIC Institutos de Investigaci\'on de Paterna}\cr
\textit{Aptdo. 22085, 46071 Valencia, Spain \&}\cr
\textit{Institute of Modern Physics}\cr
\textit{Chinese Academy of Sciences}\cr
\textit{Lanzhou 730000, China}\cr\cr
\large{Ju-Jun~Xie}\cr
\textit{Institute of Modern Physics}\cr
\textit{Chinese Academy of Sciences}\cr
\textit{Lanzhou 730000, China \&}\cr
\textit{State Key Laboratory of Theoretical Physics}\cr
\textit{Institute of Theoretical Physics}\cr
\textit{Chinese Academy of Sciences}\cr
\textit{Beijing 100190, China}\cr\cr
\large{Wei-Hong~Liang}\cr
\textit{Department of Physics}\cr
\textit{Guangxi Normal University}\cr
\textit{Guilin 541004, China}\cr}

\begin{abstract}
Results are shown for the cross section of several reactions
induced by $\bar K$ scattering on protons. The reactions studied 
are $K^-p \to \Lambda \pi^+ \pi^-$, $K^-p \to \Sigma^0 \pi^+ \pi^-$,
$K^-p \to \Lambda \pi^0 \eta$, $K^-p \to \Sigma^0 \pi^0 \eta$, $K^-p
\to \Sigma^+ \pi^- \eta$, $\bar K^0 p \to \Lambda \pi^+ \eta$, $\bar
K^0 p \to \Sigma^0 \pi^+ \eta$, $\bar K^0 p \to \Sigma^+ \pi^+\pi^-$, 
$\bar K^0 p \to \Sigma^+ \pi^0 \eta$. In the reactions with a final 
$\pi^+\pi^-$ a clear peak is seen for the $f_0(980)$ formation, with 
no trace of the $f_0(500)$. In the cases of $\pi \eta$ production 
the $a_0(980)$ resonance shows up, with a characteristic strong cusp 
shape.
\end{abstract}

\begin{enumerate}
\item \textbf{Introduction}

Kaon beams are increasingly becoming a good source for new information 
in hadron physics. At intermediate energies J-PARC offers good
intensity secondary Kaon beams up to about 2~GeV/$c$~\cite{Sato:2009zzeE,
Kumano:2015gnaE}. DAPHNE at Frascati provides low energy Kaon 
beams~\cite{Okada:2010oxaE,Bazzi:2013vftE}. Plans are been made for a 
secondary meson beam Facility at Jefferson Lab, which would include 
Kaons, both charged and neutral~\cite{whitebookE}. One of the purposes 
is to produce hyperons ($\equiv Y$) \cite{Zhang:2013cuaE}, and cascade 
states, which are poorly known~\cite{Jackson:2013bbaE,Jackson:2015dvaE}. 
Here we address a different problem using Kaon beams: the Kaon induced 
production of the $f_0(980)$ and $a_0(980)$ resonances. We show some 
results for the reactions $\bar{K} p\to\pi\pi Y$ and $\bar{K} p\to\pi
\eta Y$, which produce the $f_0(980)$ and $a_0(980)$ resonances, 
respectively. These two scalar resonances have generated an intense 
debate as to their nature, as $q \bar q$, tetraquarks, meson
molecules, glueballs, dynamically generated states, 
\textit{etc.}~\cite{Klempt:2007cpE}. By now it is commonly accepted 
that these mesons are not standard $q \bar q$ states but 
``extraordinary" states~\cite{jaffeE}. The advent of chiral dynamics 
in its unitarized form in coupled channels, the chiral unitary 
approach, has shown how these resonances appear from the interaction 
of pseudoscalar mesons,  with a kernel, or potential, extracted from 
the chiral Lagrangians~\cite{gasserE}, and using the coupled channels 
Bethe Salpeter equation~\cite{npaE,kaiserE,markushinE,juanitoE}, or 
similar methods, like the inverse amplitude method~\cite{ramonetE,riosE}.  
A recent review on this issue is given in~\cite{sigmaE}.

New information on these states  has come from the study of $B$ and 
$D$ decays~\cite{Aaij:2011fxE,Muramatsu:2002jpE}, which has stimulated 
much theoretical work~\cite{liangE,daiE,Dedonder:2014xpaE,Daub:2015xjaE,
Doring:2013wkaE,Wang:2015ueaE,Sekihara:2015ihaE,Liang:2014amaE}. However, 
not much has been done in reactions involving baryons, with the
exception of $f_0(980)$ photoproduction, measured in 
Refs.~\cite{Battaglieri:2008psE,Battaglieri:2009aaE}, which had been 
addressed theoretically earlier in Ref.~\cite{Marco:1999nxE}. Other 
theoretical studies have been done after the experiment 
in~\cite{daSilva:2013ykaE,Donnachie:2015jaaE}. With this perspective, 
the use of Kaon induced reactions on proton targets to produce these 
states promises to be a new good source of information which should 
help us understand better the nature of these resonances.

In chiral unitary theories the $f_0(980)$ couples strongly to $K\bar 
K$, and much less to $\pi\pi$ which becomes the decay channel. 
Similarly, the $a_0(980)$ couples both to $K \bar K$ and to $\pi\eta$, 
the latter being the decay channel. This is why the use of Kaon beams 
to produce these resonances provides a new method to test these ideas.

\item \textbf{The Chiral Unitary Approach for the $f_0(980)$ and 
	$a_0(980)$ Resonances}

Following Refs.~\cite{npaE,daniE}, The first step is the use of the 
transition potentials from the lowest order chiral Lagrangians of 
Ref.~\cite{gasserE} with the coupled channels, $\pi^+\pi^-, \pi^0\pi^0, 
\pi^0\eta, \eta\eta, K^+K^-$, $K^0\bar K^0$, for which explicit 
expressions in $s$ wave can be seen in Refs.~\cite{liangE,daiE}. By 
using the on shell factorization of the Bethe-Salpeter equation in 
coupled channels~\cite{nsdE,ollerulfE}, one finds in matrix form
\begin{equation}\label{eq:BSeq}
	 T=V+VGT; ~~~~~~~T=[1-VG]^{-1} V,
\end{equation}
where $V$ is the transition potential and $G$ the loop function for
two intermediate meson propagators. This function is regularized in 
dimensional regularization or  taking a cut off in three momenta. 
For the reactions studied on needs the $t_{K^+K^-\to\pi^+\pi^-}$, 
$t_{K^0\bar K^0\to\pi^+\pi^-}$, $t_{K^+K^-\to\pi^0\eta}$, $t_{K^0 
\bar K^0 \to \pi^0\eta}$ matrix elements, which contain a pole
associated to the $f_0(980)$ (the first two), or to the $a_0(980)$ 
(the last two).  The $a_0(980)$ appears usually as a big cusp around 
the $K\bar K$ threshold, both in the theory as in 
experiments~\cite{rubinE,adamsE}. 

\item \textbf{Formalism}

In our formalism the picture for $f_0(980)$ and $a_0(980)$ anki-Kaon 
induced production proceeds via the creation of one $K$ by the 
$\bar{K} p$ initial state in a first step followed by the 
interaction of the $K$ and $\bar{K}$ which generates the resonances. 
This is shown by the Feynman diagram  depicted in 
Fig.~\ref{fig:FeynmanDiag1}.
\begin{figure}[htbp]
\begin{center}
\includegraphics[scale=0.7]{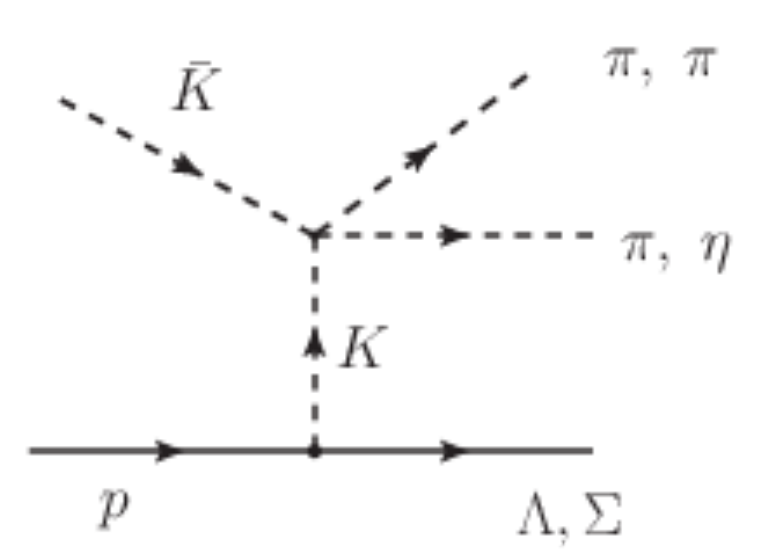}
\centerline{\parbox{0.80\textwidth}{
 \caption{Feynman diagram for the $\bar{K} p\to\pi\pi (\pi^0\eta)
	Y$ reaction. \label{fig:FeynmanDiag1}} } }
\end{center}
\end{figure}

We first take the $K^-p\to\Lambda (\Sigma^0) \pi^+\pi^- (\pi^0
\eta)$ as a reference and from this reaction we construct the 
other five reactions with minimal changes. In this case, the 
$K^-$ must couple with another $K^+$ to form the resonances. 
In this case one of the Kaons (the $K^+$) is necessarily off 
shell, which would require the use of the $K^+K^-\to\pi^+\pi^- 
(\pi^0\eta)$ amplitude with the $K^+$ leg off shell, which 
readily comes from the chiral Lagrangians. Yet, the structure 
of these Lagrangians is such that the potential can be written 
as~\cite{npaE}
\begin{eqnarray}\label{eq:potentialV}
	V_{K^+K^-\to\pi^+\pi^-} (p_{K^-}, q) &=& V^{\rm on}_{K^+K^- 
	\to\pi^+\pi^-} (M_{\rm inv}) \nonumber \\
	&& + b (q^2-m_{K^+}^2),
\end{eqnarray}
where $p_{K^-}$ and $q$ are the four momenta of $K^-$ and $K^+$
mesons, respectively, with $M_{\rm inv} = \sqrt{(p_{K^-} + q)^2}$
the invariant mass of the $K^+ K^-$ system. The off shell term 
with $b$ is unphysical while $V^{\rm on}$ is a physical quantity. 
Then, the term $b(q^2-m^2_{K^+})$ multiplied by the $K^+$ 
propagator of Fig.~\ref{fig:FeynmanDiag1} leads to a contact term 
as depicted in Fig.~\ref{fig:FeynmanDiag2}.
\begin{figure}[htbp]
\begin{center}
\includegraphics[scale=0.8]{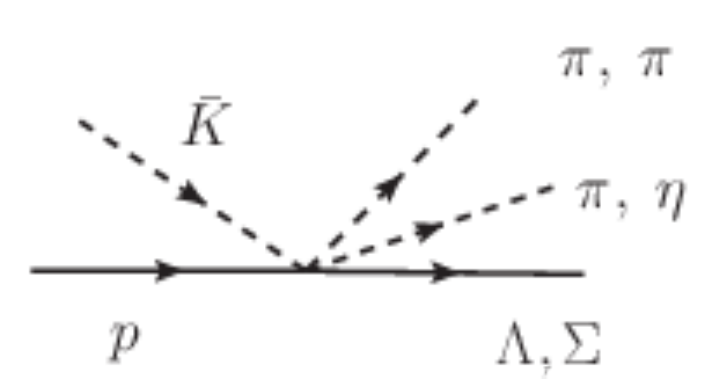}
\centerline{\parbox{0.80\textwidth}{
 \caption{Contact term stemming from the Feynman diagram of
	Fig.~\protect\ref{fig:FeynmanDiag1} from the off shell 
	part of the $K^+K^-\to\pi^+\pi^- (\pi^0 \eta)$ 
	transition potential. \label{fig:FeynmanDiag2}} } }
\end{center}
\end{figure}

The interesting things is that the same chiral Lagrangian for 
meson baryon~\cite{eckerE,ulfrepE}, expanding on the number of 
pion fields, contains a contact terms with the same topology 
as the one obtained from the off shell part of the 
amplitude~\cite{albertoE} and it cancels this off shell term.
Hence, only the physical on shell $K\bar K\to\pi\pi (\pi\eta)$ 
amplitude is needed in Fig.~\ref{fig:FeynmanDiag1} for the 
evaluation of the diagrams. These cancellations were already  
observed in Ref.~\cite{manoloE} in the study of the $\pi N\to 
\pi\pi N$ reaction and in Ref.~\cite{carmenE} for the study of 
the pion cloud contribution to the Kaon nucleus optical 
potential.

One also needs the Lagrangians for the Yukawa 
meson-baryon-baryon vertex of Fig.~\ref{fig:FeynmanDiag1} which 
is given by
\begin{eqnarray}\label{eq:LagrangianMBB}
	\mathcal{L} &=& \frac{D}{2} \langle \bar B \gamma^{\mu} 
	\gamma_5 \{u_{\mu}, B\} \rangle + \frac{F}{2} \langle 
	\bar B \gamma^{\mu} \gamma_5 [u_{\mu}, B] \rangle \nonumber \\
	&=& \frac{D+F}{2} \langle \bar B \gamma^{\mu} \gamma_5 
	u_{\mu} B \rangle + \frac{D-F}{2} \langle \bar B 
	\gamma^{\mu} \gamma_5 B u_{\mu}  \rangle,
\end{eqnarray}
where the symbol $<>$ stands for the trace of SU(3). The term 
linear in meson field gives
\begin{equation}\label{eq:umu}
	u_{\mu}\simeq -\sqrt{2} \frac{\partial_{\mu}\Phi}{f}
\end{equation}
with $f$ the pion decay constant, $f = f_{\pi} = 93$ MeV, and
$\Phi$, $B$ the meson and baryon SU(3) field matrices given by
\begin{eqnarray}
	\Phi &=& \left(
\begin{array}{ccc}
	\frac{1}{\sqrt{2}} \pi^0 + \frac{1}{\sqrt{6}} \eta & \pi^+ 
	& K^+ \\
	\pi^- & - \frac{1}{\sqrt{2}} \pi^0 + \frac{1}{\sqrt{6}} 
	\eta & K^0 \\
	K^- & \bar{K}^0 & - \frac{2}{\sqrt{6}} \eta
\end{array}
	\right), \\
	B &=& \left(
\begin{array}{ccc}
	\frac{1}{\sqrt{2}} \Sigma^0 + \frac{1}{\sqrt{6}} \Lambda &
	\Sigma^+ & p \\
	\Sigma^- & - \frac{1}{\sqrt{2}} \Sigma^0 + \frac{1}{\sqrt{6}} 
	\Lambda & n \\
	\Xi^- & \Xi^0 & - \frac{2}{\sqrt{6}} \Lambda
\end{array}
	\right).
\end{eqnarray}
We take $D = 0.795$, $F = 0.465$ ~\cite{borasoyE}. The explicit
evaluation of the SU(3) matrix elements of 
Eq.~\eqref{eq:LagrangianMBB} leads to the following expression
\begin{equation}\label{eq:vertexMBB}
	\mathcal{L}\to i \left( \alpha \frac{D+F}{2f} + \beta
	\frac{D-F}{2f} \right)\bar u(p',s'_B) \Slash q \gamma_5 
	u(p,s_B),
\end{equation}
where $u(p,s_B)$ and $\bar u(p',s'_B)$ are the ordinary Dirac
spinors of the initial and final baryons, respectively, and $p$,
$s_B$ and $p'$, $s'_B$ are the four-momenta and spins of the
baryons, while $q = p - p'$ is the four momentum of the meson. 
The values of $\alpha$ and $\beta$ are tabulated in
Table~\ref{table:alpha_beta}.
\begin{table}[htbp]
\centering 
 \caption{\small Coefficients for the $\bar{K}NY$
	couplings of Eq. (\protect\ref{eq:vertexMBB}).} 
\vspace{0.5cm}
\begin{tabular}{c|ccc}
\hline
	 & ~~~$K^-p\to\Lambda$~~~~       & $K^-p\to \Sigma^0$ ~~~~     & $K^- n\to \Sigma^-$ \\
\hline 
$\alpha$ & ~~~$-\frac{2}{\sqrt{3}}$      & 0                           & 0 \\
$\beta$  & ~~~$\frac{1}{\sqrt{3}}$       & 1                           & $\sqrt{2}$ \\
\hline\hline
         & ~~~$\bar{K}^0n\to\Lambda$~~~~ & $\bar{K}^0n\to\Sigma^0$~~~~ & $\bar{K}^0 p \to \Sigma^+$ \\
\hline
$\alpha$ & ~~~$-\frac{2}{\sqrt{3}}$      & 0                           & 0 \\
$\beta$  &~~~ $\frac{1}{\sqrt{3}}$       & $-1$                        & $\sqrt{2}$ \\
\hline
\end{tabular}
\label{table:alpha_beta}
\end{table}

Finally  the amplitude for the diagram of Fig.~\ref{fig:FeynmanDiag1} 
can be written as
\begin{equation}
	T= -i t_{K\bar K \to MM} ~\frac{1}{q^2-m_K^2} \left( 
	\alpha\frac{D+F}{2f} + \beta\frac{D-F}{2f} \right) \nonumber \\
\end{equation}
\begin{equation}\label{eq:amplitude_T}
	~\times \bar u(p',s'_{\Lambda/\Sigma}) \Slash q \gamma_5 
	u(p,s_p) F(q^2),
\end{equation}
where we have added the usual Yukawa form factor that we take 
of the form
\begin{equation}\label{eq:FormFactor}
	F(q^2) = \frac{\Lambda^2}{\Lambda^2-q^2}
\end{equation}
with typical values of $\Lambda$ of the order of 1~GeV.

The sum and average of $|T|^2$ over final and initial polarization
of the baryons is easily done and the full expression can be seen 
in~\cite{Xie:2015mzpE}.

We can write $q^2$ in terms of the variables of the external
particles and have
\begin{equation}\label{eq:q2}
	q^2 = M_p^2 + M'^2-2E E' + 2|\vec{p}| |\vec{p'}| \cos\theta,
\end{equation}
where $\vec{p},\vec{p'}$ and $E, E'$ are the momenta and energies 
of the proton and the final baryon, and $\theta$ is the angle 
between the direction of the initial and final baryon, all of 
them evaluated in the global center of mass frame (CM).

We can write the differential cross section as
\begin{equation}\label{eq:CrossSection}
	\frac{{\rm d}^2 \sigma}{{\rm d} M_{\rm inv} {\rm d}\cos 
	\theta} =
	\frac{M_p M'}{32\pi^3}\frac{|\vec{p'}|}{|\vec{p}|}
	\frac{|\vec{\tilde{p}}|}{s}
	{\overline {\sum_{s_p}}} \sum_{s'_{\Lambda/\Sigma}} |T|^2 ,
\end{equation}
with $|\vec {\tilde p}|$ the momentum of one of the mesons in the
frame where the two final mesons are at rest,

We study nine reactions:
\begin{equation}
	K^-p\to\Lambda\pi^+\pi^-, ~~ K^-p\to\Sigma^0\pi^+\pi^-, 
	~~K^-p\to\Lambda\pi^0\eta, \nonumber \\
\end{equation}
\begin{equation}\label{eq:9Reactions}
	K^-p\to\Sigma^0\pi^0\eta, ~~ K^-p\to\Sigma^+\pi^-\eta, 
	~~\bar K^0p\to\Lambda\pi^+\eta, \\
\end{equation}
\begin{equation}
	\bar K^0p\to\Sigma^0\pi^+\eta,~~\bar K^0p\to\Sigma^+ 
	\pi^+\pi^-,~~\bar K^0p\to\Sigma^+\pi^0\eta. \nonumber
\end{equation}

The Yukawa vertices for $K B B $ are summarized in
Table~\ref{table:alpha_beta}. For the $K \bar K\to MM$ 
amplitudes only the $I_3 = 0$ components, corresponding to 
zero charge, are needed here.  We have three cases with 
$\pi\eta$ where the charge is non zero, $K^-p\to\Sigma^+\pi^- 
\eta$, $\bar K^0p\to\Lambda\pi^+\eta$ and $\bar K^0p\to\Sigma^0
\pi^+\eta$, which  can be easily related to the $K^+K^-\to\pi^0 
\eta$ using isospin symmetry~\cite{Xie:2015mzpE}. 

With these ingredients, we use Eq. (\ref{eq:CrossSection}) to
evaluate the cross section in each case, changing the $t_{K\bar 
K, MM}$ in each case and the values of $\alpha$ and $\beta$. 
These magnitudes are summarized in Table~\ref{table:t_alpha_beta}.
\begin{table}[htbp]
\renewcommand{\arraystretch}{1.5}
\setlength{\tabcolsep}{0.2cm}
\centering
   \caption{Matrices $t_{K\bar K\to MM}$, $\alpha$, $\beta$ 
	used in each reaction and resonance obtained.}
\vspace{0.5cm}
\begin{tabular}{llccc}
\hline
Reaction                         & $t_{K\bar K \to MM}$ & $\alpha$ &  $\beta$  & Resonance\\
\hline
$K^-p\to\Lambda\pi^+\pi^-$       & $t_{K^+K^-\to\pi^+\pi^-}$         & $-\frac{2}{\sqrt{3}}$ & $\frac{1}{\sqrt{3}}$ & $f_0(980)$  \\
$K^-p\to\Sigma^0\pi^+\pi^-$      & $t_{K^+K^-\to\pi^+\pi^-}$         & $0$                   & $1$                  & $f_0(980)$  \\
$K^-p\to\Lambda\pi^0\eta$        & $t_{K^+K^-\to\pi^0\eta}$          & $-\frac{2}{\sqrt{3}}$ & $\frac{1}{\sqrt{3}}$ & $a_0(980)$  \\
$K^-p\to\Sigma^0\pi^0\eta$       & $t_{K^+K^-\to\pi^0\eta}$          & $0$                   & $1$                  & $a_0(980)$  \\
$K^-p\to\Sigma^+\pi^-\eta$       & $\sqrt{2}~t_{K^+K^-\to\pi^0\eta}$ & $0$                   & $\sqrt{2}$           & $a_0(980)$  \\
$\bar K^0p\to\Lambda\pi^+\eta$   & $\sqrt{2}~t_{K^+K^-\to\pi^0\eta}$ & $-\frac{2}{\sqrt{3}}$ & $\frac{1}{\sqrt{3}}$ & $a_0(980)$  \\
$\bar K^0p\to\Sigma^0\pi^+\eta$  & $\sqrt{2}~t_{K^+K^-\to\pi^0\eta}$ & $0$                   & $1$                  & $a_0(980)$  \\
$\bar K^0p\to\Sigma^+\pi^+\pi^-$ & $t_{K^0\bar K^0\to\pi^+\pi^-}$    & $0$                   & $\sqrt{2}$           & $f_0(980)$  \\
$\bar K^0p\to\Sigma^+\pi^0\eta$  & $t_{K^0\bar K^0\to\pi^0\eta}$     & $0$                   & $\sqrt{2}$           & $a_0(980)$  \\
\hline
\end{tabular}
\label{table:t_alpha_beta}
\end{table}

\item \textbf{Results}

The cross section depend on the energy, $M_{\rm inv}$, and 
the scattering angle $\theta$ given by Eq.~(\ref{eq:q2}). We
first evaluate the cross section for $\theta=0$, in the 
forward direction. In Fig.~\ref{Fig:dsigdm-kmp-f0}, we show 
the numerical results of ${\rm d} \sigma / {\rm d} M_ {\rm 
inv} {\rm d} \cos\theta$ for $\cos (\theta) =1$ as a function 
of $M_{\rm inv}$ of the $\pi^+\pi^-$ for $K^-p\to\Lambda 
(\Sigma^0) \pi^+\pi^-$ reactions. We take $\sqrt s= 2.4 
~{\rm GeV}$, which corresponds to the $K^-$ momentum 
$p_{K^-} = 2.42 ~{\rm GeV}$ in the laboratory frame. One 
can see  a clear peak around $M_{\rm inv} = 980$ MeV which 
is the signal of the $f_0(980)$ resonance  produced by the 
initial $K^+K^-$ coupled channel interactions and decaying 
into $\pi^+\pi^-$ channel. We also see that the magnitude 
of the cross section for $\Lambda$ production is about ten 
times larger than for $\Sigma^0$ production, because the 
coupling of $KN\Lambda$ is bigger than the $KN\Sigma$ one.
\begin{figure}[htbp]
\begin{center}
\includegraphics[scale=0.5]{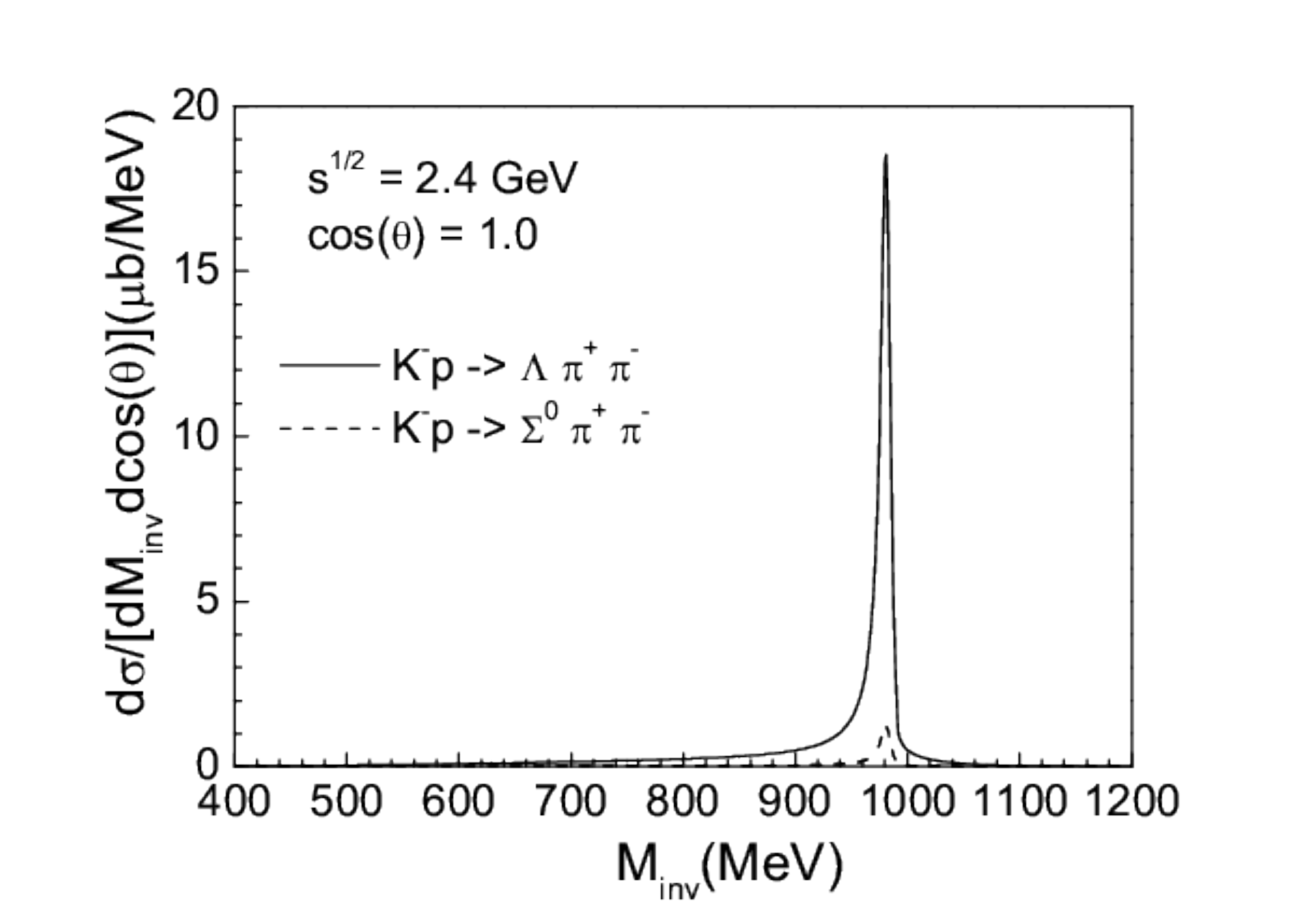}
\centerline{\parbox{0.80\textwidth}{
 \caption{Theoretical predictions for $S$ wave $\pi^+
	\pi^-$ mass distributions for $K^-p\to\Lambda 
	(\Sigma^0) \pi^+\pi^-$ reactions at $\sqrt{s} 
	= 2.4$~GeV and ${\rm cos}(\theta) = 1$.}
	\label{Fig:dsigdm-kmp-f0} } }
\end{center}
\end{figure}

In Fig.~\ref{Fig:dsigdm-kmp-a0}, we can see the numerical 
results of ${\rm d} \sigma / {\rm d} M_ {\rm inv} {\rm d} 
\cos\theta$ for $\cos (\theta) =1$ as a function of $M_{\rm 
inv}$ of the $\pi \eta$ for $K^-p\to\Lambda (\Sigma^0) \pi^0
\eta$ and $K^-p\to\Sigma^+\pi^-\eta$ reactions. In this case 
one also observes a clear peak/cusp around $M_{\rm inv} = 
980$~MeV, corresponding to the $a_0(980)$ state.
\begin{figure}[htbp]
\begin{center}
\includegraphics[scale=0.5]{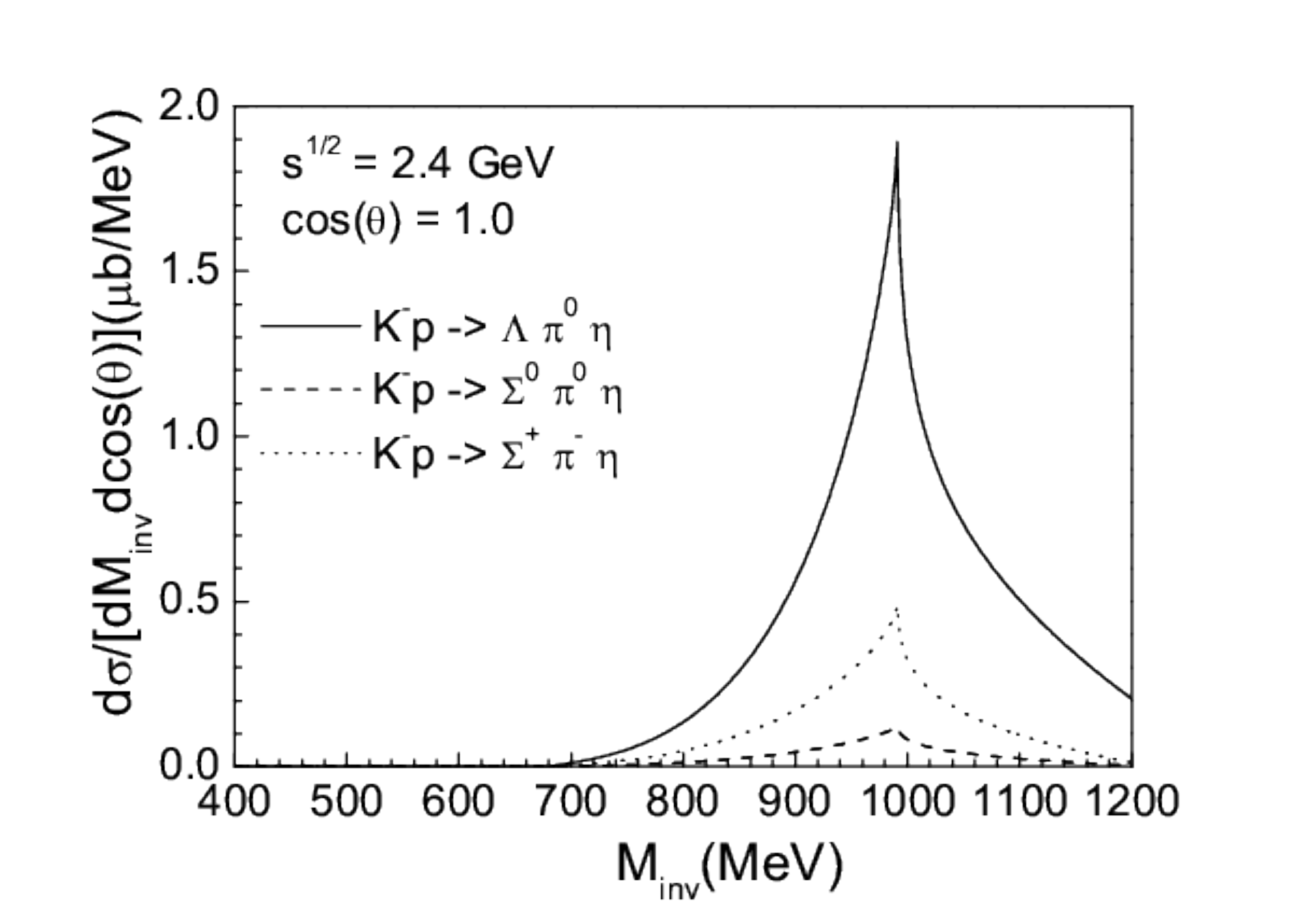}
\centerline{\parbox{0.80\textwidth}{
 \caption{Theoretical predictions for $S$ wave $\pi 
	\eta$ mass distributions for $K^-p\to\Lambda 
	(\Sigma^0)\pi^0\eta$ and $K^-p\to\Sigma^+\pi^- 
	\eta$ reactions at $\sqrt{s} = 2.4$~GeV and
	${\rm cos}(\theta) = 1$.} 
	\label{Fig:dsigdm-kmp-a0} } }
\end{center}
\end{figure}

The results for $\bar{K}^0p$ reactions are shown in
Fig.~\ref{Fig:dsigdm-k0barp}. One observes again the 
clear peaks for $a_0(980)$ and $f_0(980)$ resonances 
around $M_{\rm inv} = 980$~MeV.
\begin{figure}[htbp]
\begin{center}
\includegraphics[scale=0.5]{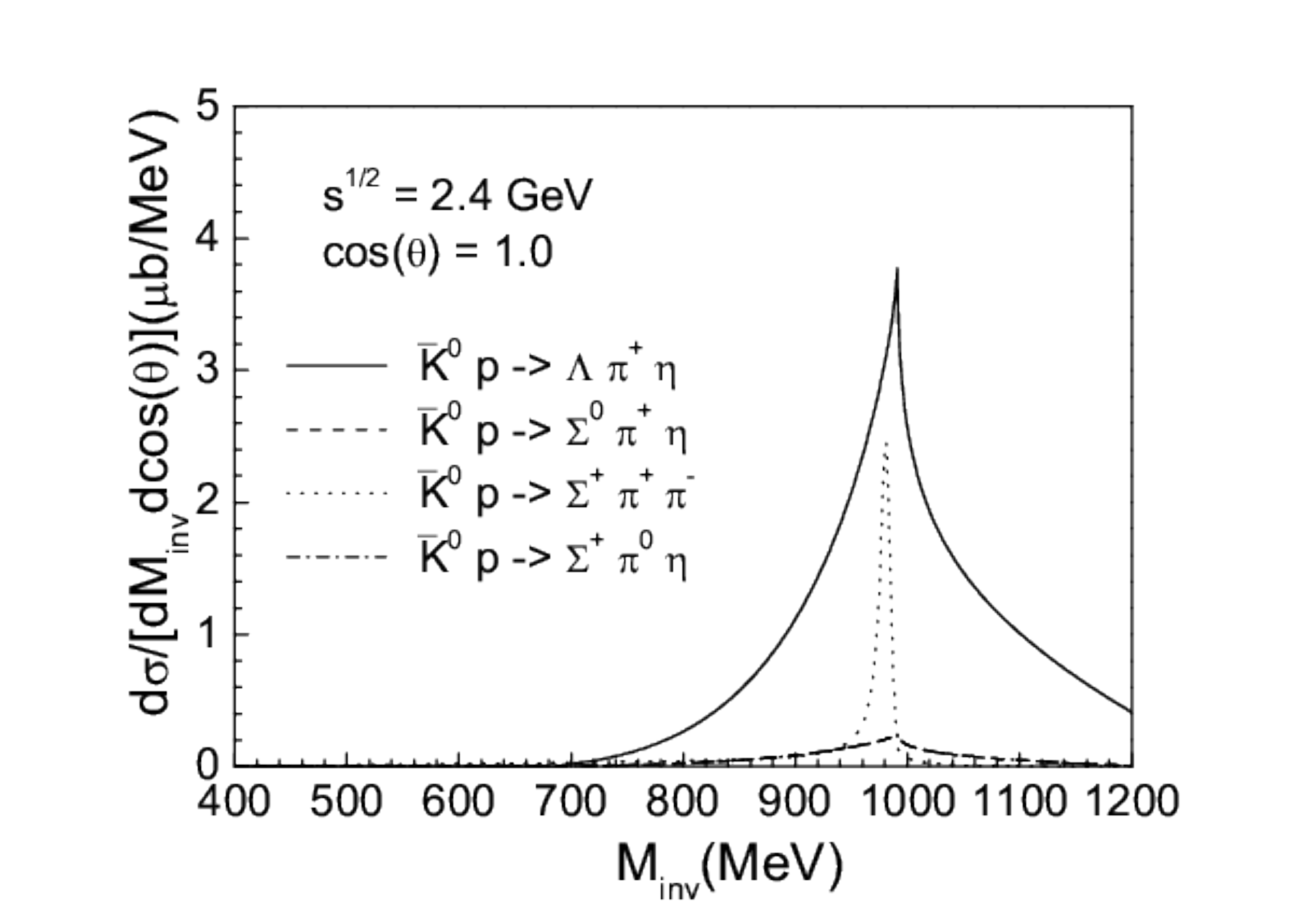}
\centerline{\parbox{0.80\textwidth}{
 \caption{Theoretical predictions for $S$ wave $\pi 
	\eta$ and $\pi^+\pi^-$ mass distributions for 
	$\bar{K}^0p\to\Lambda (\Sigma^0) \pi^+\eta$ 
	and $K^-p\to\Sigma^+\pi^0\eta (\pi^+\pi^-)$
	reactions at $\sqrt{s} = 2.4$~GeV and ${\rm 
	cos}(\theta) = 1$.} \label{Fig:dsigdm-k0barp} } }
\end{center}
\end{figure}

In all the reactions mentioned above, one observes clear 
peaks for the $f_0(980)$ with the $\pi^+\pi^-$ production 
or for the $a_0(980)$ with $\pi\eta$ production. One 
should note that in the case of the $f_0(980)$ production 
there is no signal for $f_0(500)$ ($\sigma$) production. 
This is similar to what was found in $B^0_s\to J/\psi\pi^+ 
\pi^-$, where a clear peak was seen for the $f_0(980)$ but 
no trace was observed of the $f_0(500)$~\cite{Aaij:2011fxE}. 
An explanation for this fact was given in Ref.~\cite{liangE} 
using the chiral unitary approach.

The reactions with $\pi\eta$ in the final state produce the
$a_0(980)$ resonance, with a cusp around the $K \bar K$ 
threshold, but with a large strength. 

In Figs.~\ref{Fig:dsigdm-kmp-f0-costheta} 
to~\ref{Fig:dsigdm-k0barp-costheta}, we show the results for 
${\rm d}\sigma / {\rm d} M_ {\rm inv} {\rm d} \cos\theta$ for 
the $\bar{K}p$ reactions at the peak of the invariant mass 
of, $f_0(980)$, $a_0(980)$ respectively, as a function of 
$\cos\theta$.  The reactions peak forward because we
considered only the contributions from the $t$ channel $K$ 
exchange. One can see that the cross section falls down from 
forward to backward angles by about a factor ten.
\begin{figure}[htbp]
\begin{center}
\includegraphics[scale=0.5]{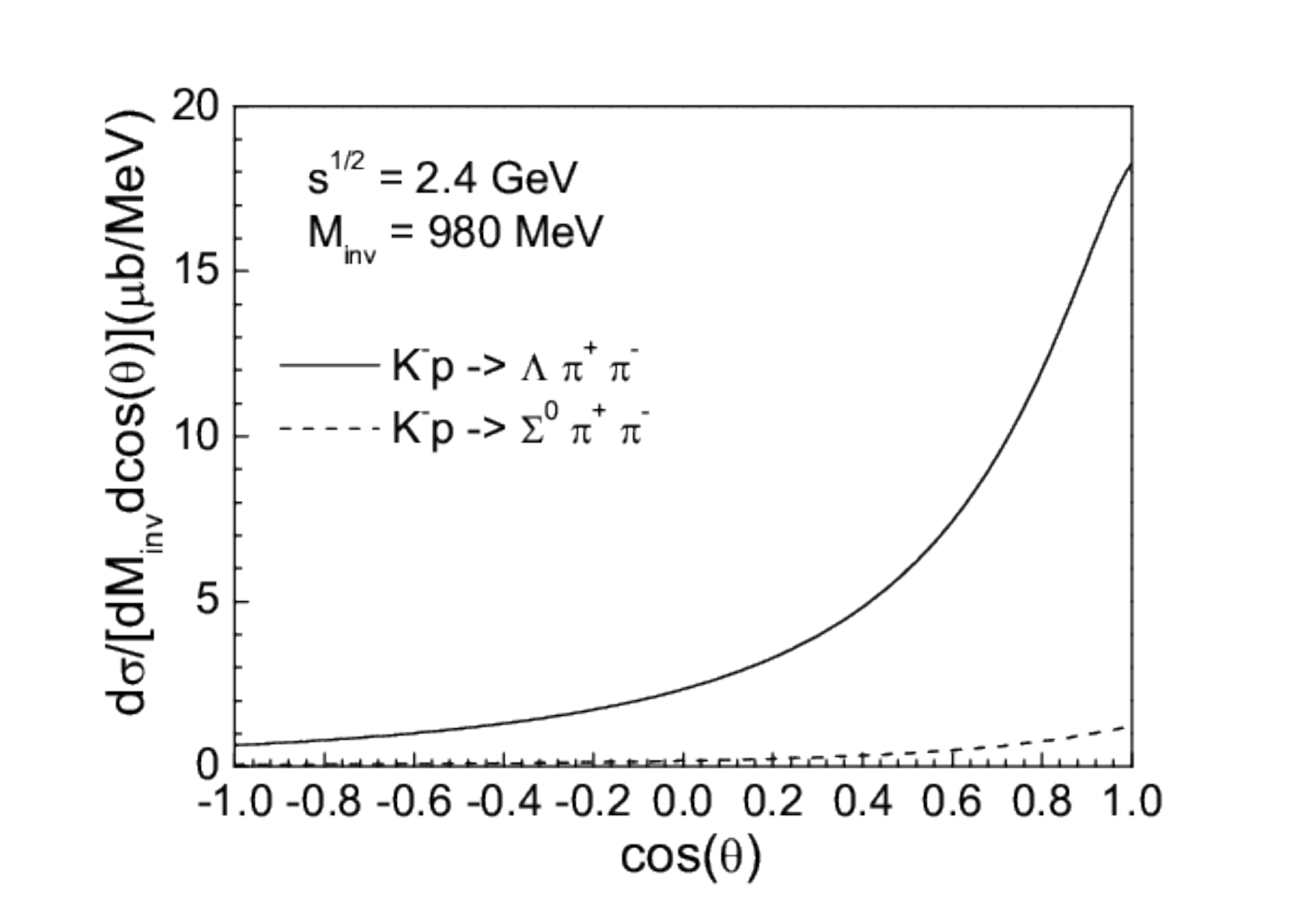}
\centerline{\parbox{0.80\textwidth}{
 \caption{Theoretical predictions for ${\rm d}\sigma / 
	{\rm d} M_{\rm inv} {\rm d} \cos \theta$ as a 
	function of ${\rm cos}(\theta)$ for $K^-p\to
	\Lambda (\Sigma^0) \pi^+\pi^-$ reactions at
	$\sqrt{s} = 2.4$~GeV and $M_{\rm inv} = 980$~MeV.}
	\label{Fig:dsigdm-kmp-f0-costheta} } }
\end{center}
\end{figure}
\begin{figure}[htbp]
\begin{center}
\includegraphics[scale=0.5]{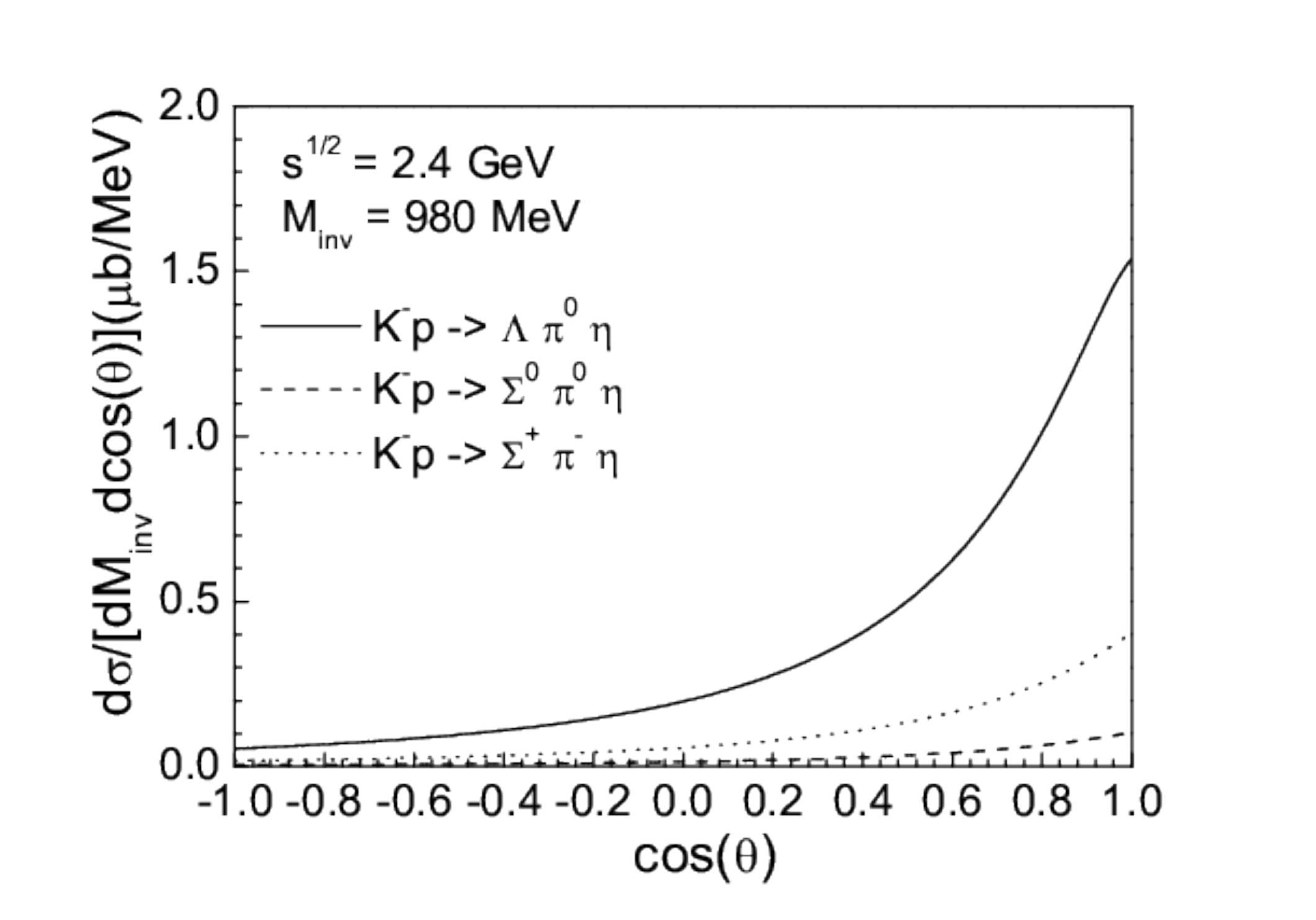}
\centerline{\parbox{0.80\textwidth}{
 \caption{Theoretical predictions for ${\rm d} \sigma / 
	{\rm d} M_{\rm inv} {\rm d} \cos \theta$ as a 
	function of ${\rm cos}(\theta)$ for $K^-p\to 
	\Lambda (\Sigma^0)\pi^0\eta$ and $K^-p\to
	\Sigma^+\pi^-\eta$ reactions at $\sqrt{s} = 
	2.4$~GeV and $M_{\rm inv} = 980$~MeV.} 
	\label{Fig:dsigdm-kmp-a0-costheta} } }
\end{center}
\end{figure}
\begin{figure}[htbp]
\begin{center}
\includegraphics[scale=0.5]{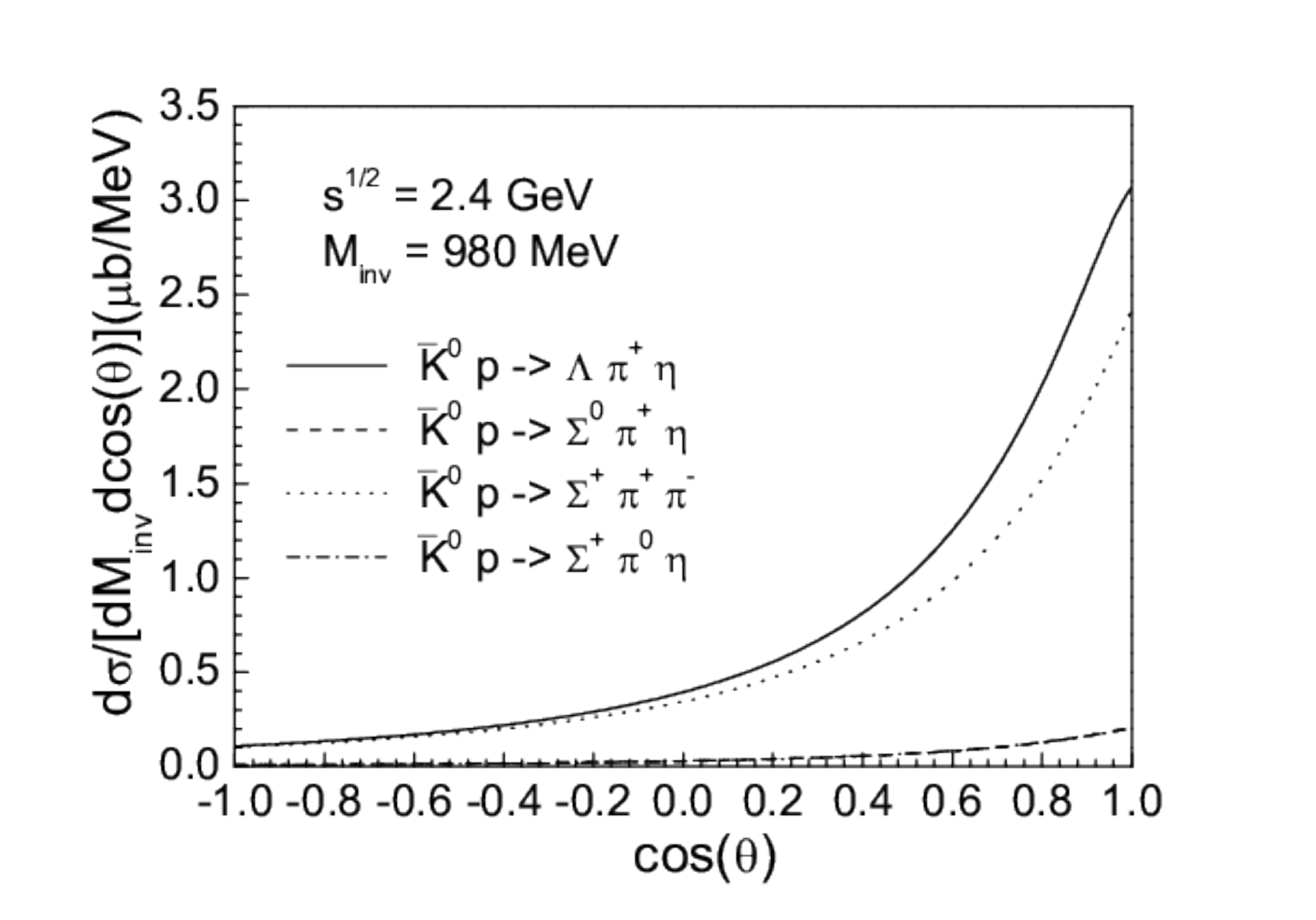}
\centerline{\parbox{0.80\textwidth}{
 \caption{Theoretical predictions for ${\rm d}\sigma / 
	{\rm d} M_{\rm inv} {\rm d} \cos \theta$ as a 
	function of ${\rm cos}(\theta)$ for $\bar{K}^0p 
	\to\Lambda (\Sigma^0) \pi^+\eta$ and $\bar{K}^0p
	\to\Sigma^+\pi^0\eta (\pi^+\pi^-)$ reactions at 
	$\sqrt{s} = 2.4$~GeV and $M_{\rm inv} = 980$~MeV.} 
	\label{Fig:dsigdm-k0barp-costheta} } }
\end{center}
\end{figure}

Finally, by fixing $M_{\rm inv} = 980$~MeV at the peak of 
the resonance and $\cos\theta =1$ we show the results by
looking at the dependence of the cross section with the 
energy of the $\bar K $ beam.  We only show the results 
for the $\Lambda$ production in 
Fig.~\ref{Fig:dsigdm-lambda-plab} because the $\Lambda$ 
production is larger than the $\Sigma$ production. One can 
observe that the cross section grows fast from the reaction 
threshold and reaches a peak around $p_{\bar K} = 2.5$~GeV.
\begin{figure}[htbp]
\begin{center}
\includegraphics[scale=0.5]{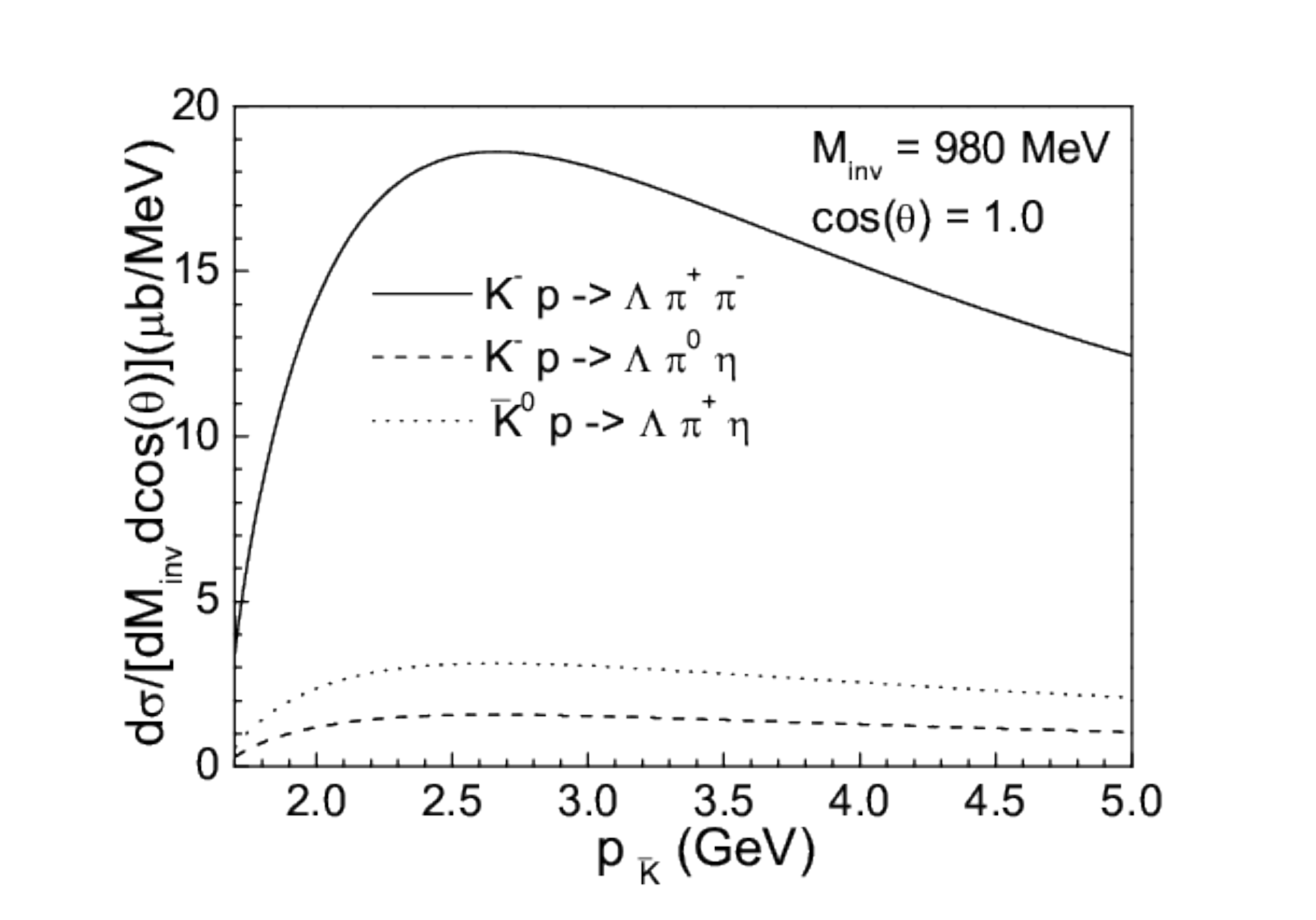}
\centerline{\parbox{0.80\textwidth}{
 \caption{Theoretical predictions for ${\rm d} \sigma / 
	{\rm d} M_{\rm inv} {\rm d}\cos\theta$ as a 
	function of $p_{\bar K}$ for $K^-p\to\Lambda\pi^+ 
	\pi^- (\pi^0\eta)$ and $\bar{K}^0p\to\Lambda\pi^+ 
	\eta$ reactions at ${\rm cos} \theta = 1$ and 
	$M_{\rm inv} = 980$~MeV.} 
	\label{Fig:dsigdm-lambda-plab} } }
\end{center}
\end{figure}

\item \textbf{Conclusions}

We have shown results of  the cross sections for the 
production of $f_0(980)$ and $a_0(980)$ resonances in 
$\bar{K} p$ reactions from the perspective that these 
two resonances are dynamically generated from the coupled
pseudoscalar-pseudoscalar channels interaction in $I = 0$ 
and $1$, respectively. These results provide the first 
evaluation of the cross section for these reactions. In 
the cases of $\pi^+\pi^-$ final states, we find a neat 
peak for the $f_0(980)$ production and no trace of the
$f_0(500)$. This feature is related to the fact that 
the resonance is produced from $K \bar K$ and the 
$f_0(980)$ has a strong coupling $K \bar K$ while the 
$f_0(500)$ has a smaller coupling to this component.  
This feature was also observed in the $B_s\to J/\psi\pi^+ 
\pi^-$ reaction and one finds a natural explanation of 
both reactions within the chiral unitary approach. It 
would be most interesting to have the reactions proposed
measured in actual experiments to bring further light on 
possible interpretations of the nature of these resonances. 

The reactions with the $\pi\eta$ production give also rise 
to a clear peak corresponding to the $a_0(980)$. This 
resonance appears as a limit of a resonance in the chiral 
unitary approach, corresponding to a state slightly unbound, 
or barely bound. Consequently, it shows up in form of a 
strong cusp around the $K\bar K$ threshold, a feature which 
is observed in recent experiments with large statistics. We 
should also note that our theoretical results provide the 
absolute strength for both the $f_0(980)$ and $a_0(980)$ 
production as a consequence of the theoretical framework 
that generates dynamically these two resonances. Comparison 
of the strength of these reactions, when measured, could 
serve to assert the accuracy of the production model that 
we have considered, and help us narrow the scope on 
pictures for the nature of the $f_0(980)$ and $a_0(980)$ 
resonances. 

\newpage
\item \textbf{Acknowledgments}

One of us, E.O., wishes to acknowledge support from the Chinese
Academy of Sciences in the Program of Visiting Professorship for
Senior International Scientists (Grant No. 2013T2J0012). This 
work is partly supported by the Spanish Ministerio de Economia 
y Competitividad and European FEDER funds under the contract 
number FIS2011--28853--C02--01 and FIS2011--28853--C02--02, and 
the Generalitat Valenciana in the program Prometeo II-2014/068. 
This work is also partly supported by the National Natural 
Science Foundation of China under Grant Nos. 11165005, 11565007 
and 11475227. This work is also supported by the Open Project 
Program of State Key Laboratory of Theoretical Physics,
Institute of Theoretical Physics, Chinese Academy of Sciences, 
China (No.~Y5KF151CJ1).
\end{enumerate}


\newpage
\subsection{TREK $@$ J-PARC: Beyond the Standard Model with Stopped 
	$K^+$}
\addtocontents{toc}{\hspace{2cm}{\sl M.~Kohl}\par}
\setcounter{figure}{0}
\setcounter{equation}{0}
\halign{#\hfil&\quad#\hfil\cr
\large{Michael Kohl}\cr
\textit{Hampton University}\cr
\textit{Hampton, VA 23668, U.S.A. \&}\cr
\textit{Thomas Jefferson National Accelerator Facility}\cr
\textit{Newport News, VA 23606, U.S.A.}\cr
\textit{(TREK/E36 Collaboration)}\cr}

\begin{abstract}
The TREK/E36 experiment has been carried out at J-PARC to 
provide a precision test of lepton universality in the 
$K_{e2}/K_{\mu2}$ ratio to search for new physics beyond 
the Standard Model. Simultaneously it will be sensitive 
to light U(1) gauge bosons and sterile neutrinos below 
300~MeV/$c^2$, which could be associated with dark matter 
or explain established muon-related anomalies such as the 
muon $g-2$ and the proton radius puzzle. The experiment 
has been set up at the J-PARC K1.1BR kaon beamline since 
fall 2014, it has been fully commissioned in spring 2015, 
and completed the accumulation of production data in fall 
2015. The experiment has used a scintillating fiber target 
to stop a beam of up to 1.2 Million $K^+$ per spill. The 
kaon decay products were detected with a large-acceptance 
toroidal spectrometer capable of tracking charged particles 
with high resolution, combined with a photon calorimeter 
with large solid angle and particle identification systems. 
The status and recent progress of the experiment will be 
presented.
\end{abstract}

\begin{enumerate}
\item \textbf{Introduction}

High precision electroweak tests represent a powerful tool 
to test the Standard Model (SM) and to obtain indirect 
hints of new physics. Lepton universality is a central 
characteristic of the SM, describing the flavor independence 
of the electroweak couplings of the charged leptons. 
Experimentally, lepton universality has been tested and 
established rather well in many processes -- for an overview 
see, \textit{e.g.}, Ref.~\cite{pich2012QQ}.  However, there 
have also been a few exceptions. Small deviations from lepton 
universality at the 2-3 $\sigma$ level have been observed in 
the $\tau$ sector~\cite{pdg2010QQ,babar2013QQ,belle2010QQ}. 
More recently, universality violation has also been reported 
in the $\mu$ sector with a 2.6 $\sigma$ deviation from lepton 
universality in the LHCb result of $B^+$ mesons decaying to 
$K^+ l^+ l^-$~\cite{LHCb2014QQ}. 

Although not strictly referring to the electroweak couplings, 
the unresolved proton radius puzzle~\cite{pohl2013QQ} can also 
be interpreted as a violation of lepton universality. The 
proton radius puzzle is the seven standard deviation 
difference in Lamb shift measurements of the proton radius 
with muonic hydrogen~\cite{pohl2010QQ,antognini2013QQ} and radius 
measurements with electrons~\cite{bernauer2010QQ,bernauer2014QQ,
codata2010QQ}.

The observation of any non-universal behavior could imply new 
physics beyond the SM. Lepton universality can be tested with 
high precision experiments that aim at measuring certain 
observables very precisely which can be calculated in the SM 
very accurately and where new physics effects would be enhanced. 
Such an observable is the ratio of leptonic decay widths of the 
charged kaon, $R_K$, defined as
\begin{equation}\label{eq:decayratio}
	R_K = \frac{\Gamma(K^+ \rightarrow e^+ \nu_e)}{\Gamma(K^+ 
	\rightarrow\mu^+ \nu_\mu)}\;.
\end{equation}  
\begin{figure}[ht!]
\includegraphics[width=\textwidth]{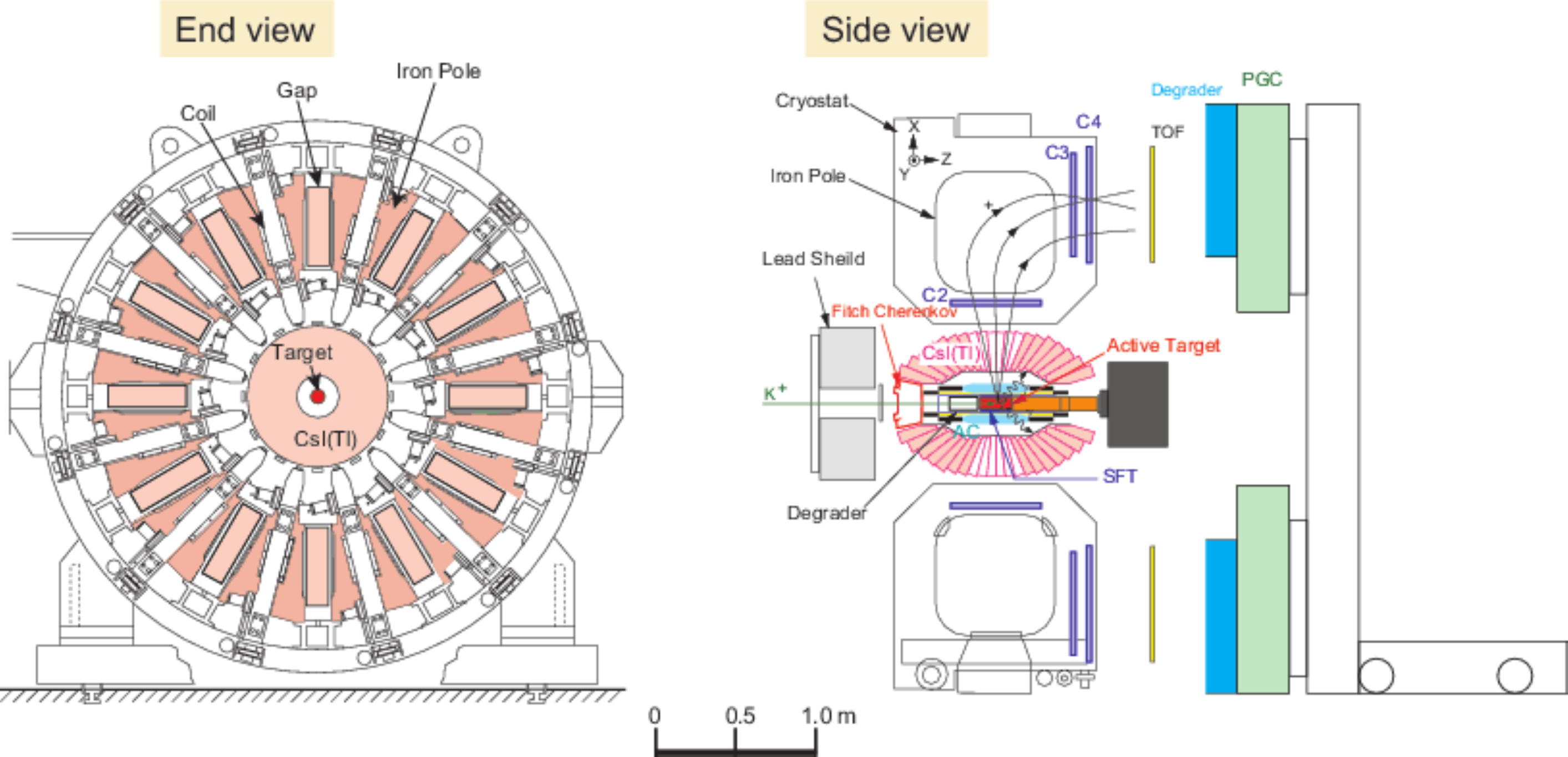}
\centerline{\parbox{0.80\textwidth}{
 \caption{\label{e36_setup}Schematic end and side views of 
	the E36 setup. See text for description.} } }
\end{figure}

Since the hadronic uncertainties associated with the form 
factor of the decay drops out in leading order from the ratio 
of decay widths, the SM prediction is very precise, as small 
as 0.04\%, giving a value of 
\begin{equation}\label{eq:rk_sm}
	R_K^{SM} = \frac{m_e^2}{m_\mu^2} \left( \frac{m_K^2-
	m_e^2}{m_K^2-m_\mu^2}
	\right ) ^2 (1 + \delta_r) =  (2.477 \pm 0.001) 
	\times 10^{-5},
\end{equation}
with a radiative correction of $\delta_r = 
-0.036$~\cite{cirigliano2007QQ}.  Moreover, the value of $R_K$ 
is helicity-suppressed, which enhances the relative size of 
any effect due to new physics.  New physics effects in the 
ratio $R_K$ have been predicted in the minimal SUSY extension 
of the SM (MSSM)~\cite{masiero06QQ,masiero08QQ,girrbach2012QQ}. In 
addition, $R_K$ is also sensitive to the neutrino mixing 
parameters within SM extensions involving a fourth generation 
of quarks and leptons~\cite{lacker2010QQ} or sterile 
neutrinos~\cite{abada2013QQ}.

The ratio $R_K$ has been measured previously with 
KLOE~\cite{kloe2009QQ} and NA62~\cite{na62_2012QQ}, averaging to 
$R_K = (2.488 \pm 0.009) \times 10^{-5}$, consistent with the 
SM value.

Experiment E36 at J-PARC has been designed to measure the ratio 
$R_K$ to test lepton universality with a total uncertainty of 
0.25\% using stopped kaons~\cite{E36-ProposalQQ}. As such it 
provides a complementary method to the in-flight decay 
measurements of NA62 and KLOE, governed by different systematics.

In addition, E36 is sensitive to byproduct searches for light 
neutral particles in the exotic decay modes $K^+\rightarrow 
\mu^+\nu V$ or $K^+\rightarrow\pi^+V$, with $V\rightarrow 
e^+e^-$. Such a particle, also known as the dark or hidden 
photon $A'$ or also dark $Z'$, could represent the hidden force 
carrier of a dark sector associated with dark 
matter~\cite{bjorken2009QQ,pospelov2009QQ}. If it is sufficiently 
light and weakly coupled to the SM, it could lead to observable 
decays into electron-positron pairs, which can be used to 
reconstruct the invariant mass of the hypothetic dark particle. 
Further, a possible resolution of the proton radius could be 
due to the existence of light neutral particles~\cite{barger2011QQ,
tucker-smith2011QQ,carlson_rislow2012QQ,carlson_rislow2014QQ,
batell2011QQ}. Such hypothetic particles can be conceived without 
violating existing constraints if they are fine-tuned and 
non-universally coupled~\cite{carlson_rislow2012QQ,
carlson_rislow2014QQ,batell2011QQ}.  In this case there would be 
a prediction of a strong observable signals in kaon decays, in 
particular in the caclulable leptonic radiative mode $K^+
\rightarrow\mu^+\nu e^+e^-$~\cite{carlson_rislow2014QQ}.

\item \textbf{Experimental Setup}

Experiment E36~\cite{E36-ProposalQQ} has been part of the TREK 
program at J-PARC, where TREK stands for Time Reversal Experiment 
with Kaons.  E36 has been mounted from fall 2014 -- spring 2015 
and completed data taking in fall 2015 at the K1.1BR beamline at 
J-PARC, using the existing E-246 apparatus~\cite{E246_setupQQ} from 
a previous T-violation search via transverse polarization of muons 
in $K^+ \rightarrow \mu^+ \pi^0 \nu_\mu$ ($K_{\mu3}$) decays at 
KEK~\cite{abe2006QQ,abe2004QQ,abe1999QQ}. A next-generation T-violation 
search in $K_{\mu3}$ decays (E06) has been proposed at J-PARC to 
take place when sufficient primary proton beam power of 100-300~kW 
becomes available~\cite{E06-ProposalQQ}. The E-246 apparatus has 
been upgraded for E36 and features
\begin{itemize}
\item a smaller-diameter scintillating fiber target to 
	stop the kaon beam to minimize multiple scattering 
	and energy loss of the outgoing decay particles,
\item redundant particle ID systems to distinguish $e$ and 
	$\mu$ with high efficiency and low misidentification 
	probability,
\item improved near-target tracking with a Spiraling Fiber 
	Tracker (SFT), and
\item a faster readout of the CsI(Tl) calorimeter with a pile-up 
	capable data acquisition system using FPGA based wave 
	form digitization.
\end{itemize}

The E36 setup is shown schematically in Fig.~\ref{e36_setup}. 
The incoming $K^+$ is tagged with the Fitch Cherenkov and 
moderated down to range out inside the active volume of the 
stopping target, a matrix of 256 scintillating fibers oriented 
longitudinally along the beam, which determines the location 
of the kaon stop in the transverse plane. The target is 
surrounded by a Spiraling Fiber Tracker (SFT), consisting of 
two pairs of fiber layers spiraling in either helicity around 
the target, providing a longitudinal coordinate of the outgoing 
decay particle~\cite{sftQQ}. The target+SFT assembly is further 
surrounded by 12 time-of-flight scintillators (TOF1) and 12 
aerogel (AC) counters~\cite{acQQ} aligned with the 12 sectors 
of the toroidal spectrometer. Photons and positron-electron 
pairs from $\pi^0$ or directly from $K^+$ decays are registered 
in the highly segmented large-acceptance CsI(Tl) calorimeter
barrel covering a solid angle of about $3\pi$. The calorimeter 
features 12 gaps aligned with the sectors of the toroid,
allowing energetic charged $\pi^+, \mu^+$, and $e^+$ to be 
momentum-analyzed through the magnetic field and tracked with 
Multi-Wire Proportional Chambers C2-C4 in each of 12 gaps of 
the magnetic toroid. At the exit of each magnet gap, another 
set of fast scintillators (TTC and TOF2), and leadglass (PGC) 
counters~\cite{pgcQQ} are providing trigger signals and particle 
identification to discriminate between $\mu$ and $e$.

Three particle identification systems allow to redundantly 
distinguish between positrons, and muons or pions: The 
threshold Aerogel Cerenkov (AC) counters sensitive to positrons 
surround the target bundle, the time of flight (TOF) is measured 
between scintillators near the target (TOF1) and in each gap 
(TTC and TOF2), and leadglass counters (PGC) are located at the 
end of each gap to identify positrons by their shower.

The data acquisition was set up for three different trigger 
types. All trigger types required a good kaon stop defined by 
the Fitch counter and minimal signals in the target.  The 
``Positron" trigger required a good kaon stop in combination 
with a good gap trigger and an additional sector aerogel hit 
for positron candidates.  The ``Muon" trigger did not have any 
gap positron requirement and was prescaled by an order of 
magnitude. The third trigger type was dedicated to the light
neutral boson search, by requiring at least three TOF1 
counters for three charged particle to be observed, in 
coincidence with a good gap trigger, but no PID constraints 
otherwise, resulting in a tolerable count rate.

\item \textbf{Status of E36}

The E36 experiment has been performed with the TREK apparatus 
at J-PARC employing a $K^+$ beam stopped in an active target 
consisting of scintillating fibers. The technique is different 
from the NA62 and KLOE experiments which used the in-flight-kaon 
decay method.  The $K_{e2}$ ($p_{e+} = 247$~MeV/{\it c}) and 
$K_{\mu 2}$ ($p_{\mu^+} = 236$~MeV/{\it c}) events were detected 
using the TREK toroidal spectrometer. In order to compare the 
experimental $R_K$ value with the SM prediction, the internal 
bremsstrahlung process in radiative $K^+ \rightarrow e^+\nu
\gamma$ ($K_{e2\gamma}^{IB}$) and $K^+\rightarrow\mu^+\nu 
\gamma$ ($K_{\mu 2\gamma}^{IB}$) decays is included into the 
$K_{e2}$ and $K_{\mu 2}$ samples, respectively.  The $R_K$ 
value is derived from the accepted $K_{e2}$ and $K_{\mu 2}$ 
events after correcting for the detector acceptance. Charged 
particles from the kaon stopping target are tracked and 
momentum-analyzed using four-point tracking with the target+SFT 
and three multi-wire proportional chambers in each toroidal 
sector. The tracking redundancy allows to determine the 
efficiency of each tracking element. The $K_{e2}$, $K_{\mu 2}$, 
and their radiative decays were collected for a central magnetic 
field of the spectrometer, $B = 1.4$~T. In order to remove 
$K^+\rightarrow\pi^0e^+\nu (K_{e3})$ and $K^+\rightarrow\pi^0 
\mu^+\nu (K_{\mu 3})$ backgrounds, the $K_{e2}$ and $K_{\mu 
2}$ events are identified by requiring the $e^+$ and $\mu^+$ 
momentum to be higher than the $K_{e3}$ and $K_{\mu 3}$ 
endpoints ($p_{max} = 228$ and 215~MeV/{\it c}). Particle 
discrimination between $e^+$ and $\mu^+$ is carried out 
using aerogel Cherenkov (AC) counters surrounding the target, 
by measuring the time-of-flight (TOF) between the TOF1 and TOF2 
scintillation counters, and by a lead glass shower calorimeter 
(PGC). The TOF1 and AC counters surround the fiber target, and 
the TOF2 and PGC counters are located at the exit of the 
spectrometer. Simulations have shown that a muon 
misidentification probability $< 10^{-6}$ is adequate and 
achievable by using the particle ID systems in combination with 
the momentum selection.
...
The $R_K = \Gamma (K_{e2}) / \Gamma (K_{\mu 2})$ ratio can be 
obtained from the number of accepted events (N), $\tilde{K}_{e 2} 
= K_{e 2} + K_{e 2}^{IB}$ and $\tilde{K}_{\mu 2} = K_{\mu 2} +
K_{\mu 2}^{IB}$, corrected for the detector acceptance. The 
acceptance ratio can be calculated by a Monte Carlo simulation, 
and dedicated calibration datasets have been taken to validate 
the simulations. It should be noted that the analysis procedure 
is identical for both $K_{e 2}$ and $K_{\mu 2}$ except for the 
particle identification in order to reduce the systematic error 
due to the analysis. The statistical error of the $R_K$ value 
will be dominated by that of the accepted $K_{e 2}$ events 
because the BR($K_{e 2})/BR(K_{\mu 2}) \approx 10^{-5}$. The 
number of $K_{e 2}$ events has been estimated to be 
$\approx 2.5 \times 10^5$ assuming 1,500 kWdays of data 
collection, corresponding to a statistical error of $\Delta R_K
 = 0.0054$ ($\Delta R_K /R_K = 0.2$\%).  At the end of data 
taking, the actually delivered integrated beam amounted to
about 1,000~kWdays.

Systematic errors have been considered due to (1) uncertainty 
of the detector acceptance ratio, (2) imperfect reproducibility 
of the experimental conditions by a Monte Carlo simulation, (3) 
performance of particle identification, and (4) background 
contamination. The total systematic error has been estimated 
with detailed simulation to be $\Delta R_K /R_K = 0.15$\% by 
adding all items in quadrature. 

For the byproduct search of light neutral bosons $A'$ in the 
processes $K^+\to\mu^+\nu_\mu A'$ or $K^+\to\pi^+ A'$ as 
possible signals from the dark sector, the $A'$ particle would 
be identified as a narrow peak in the $e^+e^-$ invariant-mass 
distribution through its decay, $A'\to e^+e^-$.  In the E36 
experiment the $e^+e^-$ pair has been detected in the CsI(Tl) 
calorimeter in coincidence with the charged muon or pion 
tracked in the toroidal spectrometer. A multiplicity of at 
least three TOF1 elements has been required for the dedicated 
trigger type, while the less biased muon trigger type was 
prescaled during data taking.

Data taking of E36 has recently been completed at J-PARC by the 
end of 2015. Two independent analyses have been started in two 
teams, one based in Japan and another in North America.

\item \textbf{Acknowledgments}

This work has been supported by DOE Early Career Award 
DE--SC0003884 and DE--SC0013941 in the US, NSERC in Canada 
and Kaken-hi in Japan.
\end{enumerate}


\newpage
\subsection{Simulation Study of $K_L$ Beam: $K_L$ Rates and 
	Background}
\setcounter{figure}{0}
\addtocontents{toc}{\hspace{2cm}{\sl I.~Larin}\par}
\halign{#\hfil&\quad#\hfil\cr
\large{Ilya Larin}\cr
\textit{Department of Physics}\cr
\textit{Old Dominion University}\cr
\textit{Norfolk, VA 23529, U.S.A.}\cr}

\begin{abstract}
We report our simulation results for $K_L$-beam and neutron 
background production, estimated rates for certain run conditions 
and resolution for  $K_L$-beam momentum.
\end{abstract}

\begin{enumerate}
\item \textbf{$K_L$ Beam Line}

Our calculations have been performed for Jefferson Lab Hall~D 
setup geometry. Primary $K_L$-production target has been placed 
in Hall~D collimator cave. For the target material, we selected 
beryllium as for thick targets $K_L$-yield roughly proportional 
to the radiation length and density, which gives beryllium as 
the best candidate. Beam plug and sweeping magnet are placed 
right after the target. For our calculations we took a simple 
beam plug: 15~cm thick piece of lead. Sweeping magnet is
cleaning up charged component and has a field integral 
2$\,$Tesla$\,$.$\,$meter, which is enough to remove all charged 
background coming out of the beam plug. Vacuum beam pipe has 
7~cm diameter and preventing neutron rescattering in air.
Where are two collimators: one placed before the wall between 
collimator cave and experimental hall, another - in front of the 
Hall~D detector. Distance between primary Be~target and liquid 
hydrogen (LH$_2$) target (located inside Hall~D detector) has 
been taken 16~m in our calculations, it can be increased upto 
20~m.

\item \textbf{$K_L$ Production}

We simulated $K_L$-production in photon bremstruhlung beam produced 
by 12~GeV electron beam in Hall D tagger amorphous radiator. We 
analyzed $K_L$-production via $\phi$-meson photoproduction in 
detail. This is one of the main mechanisms of $K_L$-production at 
our energy range. It gives the same number of $K^0$ and $\bar{K}^0$. 
Another mechanism is hyperon photoproduction (which gives only 
$K^0$) was not studied in our simulations separately. Instead, we 
have taken as an alternative model Pythia generator~\cite{pythiaL}, 
which includes hyperon production. $\phi$-meson photoproduction total 
and differential cross sections on proton and complex nuclei (coherent 
and incoherent) were taken from Refs.~\cite{phi_prod_totL,phi_prod_mecL}. 
Angular distributions for $\phi\to K_L K_S$ decay, we used are 
from Ref.~\cite{phi_prod_totL,phi_prod_ang1L,phi_prod_ang2L}. These 
calculations show that $\phi$ decay in its rest frame is going 
mostly perpendicular to the axis of $\phi$ momentum. Since $K_L$s 
need to stay on original photon beam direction to get LH$_2$ target, 
this condition requires that $\phi$ production and decay angles in 
laboratory frame should be about the same. That means we will have 
in the LH$_2$ only $K_L$s from $\phi$-mesons produced at relatively 
high $t$. It suppresses the number of "useful" $K_L$s by factor of 
$\sim$3 or more (in comparison with the case if $K_L$ and $K_S$
momenta are parallel to $\phi$ momentum). $K_L$ absorption used in 
our calculations was studied in Ref.~\cite{kl_absL} very well. About 
80\% of produced $K_L$s will be absorbed in Be target itself and 
beam plug. This value of absorbed $K_L$s can be reduced by optimizing 
beam plug setup.

\item \textbf{$K_L$ Beam Parameters}

One of the main $K_L$-beam parameters is momentum distribution.
Momentum spectrum is a function of the distance and angle. We are 
giving here resulting spectra for $K_L$ reaching LH$_2$ target. 
Results of our simulations for $K_L$ momentum spectrum is shown 
on Fig.~\ref{klcomp}. The spectrum first has increasing 
shape since $\phi$ decay cone angle decreasing at higher 
$\gamma$-beam and $K_L$ momentum. This selecting lower $\phi$ 
production $t$ values, which are more favorable according to 
$\phi$ differential cross section. At certain point highest 
possible $\gamma$-beam momentum is reached and $K_L$ momentum 
spectrum is dying out pretty fast. For comparison, we selected 
part of $K_L$ spectrum from Pythia generator originated only 
from $\phi$ decays and showed it on the same plot (red 
histogram).
\begin{figure}[ht!]
\begin{center}
\includegraphics[clip,width=0.65\textwidth]{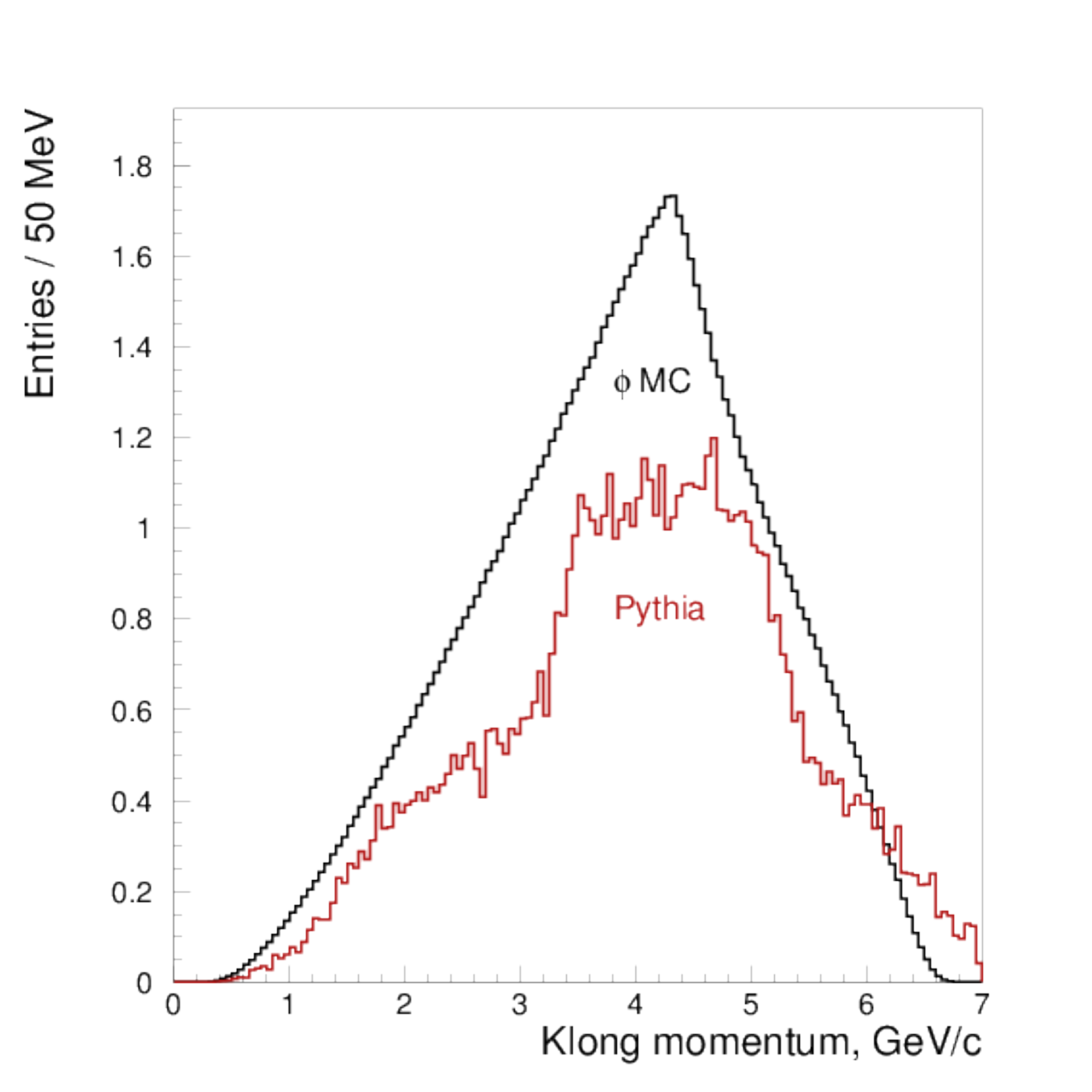}
\end{center}
\centerline{\parbox{0.80\textwidth}{
 \caption{$K_L$ momentum spectra originated from $\phi$ 
	decays: black histogram - our simulation using 
	GEANT~\protect\cite{geantL}, red - Pythia generator 
	result~\protect\cite{pythiaL}.} \label{klcomp} } }
\end{figure}

Pythia shows, that $\phi$ decays give about 30\% of $K_L$s. 
Number of $K^0$ exceeds number of $\bar{K}^0$ by 30\% 
according to this generator for our conditions. Their 
momentum spectra are shown on Fig.~\ref{k0_k0bar} separetly.
\begin{figure}[ht!]
\begin{center}
\includegraphics[clip,width=0.65\textwidth]{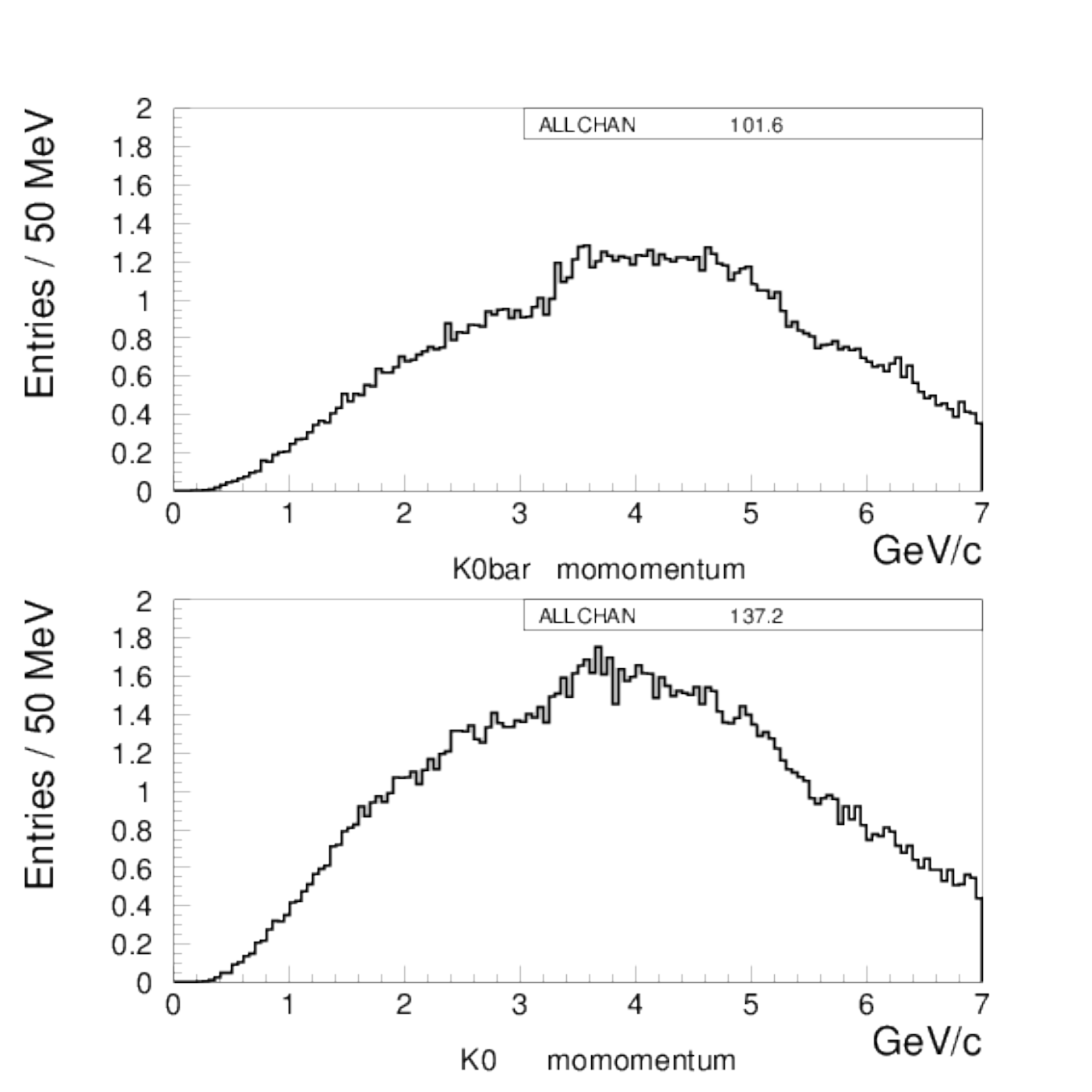}
\end{center}
\centerline{\parbox{0.80\textwidth}{
 \caption{$\bar{K}^0$ (top plot) and $K^0$ (bottom plot) 
	momentum spectra from Pythia generator.} 
	\label{k0_k0bar} } }
\end{figure}

To estimate expected rate of $K_L$ at LH$_2$ target we used 
the following conditions: electron beam current is 
3.2~$\mu$A, tagger radiator thickness is 1\% of radiation 
length, Be target thickness is 40~cm, distance Be to 
LH$_2$ target is 16~m, radius of LH$_2$ target is 2~cm. Our 
calculations are related to the $K_L$ flux at that distance 
and solid angle. For $K_L$-beam intensity under the above 
condition, we got 100 $K_L$s per second for our $\phi$ 
photoproduction simulation and 240 $K_L$s per second from 
all sources from Pythia. There are ways to increase the 
$K_L$-beam intensity by increasing tagger radiator thickness, 
electron beam current and other parameters. Increasing LH$_2$ 
target radius will increase number of $K_L$s reaching it 
proportionally to the solid angle. For example for LH$_2$ 
target radius 4~cm, electron beam current 5~$\mu$A, 5\% rad. 
length radiator and increased Be target sizes we shall be 
able to obtain beam rate about 7,000 $K_L$s per second from 
all production mechanisms at LH$_2$ target face. For 
comparison this value corresponds to  $\sim$10 million of 
produced $K_L$s in Be target per second.

\item \textbf{$K_L$ Beam Resolution}

$K_L$-beam momentum can be measured using TOF - time between 
accelerator bunch and reaction in LH$_2$ target detected by 
start counter.  Hall~D tagger timing can not be used at such 
high intensity conditions. Thus TOF resolution is a quadratic 
sum of accelerator time and start counter time resolutions.
Since accelerator signal has very good time resolution 
($\sim$0.1$\,$ns or better), TOF resolution will be defined 
by start counter. Hall~D start counter has resolution 
$\sim$0.35~ns. This value can be hopefully improved with 
upgrading counter design and parameters. In our calculations
we used an optimistic value of 0.25~ns start counter time 
resolution. Of course to get TOF information electron beam
needs to have narrow bunch time structure with the distance
between bunches at least 30~ns. At low ($<1$~GeV/$c$) 
$K_L$ momenta uncertainty in $K_L$ production point position 
within Be target will also affect TOF calculation precision. 
Fig.~\ref{res_p} shows TOF and beam momentum resolution as a 
function of $K_L$ beam momentum.
\begin{figure}[ht!]
\begin{center}
\includegraphics[clip,width=0.45 \textwidth]{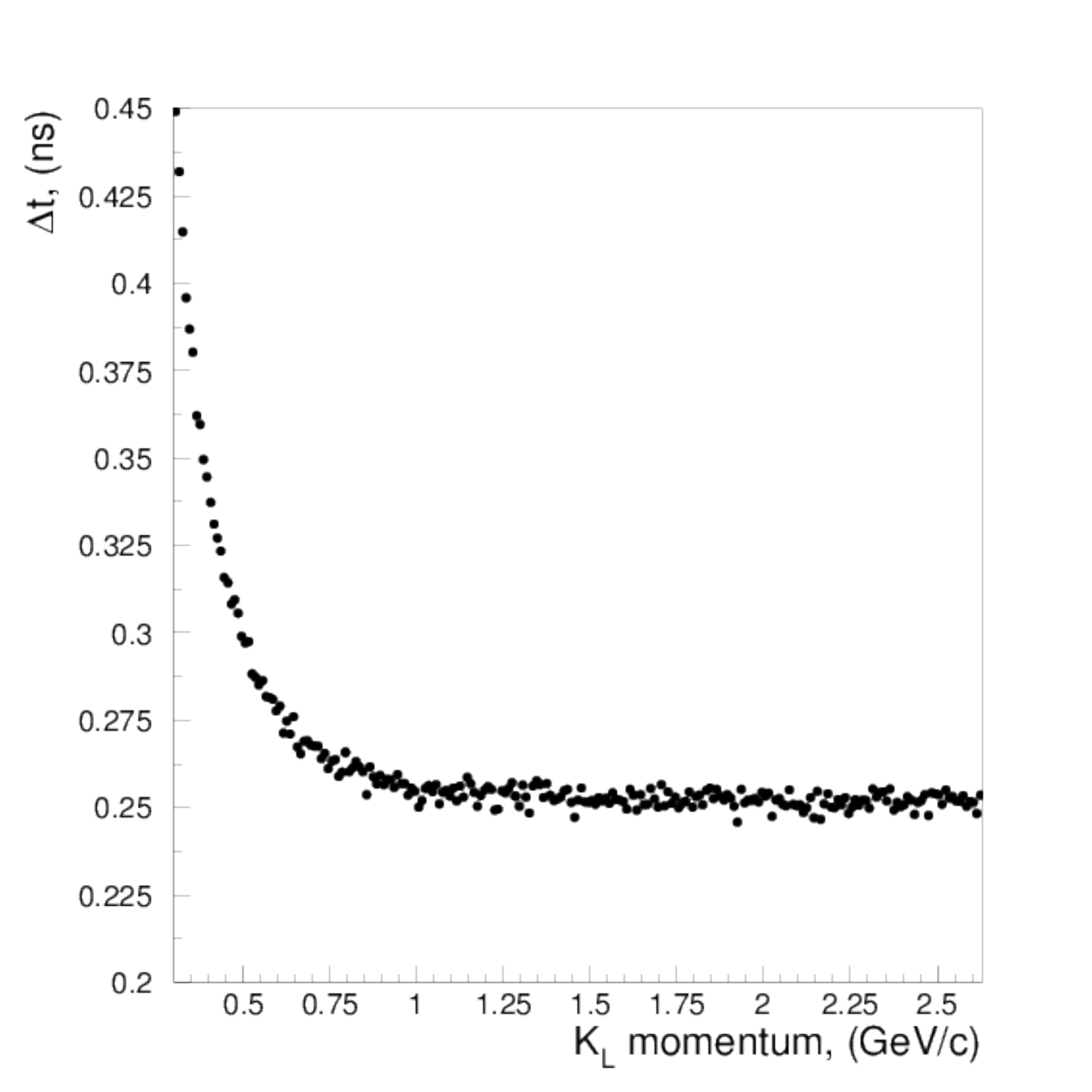}
\includegraphics[clip,width=0.45 \textwidth]{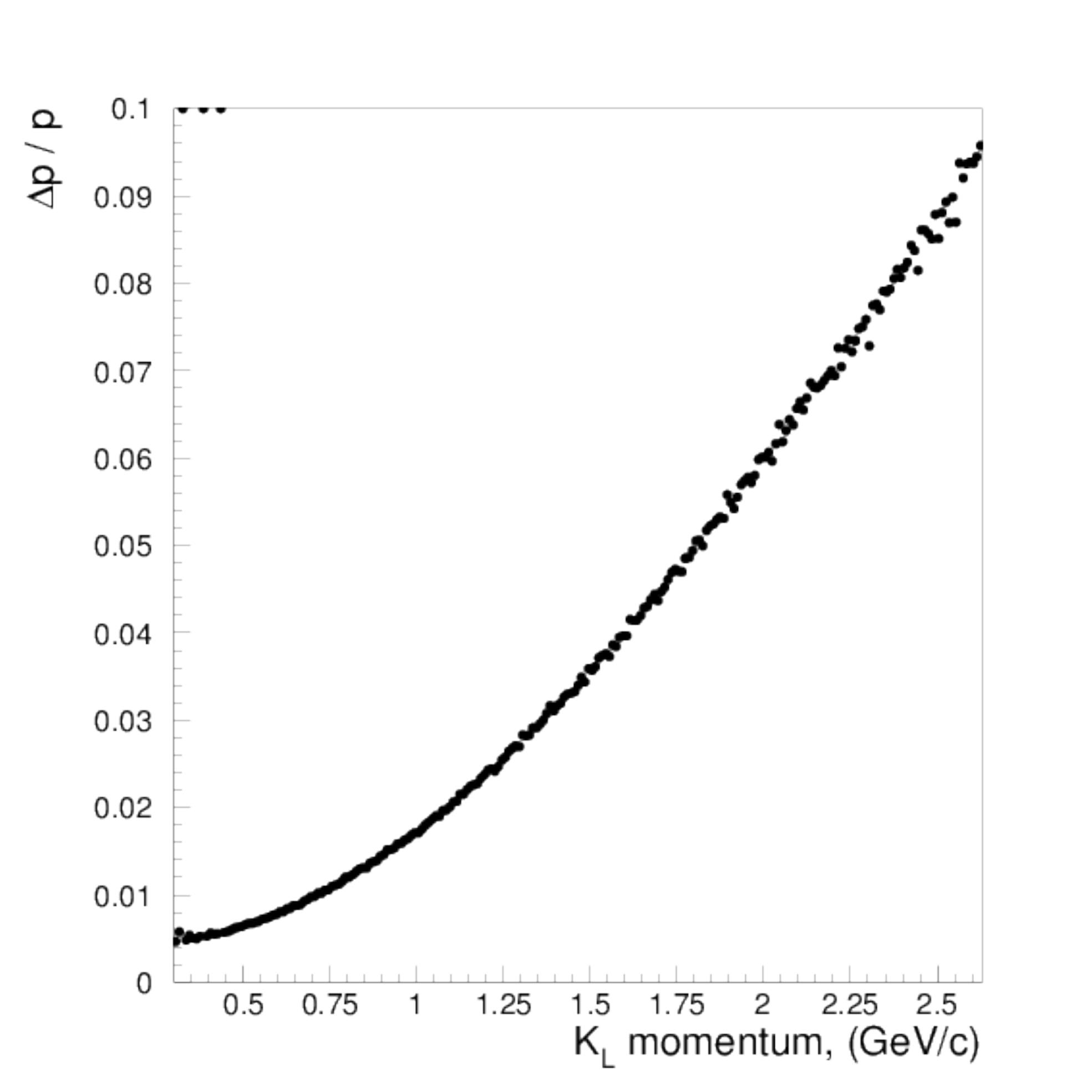}
\end{center}
\centerline{\parbox{0.80\textwidth}{
 \caption{$K_L$-beam TOF (left plot) and momentum (right 
	plot) resolution as a function of momentum.}
	\label{res_p} } }
\end{figure}

TOF resolution is flat for momenta higher that 1~GeV/$c$.
Momentum resolution is growing with momentum value, for 
1~GeV/$c$ it is $\sim$1.7\%, for 2~GeV/$c$ $\sim$6\%.

\item \textbf{$K_L$ Beam Background}

Background conditions is one of the most important parameter
of the beam. After passing through 30 radiation length beam 
plug and swiping out charged background component, we will 
have some residual $\gamma$ background and neutrons produced 
by electromagnetic showers. Momentum spectrum of residual 
$\gamma$s shown on Fig.~\ref{egam_residual_layers}.
\begin{figure}[ht!]
\begin{center}
\includegraphics[clip,width=0.65 \textwidth]{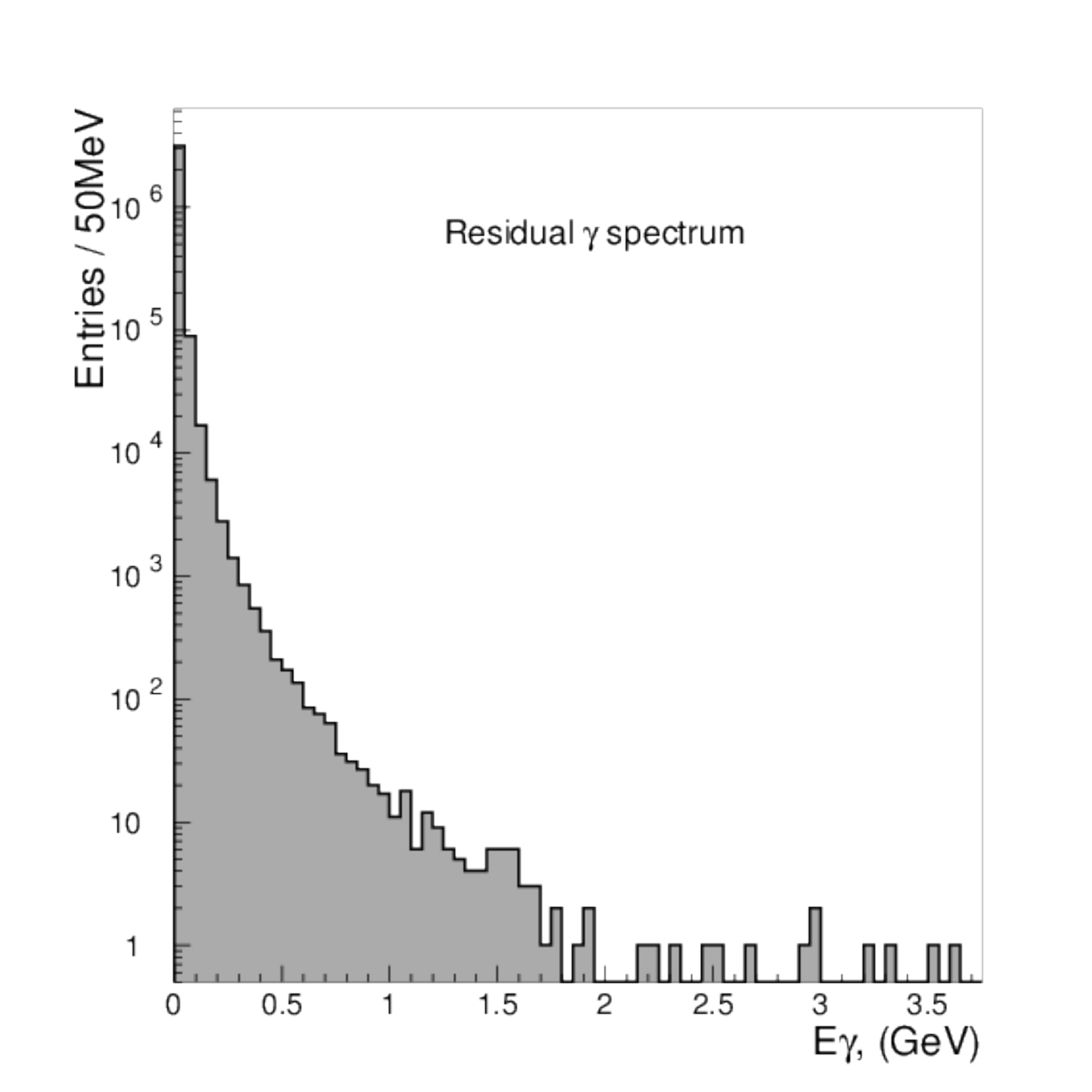}
\end{center}
\centerline{\parbox{0.80\textwidth}{
 \caption{Momentum spectrum of residual $\gamma$s.}
	\label{egam_residual_layers} } }
\end{figure}

It decreases exponentially with increasing $\gamma$ energy.
For the rates we obtained $\sim$100,000 per second for 
$\gamma$s with energy above 50~MeV and $\sim$1,000 per second 
for $\gamma$s above 500~MeV.

The most important and unpleasant background for $K_L$-beam is 
neutron background. Special care needs to be taken to estimate 
and if possible to eliminate this kind of background. In our 
calculations to estimate neutron background we used two 
independent program packages: Pythia~\cite{pythiaL} and 
DINREG~\cite{dinregL}. Both packages give the same order 
of magnitude neutron background level. At our condition it is
$\sim$140 neutrons per second at LH$_2$ target for neutrons with 
momenta higher than 500~MeV/c. These spectra along with $K_L$ 
momentum spectrum are shown on Fig.~\ref{klvsn}.
\begin{figure}[ht!]
\begin{center}
\includegraphics[clip,width=0.65 \textwidth]{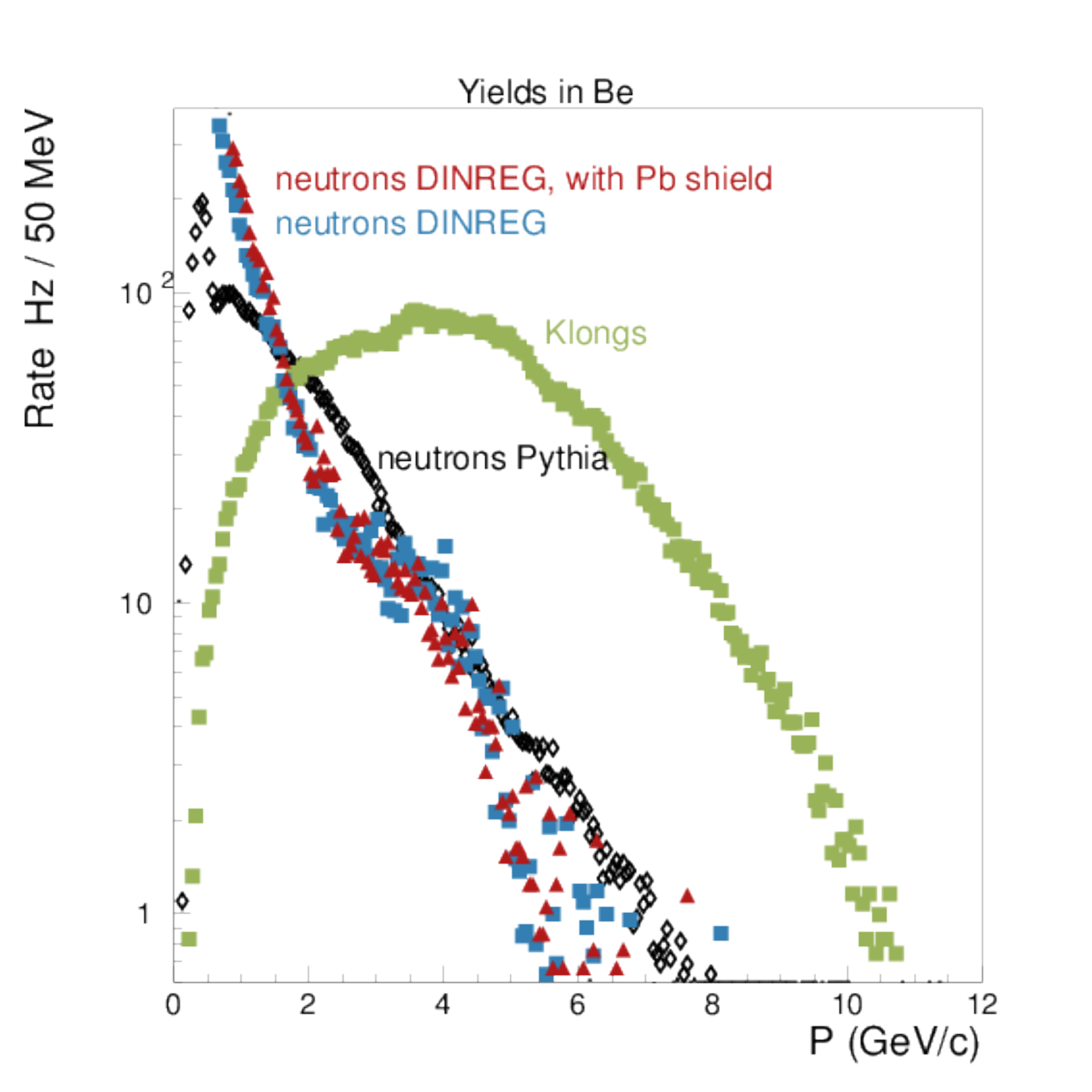}
\end{center}
\centerline{\parbox{0.80\textwidth}{
 \caption{$K_L$ and neutron momentum spectra obtained with 
	different packages.} \label{klvsn} } }
\end{figure}

Additionally we calculated muon production level. Muon will be 
swiped out of the beam line thus they are not our background. 
But since their high penetration ability it might be important 
the for purposes of the shielding. We taken into account only 
Bethe-Heitler muon production process. Muons from pion decays 
and other production mechanisms will increase total muon yield 
only slightly. They were not included in our model. Number of 
produced muon in Be target and lead beam plug is about the 
same, lead originating muons have much softer momentum spectrum. 
Estimated number of produced muons $\sim$6 million per second. 
Their momentum spectrum is shown on Fig.~\ref{mu_mom}.
\begin{figure}[ht!]
\begin{center}
\includegraphics[clip,width=0.65 \textwidth]{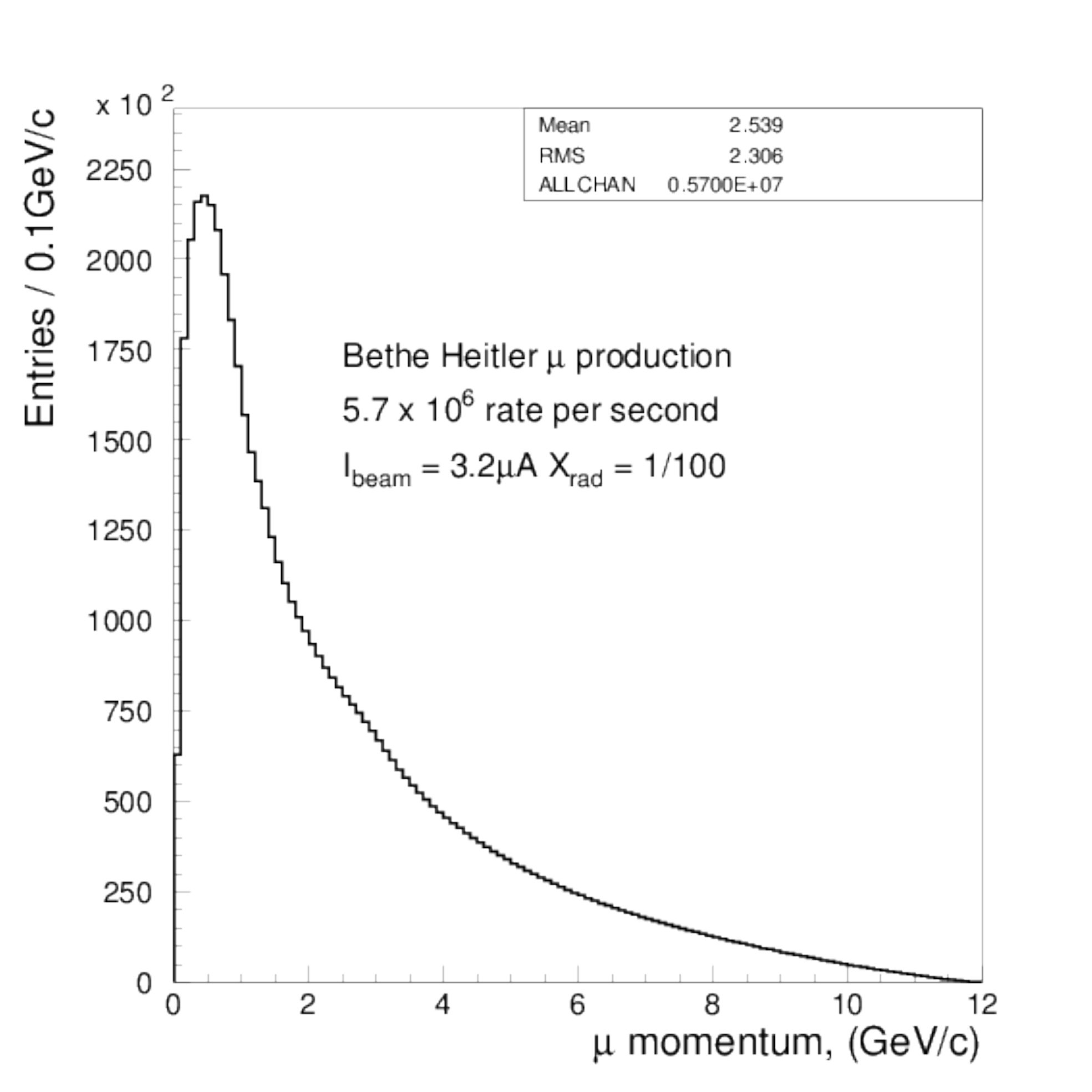}
\end{center}
\centerline{\parbox{0.80\textwidth}{
 \caption{Muon momentum spectrum for Bethe-Heitler 
	production.} \label{mu_mom} } }
\end{figure}

Half of muons have momentum higher than 2~GeV/$c$, $\sim$10\% 
of muons have momentum higher than 6~GeV/$c$ and $\sim$1\% of 
muons - momentum above 10~GeV/$c$.

\item \textbf{Summary}

In the summary part, we want to emphasize that $K_L$-beam 
facility opens horizons for new rich physics. Jefferson Lab 
GlueX spectrometer has very good acceptance and resolution 
parameters~\cite{simonL}, which perfectly fit $K_L$-beam 
facility requirements. Expected rates for $K_L$-beam with
increased $\gamma$-beam luminosity will allow to collect 
statistics order of magnitude higher than other facilities 
can provide. One of the main advantage of such facility is 
that $K_L$-beam is produced by $\gamma$-beam which provides 
low neutron background level comparing with hadron produced 
$K_L$-beam. To get more precise $K_L$-beam rates and neutron 
background estimations as well as radiation levels induced, 
it is important to conduct a few days measurements on low 
intensity test beam.

\newpage
\item \textbf{Acknowledgments}

I thank the organizing committee for the possibility to 
participate in this Workshop. I also want to express my 
special thanks to Igor Strakovsky and Moskov Amaryan for 
fruitful discussions which help me a lot in $K_L$-beam 
analysis.  This work is supported, in part, by the U.S. 
Department of Energy, Office of Science, Office of Nuclear 
Physics, under Award Number DE--FG02--96ER40960.
\end{enumerate}


\newpage
\subsection{$K_L$ Simulation Studies with the GlueX Detector}
\setcounter{figure}{0}
\setcounter{equation}{0}
\addtocontents{toc}{\hspace{2cm}{\sl S.~Taylor}\par}
\halign{#\hfil&\quad#\hfil\cr
\large{Simon Taylor}\cr
\textit{Thomas Jefferson National Accelerator Facility}\cr
\textit{Newport News, VA 23606, U.S.A.}\cr}

\begin{abstract}
Results of simulations of three reactions of interest for a
potential $K_L$ program in Hall~D at Jefferson Lab, namely
$K_Lp\rightarrow~pK_S$, $\pi^+\Lambda$, and $K^+\Xi^0$,
are presented.
\end{abstract}

\begin{enumerate}
\item \textbf{Introduction}

The GlueX detector is a large acceptance detector based
on a solenoid design with good coverage for both neutral
and charged particles.  This article describes some
simulations of events generated by $K_L$ beam particles
interacting with a liquid hydrogen target at the center
of the solenoid.  The GlueX detector is used to detect
one or all of the final state particles. I will be
focusing on a few of the simplest two-body reactions,
namely $K_Lp\rightarrow pK_S$, $K_Lp\rightarrow\Lambda
\pi^+$, and $K_Lp\rightarrow K^+\Xi^0$.

\item \textbf{Event Generation, Simulation and
        Reconstruction}

The $K_L$ beam is generated by sampling the momentum
distribution of $K_L$ particles coming from the decays of
$\phi$ mesons produced by interactions of a photon beam
with a beryllium target 16 meters upstream of the liquid 
hydrogen target.  The $K_L$ beam profile was assumed to
be uniform within a 2~cm radius at the hydrogen target.

The cross section model for the $pK_S$ channel was
determined by parameterizing fits to the existing data
for $W\le2.17$~GeV and connecting the cross section at
$W=2.17$~GeV to a power-law approximation to the cross
section for higher W.  The parametrization for low W took
the form
\begin{equation}
        \frac{d\sigma}{d\Omega}=f_0(W)P_0(\cos\theta)+f_1(W)
        P_1(\cos\theta)+f_2(W)P_2(\cos\theta),
\end{equation}
where $P_0$, $P_1$, and $P_2$ are the first three Legendre
polynomials and $f_0$, $f_1$, and $f_2$ were determined
empirically.  The high-$W$ behavior was modeled according
to the results reported in Brandenburg \textit{et al.},
\cite{Brandenburg:1973ntD}: the total cross section falls
off as function of the $K_L$ momentum $p_K$ according to
$\sigma\propto p_K^{-2.1}$ and the angular dependence for
high $W$ depended on $u^\prime=u-u_{max}$, $s$ and $t$
according to
\begin{eqnarray}
        \frac{d\sigma}{dt} & \propto & p_K^{-1.33}
        e^{\left(3.1+2.8\log{s}\right)t} ,\\
        \frac{d\sigma}{du^\prime} &\propto &  p_K^{-5.24}
        e^{5.4u^\prime}.
\end{eqnarray}
The cross section model for the $\Lambda\pi^+$ channel was
based on distributions from Yamartino~\cite{Yamartino:1974ybD}.

The cross section model for the $K^+\Xi^0$ channel was based
on parametrizations of functions from Jackson
\textit{et al.}~\cite{Jackson:2015dvaD}.

The generated events were passed through a full GEANT3-based
Monte Carlo of the GlueX detector.  The detector consists of
a solenoid magnet enclosing devices for tracking charged
particles and detecting neutral particles and a forward
region consisting of two layers of scintillators (TOF) and a
lead-glass electromagnetic calorimeter (FCAL).  A schematic
view of the detector is shown in Fig.~\ref{Fig:GlueX}. The
magnetic field at the center of the bore of the magnet for
standard running conditions is about 2~T.  The trajectories
of charged particles produced by interactions of the beam
with the 30-cm liquid hydrogen target at the center of the
bore of the magnet are measured using the Central Drift
Chamber for angles greater than $\sim20^\circ$ with respect
to the beam line.  Forward-going tracks are reconstructed
using the Forward Drift Chambers. The timing of the
interaction of the Kaon beam with the hydrogen target is
determined using signals from the Start Counter, an array of
30 thin (3~mm thick) scintillators enclosing the target
region.  Photons are registered in the central region by the
Barrel Calorimeter (BCAL).
\begin{figure}[ht!]
\begin{center}
\epsfig{file=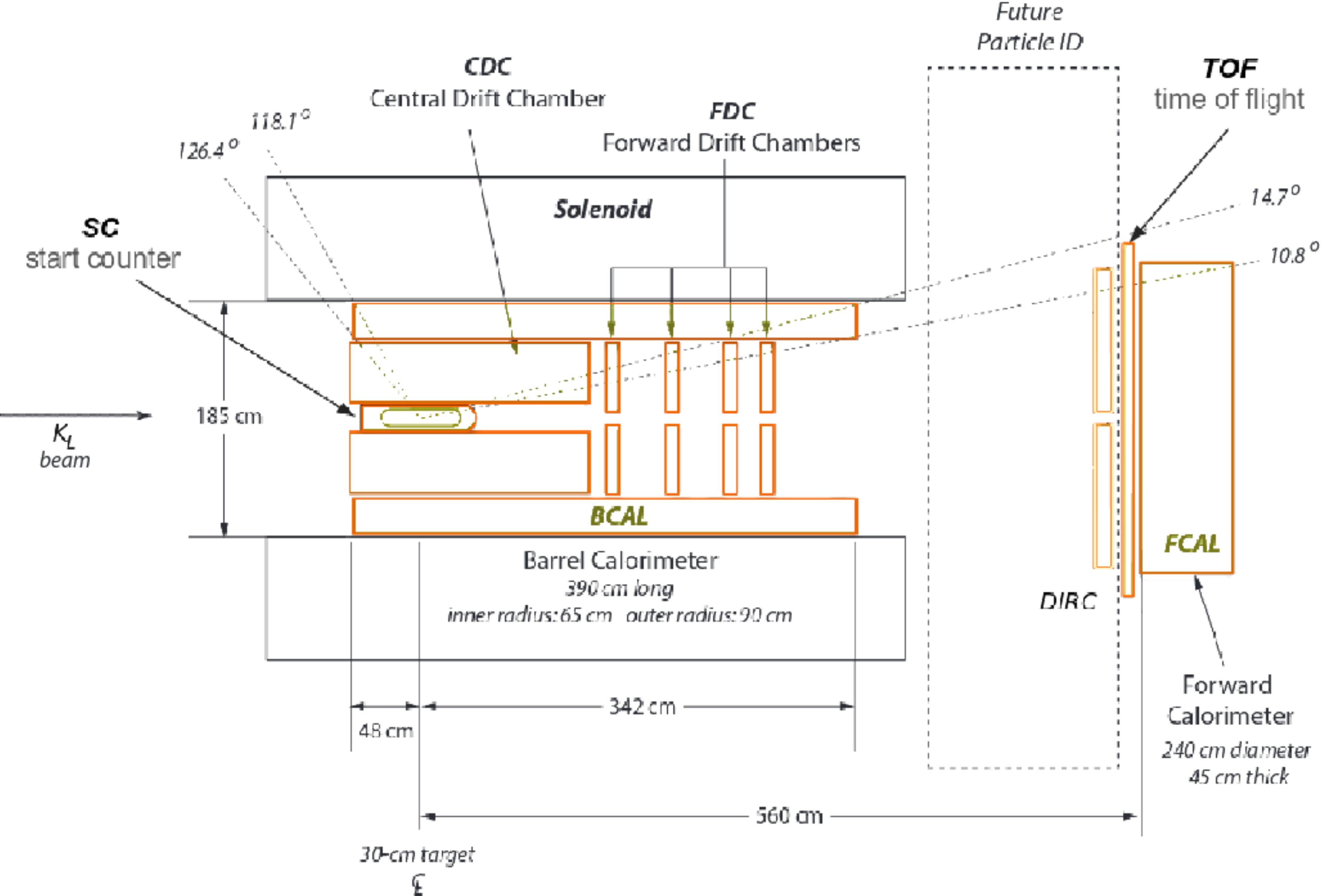,width=\linewidth}
\end{center}
\centerline{\parbox{0.80\textwidth}{
 \caption{Schematic view of the GlueX detector.}
        \label{Fig:GlueX} } }
\end{figure}

For each topology, one particle (the proton for the $pK_S$
channel, the $\pi^+$ for the $\Lambda\pi^+$ channel and the
$K^+$ for the $K^+\Xi^0$ channel) provides a rough
determination for the position of the primary vertex along
the beam line that is used in conjunction with the start
counter to determine the flight time of the $K_L$ from the
beryllium target to the hydrogen target. Protons, pions,
and Kaons are distinguished using a combination of $dE/dx$
in the chambers and time-of-flight to the outer detectors
(BCAL and TOF).  The energy loss and timing distributions
for the $pK_S$ channel are shown in Fig.~\ref{Fig:PID}; the
distributions are similar for the $\Lambda\pi^+$ channel,
where a proton band arises from the $\Lambda\rightarrow p
\pi^-$ decay channel.  Also shown is the $dE/dx$
distribution for the $K^+\Xi^0$ channel, where a prominent
Kaon band can be seen, along with pion and proton bands
arising from $\Lambda$ decays.
\begin{figure}[ht!]
\begin{center}
\epsfig{file=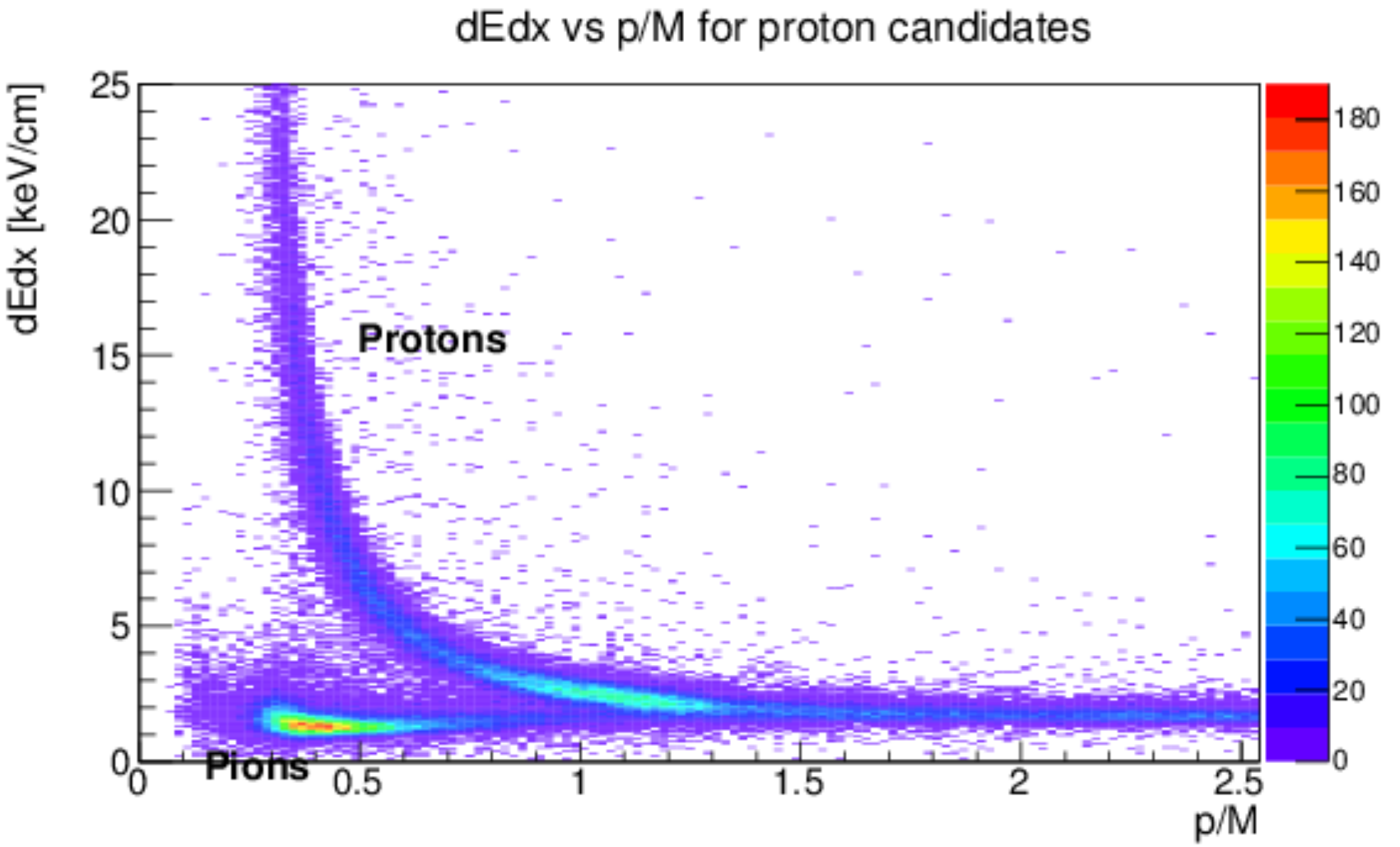,width=3.0in}
\epsfig{file=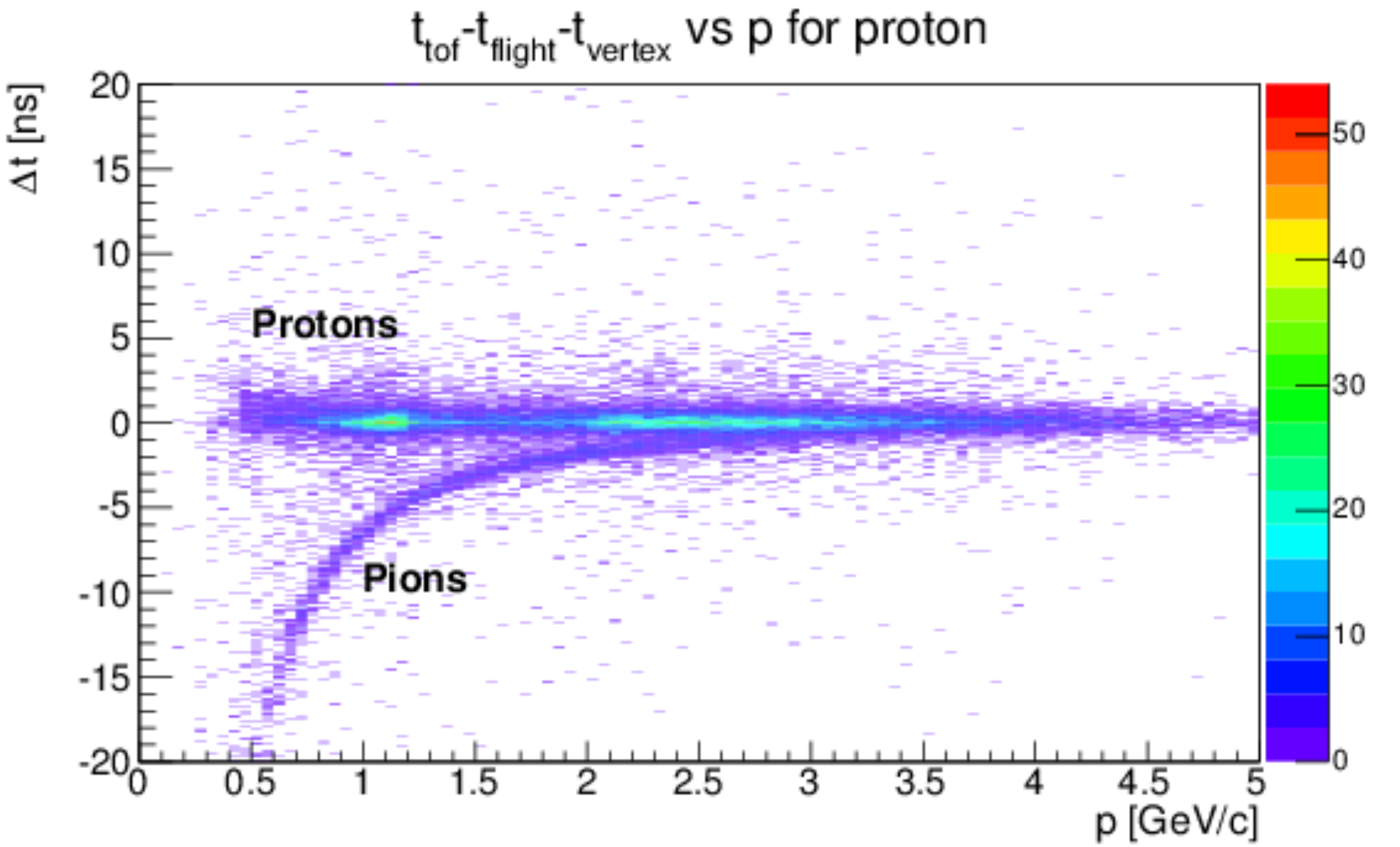,width=3.0in}
\epsfig{file=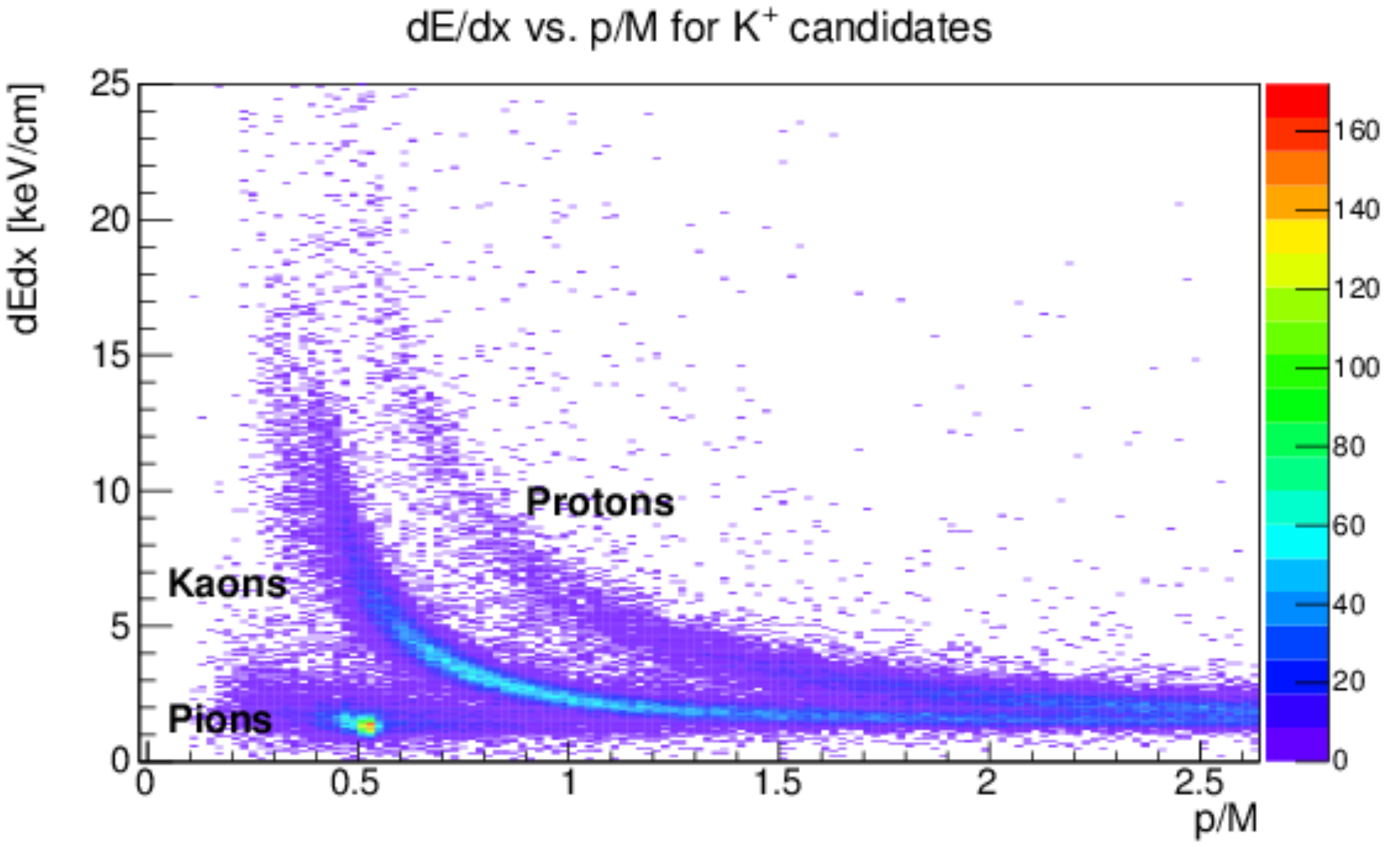,width=3.0in}
\end{center}
\centerline{\parbox{0.80\textwidth}{
 \caption{Particle identification:  (top left) $dE/dx$ for
        the $pK_S$ channel; (top right) time difference at
        the primary ``vertex'' for the proton hypothesis 
        for the $pK_S$ channel using the TOF; (bottom)
        $dE/dx$ for the $K^+\Xi^0$ channel.  The
        proton and pion bands arise from the decay of the
        $\Lambda$. } \label{Fig:PID} } }
\end{figure}

\item \textbf{Results for each Topology}

\begin{enumerate}
\item \textbf{$K_Lp\rightarrow pK_S$}

The $K_L$ momentum distribution for the $pK_S$ channel and
the mass distribution for the $K_S$ recoiling against the
proton are shown in Fig.~\ref{Fig:pks klong}.  The missing
mass distribution suffers from long non-Gaussian tails.
\begin{figure}[ht!]
\begin{center}
\epsfig{file=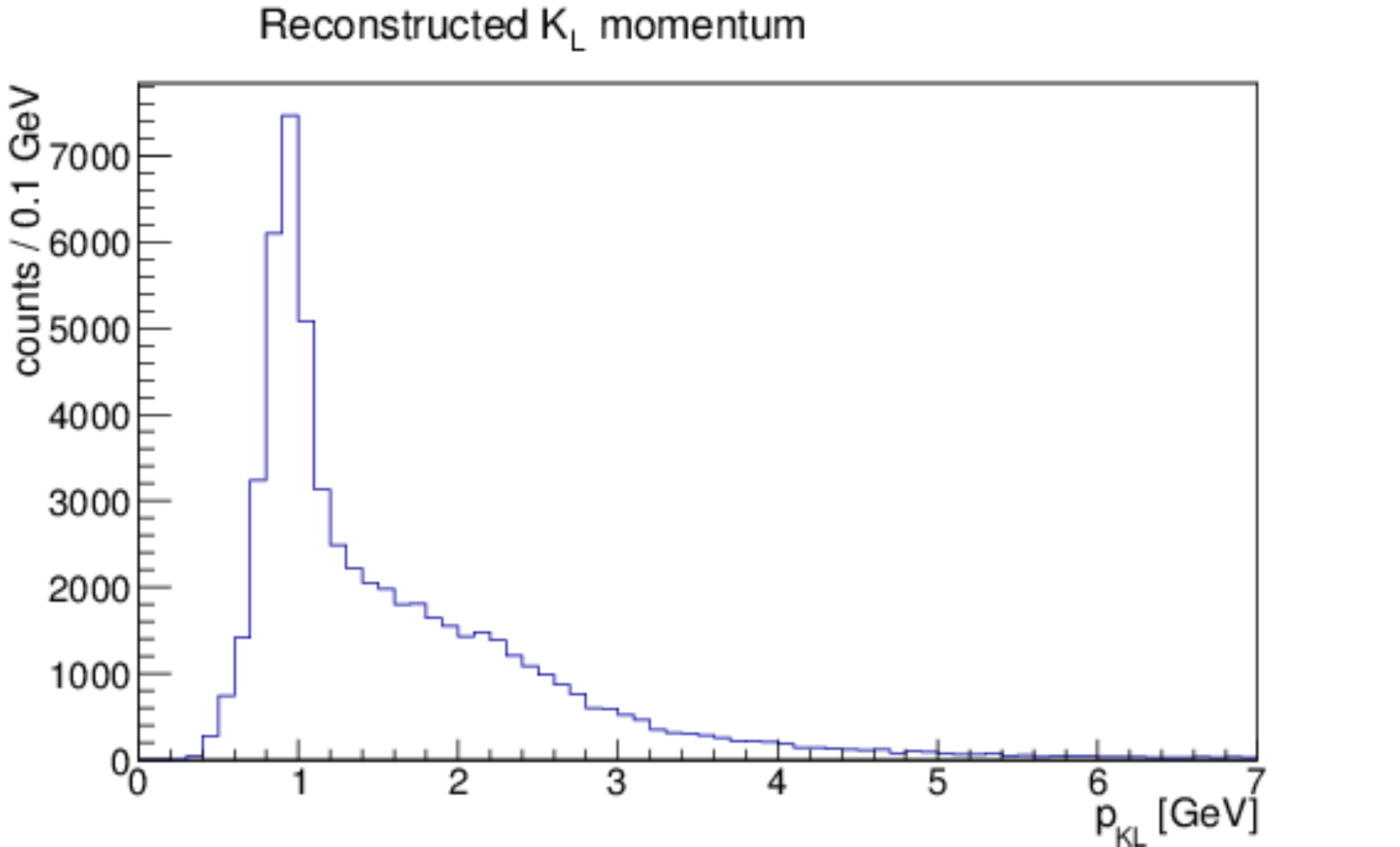,width=3.0in}
\epsfig{file=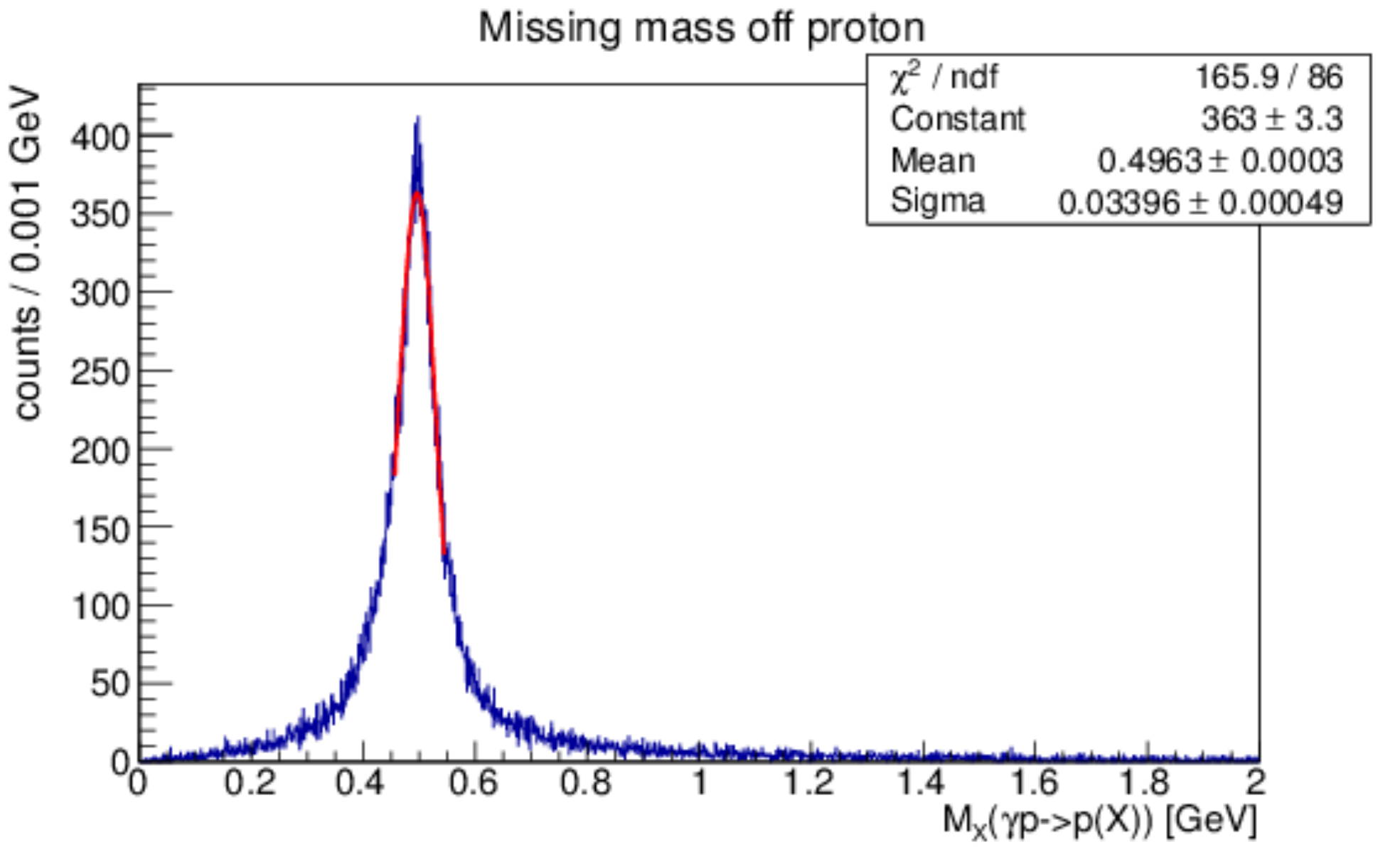,width=3.0in}
\end{center}
\centerline{\parbox{0.80\textwidth}{
 \caption{$K_L$ momentum distribution (left) and missing
        mass off the proton (right) for the $pK_S$ channel.}
        \label{Fig:pks klong} } }
\end{figure}

Since the GlueX detector has full acceptance in $\phi$ for
charged particles and large acceptance in $\theta$ (roughly
$1^\circ\sim140^\circ$), reconstruction of full events is
feasible.  For the $pK_S$ channel, I take advantage of the
branching ratio of 69.2\% for $K_S\rightarrow\pi^+
\pi^-$~\cite{Agashe:2014kdaD}: the invariant mass of the 
$\pi^+\pi^-$ pair and $W$ as computed from the four-momenta
of the proton and the two pions is shown in
Fig.~\ref{Fig:pks full reaction}.  After combining the
four-momenta of the final state particles  with the
four-momenta of the beam and the target, the missing mass
squared for the full reaction should be zero, which is also
shown in Fig.~\ref{Fig:pks full reaction}.  A comparison
between two methods for computing $W$, one using the $K_L$
momentum and the other using the final state particles, is
shown in Fig.~\ref{Fig:pks w resolution}.  Finally, I
require conservation of energy and momentum in the reaction
by applying a kinematic fit to the data. After applying a
0.1 cut on the confidence level of the fit, I computed an
 estimate for the reconstruction efficiency as a function
of $W$ as shown in Fig.~\ref{Fig:pks efficiency}.  The
efficiency is $\varepsilon=N(W,reconstructed)/N(W,thrown)$.
Here the efficiency includes the branching ratio for
$K_S\rightarrow\pi^+\pi^-$.  The average reconstruction
efficiency is about 7\%.
\begin{figure}[ht!]
\begin{center}
\epsfig{file=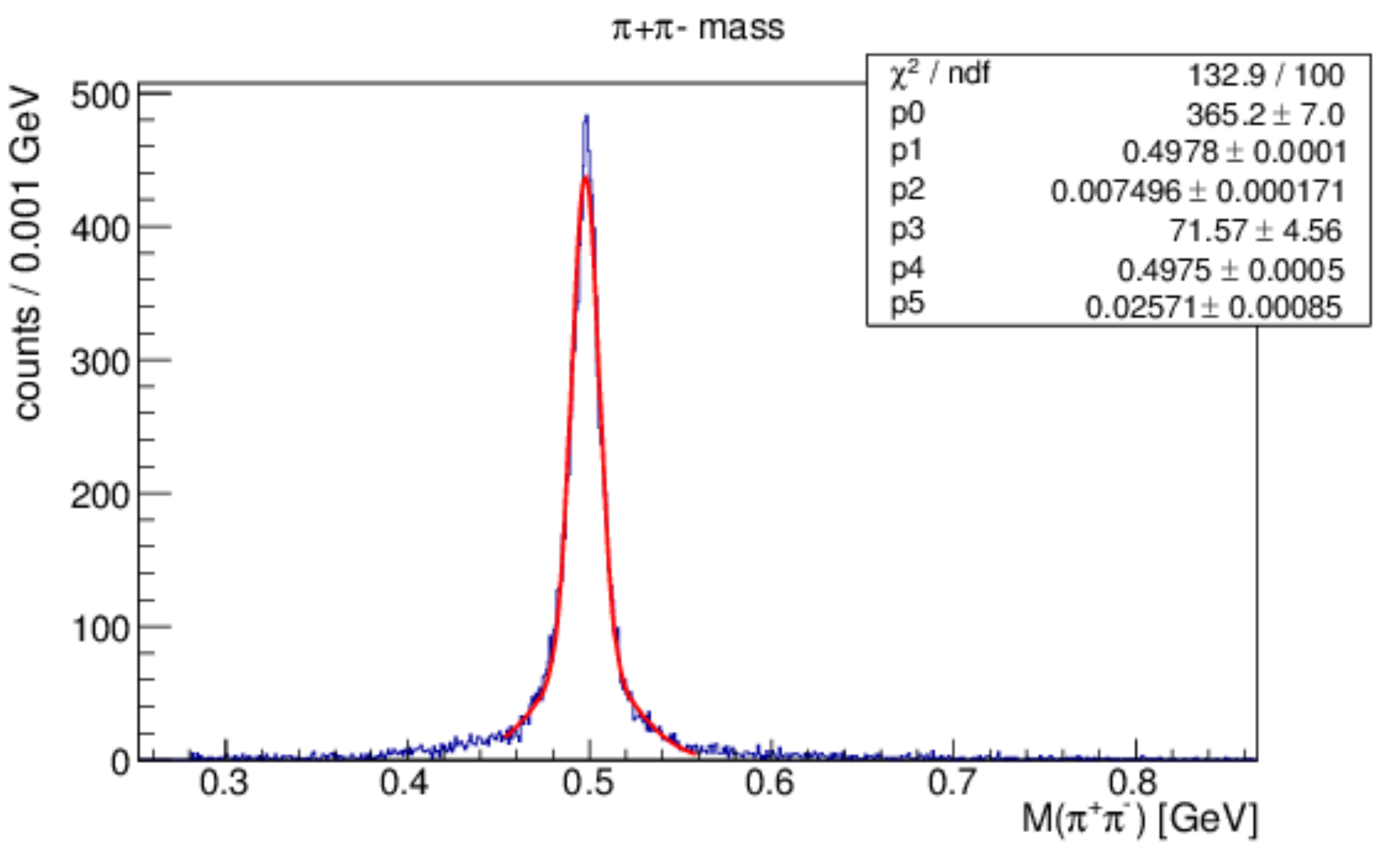,width=3.0in}
\epsfig{file=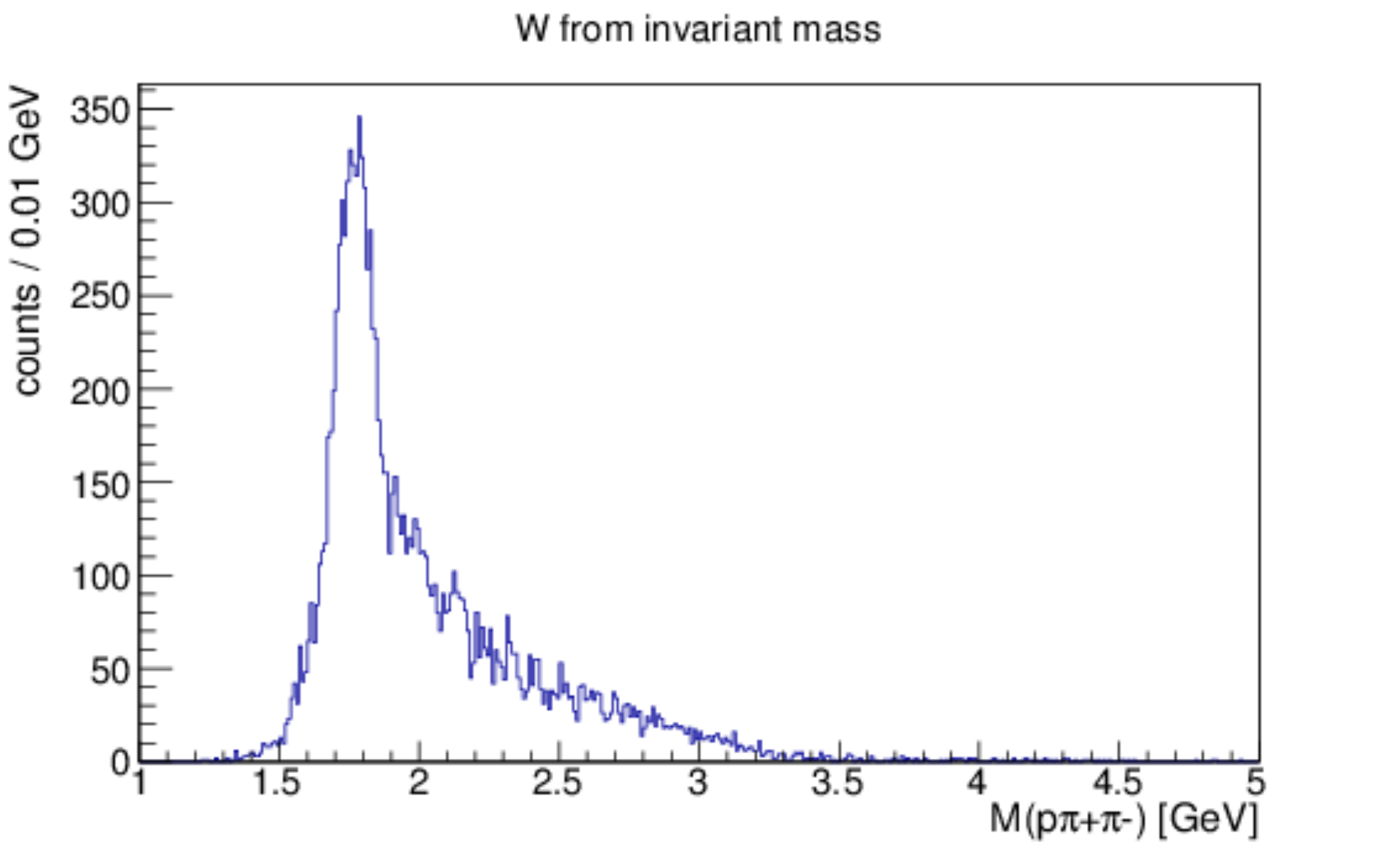,width=3.0in}
\epsfig{file=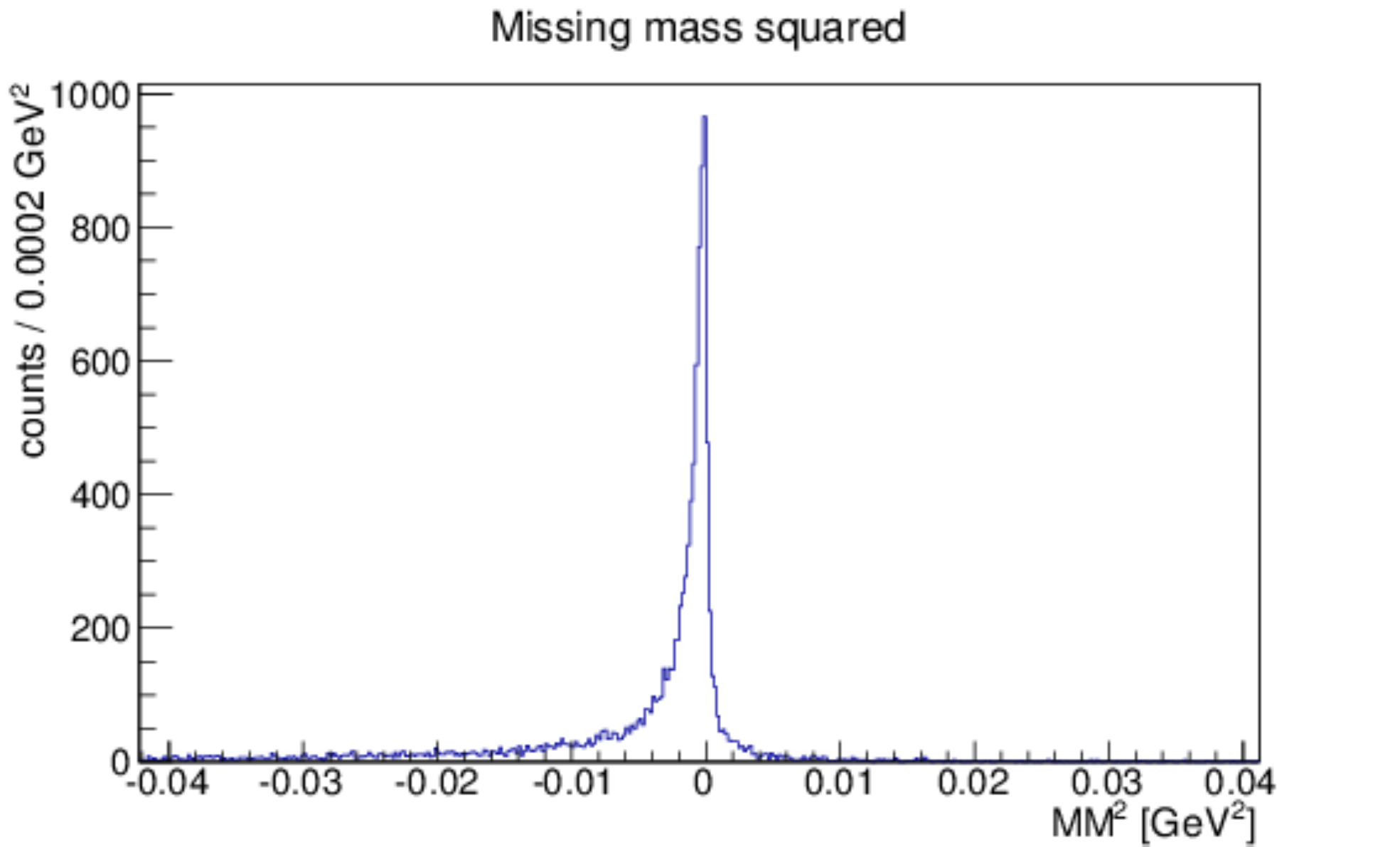,width=3.0in}
\end{center}
\centerline{\parbox{0.80\textwidth}{
 \caption{Full reconstruction for $K_Lp\rightarrow pK_S$,
        $K_S\rightarrow\pi^+\pi^-$:  (top left) $\pi^+\pi^-$
        invariant mass; (top right) W computed from $p\pi^+
        \pi^-$ invariant mass; (bottom) missing mass squared
        for the full reaction.} \label{Fig:pks full reaction} } }
\end{figure}
\begin{figure}[ht!]
\begin{center}
\epsfig{file=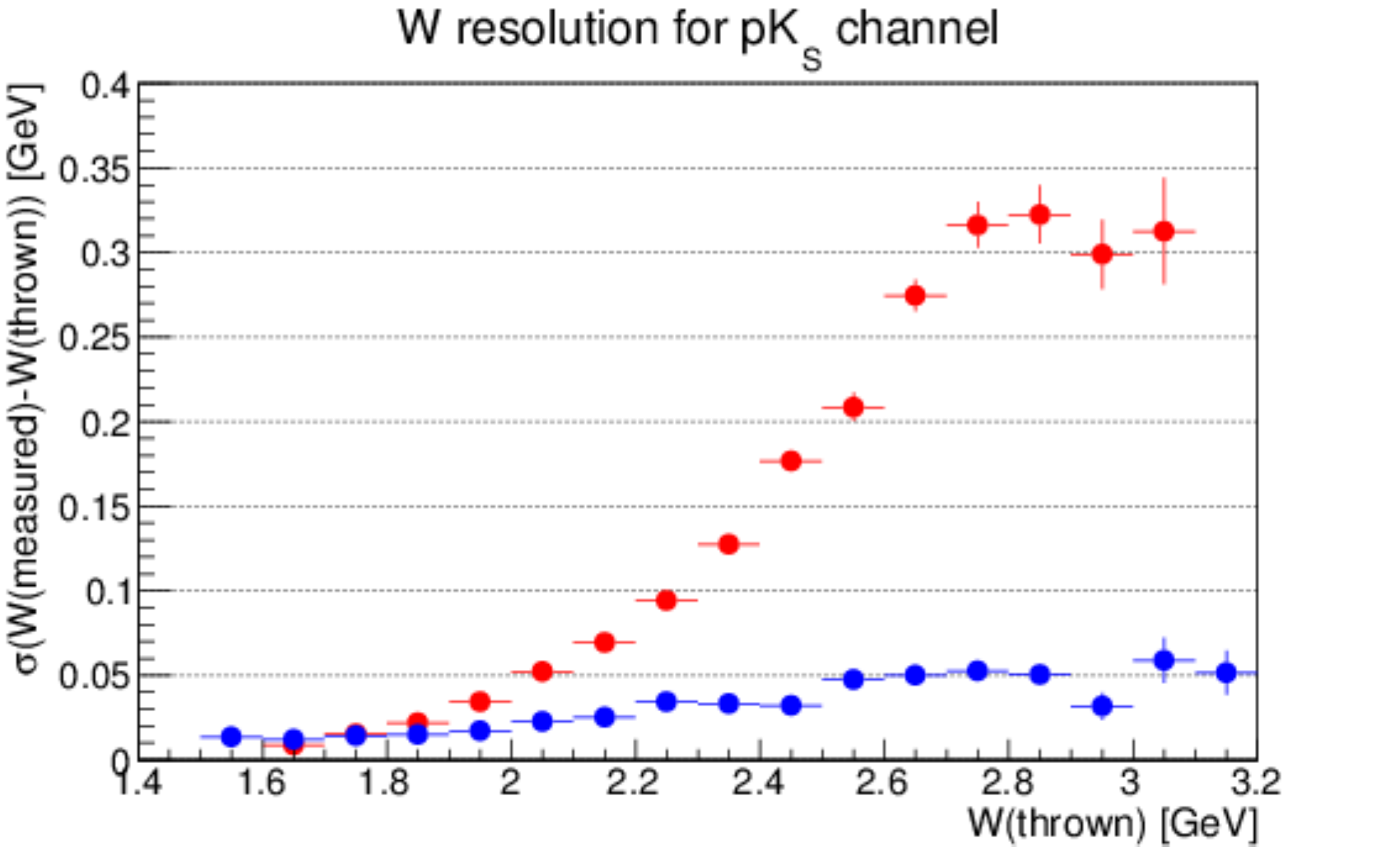,width=3.0in}
\end{center}
\centerline{\parbox{0.80\textwidth}{
 \caption{W resolution for the $pK_S$ channel.}
        \label{Fig:pks w resolution} } }
\end{figure}
\begin{figure}[ht!]
\begin{center}
\epsfig{file=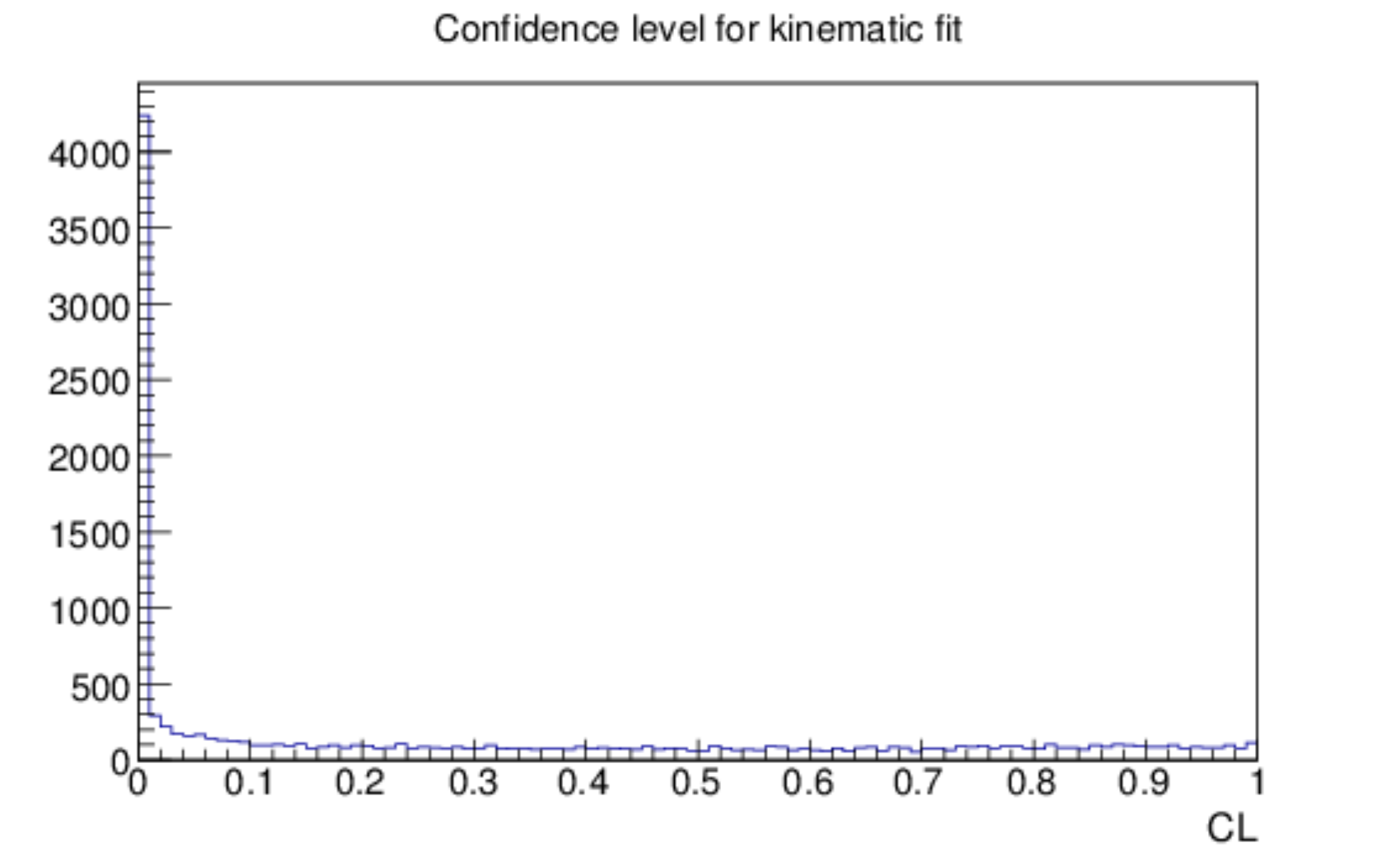,width=3.0in}
\epsfig{file=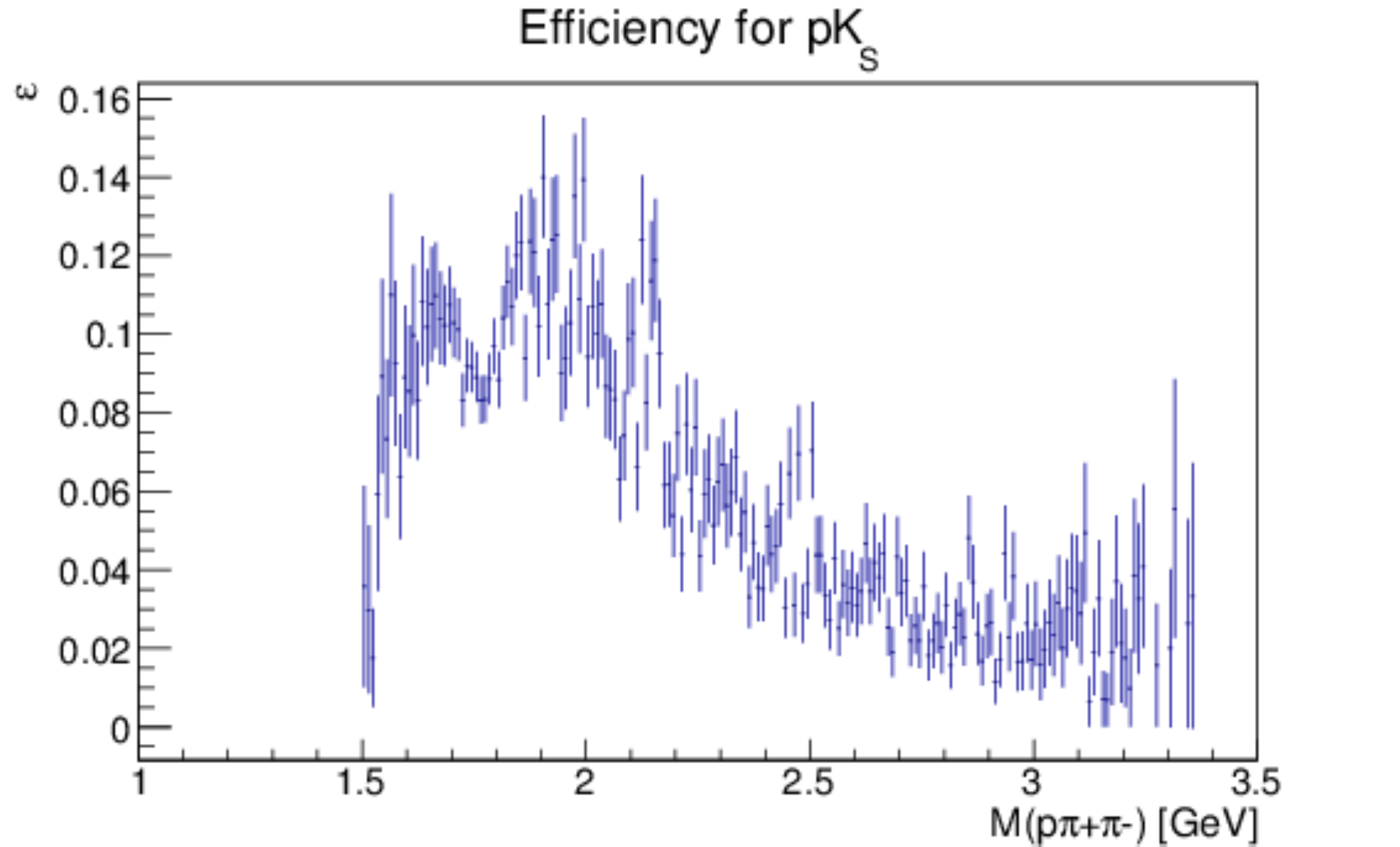,width=3.0in}
\end{center}
\centerline{\parbox{0.80\textwidth}{
 \caption{(left) Confidence level distribution for
        kinematic fit for the $pK_S$ channel. (right)
        Estimate for efficiency for full reconstruction
        of the $K_Lp\rightarrow pK_S$, $K_S\rightarrow
        \pi^+\pi^-$ reaction chain as a function of W.}
        \label{Fig:pks efficiency} } }
\end{figure}

\item \textbf{$K_Lp\rightarrow \Lambda\pi^+$}

The reconstructed $K_L$ momentum distribution and the missing
mass off the $\pi^+$ for the $\gamma p\rightarrow\Lambda\pi^+$
simulation are shown in Fig.~\ref{Fig:lampi pkl}. As with the
previous topology, the missing mass distribution has very
long non-Gaussian tails.

Taking advantage of the large (63.9\%) branching ratio for
$\Lambda\rightarrow p\pi^-$~\cite{Agashe:2014kdaD}, the full
final state can be reconstructed.  The mass distributions for
this reaction chain are shown in Fig.~\ref{Fig:lampi full
reaction}.  A comparison of the $W$ resolution using two 
methods (one using the $K_L$ momentum and the other using the
four-momenta of the final state particles) is shown in
Fig.~\ref{Fig:lampi w res}.  After applying a kinematic fit
to the data and cutting at a confidence level of 0.1, I
determined the efficiency as a function of $W$, as shown in
Fig.~\ref{Fig:lampi eff}.  Here the efficiency includes the
branching ratio for $\Lambda\rightarrow p \pi^{-}$.  The
average efficiency is about 2\%.
\begin{figure}[ht!]
\begin{center}
\epsfig{file=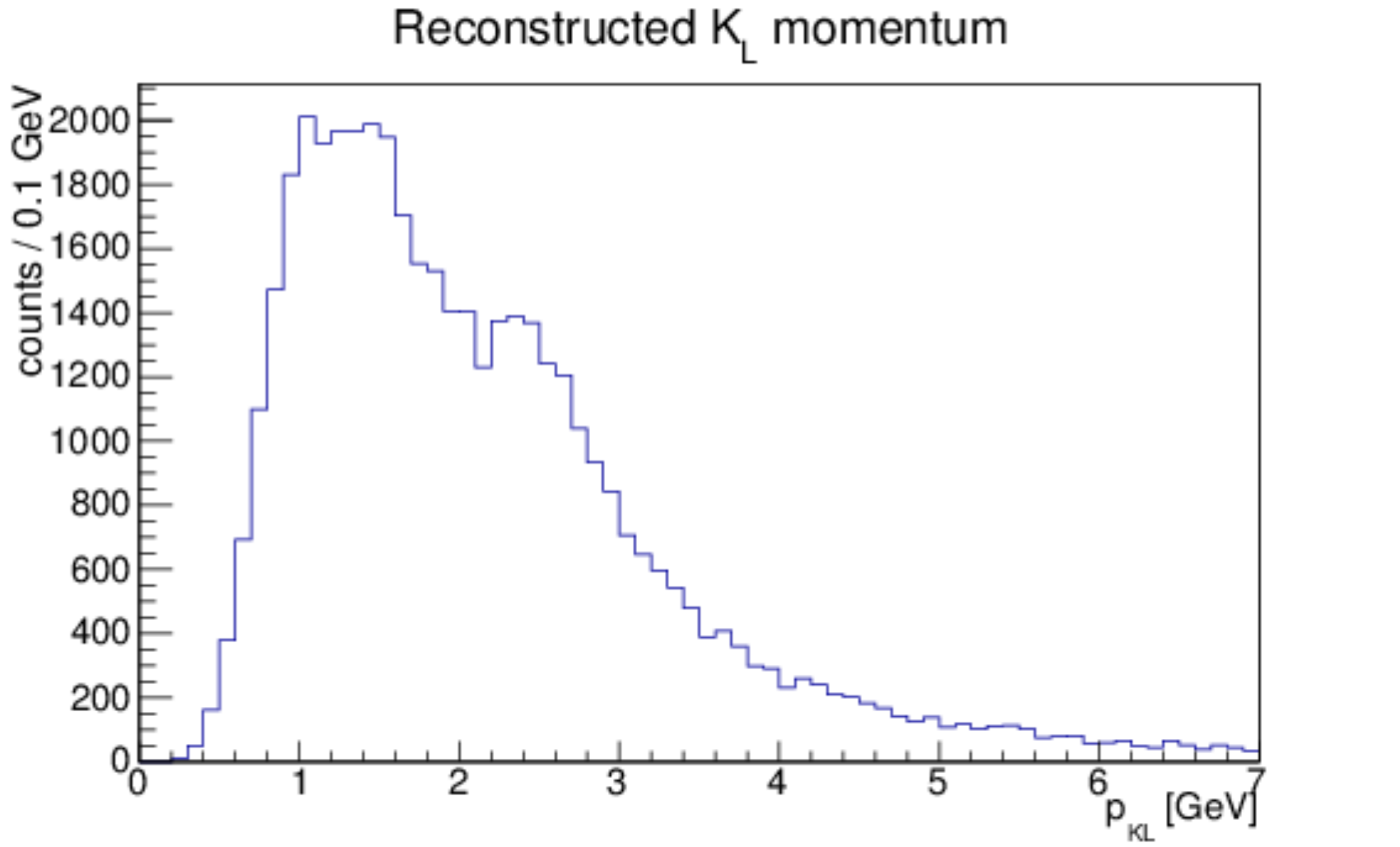,width=3.0in}
\epsfig{file=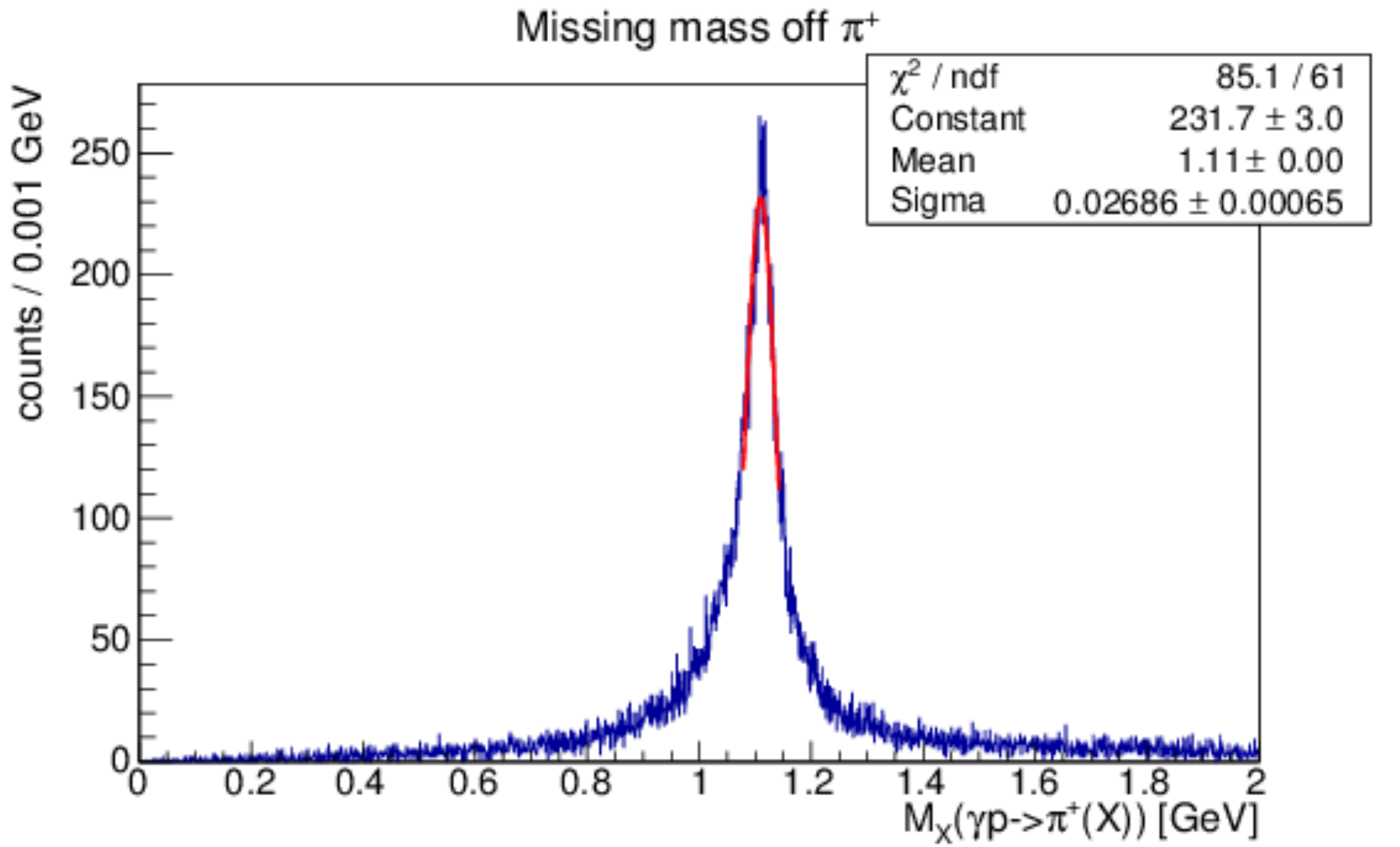,width=3.0in}
\end{center}
\centerline{\parbox{0.80\textwidth}{
 \caption{{$K_L$ momentum distribution (left) and missing
        mass off the proton (right) for the $\Lambda\pi^+$
        channel.}} \label{Fig:lampi pkl} } }
\end{figure}
\begin{figure}[ht!]
\begin{center}
\epsfig{file=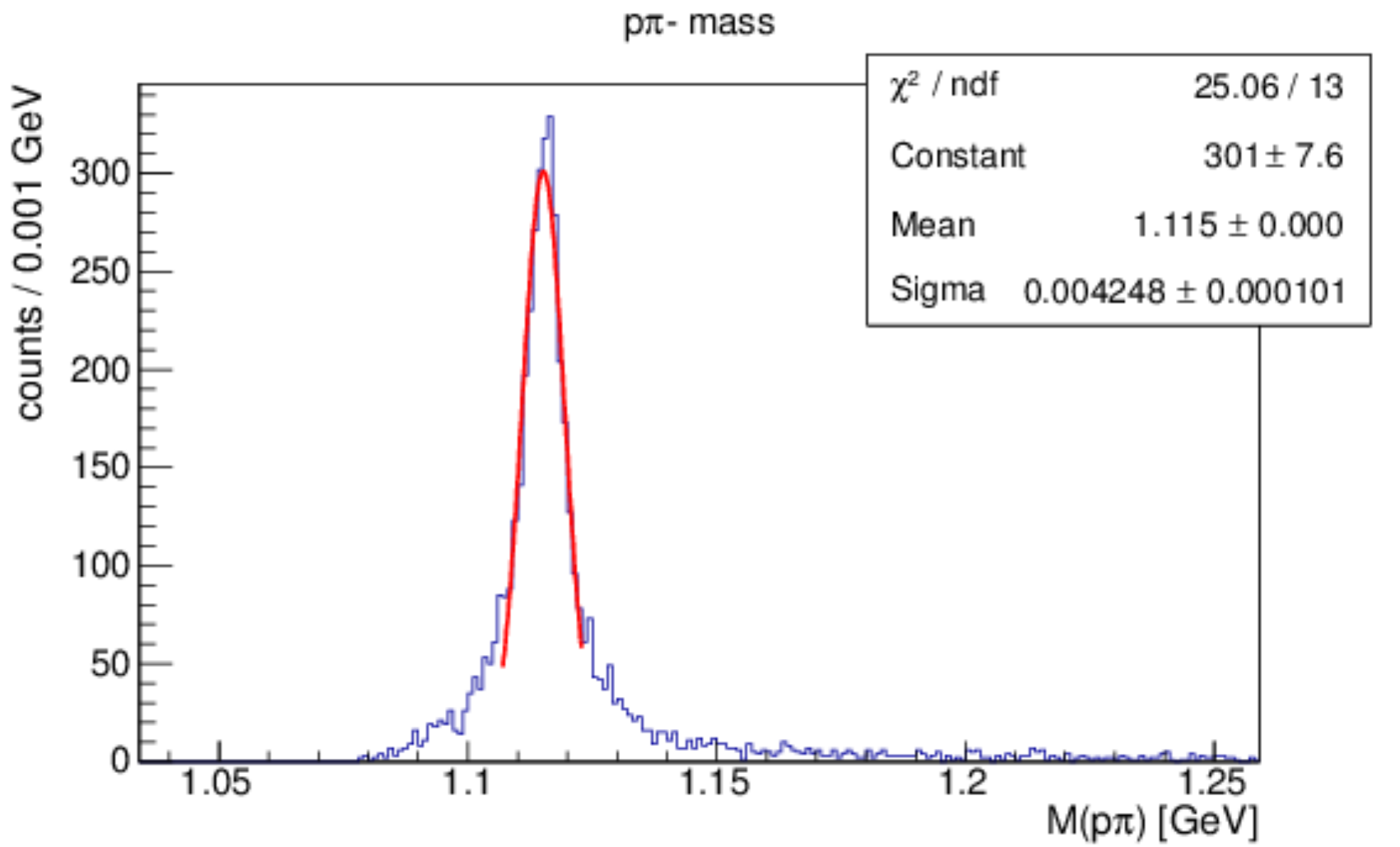,width=3.0in}
\epsfig{file=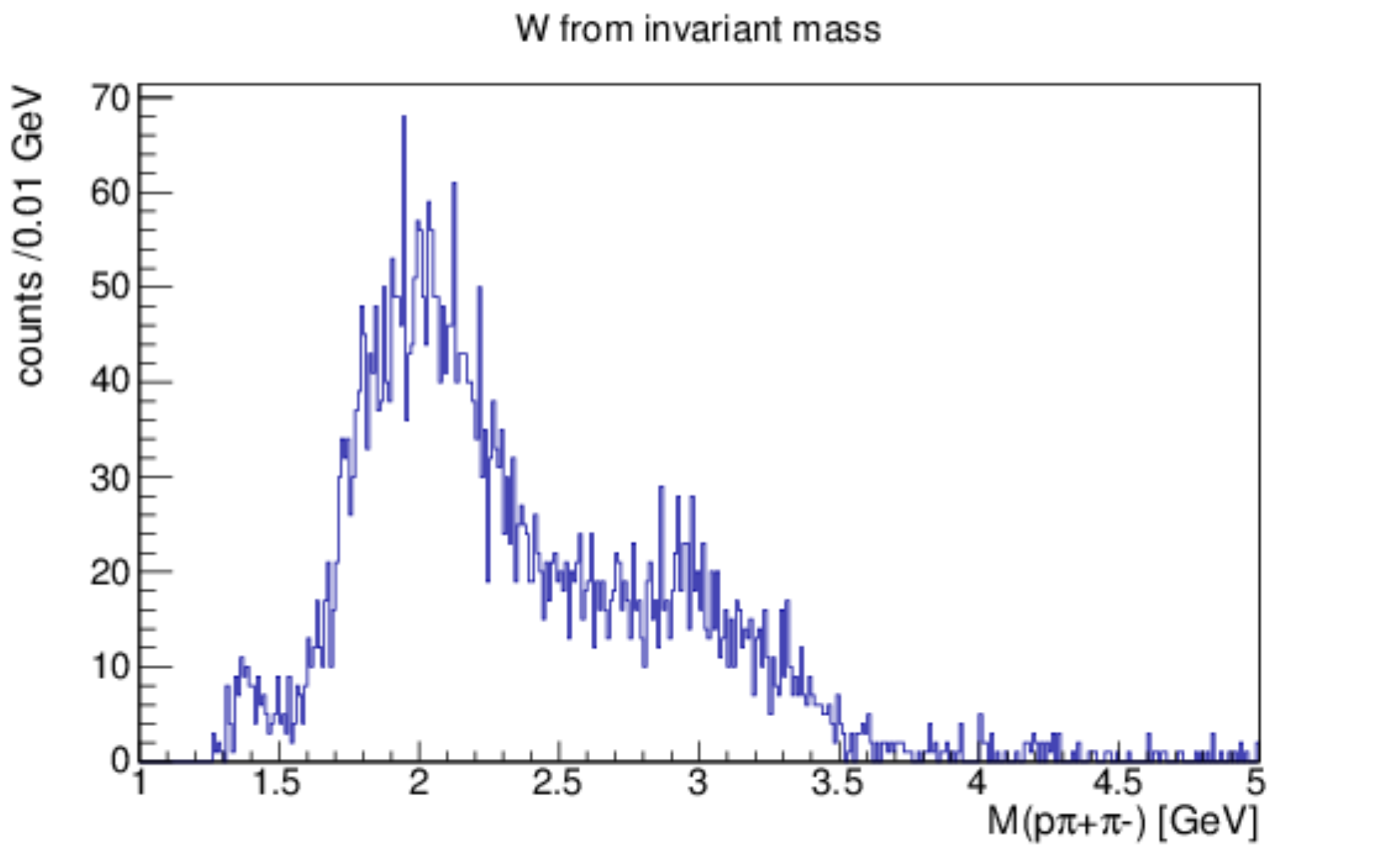,width=3.0in}
\epsfig{file=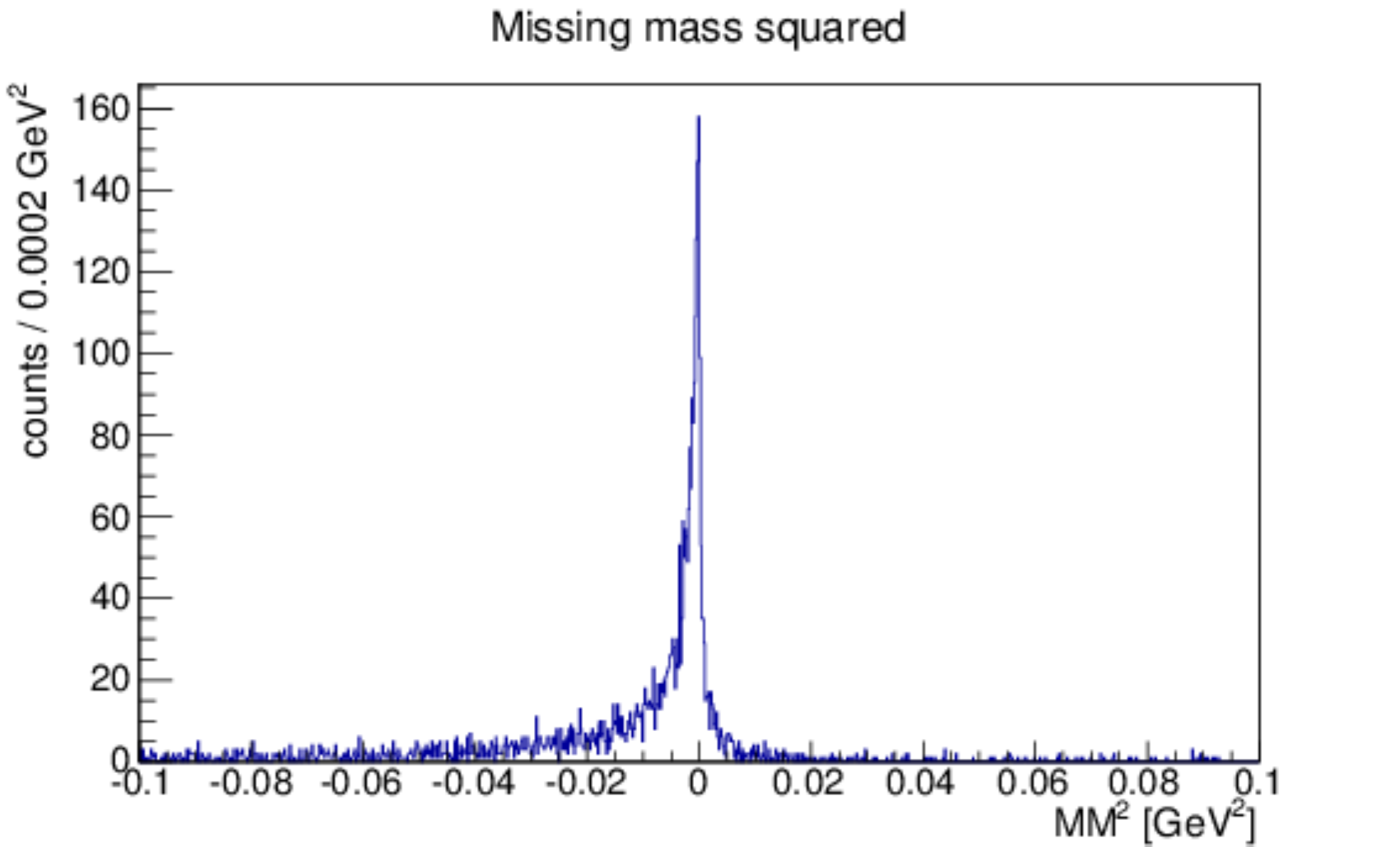,width=3.0in}
\end{center}
\centerline{\parbox{0.80\textwidth}{
 \caption{Full reconstruction for $K_Lp\rightarrow\Lambda
        \pi^+$, $\Lambda\rightarrow p\pi^-$:  (top left)
        $p\pi^-$ invariant mass; (top right) W computed
        from $p\pi^+\pi^-$ invariant mass; (bottom) missing
        mass squared for the full reaction.}
        \label{Fig:lampi full reaction} } }
\end{figure}
\begin{figure}[ht!]
\begin{center}
\epsfig{file=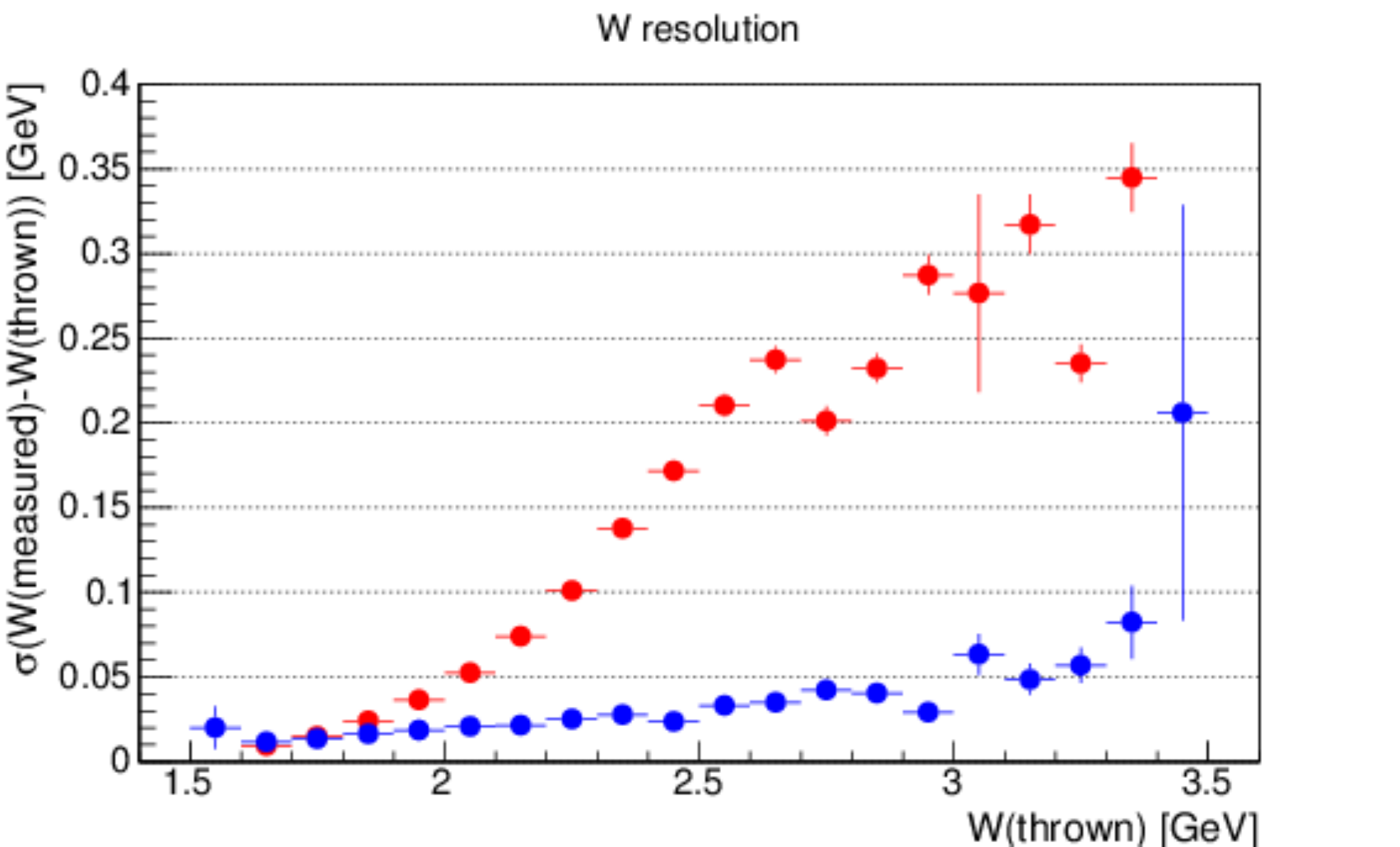,width=3.0in}
\end{center}
\centerline{\parbox{0.80\textwidth}{
 \caption{W resolution for the $\Lambda\pi^+$ channel.}
        \label{Fig:lampi w res} } }
\end{figure}
\begin{figure}[ht!]
\begin{center}
\epsfig{file=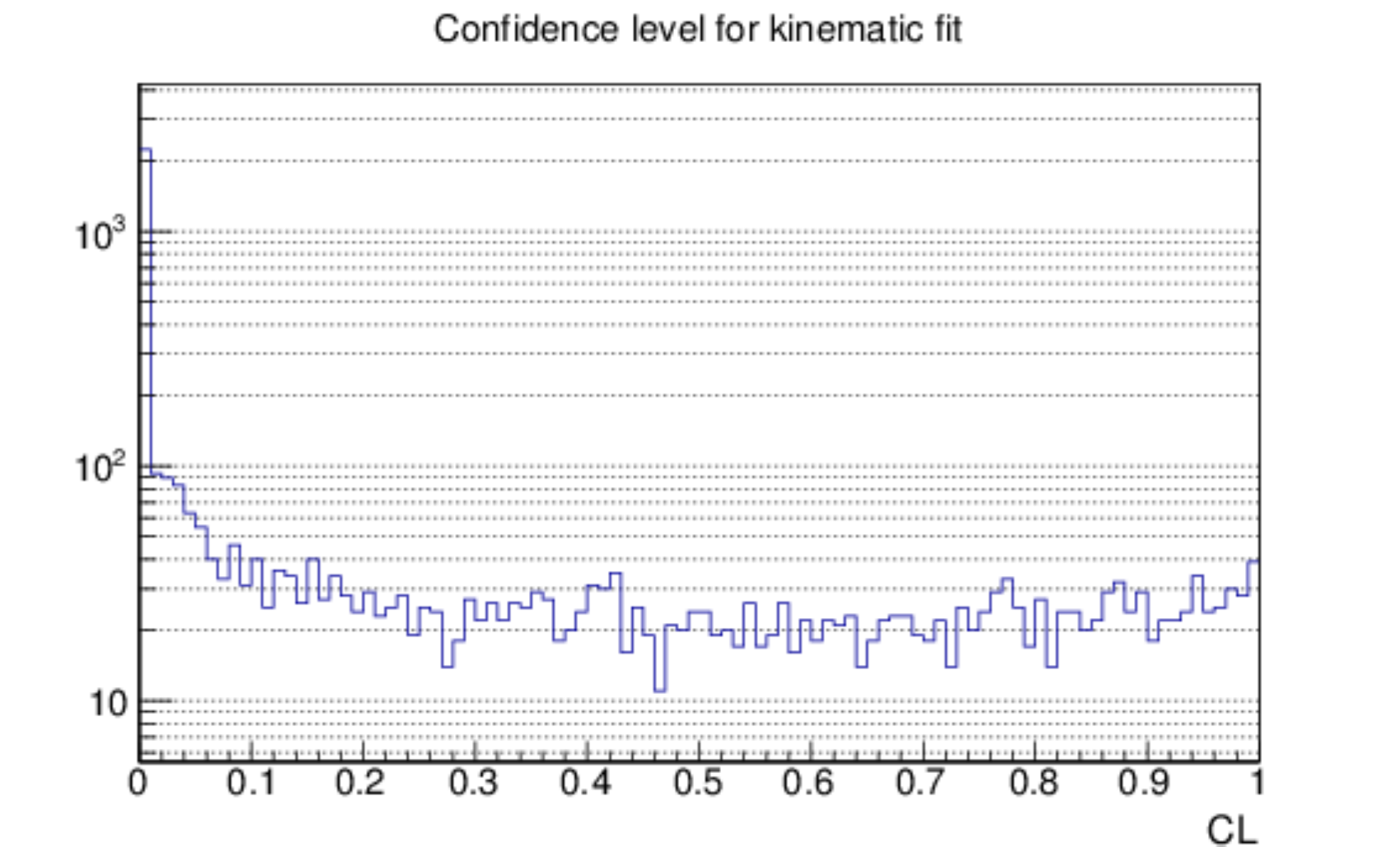,width=3.0in}
\epsfig{file=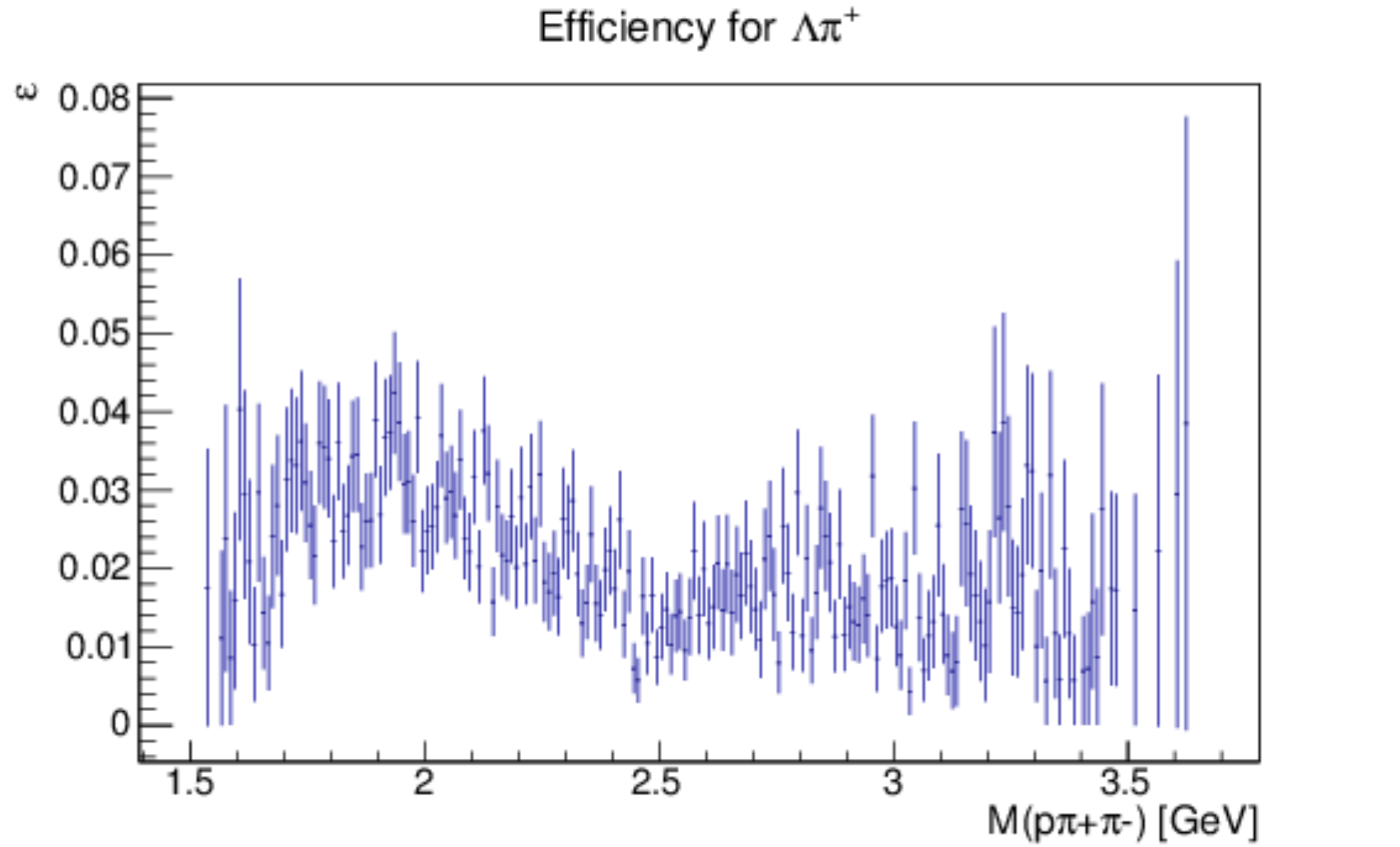,width=3.0in}
\end{center}
\centerline{\parbox{0.80\textwidth}{
 \caption{(left) Confidence level distribution for kinematic
        fit for the $\Lambda\pi^+$ channel.  (right) Estimate
        for efficiency for full reconstruction of the $K_L p
        \rightarrow\Lambda\pi^+$, $\Lambda\rightarrow p\pi^-$
        reaction chain as a function of W.}
        \label{Fig:lampi eff} } }
\end{figure}

\item \textbf{$K_Lp \rightarrow K^+\Xi^0$}

The reconstructed $K_L$ momentum distribution and the missing
mass off the $K^+$ for the $\gamma p \rightarrow K^+\Xi^0$
simulation are shown in Fig.~\ref{Fig:kxi pkl}.  The $\Xi^0$
mass distribution reconstructed using the missing mass
technique is broad (full-width-at-half-maximum =$\sim$300~MeV).

The $\Xi^0$ decays almost 100\% of the time to $\Lambda
\pi^0$~\cite{Agashe:2014kdaD}. Here we take advantage of the
large branches for $\Lambda\rightarrow p\pi^-$ and $\pi^0
\rightarrow\gamma\gamma$ to reconstruct $\Xi^0$'s using the
four-momenta for all of the final state particles. The 
reconstructed mass distributions are shown in
Fig.~\ref{Fig:kxi full reconstruction}.  A comparison of the
two methods for computing $W$ is shown in Fig.~\ref{Fig:kxi
W res}.  An estimate for the efficiency as a function of $W$
after applying a kinematic fit to the data and cutting at a
confidence level of 0.1 is shown in Fig.~\ref{Fig:kxi eff}.
The average efficiency is about 0.4\%.   This efficiency
includes the branching ratios for the $\Lambda$ and $\pi^0$
decays.  Also shown is the invariant mass of the $\Xi^0$
constructed in $p\pi^-2\gamma$.  The peak is much narrower
than the $\Xi^0$ peak seen in missing mass.
\begin{figure}[ht!]
\begin{center}
\epsfig{file=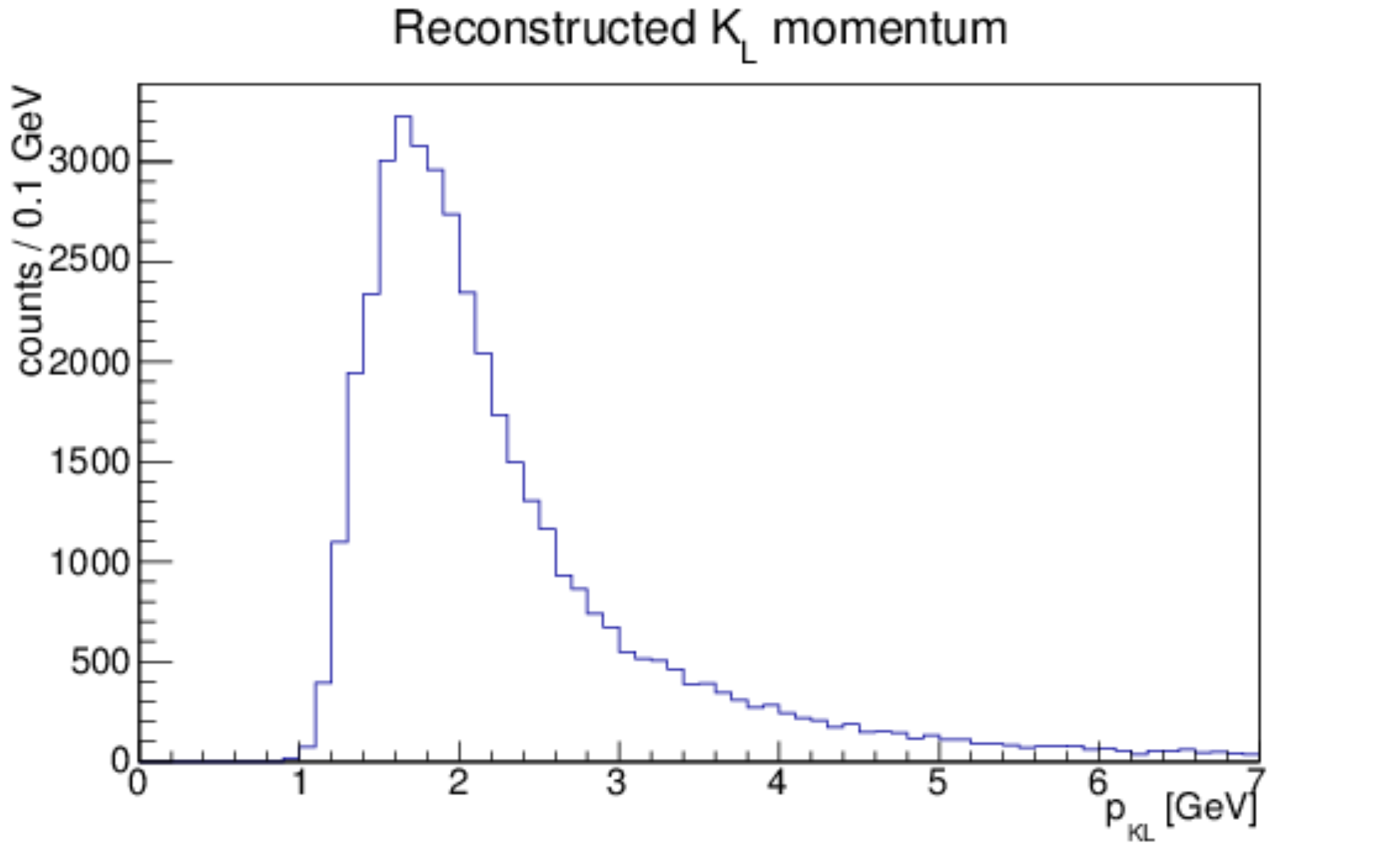,width=3.0in}
\epsfig{file=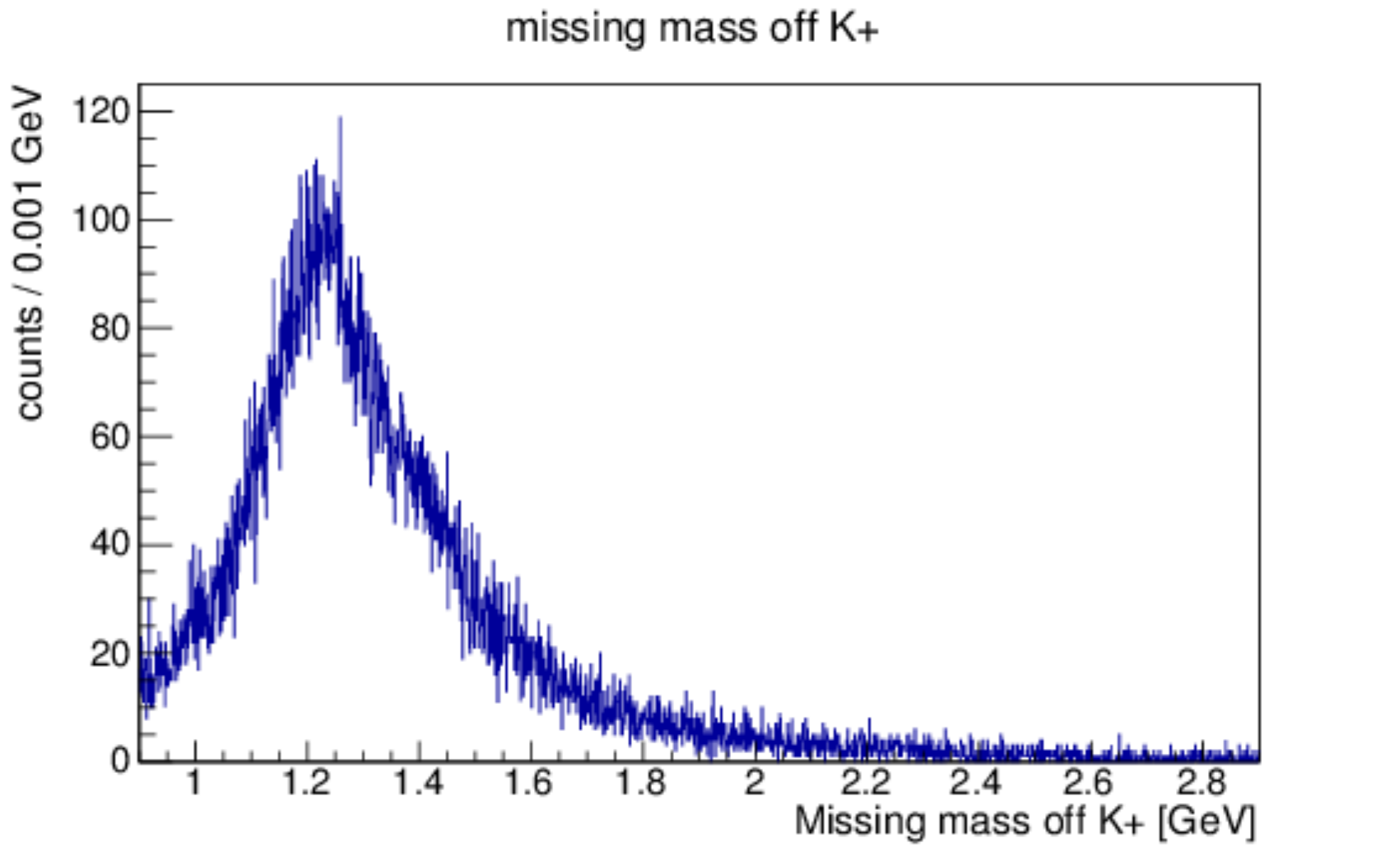,width=3.0in}
\end{center}
\centerline{\parbox{0.80\textwidth}{
 \caption{{$K_L$ momentum distribution (left) and missing
        mass off the proton (right) for the $K^+\Xi^0$
        channel.}} \label{Fig:kxi pkl} } }
\end{figure}
\begin{figure}[ht!]
\begin{center}
\epsfig{file=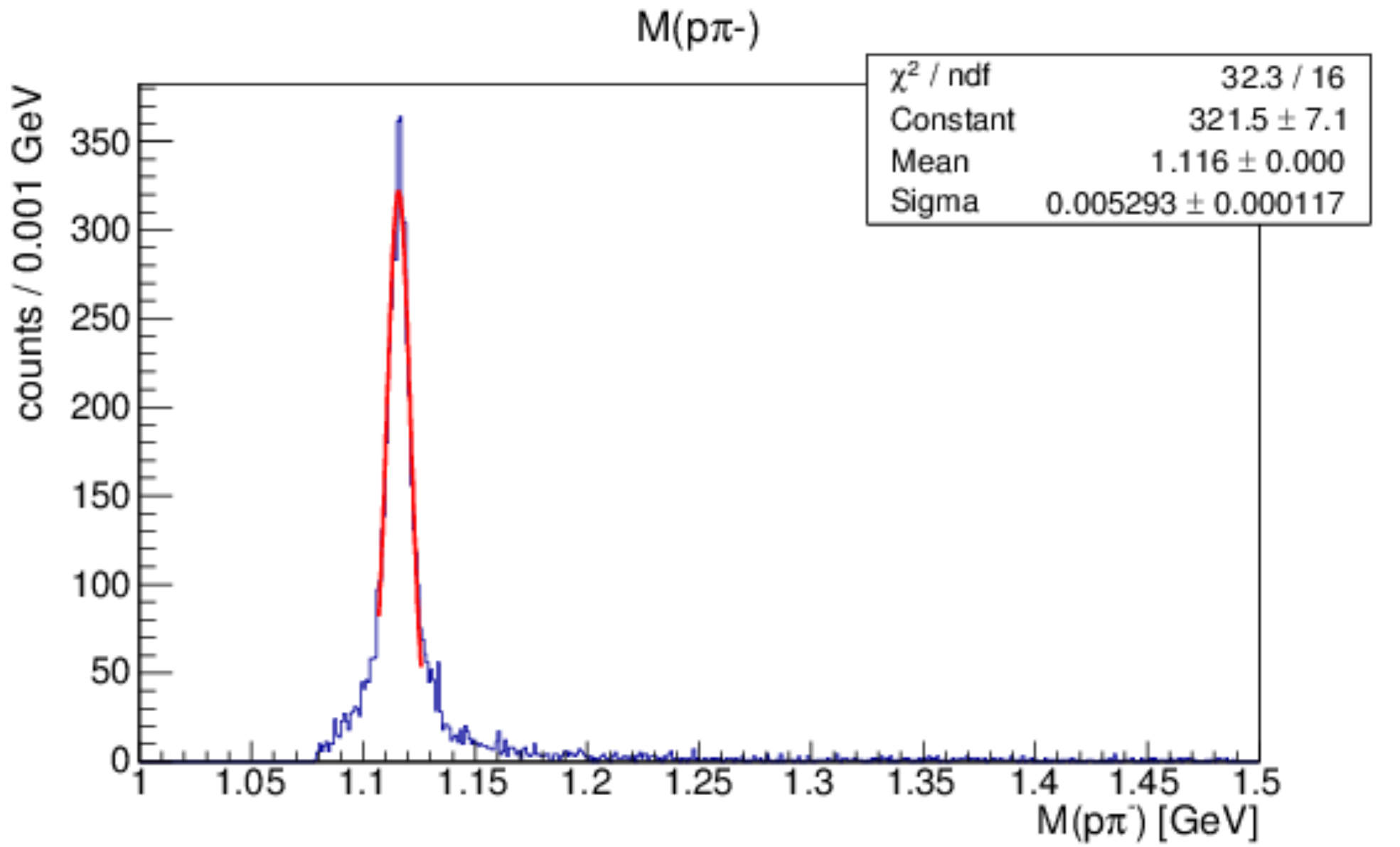,width=3.0in}
\epsfig{file=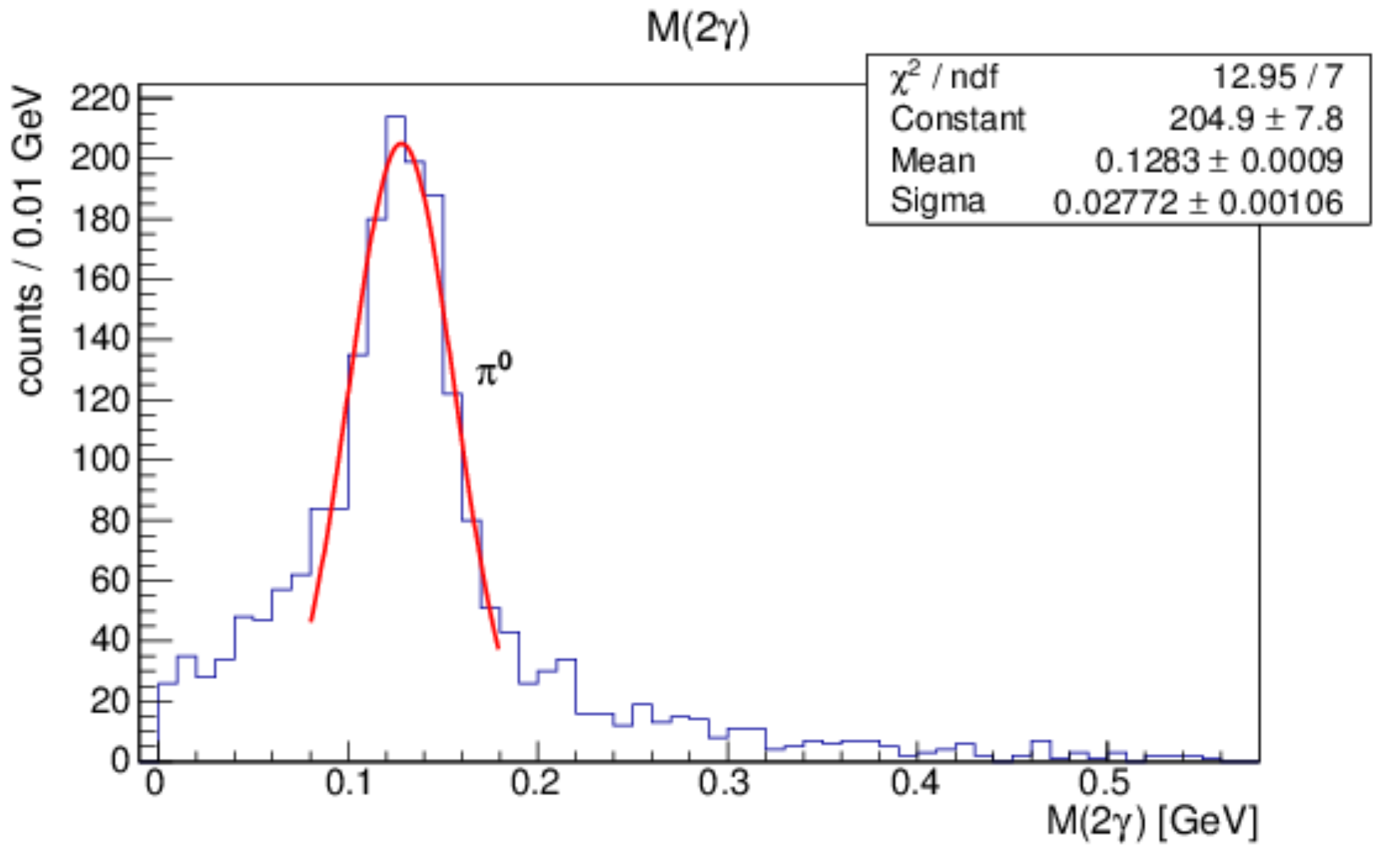,width=3.0in}
\epsfig{file=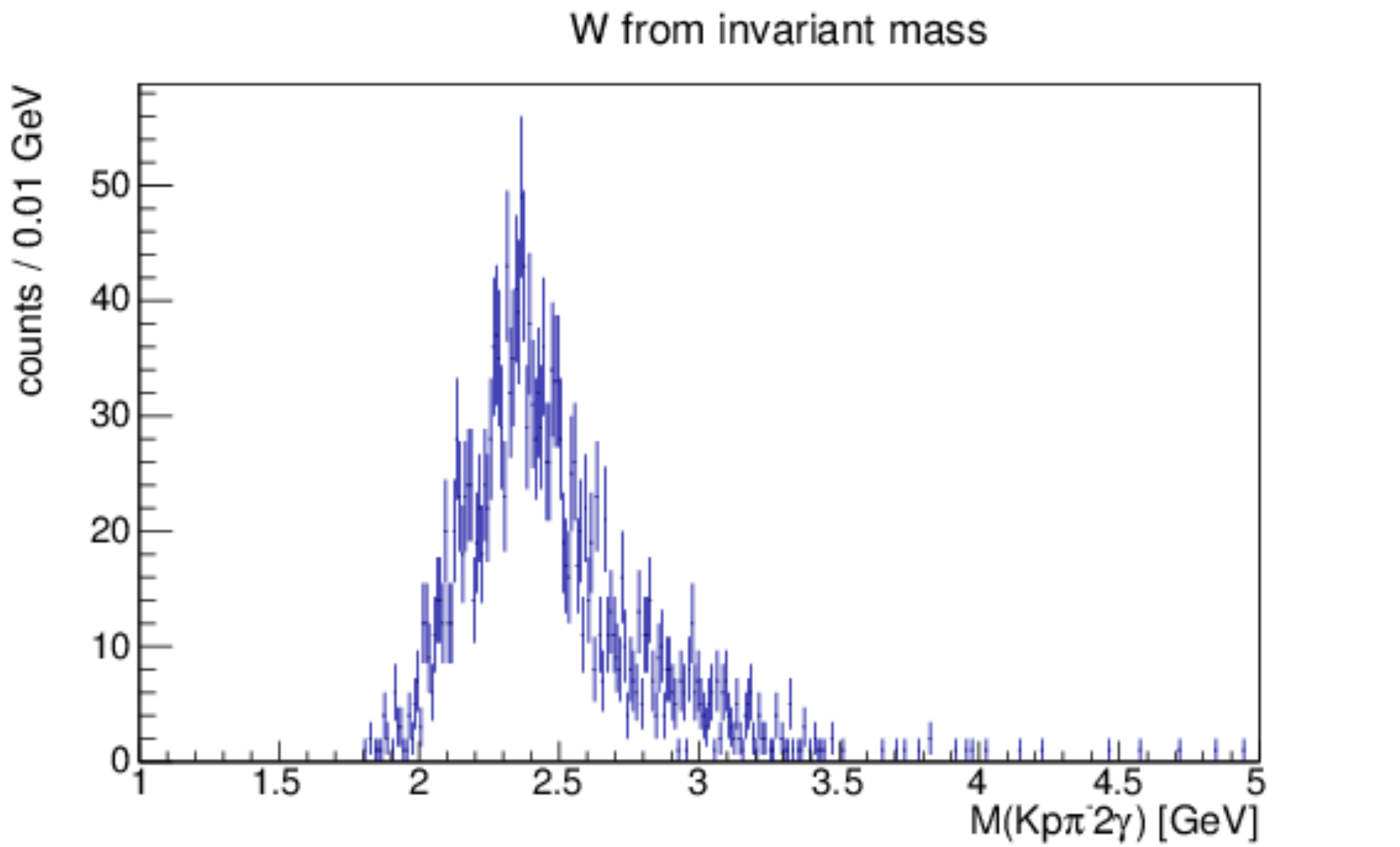,width=3.0in}
\epsfig{file=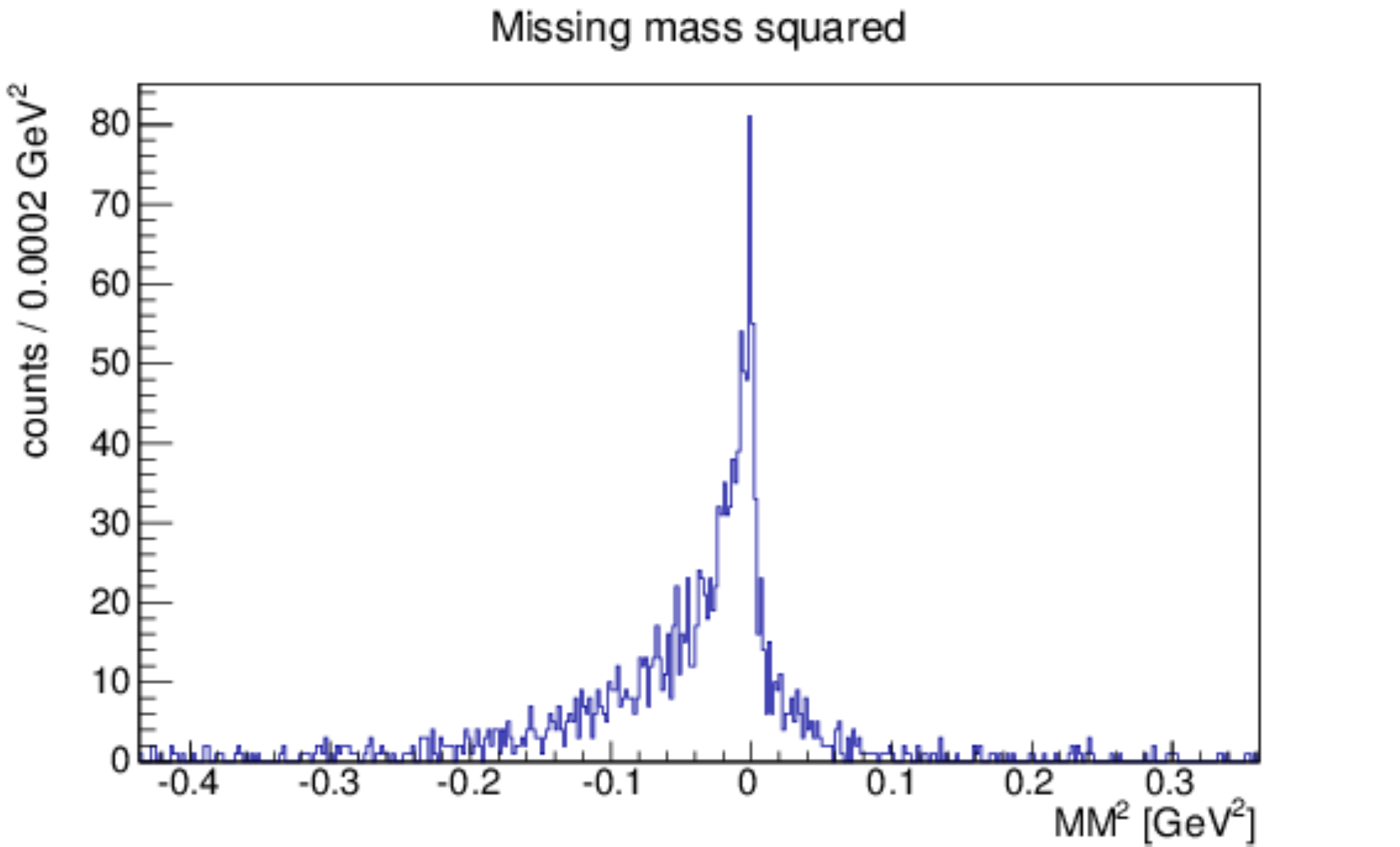,width=3.0in}
\end{center}
\centerline{\parbox{0.80\textwidth}{
 \caption{Full reconstruction for $K_Lp\rightarrow K^+\Xi^0$,
        $\Xi^0\rightarrow\Lambda\pi^0$, $\Lambda\rightarrow
        p\pi^-$, $\pi^0\rightarrow\gamma\gamma$:  (top left)
        $p\pi^-$ invariant mass; (top right) two photon
        invariant mass; (bottom left) W computed from $K^+p
        \pi^-2\gamma$ invariant mass; (bottom right) missing
        mass squared for the full reaction.}
        \label{Fig:kxi full reconstruction} } }
\end{figure}
\begin{figure}[ht!]
\begin{center}
\epsfig{file=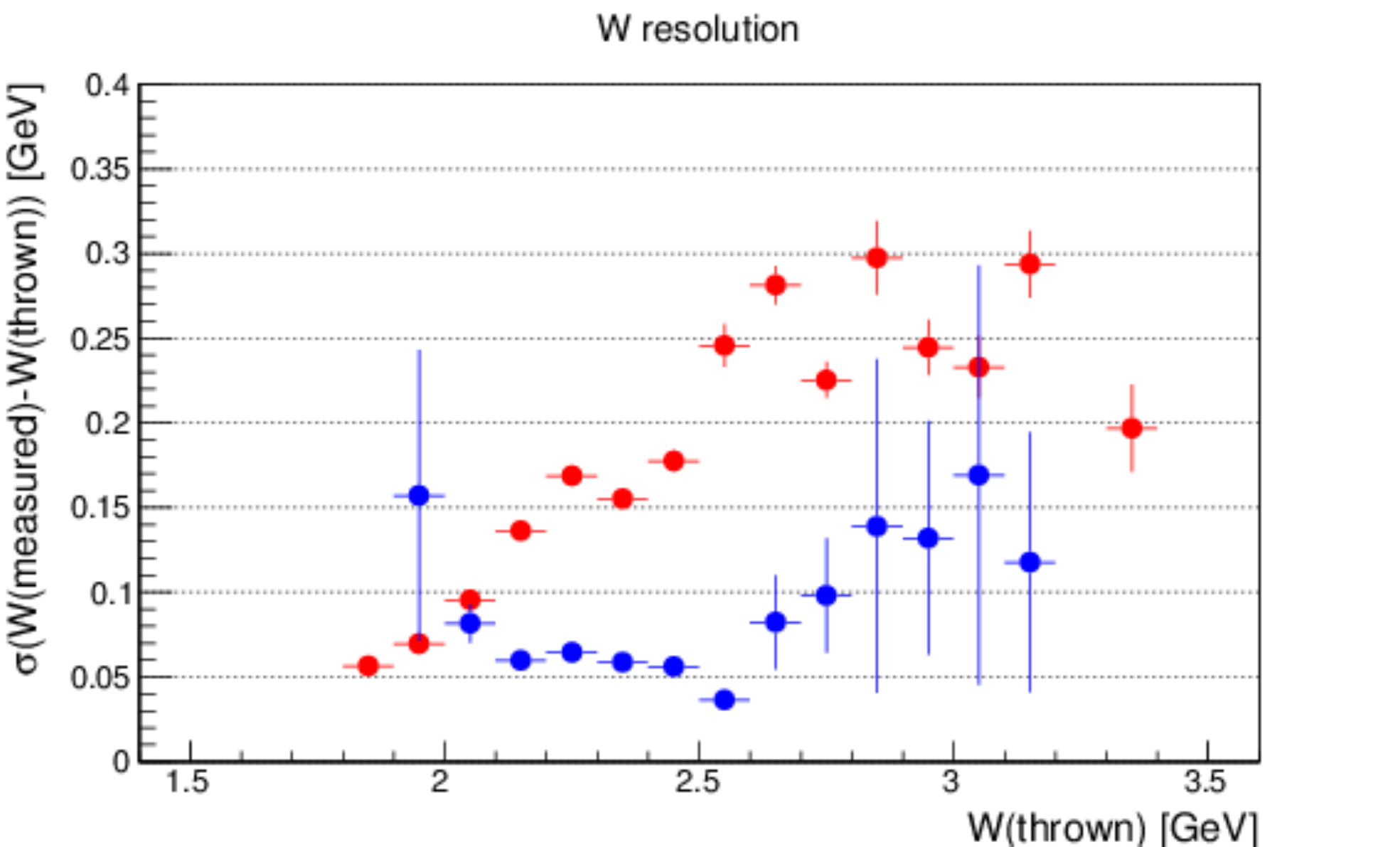,width=3.0in}
\end{center}
\centerline{\parbox{0.80\textwidth}{
 \caption{W resolution for the $K^+\Xi^0$ channel.}
        \label{Fig:kxi W res} } }
\end{figure}
\begin{figure}[ht!]
\begin{center}
\epsfig{file=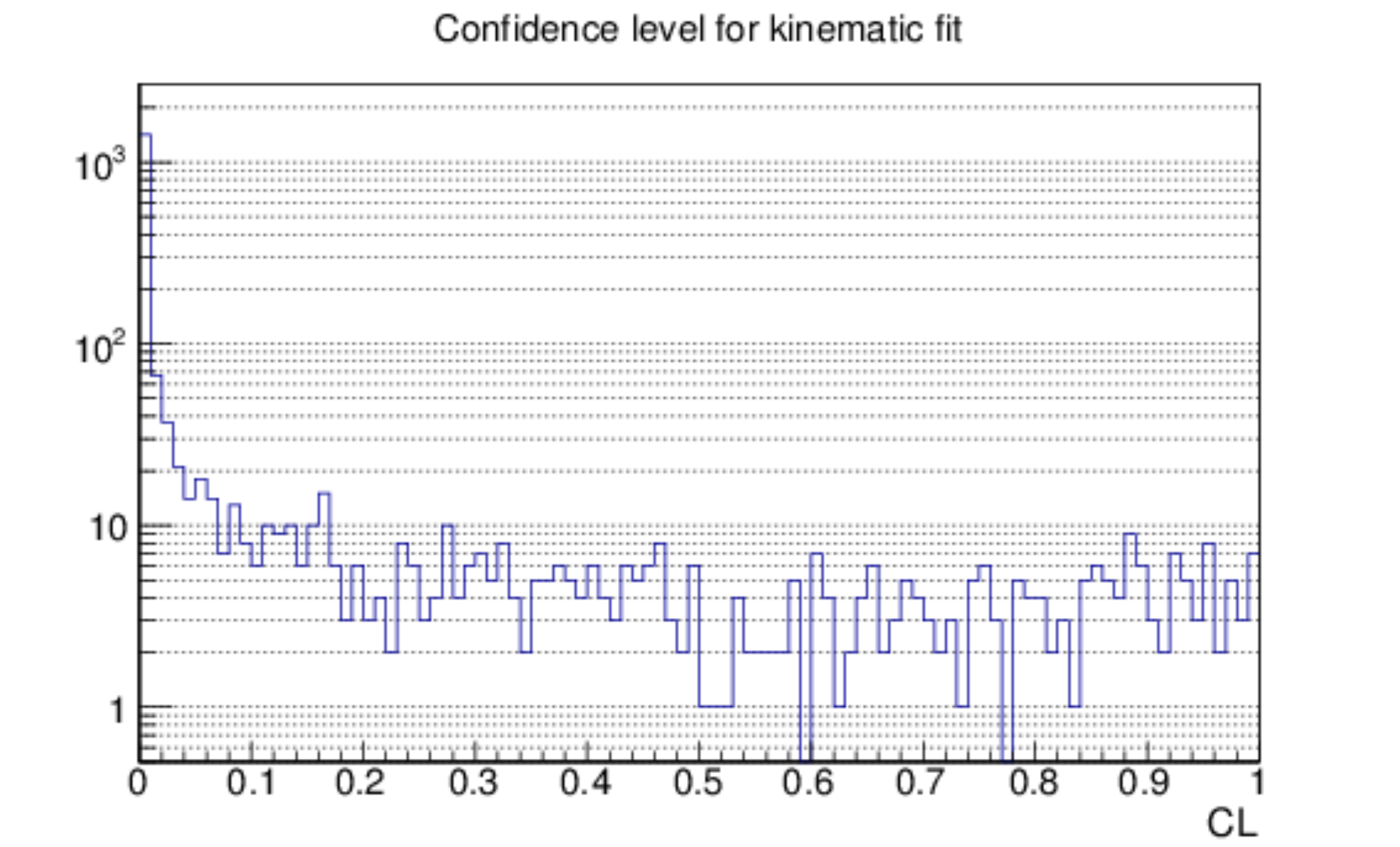,width=3.0in}
\epsfig{file=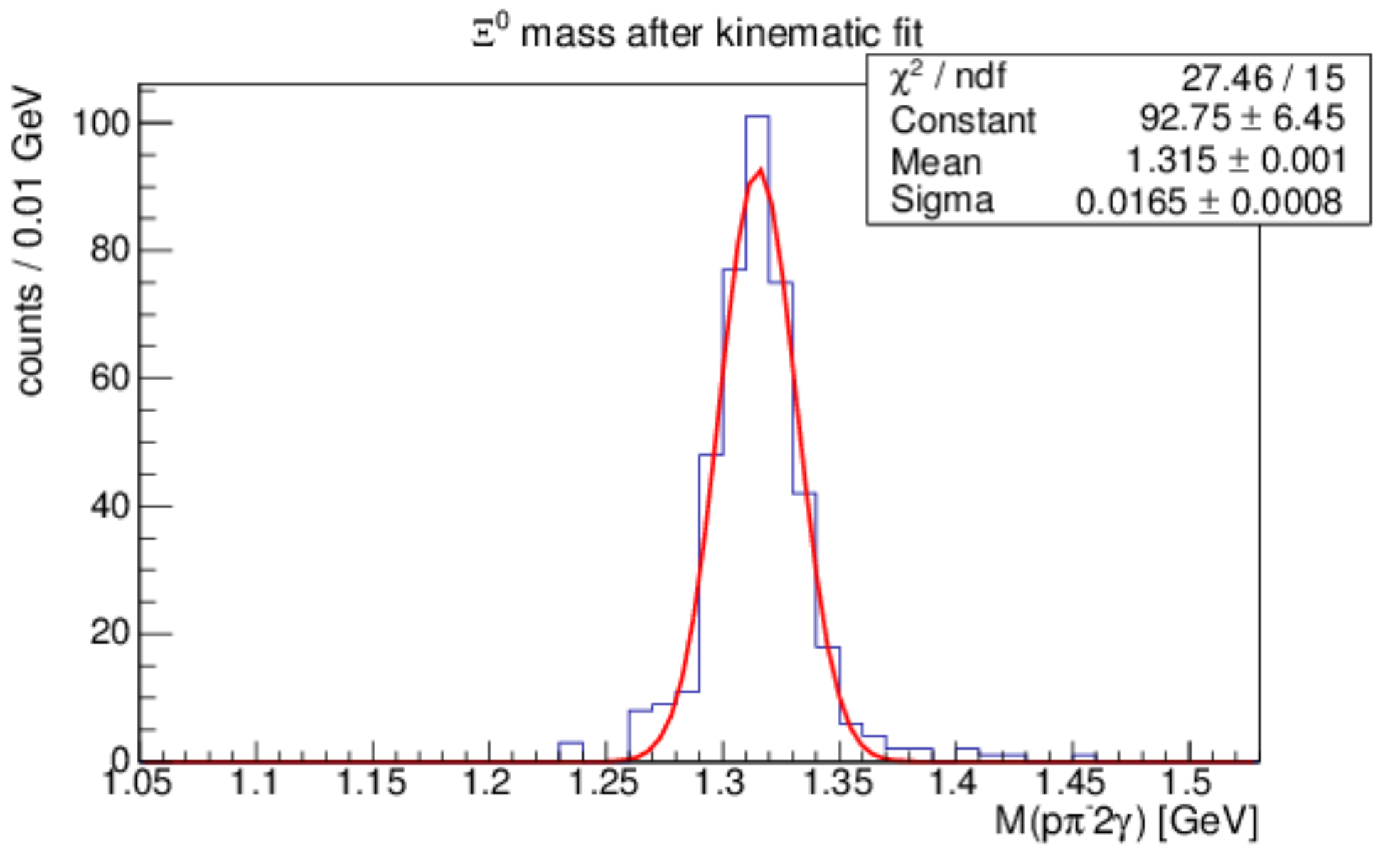,width=3.0in}
\epsfig{file=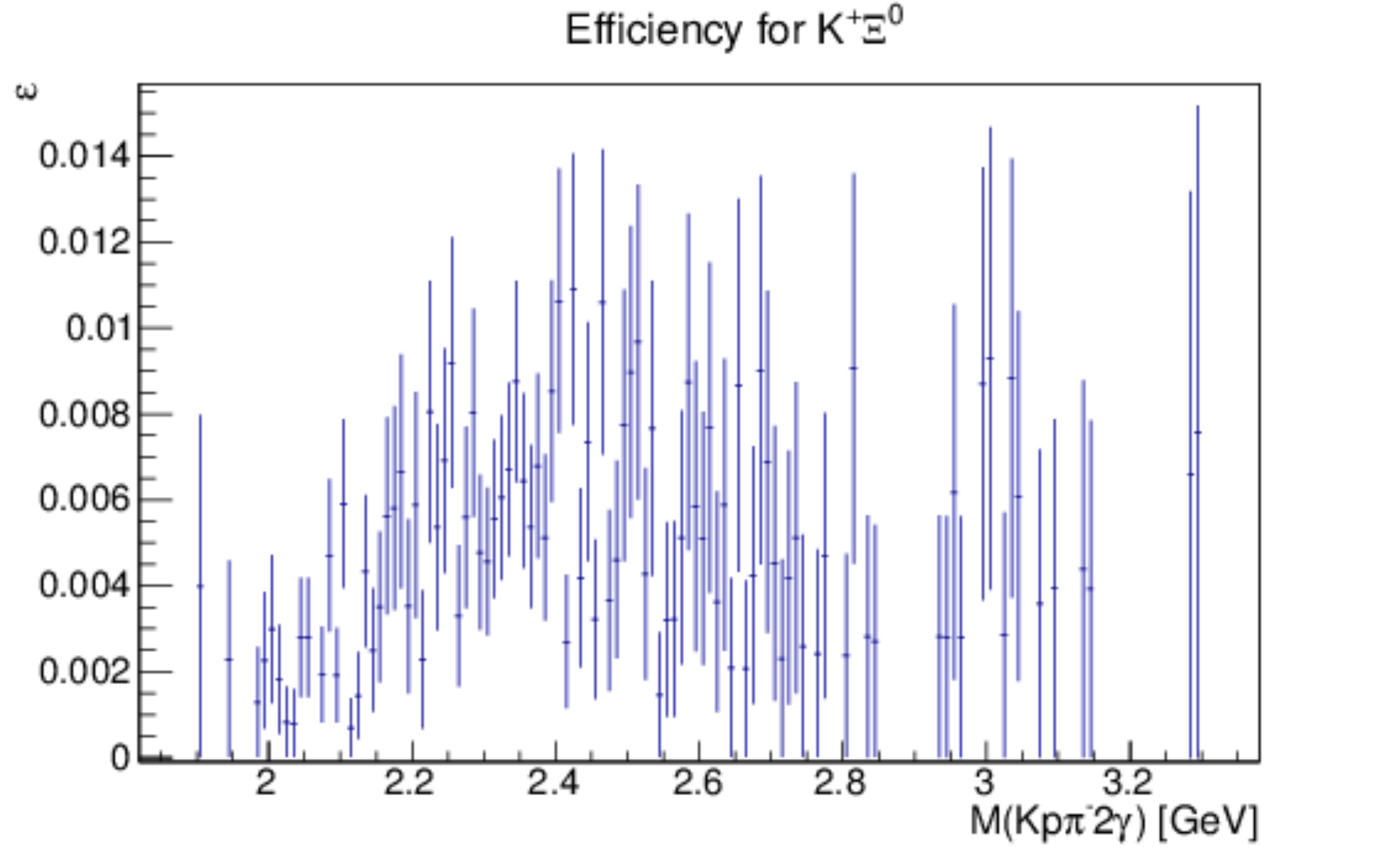,width=3.0in}
\end{center}
\centerline{\parbox{0.80\textwidth}{
 \caption{(top left) Confidence level distribution for
        kinematic fit for the $K^+\Xi^0$ channel. (top right)
        $\Xi^0$ mass distribution after kinematic fit with a
        0.1 confidence level cut.  (bottom) Estimate for
        efficiency for full reconstruction of the $K_L p
        \rightarrow K^+\Xi^0$, $\Xi^0\rightarrow\Lambda
	\pi^0$, $\Lambda\rightarrow p\pi^-$, $\pi^0
	\rightarrow\gamma\gamma$ reaction chain as a 
	function of W.} \label{Fig:kxi eff} } }
\end{figure}
\end{enumerate}

\newpage
\item \textbf{Remarks}

For all topologies under consideration here, the $W$ 
resolution using the measured $K_L$ momentum worsens
considerably as $W$ increases, whereas the $W$
resolution using the final state particles is flatter
as a function of $W$.  Resolution at the level of
50~MeV or less can be achieved with the latter
technique, at the expense of efficiency.  One source
for the inefficiency is the spiraling of low-momentum
pions in the magnetic field of the GlueX solenoid,
which makes track finding and fitting difficult.
This can be mitigated by running the solenoid at a
lower current than the standard current for regular
GlueX runs.  The preliminary kinematic fitting results
are encouraging and further improvement in the $W$
resolution can be expected for the case where events
are fully reconstructed.

\item \textbf{Acknowledgments}

The author would like to thank Ilya Larin for providing
event generation code and Mark Manley for providing plots
of cross section data for the $K_L p\rightarrow p K_S$
reaction.  This material is based upon work supported by
the U.S. Department of Energy, Office of Science, Office
of Nuclear Physics under contract DE--AC05--06OR23177.
\end{enumerate}


\newpage
\subsection{Compact Photon Source Conceptual Design}
\addtocontents{toc}{\hspace{2cm}{\sl P.~Degtyarenko and B.~Wojtsekhowski}\par}
\setcounter{figure}{0}
\halign{#\hfil&\quad#\hfil\cr
\large{Pavel Degtyarenko and Bogdan Wojtsekhowski}\cr
\textit{Thomas Jefferson National Accelerator Facility}\cr
\textit{Newport News, VA 23606, U.S.A.}\cr}

\begin{abstract}
We describe options for the production of an intense photon
beam at the CEBAF Hall~D Tagger facility, needed for creating a
high-quality secondary $K^0_L$ delivered to the Hall D detector.
The conceptual design for the Compact Photon Source apparatus
is presented.
\end{abstract}

\begin{enumerate}
\item \textbf{Introduction}

An intense high energy gamma source is a pre-requisite for the
production of the $K_L^0$ beams needed for the new proposed
experiments at Hall~D.  Here we describe a new approach to
designing such photon sources.  We will discuss possible
practical implementation of the new approach in the design,
adjusted to the parameters and limitations of the available
infrastructure.  The plan view of the present
Tagger vault area is shown in Fig.~\ref{fig:CPS_plot1}.

\item \textbf{Available facilities and Options at the Tagger
        Area}

An electron beam at the nominal energy of 12~GeV enters the
Tagger vault through the Hall~D beamline tunnel and is
directed into a typically thin radiator in front of the
Tagger magnet. The bulk of the incident beam goes through
the gap in the magnet without interaction and is then
directed into the exit beam pipe, leading to the beam dump
at the end of the beam dump alcove, behind the labyrinth
shielding walls.  Electron interactions in the radiator
produce bremmstrahlung photons going straight through the
opening in the magnet, and then through the long pipe
leading to the Hall~D beam entry port. Electrons that have
lost a portion of their original energy to the photon
production are deflected in the magnet and exit through
the row of position-sensitive detectors at the magnet side.
The energy of each corresponding photon is thus determined.
The need to count and the need to determine the energy of
each produced photon imposes major limitations on the
maximum beam current (5~$\mu$A at 12~GeV) and on the
maximum radiator thickness (about 0.0005~radiation lengths
(R.L.) for the maximum beam current). The whole area layout
and design parameters, including the radiation shielding
requirements, were chosen with these limitations taken into
account.
\begin{figure}[ht!]
\begin{center}
\includegraphics[width=\linewidth]{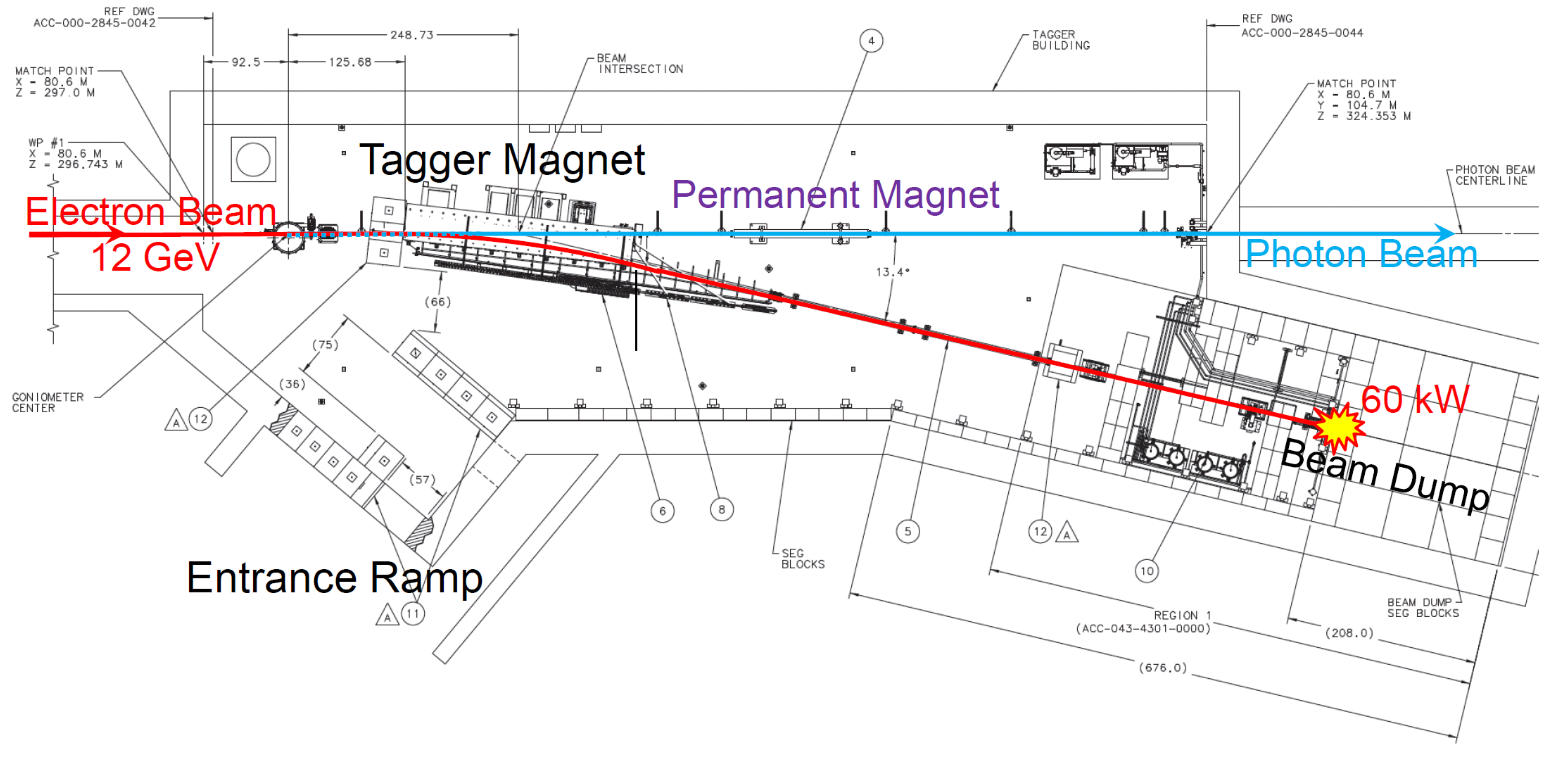}
\centerline{\parbox{0.80\textwidth}{
 \caption{Tagger vault area in the Hall~D complex at CEBAF.}
        \label{fig:CPS_plot1} } }
\end{center}
\end{figure}
\begin{figure}[ht!]
\begin{center}
\includegraphics[width=\linewidth]{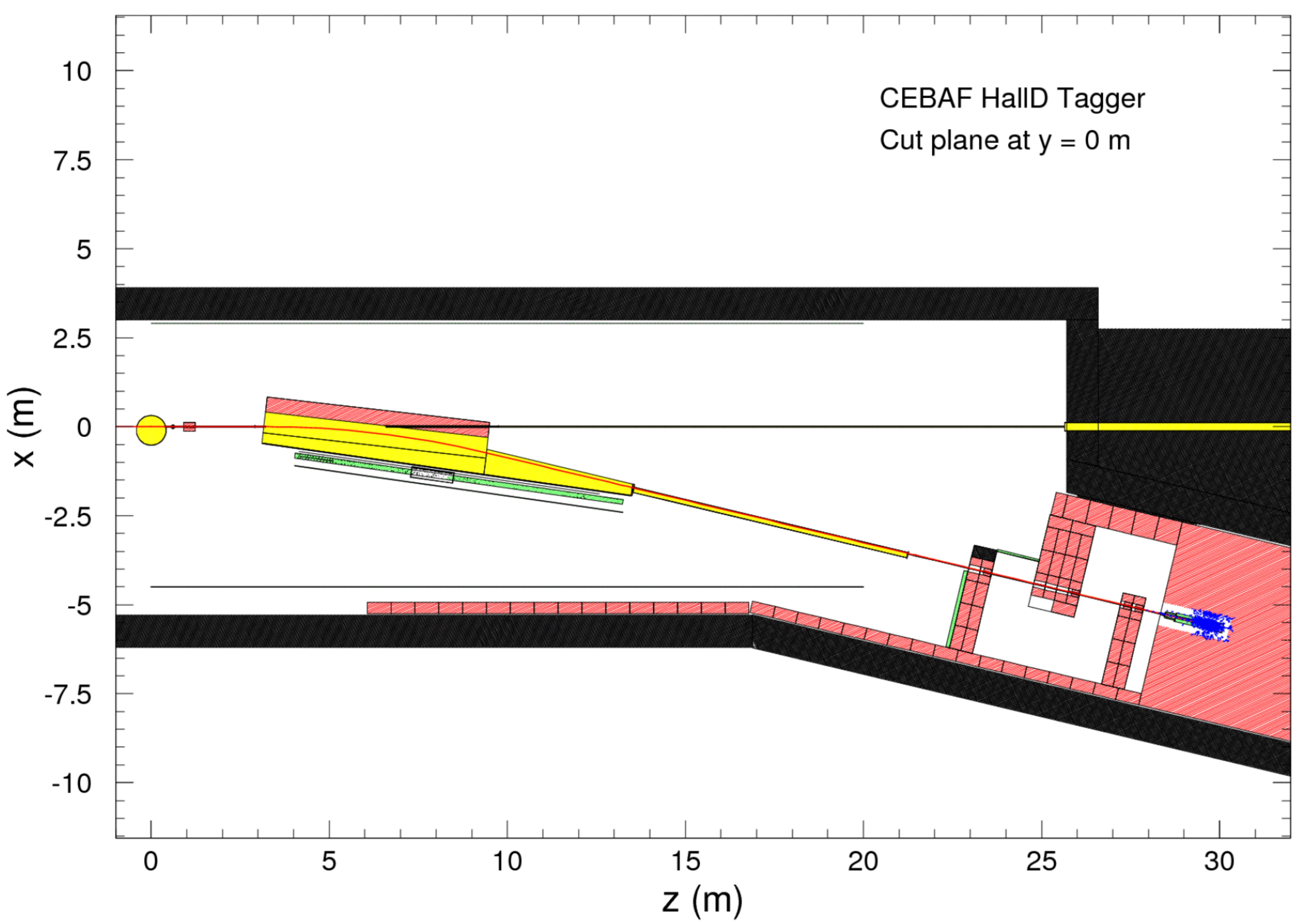}
\centerline{\parbox{0.80\textwidth}{
 \caption{Plane cut of the Tagger vault model built using the
        GEANT3 detector simulation package.  Black areas
        correspond to the concrete walls. Red hatch style
        is used for iron shielding blocks. Yellow areas
        correspond to the beam vacuum.} \label{fig:CPS_plot2} } }
\end{center}
\end{figure}

\item \textbf{Consequences of Intensity Increase by ``Brute
        Force"}

The proposal for the production of the $K_L^0$ beams at the
Hall~D complex would require photon beams which are orders
of magnitude more intensive than those available currently,
corresponding approximately to the effective radiator
thickness of 0.05 -- 0.1~R.L. Simply replacing the present
radiator with a much thicker one (a ``brute force"
solution) is possible in principle. However, such a solution
would lead to several consequences which would make it
practically unacceptable. The radiation levels at the vault,
both prompt and post-operational due to the beam line
elements' radioactivation, are evaluated to be too high. 
Mitigation would require removal of sensitive electronic
components around the vault, building of new temporary
shielding walls and disposal of radioactive beam line
components after the operations.  Dose rates and activation
evaluation would require complex simulations and quality and
reliability control.  That is all deemed to be expensive and
risky, and it is not clear if the necessary photon beam
intensity can be reached.

\item \textbf{Simulation of the Radiation Environment}

GEANT3~\cite{GEANT3X} detector simulation package was used
originally for modeling and optimizing the Tagger vault
setup and finding acceptable radiation shielding solutions.
An example of the GEANT3 geometry may be seen in
Fig.~\ref{fig:CPS_plot2}.  One 12~GeV beam electron trace
is shown in Fig.~\ref{fig:CPS_plot2} by the red curve,
starting in the beam tunnel and going through the radiator,
the tagger magnet, the exit vacuum volumes and beam pipes.
Upon arriving at the beam dump, it cascades inside the copper
core of the dump, with some blue gamma and black neutron
tracks shown exiting the core and stopping in the iron
shielding blocks.
\begin{figure}[ht!]
\begin{center}
\includegraphics[width=\linewidth]{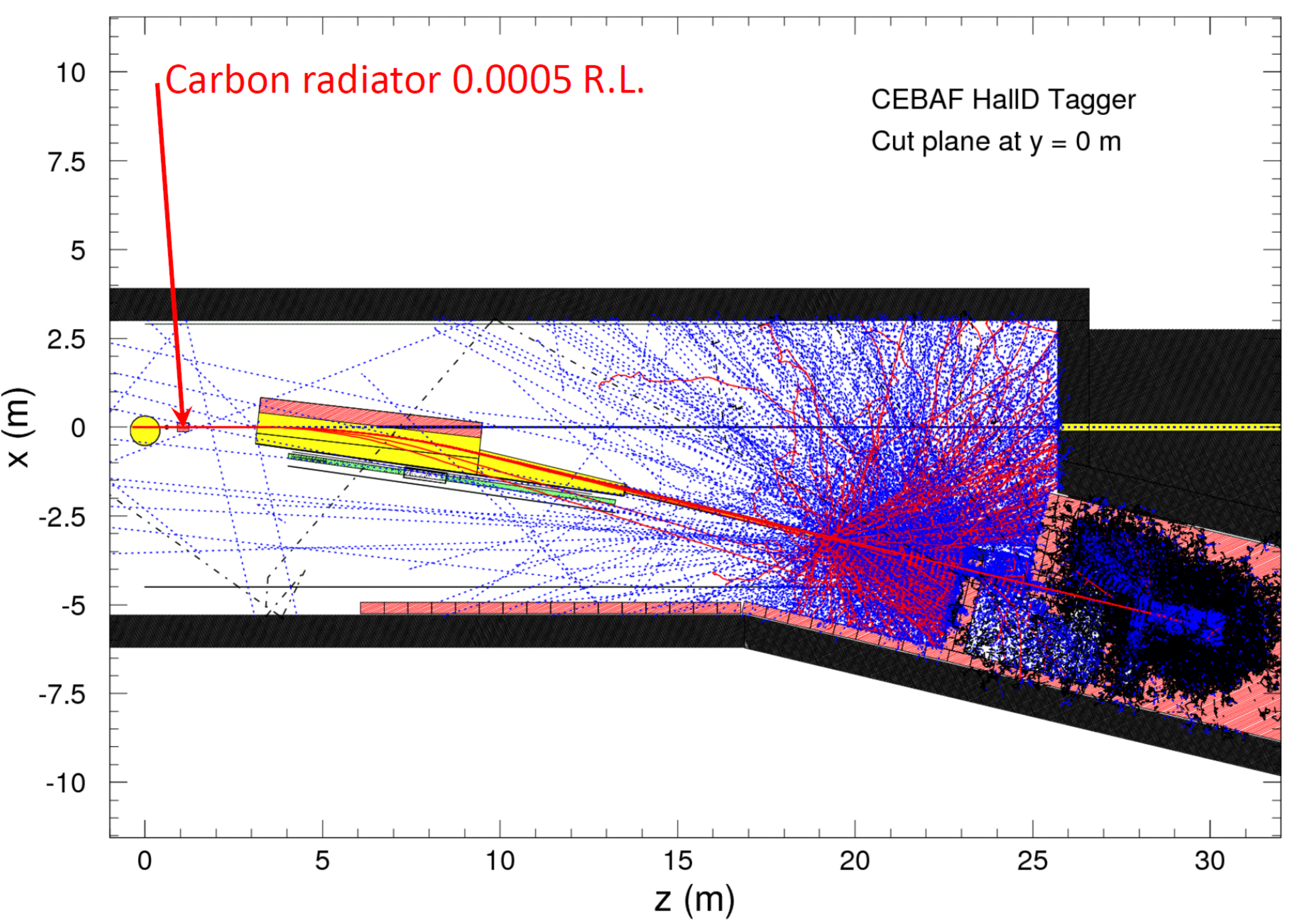}
\centerline{\parbox{0.80\textwidth}{
 \caption{Same plot as in Fig.~\protect\ref{fig:CPS_plot2},
        but showing the simulation of 2000 beam electrons
        at 12~GeV. Red tracks show charged particles, mostly
        electrons, blue tracks are gammas, and neutrons are
        tracked in black.} \label{fig:CPS_plot3} } }
\end{center}
\end{figure}
\begin{figure}[ht!]
\begin{center}
\includegraphics[width=\linewidth]{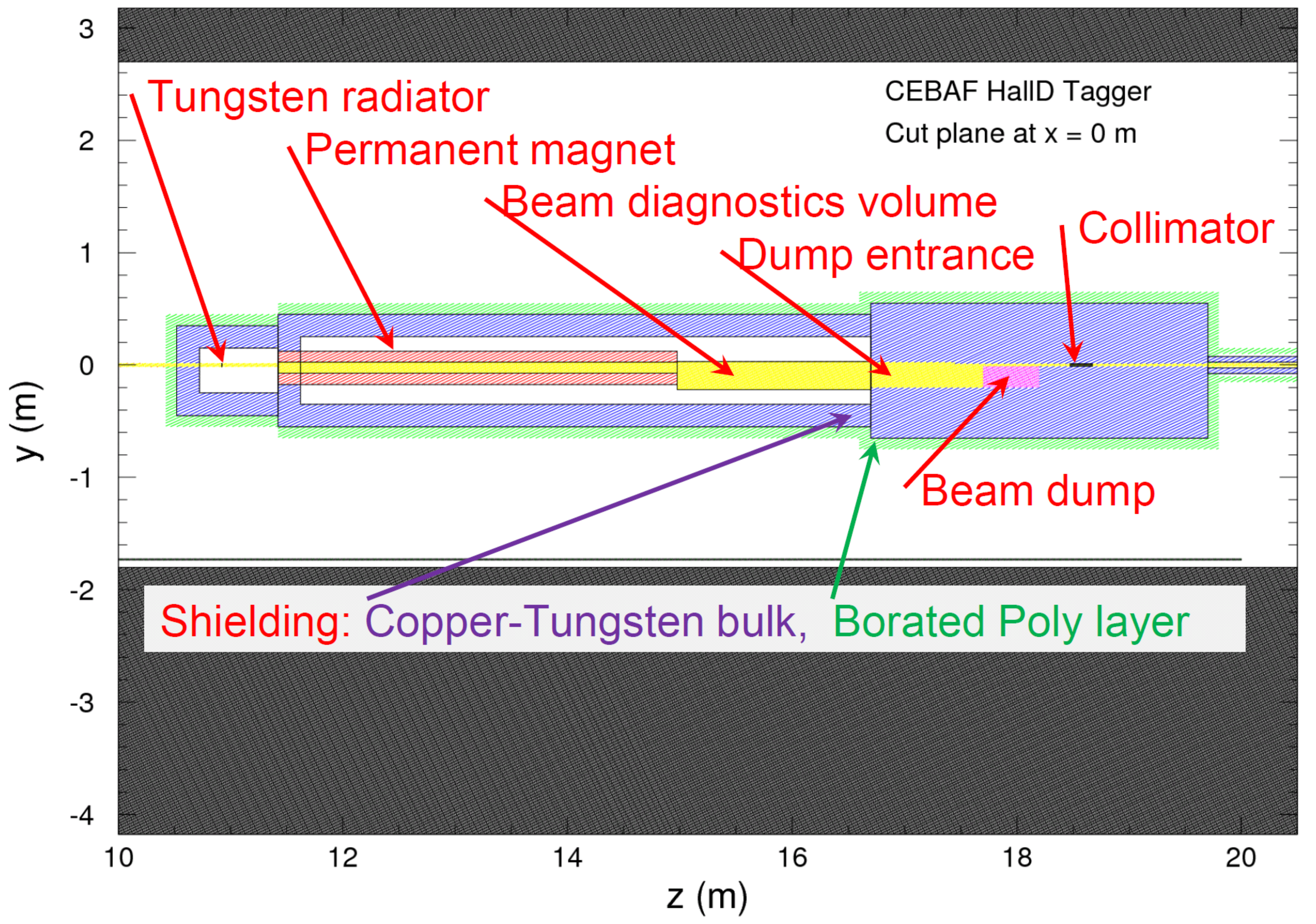}
\centerline{\parbox{0.80\textwidth}{
 \caption{Vertical cut plane of the GEANT3 model of the CPS
        device. Elements of the design are indicated in the
        plot. See discussion in the text.}
        \label{fig:CPS_plot4} } }
\end{center}
\end{figure}

Fig.~\ref{fig:CPS_plot3} illustrates the major causes of the
prompt radiation inside the vault during normal operations.
Most of the beam electrons do not interact in the radiator
and end up depositing their energy in the dump, producing a
cascade of electrons and photons which are mainly absorbed in
the core of the dump. The secondary neutrons escape the core
but mostly stop in the surrounding iron shielding and the
labyrinth walls. A couple of the beam electrons shown in the
plot have produced bremsstrahlung gammas energetic enough to
kick the electrons visibly away from the beam line.  They
end up cascading and depositing their energy in the first
labyrinth wall and spraying some of the cascading electrons
and gammas back into the vault. More beam electrons produce
lower energy gammas in the radiator and get deflected at
smaller angles, hitting exit flanges and narrower portions
of the exit beam line and producing essentially full 12~GeV
electromagnetic cascades in the vault. The bremsstrahlung
gammas produced at the radiator go straight along the exit
beam line through the magnet yoke and forward to Hall~D
proper.

The model allowed us to simulate the operational radiation
environment and optimize the shielding design at the vault,
corresponding to the nominal operating conditions in the
Hall~D complex. The first commissioning runs in 2015
indicated that the observed radiation fields in the Tagger
vault are in agreement with the calculations within a factor
2 or 3, which is acceptable given all uncertainties in the
model and in the measurements. The estimates for the
``brute force" solution of simply increasing the
radiator thickness to produce more intensive photon beams
at the vault indicate that that an increase of the prompt
radiation would be way beyond the present design
limitations and is indeed problematic.

\item \textbf{New ``Compact Photon Source" Solution}

As a solution to this apparent problem, we suggest
designing and implementing the ``Compact Photon Source"
(CPS) device.  The new conceptual design combines in a
single properly shielded assembly all elements necessary
for the production of the intense photon beam, such that
the overall dimensions of the setup are limited and the
operational radiation dose rates around it are acceptable.
The experiment does not require tagging of the produced
photons, so the new design could be compact and
hermetically closed by the shielding, as opposed to the
present Tagger Magnet concept. One of the earlier CPS
concepts was published recently~\cite{CPS_BogdanX} and
proposed for use in the WACS experiment at 
JLab~\cite{CPS_WACSX}.

Fig.~\ref{fig:CPS_plot4} illustrates the GEANT3 model of
such a device. The assembly is small enough to fit in the
Tagger vault immediately after the Tagger magnet. We propose
to extend the path of the incoming 12~GeV electron beam
beyond the Tagger magnet by removing the standard radiator
from the beam and by switching the magnetic field off. After
exiting the Tagger magnet through the currently available
photon beam exit pipe, the electron beam enters the well
shielded CPS device. First, the beam electrons see the new
radiator, as thick as necessary to produce photon beam
optimized for $K_L^0$ production downstream. After the
radiator the ``spent" electrons are deflected by the region
of a strong magnetic field sufficient to deflect the main
12~GeV beam through the vacuum volume with the beam
diagnostics equipment and into the high power beam dump,
away from the straight photon beam axis. A long-bore
straight collimator lets the photon beam through the
assembly.  An illustration of the beam propagation and
absorption in the CPS device is given in
Fig.~\ref{fig:CPS_plot7}.
\begin{figure}[t]
\begin{center}
\includegraphics[width=\linewidth]{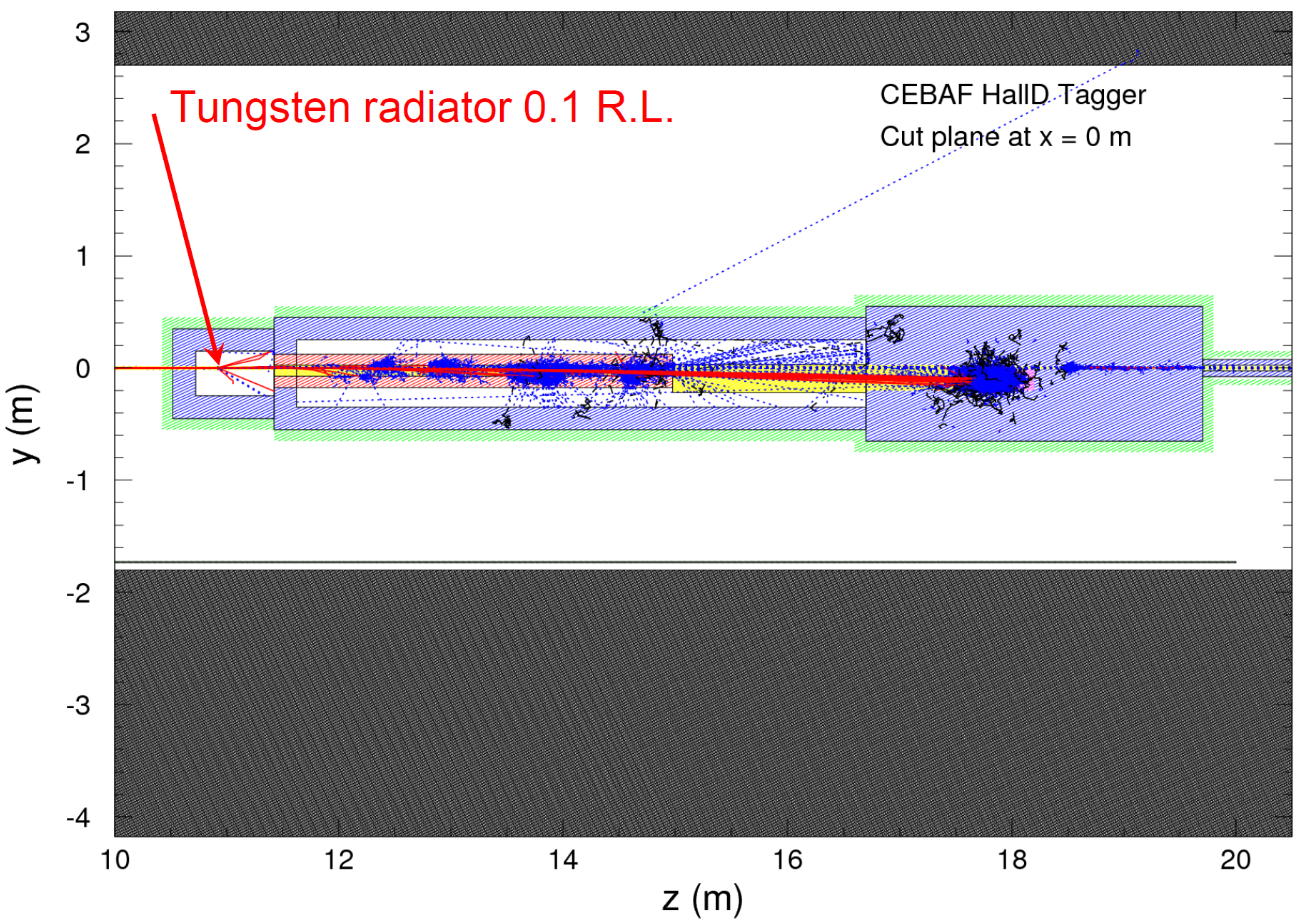}
\centerline{\parbox{0.80\textwidth}{
 \caption{The same plot as in Fig.~\protect\ref{fig:CPS_plot4},
        but showing the simulation of 50 beam electrons at
        12~GeV, with a 0.1~R.L. Tungsten radiator installed.
        Beam electrons interact in the radiator frequently,
        lose energy, and get deflected in the 0.8~Tesla
        magnetic field. Most of them go down to the core
        of the dump and deposit their energy there. The
        photon beam goes straight to (and through) the
        aperture collimator.} \label{fig:CPS_plot7} } }
\end{center}
\end{figure}

The dump including the shielding is by necessity the most
massive element of the design, consisting of a water-cooled
thermoconductive copper or silver core, surrounded by thick,
high-Z, and high density material, such as Tungsten alloy
or Lead. The dump design includes the entrance slit for the
main electron beam and for the tail electrons that lost
their energy in the radiator and are deflected farther down.
The slit provides the condition for the main
electron-produced cascade to originate roughly in the middle
of the bulk dump volume, such that there was sufficient
shielding in all directions and the products of the cascade
were contained in the dump as much as possible.

The dump will also provide a long-bore channel for the main
bremsstrahlung gamma beam to go through, with the critical
collimation aperture placed in the middle of the bulk volume
to contain most of the secondary products generated in the
collimator. A Borated Poly outer layer is useful for slowing,
thermalizing, and absorbing fast neutrons still exiting the
bulk shielding.  We suggest that the exit photon channel
and beam line in the vault also be shielded properly.
\begin{figure}[t]
\begin{center}
\includegraphics[width=\linewidth]{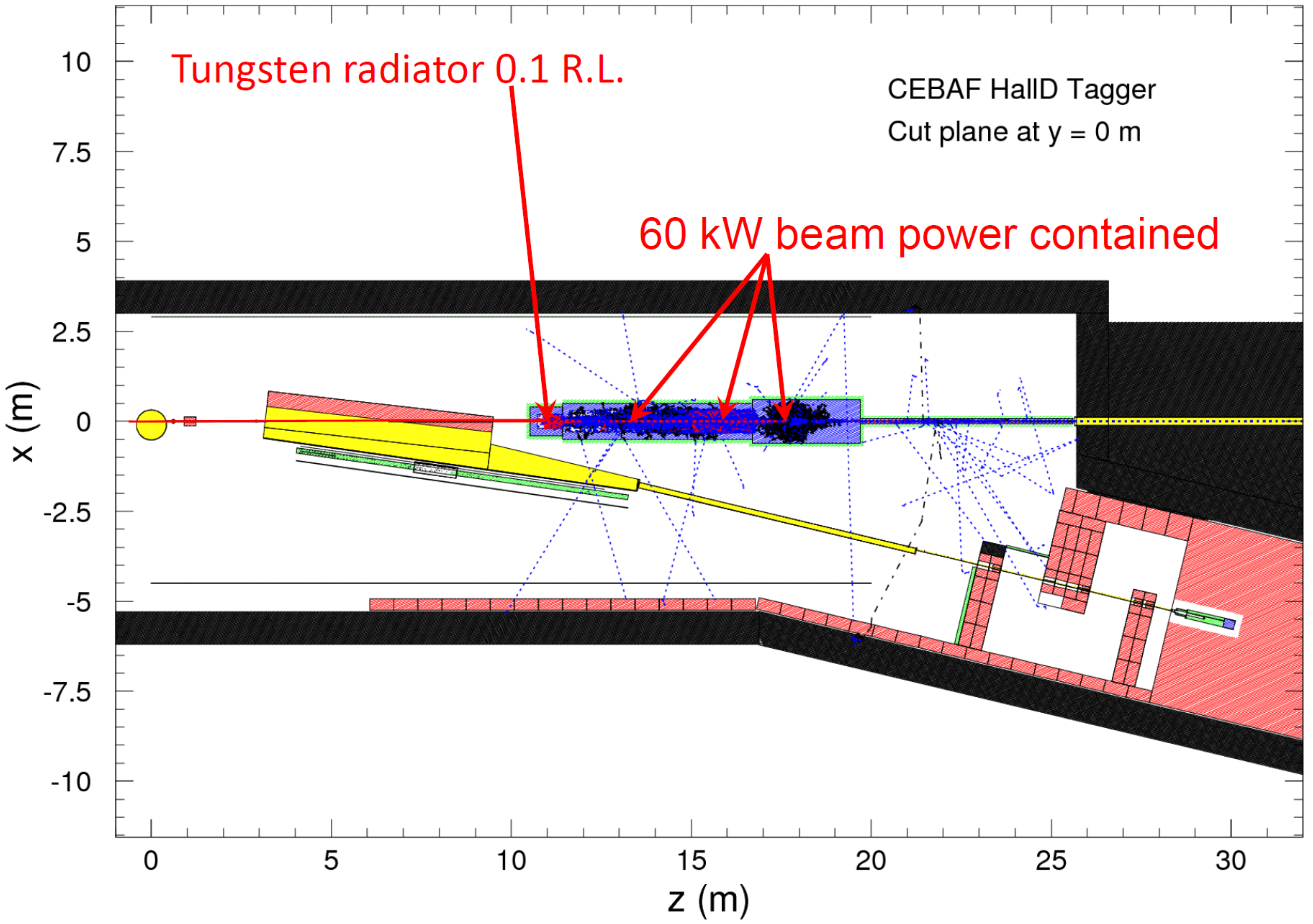}
\centerline{\parbox{0.80\textwidth}{
 \caption{A plane cut of the Tagger vault model built
        using the GEANT3 package, showing the CPS 
        assembly and the simulation of 2000 beam
        electrons at 12~GeV.} \label{fig:CPS_plot5} } }
\end{center}
\end{figure}

The GEANT3 model calculations show that the overall
dimensions of the CPS assembly could stay reasonably
small (see Fig.~\ref{fig:CPS_plot5}), while achieving
the overall dose rates at the Tagger vault comparable
to nominal Hall~D operations. The dense high-Z material
covering the core of the dump from all sides, together
with a 2-inch layer of Borated Poly material all around,
work rather effectively. The estimate shows that, for
the ultimate new setup with 10\% R.L. radiator, the dose
rates in the vault during full 60~kW beam operations are
comparable to the nominal running conditions in the vault,
as shown in Fig.~\ref{fig:CPS_plot6}.  The radiation
spectral composition is different; most of the dose rate
contribution in the CPS setup is from higher energy
neutrons. The comparison indicates that at equal beam
currents, gamma radiation dose rates are much smaller for
the CPS run (an order of magnitude), and neutron dose
rates in the area are comparable.
\begin{figure}[t]
\begin{center}
\includegraphics[width=\linewidth]{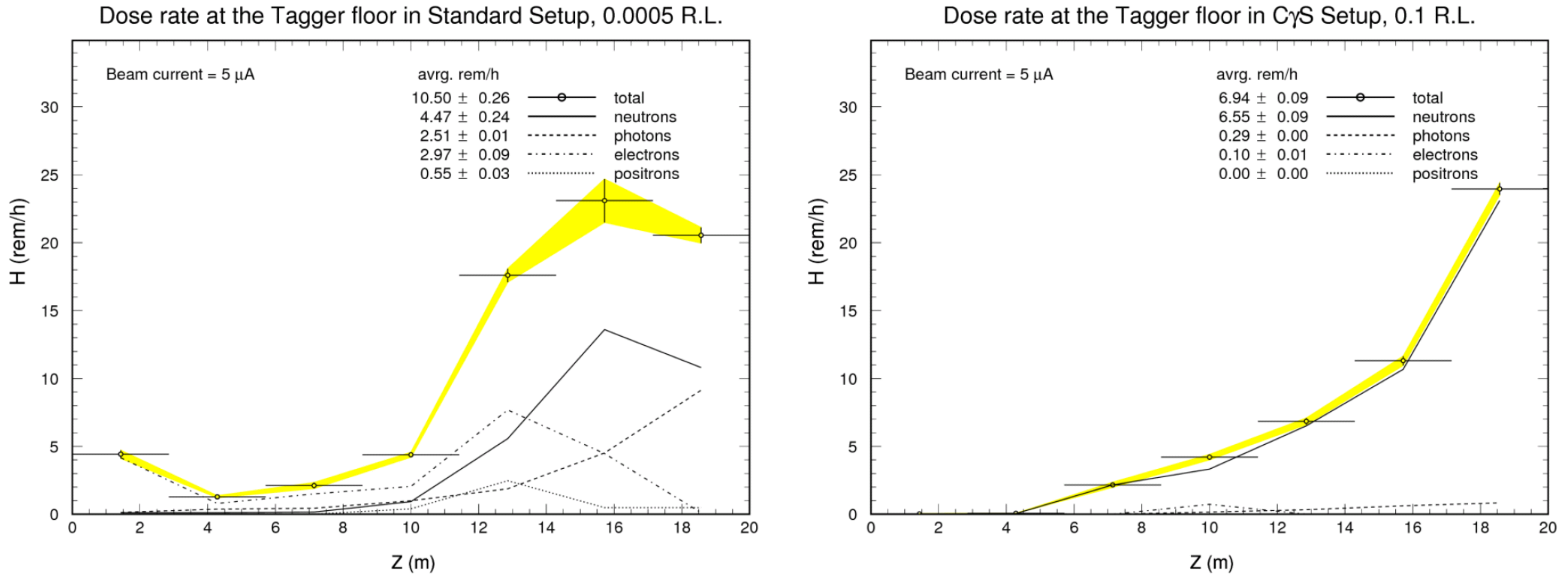}
\centerline{\parbox{0.80\textwidth}{
 \caption{The plots show a comparison of dose rate estimates
        in the Tagger Area in the two conditions: (left panel)
        nominal Hall~D operation with the standard amorphous
        radiator at 0.0005~R.L., - with (right panel) radiator
        at 0.1~R.L., used as part of the Compact Photon Source
        setup.} \label{fig:CPS_plot6} } }
\end{center}
\end{figure}

The dramatic difference between the overall dimensions of 
the present dump package and the new proposed compact 
solution is explained by using high density and high-Z 
material in the new design, and also by making it hermetic, 
that is, covering the beam dump entrance volume almost as 
solidly as in all other directions. The present Hall~D 
Tagger Dump design was opportunistic as it used the standard 
copper beam dump device and standard iron shielding blocks 
available at that time. The design required dump accessibility, 
so it was not possible to make the shielding hermetic, and 
heavy labyrinth walls were needed. Using a Tungsten alloy 
with a density about 2.5 times higher than Iron, all linear 
dimensions can be made correspondingly smaller. Another 
factor was a certain degree of conservatism required for a 
permanent design, which can be avoided in the design dedicated 
for a limited-period experiment.

One of the special features of the new concept design is the
difficulty of reaching its innards once operations have
started, because of the high levels of induced radioactivity
inside.  Thus, special attention should be devoted to the
reliability of all inner elements of the CPS.

\item \textbf{Implementation Features and Cost Estimate}

We do not see principal obstacles to implementation of the
CPS concept in the experiment.

The new dump core may have characteristics close to the one
installed already, such that the dump cooling system can be
re-used (maximum 60~kW cooling power).

To make long and narrow photon beam collimation we propose
to build the core using two symmetric flat plates, left and
right, and make matching grooves in them for the beam entry
cones, beam line, and the aperture collimator.

Most of the present Tagger Area equipment may remain in place;
the CPS will be assembled around the modified gamma line.

The available permanent magnet may be used in the CPS assembly
(pending thermal engineering analysis, as there will be a need
to have it cooled, dissipating approximately 1.5~kW of power
deposited in it). Available identical spare magnet can be
installed at the end of the beam line in the Tagger as a
protective measure.  Alternatively, a new powered magnet
could be designed for the CPS, with a comparable field
integral (it does not have to be highly precise). In such a
case, the present permanent magnet will be moved downstream.

The CPS solution for the new intense high energy gamma source
in the Tagger vault will be characterized by the absence of
extra prompt irradiation or extra beam line activation for
existing structures in the area during and after the
operations. The accumulation of radioactivity inside the CPS
will be significant, to a large degree preventing access to
the inner parts immediately after operations. However, such
activation will not present a problem while the CPS stays
assembled, due to a very strong self-shielding.

There will be a possibility of switching between the two
modes of Hall~D operations: a low intensity tagged photon
beam, and high intensity photon beam from the CPS.

Disassembly and decommissioning could be postponed until
radioactive isotopes decay inside to manageable levels.

Cost would include detailed iterative modeling and simulation
to optimize operation parameters, design, engineering and
production, plus the choice and cost of bulk shielding
material.

Rough cost expectation: within \$0.5~M.

\newpage
\item \textbf{Conclusions}

Compared to the alternative, the proposed CPS solution
presents several advantages, including much less
disturbance of the available infrastructure at the Tagger
Area and better flexibility in achieving high-intensity
photon beam delivery to Hall~D.

The proposed CPS solution will satisfy proposed $K_L^0$
beam production parameters

We do not envision big technical or organizational
difficulties in the implementation of the conceptual
design.

\item \textbf{Acknowledgments}

This work is authored by The Southeastern Universities Research
Association, Inc. under U.S. DOE Contract No. DEAC05--84150. 
The U.S. Government retains a non-exclusive, paid-up, 
irrevocable, world-wide license to publish or reproduce this 
manuscript for U.S. Government purposes.
\end{enumerate}


\newpage
\subsection{Targets for a Neutral Kaon Beam}
\addtocontents{toc}{\hspace{2cm}{\sl C.~Keith}\par}
\setcounter{figure}{0}
\halign{#\hfil&\quad#\hfil\cr
\large{Christopher Keith}\cr
\textit{Thomas Jefferson National Accelerator Facility}\cr
\textit{Newport News, VA 23606, U.S.A.}\cr}

\begin{abstract}
A secondary beam of neutral Kaons is under consideration for 
Hall~D at Jefferson Lab to perform spectroscopic studies of 
hyperons produced by $K^0_L$ particles scattering from proton 
and deuteron targets. The proposed physics program would utilize 
the GlueX detector package currently installed in Hall~D.  This 
contribution looks at potential targets for use in the new 
facility, paying close attention to the existing infrastructure 
of GlueX and Hall~D. Unpolarized cryotargets of liquid hydrogen 
and deuerium, as well as polarized solid targets of protons and 
deuterons are examined.
\end{abstract}

\begin{enumerate}
\item \textbf{Introduction}

A proposal is currently under consideration to expand
Jefferson Lab's program of hadron spectroscopy and
develop a secondary $K^0_L$ beam in experimental Hall~D~\cite{LOIY}.
The Kaon beam will be produced from photoproduction on a beryllium
target located about 85~m downstream from the Hall~D radiator for
Bremsstrahlung photons.  Photons escaping the beryllium target will
be absorbed by a lead shield, while charged particles will be
removed by a sweeper magnet.  The $K^0_L$ flux, collimated into a
6~cm diameter beam, is expected to be of order $10^{4}$~s$^{-1}$,
along with a similar rate of high energy neutrons.  The existing
GlueX detector package and its 1.8~T superconducting solenoid would
be utilized for the program.

This contribution examines possible targets for the neutral Kaon
beam facility, both unpolarized and polarized hydrogen and
deuterium.  Emphasis is placed on the former, and in particular
on straightforward modifications to the existing GlueX cryotarget
that will make it suitable for a large diameter beam of $K^0_L$.

\item \textbf{Liquid Hydrogen Target}

If possible, the proposed experimental program will utilize the
existing GlueX liquid hydrogen cryotarget (Fig.~\ref{fig:Cart}),
modified to accept a larger diameter target cell.  The GlueX
target comprises a kapton cell containing liquid hydrogen (LH$_2$)
at a temperature and pressure of about 20~K and 19~psia. The
100~ml cell is filled through a pair of 1.5~m long stainless
steel tubes (fill and return) connected to a small vessel where
hydrogen gas is condensed from two room temperature storage tanks.
Inside the vessel is a large condensation surface for the gas,
consisting of numerous copper fins that are cooled by a pulse
tube refrigerator (PTR) with a base temperature of 3~K and
cooling power of about 20~W at 20~K.  A 100~W temperature
controller regulates the condenser at 18~K.
\begin{figure}
\begin{center}
\includegraphics[width=4.0 in]{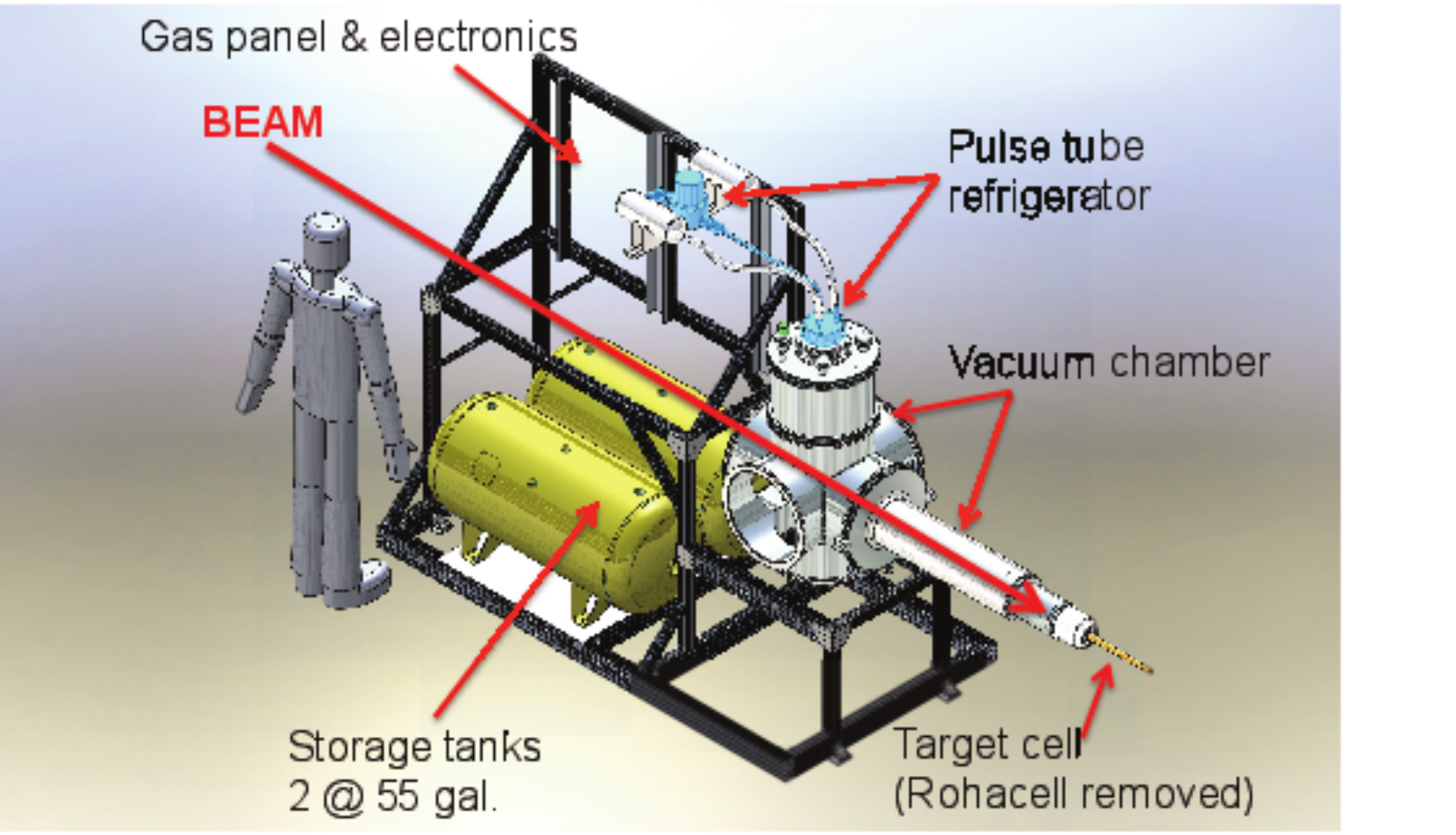}
\end{center}
\centerline{\parbox{0.80\textwidth}{
 \caption{The GlueX liquid hydrogen target.}
        \label{fig:Cart} } }
\end{figure}

The target assembly is contained within an ``L"-shaped, stainless
steel and aluminum vacuum chamber with a 1~cm thick Rohacell
extension surrounding the target cell.  The start counter for the
GlueX experiment fits snugly over this extension. The vacuum
chamber, along with the hydrogen storage tanks, gas handling
system, and control electronics, is mounted on a custom-built beam
line cart for easy insertion into the Hall~D solenoid. A compact
I/O system monitors and controls the performance of the target,
while hardware interlocks on the target temperature and pressure
and on the chamber vacuum ensure the system's safety and integrity.
The target can be cooled from room temperature and filled with
liquid hydrogen in about 5 hours.  For empty target runs, the
liquid can be boiled from the cell in less than twenty minutes.
The cell remains filled with cold hydrogen gas for these runs and
can refilled with liquid in about forty minutes.

The GlueX cell (Fig.~\ref{fig:TwoCells}) is closely modeled on
those used for experiments in Hall~B at Jefferson Lab for more
than a decade~\cite{Mec2003Y}. It is a horizontal, tapered
cylinder about 38~cm long with a mean diameter of 2~cm.
A 2~cm diameter reentrant beam window defines the length of
LH$_2$ in the beam to be about 30~cm.  Both entrance and exit
windows on the cell are 75~$\mu$m kapton while the cylindrical
walls are 130 $\mu$m kapton glued to an aluminum base.  In normal
operation the cell, the condenser, and the pipes between them are
all filled with liquid hydrogen.  In this manner the liquid can 
be subcooled a few degrees below the vapor pressure curve,
greatly suppressing the formation of bubbles in the cell.
In total, about 0.4~liter of LH$_2$ is condensed from the storage
tanks, and the system is engineered to safely recover this
quantity of hydrogen back into the tanks during a sudden loss of
insulating vacuum, with a maximum allowed pressure of
49~psia~\cite{MeekinsY}.
\begin{figure}
\begin{center}
\includegraphics[width=3.5 in]{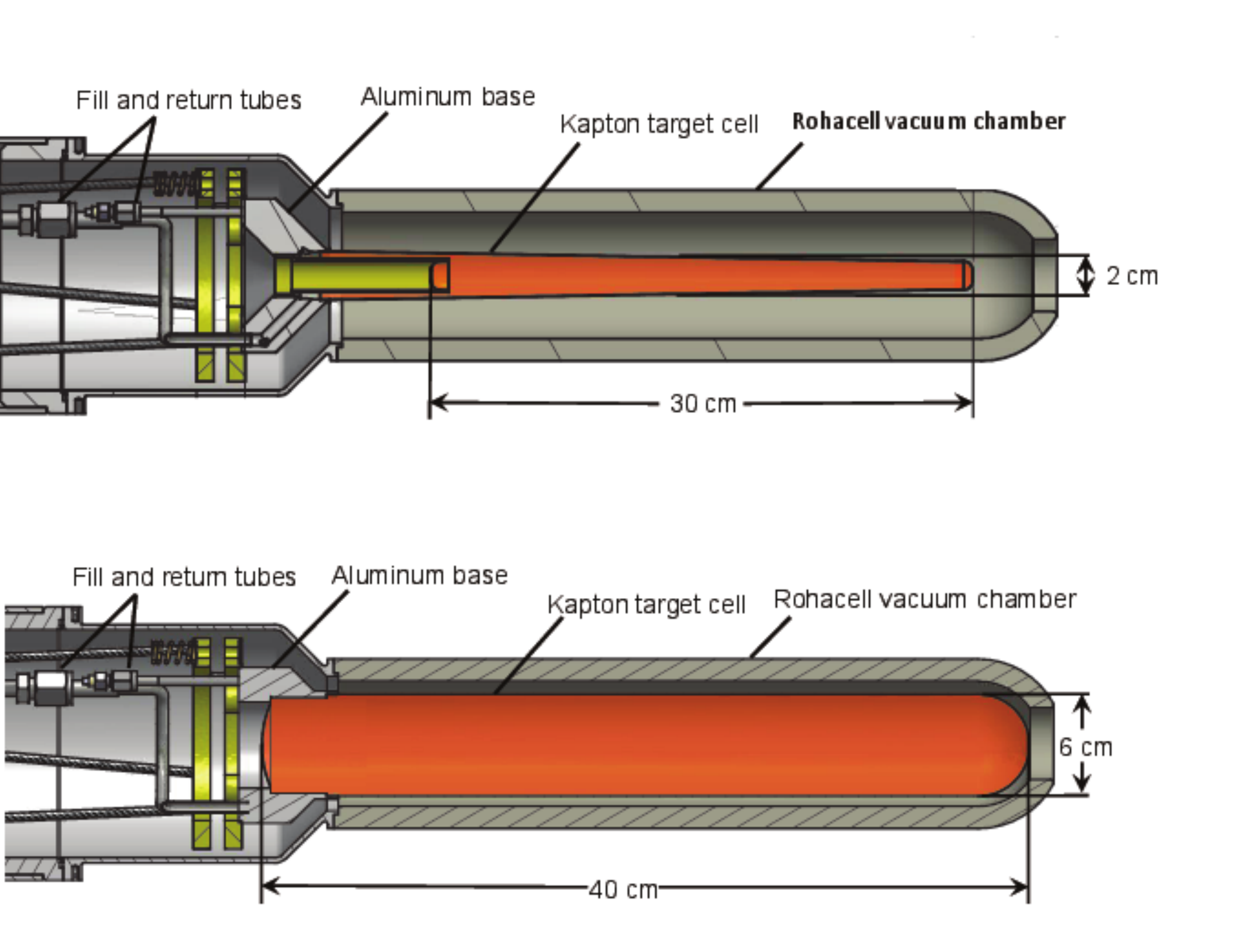}
\end{center}
\centerline{\parbox{0.80\textwidth}{
 \caption{Top: Kapton target cell for the GlueX LH$_2$ target.
        Bottom: Conceptual design for a larger target cell for
        the proposed $K^0_L$ beam in Hall~D.} 
	\label{fig:TwoCells} } }
\end{figure}

A conceptual design for the neutral Kaon beam target is also
shown in Fig.~\ref{fig:TwoCells}.  The proposed target cell has
a diameter of 6~cm and a 40~cm length from entrance to exit
windows, corresponding to a volume of about 1.1~liter. The
inventory of gas required to operate the target with this cell
will be about 1500 STP liter, which can be stored in the existing
tanks at about 50~psia. The JLab Target Group will investigate
alternative materials and construction techniques to
increase the strength of the cell.

The GlueX target system is expected to operate equally well with
liquid deuterium (LD$_2$), which condenses at a slightly higher
temperature than hydrogen: 23.3~K versus 20.3~K at atmospheric
pressure.  Because the expansion ratio of LD$_2$ is 13\% higher,
the storage pressure will about 60~psia.  The new target cell
will therefore be engineered and constructed to accomodate either
H$_2$ or D$_2$.

\item \textbf{Solid Polarized Target}

Dynamically polarized targets were successfully utilized in each
of Jefferson Lab's three experimental halls (A, B, and C) during
its 6~GeV era~\cite{KeithPSTP2015Y}.  It is natural then to
contemplate their use in Hall~D as well.  We can expect the
Hall~D solenoid and its accompanying cryogenic facility to play
significant roles in the target's design and operation.

To realize dynamic nuclear polarization (DNP), a solid
dielectric material is doped with paramagnetic radicals.  The
unpaired electrons in these radicals are polarized at low
temperature and high magnetic field, and microwave-driven spin
flip transitions transfer the electron's polarization to nearby
nuclei in the material.  The nuclear polarization is then
transported through the bulk via spin diffusion.

DNP targets generally fall into one of two categories:
continuously polarized and frozen spin.  In the former case the
DNP process is maintained continuously throughout the 
scattering experiment, while frozen spin targets are polarized 
intermittently, and the scattering data is acquired  while the 
polarization slowly decays.  Continuously polarized targets 
require a highly uniform polarizing magnet of 2.5--5~T whose 
geometry limits the acceptance of scattered particles.  A 
similar magnet is also required to polarize a frozen spin 
target, but the target sample can then be removed from the high 
field and maintained by a much less massive ``holding" magnet 
during data acquisition, provided it is cooled to a temperature 
of 50~mK or less.  For this reason frozen spin targets are 
limited to particle beams no greater than about 10$^8$~s$^{-1}$, 
while continuously polarized targets have operated up to 
$\sim10^{12}$~s$^{-1}$. Because of its high resistance to 
radiation damage, irradiated ammonia (NH$_3$ or ND$_3$) 
continuously polarized at 1~K and 5~T is the usual choice for 
electron beams up to 140~nA.  Chemically doped butanol
(C$_4$H$_{10}$O) has become the material of choice for frozen
spin targets, thanks to its ease of production and handling, 
and its lack of polarizable background nuclei other than 
hydrogen.  Protons in either ammonia or butanol can be 
dynamically polarized in excess of 90\%.  Deuterated ammonia 
can be polarized to about 60\%, and deuterated butanol to 
80--90\%.

At the expected $K^0_L$ rate of the proposal, either type of
polarized target is suitable. However, the 1.8~T field of the
Hall~D solenoid is too weak and too inhomogeneous (0.25~T/m)
to act as an effective polarizing magnet for DNP.  It would
serve as an excellent holding magnet for a frozen spin target
though.  Polarization decay times up to 4000 hours were
observed with the FROST target in Hall B using a 0.5~T
holding field~\cite{FROSTY}. From these results we anticipate
relaxation times exceeding 10,000 hours at 1.8~T. The target
would be polarized outside the Hall~D solenoid using a
warm-bore magnet similar to the one used for the FROST target
and moved to the GlueX solenoid for data acquisition. A small
transfer coil would be incorporated inside the target cryostat
to maintain the polarization while the target is moving.

The size of the polarized target sample will be critical.  For
best results, the polarizing field should be uniform to about
100~ppm over the sample volume.  The cost of a magnet suitable
for a 6~cm sample diameter will be significant, so smaller
diameters should be considered.  Approximately 2~mW/g of
microwave  power is necessary for DNP at 2.5~T.  Thus the
sample volume will also determine the refrigeration capacity
of a frozen spin target.  Frozen butanol consisting of 1--2~mm
beads has a density of 1.1~g/cm$^3$, a packing fraction of
about 0.6, and a dilution factor of 0.135.  A target sample
2~cm in diameter and 23~cm long would provide a similar
proton luminosity as the 30~cm long LH$_2$ target.  Dynamic
polarization of the 50~g sample would require about 0.1~W of
microwave power at 2.5~T and 0.3~K.   A $^3$He-$^4$He
dilution refrigerator similar to FROST's would be suitable
for this application, operating at a $^3$He circulation rate
of 30~mmol/s during polarization and 1--2~mmol/s during
frozen spin mode.

A frozen spin target consumes liquid helium at a rate of a few
liters per hour to operate the dilution refrigerator.  The Hall~D
cryogenic plant may not be able to provide this volume of LHe to
a polarized target and maintain the GlueX solenoid at the same
time.  In this case the polarized target will require a separate
source of LHe, and should be designed to economize $^4$He
consumption.  Alternatively, one may consider a so-called
``cryogen-free" dilution refrigerator (CFDR), where the 
circulating $^3$He is condensed by a small cryocooler such as the 
PTR utilized for the GlueX cryotarget~\cite{DryDRY}.  Unfortunately, 
present day CFDRs cannot provide the cooling power necessary to 
polarize a 50~g target sample.  In its place, we can consider a 
hybrid system using two pulse tubes (Fig.~\ref{fig:CFPT}).  
Together these can condense $^3$He at a rate sufficient for frozen 
spin operation and simultaneously condense 5--10~l/day of $^4$He 
into a 50~liter reservoir within the target cryostat.  Once 
sufficient liquid is accumulated, it would support an increased 
$^3$He circulation rate long enough to polarize the sample, about 
8~hours.  The $^4$He level in the reservoir would naturally 
decrease during this time but recover during the week or more of 
frozen spin operation.
\begin{figure}
\begin{center}
\includegraphics[width=6 in]{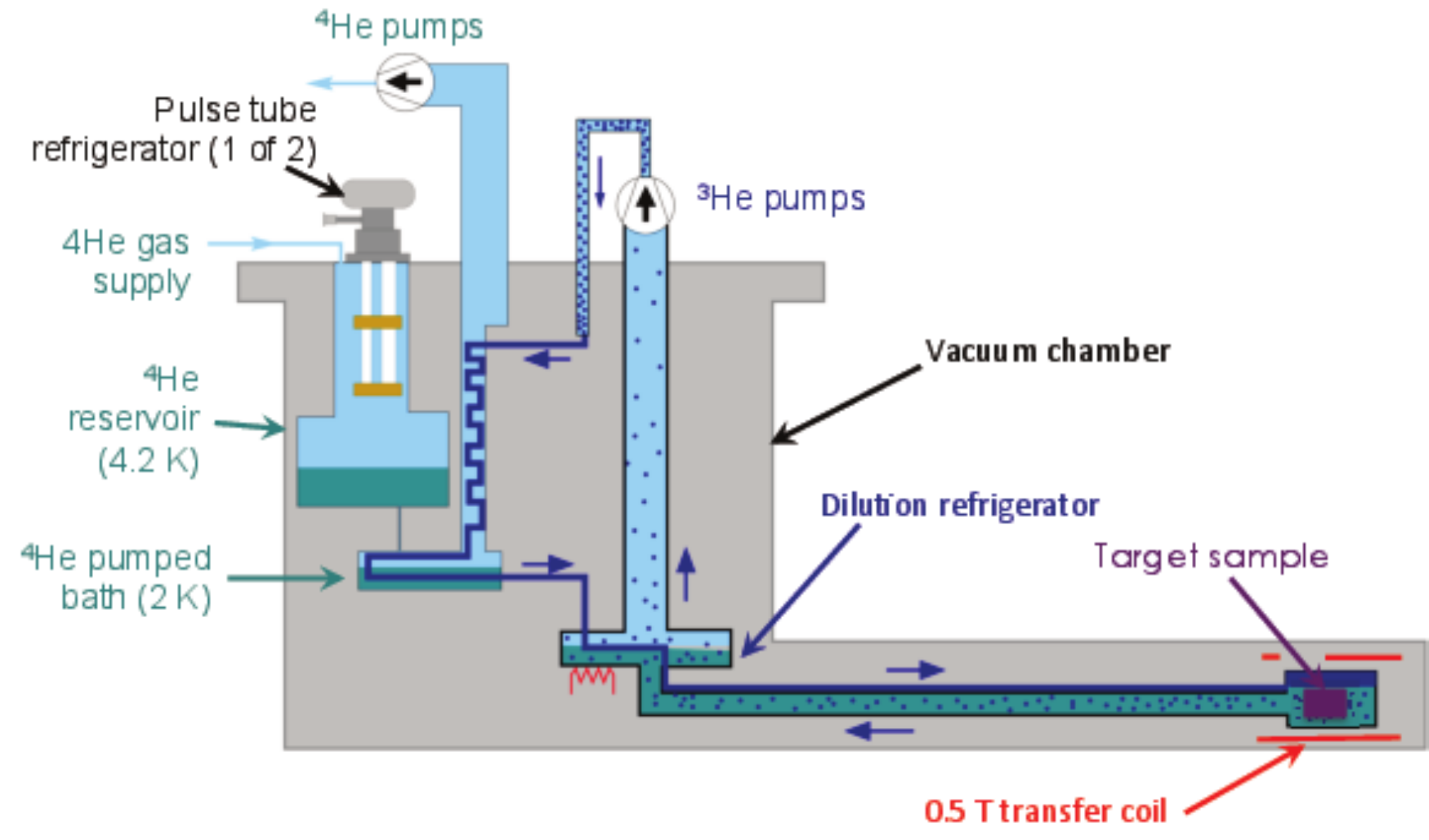}
\end{center}
\centerline{\parbox{0.80\textwidth}{
\caption{Schematic drawing of a hybrid cryogen-free frozen spin
        target. See text for details.} \label{fig:CFPT} } }
\end{figure}

\item \textbf{Summary}

Possible targets for a neutral Kaon beam in Hall~D have been
examined.  It is found that the existing GlueX cryotarget can
be modified to accept liquid hydrogen or deuterium cell
diameters up to 6~cm, with some R\&D required to increase the
working pressure of the current GlueX cells.  For polarized
target experiments a frozen spin target of butanol is 
indicated.  It will be more difficult to realize a 6~cm 
diameter sample in this case, due to the magnetic field and 
cooling power requirements necessary for dynamic polarization.  
Instead a 23~cm long sample with 2~cm diameter is considered.

\item \textbf{Acknowledgments}

This material is based upon work supported by the U.S. Department
of Energy, Office of Science, Office of Nuclear Physics under
contract DE--AC05--06OR23177.

\end{enumerate}


\newpage
\section{List of Participants of KL2016 Workshop}

\begin{itemize}
\item Mohammad (Saif) Ahmad, GWU       		  <msahmad@gwu.edu>
\item Michael Albrow, FNAL      		  <albrow@fnal.gov>
\item Moskov Amaryan, ODU       		  <mamaryan@odu.edu>
\item Frank (Ted) Barnes, DOE-NP	          <ted.barnes@science.doe.gov>
\item William J. Briscoe, GWU                     <briscoe@gwu.edu>
\item Daniel Carman, JLab 	  		  <carman@jlab.org>
\item Shloka Chandavar, Ohio U. 		  <cshloka@jlab.org>
\item Eugene Chudakov, JLab	  		  <gen@jlab.org>
\item Pavel Degtyarenko, JLab	  		  <pavel@jlab.org>
\item Michael D\"oring, GWU	  		  <doring@gwu.edu>
\item Robert Edwards, JLab			  <edwards@jlab.org>
\item Ishara Fernando, Hampton U. 		  <ishara@jlab.org>
\item Igor Filikhin, NCCU			  <ifilikhin@nccu.edu>
\item Alessandra Filippi, I.N.F.N. Torino	  <filippi@to.infn.it>
\item Jos\'e~L.~Goity, Hampton U./JLab 		  <goity@jlab.org>
\item Helmut Haberzettl, GWU 			  <helmut@gwu.edu>
\item Avetik Hayrapetyan, JLU Giessen		  <Avetik.Hayrapetyan@uni-giessen.de>
\item Charles Hyde, ODU			  	  <chyde@odu.edu>
\item Hiroyuki Kamano, RCNP, Osaka U.		  <kamano@rcnp.osaka-u.ac.jp>
\item Christopher Keith, JLab                     <ckeith@jlab.org>
\item Roman Kezerashvili, NY City College of Tech., CUNY   <rkezerashvili@citytech.cuny.edu>
\item Franz Klein, GWU			  	  <fklein@jlab.org>
\item Michael Kohl, Hampton U.		  	  <kohlm@jlab.org>
\item Valery Kubarovsky, JLab			  <vpk@jlab.org>
\item Ilya Larin, ODU	 			  <ilarin@jlab.org>
\item Haiyun Lu, U. of Iowa           	  	  <hlu@jlab.org>
\item David Mack, JLab				  <mack@jlab.org>
\item Maxim Mai, HISKP Bonn U. 			  <mai@hiskp.uni-bonn.de>
\item D. (Mark) Manley, KSU 			  <manley@kent.edu>
\item Vincent Mathieu, IU 			  <mathieuv@indiana.edu>
\item Georgie Mbianda Njencheu, ODU 		  <gmbia001@odu.edu>
\item Robert McKeown, JLab			  <bmck@jlab.org>
\item Curtis Meyer, CMU                           <cmeyer@cmu.edu>
\item Victor Mokeev, JLab			  <mokeev@jlab.org>
\item Hugh Montgomery, JLab			  <mont@jlab.org>
\item Fred Myhrer, USC			  	  <myhrer@sc.edu>
\item Kanzo Nakayama, UGa			  <nakayama@uga.edu>
\item James Napolitano, Temple U.		  <napolj@temple.edu>
\item Hiroyuki Noumi, RCNP, Osaka U.		  <noumi@rcnp.osaka-u.ac.jp>
\item Yongseok Oh, Kyungpook Nat. U.		  <yohphy@knu.ac.kr>
\item Hiroaki Ohnishi, RIKEN/RCNP Osaka U. 	  <h-ohnishi@riken.jp>
\item Eulogio Oset, U. de Valencia 		  <eulogio.oset@ific.uv.es>
\item Emilie Passemar, IU/JLab 			  <epassema@indiana.edu>
\item Eugene Pasyuk, JLab			  <pasyuk@jlab.org>
\item Michael Pennington, JLab		  	  <michaelp@jlab.org>
\item Angels Ramos, U. of Barcelona		  <ramos@ecm.ub.edu>
\item David Richards, JLab  			  <dgr@jlab.org>
\item James Ritman, FZJ  			  <j.ritman@fz-juelich.de>
\item Torri Roark, ODU 			  	  <troar001@odu.edu>
\item Patrizia Rossi, JLab			  <rossi@jlab.org>
\item Elena Santopinto, I.N.F.N. Genova		  <elena.santopinto@ge.infn.it>
\item Reinhard Schumacher, CMU		  	  <schumacher@cmu.edu>
\item Elton Smith, JLab			  	  <elton@jlab.org>
\item Alexander Somov, JLab			  <somov@jlab.org>
\item Justin Stevens, JLab			  <jrsteven@jlab.org>
\item Igor Strakovsky, GWU                        <igor@gwu.edu>
\item Adam Szczepaniak, IU/JLab		  	  <aszczepa@indiana.edu>
\item Simon Taylor, JLab 			  <staylor@jlab.org>
\item Dominik Werthmueller, Glasgow U.	  	  <dominik.werthmueller@glasgow.ac.uk>
\item Bogdan Wojtsekhowski, JLab   		  <bogdanw@jlab.org>
\item Veronique Ziegler, JLab 		  	  <ziegler@jlab.org>
\item Bingsong Zou, ITP/CAS			  <zoubs@itp.ac.cn>
\end{itemize}
\end{document}